\documentclass[fleqn,usenatbib]{mnras}

\usepackage{booktabs}

\usepackage{times}
\usepackage{mathptmx}
\usepackage{subfig}
\usepackage{color}

\usepackage{graphicx}	
\usepackage{amsmath}	
\usepackage{amssymb}	

\usepackage{longtable}
\usepackage{lscape}
\newcommand{\myaffil}[1]{$^{\rm #1}$}
\newcounter{inst}
\newcommand{\inst}[1]{\noindent%
  \refstepcounter{inst}\myaffil{\arabic{inst}\label{#1}}
}

\title[Spectral Variability at Low Frequencies]{Spectral Variability of Radio Sources at Low Frequencies}

\author[K. Ross et al.]{K.~Ross\myaffil{\ref{ICRAR}}\thanks{E-mail: kathryn.ross@icrar.org},
J.~R.~Callingham\myaffil{\ref{Leiden},\ref{ASTRON}},
N.~Hurley-Walker\myaffil{\ref{ICRAR}},
N.~Seymour\myaffil{\ref{ICRAR}},
P.~Hancock\myaffil{\ref{ICRAR}},
\newauthor T.~M.~O.~Franzen\myaffil{\ref{ASTRON}},
J. Morgan \myaffil{\ref{ICRAR}},
S.~V.~White\myaffil{\ref{ICRAR},\ref{Rhodes}},
M.~E.~Bell\myaffil{\ref{UTS}},
P.~Patil\myaffil{\ref{Virginia},\ref{NRAO}}
\\
{\small\inst{ICRAR}\,International Centre for Radio Astronomy Research, Curtin University, Bentley, WA 6102, Australia}\\
{\small\inst{Leiden}\,Leiden Observatory, Leiden University, PO Box 9513, Leiden, 2300\,RA, The Netherlands} \\
{\small\inst{ASTRON}\,ASTRON, Netherlands Institute for Radio Astronomy, Oude Hoogeveensedijk 4, Dwingeloo, 7991\,PD, The Netherlands}\\
{\small\inst{Rhodes}\,Department of Physics and Electronics, Rhodes University, PO Box 94, Makhanda, 6140, South Africa}\\
{\small\inst{Virginia}\,Department of Astronomy, University of Virginia, 530 McCormick Road, Charlottesville, VA 22903, USA}\\
{\small\inst{NRAO}\,National Radio Astronomy Observatory, 520 Edgemont Road, Charlottesville, VA 22903, USA}\\
{\small\inst{UTS}\,University of Technology Sydney, 15 Broadway, Ultimo NSW, 2007, Australia}
}

\date{Accepted 2020 November 30. Received 2020 November 29; in original form 2020 October 12}

\pubyear{2020}

\begin{document}
\label{firstpage}
\pagerange{\pageref{firstpage}--\pageref{lastpage}}
\maketitle

\begin{abstract}
Spectral variability of radio sources encodes information about the conditions of intervening media, source structure, and emission processes. With new low-frequency radio interferometers observing over wide fractional bandwidths, studies of spectral variability for a large population of extragalactic radio sources are now possible. Using two epochs of observations from the GaLactic and Extragalactic All-sky Murchison Widefield Array (GLEAM) survey that were taken one year apart, we search for spectral variability across 100--230\,MHz for 21,558 sources. We present methodologies for detecting variability in the spectrum between epochs and for classifying the type of variability: either as a change in spectral shape or as a uniform change in flux density across the bandwidth. We identify 323 sources with significant spectral variability over a year-long timescale. Of the 323 variable sources, we classify 51 of these as showing a significant change in spectral shape. Variability is more prevalent in peaked-spectrum sources, analogous to gigahertz-peaked spectrum and compact steep-spectrum sources, compared to typical radio galaxies. We discuss the viability of several potential explanations of the observed spectral variability, such as interstellar scintillation and jet evolution. Our results suggest that the radio sky in the megahertz regime is more dynamic than previously suggested.
\end{abstract}




\section{Introduction}
\label{sec:intro}
Radio source variability is a powerful resource for studying extragalactic source structure and the physics of the environmental interaction of a radio galaxy. The two main categories of variability, intrinsic and extrinsic, provide information about the source itself or the intervening media along the line of sight, respectively. For example, radio variability can inform us about adiabatic expansion from changes in optical depth with time \citep{tingay2015spectral} or changes in accretion state and jet evolution \citep{tetarenko2019cygnus}. Extrinsic variability induced by scintillation provides information about the electron density variations between the source and observer. It can also give detailed information on the instrnsic structure of the source, particularly on the smallest angular scales \citep[e.g.][]{2007MNRAS.380L..20M}. The majority of previous studies of spectral variability have been conducted at gigahertz frequencies, which have shown to be dominated by the contributions from the core and jets \citep{hardcastle2008properties}, and thus detections of variability in the gigahertz regime have been common \citep[e.g.][]{quirrenbach1989rapid,quirrenbach1992variability,Fan2007variabilityghz,Bower_2011}. 

Intrinsic variability of synchrotron radiation allows an observer to place a strict upper limit on the brightness temperature of the emission \citep[e.g.][]{JMJ2008v404}. Brightness temperatures for all sources emitting synchrotron radiation are subject to the strict upper limit of $10^{12}$\,K due to the Compton scattering limit \citep{kellerman1969opaque}. Sources with temperatures which exceed this limit indicate that their emission is coherent, as in the case of pulsars, or beamed towards the observer, as in the case of blazars. 

The magnitude of the brightness modulation and timescales of extrinsic variability are dependent on which intervening medium is causing the scintillation \citep{hancock2019refractive} and the source size \citep{narayan1992physics}. Depending on the frequency of the radiation, interstellar scintillation (ISS) typically varies source brightness on timescales of months to years \citep{Coles1987RISS}, and is a result of the intervening electron density in the interstellar medium (ISM). ISS has two subcategories, refractive and diffractive, which produce slow (months--years) and short (days--weeks) timescale variability, respectively \citep{Rickett1986RISS}. Interplanetary scintillation (IPS) occurs when radio waves are distorted as they travel through the Solar wind \citep{Clarke:phdthesis,burnell1970ips}. Typically IPS has timescales of seconds or shorter, and sources with a larger angular size can vary due to IPS compared to ISS.

Previous studies of variability and transients at low frequencies ($<1$\,GHz) have searched a wide range of timescales and types of radio sources since the first discovery of variability due to refractive interstellar scintillation \citep[RISS; ][]{hunstead_1972,Rickett1986RISS,fanti1990nature, Riley_1993, hancock2019refractive}. However, such searches have only identified a small population. \citet{ipsII} searched for variability in a sample of compact 37~extragalactic radio sources and identified only one source as showing significant variability; J013243-165444, a known blazar with a peaked radio spectrum. 

A comparison of TIFR GMRT 150\,MHz Sky Survey Alternative Data Release 1 \citep[TGSS-ADR1;][]{intema2016TGSS} and the GaLactic and Extragalactic All-Sky Murchison Widefield Array \citep[GLEAM;][]{wayth2015gleam,2017MNRAS.464.1146H} surveys at $\sim150$\,MHz to search for transients between the two surveys yielded only one candidate that had no detectable spectral curvature \citep{Murphy2016Transients}. \cite{lofar2015transients} conducted a search for transients at 60\,MHz using the Low-Frequency Array \citep[LOFAR;][]{2013lofar} and also found only one candidate, showing that bright transient radio sources at low frequencies are fairly uncommon. The Murchison Widefield Array Transients Survey \citep[MWATS;][]{mwats} surveyed $\sim$1,000 sources for almost three years at a cadence of $\approx$3\,months. MWATS found 15 variable sources with significant flux-density modulation at 154\,MHz, seven of which were identified as having a curved spectrum by \citet{Callingham_2017}, and detected no transients.

Previously, it has been suggested that surveys of variability at low frequencies ($\leq\,500$\,MHz) have found few variable extragalactic sources because emission at megahertz frequencies is expected to be dominated by the emission from the lobes of the radio galaxies \citep{mwats}. Such radio lobes are $\sim$10--1000\,kpc in size \citep{hardcastle2008properties}, and thus are often too large for ISS to be significant, which requires angular sizes $\lesssim5$\,milliarcseconds. However, the radio sources that have previously been identified as low-frequency variables are more likely to also have a peaked spectrum \citep{mwats,ipsII}. It still remains unclear whether peaked-spectrum sources dominate the low-frequency variable population due to intrinsic effects, such as source evolution, or due to their potentially small spatial structures causing them to be more susceptible to scintillation. 

Peaked spectrum sources (PSS), analogous to gigahertz-peaked spectrum (GPS), high-frequency peaked (HFP) and compact steep-spectrum (CSS) sources \citep{o1998compact,kunert2010compact,2020arXiv200902750O}, are a unique subset of AGN that can display far more compact double-lobe morphology than typical radio-loud AGN \citep{phillips1982symmetric, Tzioumis_2010}. GPS and HFP radio sources are categorised by their notable peak at gigahertz frequencies, and CSS radio sources are expected have a peak at a lower radio frequency ($<200$\,MHz) and display a compact double structure. A subclass of PSS were identified by \cite{Callingham_2017} that display the same identifiable peak but in the megahertz regime that are believed to be the same class of object as GPS and HFP sources \citep{callingham2015,coppejans2015mps,coppejans2016mps}.

Previous studies of the variability of PSS at gigahertz frequencies have yielded several sources that show flux density variability across their radio spectra while maintaining their PSS classification (i.e., retain a clear peak in their radio spectra at each epoch). However, it has also been observed that some sources can lose their PSS classification over time \citep{torniainen2005long}. Several sources displayed a temporary peaked spectrum which over time smoothed to a flat spectrum. Such sources with a temporary peaked spectrum at gigahertz frequencies are believed to be blazars \citep{2005A&A...432...31T}, where the features in the small core-jet structure are likely also scintillating at gigahertz frequencies.
In contrast, one known peaked-spectrum source, PKS\,B1718-649, which has a double-lobe morphology on parsec scales, has been observed to show variability both above and below the spectral peak at gigahertz frequencies over an approximately two-year period \citep{tingay2015spectral}. The spectral variability of the spectral energy distribution (SED) of PKS\,B1718--649 was best modelled by variations in the optical depth and adiabatic expansion of the source. 

Despite the plethora of radio frequency variability research, the majority of previous studies have been limited to identifying variability at a single frequency \citep{lofar2015transients,Murphy2016Transients,ipsII,mwats}, small sample size \citep{tingay2015spectral}, or spectral variability at gigahertz-frequencies formed from non-contemporaneous data \citep{torniainen2005long}. Such shortcomings have limited our understanding about the cause of the identified variability since, for example, scintillation is a broadband effect producing unique variability across the entire radio spectrum. Likewise, intrinsic variability produces frequency-dependent effects depending on the emission or absorption mechanism. Distinguishing between intrinsic or extrinsic processes as the cause of variability requires simultaneous multi-frequency spectral coverage, but this has been hard to achieve. 

Large population studies with significant spectral and temporal coverage have only recently become available with the development of radio telescopes like the Murchison Widefield Array \citep[MWA;][]{tingay2013MWA} and LOFAR \citep{2013lofar}. The MWA has a large field of view ($\sim600$\,$\mathrm{deg}^{2}$) and operates over a wide frequency range ($\sim$80-300\,MHz) with an instantaneous bandwidth of 30.72\,MHz. As we move into the Square Kilometre Array (SKA) era, low-frequency surveys with wide spectral coverage of large populations will become more readily available, permitting us to discern the origins of radio variability. Consequently, it is imperative that we derive appropriate methodologies and robust statistical techniques in order to produce insightful results from these future surveys. 

This paper presents the first large population survey of low-frequency spectral variability using two epochs of the GLEAM survey. In Section~\ref{sec:data} we outline the sample selection process and data used in this analysis. Section~\ref{sec:var-analysis} describes the methodology for detecting and classifying spectral variability. We present the catalogue of variable candidates in Section~\ref{sec:results}, in particular the sources with persistent PSS in Section \ref{sec:results_VIPPSS} and variable spectral shape in Section~\ref{sec:results_changingshape}. The potential mechanisms for the observed variability of each class are discussed in Section~\ref{sec:discussion}. We adopt the standard $\Lambda$-cold dark matter cosmological model, with $\Omega_{\rm M} = 0.28$, $\Omega_\Lambda = 0.72$, and the Hubble constant $H_0 = 70$\,km\,s$^{-1}$\,Mpc$^{-1}$ \citep{lamdacdm}.

\section{Data}
\label{sec:data}

\subsection{GLEAM Year 1 and Year 2}
\label{sec:data_GLEAM}
In the first year of GLEAM observations (`Year 1': Aug~2013 -- Jun~2014), the entire sky south of Dec $+$30\,deg was surveyed at 72--231\,MHz using meridian drift scan observations at different declination stripes, taken at night in week-long runs spaced about three months apart. Due to the observing strategy, 8--10\,hour scans taken at night, there is little crossover in surveyed sky area in the observations separated by three months, making a variability search on a three month timescale feasible only for small areas of sky. Quasi-simultaneous spectral coverage was ensured by cycling through a sequence of two minute scans in five frequency bands; the frequency bands covered 72--103, 103--134, 139--170, 170--200, 200--231\,MHz. \citet{2017MNRAS.464.1146H} published the GLEAM Extragalactic Catalogue, which excludes Galactic latitudes, |$b$|$<$|10$^{\circ}$|, and a few areas around bright sources, based on these observations. The catalogue provides 20 almost independent flux density measurements across 72--231\,MHz and covers 24,000\,deg$^{2}$ of sky, making it the widest fractional bandwidth radio survey to date.

An additional epoch (`Year 2', Aug to Dec 2014) was conducted with minimal changes to the observing strategy. In Year 2 the observations of year one were repeated twice with hour angles of $\pm1\,$hour. This second epoch of GLEAM provides a unique opportunity to search for low-frequency variability in the flux density over a large fractional bandwidth over a one year timescale (hereafter referred to as Year 1 and Year 2 observations). 

We independently processed a subset of the Year 1 and Year 2 data from the highest four frequency bands over an $\sim8000$\,deg$^{2}$ region of sky centred on the South Galactic Pole and covering $300^{\circ}\leq\mathrm{RA}\leq 100^{\circ}$ and $0^{\circ}\leq \mathrm{Dec.}\leq -60^{\circ}$. In this region of sky, the Year 1 data were taken almost entirely over the period Aug--Nov~2013 and the Year 2 data were taken over the period Aug--Dec~2014. We used an improved pipeline and beam model for the MWA, and used the published GLEAM catalogue as a sky model for calibration. 

The lowest MWA sub-band (72--103\,MHz) was not used as the presence of the bright sources Fornax~A and Pictor~A in the sidelobes of the primary beam prevented maps with sufficient quality being produced for the Year 2 data. A more detailed explanation of this enhanced processing is provided by \citet{FranzenGLEAM2020}. We use these two epochs of data, each composed of sixteen 7.68\,MHz mosaics spanning 103--231\,MHz and a wide-band mosaic covering 200--231\,MHz, to search for variability.

\subsection{Source Extraction}
We used \textsc{Aegean}\footnote{\url{https://github.com/PaulHancock/Aegean}}, a source finding algorithm \citep{2012MNRAS.422.1812H,2018PASA...35...11H}, to create the catalogue of sources for each epoch. Firstly, \textsc{Aegean} was run blindly (i.e. without prior positional constraints) over the most sensitive and highest resolution mosaic, the 200--231\,MHz mosaic for the Year 2 data. The background and noise mosaics were generated using the Background and Noise Estimation tool (\textsc{BANE}). The resulting catalogue was used to provide the positions for a priorized fit measurement in each of the other mosaics. \textsc{Aegean} also used the point spread function (PSF) maps as well as the background and noise images for source characterisation. 

\textsc{Aegean} measures the peak flux density of a source by a Gaussian fit to its brightness distribution, with the reported total flux density representing the integrated flux density under the fit. We used the priorized fitting mode of \textsc{Aegean}, which took the known position and size of the source and only fit for the flux density. Since we are only investigating high signal-to-noise sources, small variations induced by the fitting algorithm have a negligible impact on the spectra of the sources and were accurately captured in our flux density uncertainties. For thoroughness, we checked for any significant changes in source shapes or the PSF between epochs and found no such changes. Consequently, by assuming a consistent shape and position and by using the priorized mode to fit the flux, we reduced the possibility of induced artificial variability due to differences in the source finder fitting. 

An accurate estimate of the uncertainty for individual flux density measurements for each source at each frequency is necessary to evaluate the reliability of any observed variability. For sources detected above 100\,$\sigma$ in the 200--231\,MHz Year 1 mosaic and 200--231\,MHz Year 2 mosaic, we examined the distribution of the flux density ratios of the Year 1 to Year 2 integrated flux densities. Note that since the variability is rare, the width of the distribution is dominated by systematic and random noise. We used this distribution to confirm that the flux density scales are consistent. The measured FWHM of this distribution was used to determine the random flux density uncertainty at each frequency. The percentage uncertainty for each of the 16 frequency bands was found to be $\sim0.5$--$1\%$, consistent with the internal uncertainties reported by \cite{2017MNRAS.464.1146H}.

Near the edges of the mosaic there is correlated noise in some bands. As a result, we increased the error for sources within roughly five degrees of the edge of the mosaics by $\sim3$\%. Likewise, in several regions, poor calibration due to bright sources in the sidelobe of the MWA or bright nearby sources could result in correlated variability. The sources we account for are Pictor A (both for nearby sources and when Pictor~A is in the sidelobe), Fornax~A, Cassiopeia~A in the sidelobe and the Crab nebula in the sidelobe. We increased the error in these regions to $\sim2$--$5\%$ to account for this. The percentage increase for the error was calculated to ensure there was no structure in the VIP according to RA or Dec. Firstly, regions with a higher density of sources with large values for the VIP were identified. The error was increased in these regions until there was no discernible structure in the VIP across the entire mosaic. The central coordinates of these regions are presented in Table~\ref{tab:coords}. Furthermore, any sources with $18^{\circ}\leq \mathrm{RA} \leq 36^{\circ}$ and $-35^{\circ} \leq \mathrm{Dec} \leq -20^{\circ}$ or $54^{\circ}\leq \mathrm{RA} \leq 95^{\circ}$ and $-35^{\circ} \leq \mathrm{Dec} \leq -22^{\circ}$ were excluded entirely as the quality of the mosaics was lower in these sky regions due to issues with calibration or bright sources in the primary beam in the Year 2 data.

\begin{table}
\centering
\begin{tabular}{|p{3cm}|p{3.5cm}|}
\hline
Source Name     & Coordinates, $\mathrm{RA}$, $\mathrm{Dec}$ (deg) \\
\hline
Pictor~A & $80$, $-46$ \\
Pictor~A sidelobe & $62$, $-15$ \\
Fornax~A & $51$, $-37$ \\
Cassiopeia~A sidelobe & $350$, $-25$ \\
Crab Nebula sidelobe & $31$, $-42$ \\
\hline
\end{tabular}%
\caption{Bright sources where correlated variability was noticed. We assign higher error when calculating the VIP and MOSS to reduce the measured variability making it in line with other regions. }
\label{tab:coords}
\end{table}

The final catalogue used for this project contains 93,928 sources (selected from the Year 2 200--231\,MHz mosaic). Each source has 16 individual flux density measurements across 103--231\,MHz, and a wide-band (200--231\,MHz) flux density for both Year 1 and Year 2. 
Noise levels were measured from the local root-mean-squared (rms) of the initial mosaics. For the 200--231\,MHz wide-band mosaics, the mean and standard deviation of the rms noise was found to be $7\pm 5$\,mJy beam$^{-1}$.

\subsection{Sample Selection}
\label{sec:methods_sampleselection}
Several quality cuts were applied to the catalogue to select reliable sources. These cuts are based on those outlined by \cite{Callingham_2017}. In summary, the quality cuts ensure that unresolved GLEAM sources are bright enough to form high signal-to-noise spectra to reliably search for spectral variability. The selection criteria, applied to both years, are presented in Table~\ref{tab:QualityCuts}. 
Sources that met the first five criteria in Table~\ref{tab:QualityCuts} (in both years) are classified as the ``master sample'', which is composed of a total of 21,558 sources.

\subsection{Description of Additional Radio Data}
\label{sec:extra_surveys}
We use the Sydney University Molonglo Sky Survey \citep[SUMSS;][]{2003MNRAS.342.1117M} and the NRAO VLA Sky Survey \citep[NVSS; ][]{condon+98}, as part of our spectral modelling to estimate the spectral index for the high frequency section of the SED (180\,MHz -- 843\,MHz/1.4\,GHz). We considered the possibility that the higher resolution of SUMSS and NVSS would mean some of our MWA sources are resolved into multiple components. After visually inspecting the NVSS and SUMSS counterparts to our PSS candidates, no source was found to be heavily resolved. We also note that the SUMSS and NVSS data is only used in the PSS classification, not in the variability analysis.

The Australia Telescope 20\,GHz (AT20G) Survey was used to test if there was the presence of compact features and the detections of core components \citep{at20g}, see Section~\ref{sec:results_at20g} for more details. All other radio surveys used in this paper were not explicitly used in any analysis, but are included in the SEDs for completeness. The additional radio surveys are the Very Large Array Low-frequency Sky Survey Redux \citep[VLSSr; ][]{2014MNRAS.440..327L}, TIFR GMRT 150\,MHz Sky Survey Alternative Data Release 1 \citep[TGSS-ADR1; ][]{intema2016TGSS}, and the Molonglo Reference Catalogue \citep[MRC: ][]{1981MNRAS.194..693L,1991Obs...111...72L}. All catalogues were cross-matched using Topcat's \citep{2005ASPC..347...29T} nearest neighbour routine with a 2\,arcmin radius. A 2\,arcmin radius was chosen as it is comparable to the resolution of GLEAM. 

\subsubsection{SUMSS}
SUMSS is a continuum survey at 843\,GHz with observations taken between 1997 and 2003 \citep{2003MNRAS.342.1117M}. SUMSS was conducted by the Molonglo Observatory Synthesis Telescope \citep[MOST;][]{1981PASAu...4..156M,1991AuJPh..44..729R} covering the southern sky up to a declination of $-30^{\circ}$, excluding Galactic latitudes below $10^{\circ}$. The published catalogue has a total of 211,063 sources and the resolution of the survey varied with declination $\delta$ as $45^{\prime\prime} \,\times \,45^{\prime\prime} \,\mathrm{cosec}| \delta | $. SUMSS is 100\% complete above $\approx8$\,mJy south of a declination of $-50^{\circ}$, and above $\approx18$\,mJy for sources with a declination between $-50^{\circ}$ and $-30^{\circ}$. 

\subsubsection{NVSS}
NVSS is a continuum survey at 1.4\,GHz with observations taken between 1993 and 1996 \citep{condon+98}. NVSS was conducted by the Very Large Array (VLA) covering the northern sky down to a declination of $-40^{\circ}$ with a resolution of $\approx45$\,arcseconds. The published catalogue has a total of 1,810,672 sources, and is 100\% complete above 4\,mJy. 

\subsubsection{AT20G}
AT20G is a blind search for radio sources at 20\,GHz with observations taken between 2004 and 2008 \citep{at20g}. AT20G was conducted by the Australia Telescope Compact Array (ATCA) covering the southern sky up to a declination of $0^{\circ}$ with a resolution of $\approx10$\,arcseconds. The published catalogue has a total of 5,890 sources, and is 91\% complete above 100\,mJy in regions south of declination $-15^{\circ}$.

\begin{table*}
\centering
\begin{tabular}{|p{0.5cm}|p{5cm}|p{7cm}|l|}
\hline
& Step     &  Criteria & Number of Sources \\
\hline
0 & Total Sources in Field & & 93,928 \\
1 & Unresolved in wide-band mosaic& $\frac{ab}{a_{\mathrm{psf}}b_{\mathrm{psf}}}\leq 1.1$ & 77,916 \\
2 & Bright in wide-band mosaic  & $S_\mathrm{200-231MHz}\geq 160\,\mathrm{mJy}$ & 24,089\\
3 & High signal-to-noise (S/N) & eight or more flux density measurements with S/N $\geq 3$ & 24,624\\
4 & NVSS and/or SUMSS counterparts & Cross-match within 2\,arcmin counterpart &  24,619\\
5 & Cut bad RA and Dec regions & Source outside regions with $18^{\circ}\leq \mathrm{RA} \leq 36^{\circ}$ and $-35^{\circ} \leq \mathrm{Dec} \leq -20^{\circ}$ or $54^{\circ}\leq \mathrm{RA} \leq 95^{\circ}$ and $-35^{\circ} \leq \mathrm{Dec} \leq -22^{\circ}$ & 21,558 \\
 & \emph{Total Master Population} & \emph{ Sources that passed steps 0--5} & \emph{21,558}\\
6 & Variable & VIP $\geq 58.3$ & 340 \\
7 & Manual Check & Pass manual inspection & 323 \\
8 & Uniform Spectral Change & MOSS $< 36.7$ & 272 \\
9 & Changing Spectral Shape & MOSS $\geq 36.7$ & 51 \\
\hline
\end{tabular}%
\caption{Quality cuts (from Section~\ref{sec:methods_sampleselection}) applied to the catalogue of sources to derive the master population and the criteria for variability classification. Any source that did not pass criteria 1--5 in each year of data was discarded. $a$, $b$, $a_{\mathrm{psf}}$ and $b_{\mathrm{psf}}$ are the semi-major and semi minor axes of the source and the point spread function, respectively. $S_\mathrm{200-231MHz}$ is the measured flux density in the wide band mosaic covering 200--231\,MHz. NVSS is the NRAO VLA Sky Survey \citep{condon+98} and SUMSS is the Sydney University Molonglo Sky Survey \citep{2003MNRAS.342.1117M}. These quality cuts are identical to those implemented by \citet{Callingham_2017}. Sources which pass steps 1--5, in both years, are classified as the ``master population''. Variable sources are selected as described in Section~\ref{sec:var-analysis}. The VIP is a measure of variability and is calculated according to Equation~\ref{eq:varparam}. The MOSS parameter is presented in Equation~\ref{eq:MOSS} and measures the change in spectral shape. }
\label{tab:QualityCuts}
\end{table*}

\section{Variability Analysis}
\label{sec:var-analysis}
In this section, we present methods to determine if a source is variable and if it changes spectral shape. The classification steps are presented in Table~\ref{tab:QualityCuts}. 

\subsection{Variability Index Parameter}
\label{sec:var-analysis_VIPdescription}
In order to identify true source variability, instead of instrumental noise, we define the variability index parameter (VIP). The VIP is adopted from the $\chi^2$ statistic, 
\begin{equation}
    \label{eq:varparam}
    \mathrm{VIP} = \sum^{n}_{i=1} \frac{\big(S_{\mathrm{yr1}}(i)- S_{\mathrm{yr2}}(i)\big)^2}{\sigma_i^2},
\end{equation}
where $S_{\mathrm{yr1}}(i)$ and $S_{\mathrm{yr2}}(i)$ are the flux densities in Year 1 and Year 2, respectively, in a given sub-band, $i$. $\sigma_i$ is the combined uncertainty of each flux density added in quadrature. The VIP is calculated entirely from the raw flux density measurements to avoid biasing variability estimates induced when fitting spectral models to the data.

The VIP was calculated for each source in the master sample, including the 422 sources previously identified as a PSS by \cite{Callingham_2017}.

\begin{figure}
    \centering
    \includegraphics[width=\linewidth]{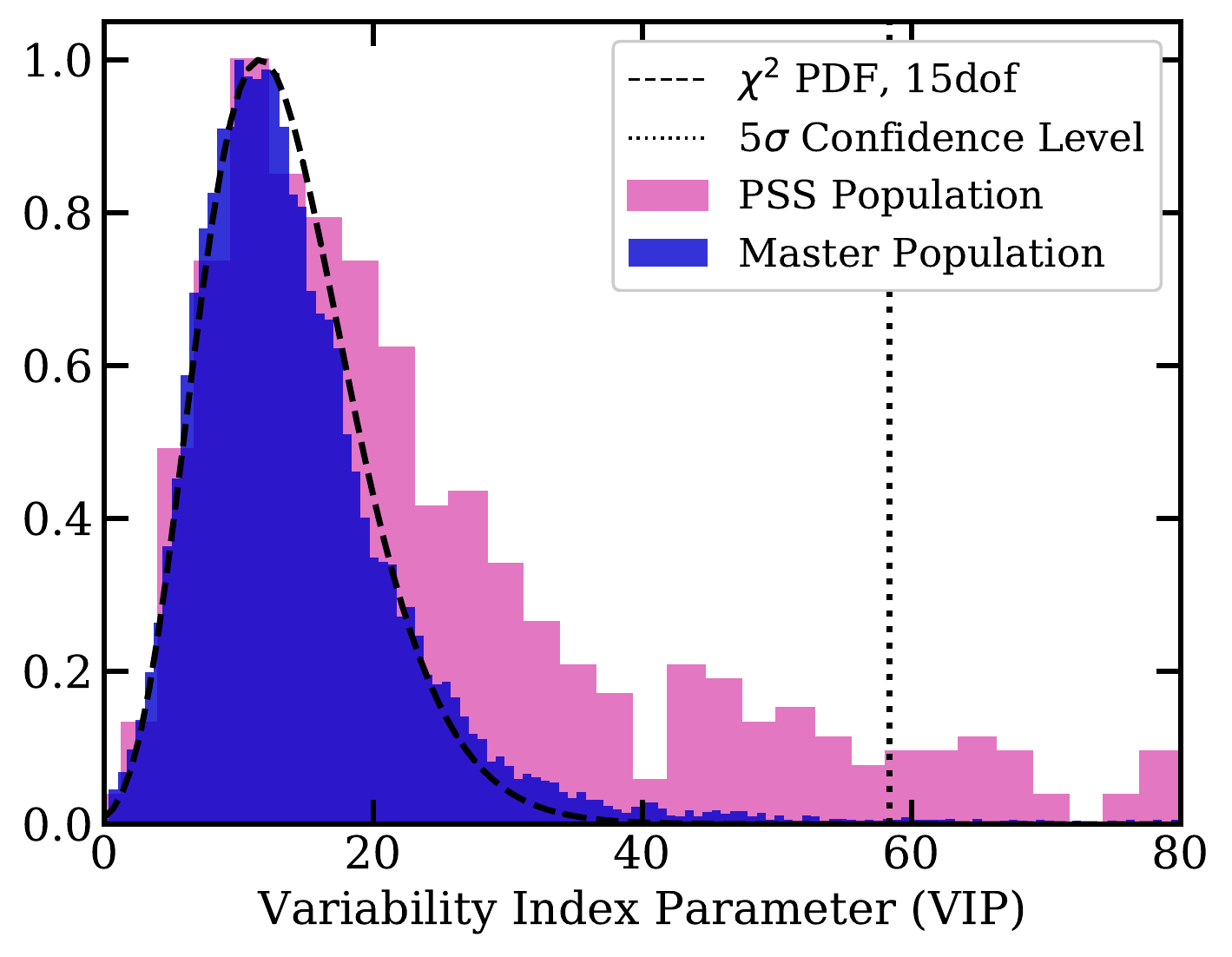}
    \caption{Histogram of the VIP as a measure of spectral variability for both the master population (blue) and the PSS population (pink) normalised to a maximum of one. The master population was modelled with a $\chi^2$ probability density function with 15 degrees of freedom and is consistent with a population which is mostly non-varying. Using this distribution a 99.99994\% confidence level, corresponding to a VIP of 58.3, was used to determine whether a source was variable. The PSS population is not well defined by a $\chi^2$ distribution, implying it is likely a more intrinsically variable population. }
    \label{fig:varparamhist}
\end{figure}

We plotted a $\chi^2$ probability density function (pdf) with 15 degrees of freedom using \textsc{scipy.stats.chi2} \citep{scipy}. The pdf and histogram are presented in Figure \ref{fig:varparamhist}. We observe that the shape of the VIP distribution for the master population is what we would expect for an intrinsically non-variable population, with the variation largely produced by noise in the flux density measurements. Such an interpretation is supported by the fact the vast majority of sources in the master population have a VIP within $0$--$16$, suggesting any change from Year 1 to Year 2 is entirely within 68\,per\,cent confidence limit for all 16 flux density measurements. Furthermore, the agreement with a theoretical $\chi^2$ distribution with 15 dof is consistent with a largely non-variable population. 

We chose to prioritise the reliability of the selected variable sources over completeness of the sample as we are investigating an unexplored parameter space for variability. Therefore, we have implemented a conservative cut to the VIP of $\geq58.3$ (equivalent to a confidence level of true variability of 99.99994\%, i.e. 5$\sigma$). All sources with a VIP$\geq$58.3 also had a visual inspection of the SEDs to ensure the variability was reliable. 17~sources with a VIP$\geq$58.3 were flagged as non-variable; their apparent variability is likely due to calibration errors or bright nearby sources, and is characterised by large, non-physical steps in the SED.

\begin{figure*}
    \centering
    \includegraphics[width=\linewidth]{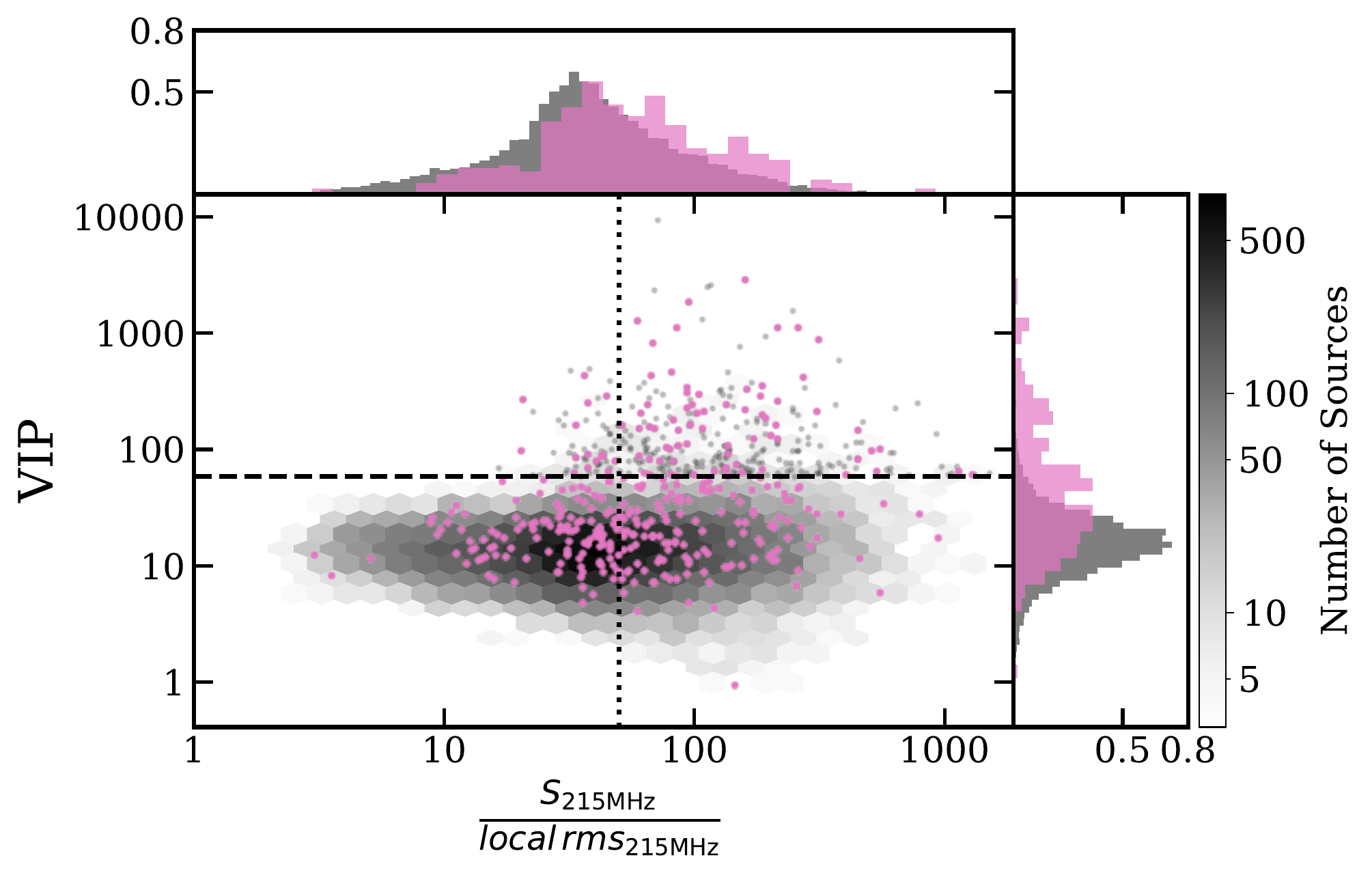}
    \caption{Distribution of the VIP as a function of the signal-to-noise ratio (S/N) in the 200--231\,MHz wide band mosaic. The grey hexagons represent the density of the master population (as indicated by the colour-bar). Pink dots are sources identified as persistent PSS and grey points are all other variable sources. The dashed horizontal line is the 99.99994\% VIP confidence level of 58.3 (equivalent to 5$\sigma$): sources with a VIP$\geq$58.3 are classified as variable. The distribution of the S/N for the master population and PSS population is shown in the grey and pink histograms in the top panel. The dotted vertical line denotes a S/N cut of 50 where the PSS population is complete, relative to the master population. The grey and pink histograms in the right panel represent the VIP distributions for the master population and PSS population respectively for sources with a S/N above 50. The PSS VIP histogram shows a significantly different distribution to that of the master population with a peak at higher VIP, a wider distribution, and longer tail towards higher VIP.}
    \label{fig:vip_snr}
\end{figure*}

We also compare the master population distribution of VIP with that of the PSS population (in Figure~\ref{fig:varparamhist}). Unlike the master population, the PSS population is not well defined by a $\chi^2$ distribution. There is an excess of sources identified as a PSS with VIP $\geq16$. This result suggests that variability is a more prominent feature of the PSS population compared to the general radio source population. To ensure this bulk property of the PSS population is not due to this population having a larger signal-to-noise ratio (S/N), we demonstrate how the VIP of the master population and PSS sources varies as a function of S/N in Figure~\ref{fig:vip_snr}.  The PSS population relative to the master population does not have completeness issues at S/N$\geq50$, as shown in the top panel of Figure~\ref{fig:vip_snr} as the PSS S/N population closely matches that of the master population. However, we find the distribution of the VIP of the PSS population much wider than the master population (for sources with S/N$\geq50$), with a larger tail towards a higher VIP, as shown in the histogram in the right panel of Figure~\ref{fig:vip_snr}. 

We take this as evidence that the PSS source \emph{population} is more variable, on average, than the general radio source population.

\subsection{Measure of Spectral Shape}
\label{sec:var-analysis_var_classification}
The large fractional bandwidth of this study enables us to detect changes in spectral shape between epochs. We classified the variable population into two classes of spectral variability:

\begin{enumerate}
    \item Uniform change: all 16 flux density measurements increase or decrease by the same absolute amount (within uncertainty);
    \item Changing shape: the shape of the spectrum has changed between epochs.
\end{enumerate}

To distinguish between these two categories, we define the measure of spectral shape (MOSS) parameter. The MOSS parameter uses the flux density measurements directly to detect changes in spectral shape in order to reduce uncertainties that accompany fitting spectral models. The MOSS parameter is also a variation of the $\chi^2$ statistic, namely: 

\begin{equation}
\label{eq:MOSS}
    \mathrm{MOSS} = \sum^{n}_{i=1} \frac{(\widetilde{\mathrm{diff}}- \mathrm{diff}(i))^2}{\sigma_{i}^2}
\end{equation}
where $\widetilde{\mathrm{diff}}$ is the median of the differences between the flux density over all frequencies, $\mathrm{diff}(i)$ is the difference of the flux densities between the two epochs at frequency $i$, and $\sigma_{i}$ is the combined uncertainty of each flux density added in quadrature. Unlike the VIP, the MOSS parameter measures how many flux density points are greater than $1\sigma$ away from the median difference value. A larger MOSS value suggests a larger spread of the difference in measurements from the median value between the two epochs and hence a change in spectral shape. 

\begin{figure}
    \centering
    \includegraphics[width=\linewidth]{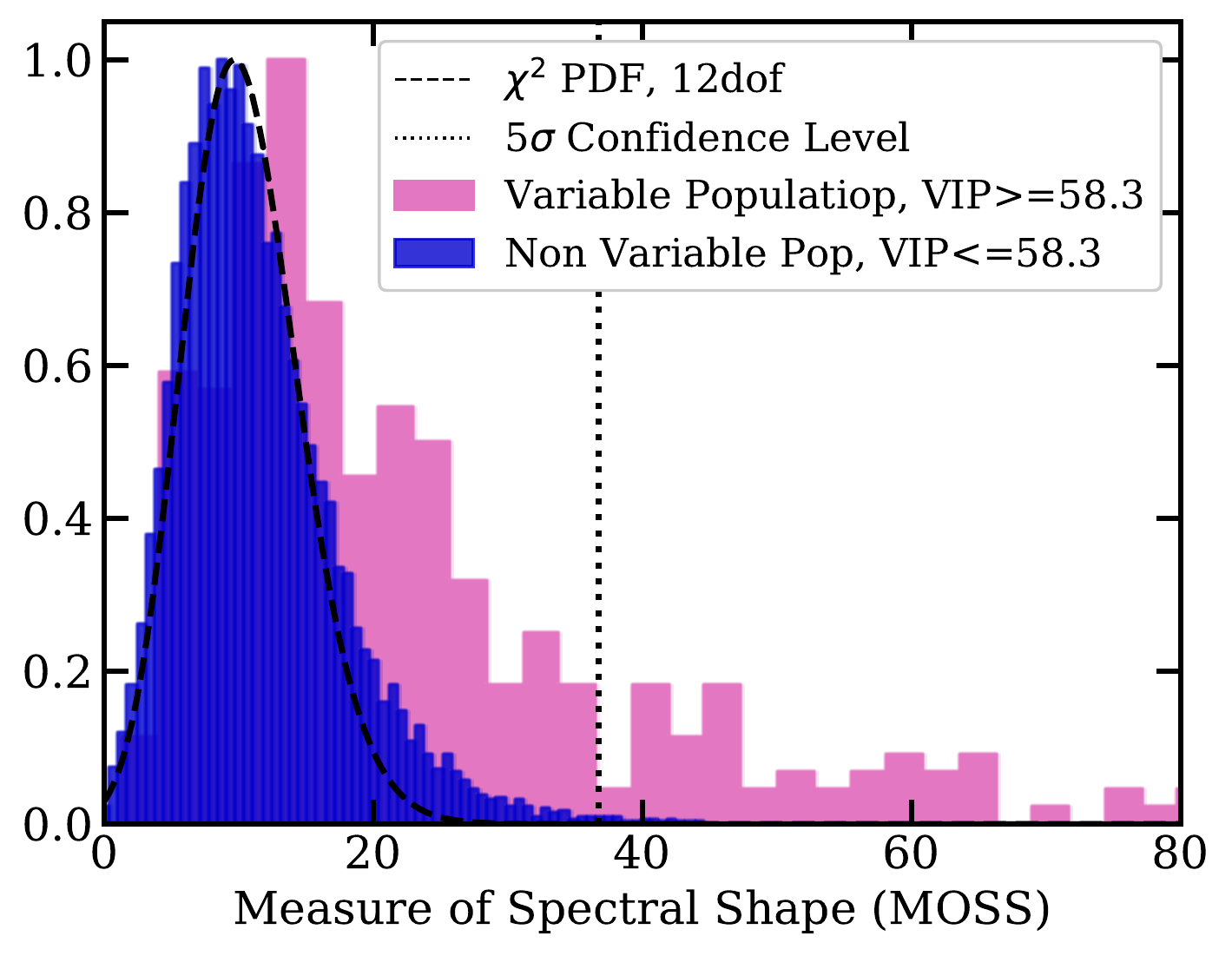}
    \caption{Histogram of the MOSS parameter, a measure of the variability of spectral shape, for the non-variable master population (blue) and the 323 variable sources (sources with a VIP$\geq58.3$) (pink) both normalised to have the max number of sources in a bin as 1. This distribution was fit with a $\chi^2$ probability density function and is consistent with the majority of variable sources not displaying changing spectral shape. Using this distribution, a 99.99994\% confidence level, corresponding to a MOSS parameter $>36.7$, was used to determine a significant change in spectral shape between Year~1 and Year~2.}
    \label{fig:moss_hist}
\end{figure}

The distribution of the MOSS parameter for the non-variable population is most consistent with a $\chi^2$ probability density function with 12 dof, as shown in Figure\,\ref{fig:moss_hist}. \citet{hurley2017galactic} noted that the errors within the 30\,MHz bands of GLEAM are correlated. Since we are measuring correlated change away from a central point, this has a stronger effect on the MOSS parameter than the VIP. We hypothesise that this is the cause of the reduced number of degrees of freedom of the optimal PDF distribution, given we expect 14 dof. Using the distribution of the non-variable population a value of $36.7$ and above for the MOSS parameter was chosen to select sources that are 99.99994\% likely to be truly changing shape (equivalent to a 5$\sigma$ confidence level). We define the \textit{uniform change population} as variable sources with no significant change in spectral shape according to the MOSS parameter (with a MOSS parameter below $36.7$). Likewise we define the \textit{changing spectral shape population} as variable sources with a significant change in spectral shape according to the MOSS parameter (with a MOSS parameter $\geq36.7$). 

\begin{figure*}
    \centering \includegraphics[width=0.45\linewidth]{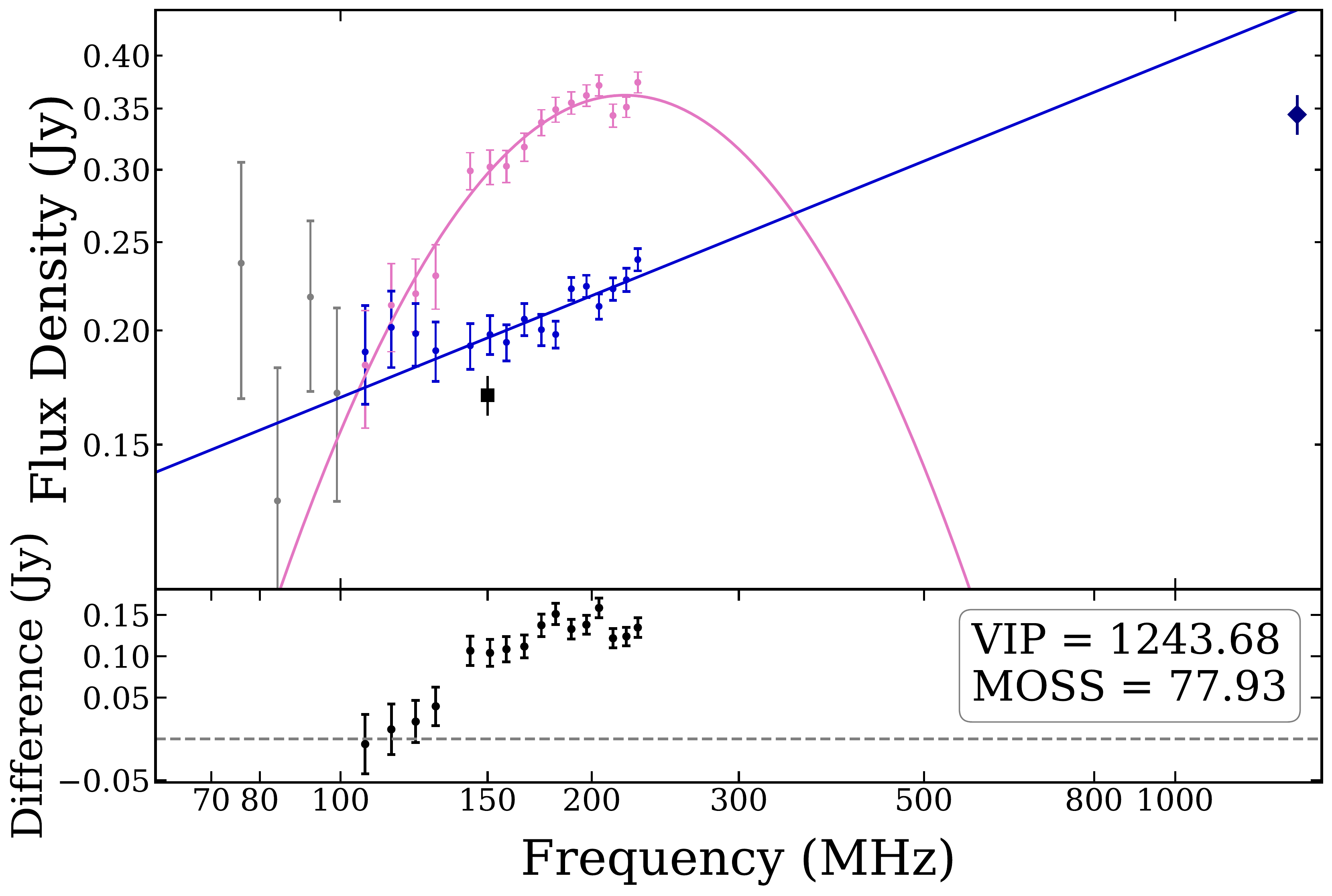}
    \label{fig:J225641}
    \centering \includegraphics[width=0.45\linewidth]{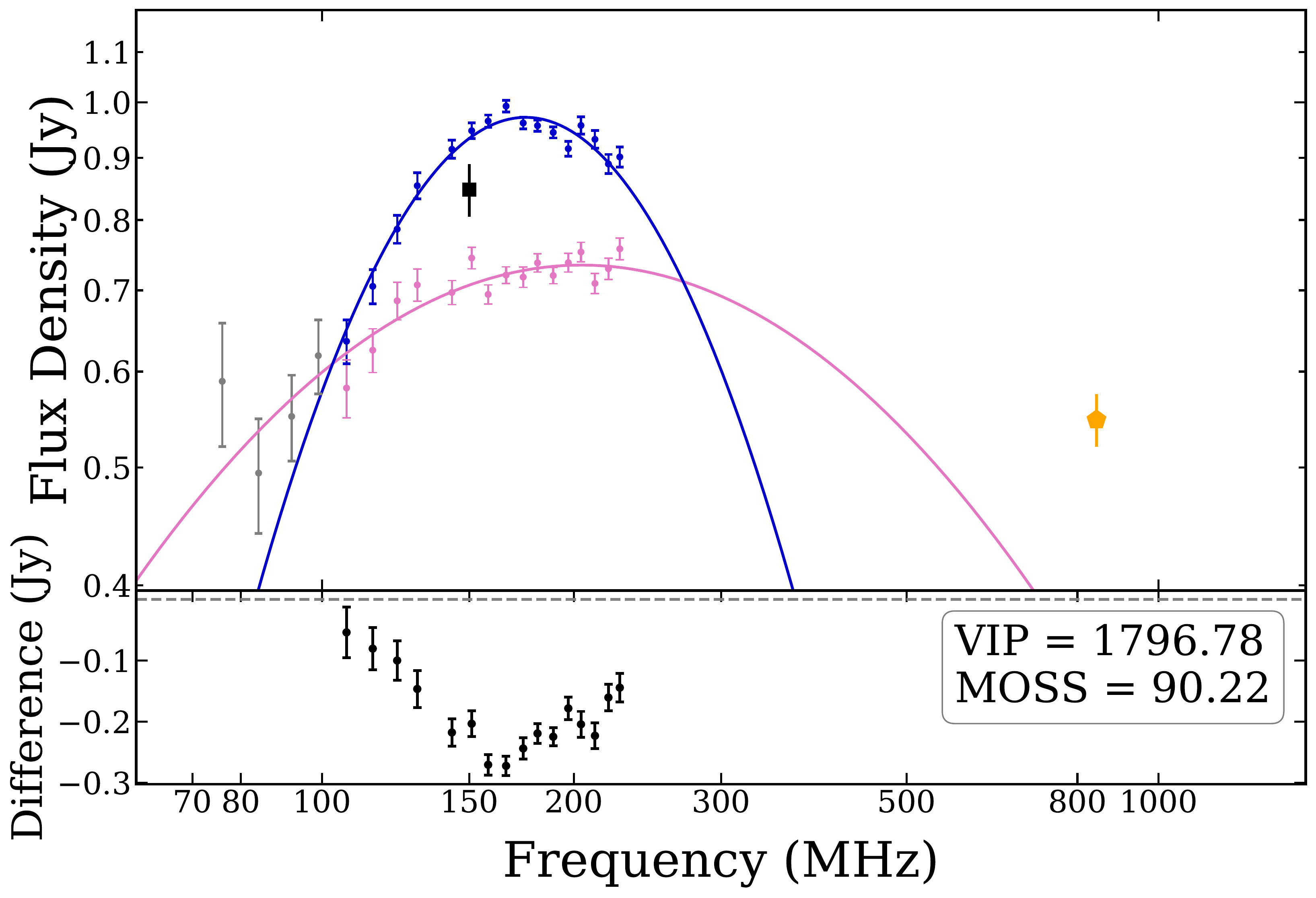}
    \label{fig:J032237}
    \caption{Examples of radio spectra of two variable sources showing a significant change in spectral shape between Years 1 and 2. One was found to be have a peaked spectrum in Year 1 and then flattened in Year 2 (GLEAM\,J225641--201140, left panel) and the other (GLEAM\,J032237--482010, right panel) showed the reverse. The points represent the following data: GLEAM low frequency (72--100\,MHz) (grey circles), Year 1 (pink circles), Year 2 (blue circles), TGSS (black square), SUMSS (yellow pentagon), and NVSS (navy diamond). The difference of the flux densities in Year 1 and 2 are plotted below. Models plotted for each year are determined by the PSS classification only. A source classified with a peak within the observed MWA band, which also satisfied the PSS criteria presented by \citet{Callingham_2017}, was modelled by a quadratic according to Equation~\ref{eq:quadratic}. Remaining sources were modelled by a power-law according to Equation~\ref{eq:plaw}, see Section~\ref{sec:methods_spectralmodels} for details.}
    \label{fig:blazars}
\end{figure*}

The spectra of two examples sources that are classified as changing spectral shape according to the MOSS parameter are presented in Figure~\ref{fig:blazars}.

\begin{figure}
    \centering
    \includegraphics[width=\linewidth]{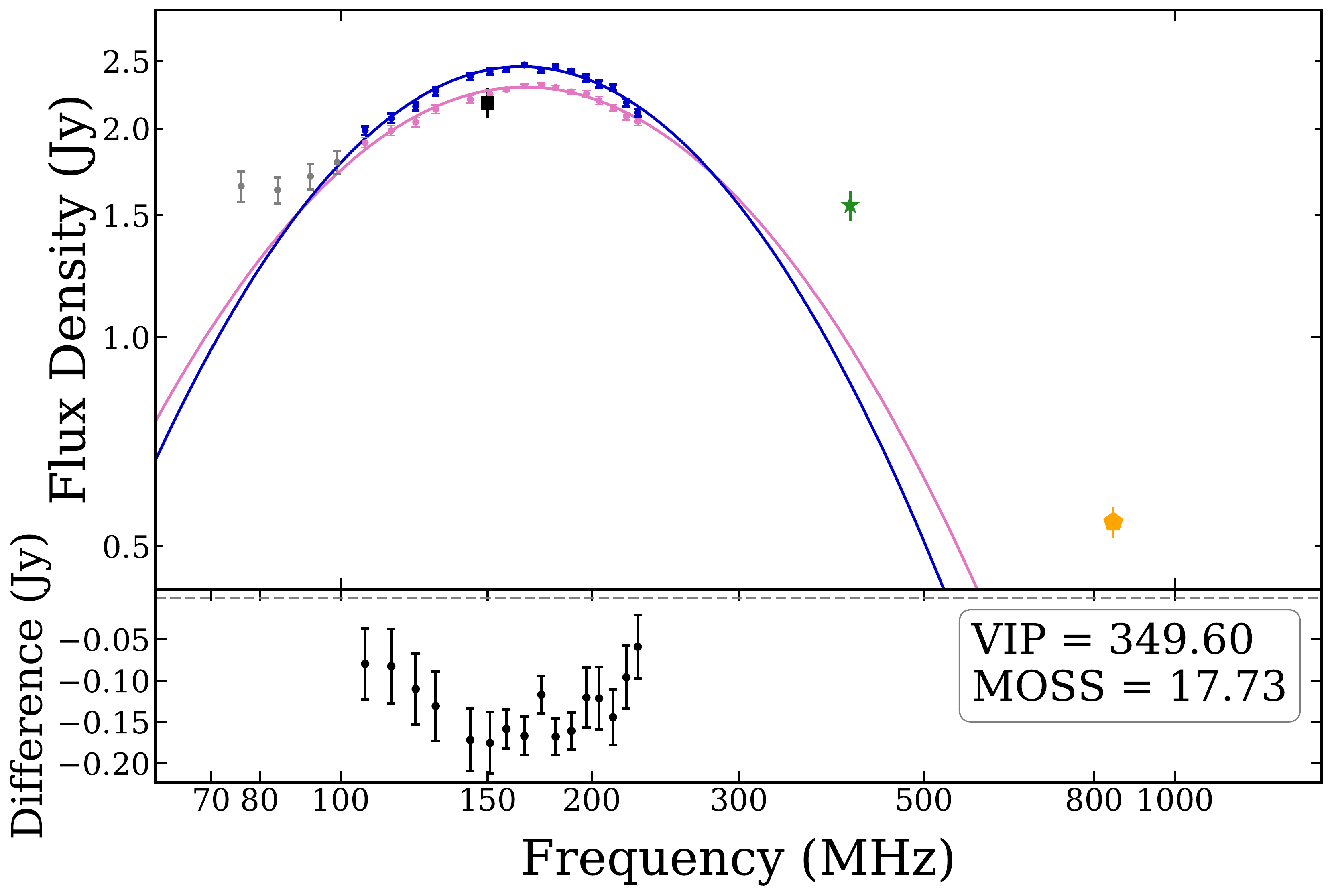}
    \caption{Example of a source for which there is insignificant change in spectral shape according to the MOSS parameter 5$\sigma$ confidence level cut despite the sharper peak in the spectra for Year 2. The data points represent the following data: GLEAM low frequency (72--100\,MHz) (grey circles), Year 1 (pink circles), Year 2 (blue circles), TIFR GMRT 150\,MHz Sky Survey Alternative Data Release 1 (TGSS) (black square), MRC (green star), and SUMSS (yellow pentagon). Residuals are calculated by differencing the flux density measurements of the two epochs of observations. The models for each year are determined by the classification of the source; We identify a peak within the observed bandwidth that is well modelled by a quadratic according to Equation~\ref{eq:quadratic} for both years.}
    \label{fig:J234312}
\end{figure}

We note that the MOSS parameter could potentially miss truly changing spectral shape sources if the difference between the two epochs are symmetric around the mean frequency. For example, GLEAM\,J234312--480659 (Figure~\ref{fig:J234312}) shows a ``tick" shape in its difference spectra. Therefore, it is classified as uniform change rather than changing shape. Furthermore, it is harder to detect a change in spectral shape for sources with lower S/N. After careful testing to balance a reliable changing shape population with completeness, sources with a MOSS $>36.7$ are classified as changing spectral shape, corresponding to a confidence level of 5$\sigma$.

\subsection{Spectral Modelling}
\label{sec:methods_spectralmodels}
Following the spectral modelling outlined by \cite{Callingham_2017}, the flux density measurements for each source for both years were fit with two different models using the non-linear least squares python \textsc{scipy} module, \verb|curve_fit| \citep{scipy}:
\begin{enumerate}
    \item Power-law: A model that fits the flux density distribution of sources whose emission is primarily non-thermal:
    
    \begin{equation}
\label{eq:plaw}
    S_\nu = a\nu^{\alpha},
\end{equation}

    where $S_{\nu}$ is the flux density at  frequency $\nu$ and $a$ is the amplitude of the spectrum.
    This model was fit to two datasets: (1) the 16-band MWA data (100--231\,MHz) to measure the spectral index for the low frequency section of the SED, $\alpha_{\mathrm{low}}$, and; (2)  two MWA flux density measurements (189 and 212\,MHz) and the flux density of the SUMSS and/or NVSS counterpart to measure the spectral index for the high frequency section of the SED, $\alpha_{\mathrm{high}}$.
    \item Quadratic: A non-physical model to detect any spectral curvature within the MWA band. This is only fit to the 16-band MWA data (100--231\,MHz):
    \begin{equation}
\label{eq:quadratic}
    S_\nu = a\nu^{\alpha}\exp^{q \left( \ln\nu \right)^{2}},
\end{equation}
\end{enumerate}
    where $q$ represents the spectral curvature and the other parameters are as defined in Equation~\ref{eq:plaw}.

It is worth noting, when classifying PSS using observations at different frequencies taken at different times, it is possible to over-estimate or under-estimate the spectral indices due to
variation in fluxes (e.g. due to scintillation or Doppler boosting at higher frequencies, i.e. the peak may seem more pronounced when combining different epochs of observation). Consequently, the best way to detect and classify PSS is with simultaneous megahertz and gigahertz-frequency spectral coverage of the SED with monitoring of a least a year to determine if the source maintains its PSS classification. In common with all other searches in the literature so far, our PSS classifications rely on measurements taken over different epochs (i.e. SUMSS/NVSS), implying there will be false positives in the population.

\section{Results}
\label{sec:results}
Firstly, we compare the distributions of $\alpha_{\mathrm{low}}$, $\alpha_{\mathrm{high}}$ and $q$ with those presented by \citet{Callingham_2017}. In both the Year 1 and 2 data, the majority of sources lie close to a $\alpha_{\mathrm{high}} = \alpha_{\mathrm{low}}$ line, consistent with no change of spectral index across the full frequency. The median and standard deviation for $\alpha_{\mathrm{low}}$ and $\alpha_{\mathrm{high}}$ are $-0.82 \pm 0.28$ and $-0.76 \pm 0.24$ in Year 1, and $-0.81 \pm 0.28$ and $-0.76 \pm 0.22$ in Year 2.

The curvature parameter distributions for Year 1 and Year 2 are both consistent with those found by \cite{Callingham_2017} with the median curvature, $q$, and standard deviation for Year 1 and Year 2 found to be $-0.12 \pm 0.50$ and $-0.12 \pm 0.41$ respectively. We compared distributions for $\alpha_{\mathrm{low}}$, $\alpha_{\mathrm{high}}$ and $q$ with those presented by \citet{Callingham_2017} and find no significant differences. 

From the 21,558~sources of the master population, we have identified 323~sources that have VIP$\geq58.3$ and classified them as variable. The majority of sources in this variable population show no significant change in spectral shape between the two MWA epochs, with 272~sources ($\sim$84\,per\,cent) showing a uniform spectral change across the observed bandwidth. The other 51~sources ($\sim$16\,per\,cent) are classified as changing spectral shape, since their spectral shapes change significantly from Year~1 to Year~2 as their MOSS parameter was $\geq36.7$.  

Of the variable population, 91~sources ($\sim28\%$) were identified as PSS by \citet{Callingham_2017}. We also classified sources in the master population as PSS according to the same criteria outlined by \cite{Callingham_2017}, and find 123~sources identified as PSS in either year in the variable population ($\sim$38\,per\,cent). We find 83~PSS in the variable population maintain a PSS classification in both years (67\,per\,cent of sources identified as PSS in either year). 

One variable source, GLEAM\,J043715--471506, shows extreme variability (with a VIP of 9,124); see Figure~\ref{app:fig:pg11} in Appendix~\ref{sec:app:fig_seds} for the SED. GLEAM\,J043715--471506 is a known pulsar, PSR\,J0437--47. Other known pulsars in the field are not included in the master population as they fail to meet the brightness cut, see step 2 in Table~\ref{tab:QualityCuts}.

We identify six sources -- GLEAM\,J001942--303118, GLEAM\,J010626--271803, GLEAM\,J033112--430208, GLEAM\,J033412--400823, GLEAM\,J041636--185102, GLEAM\,J215155-302751 -- which were classified as potential restarted galaxies by \citet{Callingham_2017} due to their ``upturned'' SEDs, each with a VIP$\geq64$. We conclude that these sources were misclassified and are likely not restarted galaxies but variable quasars with flat spectra. Of the 25 sources identified as ``upturned'' SEDs by \citet{Callingham_2017}, 19 sources (76\,per\,cent) are not identified as variable. 

Mid infra-red colour selection techniques using Wide-field Infrared Survey Explorer \citep[\textit{WISE,}][]{2010AJ....140.1868W} are widely used to distinguish between AGN and star-forming galaxies \citep[e.g.,][]{lacy+04, stern+05, wu+12, assef+13}. We compare our variable population with the \textit{WISE} infra-red colours and find all variable sources are consistent with AGN/quasar classifications, as expected based on the flux density limit of our sample. Furthermore, we searched for any trend in VIP or MOSS with Galactic latitude and find none. 

\subsubsection{Variable Persistent PSS Population}
\label{sec:results_VIPPSS}
Of the persistent PSS in the variable population, 11~sources ($\sim$13\%) are classified as showing a change in spectral shape between epochs according to the MOSS parameter, while still maintaining a PSS classification. For example, some persistent PSS may have a positive spectral index, $\alpha_{\mathrm{low}}$, in each year but the steepness changes significantly according to the MOSS parameter. The other 72~sources are classified as showing a uniform change across the MWA bandwidth. 

There are four sources that are classified as variable persistent PSS but are not classified as PSS by \cite{Callingham_2017}. The inconsistency in classification is likely due to the lack of the lowest band in this study, which prevented a robust fit in the optically thick region of the SED, making their peaked-spectrum classification less certain. These four sources are GLEAM\,J020903--495243, GLEAM\,J041913--142024, GLEAM\,J042140--152734 and GLEAM\,J234625--073042. 

\subsubsection{Changing Spectral Shape}
\label{sec:results_changingshape}
We detect 51 variable sources that have a MOSS parameter $>36.7$, and are thus classified as showing a significant change in their spectral shape between epochs. Of these changing spectral shape sources, 18 (20\,per\,cent) were selected as PSS by \cite{Callingham_2017}. The significant change in spectral shape suggests even if these sources maintain PSS classification in Year 1 and 2, it is possible this is only temporary and they will lose their peaked-spectrum classification over time. 

A change in spectral shape may allow more accurate identification of the astrophysics of sources. For instance, GLEAM\,J033023--074052 was classified peaked-spectrum by \citet{Callingham_2017} but is classified as changing spectral shape between epochs. VLBI observations of GLEAM\,J033023--074052 found it was unresolved on milliarcssecond-scales, and constrained its projected linear size to be $<45$\,pc \citep{keim2019extragalactic}. After reexamination of the optical spectral for GLEAM\,J033023--074052, we find the MgII line was a misidentified Lyman-$\alpha$ line \citep{skymapper_dr1,skymapper_dr2}\footnote{The SkyMapper ID is 21523027 and details of the object can be found here: \url{http://skymapper.anu.edu.au/object-viewer/dr3/21523027/##} and the spectra can be analysed here: \url{http://skymapper.anu.edu.au/static/sm_asvo/marz/index.html##/detailed}}, thus the redshift is likely 2.85 rather than 0.67 as previously reported by the 6df Galaxy Survey \citep{jones20096df} and used by \cite{keim2019extragalactic}. Using the updated redshift, we recalculated the upper limit for the projected linear size of J033023--074052, and the limit increases to $<50$\,pc\footnote{The small change in limit is due to the two redshifts having nearly identical angular diameter distances in the $\Lambda$-CDM cosmology.}. 

\begin{figure}
    \centering
    \includegraphics[width=\linewidth]{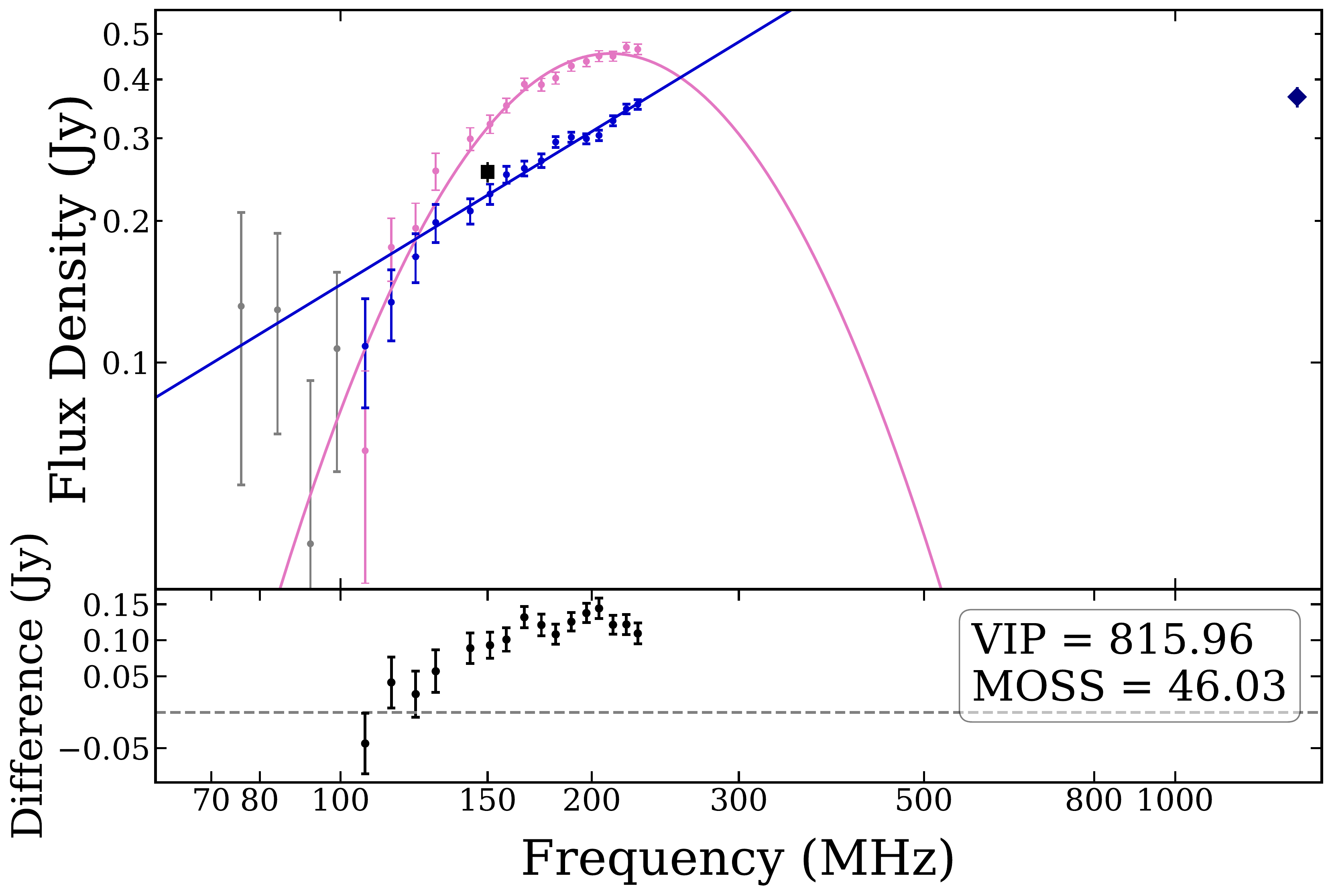}
    \caption{GLEAM\,J033023--074052, classified as changing spectral shape in this paper according to the MOSS. This source has also been found to be unresolved with VLBI with an upper limit on the projected linear size of 50\,pc \citep[from the new redshift presented here and the angular size reported in][]{keim2019extragalactic}. The points represent the following data: GLEAM low frequency (72--100\,MHz) (grey circles), Year 1 (pink circles), Year 2 (blue circles), TGSS (black square), and NVSS (navy diamond). 
    In Year 1 the spectrum is classified with a peak within the observed band and was modelled by a quadratic according to Equation~\ref{eq:quadratic}, and the Year 2 spectrum was modelled by a power-law according to Equation~\ref{eq:plaw}.}
    \label{fig:J033023}
\end{figure}

The compact linear size and changing spectral shape over a year-long period is consistent with the jets from the AGN being oriented towards the observer (i.e. a blazar), as opposed to a small double source that would be expected for a young AGN. The radio SED for GLEAM\,J033023--074052 is shown in Figure \ref{fig:J033023}. 

\subsection{AT20G Counterparts}
\label{sec:results_at20g}
\citet{at20g} present a blind search for radio sources using the Australia Telescope Compact Array (ATCA) at 20\,GHz. At 20\,GHz, the brightness of radio galaxies is more likely to be core-dominated. We cross-matched our master population with the AT20G survey to identify sources that are dominated by their core flux density and/or are more likely to be blazars. Of the 1020~sources in our master population that have a counterpart in AT20G, 116~sources are classified as variable (11\,per\,cent). 

This result contrasts with just $\sim$1.6\% of the total master population being variable and therefore supports the idea that core-dominated radio sources are more likely to be variable. Furthermore, of the variable sources with AT20G counterparts, we identify 24~sources (21\,per\,cent) as changing spectral shape, larger than the 16\,per\,cent of the total variable population showing significant change in spectral shape. 

We note one source in particular, GLEAM\,J032237--482010, has a significant change in spectral shape (MOSS$\approx$90) yet has no AT20G counterpart. GLEAM\,J032237--482010 does have a flux density of $\sim0.5$\,Jy at 840\,MHz according to SUMSS. If GLEAM\,J032237--482010 is a quasar with a flat spectrum around 0.5\,Jy or a blazar with a temporary peak in the SED, the core should be bright enough for an AT20G detection. The non-detection of AT20G suggests this source is not core dominated, contradictory to what the variability suggests. The SED for GLEAM\,J032237--482010 is presented in Figure~\ref{fig:blazars}. 

\subsection{Known blazars in the variable population}
\label{sec:results_blazarxmatch}
\citet{Massaro2015} combined optical spectra and absorption lines with the radio spectra to identify blazars, which they present in the Roma-BZCAT catalogue. We cross-match our master population with Roma-BZCAT and find 295~sources with a BZCAT counterpart. Of these sources, 64 are classified as variable by the VIP (22\,per\,cent), 18 of which (28\,per\,cent) are classified as changing spectral shape according to the MOSS parameter. This is a larger proportion of sources classified as changing spectral shape compared to the total variable population of which the changing spectral shape sources make up 16\,per\,cent.

\subsection{Comparison to literature 150\,MHz variability studies}
\label{sec:results_150var}
\subsubsection{MWA Transients Survey}
\label{sec:results_MWATS}
The MWA Transients Survey \citep[MWATS;]{mwats} was a blind search for variable sources at 150\,MHz over 3--4 years. We compare our variable population with the sources identified by the single frequency variability identified by MWATS and find no overlap. Of the 15 variable sources identified by MWATS, seven were in our field and no source had a VIP$\geq$23. 

By considering the light curves presented by \citet{mwats} for the seven sources in our field, it appears MWATS is more sensitive to changes in flux density over 3--4\,years, while this work considers only two epochs one year apart. The different parameter spaces each survey explores suggest different astrophysical explanations are driving the observed variability. MWATS predominantly attributes their observed variability to RISS. We explore the viability of RISS as the cause for the observed variability of this survey in Section~\ref{sec:discussion_extrinsic_processes}.

\subsubsection{Interplanetary Scintillation with the MWA}
\label{sec:results_IPSvar}
We performed a cross-match of the master population with the catalogue of sources displaying IPS from \cite{ipsII} and found 1,873~sources in our field, only 40 of which had a VIP$\geq58.3$. Of the variable sources presented in the IPS catalogue, only four are reported as non-scintillating while 28 are reported as highly scintillating with a normalised scintillation index above 0.9. We note, \citet{ipsII} report 12\,per\,cent of their total population were strongly scintillating due to IPS while 90\,per\,cent of the variable sources in the field are at least moderately scintillating. 

\citet{ipsII} identify 37 compact sources according to IPS and search for variability at 150\,MHz within this sample.
The one source identified as variable, GLEAM\,J013243--165444, is also in our master population. We also classify GLEAM\,J013243--165444 as variable source with significant change in spectral shape with a VIP of $\sim122$ and a MOSS parameter of $\sim85$. Despite its blazar identification and variability in both surveys, GLEAM\,J013243--165444 maintains a peaked spectrum in both years of GLEAM observations with a peak at $\sim$150\,MHz. There are several sources within the changing spectral shape population that temporarily display a peaked-spectrum classification. Such temporary spectral peaks have also been identified by \citet{torniainen2005long}. It is thus possible that the PSS classification of GLEAM\,J013243--165444 is also temporary. The SED for GLEAM\,J013243--165444 is presented in Figure~\ref{app:fig:pg4} in Appendix~\ref{sec:app:fig_seds}. 

\subsection{GMRT Search for Transients}
\label{sec:results_GMRTvar}
The VIP is calculated using only two epochs but takes advantage of the 16 individual flux density measurements. \cite{hajela2019gmrt} present a statistical method which compares two epochs on over four different timescales (4\,hours, 1\,day, 1\,month and 4\,years) but only uses one flux density measurement for each epoch. We apply the methodology presented by \cite{hajela2019gmrt} to our master population to probe variability over one year at a single frequency, and compare this to the VIP. Such a comparison helps us put our methodology of identifying variability in the context of the literature.

\begin{figure}
    \centering
    \includegraphics[width=\linewidth]{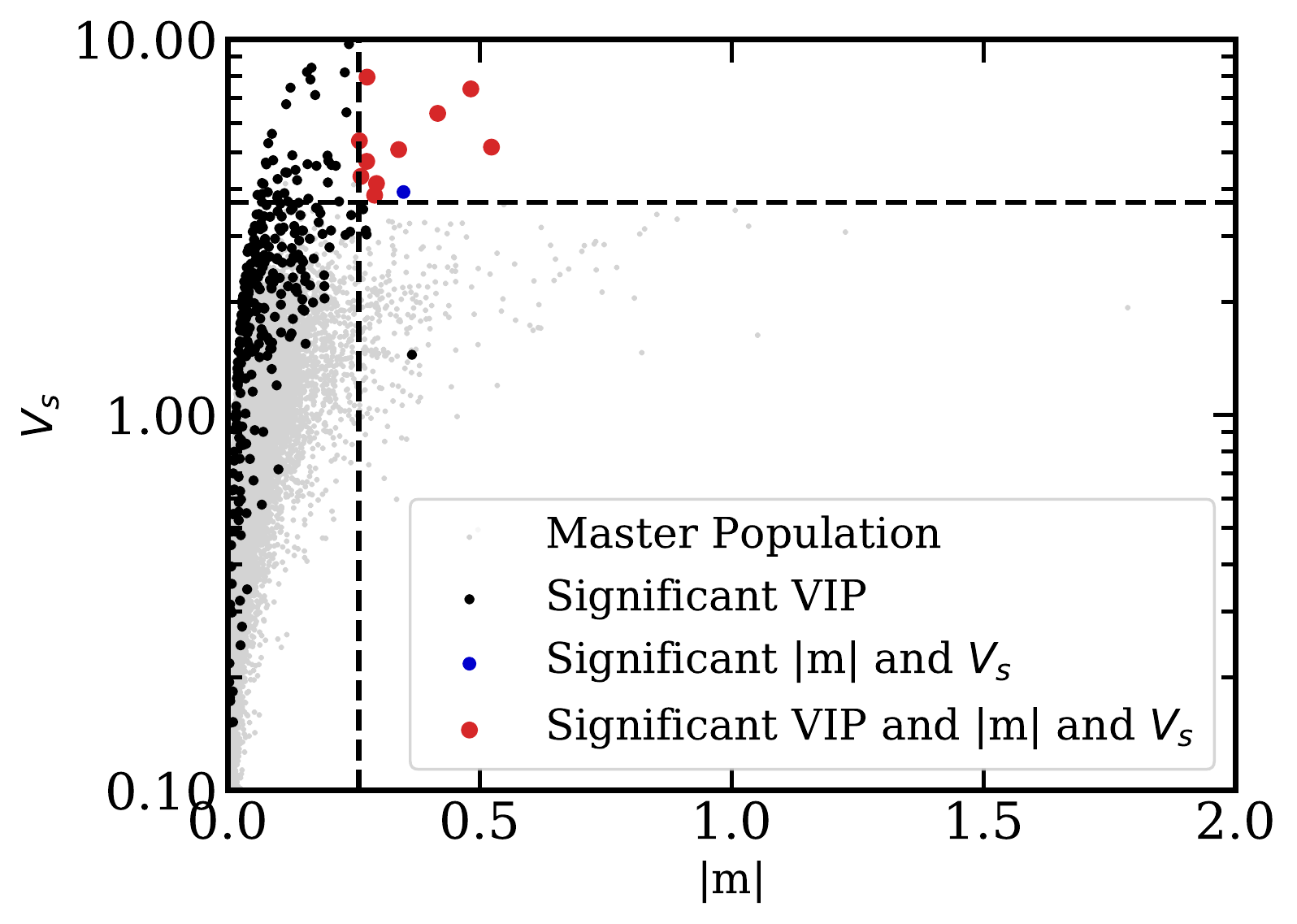}
    \caption{The distribution of the variability statistics defined by \citet{hajela2019gmrt} for our master sample; the y-axis shows $V_s$ (Equation~\ref{eq:var_stat}), as a function of the modulus of the modulation index, $m$ at 150\,MHz (Equation~\ref{eq:mod_index_GMRT}). The different populations are listed in the legend. Many sources that are variable according to the VIP are missed by the $V_s$ and $m$. Using single-frequency variability statistics seems to cause low completeness.}
    \label{fig:150MHz_variability}
\end{figure}

We calculate their variability statistic, $V_s$, at 150\,MHz: 
\begin{equation}
    V_s = \frac{\left( S_1 - S_2  \right)}{\sqrt{\sigma_1^2 +\sigma_2^2}},
    \label{eq:var_stat}
\end{equation}
and the modulation index, $m$:
\begin{equation}
    m=2\times\frac{S_1-S_2}{S_1+S_2},
    \label{eq:mod_index_GMRT}
\end{equation}
where $S_1$ and $S_2$ are the flux densities in the first and second epoch, respectively. $\sigma_1$ and $\sigma_2$ are the uncertainties on the measurements. \cite{hajela2019gmrt} state a source is \textit{truly variable} if the $V_s$ is more than four times the standard deviation of $V_s$ and $|m|>0.26$. Using this classification of variability on our MWA master sample at the 150\,MHz, we find 13 sources which would be what \cite{hajela2019gmrt} define as \textit{truly variable}. Of these 13 sources, 12 are selected by the VIP as showing significant variability as shown in Figure~\ref{fig:150MHz_variability}, all of which have a VIP$\geq$85. Additionally, of these 13 sources, only two are classified as having a change in shape, and 11 are classified as having a uniform change across the band.  The VIP takes full advantage of the multiple flux density measurements and is thus more robust to single frequency random fluctuations. The $V_s$ and $m$ presented by \cite{hajela2019gmrt} has a high reliability, but a low completeness, missing a large fraction of variable sources at low frequencies, shown by black dots in Figure~\ref{fig:150MHz_variability}. 

Separately, we cross-match our variable population with that presented by \cite{hajela2019gmrt} and find only one common source. GLEAM\,J012528--000557 \citep[referred to as J012528+000505 in][]{hajela2019gmrt}, is found to be variable in this work and by \cite{hajela2019gmrt}. This source is a known blazar (Section~\ref{sec:results_blazarxmatch}) and thus variability at these frequencies on timescales of years is not unexpected.

\section{Discussion}
\label{sec:discussion}
As we have classified the observed variability according to the \textit{type} of variability observed, and compared it to other low-frequency variability studies, we now discuss the potential physical mechanisms that could drive each variability classification.

\subsection{Extrinsic Variability}
\label{sec:discussion_extrinsic_processes}
Scintillation can cause radio sources to vary in brightness over several different timescales depending on the scattering regime.
The mosaics used in this study are composed of multiple 2 minute snapshots, and IPS and ionosphere scintillation timescales are short enough that the variations will be smoothed over in the mosaics.  Likewise the high Galactic latitude of our survey area places the transition frequency from weak scattering to the strong scattering at $\gtrsim1$\,GHz. Therefore, variability identified in our survey is probing the strong scattering regime \citep{1998MNRAS.294..307W}. 

Refractive interstellar scintillation (RISS) in the strong regime can cause variability at megahertz frequencies on year-long timescales. In comparison, diffractive interstellar scintillation (DISS) occurs on shorter timescales (seconds to minutes) and with a larger amplitude of modulation \citep{narayan1992physics}. Furthermore, DISS requires much smaller limits on source size than RISS, so more strongly influences light from extremely compact sources such as pulsars and fast radio bursts, and is a narrow band effect \citep[with the fractional decorrelation bandwidth $\ll1$][]{narayan1992physics}. We thus attribute the observed variability of the known pulsar GLEAM\,J043715--471506 (PSR\,J0437-47) to DISS, further supported by the irregularity of the SED suggesting significant frequency dependence on the modulation within the MWA bandwidth. We focus on RISS in the following sections when considering scintillation as the cause for the observed variability. 

Extended sources can still scintillate if they have point-like components embedded within the extended structure, such as hotspots in the lobe of a radio galaxy. However, for sources with an angular size far larger than the scintillation angle, and with no such compact features, the combined modulation of the smaller regions averages to a negligible total modulation. Thus, if we assume the source is point-like and find the spectral variability consistent with scintillation, it is also consistent with a point-like structure embedded within an extended source. However, for an extended structure, the point-like region is a fraction $f$ of the total flux density. The point-like region will still scintillate while the extended structure will not. Consequently, for extended sources, we measure the modulation index reduced by a factor $1-f$, \citep{hancock2019refractive}. Hence, the compact region needs to dominate the emission from the lobes of a radio galaxy for scintillation to produce the observed variability. Furthermore, if the scintillating component is larger than the Fresnel angle, the scintillation timescale increases \citep{narayan1992physics}. 

We also consider the possibility that extreme scattering events (ESEs) could cause some of the observed variability \citep[e.g.][]{Bannister2016ESEs}. While the features of some of the observed variability are consistent with an ESE, current confirmed detections of ESEs suggest they are rare events. \citet{Lazio_2001} only report finding 15~events in a survey of almost 150~sources monitored roughly once every two~days for up to 15~years). We thus discount ESEs as a likely explanation for the variability observed, but suggest a third epoch of observations on our sample is required to test the validity of this assertion.

We outline below the feasibility of RISS as the mechanism behind the observed variability for each class of variable source we observe. We note, however, that we have made several necessary assumptions regarding scintillation that may not be valid for all of the sources. For example, it possible the transition frequency for weak scattering by refractive scintillation may be much lower than expected. Consequently, this survey may be probing a transition space where many assumptions are no longer viable. However, given this study is at low (megahertz) frequencies and probing timescales of around a year, it is reasonable to assume we are well within the strong scattering regime since the transition frequency is $\gtrsim1$\,GHz.

\subsubsection{Non-PSS Uniform Change Population}
\label{sec:discussion_extrinsic_nonPPSS}
While scintillation strength is dependent on wavelength, RISS is a broadband effect at megahertz frequencies \citep[with the fractional decorrelation bandwidth $\sim1$;][]{narayan1992physics}. Thus, we expect any variability due to RISS to be approximately uniform across the $\sim$100--230\,MHz bandwidth of this study \citep{narayan1992physics,hancock2019refractive}.

Variable sources that are classified as having a uniform spectral change may have bright embedded compact features that can scintillate within their extended radio lobes, including but not limited to knots, jets, and hot spots. For such sources where the compact scintillating feature is embedded, the timescales are longer and the amount of modulation decreases as the observed scintillation is a combined average of individual regions scintillating. Sources with spectra which change uniformly can be interpreted as being compact (or their brightness being dominated by compact components) and undergoing RISS. Using estimates of the RISS of compact sources based on distribution of H$\alpha$ within the Galaxy \citep{hancock2019refractive} we confirm the observed variability of sources showing a uniform change can be explained by scintillation. Scintillation on year long timescales at megahertz frequencies requires a compact component or hot spot of size $\lesssim$5\,milliarcseconds, assuming a (Galactic) scattering screen at $D=10$\,kpc and Kolmogorov turbulence \citep{narayan1992physics,1998MNRAS.294..307W}. The preponderance of our detected variable sources having a AT20G counterpart implies that many of our sources likely have small, compact features in their morphology.

To provide confirmation of the possibility of RISS as the cause of the observed variability, we propose a long-term monitoring of these sources at centimetre wavelengths (where interstellar scintillation is negligible for sources with an angular size greater than tens of microarcseconds). Likewise, VLBI observations to obtain high resolution morphologies of these sources could search for a compact features small enough to scintillate. VLBI is performed at $\sim$gigahertz frequencies where different features may contribute more to the integrated flux density than at megahertz frequencies. Assumptions of how the morphology changes with respect to frequency would be necessary in order to estimate the morphology at megahertz frequencies.  Alternatively, IPS at megahertz frequencies could be a way to confirm the presence of a compact component without the need for VLBI. 

The morphology of the sources at different frequencies also significantly impacts the modulation due to scintillation. Megahertz frequencies are more dominated by older, likely large structures, such as radio lobes. These structures have a much larger angular size than the core or jets of an AGN. A knot or hot-spot in the lobe may not contribute as much to the overall flux density at megahertz frequencies as it does at gigahertz frequencies. Thus at low frequencies only a fraction of the flux density may be compact enough to scintillate while at higher frequencies a larger proportion (if not all) may scintillate. Long-term, multi-epoch monitoring of the entire SED paired with high resolution maps of the morphology (ideally at several frequencies) would be needed to confirm RISS as the mechanism behind the observed variability in this study. 
We have begun such a monitoring campaign combining roughly simultaneous observations (within a week) with the MWA and the ATCA with multiple epochs over a year long timescale\footnote{ATCA project code, C3333 and MWA project codes are D0025 and G0067}.

\subsubsection{Persistent PSS Uniform Change Population}
\label{sec:discussion_extrinsic_PPSS}
As PSS are a subset of AGN, all possible explanations of variability for the non-PSS uniform change population, outlined in Section~\ref{sec:discussion_extrinsic_nonPPSS}, are applicable to the persistent PSS uniform change population. However, we note that hot spots do not appear to be a dominant feature of PSS at gigahertz-frequencies \citep{keim2019extragalactic}. Sources larger than the source size limit of $\sim$5\,mas\footnote{at $z=0.5$, 5\,mas is equivalent to 30\,pc and at $z\approx1$, 5\,mas is equivalent to 40\,pc} can scintillate but have a reduced modulation index, and increased timescale of variability. If all the observed variability for the persistent PSS uniform change population is due to scintillation, at least 6\% of PSS may be dominated by core-jet structures, or have a compact component in their morphology. 
If confirmed, the variability due to scintillation can help provide milliarcsecond resolution of structures of persistent PSS at redshifts $\gtrsim0.5$ without the requirement of high resolution imaging using VLBI.  

In order to confirm scintillation as the mechanism behind the detected variability of persistent PSS, similar follow-up campaigns for the non-PSS uniform change sources are recommended. SED coverage would need to be simultaneous (within a few days) to ensure accurate estimation of the spectral peak and spectral indices.  

\subsubsection{Changing Spectral Shape Population}
\label{sec:discussion_extrinsic_changingshape}
As mentioned previously, both a point-like source and an extended source with an embedded compact feature can scintillate. In this section, we calculate the feasibility of RISS as the driving mechanism behind the observed variability for sources with a changing spectral shape. 

According to \citet{1998MNRAS.294..307W} the modulation scales with $\left(\frac{\nu}{\nu_0}\right)^{\frac{17}{30}}$, where $\nu$ is the observed frequency and the transition frequency is denoted by $\nu_0$. We note $\nu_0\gtrsim1$\,GHz, thus placing this survey well into the strong regime and RISS is a broadband effect at the frequencies probed by our survey \citep{narayan1992physics,1998MNRAS.294..307W}. At megahertz frequencies and high galactic latitudes, the expected modulation is fairly constant across the 100-230\,MHz band. Furthermore, if the source is scintillating, while scintillation strength does scale with frequency, the difference in the observed effect across 130\,MHz bandwidth is negligible according to the decoherence bandwidth for RISS ($\delta\nu_{\mathrm{dc}}\approx1$). Therefore, scintillation cannot be the sole explanation for those sources that show significant changes in their spectral shape.

\subsection{Intrinsic Variability}
\label{sec:discussion_intrinsic_processes}
The observed variability is consistent with several mechanisms of intrinsic evolutions or changes. Causes of intrinsic variability include evolution of knots in core/jet structures and changes in the immediate environment surrounding radio lobes. The megahertz frequencies of this study are probing the large scale structure and are dominated by the lobes of AGN. \cite{hardcastle2008properties} report the cores of typical AGN being four orders of magnitude fainter than the lobes at low frequencies, although it is unknown whether this holds true for compact AGN. We find it unlikely the lobes for any AGN are $\leq$1\,ly ($\sim0.3$\,pc) across, which would be necessary for light travel time to explain variability on year-long timescales. We largely focus on more plausible solutions relating to interactions within the jet/lobes and nearby surrounding absorbing media:
\begin{enumerate}
    \item Internal shocks, where shells of plasma within lobes interact and can cause a re-energisation of the electrons;
    \item Knots in the jet that are being ejected from the central core and travelling to the lobes;
    \item Changes in optical depth due to a fast-moving clumpy cloud of surrounding media;
    \item Core or jet structures being oriented towards the observer, i.e. the possibility of all sources being blazars
\end{enumerate}

As with extrinsic variability discussed in Section \ref{sec:discussion_extrinsic_processes}, we will outline the feasibility of intrinsic variability per variable class: non-PSS uniform change sources, persistent PSS, and variable spectral shape sources. 

\subsubsection{Non-PSS Uniform Change Population}
\label{sec:discussion_intrinsic_nonPPSS}
As AGN with no detectable peak in their spectra are often significantly larger than PSS, there are some intrinsic mechanisms we can eliminate as explanations for the observed variability. For example, we can rule out variability due to the evolution of a knot travelling from the core to the lobes on a year long timescale as this is unfeasible on these timescales.  

However, if the core or jet were to vary (due to changes in accretion rate, for example) we could expect to see significant changes on much shorter timescales. At low frequencies we would normally expect the lobe emission to dominate, so such variation would require either:
\begin{enumerate}
    \item The feature varying is not travelling across the entire source: this would reduce travel time to the lobes;
    \item The component of the total flux density determined by the core is greater than expected: this could result in short timescale variability from the core itself, which may be detected at low frequencies.
\end{enumerate}

In the first scenario, the timescale of variability can be reduced by introducing interaction of the plasma within the jets/lobes. This interaction creates shells of energised plasma which merge and increase the lobe brightness on much shorter timescales than energy travelling from the core \citep{Jamil_2010_iShocks}

Similarly in the second scenario, if the flux density is dominated by the core, we could detect the variability we observe via standard core or jet fuelling. This would suggest that for a small fraction of radio galaxies (and in particular PSS), current estimates of the relative brightness of the core to the lobes, ranging from $10^{-1}$--$10^{-4}$ \citep{hardcastle2008properties}, is massively underestimated at low frequencies. 

Furthermore, it is worth noting that 38 of the non-PSS uniform change sources are known blazars (of a total 175 non-PSS uniform change sources in the field). It is thus not unreasonable to see variability on short timescales from the core or jet which is oriented towards the observer providing an additional beaming effect \citep{madau1987beaming}. Any slight change of the source may be Doppler Boosted and variability on year long timescales due to the flaring or changing state of a blazar could explain the observed spectral variability of these sources. However, blazars are generally compact enough to scintillate, thus disentangling whether the observed variability is due to the intrinsic variable nature of the blazar or from scintillation in the interstellar medium is challenging. Comprehensive monitoring with high time resolution of both low frequency (radio) with high frequency (X-ray and/or Gamma) follow up is required, or VLBI observations tracing the evolution of the jet. 

\subsubsection{Persistent PSS Uniform Change Population}
\label{sec:discussion_intrinsic_PPSS}
Given PSS are a sub-population of typical AGN, all mechanisms explained in Section~\ref{sec:discussion_intrinsic_nonPPSS} are also plausible for the PSS population. But there are potentially unique intrinsic mechanisms due to the typically more compact morphologies of PSS and their interactions with the warm ISM \citep{Bicknell2018feedbackII}. 

Firstly, a subset of PSSs may have a core-jet prominence more typically associated with AGN that have a flat spectrum, even at megahertz frequencies. As such, this core variability may arise from changes in accretion rate or flaring state. If we assume an impulsive change at the core to be the cause of the variability, we would expect a uniform decrease across the entire SED of the source if the material is adiabatically expanding. In this case, the cause of the overall persistent peak in the SED remains the same. As outlined in Section \ref{sec:discussion_intrinsic_nonPPSS}, current estimates of the core prominence place core flux density at $10^{-1}$--$10^{-4}$ times fainter than the lobes at low frequencies ($\sim$150\,MHz) for typical AGN. However, measurements of the core dominance for PSS have yet to be reliably measured at megahertz-frequencies. Furthermore, for the PSS that have a compact double morphology, the core is hardly ever even detected \citep{Orienti2016properties}. Thus, the identified variable PSS in this study may have an inherently different core prominence. If a multi-epoch follow up with simultaneous observations of the broad band SED ($\sim 70$\,MHz--$\geq 10$\,GHz) showed a flare typical of a changing state in the core and/or jet, we could surmise a larger component of the flux density measured at megahertz frequencies is due to the core. Intrinsic variability due to the core may signify persistent PSS have a vastly different core prominence than their non-peaked counterparts. Only LOFAR will have the resolution in the coming decades to potentially resolve some of these sources at megahertz-frequencies.  

Secondly, similar temporary peaks in radio spectra have been observed in X-ray binary systems where ejecta from the jet has been observed \citep{Fender2009XRBreview,tetarenko2019cygnus}. Radio monitoring for Cygnus X-1 detected a lag of the radio flare from higher radio frequency to lower radio frequency  \citep[11\,GHz down to 2\,GHz][]{tetarenko2019cygnus}. Furthermore, \citet{tetarenko2019cygnus} report a decreased amplitude of modulation and increased width of the timescale of the flare at lower frequencies ($\sim 2$--3\,GHz). \citet{tetarenko2019cygnus} suggest this delay is due to the different frequencies probing further along the jet away from the core, with higher frequency observations probing younger, faster-evolving material. While black hole X-ray binary systems are orders of magnitude more compact than AGN, varying on the timescales of weeks, a similar mechanism could explain the observed variability of the persistent PSS but on year long timescales, given the typically compact ($\leq$20\,kpc) morphology of PSS. Observing the light curves at multiple radio frequencies over year long timescales for the persistent PSS may show a delay in the flare from higher frequencies to lower as the ejection from the jet travels further from the core. Furthermore, given it is not unreasonable to expect ejecta from the jets being detected on these timescales for PSS, it is also justifiable to explain the observed variability with interacting shells \citep{Jamil_2010_iShocks}, which theoretically occur on shorter timescales.

Thirdly, one potential cause of the variability below the spectral turnover could be due to changes in the free-free optical depth along the line of sight \citep[e.g.][]{tingay2015spectral}. The free-free absorption (FFA) model has been considered as the cause of absorption in the optically thick region for at least some PSS \citep{Peck_1999,Kameno2000FFA,Tremblay_2008,marr2014FFA,callingham2015,tingay2015spectral}. Attributing the variability to changes in optical depth would be evidence to support the FFA model for these sources. Recent simulations have also proposed GPS and CSS are the result of relativistic jet feedback and interactions with the surrounding warm ISM \citep{10.1093/mnras/sty070}. Several observations of absorption features linked with dust surrounding AGN have also found a strong connection with PSS \citep{Grasha_2019,10.1093/mnras/stz2452,10.1093/mnras/stz556}. However, as several studies have noted \citep{o1998compact,Callingham_2017, Bicknell2018feedbackII}, distinguishing between synchrotron self absorption (SSA) and FFA for the PSS population as a whole has thus far yielded inconclusive results as both SSA and FFA spectral models are consistent with current observations for most PSSs. Testing for prominent HI gas via would help identify galaxies that likely have a lot of intervening media between us and the radio lobes \citep{callingham2015}.

We note it is entirely possible that the majority of the persistent PSS population may be entirely composed of flaring blazars that are observed at both epochs with a peaked spectrum. While only 25\,per\,cent of the persistent PSS population were known blazars, as selected based on X-ray, optical and gigahertz-frequency characteristics, it is possible other members of the persistent PSS population happen to have a jet orientation relative to our line-of-sight that ensures it is not X-ray- or optically-bright relative to its radio luminosity. Blazars are also known to display a peaked spectrum over a prolonged period, even at gigahertz-frequencies \citep{ipsII}. For example, we find known blazar GLEAM\,J013243--165444 to be classified as a PSS in both years of observation. We thus suggest long-term observations of the light curves of the persistent PSS population at either gigahertz frequencies or, if the sources are bright enough, follow up with $\gamma$- or X--ray telescopes, like the extended ROentgen Survey with an Imaging Telescope Array \citep[eROSITA;][]{erosita}. Furthermore, follow up with multi-frequency VLBI would be critical to determine if these sources are core-jet or double-lobed objects.

\subsubsection{Changing Spectral Shape Population}
\label{sec:discussion_intrinsic_changingshape}
For intrinsic variability of a source emitting synchrotron radiation, the brightness temperature ($T_b$) has an upper limit of $10^{12}$\,K, as determined by the inverse-Compton losses. We reproduce the calculation used by \cite{mwats} to estimate if changes in the flux density are within expectations for intrinsic variability using the following equation: 
\begin{equation}
\label{eq:brightness_temp}
    \Delta S \leq \frac{2k_B \nu^2 \tau^2 T_b b^3}{D^2},
\end{equation}
where $\Delta S$ is the change in flux density, $\tau$ is the timescale (here 1\,yr), $T_b$ is the brightness temperature (set as the upper limit, $10^{12}$\,K), $b^3$ is a beaming factor for emission directed towards the observer (we assumed an upper limit of 10, a typical value for blazars \citep{Lahteenmaki_1999}), and $D$ is the distance to the source (set as 10\,Gpc, equivalent to roughly $z=1$). Using Equation \ref{eq:brightness_temp} we find that sources can not exceed 0.68\,Jy for $\Delta S$ at 150\,MHz when no beaming factor is used. By introducing the beaming factor for incoherent emission, all the observed variability for these sources can be explained via intrinsic mechanisms. We thus suggest that sources where we observe a variable spectral shape between epochs as beamed AGN. 24 of the 51 (47\,per\,cent) changing spectral shape sources have AT20G counterparts \citep{at20g}, suggesting core-dominance. Additionally, 18 sources (35\,per\,cent) are known blazars already from BZCAT \citep{Massaro2015}.

However, it is worth noting that if these sources are indeed heavily beamed, their morphology will likely be compact enough that they may also scintillate. Therefore, it is possible the observed variability is a combination of both intrinsic blazar flares and scintillation. Finally, we note that we assumed a timescale of one year due to the rough timescale of this work in searching for variability but it is likely these sources have variability that will occur on longer timescale than observed. Increasing the timescale of variability increases the $\Delta S$ limit, according to Equation~\ref{eq:brightness_temp}, for these sources to be intrinsically variable. Thus, further long-term monitoring over several years with wide spectral coverage is necessary to estimate the true timescale more accurately as well as determine the role of scintillation. 

If confirmed, spectral shape variability at low frequencies ($<1$\,GHz) has the ability to detect and classify blazars on relatively short timescales, even if they are too faint to observe at higher frequencies. 

\subsection{Summary of plausible causes of the observed variability}
\label{sec:discussion_overview}
Considering causes for intrinsic or extrinsic variability, we determined the most likely causes for the observed spectral variability for the different classes of sources. For non-PSS and PSS sources that show a uniform change in their SED between epochs, scintillation can easily explain the observed low-frequency variability. If confirmed, observing the variability due to scintillation could be paired with observations of pulsars to increase the resolution of maps of the electron column density to test models that aim to predict scintillation, such as \textsc{RISS19}\footnote{\textsc{RISS19} can be downloaded here: \url{https://github.com/PaulHancock/RISS19}}. In contrast, the sources in which we observe a changing spectral shape between epochs are more likely explained as a blazar with changing flaring states. While scintillation may be a component of the observed variability it is unlikely the sole mechanism to explain the spectral shape variability due to the wide bandwidth of RISS. We therefore present the changing spectral shape variable population as blazar candidates requiring a follow-up campaign for confirmation, for example a search for X-ray counterparts with \textit{eROSITA} \citep{erosita}.

\section{Conclusions and Outlook}
\label{sec:conclusion}
We have conducted a study of low-frequency spectral variability and devised a methodology for detecting, measuring, and classifying spectral variability using the variability index parameter (VIP) and the measure of spectral shape (MOSS) parameter. This study uses two epochs of the GLEAM survey, producing a data set that contains over 21,000~sources. Therefore, our study represents the largest survey of low-frequency spectral variability, particularly of spectra formed from contemporaneous flux density measurements over a large fractional bandwidth. 

We present 323~sources that show significant spectral variability according to the VIP, 51 of which display a significant change in shape of their spectra according to the MOSS parameter. We find that the variable sources are more likely to have a peaked spectrum, consistent with results from the MWATS and IPS surveys with the MWA. We compare the variable population with the {\it WISE} infrared survey to determine the classification of galaxy and find no variable sources to be classified as star forming. Furthermore, we conclude many of the variable sources are consistent with quasar and blazar classifications. We also compare the variable population with BZCAT to find known blazars. There is a larger proportion of variable sources that are known blazars (22\,per\,cent) compared to the master population ($\sim$1\,per\,cent), with many of the remaining variable population possessing characteristics similar to the identified blazars. Likewise, we find a larger proportion of variable sources with AT20G counterparts (11\,per\,cent) when compared to the master population (1\,per\,cent). One source in particular, GLEAM\,J\,032237--482010, shows a significant change in spectral shape, suggesting it is a core-dominated source or a blazar, yet has no AT20G counterpart. 

We discuss several sources that have particular interesting features. For reported restarted radio galaxy candidates, we find that six are variable and conclude they are misclassified quasars with a flat spectra. We compare our variable sources with several notable single frequency variability surveys conducted around 150\,MHz and only find two sources which are classified as variable in each, both of which are known blazars: GLEAM\,J012528--000557 and GLEAM\,J013243-165444. We also find one known pulsar, PSR\,J0437--47, to show significant variability and conclude this is likely due to diffractive interstellar scintillation.

We argue that the observed variability of the persistent PSS and uniform change sources are entirely consistent with refractive interstellar scintillation. The sources which show a changing spectral shape according to the MOSS parameter cannot be explained by ISS and we thus present this population as blazar candidates requiring further confirmation. 

While we suggest likely causes for each category of spectral variability, it is worth noting this is based on only two epochs of observation. Long term monitoring and specific follow up campaigns are recommended to test the presented hypotheses. Furthermore, in all cases having more epochs of observation with greater spectral coverage from megahertz to gigahertz frequencies would increase the reliability of detected variability. Thus a lower level of significance cut off for the VIP could be used to detect variability. More epochs of observation on a range of timescales could allow for the timescale of variability to be accurately estimated, this can refine viable variability mechanisms.

\subsection{SKA Era Implications} 
In the SKA era, as we gain the capability to perform large scale variability surveys with large spectral and temporal coverage, understanding how prevalent variability is at low frequencies is crucial. This paper outlines a methodology to begin dissecting this variability by detecting and classifying the low-frequency spectral variability with rigorous and reproducible statistical methods. MWATS place an upper limit prediction on the expected number of variable source at low frequencies ($\sim$150\,MHz) of 6,000 sources for a given sample of 350,000 sources. Following this trend, we would expect fewer than 400 variable sources in this study. Our results are consistent with this expectation numerically, however, the nature of spectral variability detected suggests our current understanding of low-frequency variability is not yet complete. These results highlight the insufficient understanding of the emission mechanisms at low frequencies, and AGN evolutionary scenarios. As many PSS are used for calibrators of high-frequency (gigahertz) radio telescopes, understanding the short-timescale ($\sim$years) evolution and variability of their SEDs is critical. Despite this variability being observed at low frequencies, it is important to see how this variability relates to the gigahertz regime. Furthermore, SKA\_LOW will be able to detect fainter PSS. An understanding of the current known PSS population is critical if we are to investigate the fainter population. We encourage careful monitoring of the presented sources in this paper in order to understand how they may change within the span of future surveys and studies. 

\section{Data Availability}
The data underlying this article are available in the article and in its online supplementary materia

\section*{Acknowledgements}
We acknowledge the Noongar people as the traditional owners and custodians of Wadjak boodjar, the land on which the majority of this work was completed. 
KR acknowledges a Doctoral Scholarship and an Australian Government Research Training Programme scholarship administered through Curtin University of Western Australia. JRC thanks the Nederlandse Organisatie voor Wetenschappelijk Onderzoek (NWO) for support via the Talent Programme Veni grant. NHW is supported by an Australian Research Council Future Fellowship (project number FT190100231) funded by the Australian Government. This scientific work makes use of the Murchison Radio-astronomy Observatory, operated by CSIRO. We acknowledge the Wajarri Yamatji people as the traditional owners of the Observatory site. Support for the operation of the MWA is provided by the Australian Government (NCRIS), under a contract to Curtin University administered by Astronomy Australia Limited. KR thanks Arash Bahramian, Cathryn Trott and James Ross for their statistical help and review, and Christian Wolf for their helpful discussion. We acknowledge the Pawsey Supercomputing Centre, which is supported by the Western Australian and Australian Governments. This publication makes use of data products from the Wide-field Infrared Survey Explorer, which is a joint project of the University of California, Los Angeles, and the Jet Propulsion Laboratory/California Institute of Technology, funded by the National Aeronautics and Space Administration. This research made use of NASA's Astrophysics Data System, the VizieR catalog access tool, CDS, Strasbourg, France. We also make use of the \textsc{IPython} package \citep{Ipythoncite}; SciPy \citep{2020SciPy-NMeth};  \textsc{Matplotlib}, a \textsc{Python} library for publication quality graphics \citep{Hunter:2007}; \textsc{Astropy}, a community-developed core \textsc{Python} package for astronomy \citep{astropy:2013, astropy:2018}; \textsc{pandas}, a data analysis and manipulation \textsc{Python} module \citep{reback2020pandas,mckinney-proc-scipy-2010}; and \textsc{NumPy} \citep{vaderwalt_numpy_2011}. We also made extensive use of the visualisation and analysis packages DS9\footnote{\href{ds9.si.edu}{http://ds9.si.edu/site/Home.html}} and Topcat \citep{2005ASPC..347...29T}. This work was compiled in the useful online \LaTeX{} editor Overleaf.




\bibliographystyle{mnras}
\bibliography{bibliography} 



\appendix
\section{Supplementary Online Catalogue Description}
\label{sec:app:tab}

    Column numbers, names, units and description for the supplementary online catalogue.
    Source names follow International Astronomical Union naming conventions for co-ordinate-based naming.
    Background and RMS measurements were performed by \textsc{BANE};
    the fitted spectral index parameters were derived as described in Section~\ref{sec:methods_spectralmodels}; other measurements were made using \textsc{Aegean} or measurements from additional surveys, as mentioned in Section~\ref{sec:extra_surveys}. \textsc{Aegean} incorporates a constrained fitting algorithm.
    The columns with the subscript ``wide'' are derived from the 200--230\,MHz wide-band image.
    Subsequently, the subscript indicates the central frequency of the measurement, in MHz.
    These sub-band measurements are made using the priorised fitting mode of \textsc{Aegean}, where the position and shape of the source are determined from the year 2 wide-band image, and only the flux density is fitted.
\onecolumn

    \label{tab:appendix}
    \begin{longtable*}{lllp{9cm}}
    \toprule
    Number & Name & Unit & Description \\
    \midrule
    1 & GLEAM Name & hh:mm:ss+dd:mm:ss & Name of the source in the GLEAM extragalactic catalogue  \tabularnewline 
    2 & RA\_hms & hh:mm:ss & Right ascension \tabularnewline 
    3 & Dec\_dms & dd:mm:ss & Declination \tabularnewline 
    4 & RA & $^{\circ}$ & Right ascension \tabularnewline 
    5 & err\_ra & $^{\circ}$ & Error on RA \tabularnewline 
    6 & Dec & $^{\circ}$ & Declination \tabularnewline 
    7 & err\_dec & $^{\circ}$ & Error on Dec \tabularnewline 
    8 & int\_flux\_wide\_yr1 & Jy & Integrated flux density in year 1 wideband image \tabularnewline
    9 & err\_int\_flux\_wide\_yr1 & Jy & Error on integrated flux density in year 1 wideband image \tabularnewline 
    10 & local\_rms\_wide\_yr1 & Jy\,beam$^{-1}$ & Local RMS in year 1 wideband image \tabularnewline 
    11 & a\_wide\_yr1 & $\arcsec$ & Major axis of source in year 1 wideband image \tabularnewline 
    12 & b\_wide\_yr1 & $\arcsec$ & Minor axis of source in year 1 wideband image \tabularnewline 
    13 & psf\_a\_wide\_yr1 & $^{\circ}$ & Major axis of PSF at location of source in year 1 wideband image \tabularnewline 
    14 & psf\_b\_wide\_yr1 & $^{\circ}$ & Minor axis of PSF at location of source in year 1 wideband image \tabularnewline 
    15 & int\_flux\_wide\_yr2 & Jy & Integrated flux density in year 2 wideband image \tabularnewline
    16 & err\_int\_flux\_wide\_yr2 & Jy & Error on integrated flux density in year 2 wideband image \tabularnewline 
    17 & local\_rms\_wide\_yr2 & Jy\,beam$^{-1}$ & Local RMS in year 2 wideband image \tabularnewline 
    18 & a\_wide\_yr2 & $\arcsec$ & Major axis of source in year 2 wideband image \tabularnewline 
    19 & b\_wide\_yr2 & $\arcsec$ & Minor axis of source in year 2 wideband image \tabularnewline 
    20 & psf\_a\_wide\_yr2 & $^{\circ}$ & Major axis of PSF at location of source in year 2 wideband image \tabularnewline 
    21 & psf\_b\_wide\_yr2 & $^{\circ}$ & Minor axis of PSF at location of source in year 2 wideband image \tabularnewline 
    22 & local\_rms\_107\_yr1 & Jy\,beam$^{-1}$ & Year 1 local RMS at 107\,MHz \tabularnewline
    23 & S\_107\_yr1 & Jy & Year 1 integrated flux density at 107\,MHz \tabularnewline
    24 & S\_107\_err\_yr1 & Jy & Year 1 fitting error on integrated flux density at 107\,MHz \tabularnewline
    25 & local\_rms\_115\_yr1 & Jy\,beam$^{-1}$ & Year 1 local RMS at 115\,MHz \tabularnewline
    26 & S\_115\_yr1 & Jy & Year 1 integrated flux density at 115\,MHz \tabularnewline
    27 & S\_115\_err\_yr1 & Jy & Year 1 fitting error on integrated flux density at 115\,MHz \tabularnewline
    28 & local\_rms\_122\_yr1 & Jy\,beam$^{-1}$ & Year 1 local RMS at 122\,MHz \tabularnewline
    29 & S\_122\_yr1 & Jy & Year 1 integrated flux density at 122\,MHz \tabularnewline
    30 & S\_122\_err\_yr1 & Jy & Year 1 fitting error on integrated flux density at 122\,MHz \tabularnewline
    31 & local\_rms\_130\_yr1 & Jy\,beam$^{-1}$ & Year 1 local RMS at 130\,MHz \tabularnewline
    32 & S\_130\_yr1 & Jy & Year 1 integrated flux density at 130\,MHz \tabularnewline
    33 & S\_130\_err\_yr1 & Jy & Year 1 fitting error on integrated flux density at 130\,MHz \tabularnewline
    34 & local\_rms\_143\_yr1 & Jy\,beam$^{-1}$ & Year 1 local RMS at 143\,MHz \tabularnewline
    35 & S\_143\_yr1 & Jy & Year 1 integrated flux density at 143\,MHz \tabularnewline
    36 & S\_143\_err\_yr1 & Jy & Year 1 fitting error on integrated flux density at 143\,MHz \tabularnewline
    37 & local\_rms\_151\_yr1 & Jy\,beam$^{-1}$ & Year 1 local RMS at 151\,MHz \tabularnewline
    38 & S\_151\_yr1 & Jy & Year 1 integrated flux density at 151\,MHz \tabularnewline
    39 & S\_151\_err\_yr1 & Jy & Year 1 fitting error on integrated flux density at 151\,MHz \tabularnewline
    40 & local\_rms\_158\_yr1 & Jy\,beam$^{-1}$ & Year 1 local RMS at 158\,MHz \tabularnewline
    41 & S\_158\_yr1 & Jy & Year 1 integrated flux density at 158\,MHz \tabularnewline
    42 & S\_158\_err\_yr1 & Jy & Year 1 fitting error on integrated flux density at 158\,MHz \tabularnewline
    43 & local\_rms\_166\_yr1 & Jy\,beam$^{-1}$ & Year 1 local RMS at 166\,MHz \tabularnewline
    44 & S\_166\_yr1 & Jy & Year 1 integrated flux density at 166\,MHz \tabularnewline
    45 & S\_166\_err\_yr1 & Jy & Year 1 fitting error on integrated flux density at 166\,MHz \tabularnewline
    46 & local\_rms\_174\_yr1 & Jy\,beam$^{-1}$ & Year 1 local RMS at 174\,MHz \tabularnewline
    47 & S\_174\_yr1 & Jy & Year 1 integrated flux density at 174\,MHz \tabularnewline
    48 & S\_174\_err\_yr1 & Jy & Year 1 fitting error on integrated flux density at 174\,MHz \tabularnewline
    49 & local\_rms\_181\_yr1 & Jy\,beam$^{-1}$ & Year 1 local RMS at 181\,MHz \tabularnewline
    50 & S\_181\_yr1 & Jy & Year 1 integrated flux density at 181\,MHz \tabularnewline
    51 & S\_181\_err\_yr1 & Jy & Year 1 fitting error on integrated flux density at 181\,MHz \tabularnewline
    52 & local\_rms\_189\_yr1 & Jy\,beam$^{-1}$ & Year 1 local RMS at 189\,MHz \tabularnewline
    53 & S\_189\_yr1 & Jy & Year 1 integrated flux density at 189\,MHz \tabularnewline
    54 & S\_189\_err\_yr1 & Jy & Year 1 fitting error on integrated flux density at 189\,MHz \tabularnewline
    55 & local\_rms\_197\_yr1 & Jy\,beam$^{-1}$ & Year 1 local RMS at 197\,MHz \tabularnewline
    56 & S\_197\_yr1 & Jy & Year 1 integrated flux density at 197\,MHz \tabularnewline
    57 & S\_197\_err\_yr1 & Jy & Year 1 fitting error on integrated flux density at 197\,MHz \tabularnewline
    58 & local\_rms\_204\_yr1 & Jy\,beam$^{-1}$ & Year 1 local RMS at 204\,MHz \tabularnewline
    59 & S\_204\_yr1 & Jy & Year 1 integrated flux density at 204\,MHz \tabularnewline
    60 & S\_204\_err\_yr1 & Jy & Year 1 fitting error on integrated flux density at 204\,MHz \tabularnewline
    61 & local\_rms\_212\_yr1 & Jy\,beam$^{-1}$ & Year 1 local RMS at 212\,MHz \tabularnewline
    62 & S\_212\_yr1 & Jy & Year 1 integrated flux density at 212\,MHz \tabularnewline
    63 & S\_212\_err\_yr1 & Jy & Year 1 fitting error on integrated flux density at 212\,MHz \tabularnewline
    64 & local\_rms\_220\_yr1 & Jy\,beam$^{-1}$ & Year 1 local RMS at 220\,MHz \tabularnewline
    65 & S\_220\_yr1 & Jy & Year 1 integrated flux density at 220\,MHz \tabularnewline
    66 & S\_220\_err\_yr1 & Jy & Year 1 fitting error on integrated flux density at 220\,MHz \tabularnewline
    67 & local\_rms\_227\_yr1 & Jy\,beam$^{-1}$ & Year 1 local RMS at 227\,MHz \tabularnewline
    68 & S\_227\_yr1 & Jy & Year 1 integrated flux density at 227\,MHz \tabularnewline
    69 & S\_227\_err\_yr1 & Jy & Year 1 fitting error on integrated flux density at 227\,MHz \tabularnewline
    70 & local\_rms\_107\_yr2 & Jy\,beam$^{-1}$ & Year 2 local RMS at 107\,MHz \tabularnewline
    71 & S\_107\_yr2 & Jy & Year 2 integrated flux density at 107\,MHz \tabularnewline
    72 & S\_107\_err\_yr2 & Jy & Year 2 fitting error on integrated flux density at 107\,MHz \tabularnewline
    73 & local\_rms\_115\_yr2 & Jy\,beam$^{-1}$ & Year 2 local RMS at 115\,MHz \tabularnewline
    74 & S\_115\_yr2 & Jy & Year 2 integrated flux density at 115\,MHz \tabularnewline
    75 & S\_115\_err\_yr2 & Jy & Year 2 fitting error on integrated flux density at 115\,MHz \tabularnewline
    76 & local\_rms\_122\_yr2 & Jy\,beam$^{-1}$ & Year 2 local RMS at 122\,MHz \tabularnewline
    77 & S\_122\_yr2 & Jy & Year 2 integrated flux density at 122\,MHz \tabularnewline
    78 & S\_122\_err\_yr2 & Jy & Year 2 fitting error on integrated flux density at 122\,MHz \tabularnewline
    79 & local\_rms\_130\_yr2 & Jy\,beam$^{-1}$ & Year 2 local RMS at 130\,MHz \tabularnewline
    80 & S\_130\_yr2 & Jy & Year 2 integrated flux density at 130\,MHz \tabularnewline
    81 & S\_130\_err\_yr2 & Jy & Year 2 fitting error on integrated flux density at 130\,MHz \tabularnewline
    82 & local\_rms\_143\_yr2 & Jy\,beam$^{-1}$ & Year 2 local RMS at 143\,MHz \tabularnewline
    83 & S\_143\_yr2 & Jy & Year 2 integrated flux density at 143\,MHz \tabularnewline
    84 & S\_143\_err\_yr2 & Jy & Year 2 fitting error on integrated flux density at 143\,MHz \tabularnewline
    85 & local\_rms\_151\_yr2 & Jy\,beam$^{-1}$ & Year 2 local RMS at 151\,MHz \tabularnewline
    86 & S\_151\_yr2 & Jy & Year 2 integrated flux density at 151\,MHz \tabularnewline
    87 & S\_151\_err\_yr2 & Jy & Year 2 fitting error on integrated flux density at 151\,MHz \tabularnewline
    88 & local\_rms\_158\_yr2 & Jy\,beam$^{-1}$ & Year 2 local RMS at 158\,MHz \tabularnewline
    89 & S\_158\_yr2 & Jy & Year 2 integrated flux density at 158\,MHz \tabularnewline
    90 & S\_158\_err\_yr2 & Jy & Year 2 fitting error on integrated flux density at 158\,MHz \tabularnewline
    91 & local\_rms\_166\_yr2 & Jy\,beam$^{-1}$ & Year 2 local RMS at 166\,MHz \tabularnewline
    92 & S\_166\_yr2 & Jy & Year 2 integrated flux density at 166\,MHz \tabularnewline
    93 & S\_166\_err\_yr2 & Jy & Year 2 fitting error on integrated flux density at 166\,MHz \tabularnewline
    94 & local\_rms\_174\_yr2 & Jy\,beam$^{-1}$ & Year 2 local RMS at 174\,MHz \tabularnewline
    95 & S\_174\_yr2 & Jy & Year 2 integrated flux density at 174\,MHz \tabularnewline
    96 & S\_174\_err\_yr2 & Jy & Year 2 fitting error on integrated flux density at 174\,MHz \tabularnewline
    97 & local\_rms\_181\_yr2 & Jy\,beam$^{-1}$ & Year 2 local RMS at 181\,MHz \tabularnewline
    98 & S\_181\_yr2 & Jy & Year 2 integrated flux density at 181\,MHz \tabularnewline
    99 & S\_181\_err\_yr2 & Jy & Year 2 fitting error on integrated flux density at 181\,MHz \tabularnewline
    100 & local\_rms\_189\_yr2 & Jy\,beam$^{-1}$ & Year 2 local RMS at 189\,MHz \tabularnewline
    101 & S\_189\_yr2 & Jy & Year 2 integrated flux density at 189\,MHz \tabularnewline
    102 & S\_189\_err\_yr2 & Jy & Year 2 fitting error on integrated flux density at 189\,MHz \tabularnewline
    103 & local\_rms\_197\_yr2 & Jy\,beam$^{-1}$ & Year 2 local RMS at 197\,MHz \tabularnewline
    104 & S\_197\_yr2 & Jy & Year 2 integrated flux density at 197\,MHz \tabularnewline
    105 & S\_197\_err\_yr2 & Jy & Year 2 fitting error on integrated flux density at 197\,MHz \tabularnewline
    106 & local\_rms\_204\_yr2 & Jy\,beam$^{-1}$ & Year 2 local RMS at 204\,MHz \tabularnewline
    107 & S\_204\_yr2 & Jy & Year 2 integrated flux density at 204\,MHz \tabularnewline
    108 & S\_204\_err\_yr2 & Jy & Year 2 fitting error on integrated flux density at 204\,MHz \tabularnewline
    109 & local\_rms\_212\_yr2 & Jy\,beam$^{-1}$ & Year 2 local RMS at 212\,MHz \tabularnewline
    110 & S\_212\_yr2 & Jy & Year 2 integrated flux density at 212\,MHz \tabularnewline
    111 & S\_212\_err\_yr2 & Jy & Year 2 fitting error on integrated flux density at 212\,MHz \tabularnewline
    112 & local\_rms\_220\_yr2 & Jy\,beam$^{-1}$ & Year 2 local RMS at 220\,MHz \tabularnewline
    113 & S\_220\_yr2 & Jy & Year 2 integrated flux density at 220\,MHz \tabularnewline
    114 & S\_220\_err\_yr2 & Jy & Year 2 fitting error on integrated flux density at 220\,MHz \tabularnewline
    115 & local\_rms\_227\_yr2 & Jy\,beam$^{-1}$ & Year 2 local RMS at 227\,MHz \tabularnewline
    116 & S\_227\_yr2 & Jy & Year 2 integrated flux density at 227\,MHz \tabularnewline
    117 & S\_227\_err\_yr2 & Jy & Year 2 fitting error on integrated flux density at 227\,MHz \tabularnewline
    118 & S\_076 & Jy & GLEAM integrated flux density at 76\,MHz \tabularnewline
    119 & S\_076\_err & Jy & GLEAM fitting error on integrated flux density at 76\,MHz \tabularnewline
    120 & S\_084 & Jy & GLEAM integrated flux density at 84\,MHz \tabularnewline
    121 & S\_084\_err & Jy & GLEAM fitting error on integrated flux density at 84\,MHz \tabularnewline
    122 & S\_092 & Jy & GLEAM integrated flux density at 92\,MHz \tabularnewline
    123 & S\_092\_err & Jy & GLEAM fitting error on integrated flux density at 92\,MHz \tabularnewline
    124 & S\_099 & Jy & GLEAM integrated flux density at 99\,MHz \tabularnewline
    125 & S\_099\_err & Jy & GLEAM fitting error on integrated flux density at 99\,MHz \tabularnewline
    126 & S\_tgss & Jy & TGSS flux density \tabularnewline
    127 & S\_tgss\_err & Jy & TGSS flux density error \tabularnewline
    128 & S\_mrc & Jy & MRC flux density \tabularnewline
    129 & S\_sumss & Jy & SUMSS flux density \tabularnewline
    130 & S\_nvss & Jy & NVSS flux density \tabularnewline
    131 & S\_vlssr & Jy & VLSSr flux density \tabularnewline
    132 & alpha\_low\_yr1 & -- & Low-frequency spectral index derived from fitting a power law to year 1 data between 100 and 231\,MHz \tabularnewline
    133 & alpha\_low\_error\_yr1 & -- & Uncertainty in the year 1 low-frequency spectral index \tabularnewline
    134 & alpha\_high\_yr1 & -- & High-frequency spectral index derived from fitting a power law to SUMSS and/or NVSS flux density point(s) and to year 1 data at 135 and 212\,MHz \tabularnewline
    135 & alpha\_high\_error\_yr1 & -- & Uncertainty in the year 1 high-frequency spectral index \tabularnewline
    136 & quad\_curve\_yr1 & -- & Curvature parameter for year 1 \tabularnewline
    137 & quad\_curve\_error\_yr1 & -- & Uncertainty in the year 1 curvature parameter \tabularnewline
    138 & redchisq\_low\_yr1 & -- & Reduced chi squared statistic for year 1 low-frequency power law SED fit \tabularnewline
    139 & redchisq\_high\_yr1 & -- & Reduced chi squared statistic for year 1 high-frequency power law SED fit \tabularnewline
    140 & redchisq\_quad\_yr1 & -- & Reduced chi squared statistic for year 1 quadratic SED fit \tabularnewline
    141 & alpha\_low\_yr2 & -- & Low-frequency spectral index derived from fitting a power law to year 2 data between 100 and 231\,MHz \tabularnewline
    142 & alpha\_low\_error\_yr2 & -- & Uncertainty in the year 2 low-frequency spectral index \tabularnewline
    143 & alpha\_high\_yr2 & -- & High-frequency spectral index derived from fitting a power law to SUMSS and/or NVSS flux density point(s) and to year 2 data at 152 and 212\,MHz \tabularnewline
    144 & alpha\_high\_error\_yr2 & -- & Uncertainty in the year 2 high-frequency spectral index \tabularnewline
    145 & quad\_curve\_yr2 & -- & Curvature parameter for year 2 \tabularnewline
    146 & quad\_curve\_error\_yr2 & -- & Uncertainty in the year 2 curvature parameter \tabularnewline
    147 & redchisq\_low\_yr2 & -- & Reduced chi squared statistic for year 2 low-frequency power law SED fit \tabularnewline
    148 & redchisq\_high\_yr2 & -- & Reduced chi squared statistic for year 2 high-frequency power law SED fit \tabularnewline
    149 & redchisq\_quad\_yr2 & -- & Reduced chi squared statistic for year 2 quadratic SED fit \tabularnewline
    150 & VIP & -- & Variability index parameter \tabularnewline
    151 & MOSS & -- & Measure of spectral shape parameter \tabularnewline
    152 & variable\_flag & -- & Identification as part of the variable population \tabularnewline
    153 & peaked\_spectrum\_source\_flag & -- & Previously identified as PSS by \cite{Callingham_2017} \tabularnewline
    154 & persistent\_pss flag & -- & Persistent PSS \tabularnewline
    155 & MOSS Class & -- & Variability class according to the MOSS parameter \tabularnewline
    156 & pss\_class\_yr1 & -- & PSS classification in year 1 \tabularnewline
    157 & pss\_class\_yr2 & -- & PSS classification in year 2 \tabularnewline
    158 & RAJ2000\_NVSS & -- & RA of source in NVSS \tabularnewline
    159 & DEJ2000\_NVSS & -- & Dec of source in NVSS \tabularnewline
    160 & RA\_WISE & -- & RA of source in WISE \tabularnewline
    161 & Dec\_WISE & -- & Dec of source in WISE \tabularnewline
    162 & w1mpro & mag & WISE W1 counterpart magnitude \tabularnewline
    163 & w2mpro & mag & WISE W2 counterpart magnitude \tabularnewline
    164 & w3mpro & mag & WISE W3 counterpart magnitude \tabularnewline
    165 & w4mpro & mag & WISE W4 counterpart magnitude \tabularnewline
    166 & Class & -- & Spectral class according to bzcat \tabularnewline
    167 & $z$ & -- & Reported redshift from bzcat \tabularnewline
    \hline
    \label{tab:appendix_masterpop_description}
    \end{longtable*}

\section{Spectral Energy Distributions (SEDs)}
\label{sec:app:fig_seds}
    SEDs for all sources classified as variable according to the VIP presented in order of Right Ascension. 
    Models were included to assist with visual interpretation. However, models plotted for each year are dictated by their PSS classification only. A source classified with a peak within the observed MWA band, which also satisfied the PSS criteria presented by \citet{Callingham_2017}, was modelled by a quadratic according to Equation~\ref{eq:quadratic}. Remaining sources were modelled by a power-law according to Equation~\ref{eq:plaw}.

\begin{figure*}
\begin{center}$
\begin{array}{cccccc}
\includegraphics[scale=0.15]{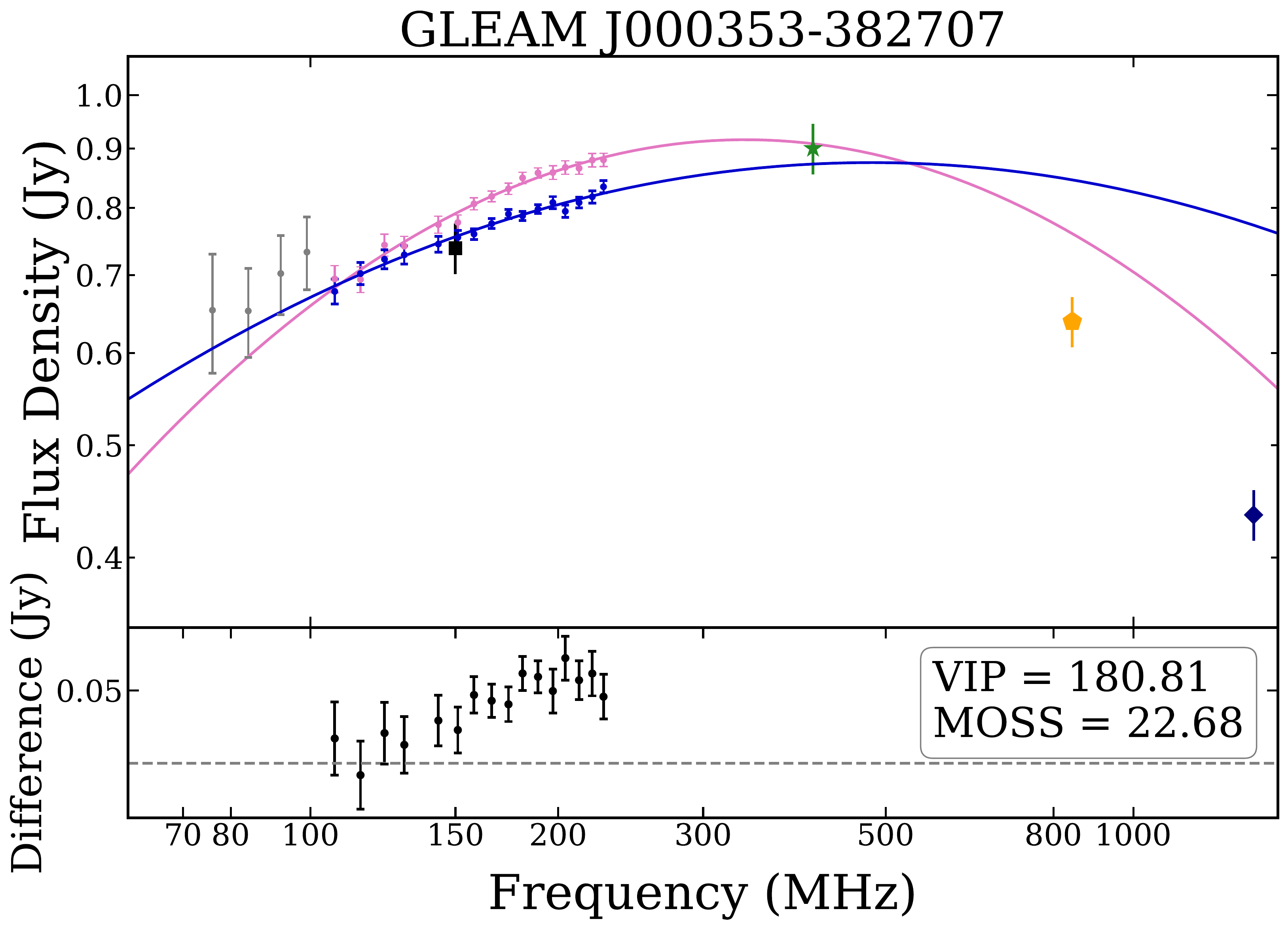} &
\includegraphics[scale=0.15]{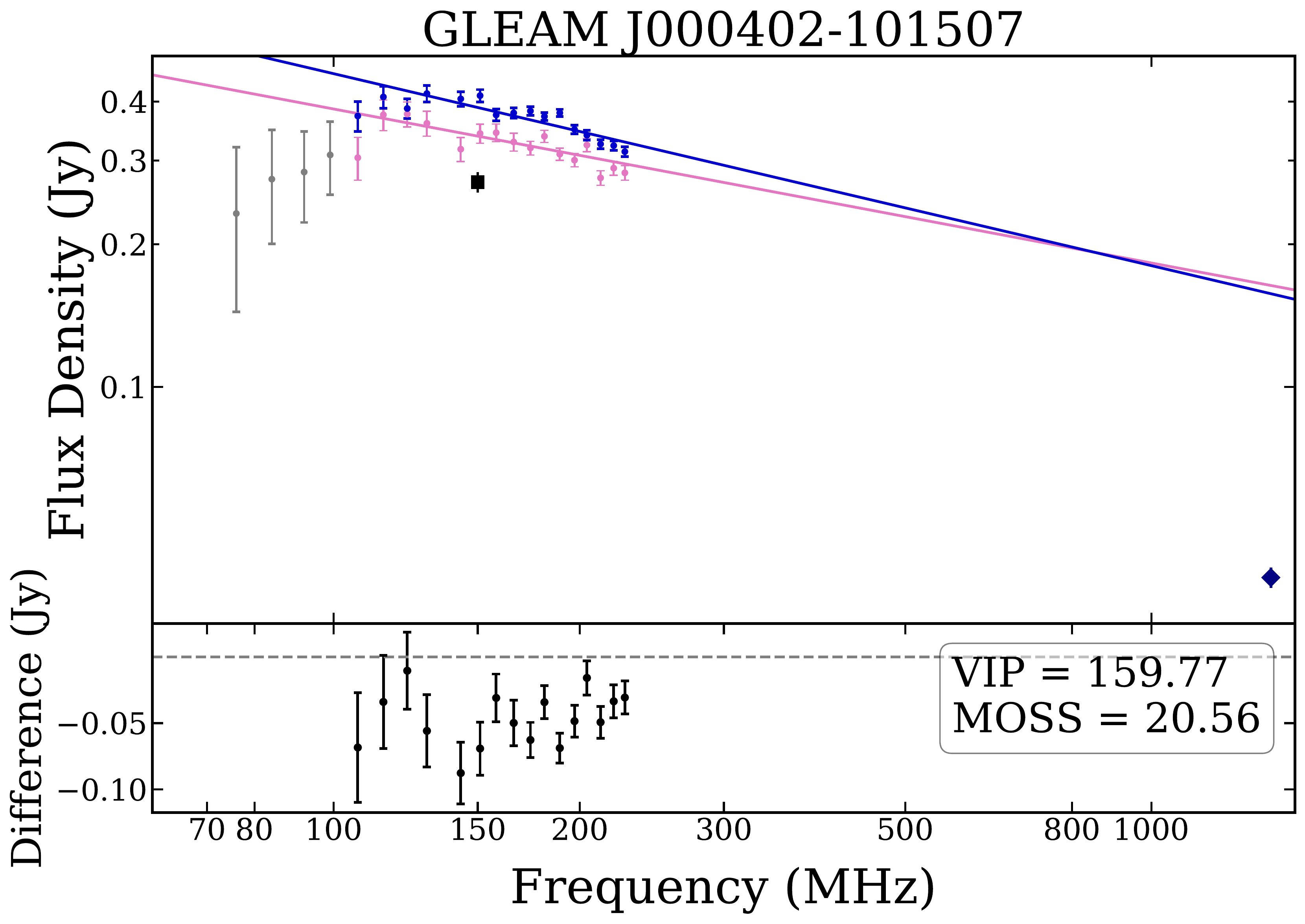} &
\includegraphics[scale=0.15]{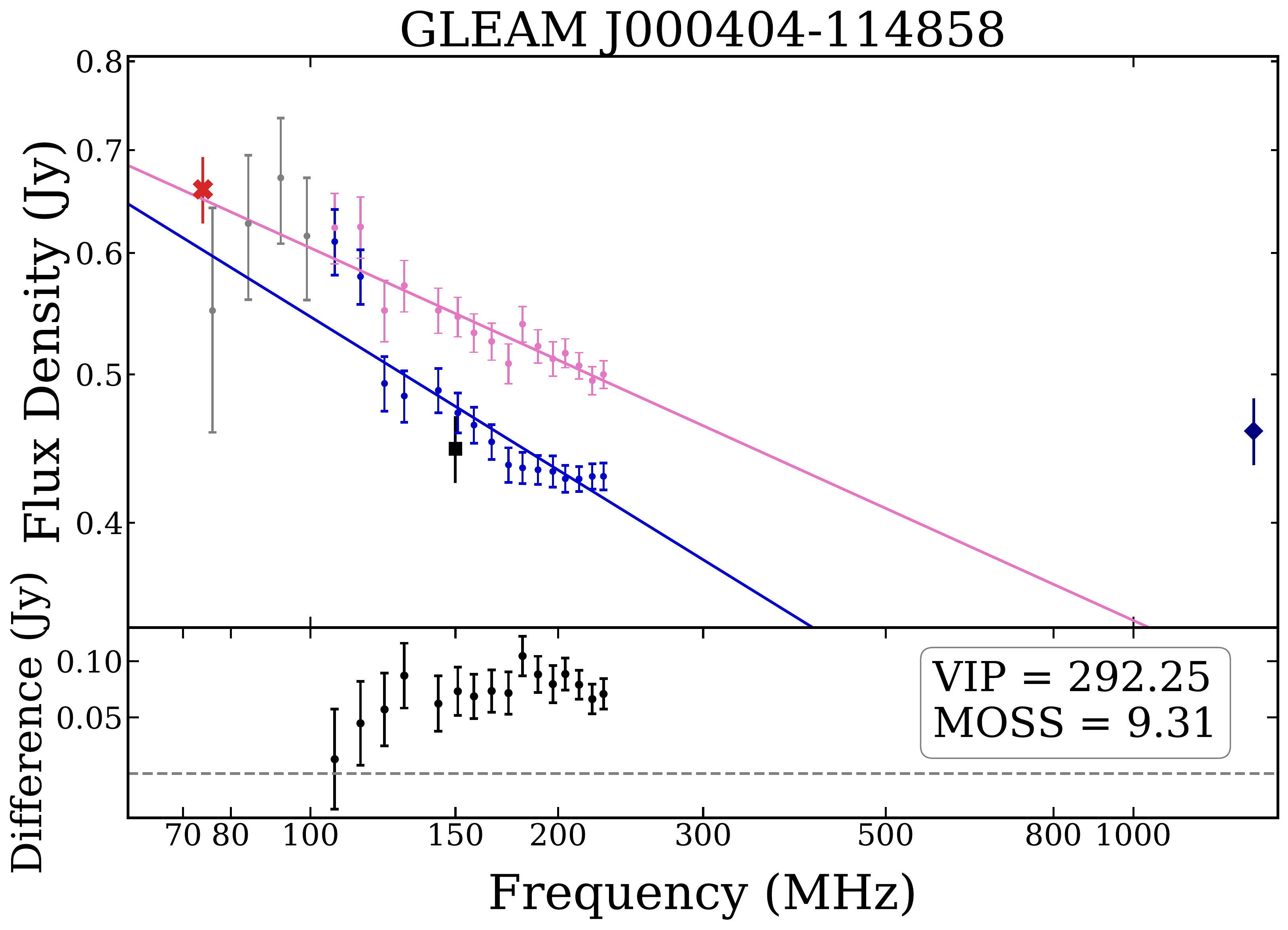} \\
\includegraphics[scale=0.15]{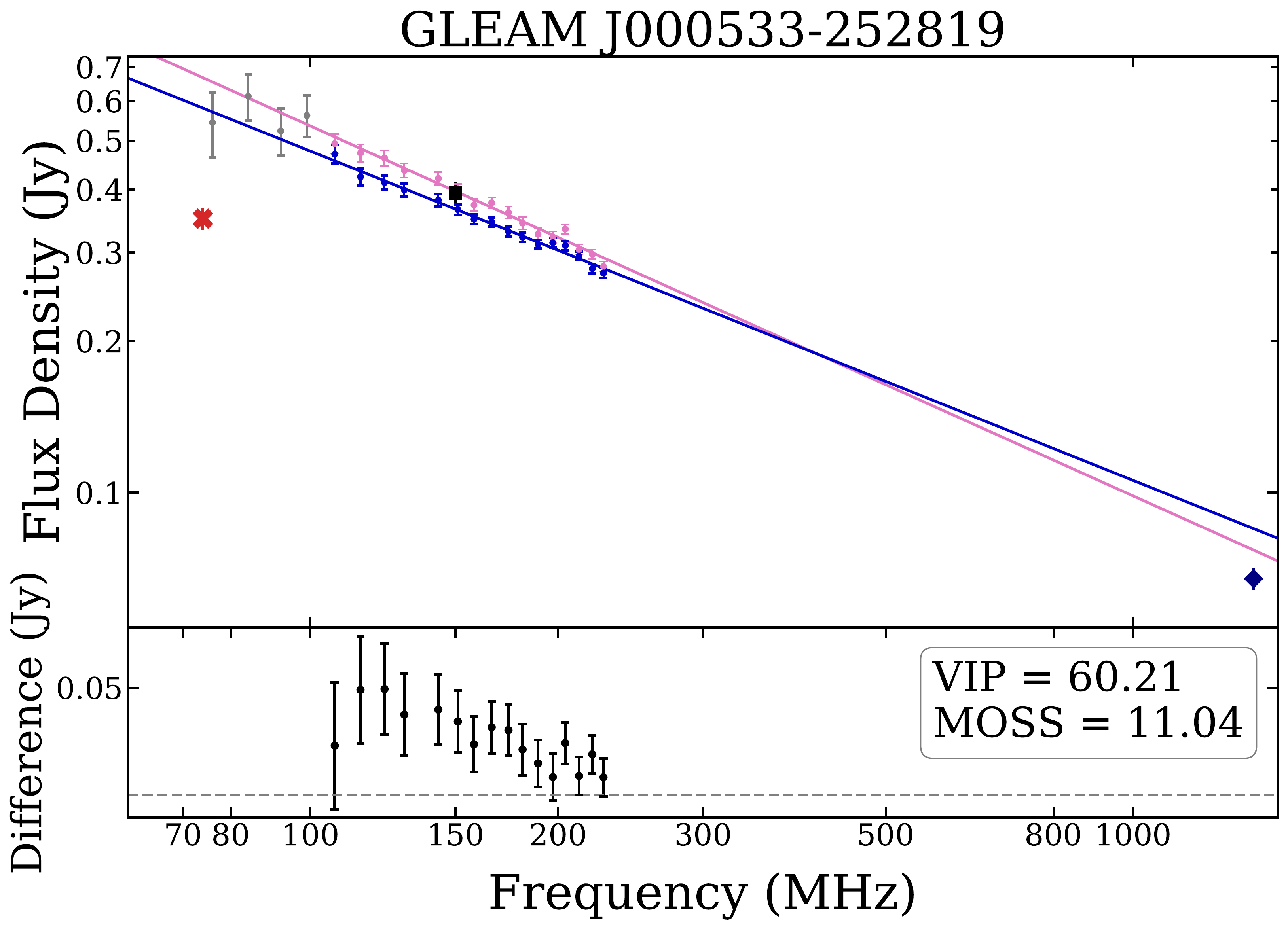} &
\includegraphics[scale=0.15]{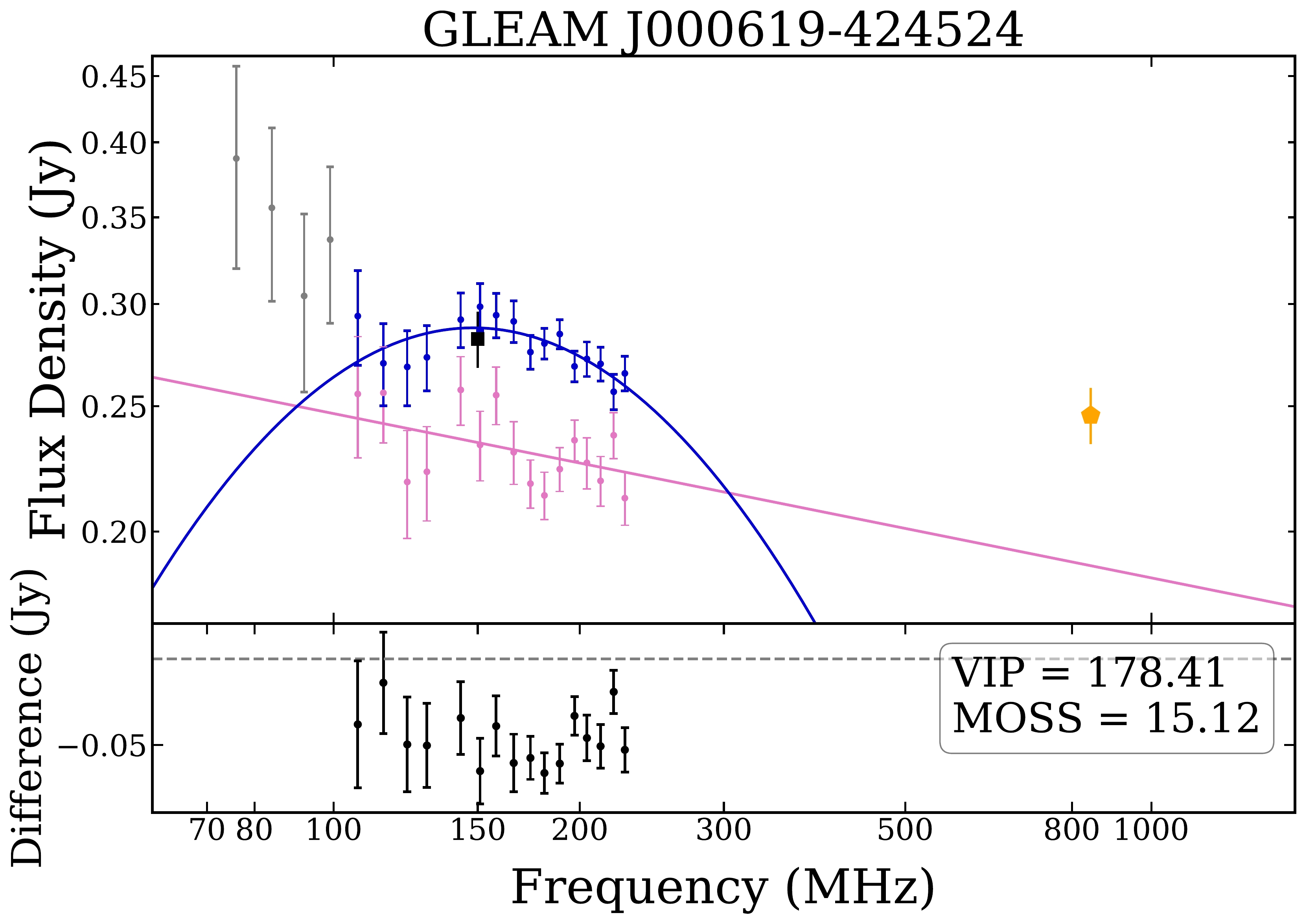} &
\includegraphics[scale=0.15]{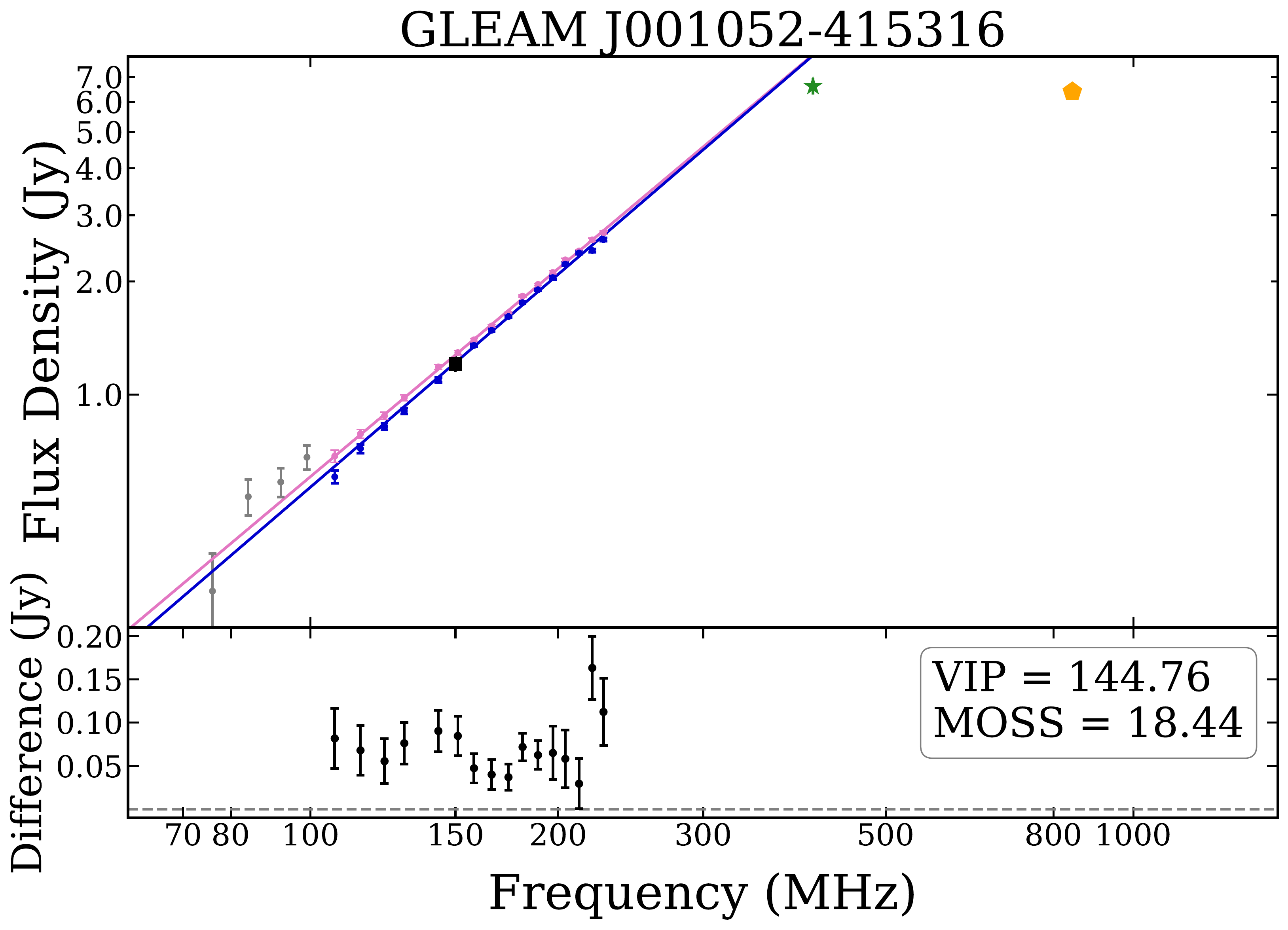} \\
\includegraphics[scale=0.15]{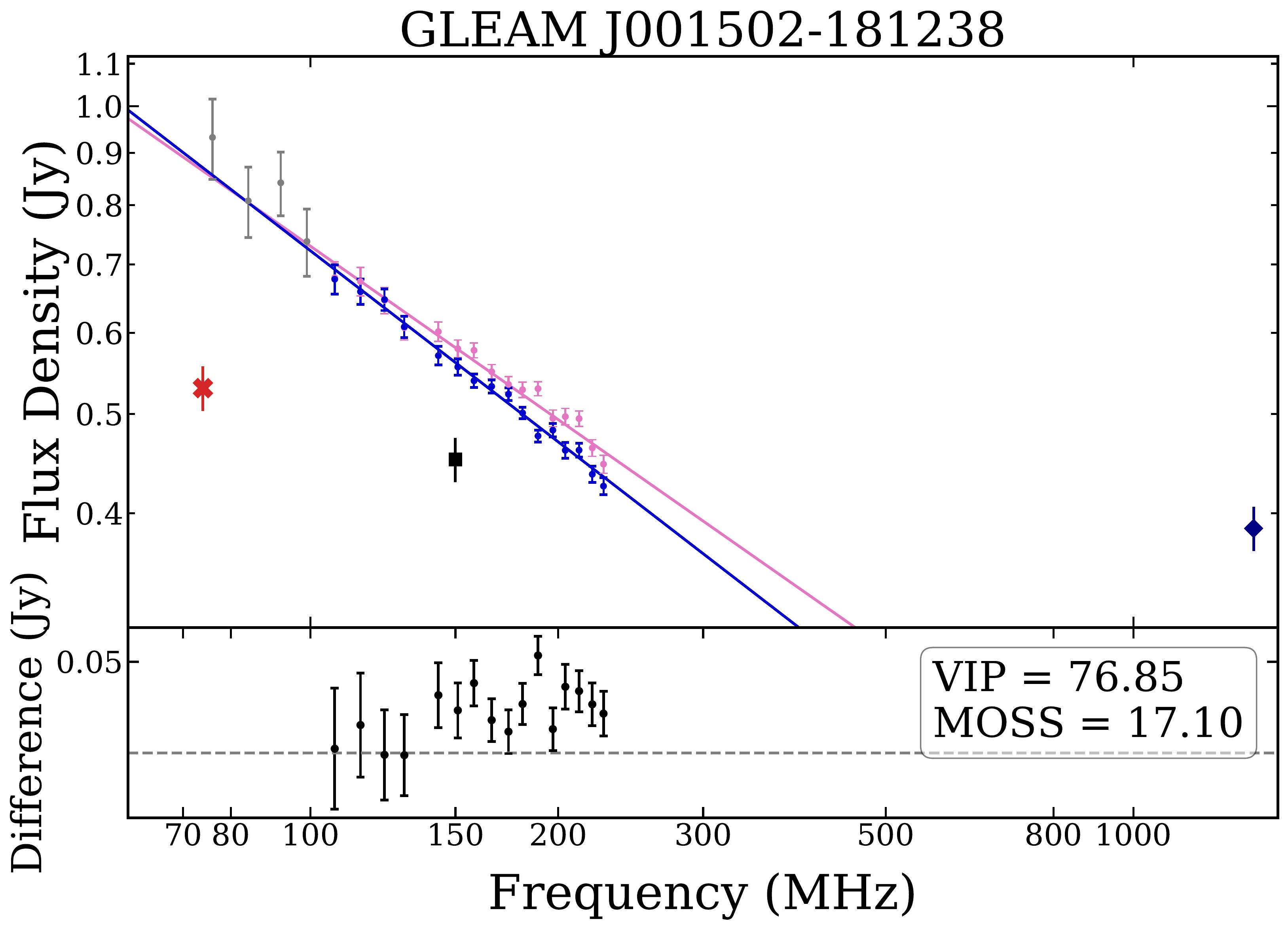} &
\includegraphics[scale=0.15]{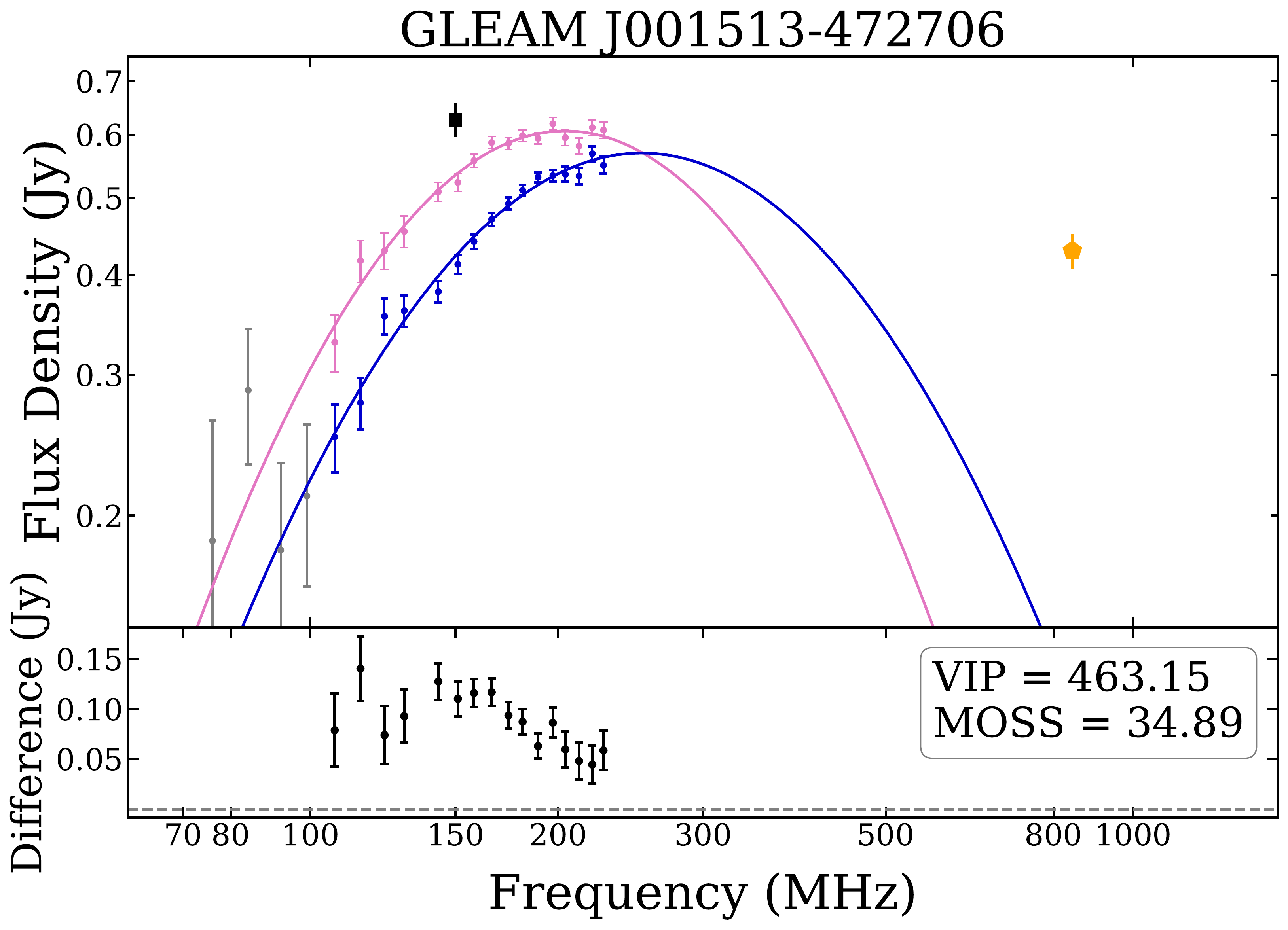} &
\includegraphics[scale=0.15]{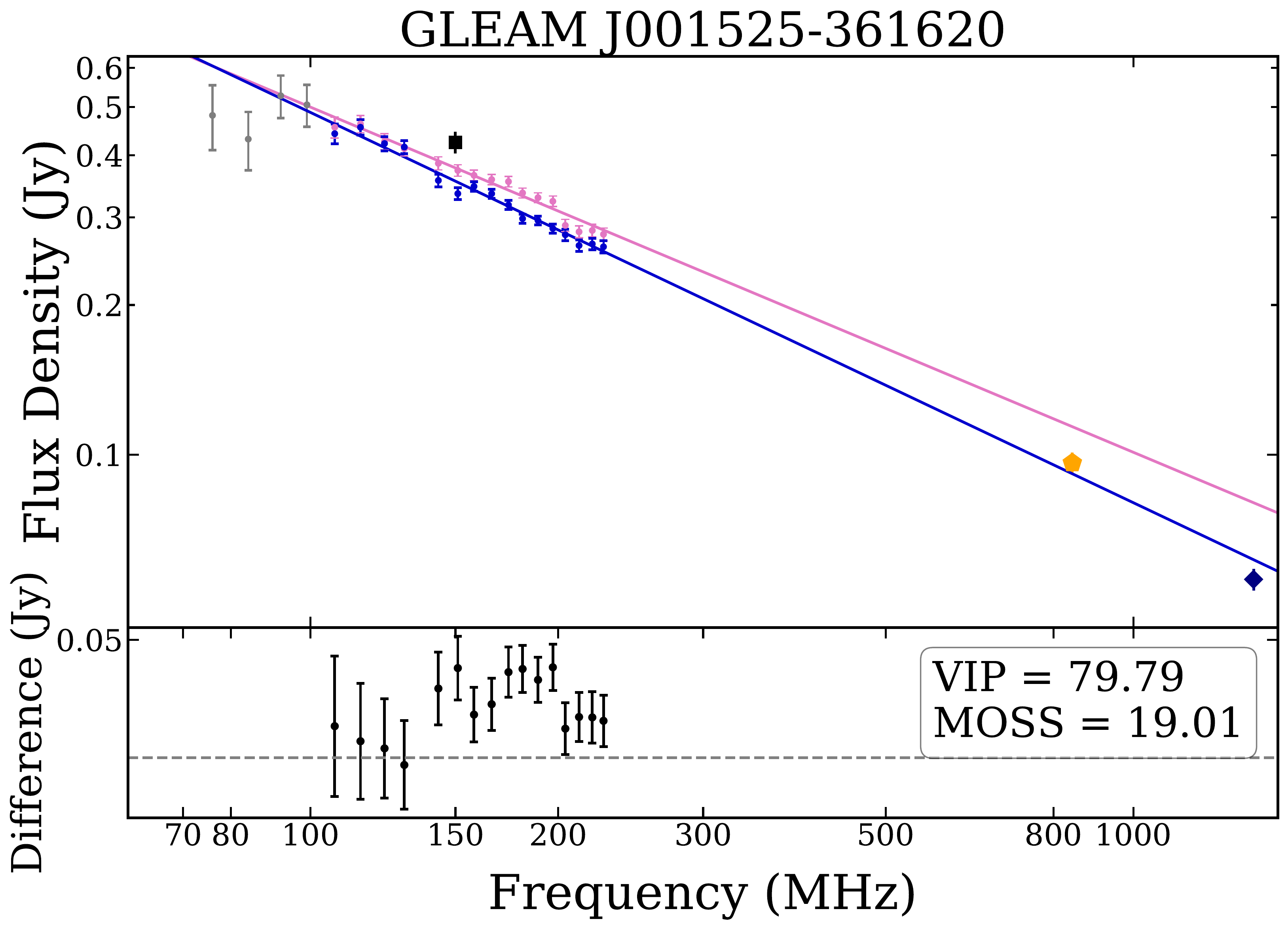} \\
\includegraphics[scale=0.15]{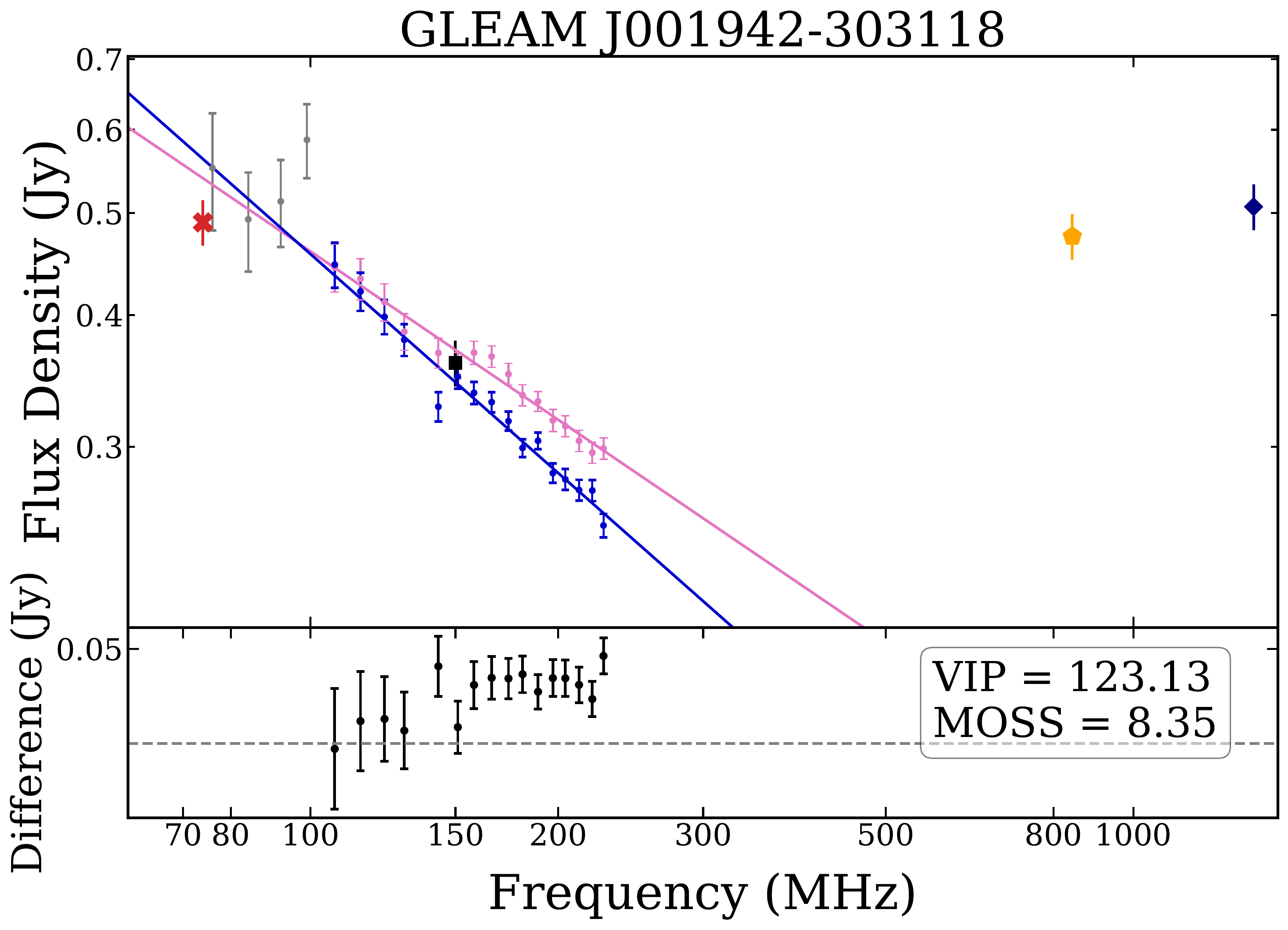} &
\includegraphics[scale=0.15]{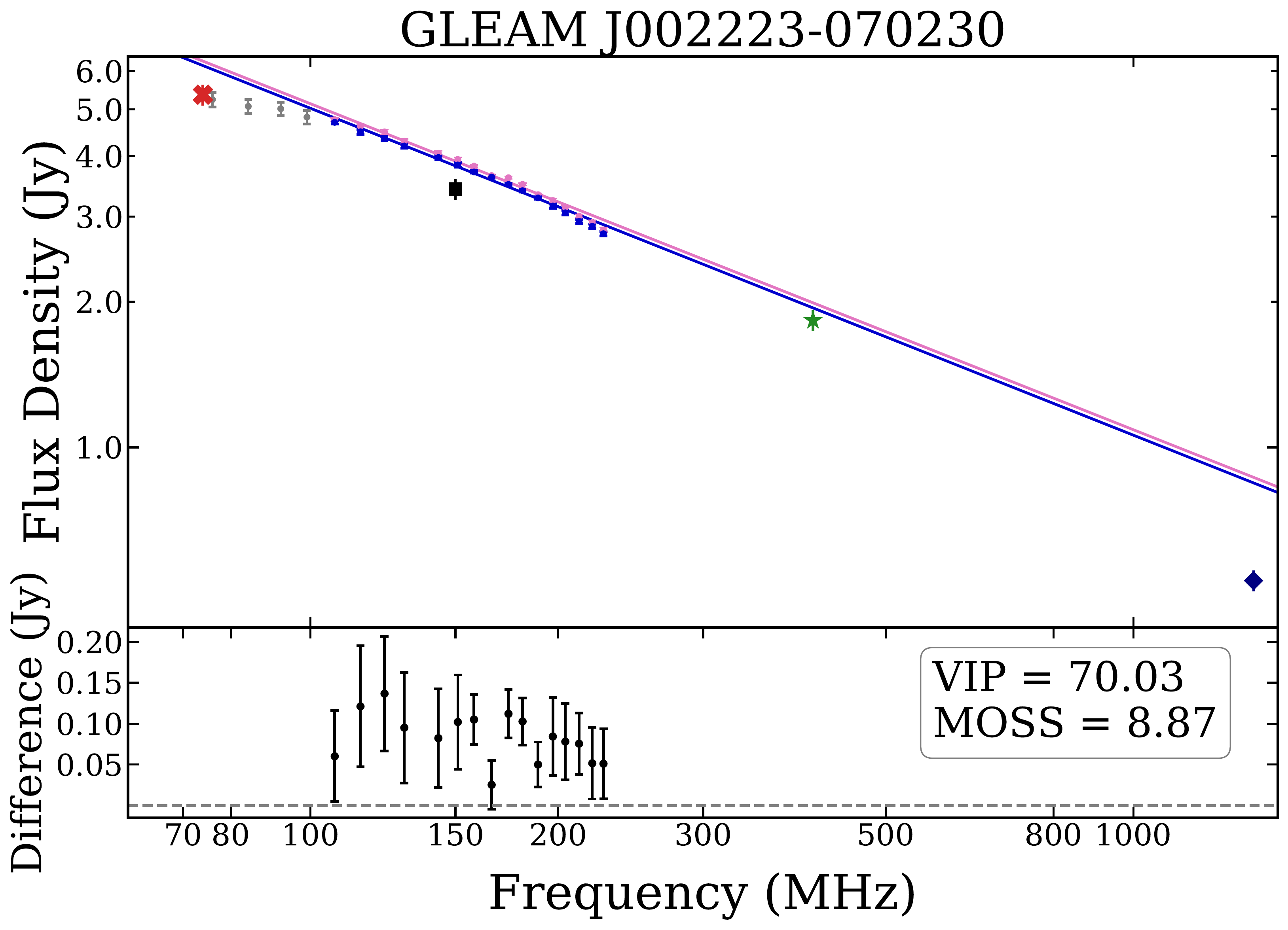} &
\includegraphics[scale=0.15]{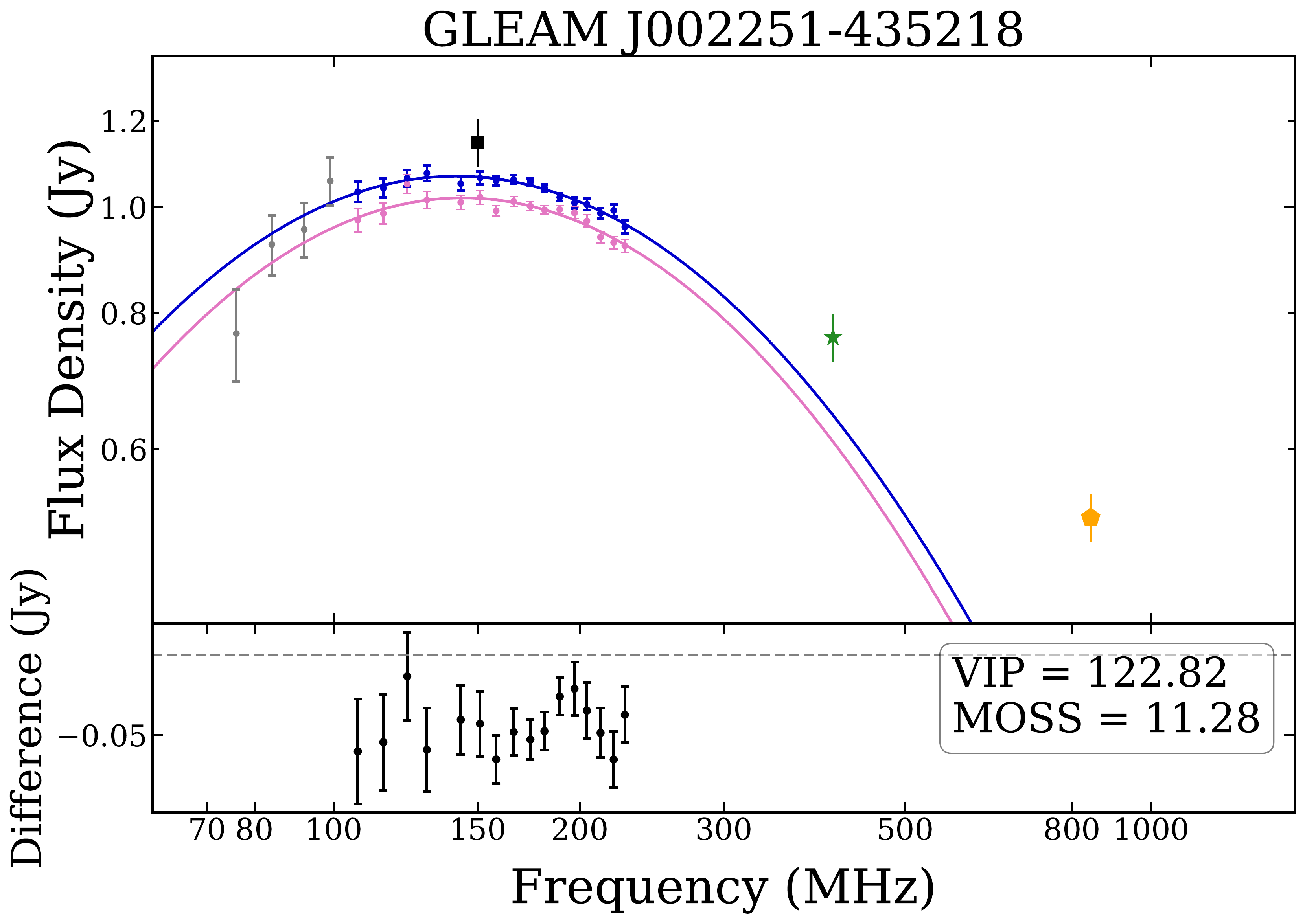} \\
\includegraphics[scale=0.15]{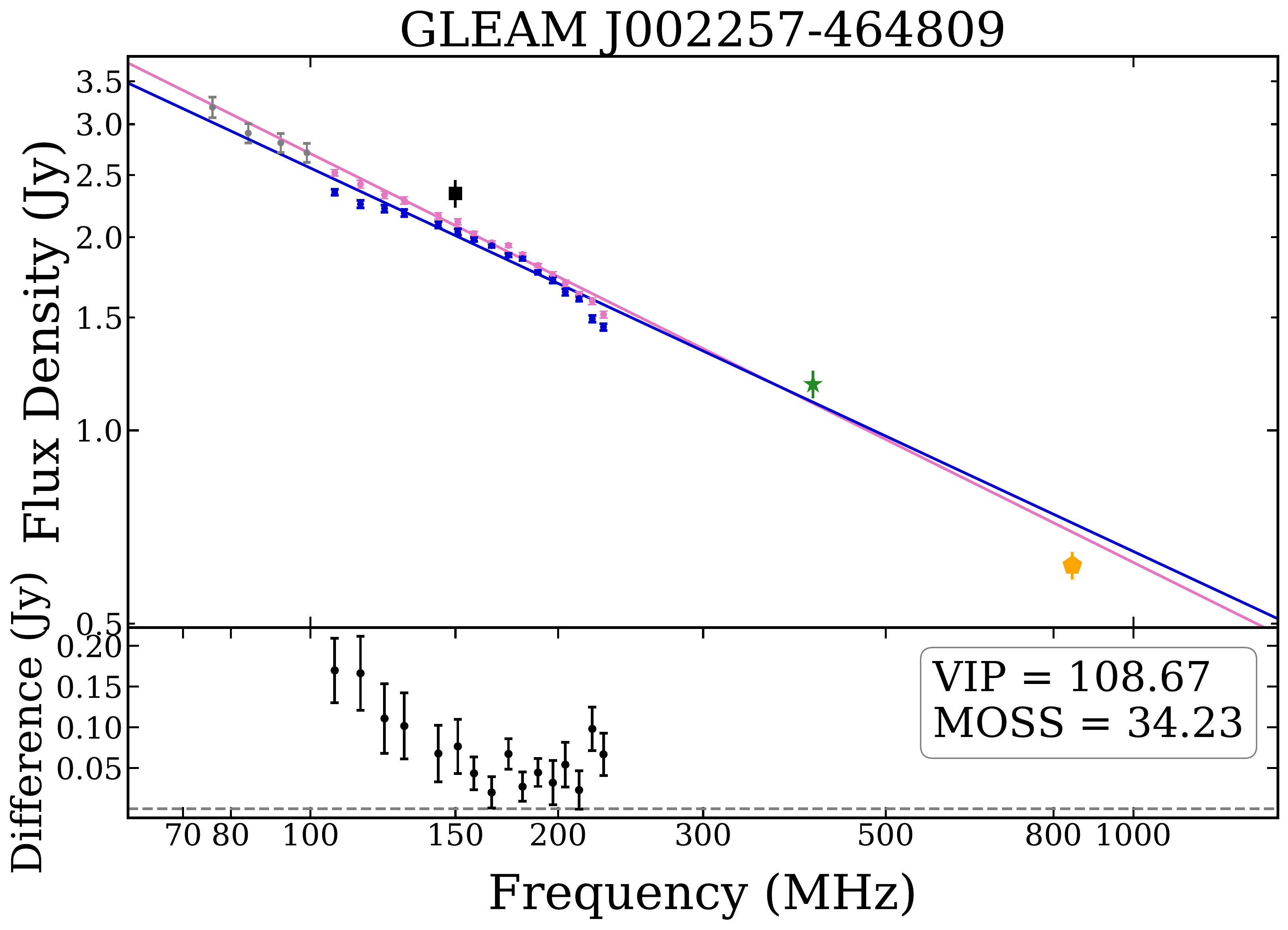} &
\includegraphics[scale=0.15]{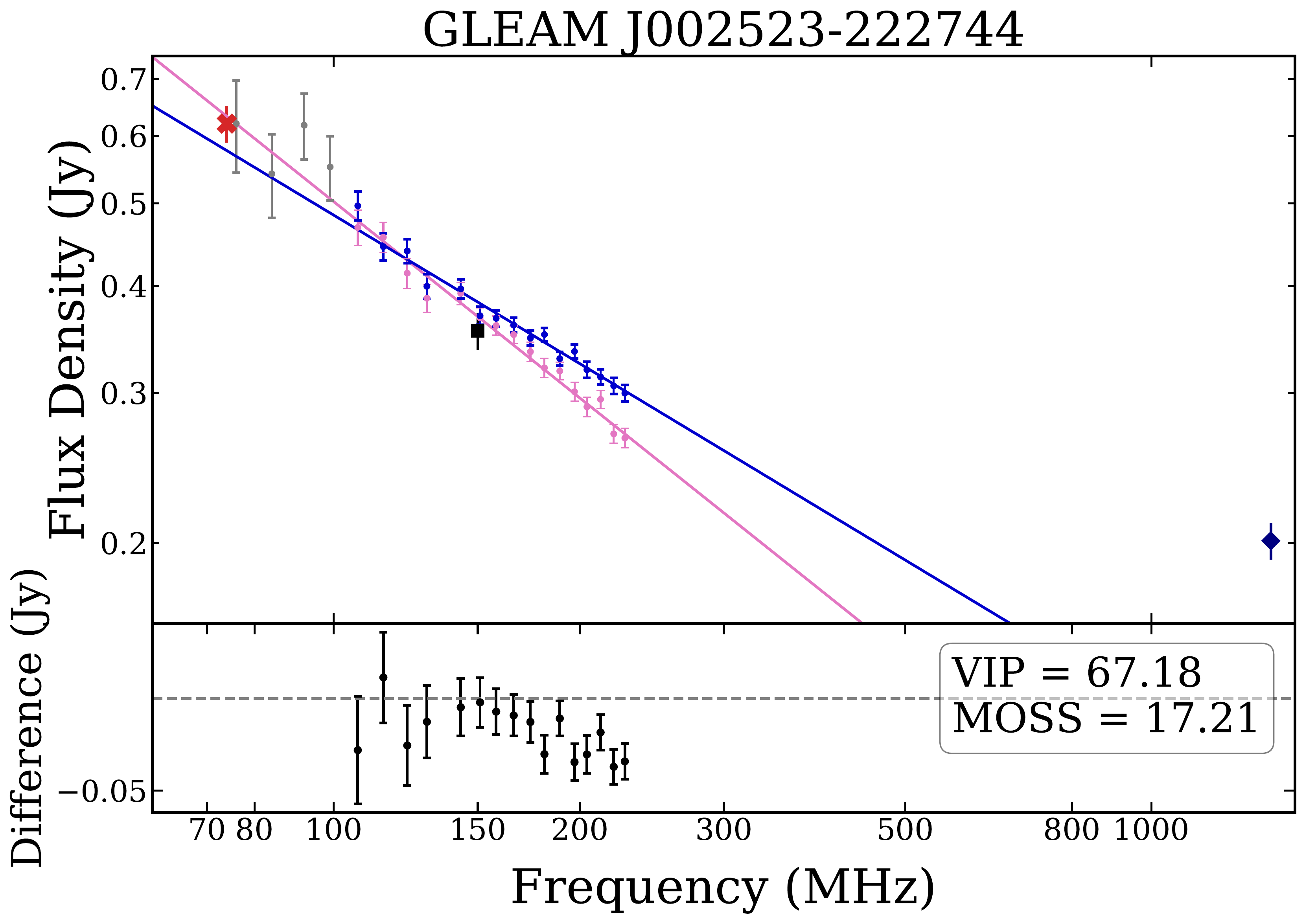} &
\includegraphics[scale=0.15]{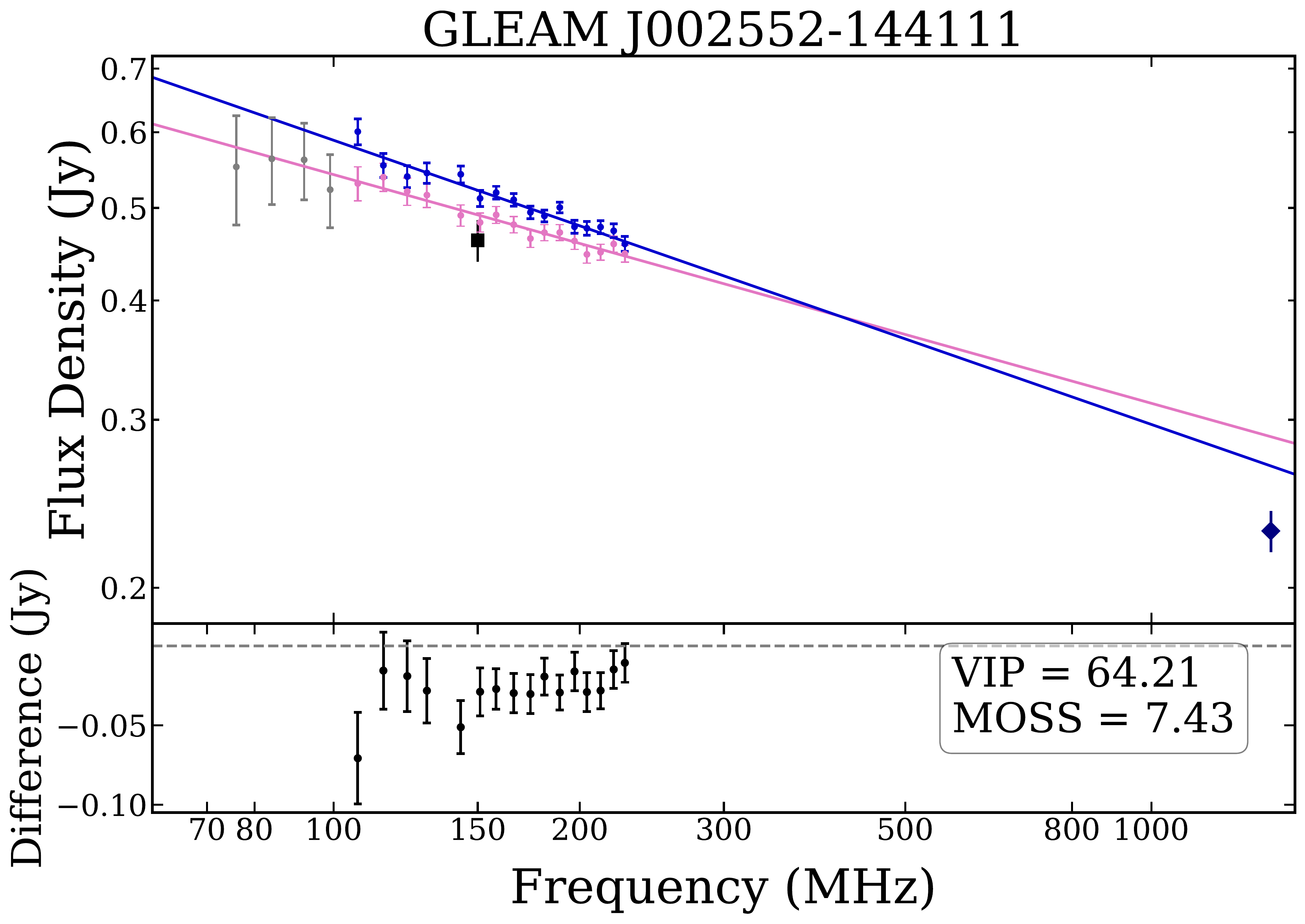} \\
\includegraphics[scale=0.15]{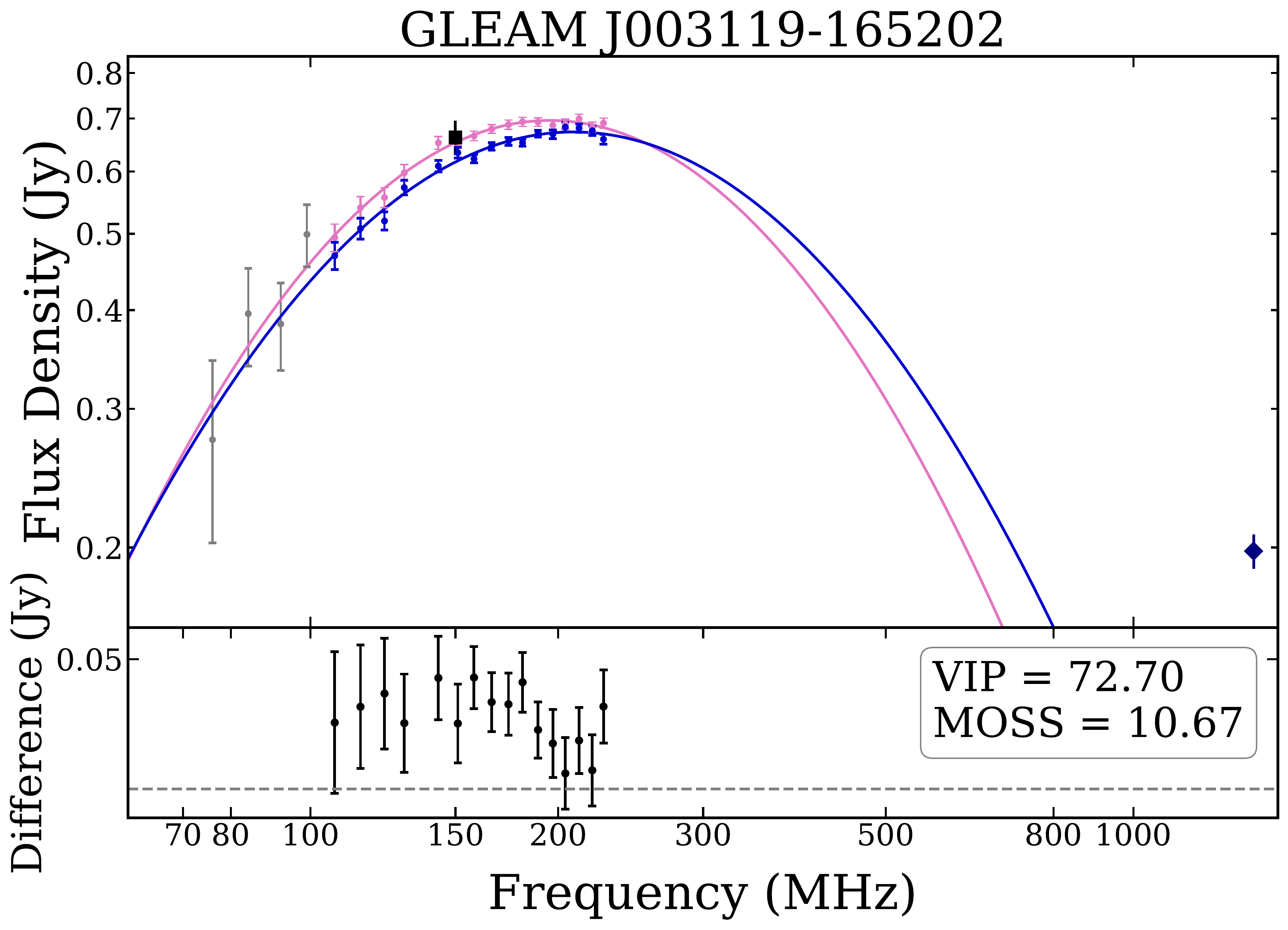} &
\includegraphics[scale=0.15]{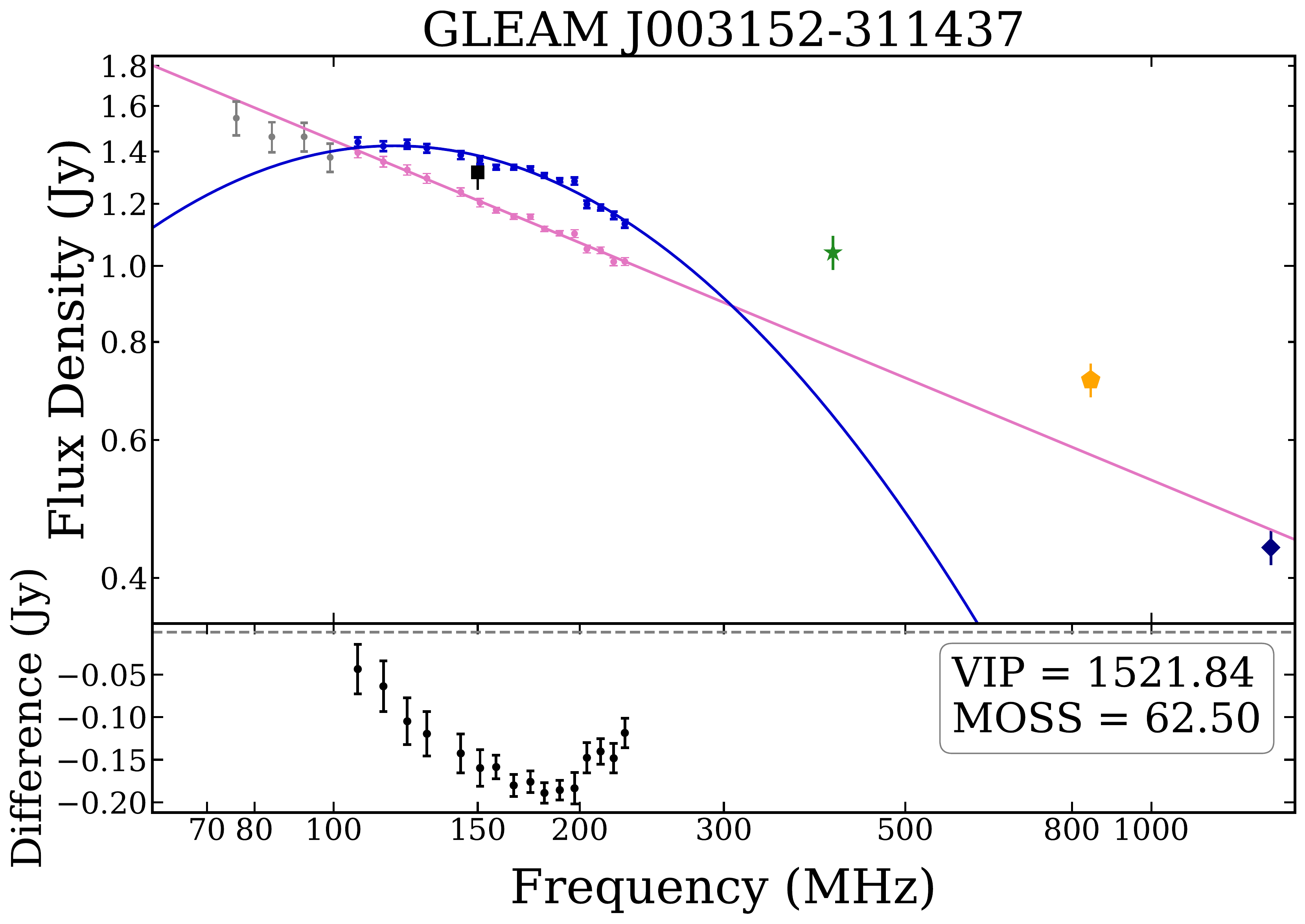} &
\includegraphics[scale=0.15]{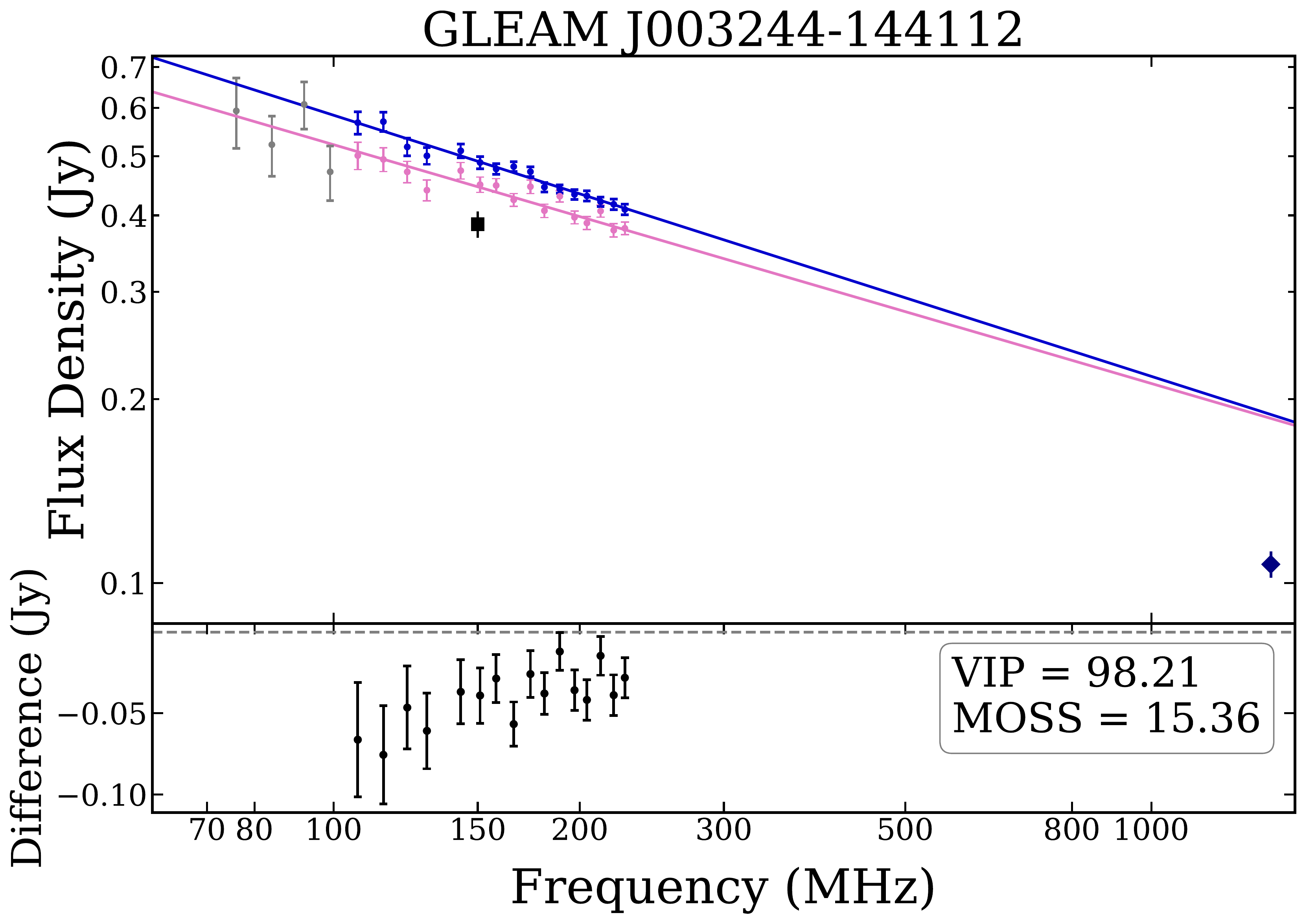} \\
\end{array}$
\caption{SEDs for all sources classified as variable according to the VIP. For each source the points represent the following data: GLEAM low frequency (72--100\,MHz) (grey circles), Year 1 (pink circles), Year 2 (blue circles), VLSSr (red cross), TGSS (black square), MRC (green star), SUMSS (yellow pentagon), and NVSS (navy diamond). The models for each year are determined by their classification; a source classified with a peak within the observed band was modelled by a quadratic according to Equation~\ref{eq:quadratic}, remaining sources were modelled by a power-law according to Equation~\ref{eq:plaw}.}
\label{app:fig:pg1}
\end{center}
\end{figure*}
\setcounter{figure}{0}
\begin{figure*}
\begin{center}$
\begin{array}{cccccc}
\includegraphics[scale=0.15]{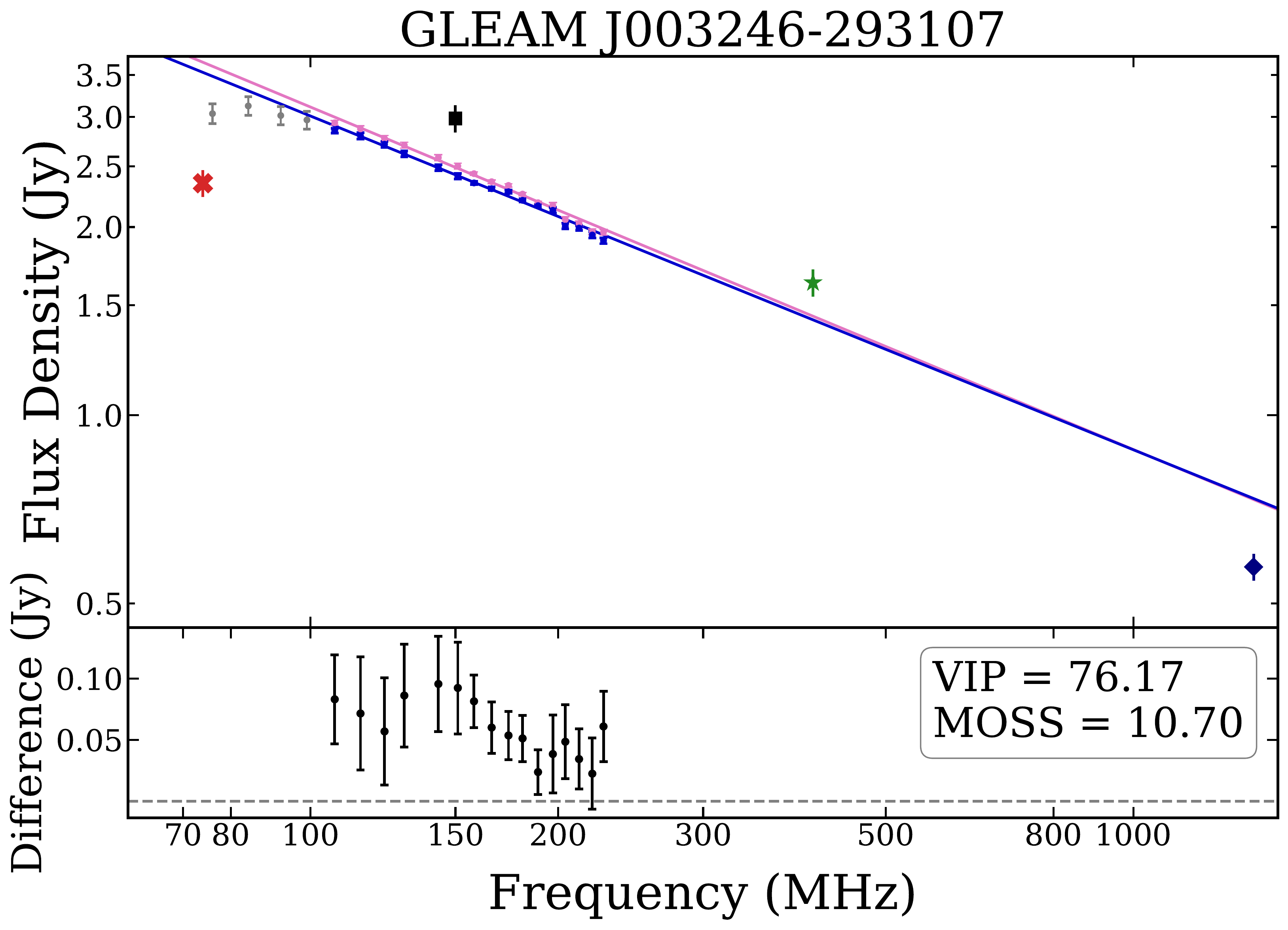} &
\includegraphics[scale=0.15]{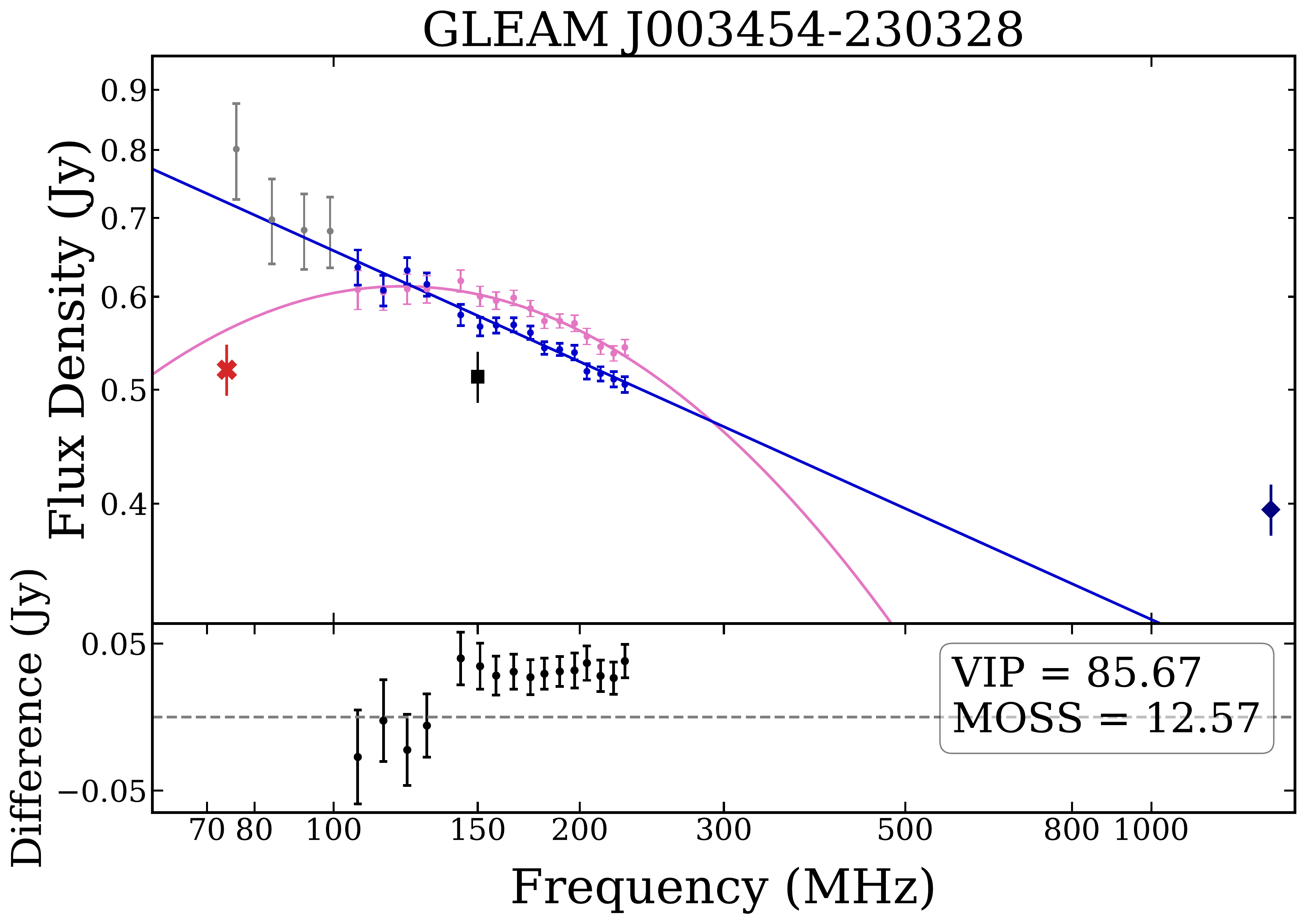} &
\includegraphics[scale=0.15]{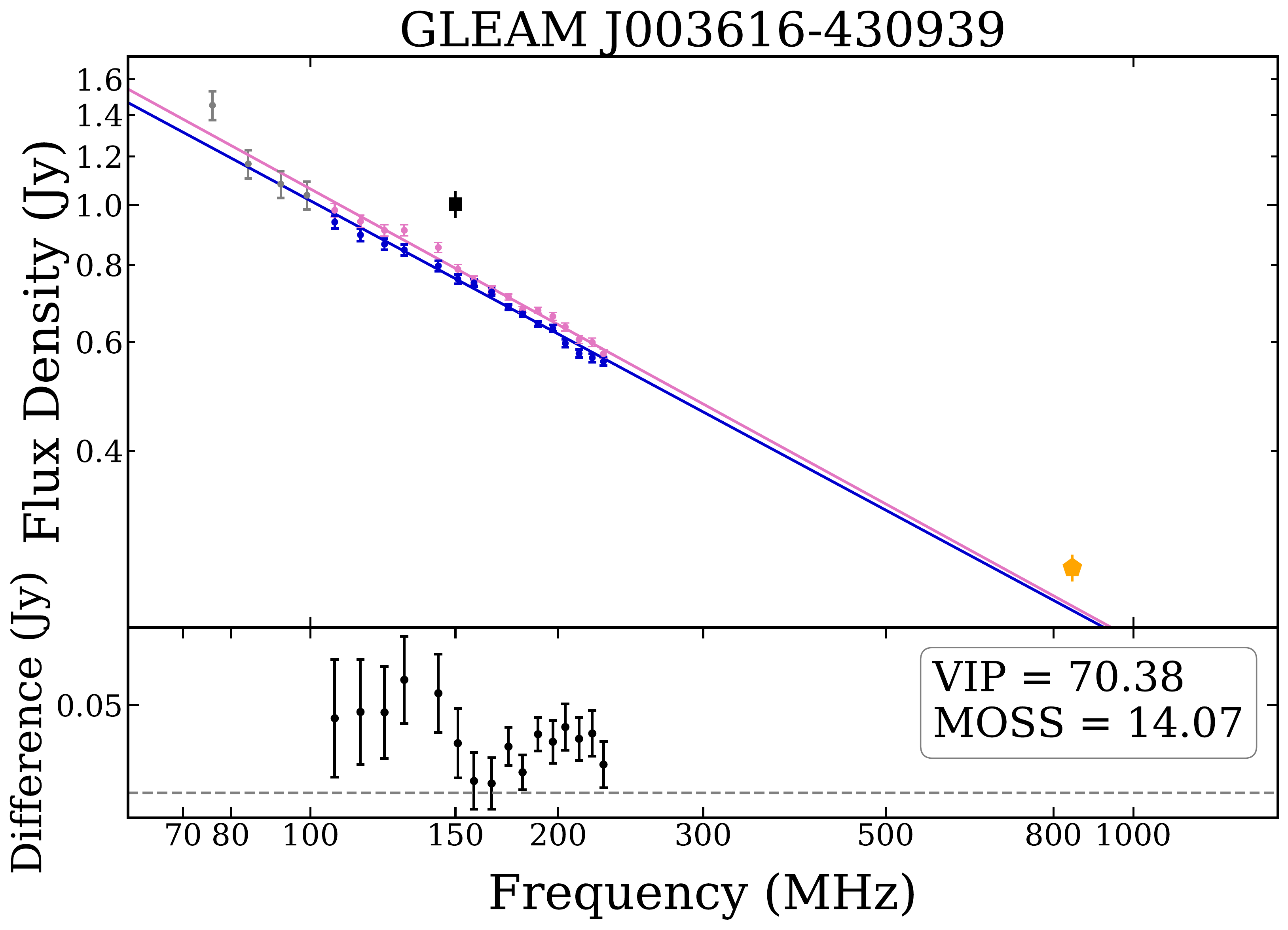} \\
\includegraphics[scale=0.15]{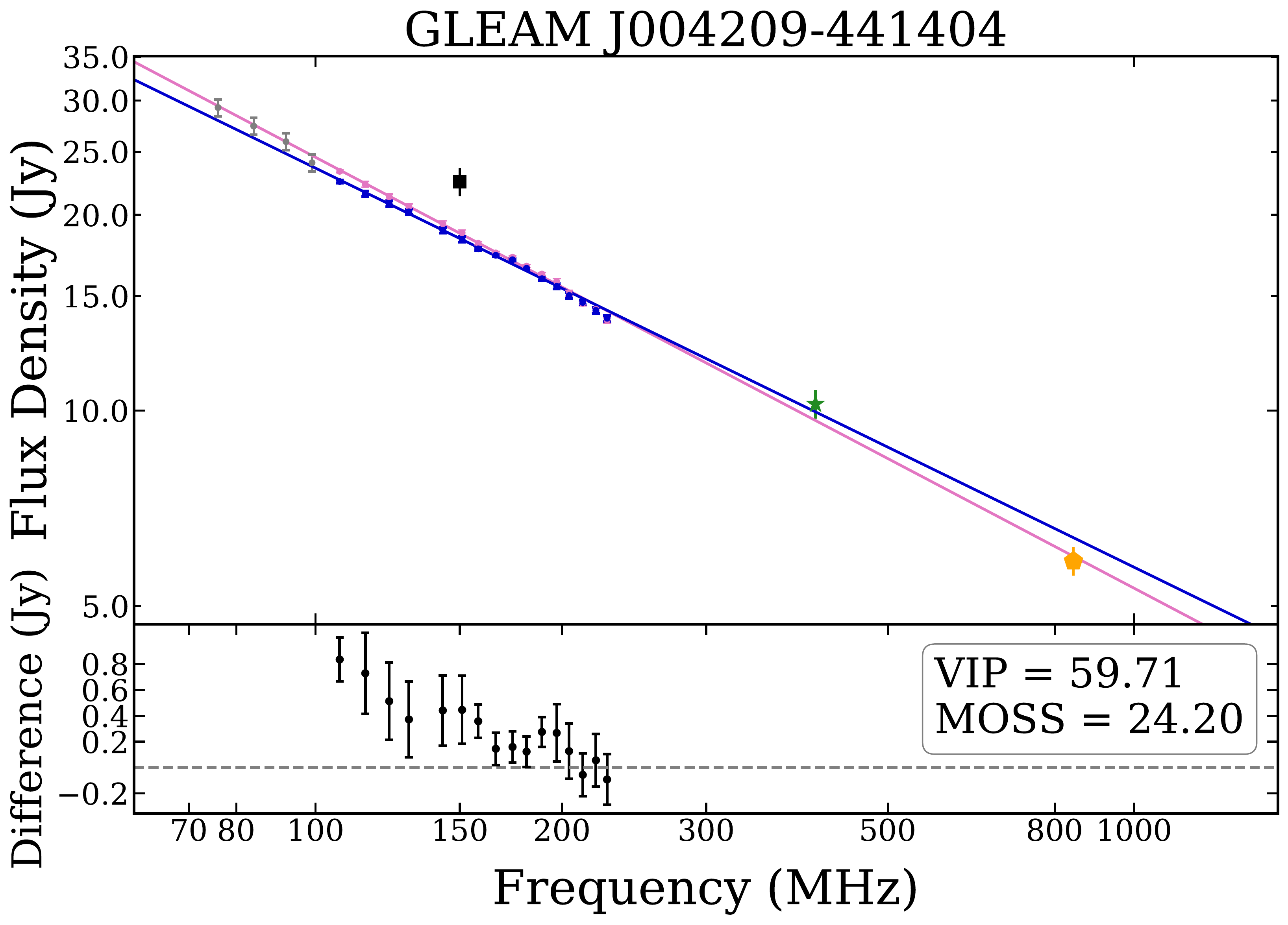} &
\includegraphics[scale=0.15]{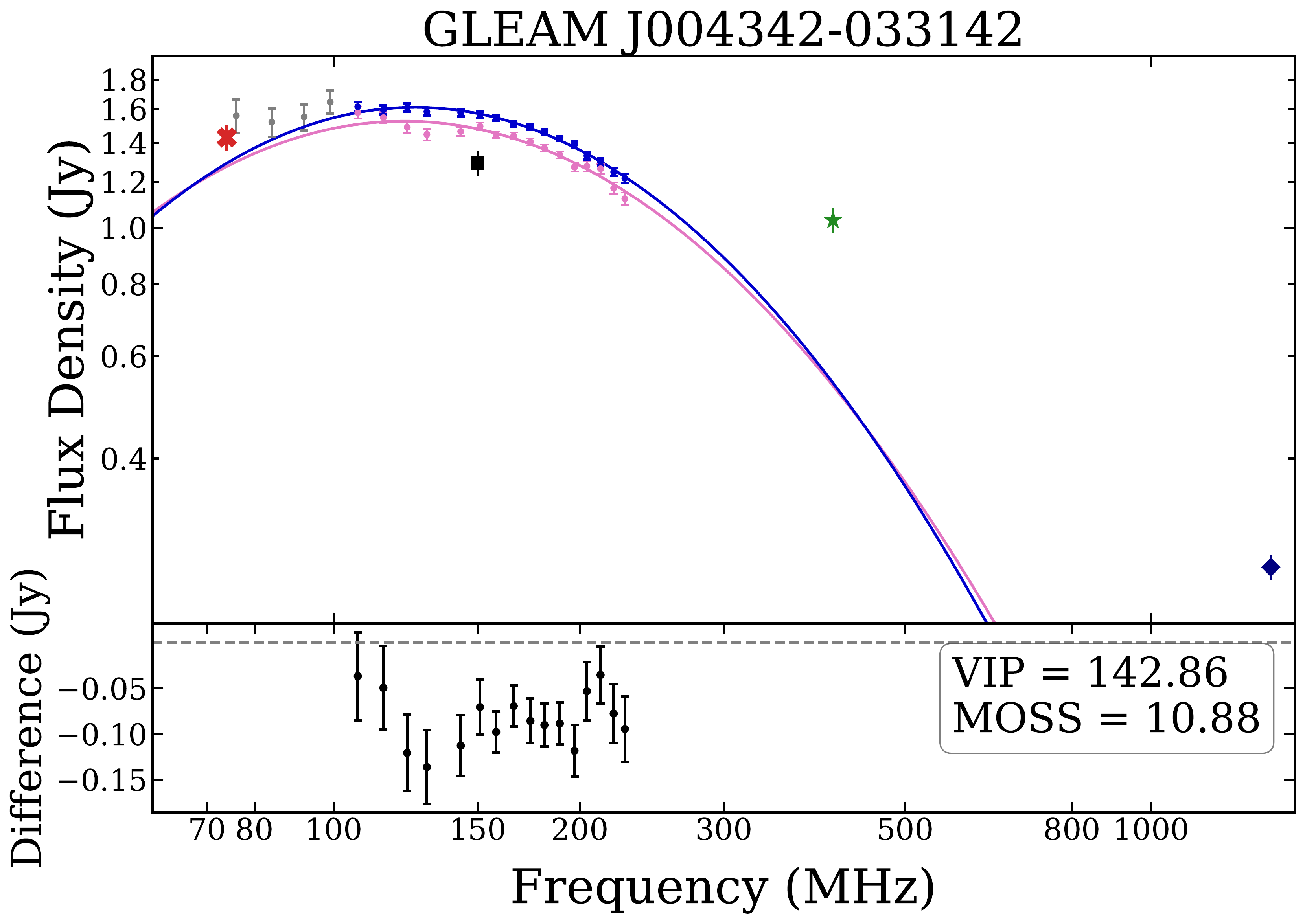} &
\includegraphics[scale=0.15]{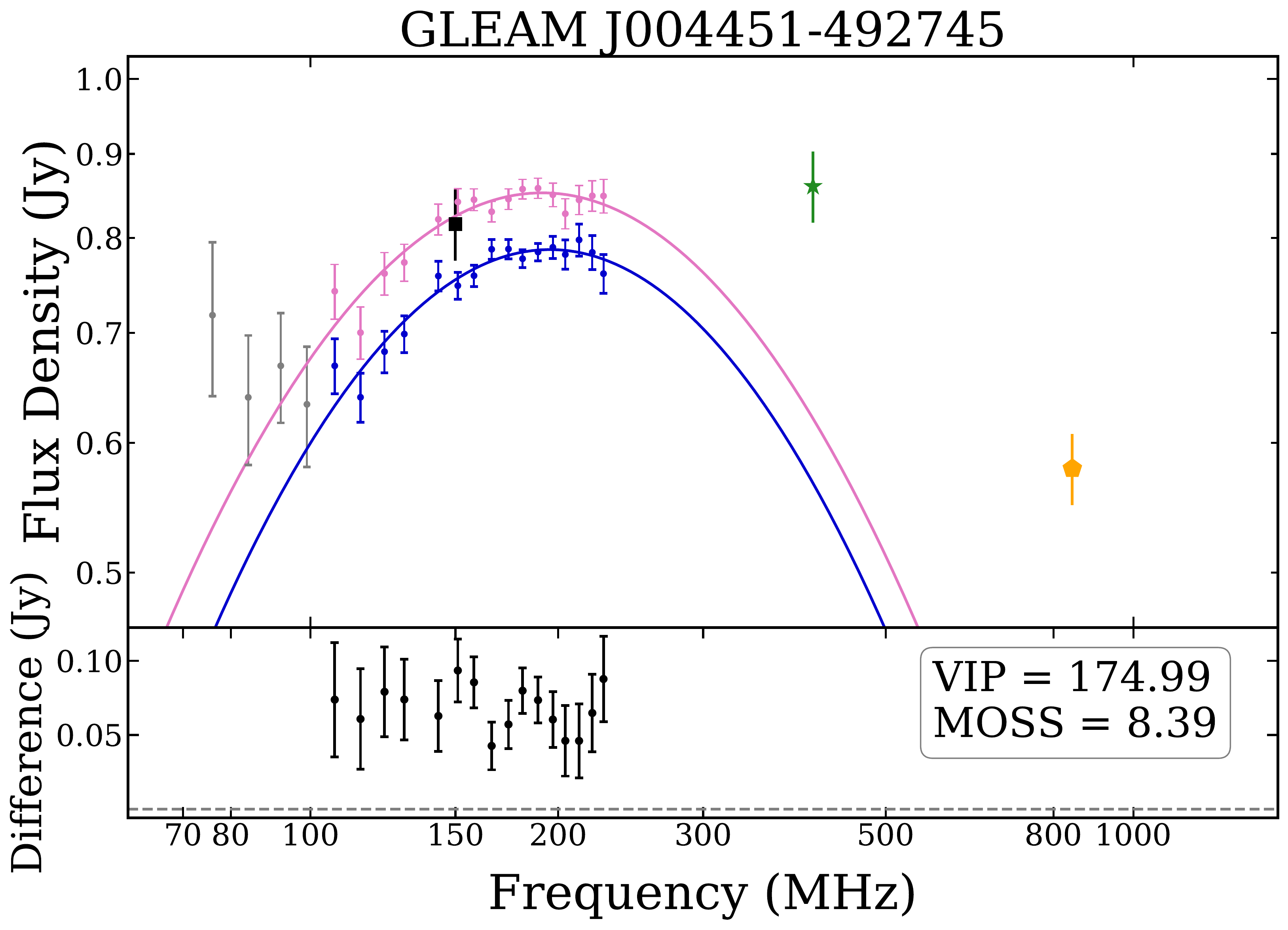} \\
\includegraphics[scale=0.15]{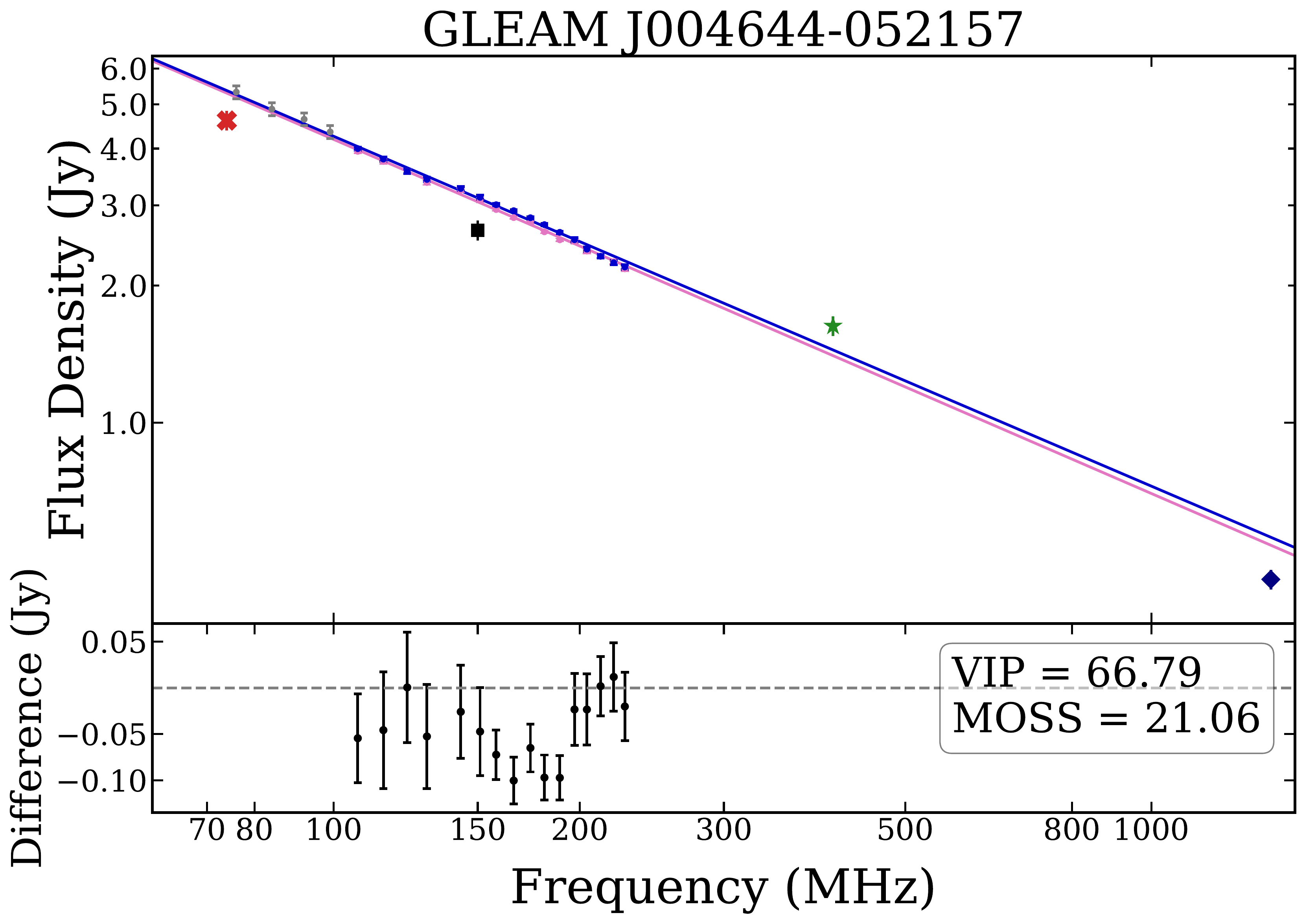} &
\includegraphics[scale=0.15]{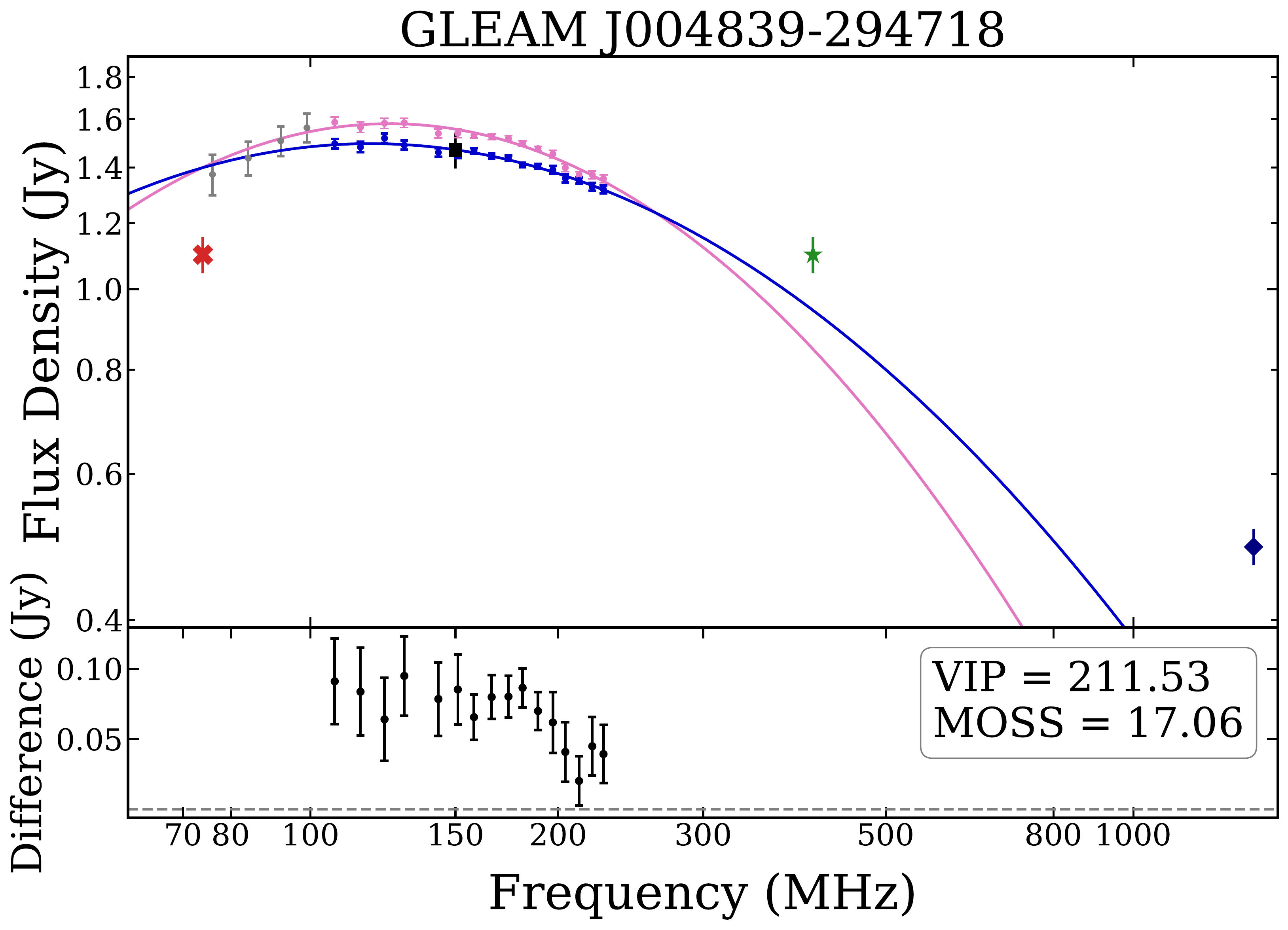} &
\includegraphics[scale=0.15]{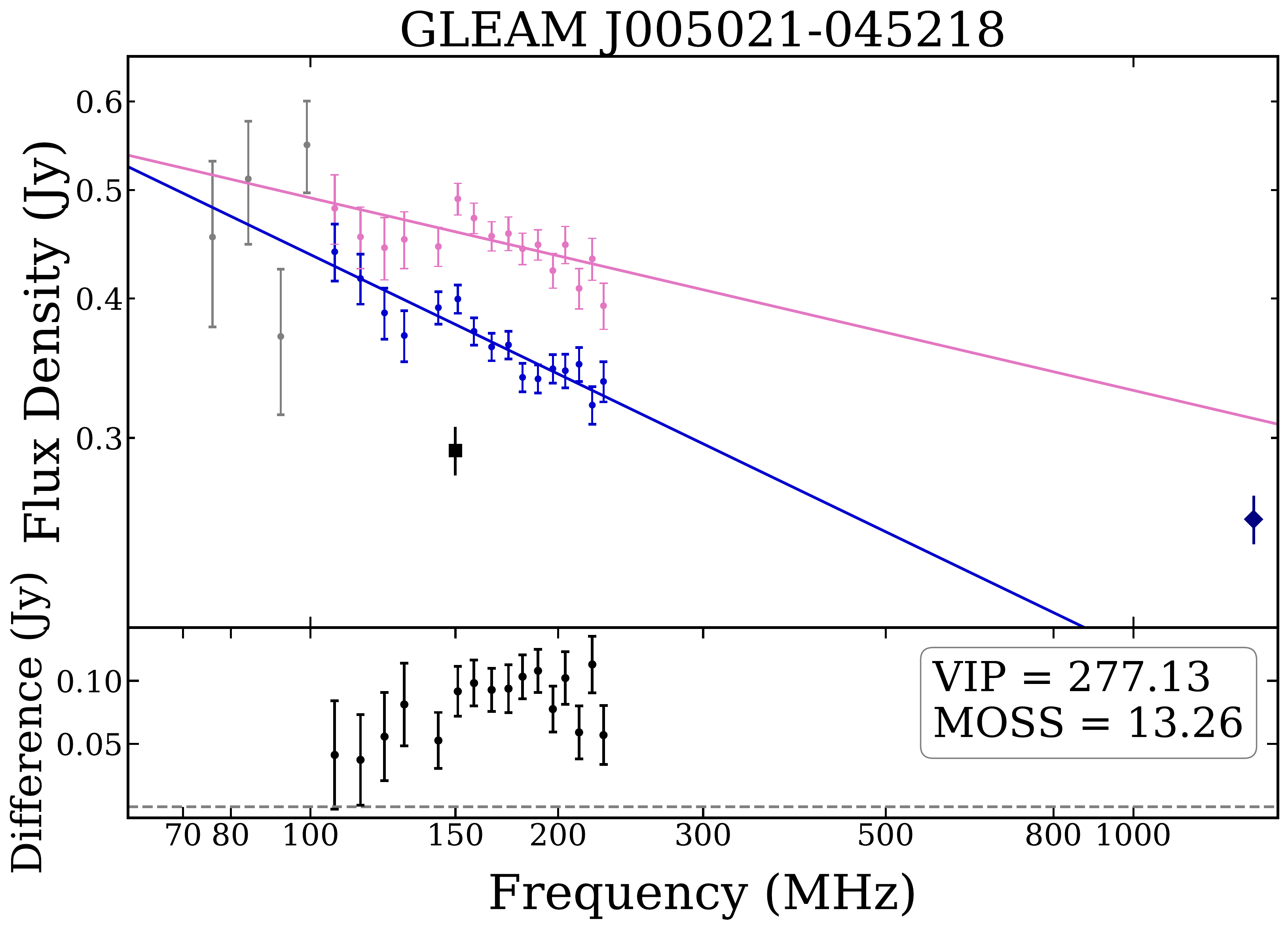} \\
\includegraphics[scale=0.15]{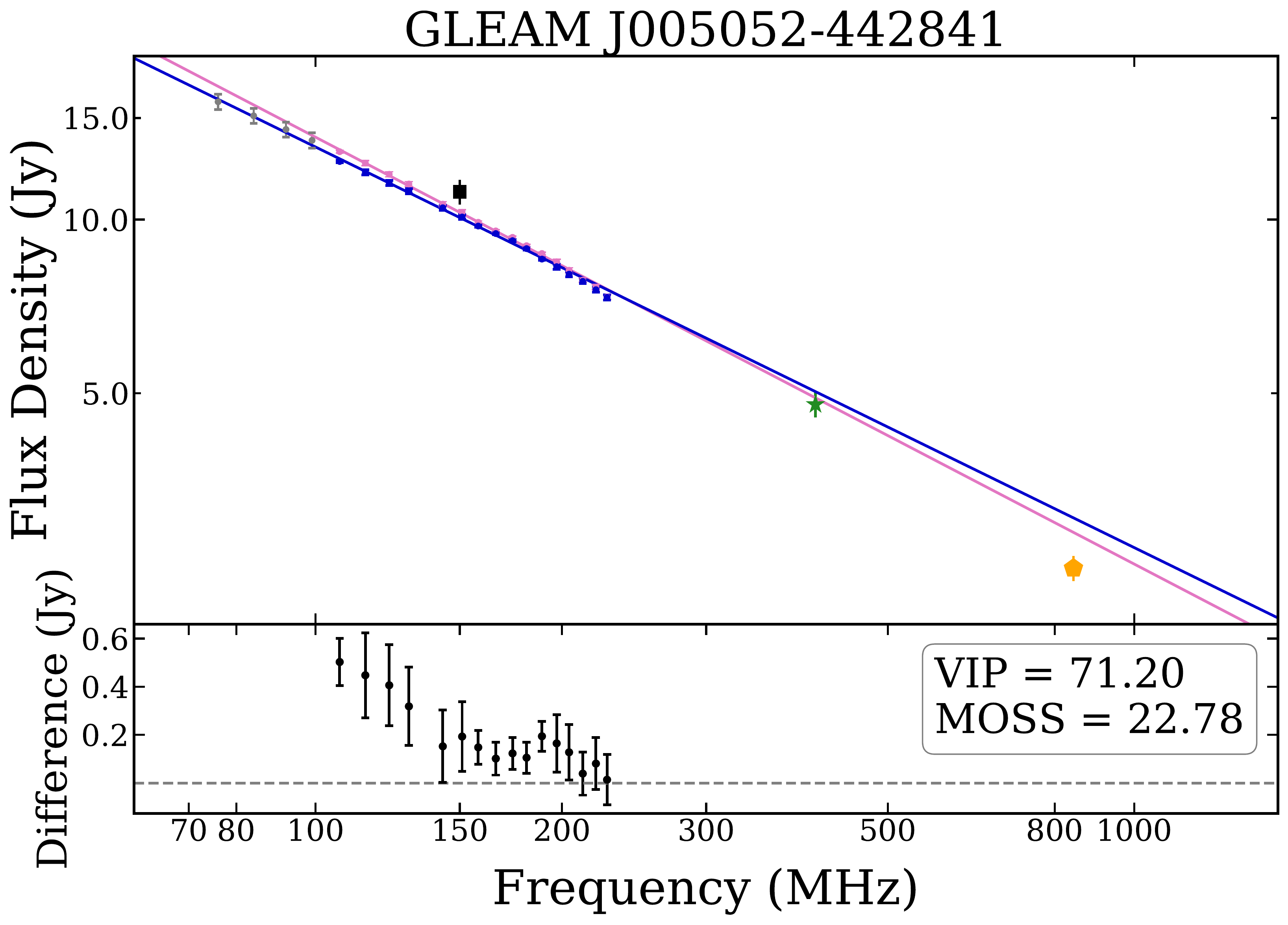} &
\includegraphics[scale=0.15]{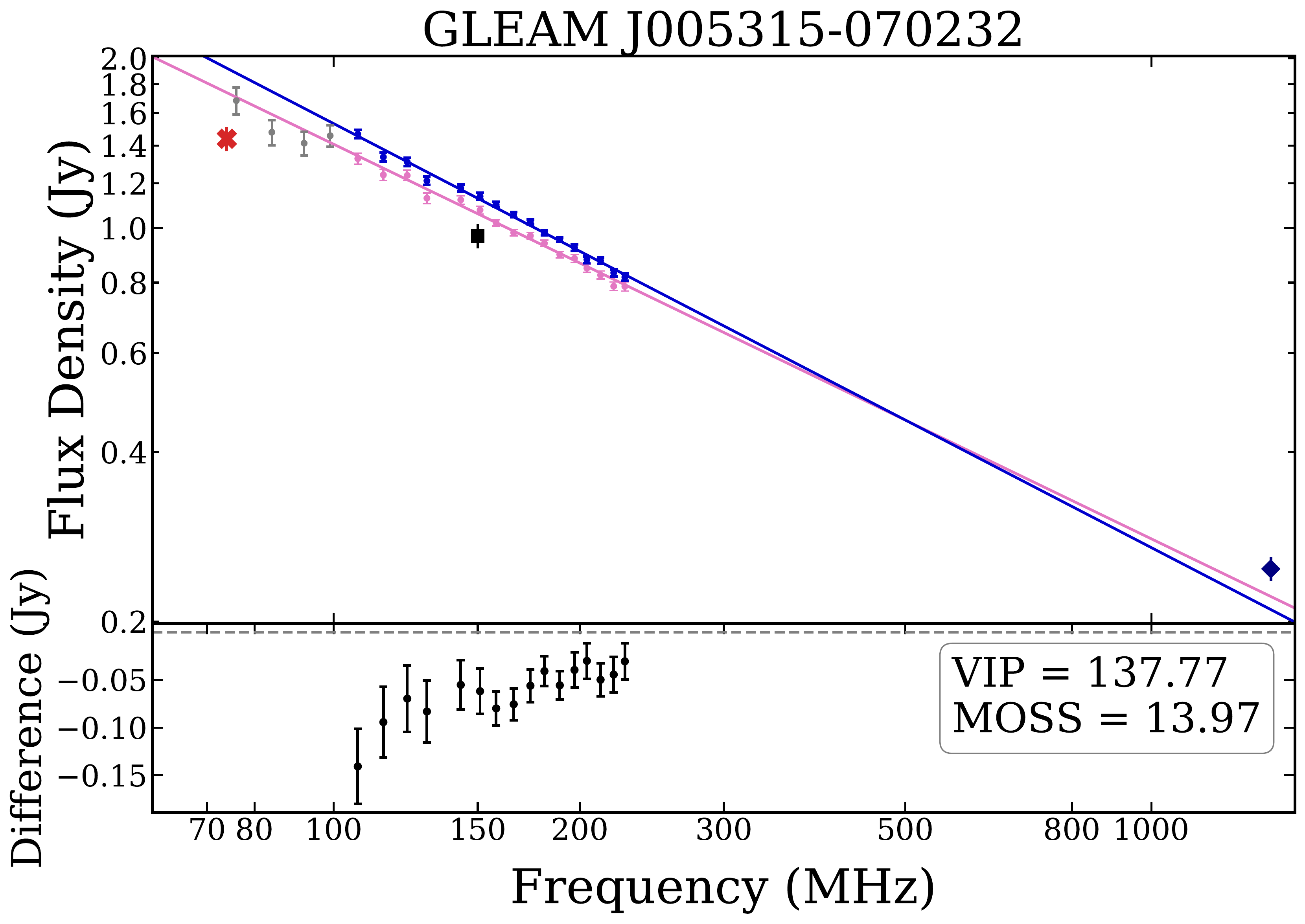} &
\includegraphics[scale=0.15]{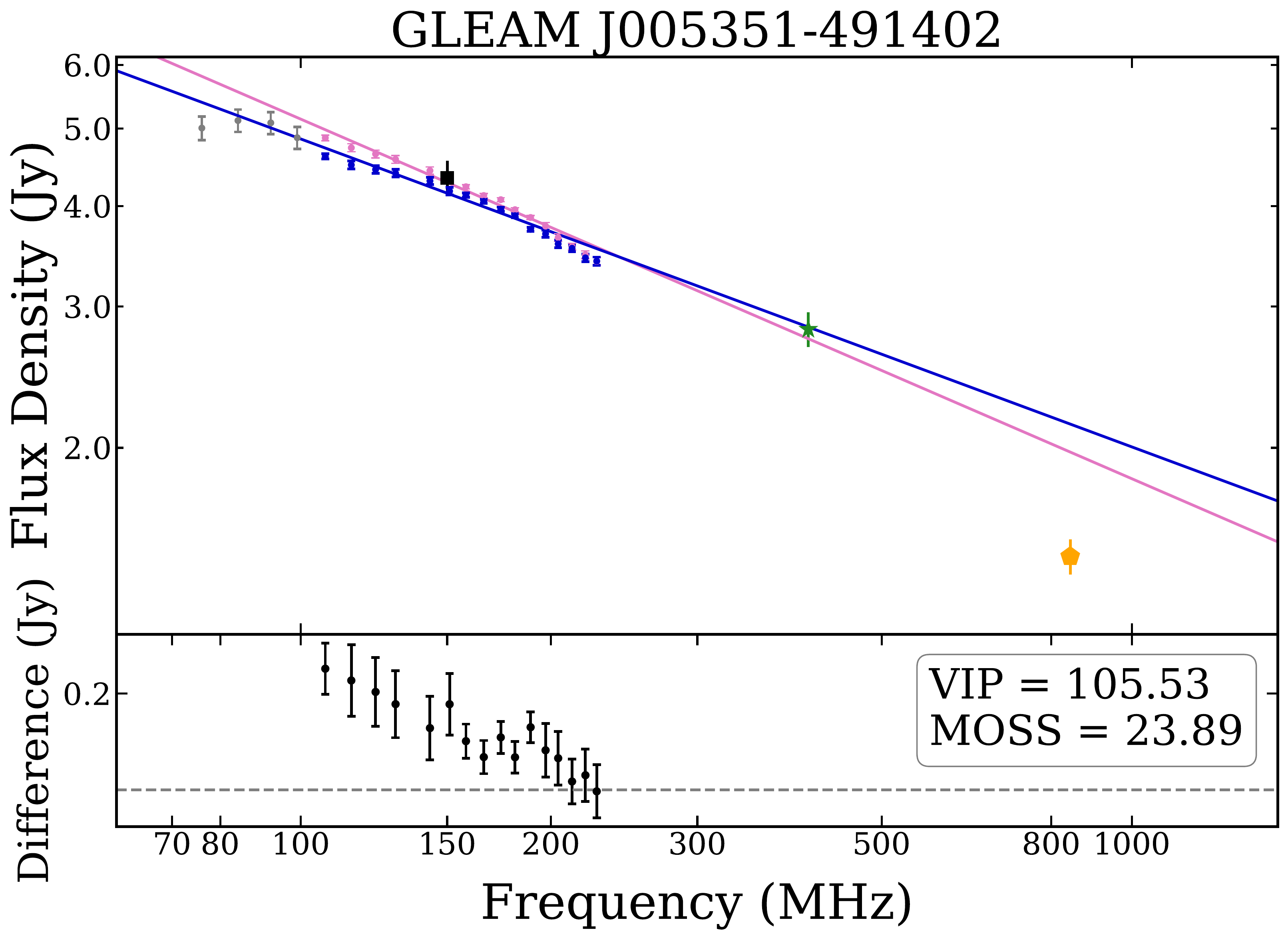} \\
\includegraphics[scale=0.15]{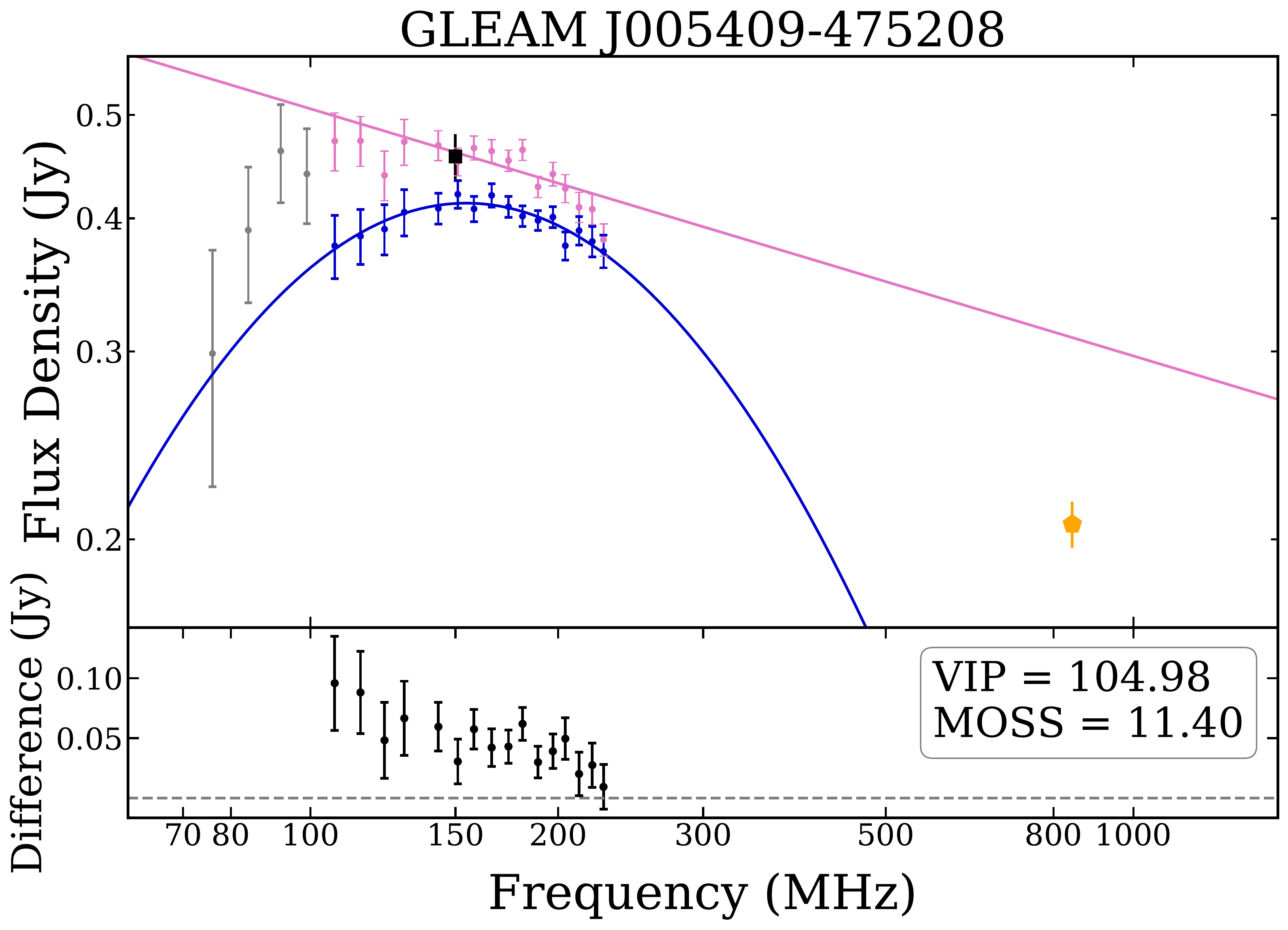} &
\includegraphics[scale=0.15]{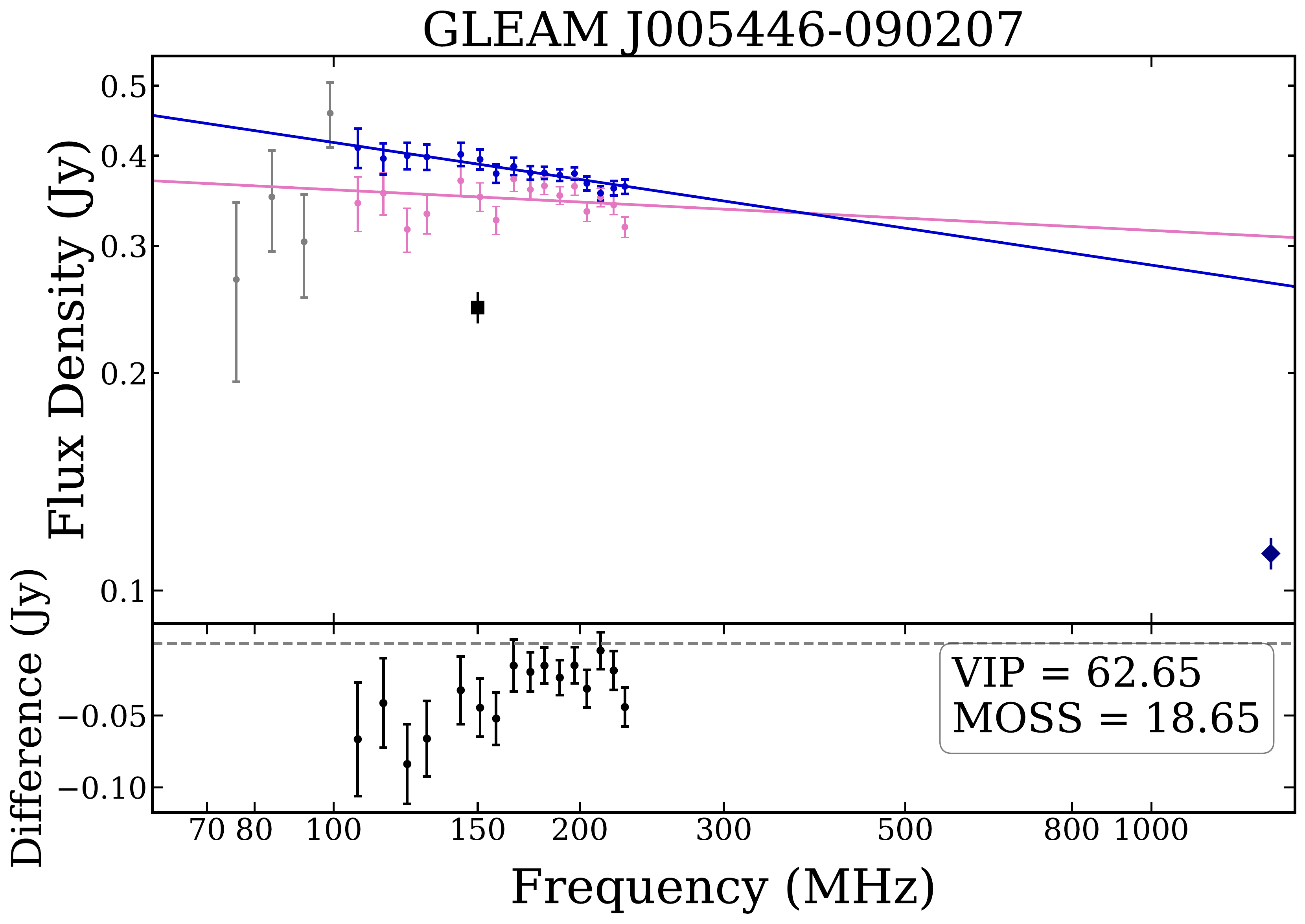} &
\includegraphics[scale=0.15]{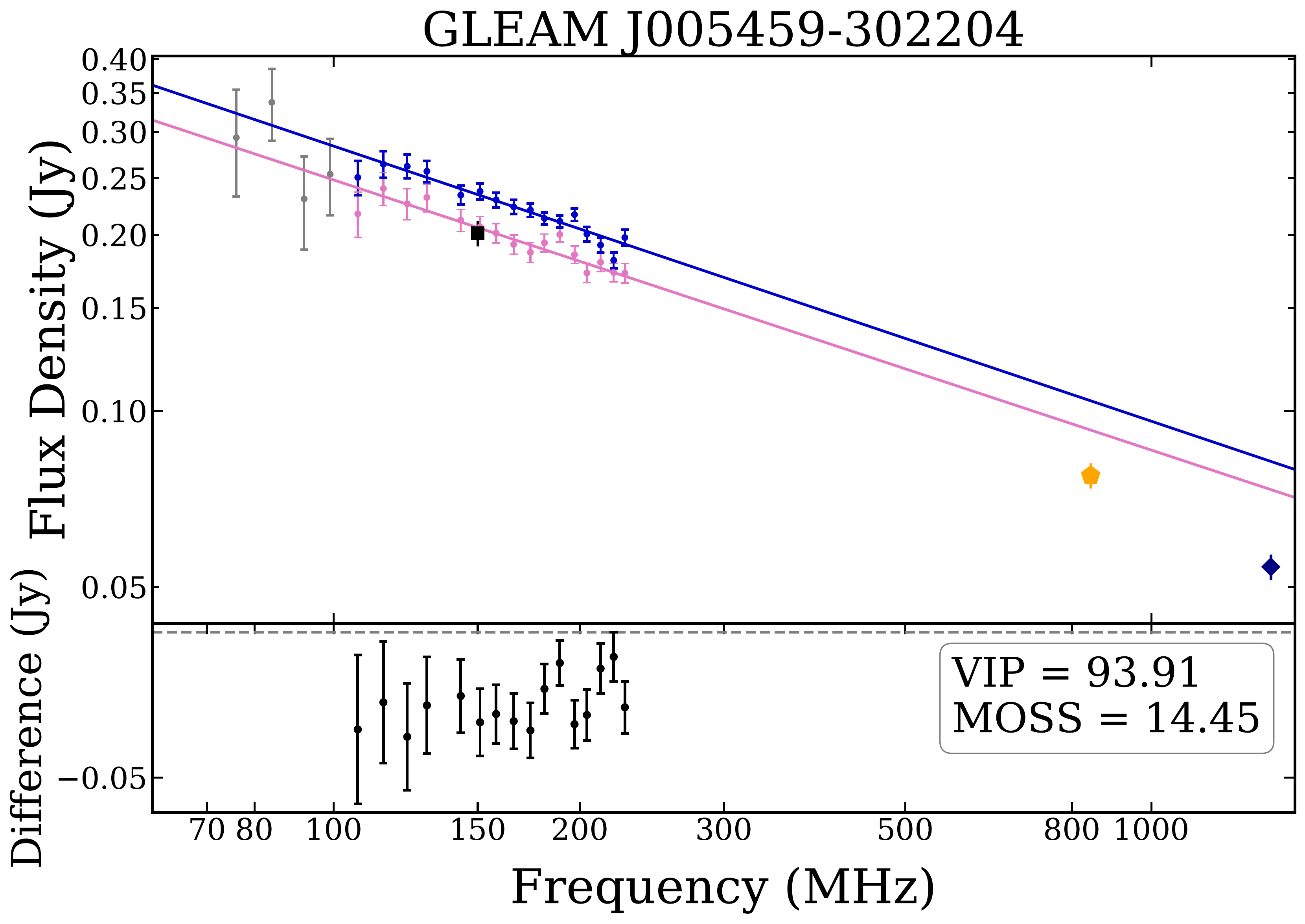} \\
\includegraphics[scale=0.15]{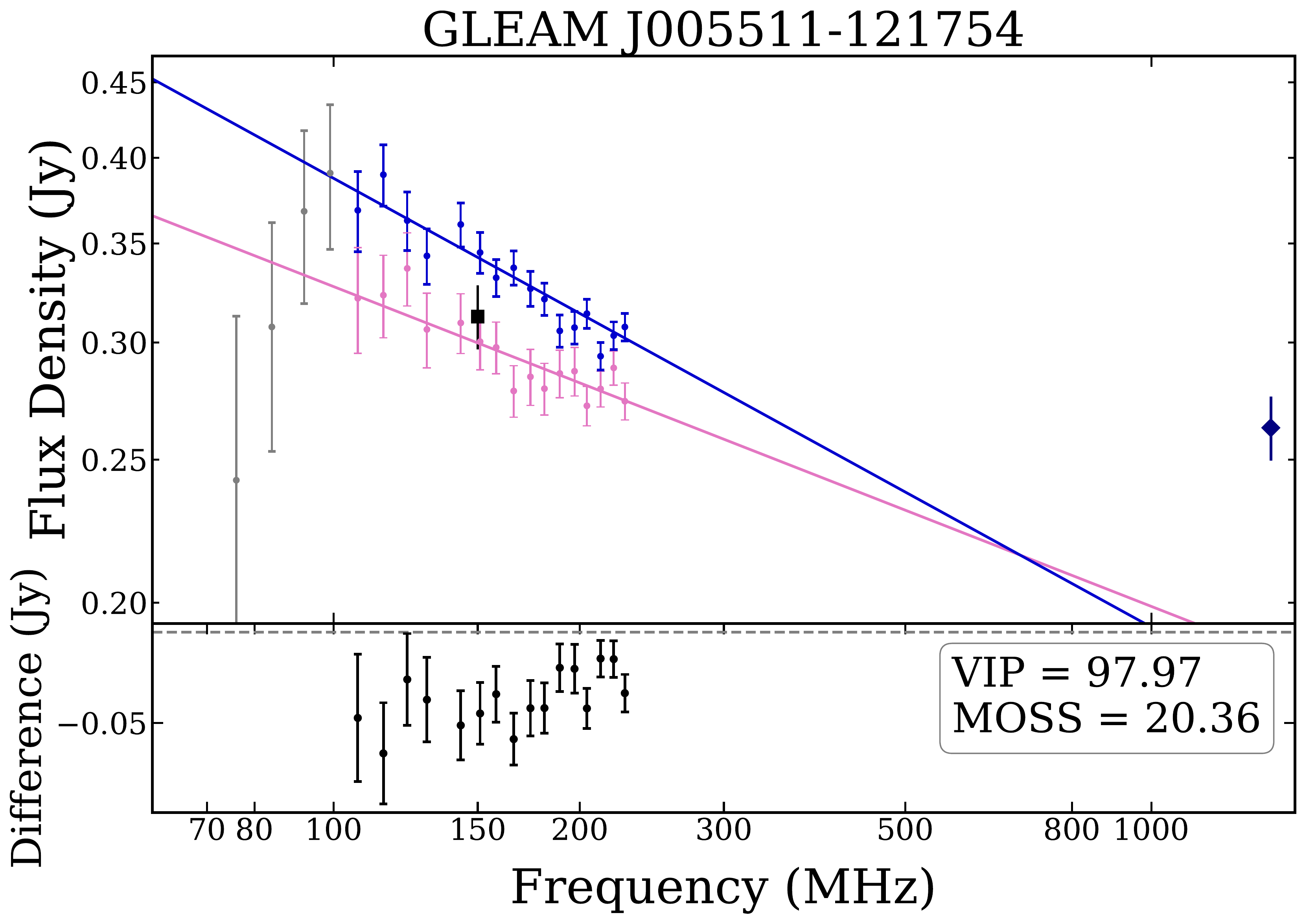} &
\includegraphics[scale=0.15]{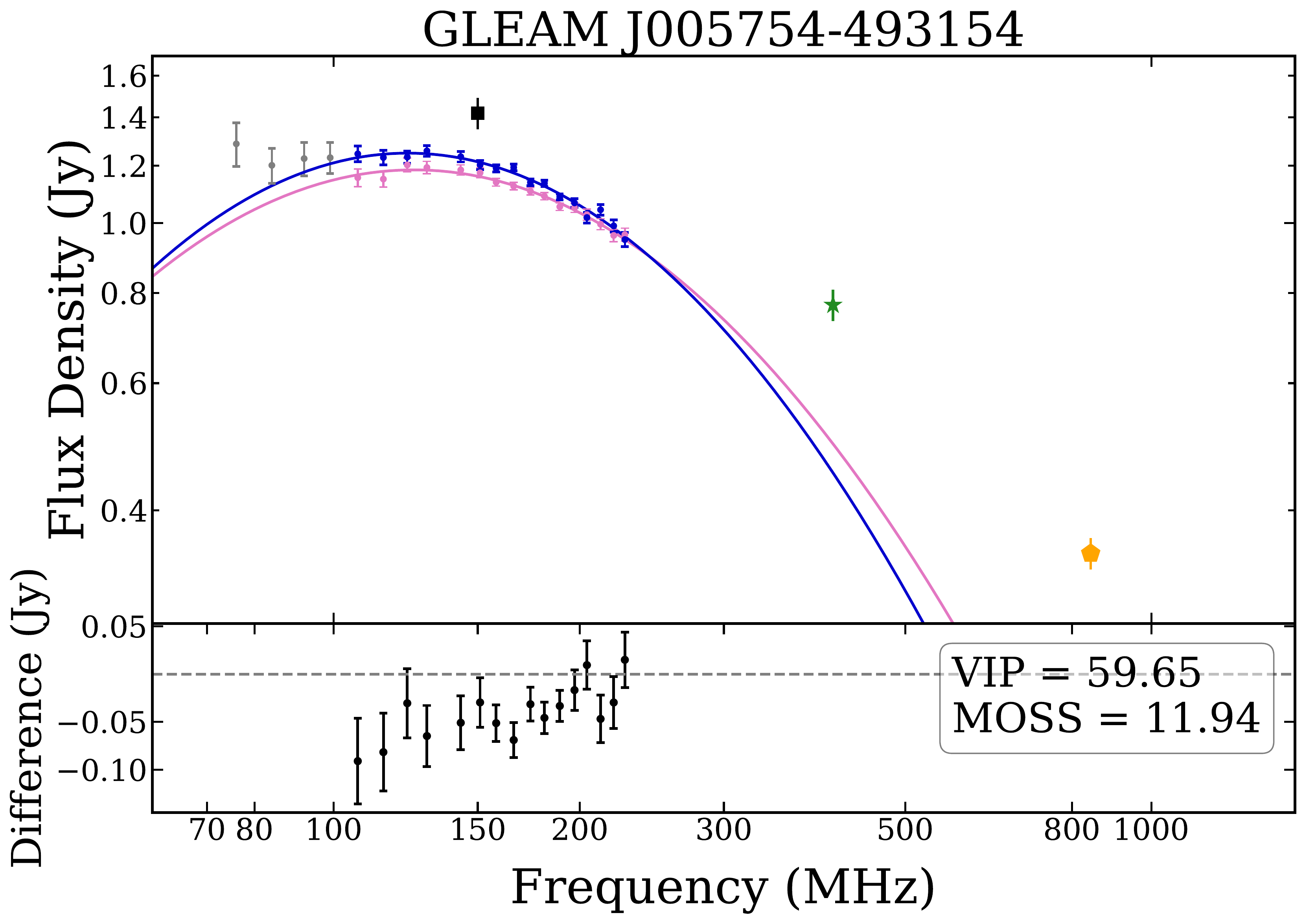} &
\includegraphics[scale=0.15]{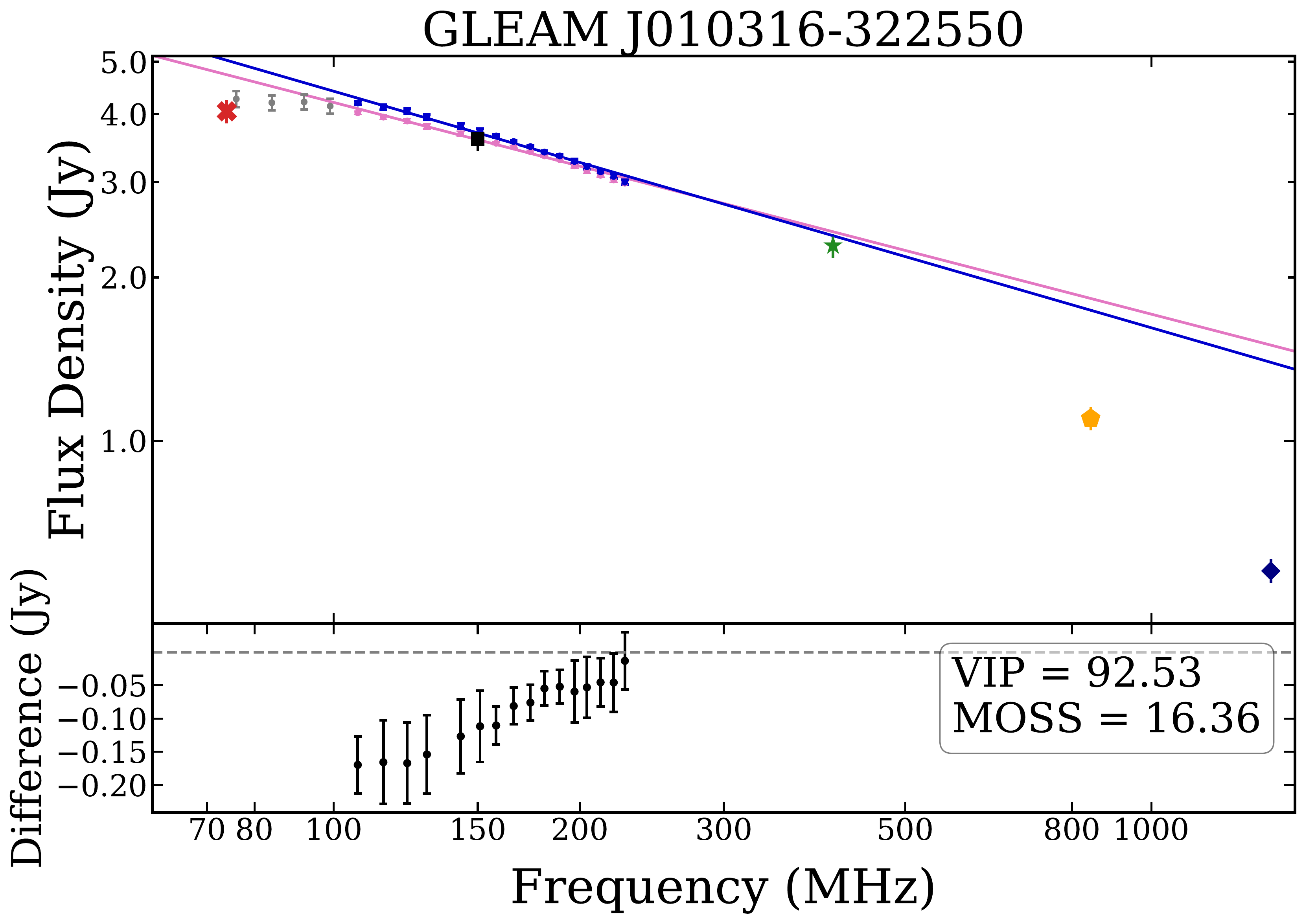} \\
\end{array}$
\caption{(continued) SEDs for all sources classified as variable according to the VIP. For each source the points represent the following data: GLEAM low frequency (72--100\,MHz) (grey circles), Year 1 (pink circles), Year 2 (blue circles), VLSSr (red cross), TGSS (black square), MRC (green star), SUMSS (yellow pentagon), and NVSS (navy diamond). The models for each year are determined by their classification; a source classified with a peak within the observed band was modelled by a quadratic according to Equation~\ref{eq:quadratic}, remaining sources were modelled by a power-law according to Equation~\ref{eq:plaw}.}
\label{app:fig:pg2}
\end{center}
\end{figure*}
\setcounter{figure}{0}
\begin{figure*}
\begin{center}$
\begin{array}{cccccc}
\includegraphics[scale=0.15]{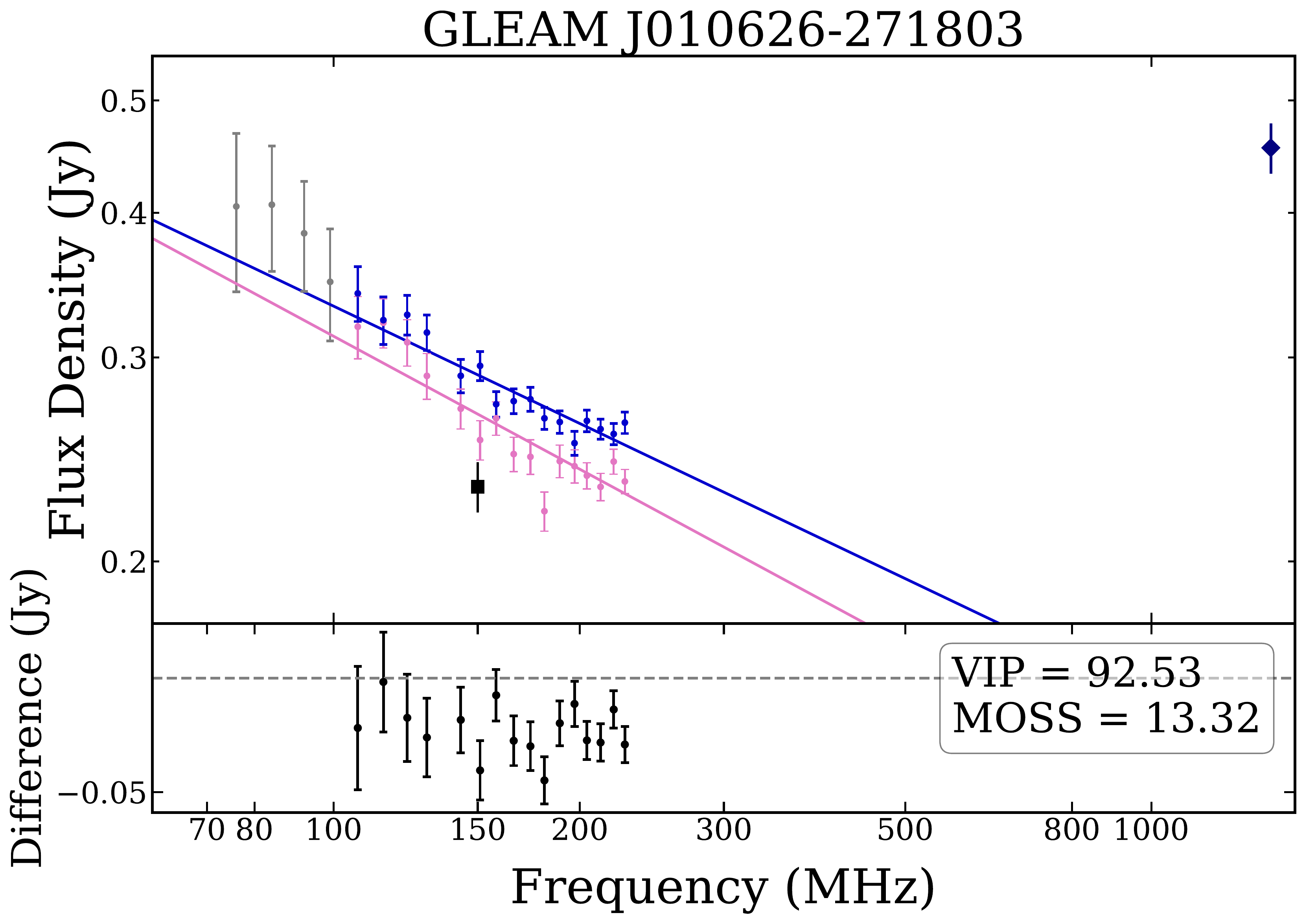} &
\includegraphics[scale=0.15]{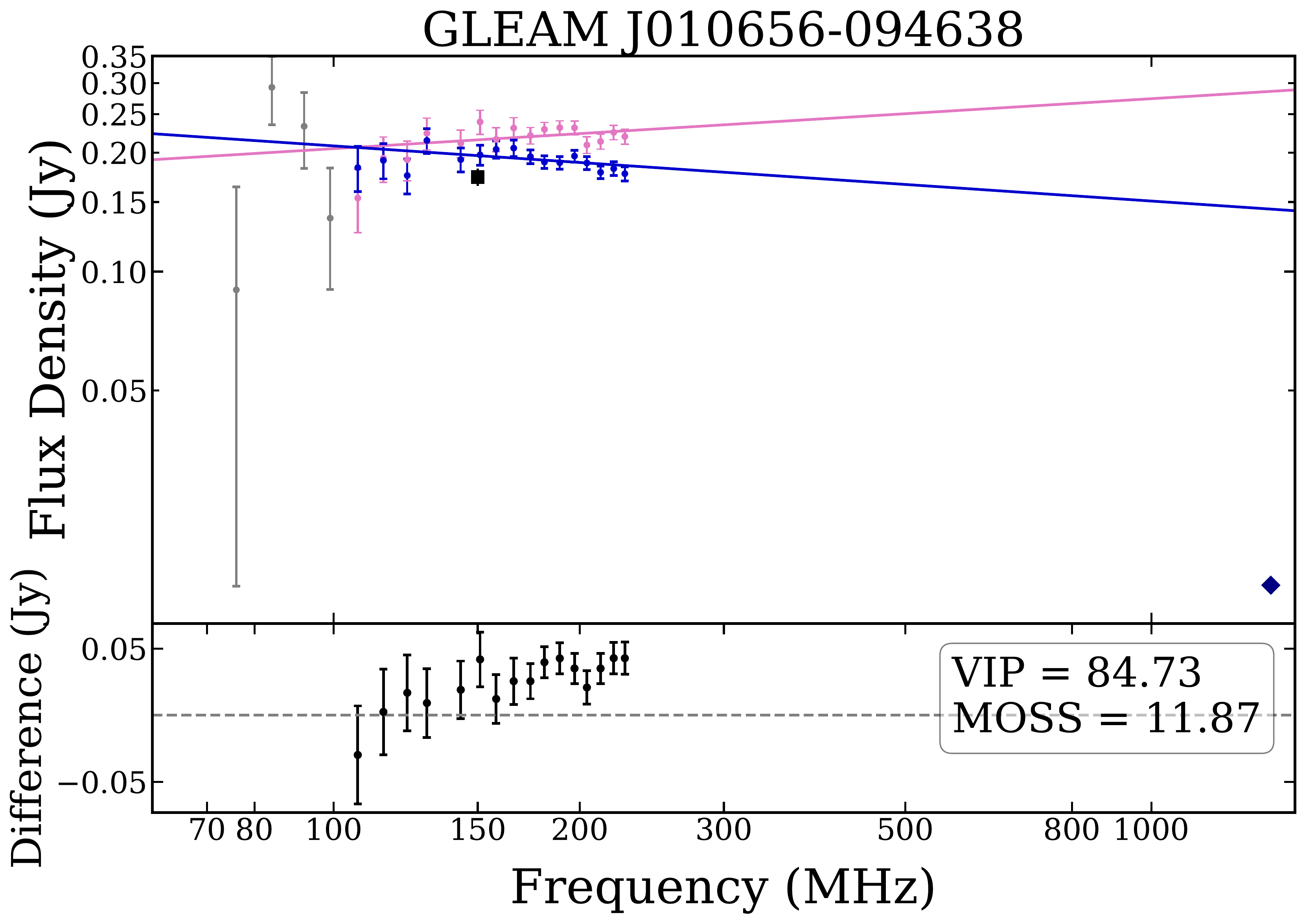} &
\includegraphics[scale=0.15]{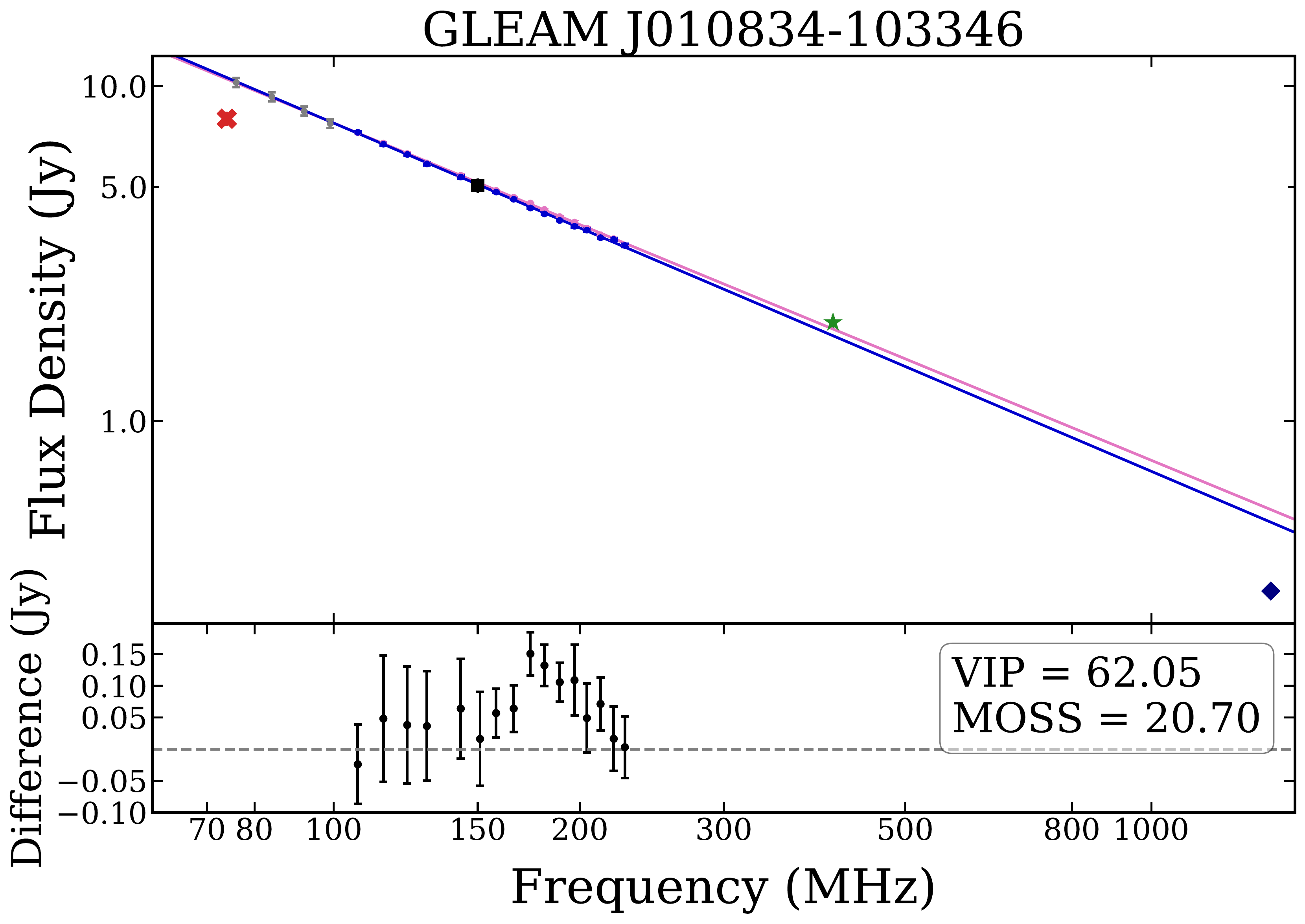} \\
\includegraphics[scale=0.15]{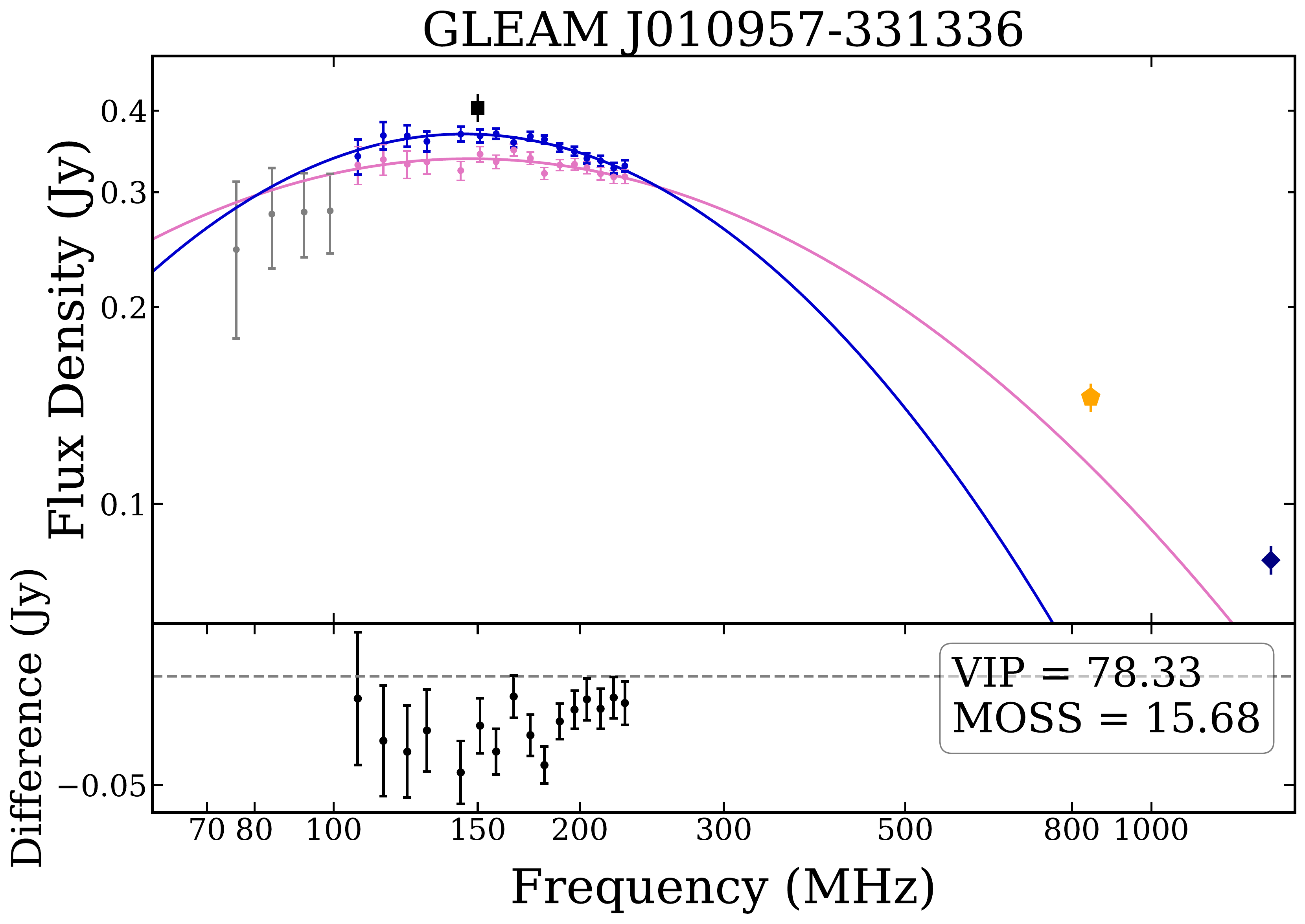} &
\includegraphics[scale=0.15]{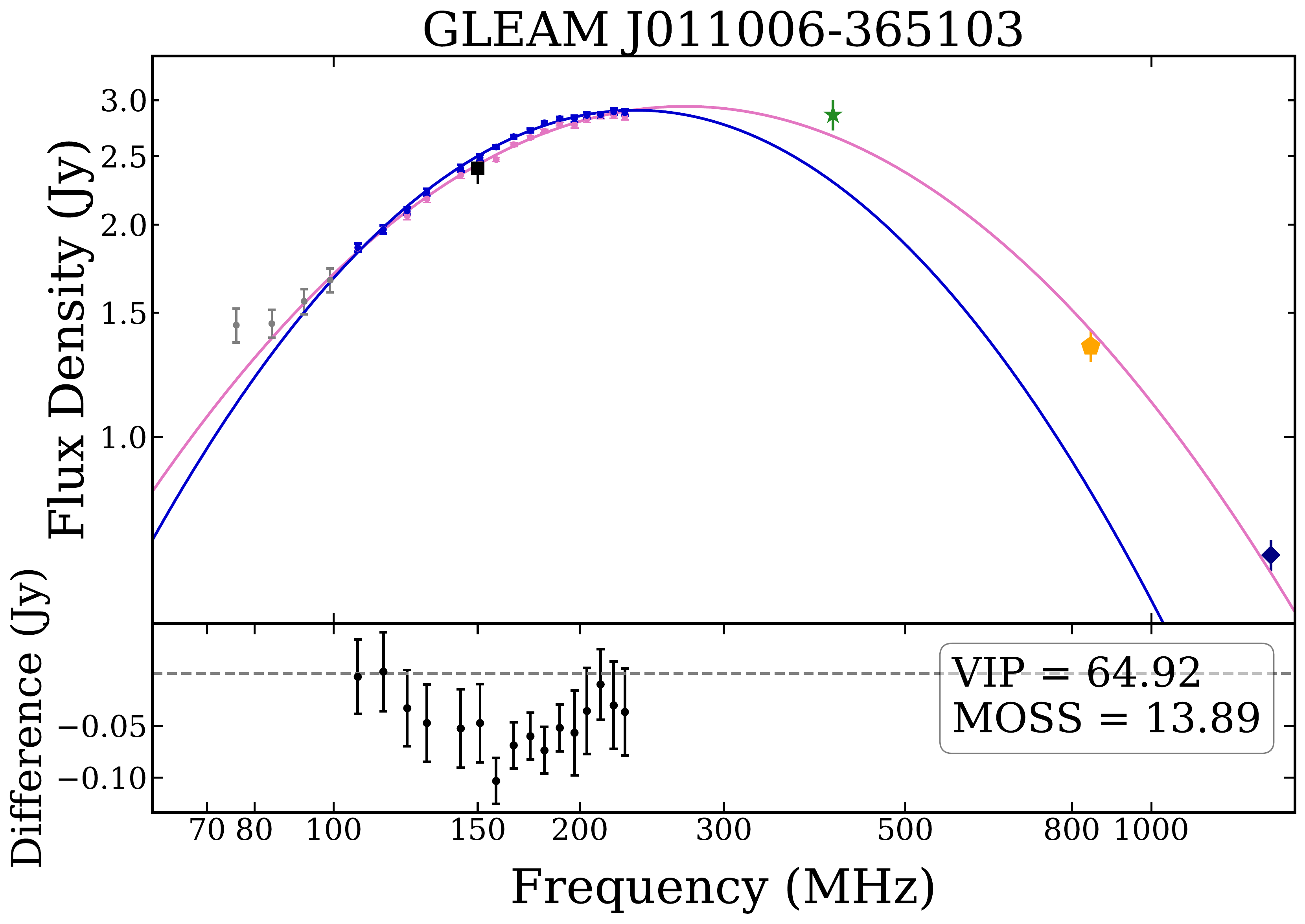} &
\includegraphics[scale=0.15]{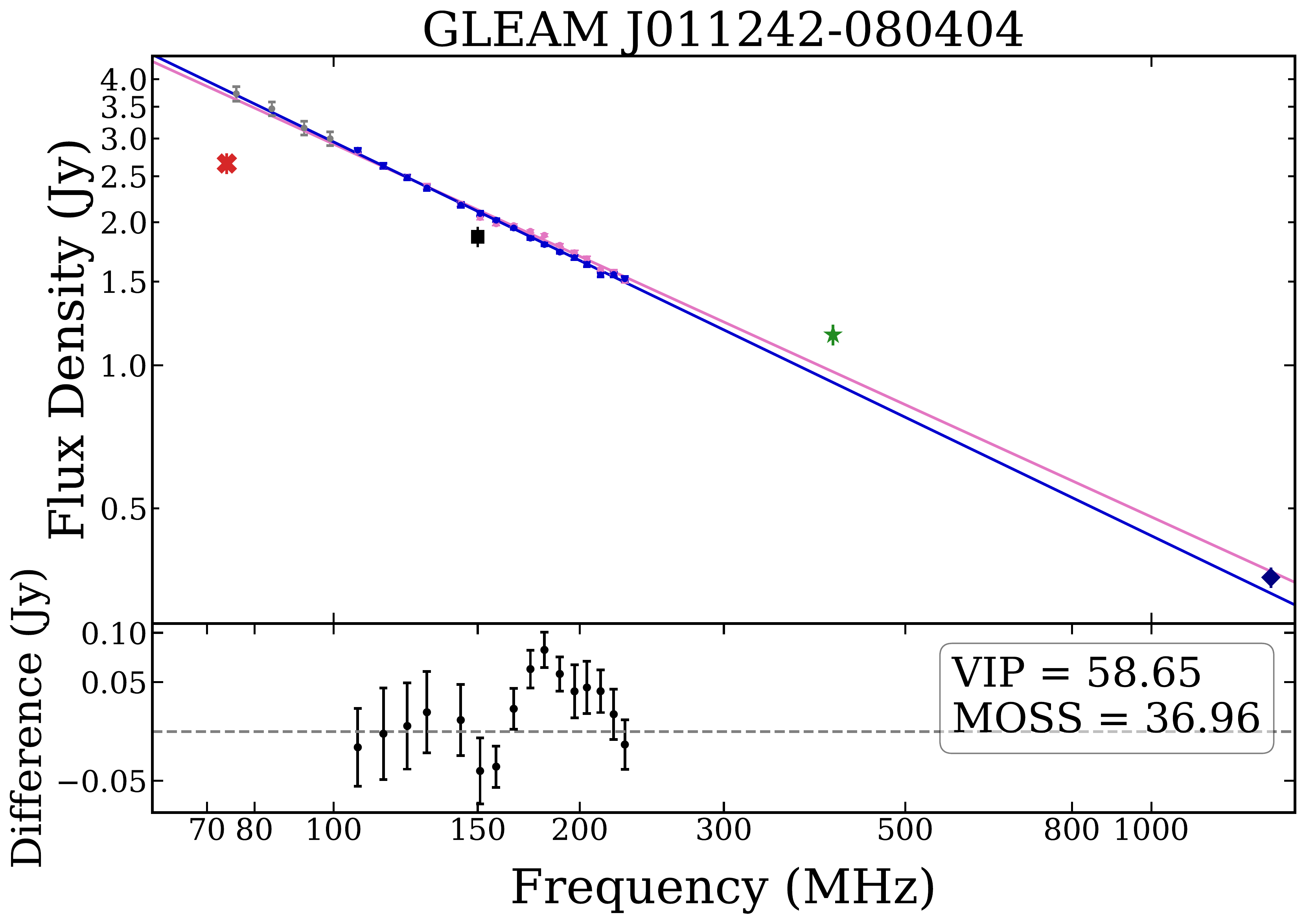} \\
\includegraphics[scale=0.15]{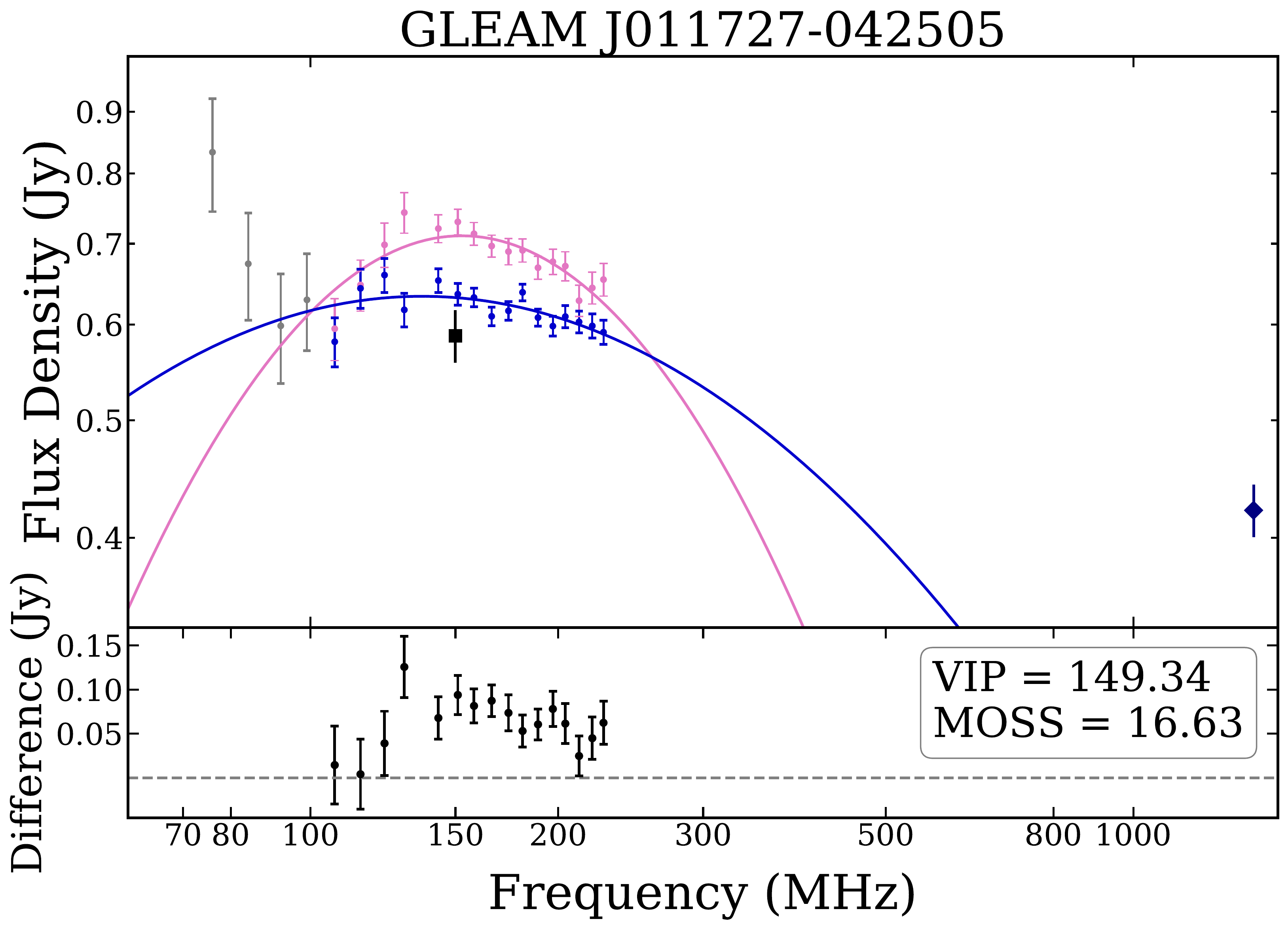} &
\includegraphics[scale=0.15]{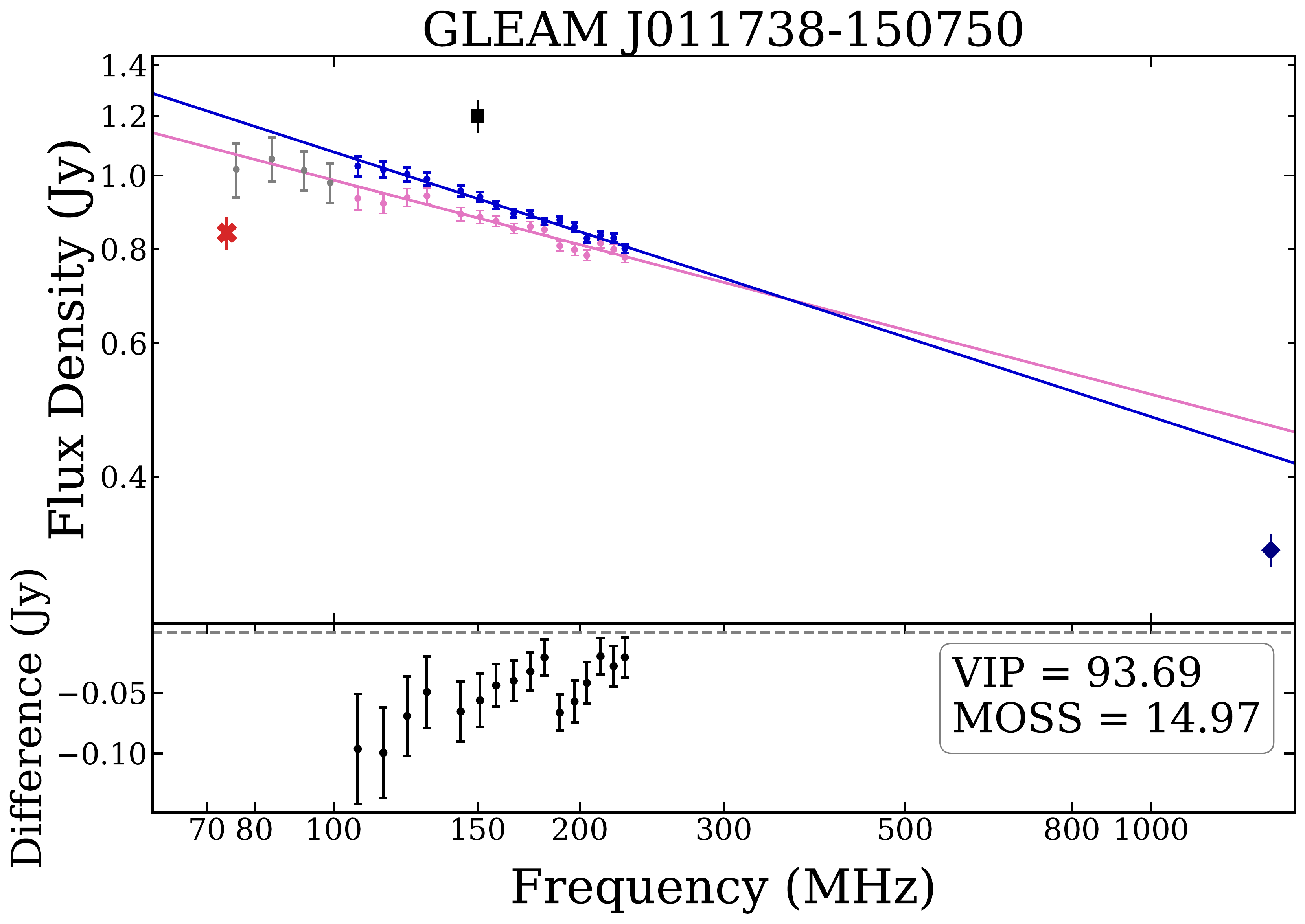} &
\includegraphics[scale=0.15]{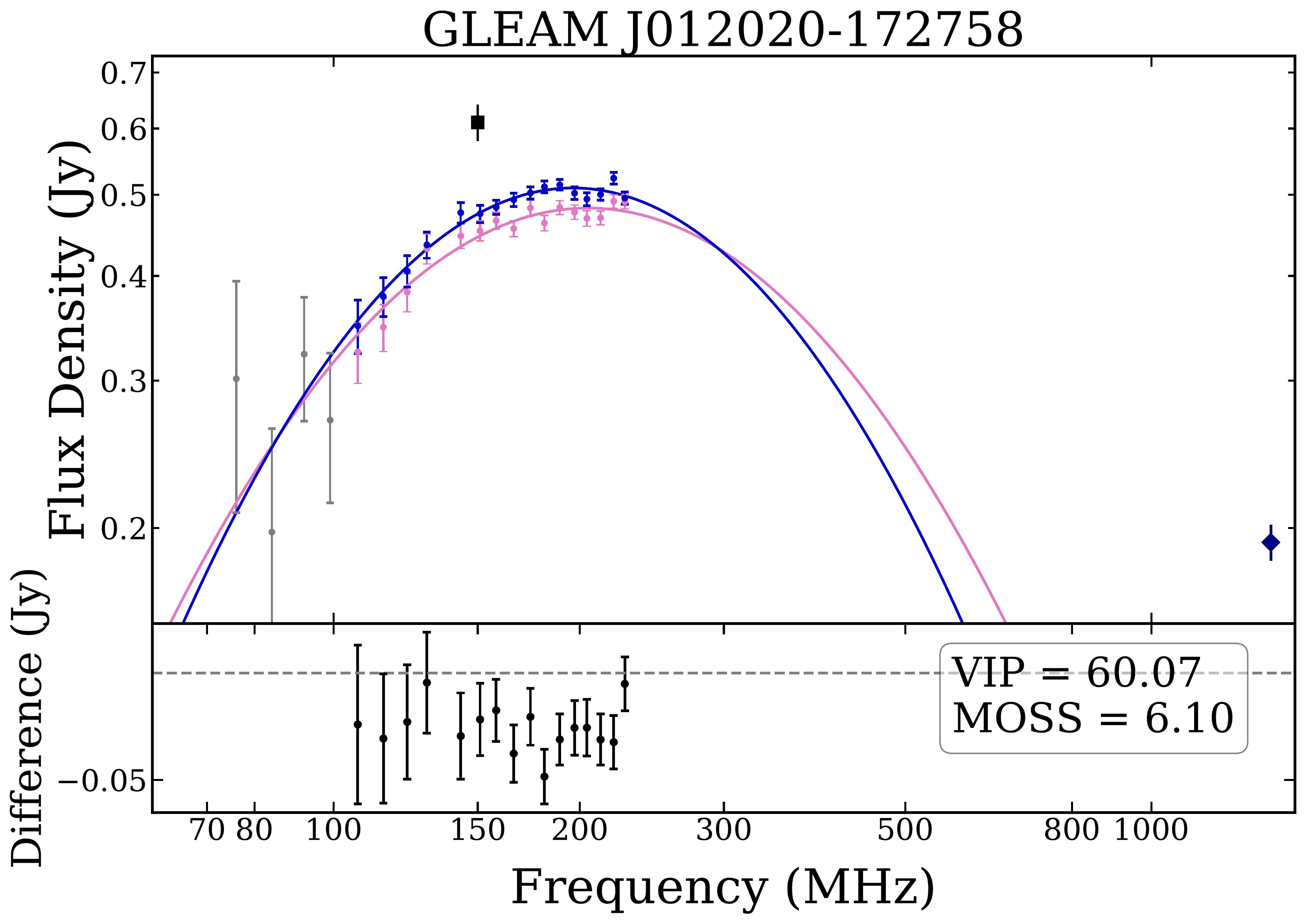} \\
\includegraphics[scale=0.15]{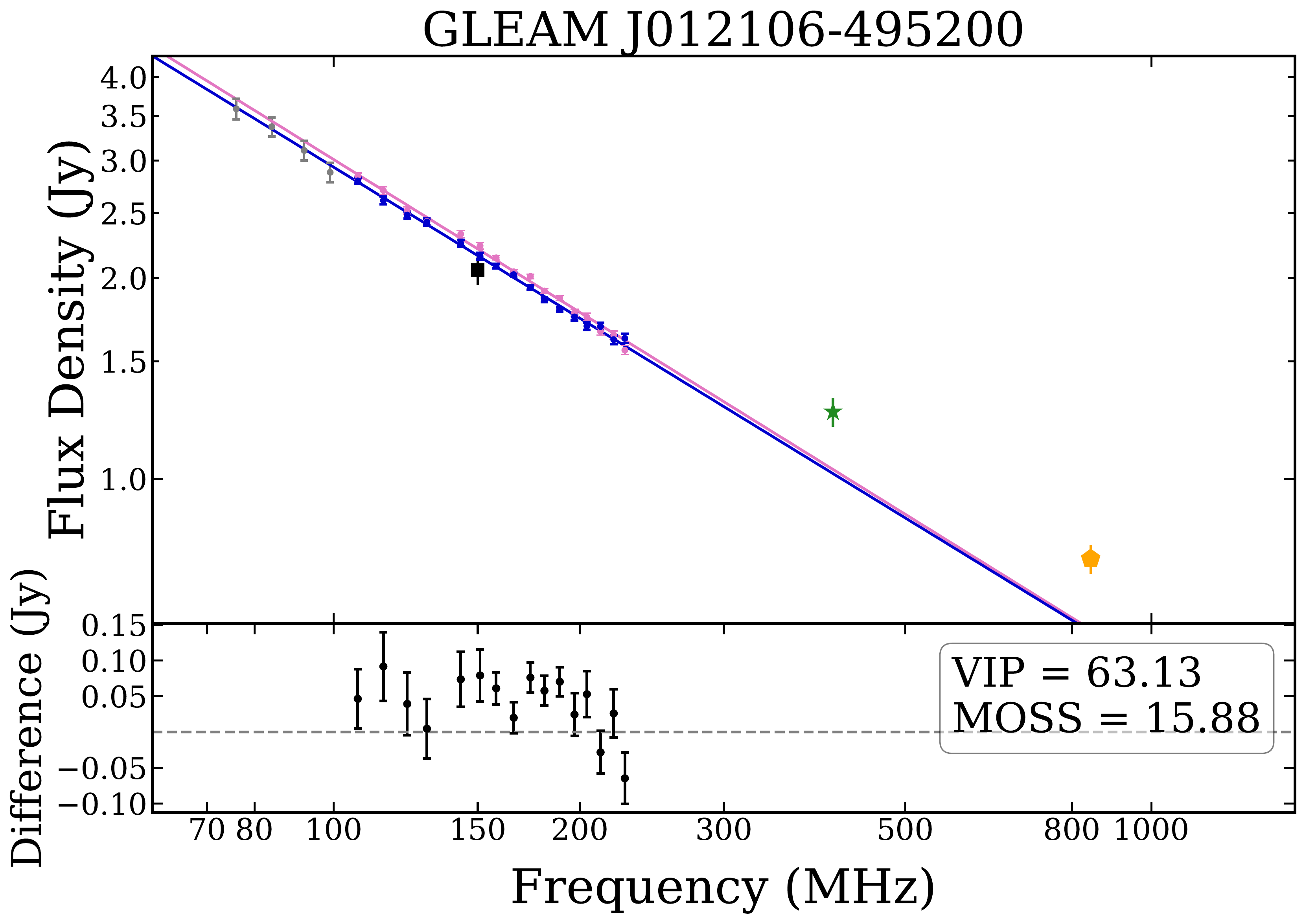} &
\includegraphics[scale=0.15]{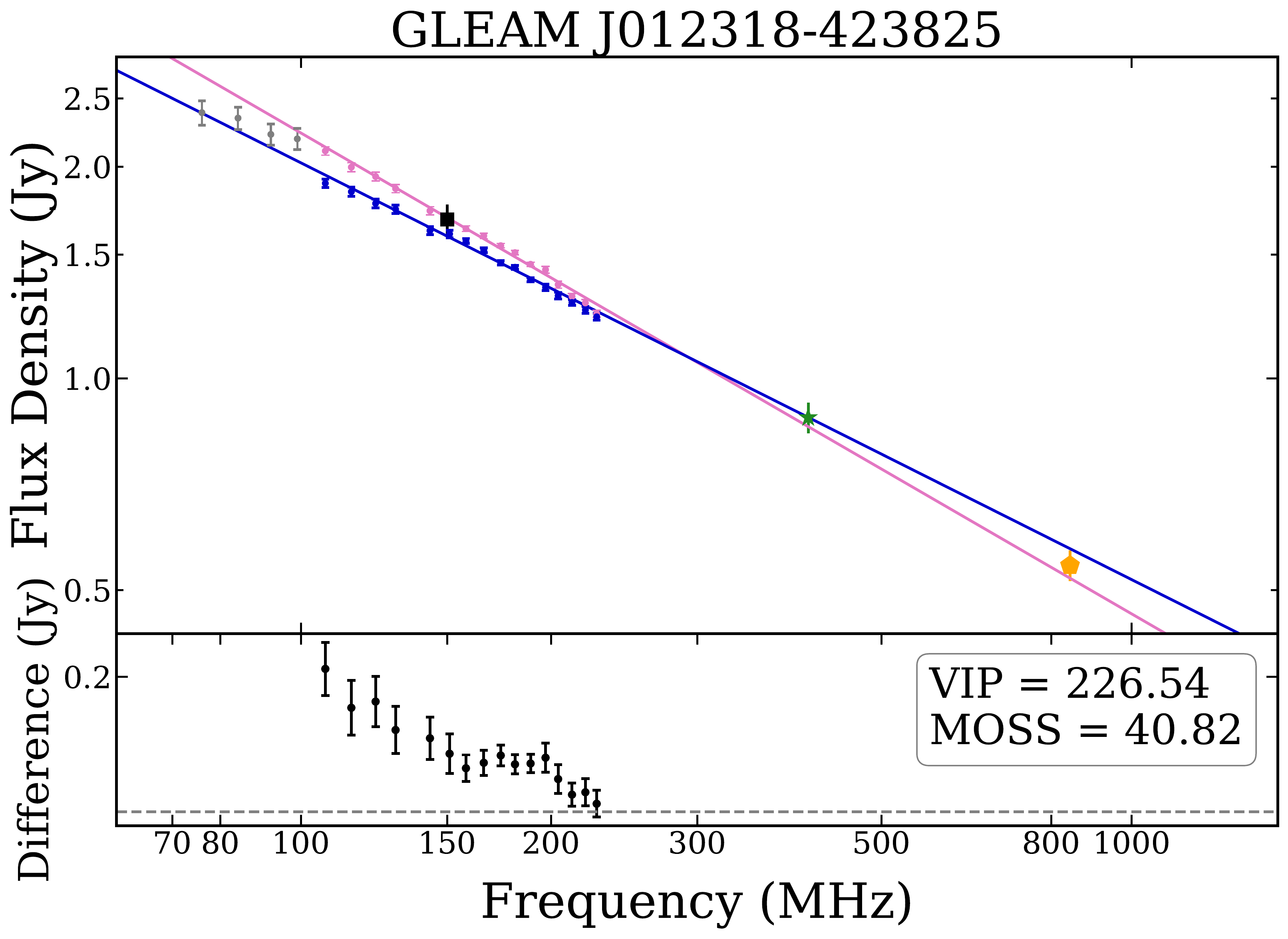} &
\includegraphics[scale=0.15]{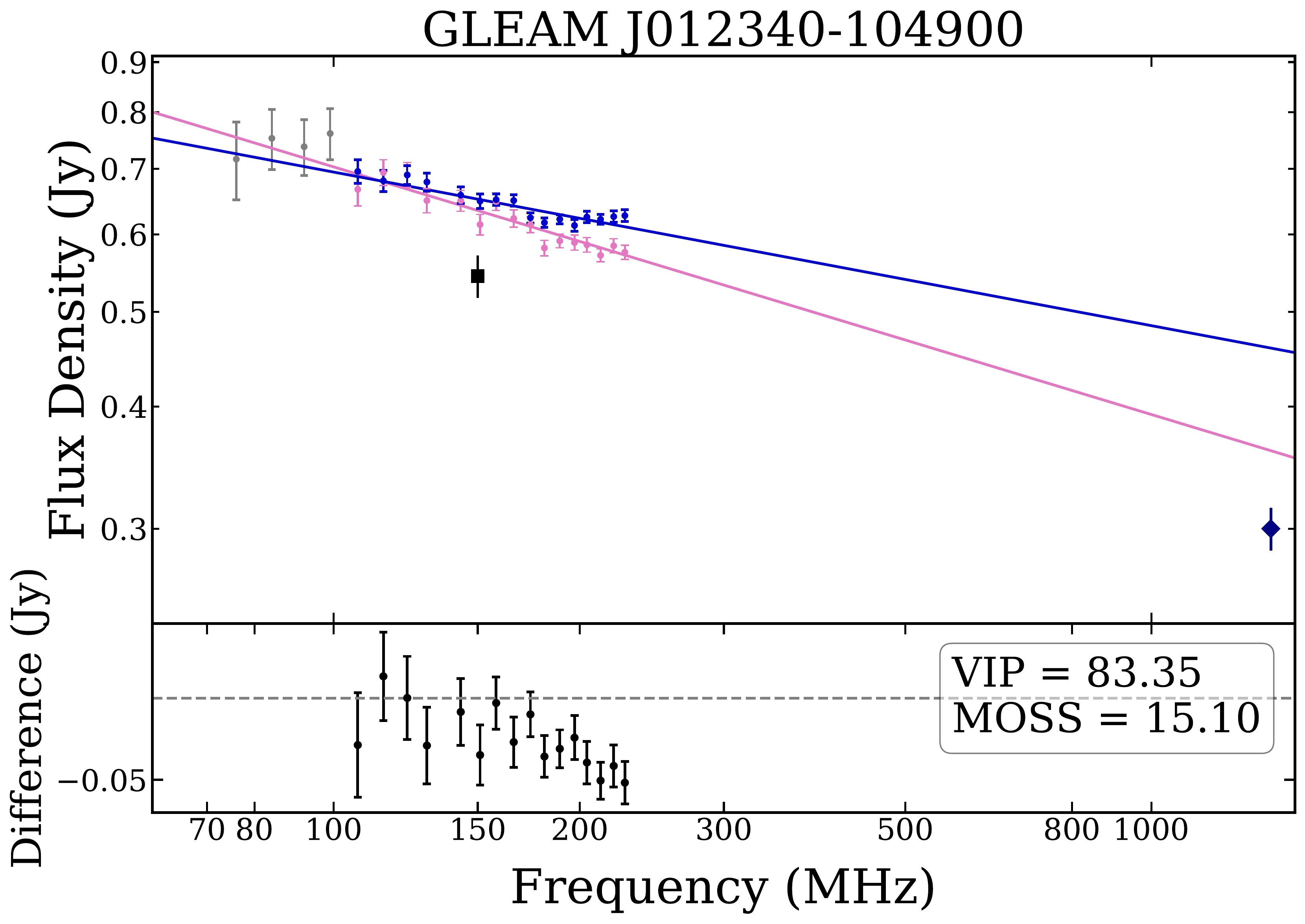} \\
\includegraphics[scale=0.15]{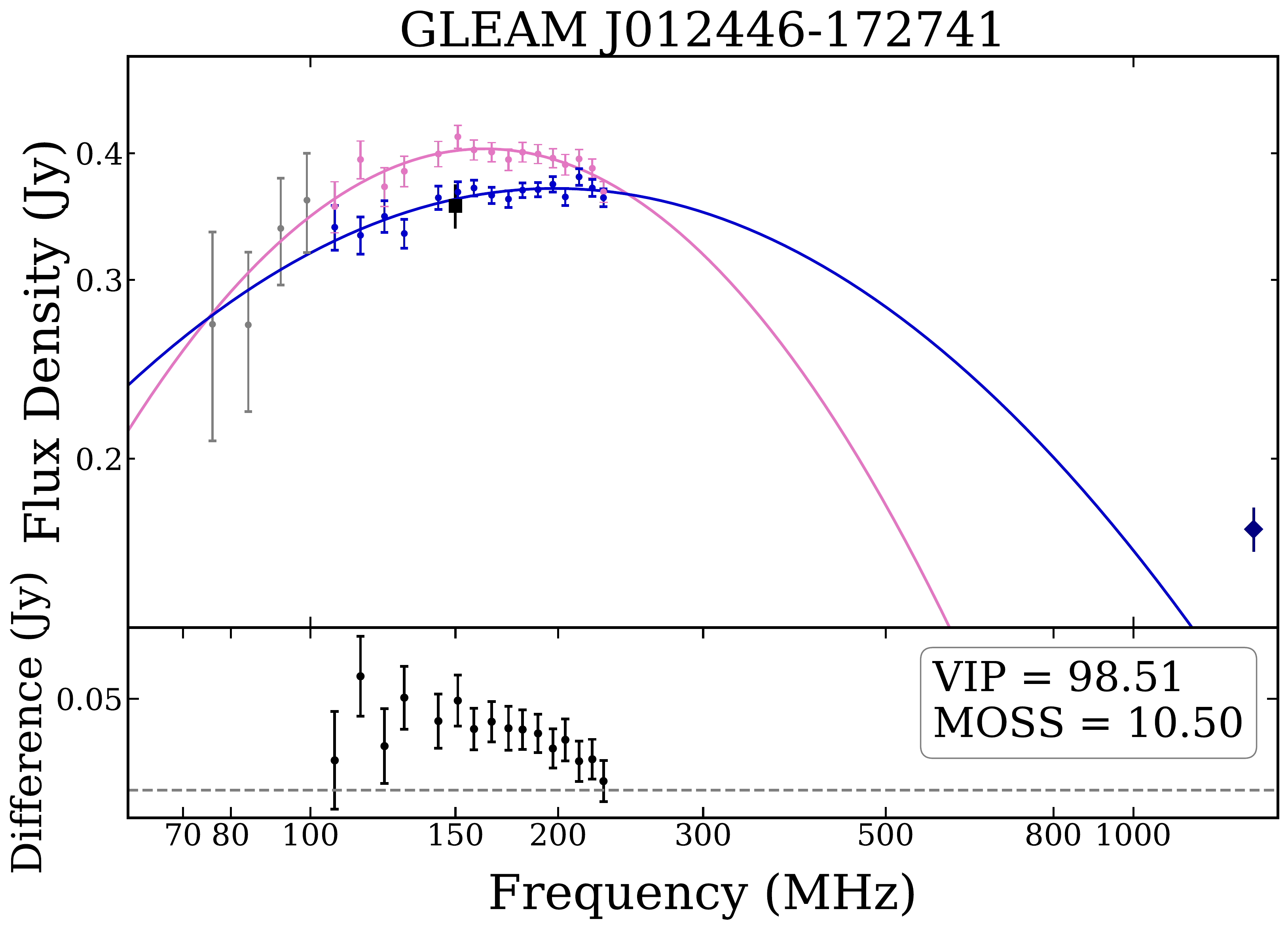} &
\includegraphics[scale=0.15]{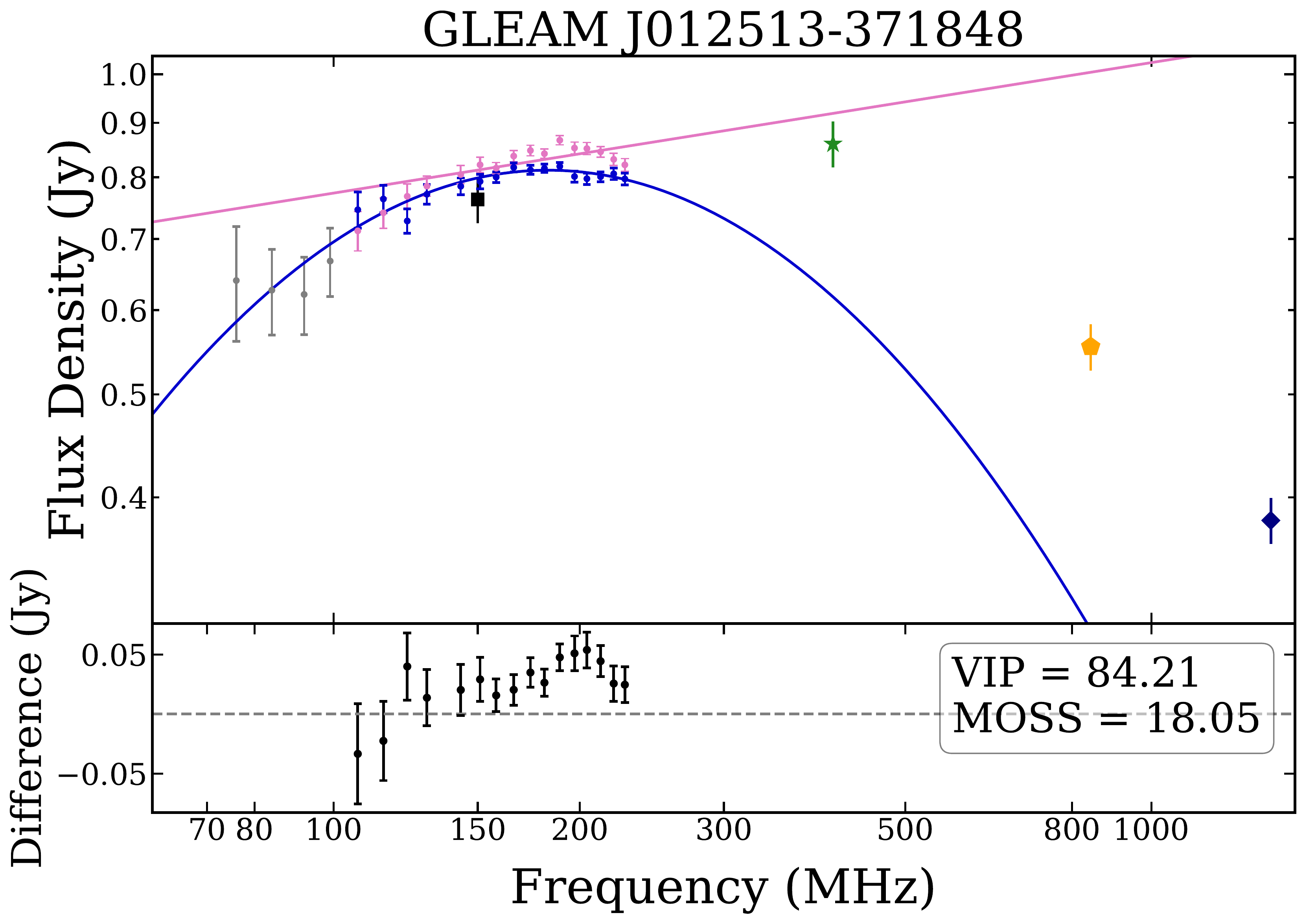} &
\includegraphics[scale=0.15]{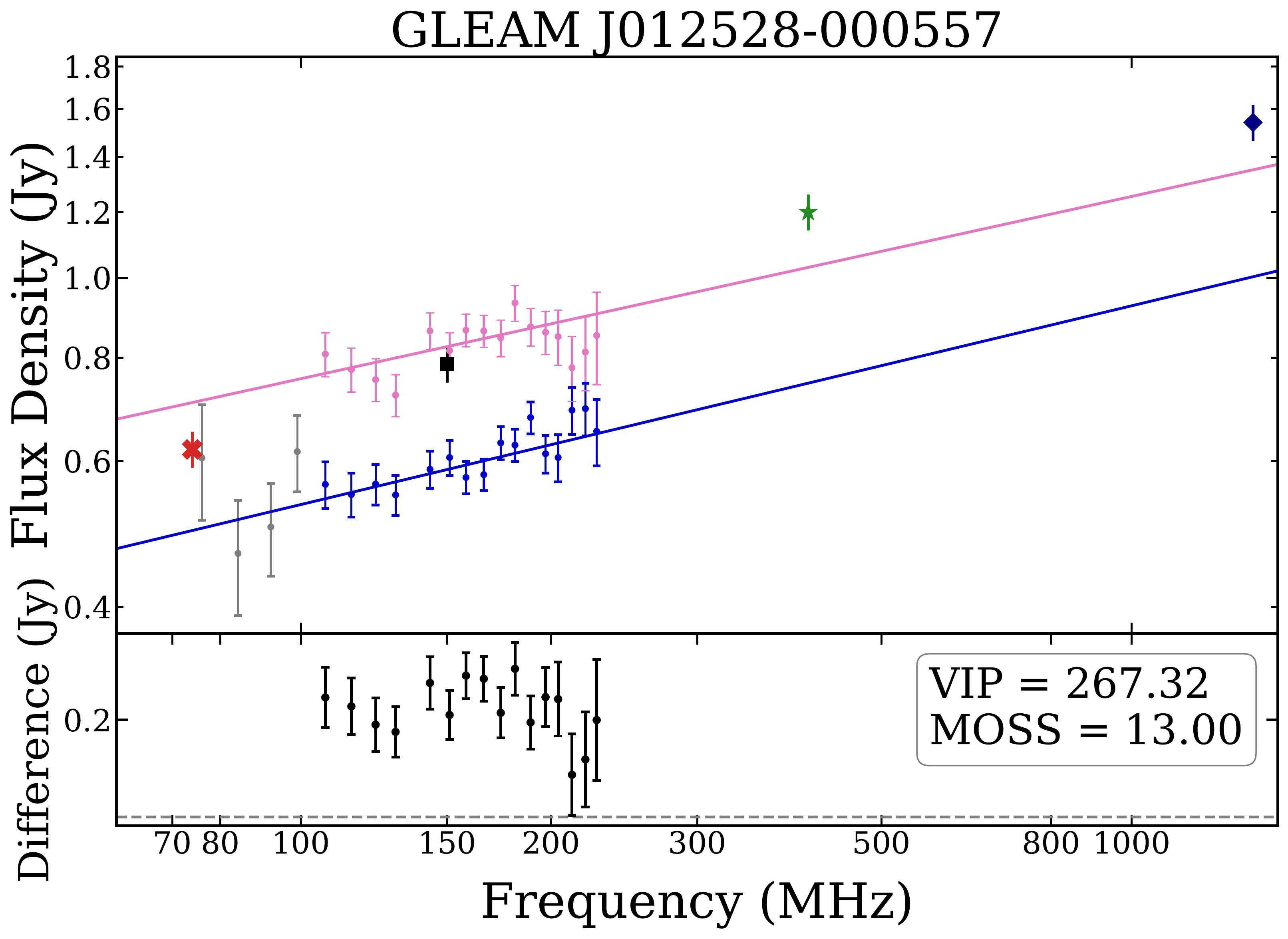} \\
\includegraphics[scale=0.15]{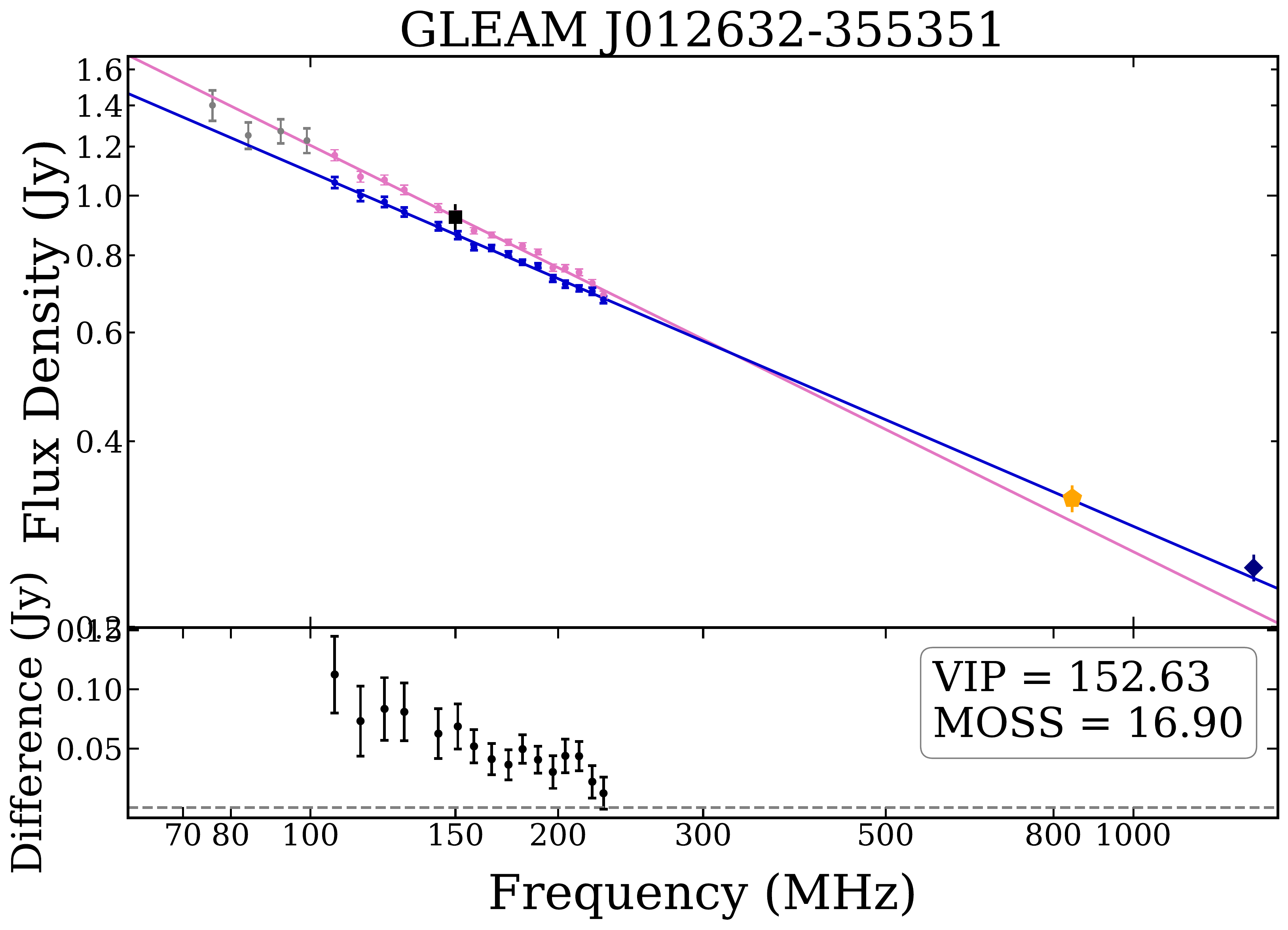} &
\includegraphics[scale=0.15]{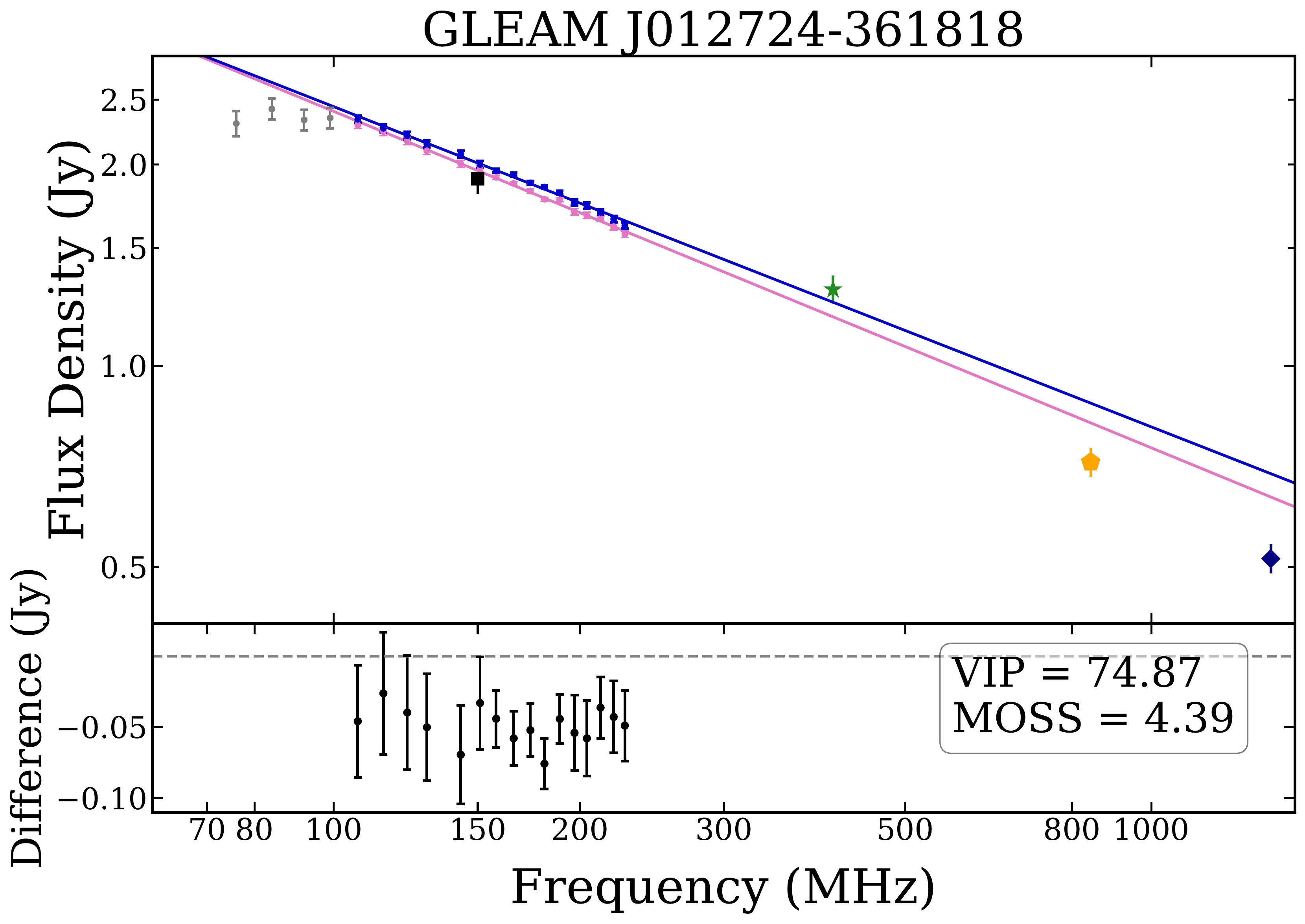} &
\includegraphics[scale=0.15]{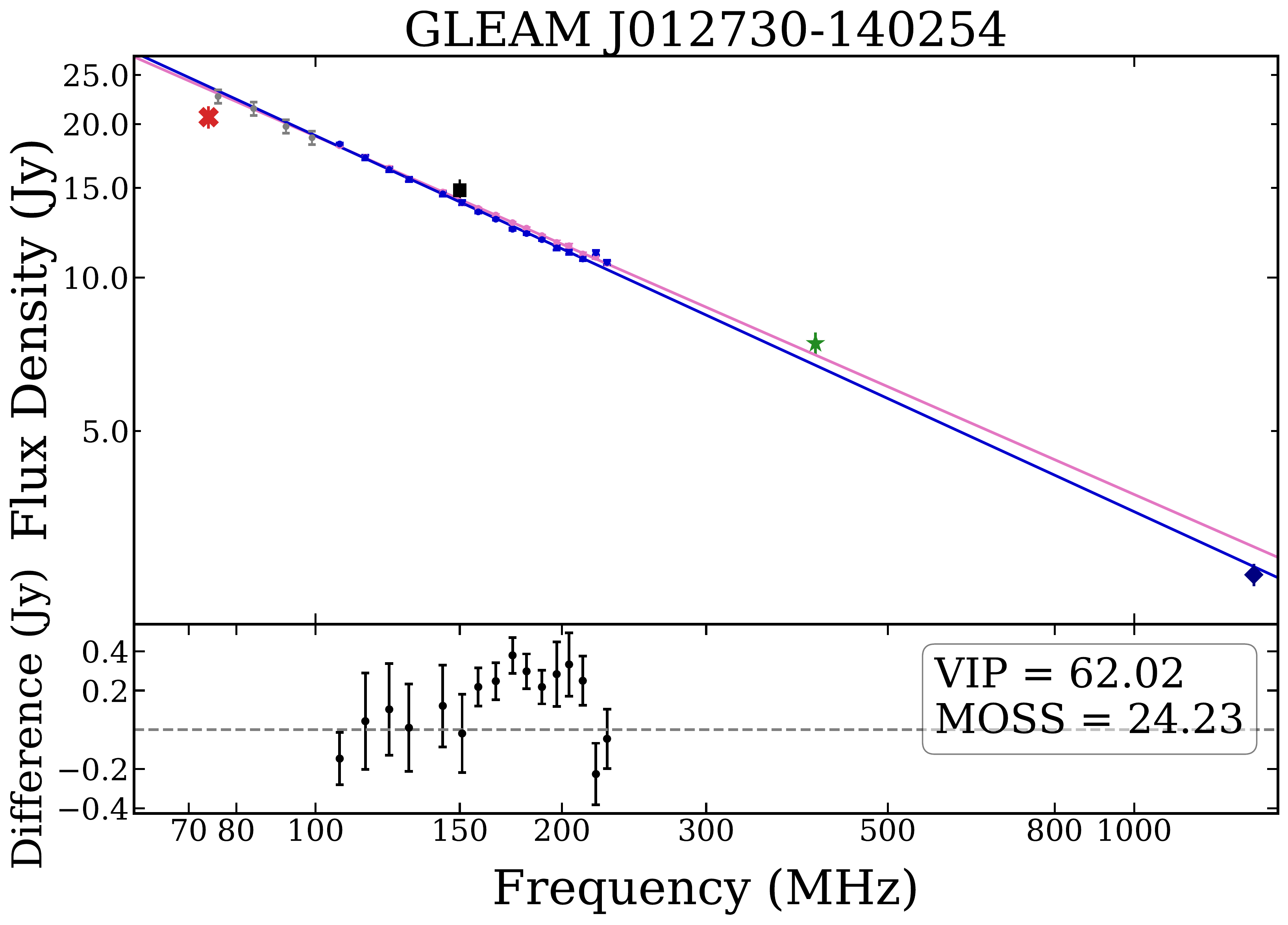} \\
\end{array}$
\caption{(continued) SEDs for all sources classified as variable according to the VIP. For each source the points represent the following data: GLEAM low frequency (72--100\,MHz) (grey circles), Year 1 (pink circles), Year 2 (blue circles), VLSSr (red cross), TGSS (black square), MRC (green star), SUMSS (yellow pentagon), and NVSS (navy diamond). The models for each year are determined by their classification; a source classified with a peak within the observed band was modelled by a quadratic according to Equation~\ref{eq:quadratic}, remaining sources were modelled by a power-law according to Equation~\ref{eq:plaw}.}
\label{app:fig:pg3}
\end{center}
\end{figure*}
\setcounter{figure}{0}
\begin{figure*}
\begin{center}$
\begin{array}{cccccc}
\includegraphics[scale=0.15]{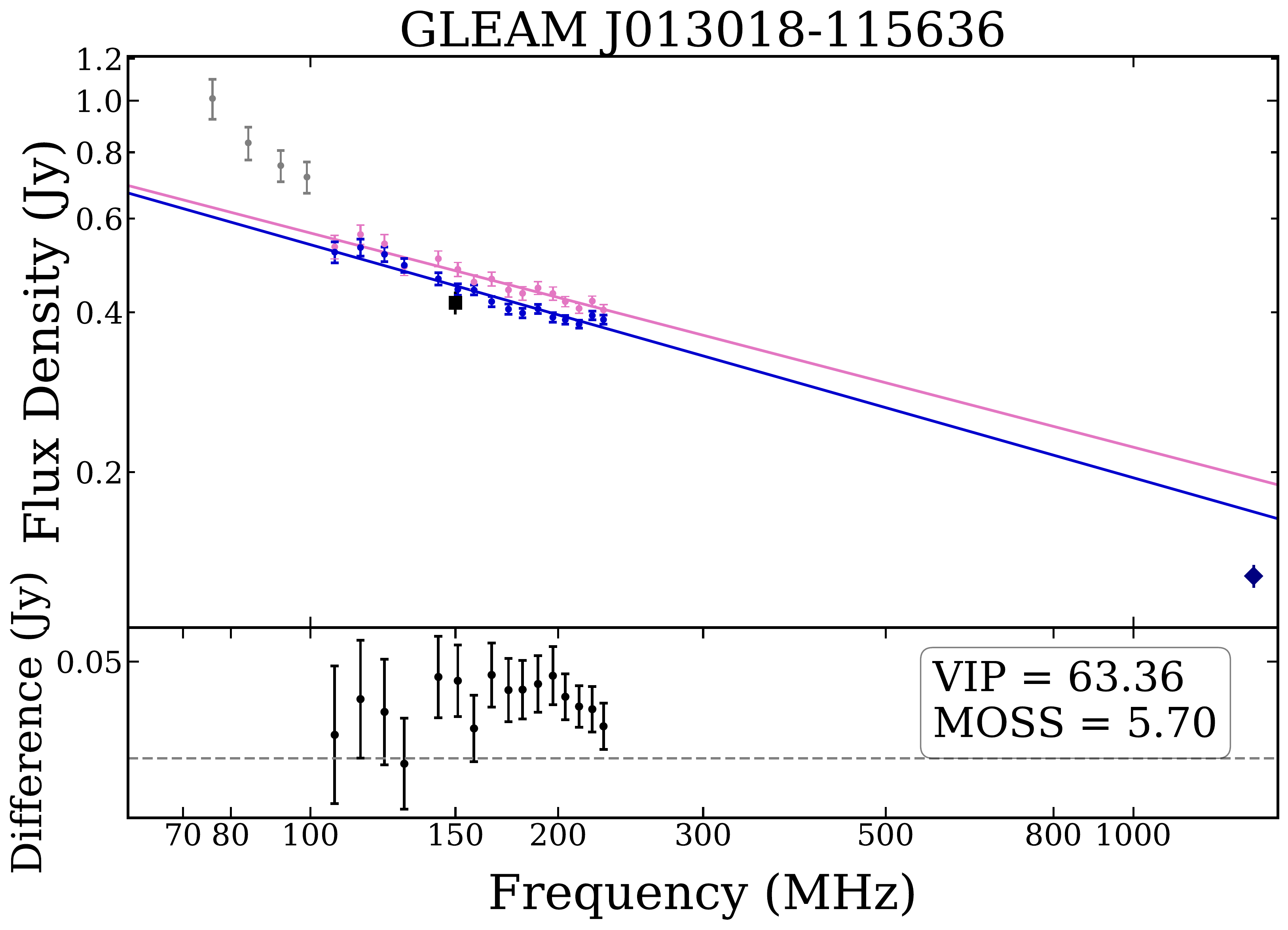} &
\includegraphics[scale=0.15]{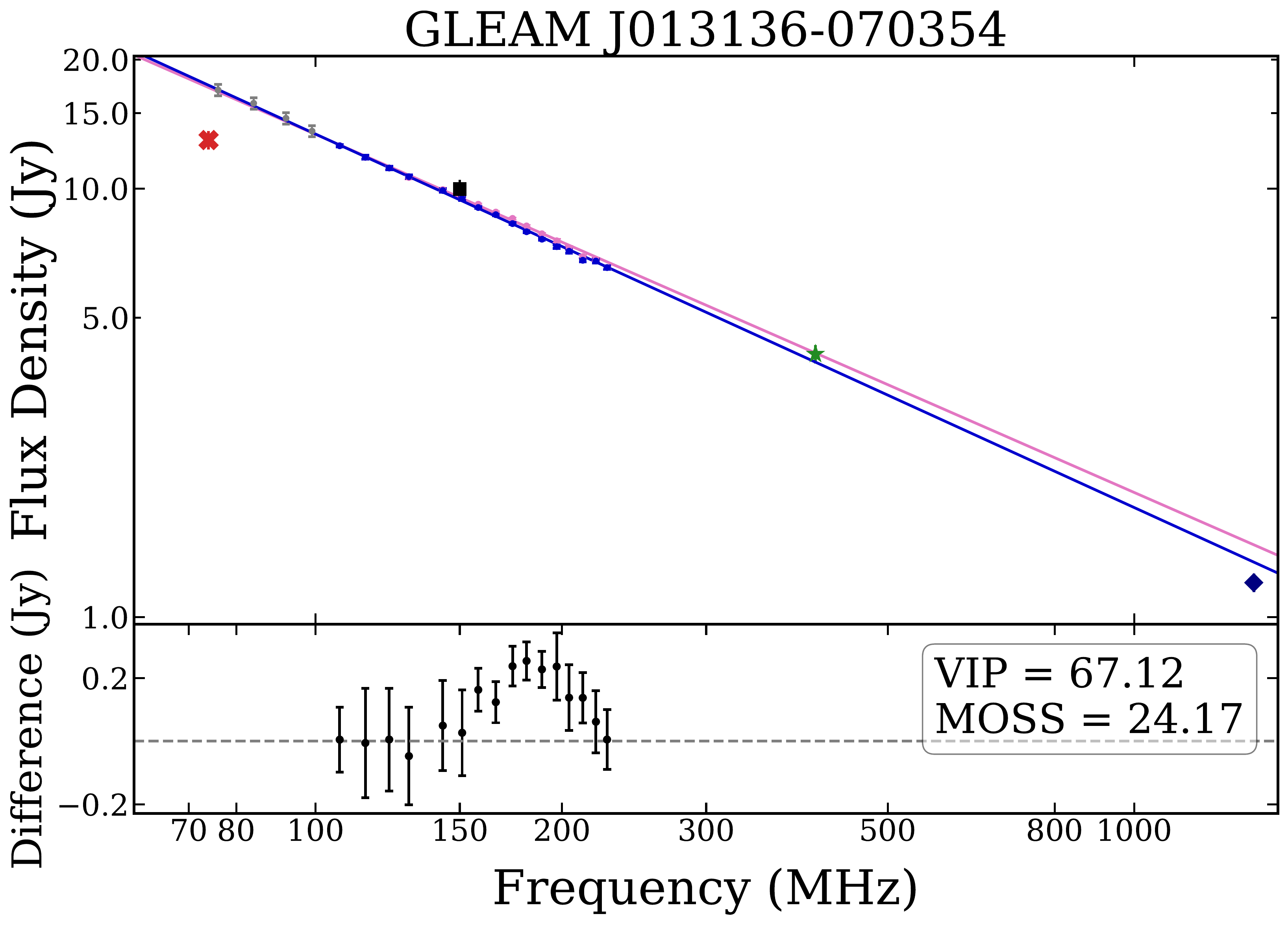} &
\includegraphics[scale=0.15]{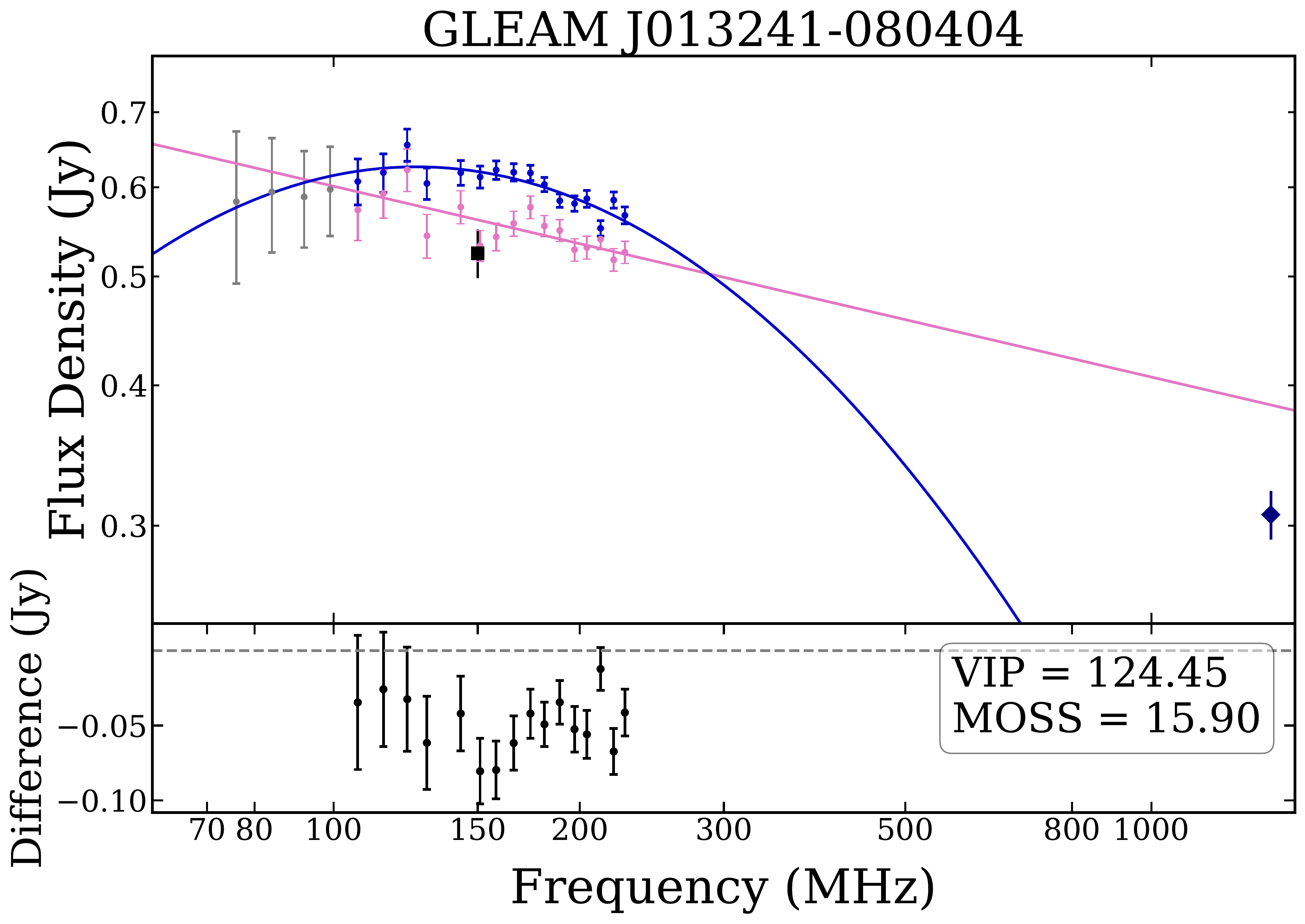} \\
\includegraphics[scale=0.15]{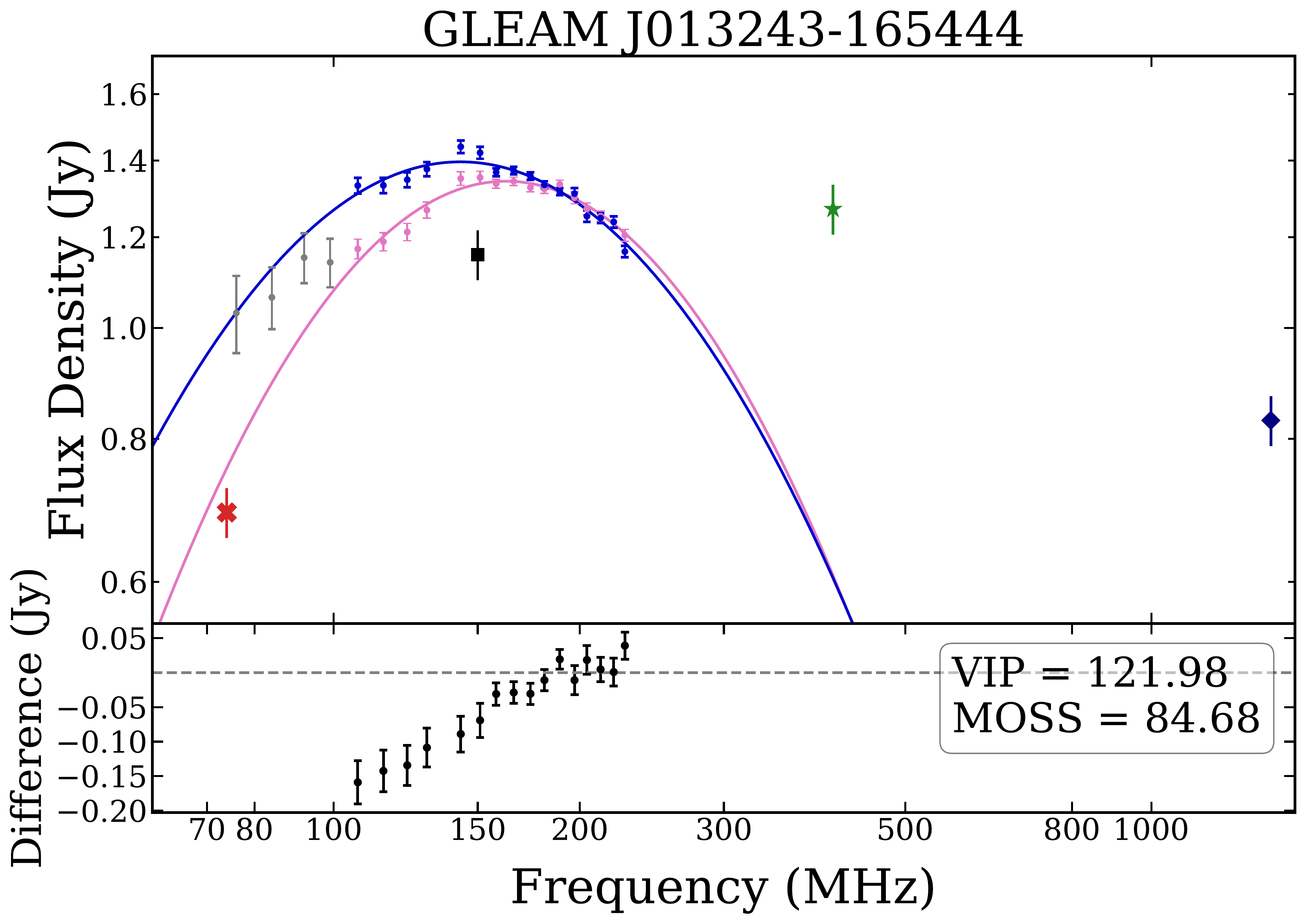} &
\includegraphics[scale=0.15]{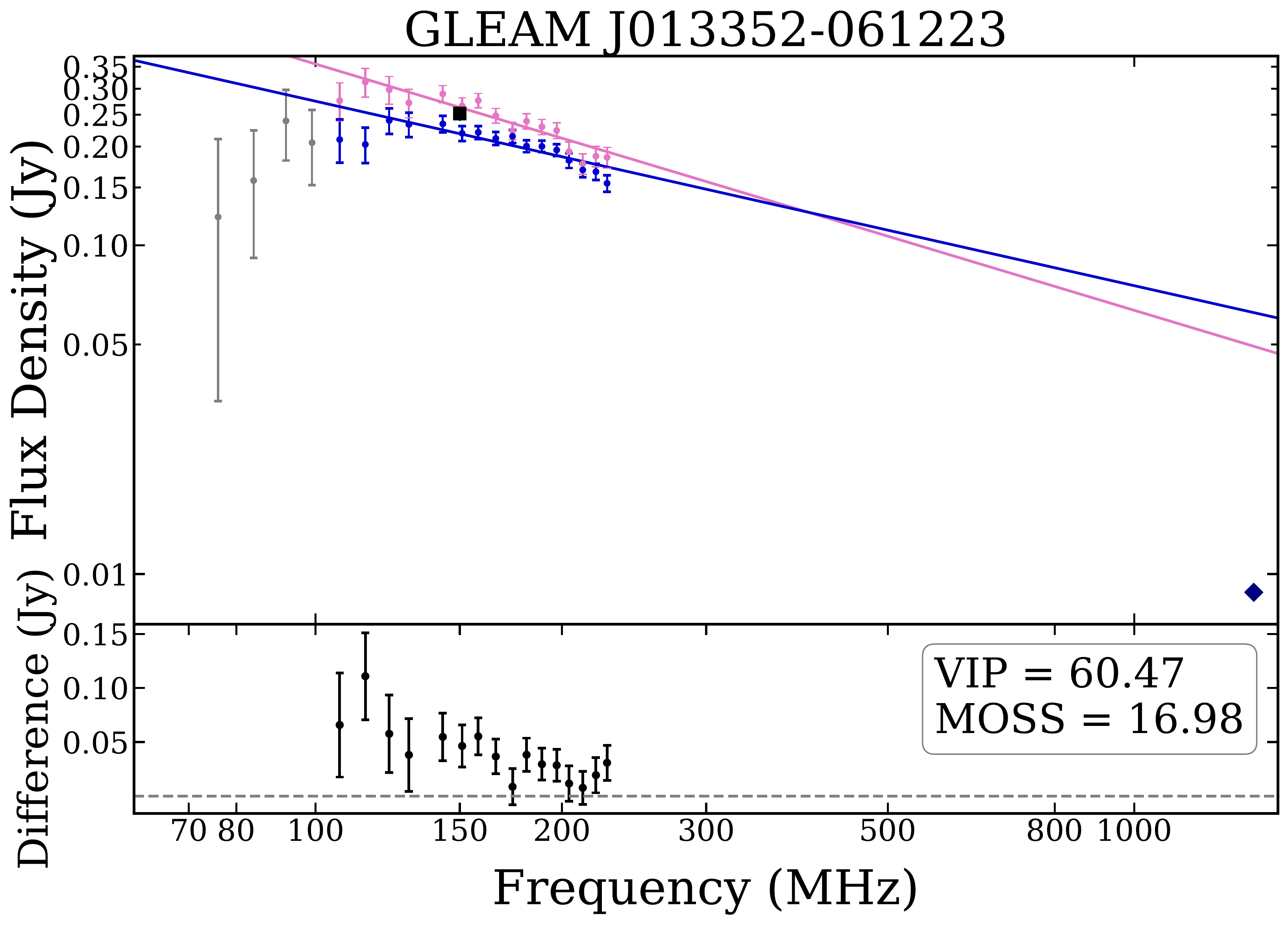} &
\includegraphics[scale=0.15]{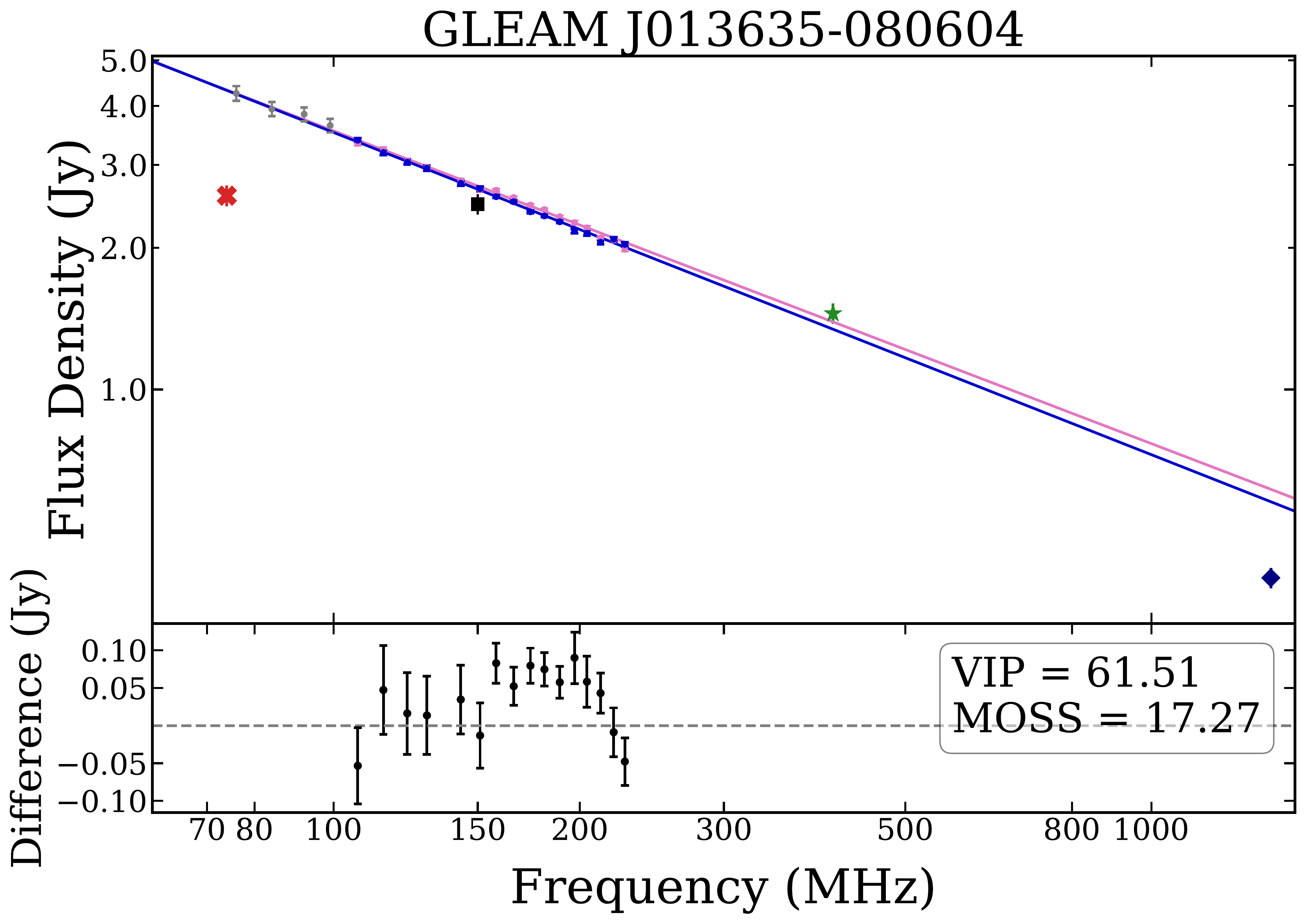} \\
\includegraphics[scale=0.15]{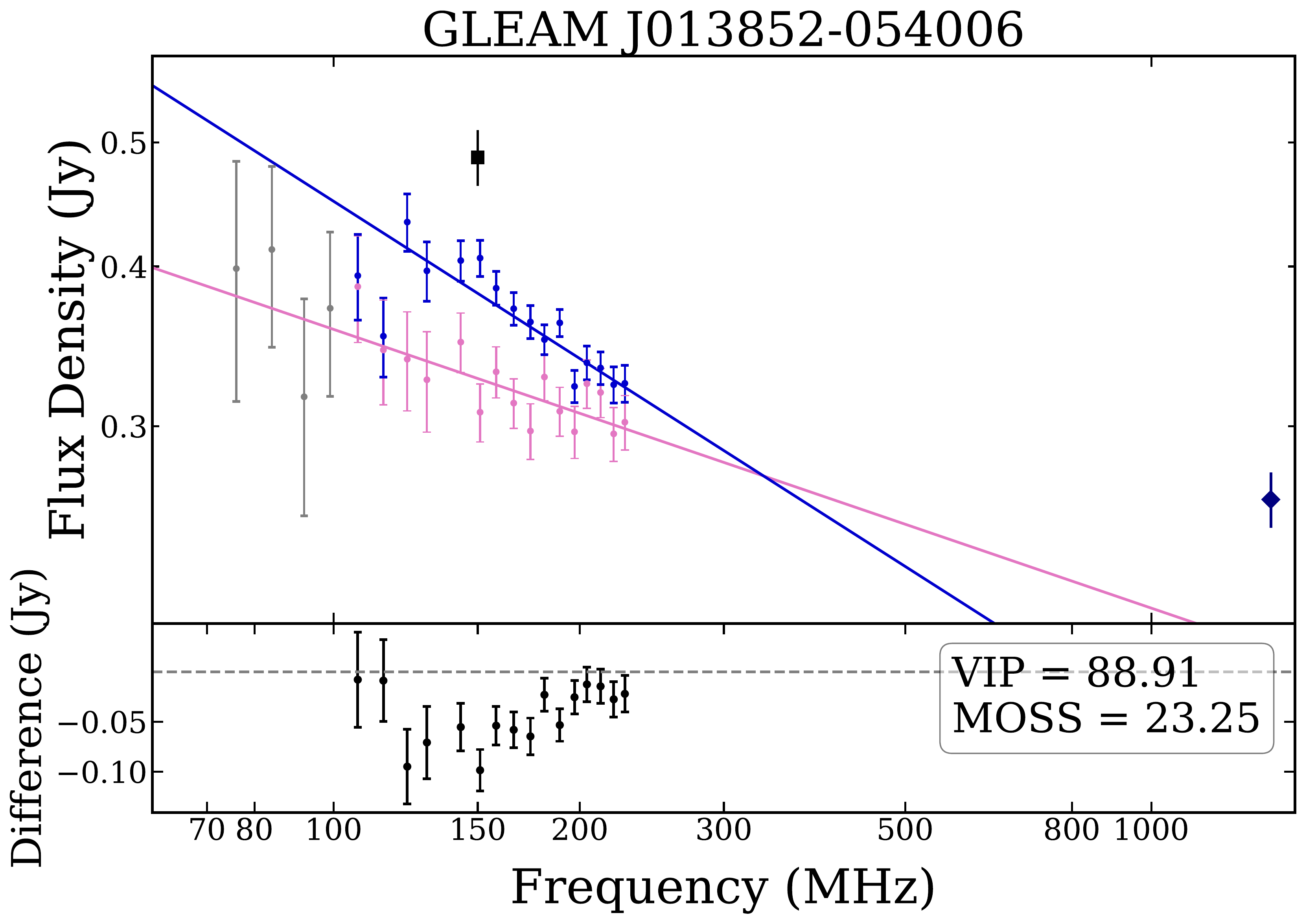} &
\includegraphics[scale=0.15]{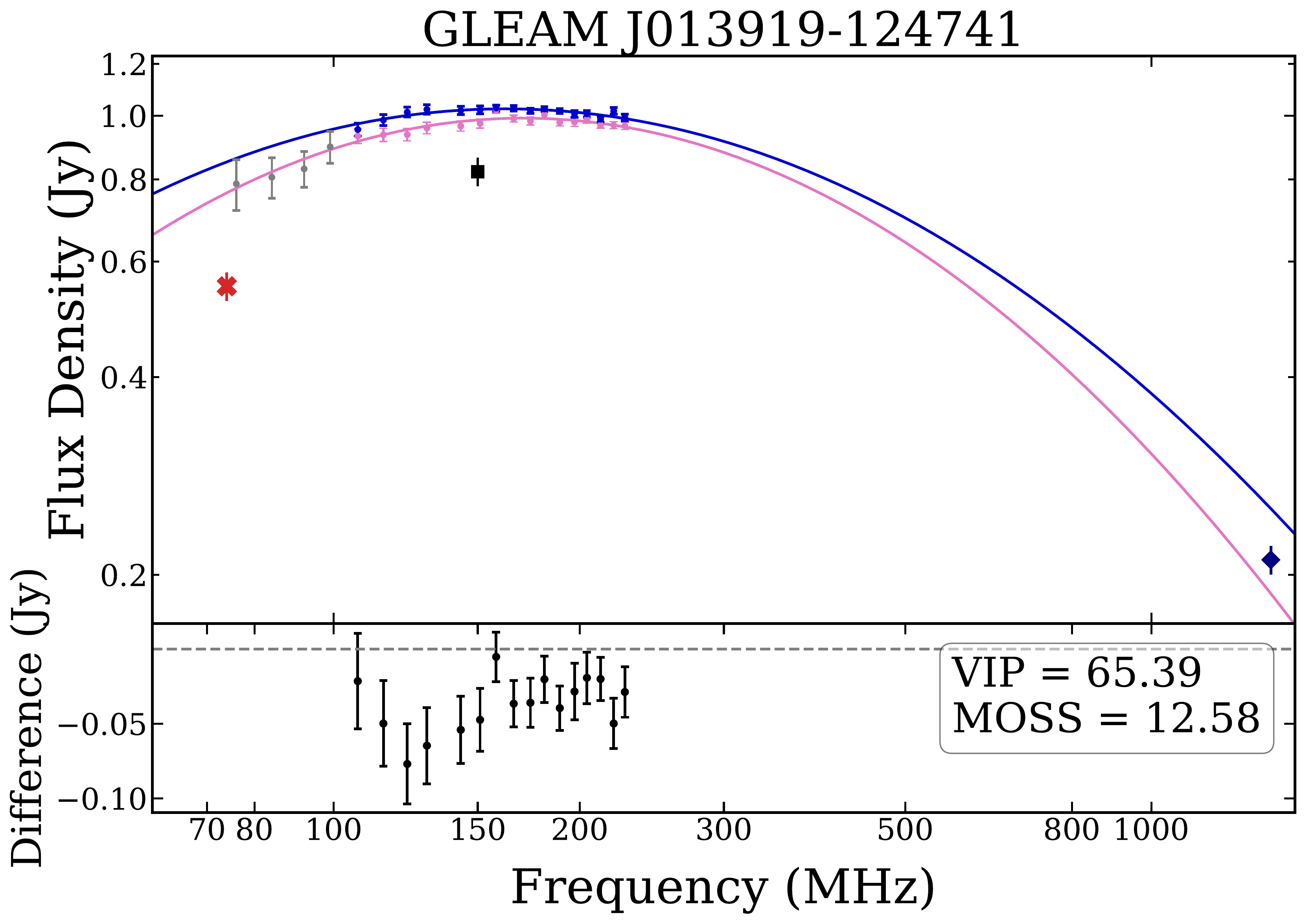} &
\includegraphics[scale=0.15]{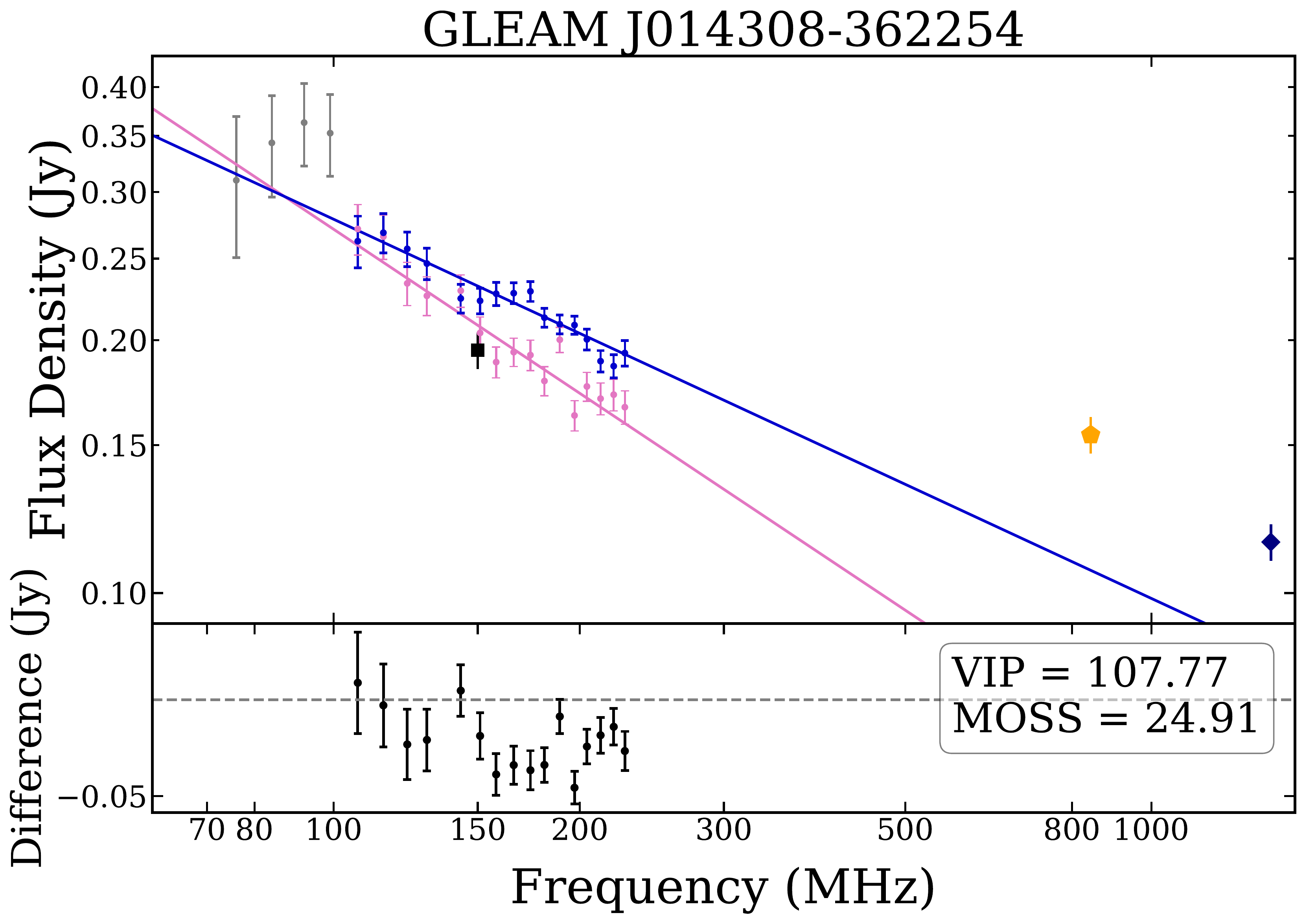} \\
\includegraphics[scale=0.15]{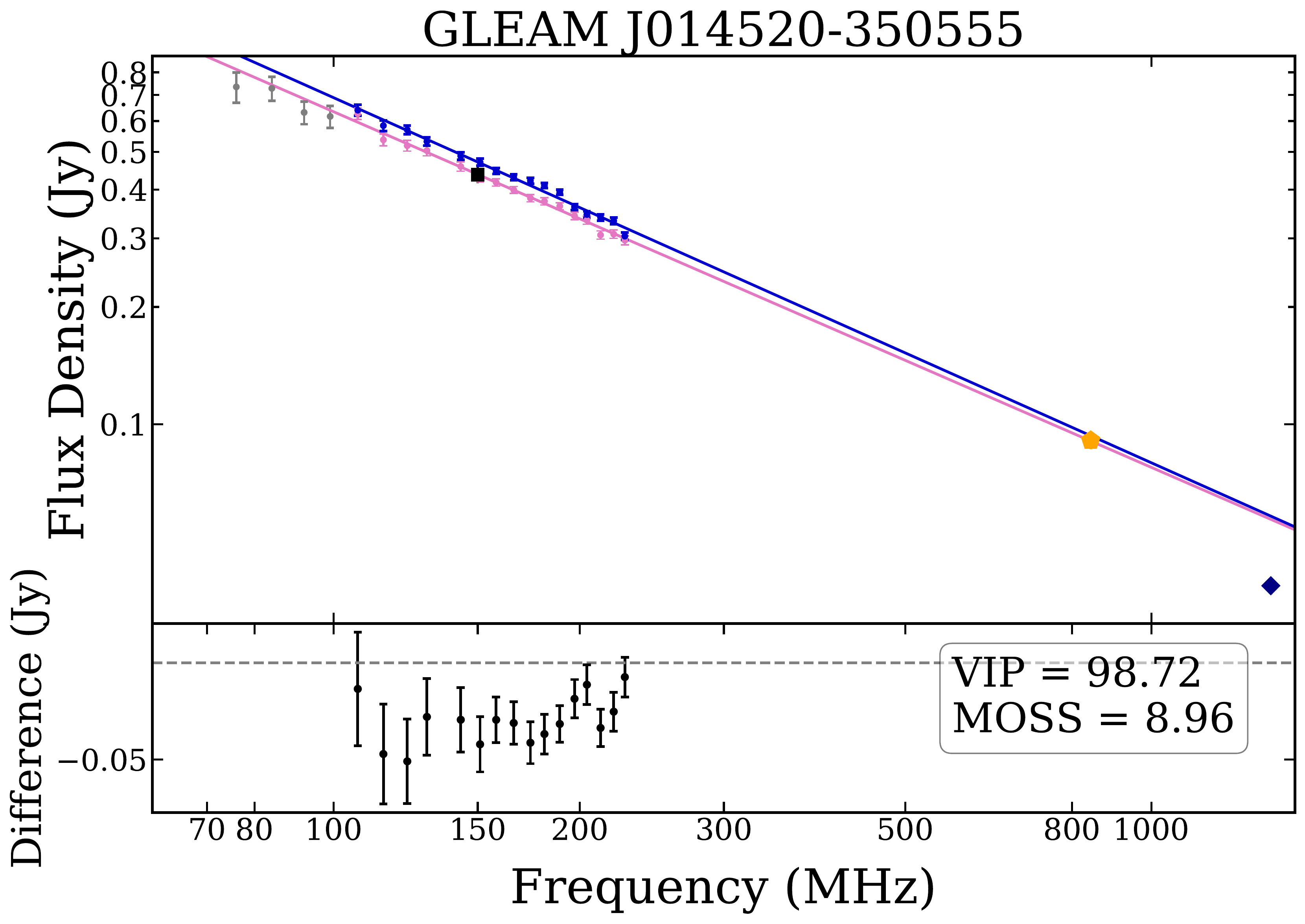} &
\includegraphics[scale=0.15]{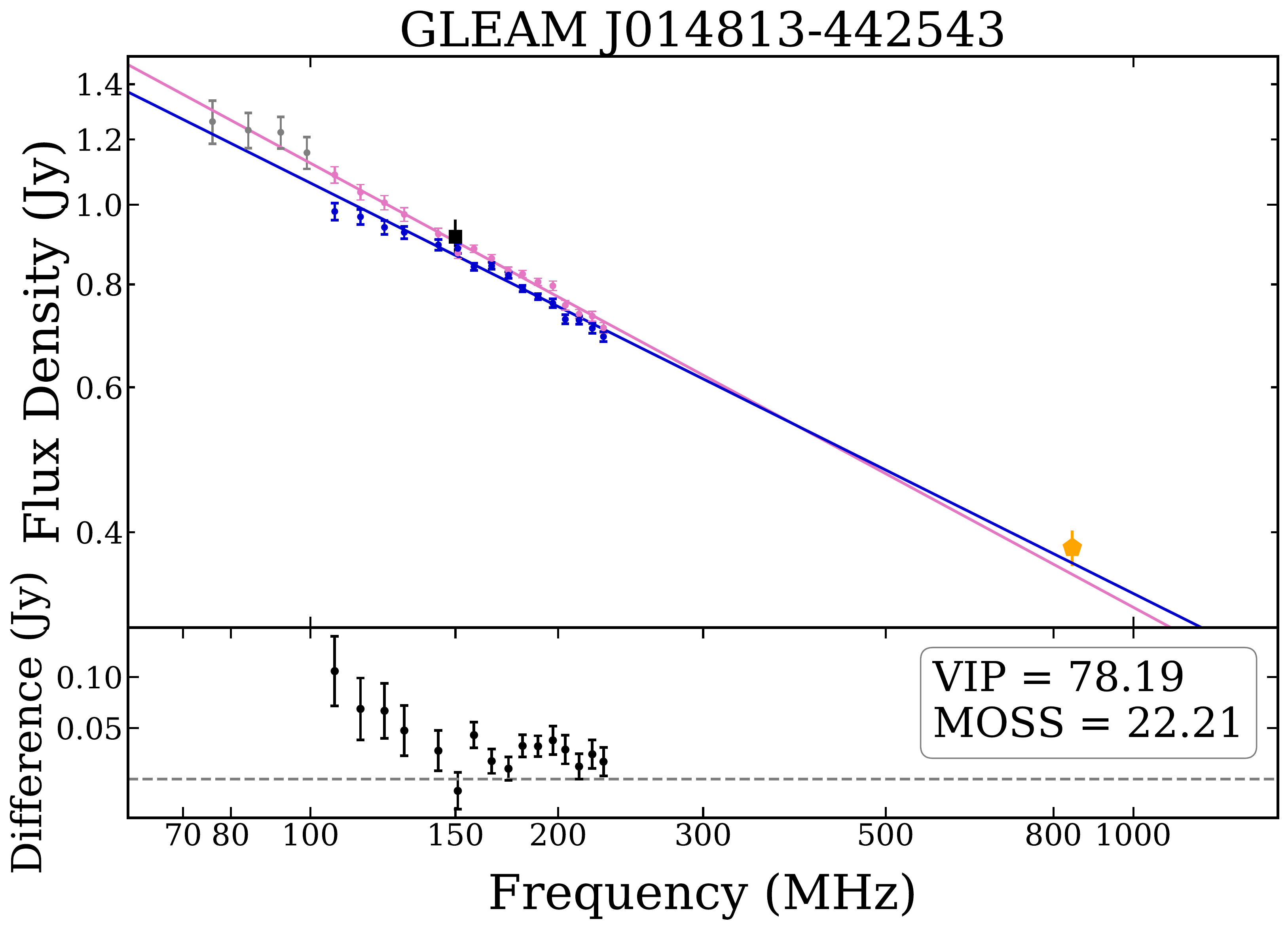} &
\includegraphics[scale=0.15]{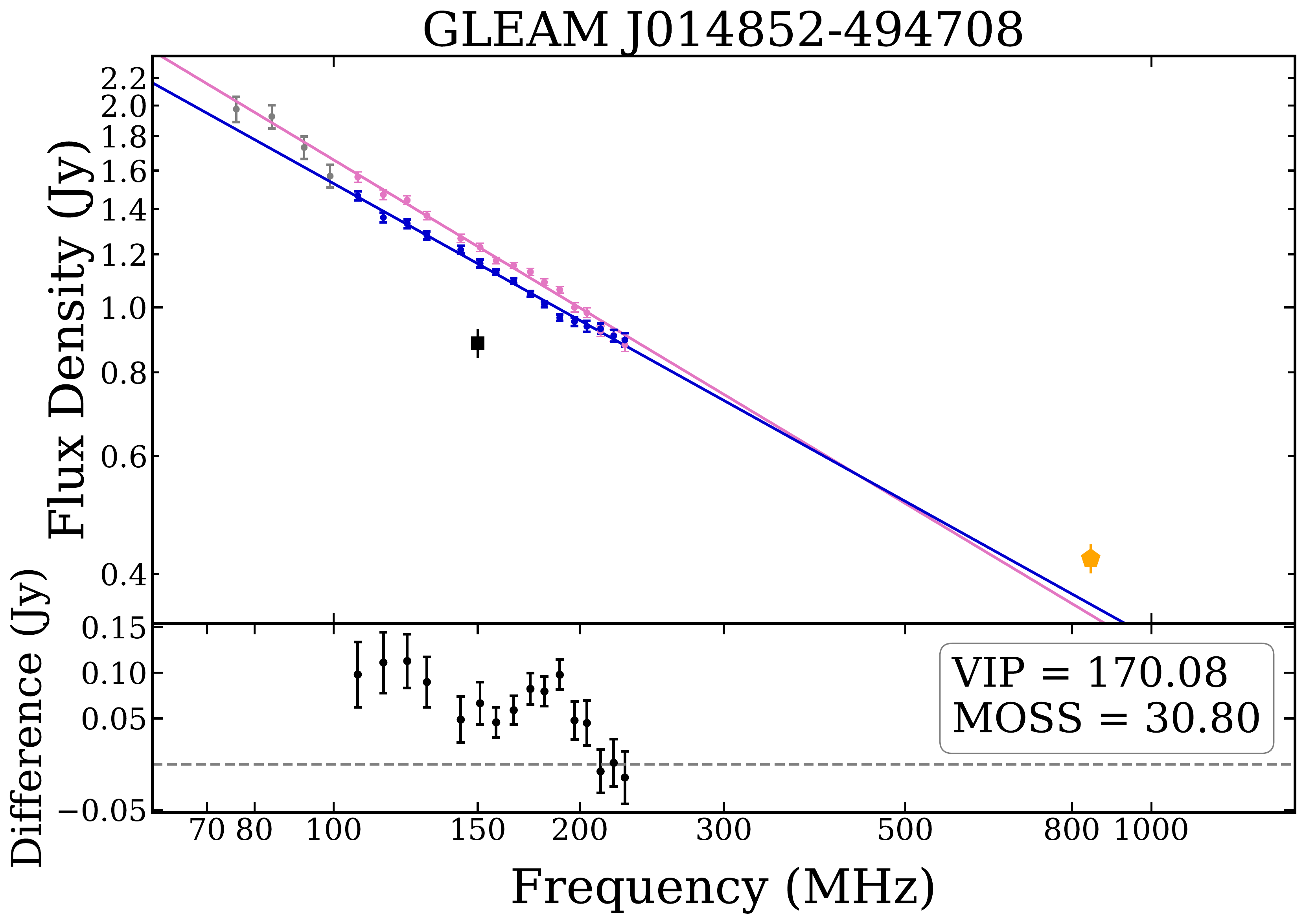} \\
\includegraphics[scale=0.15]{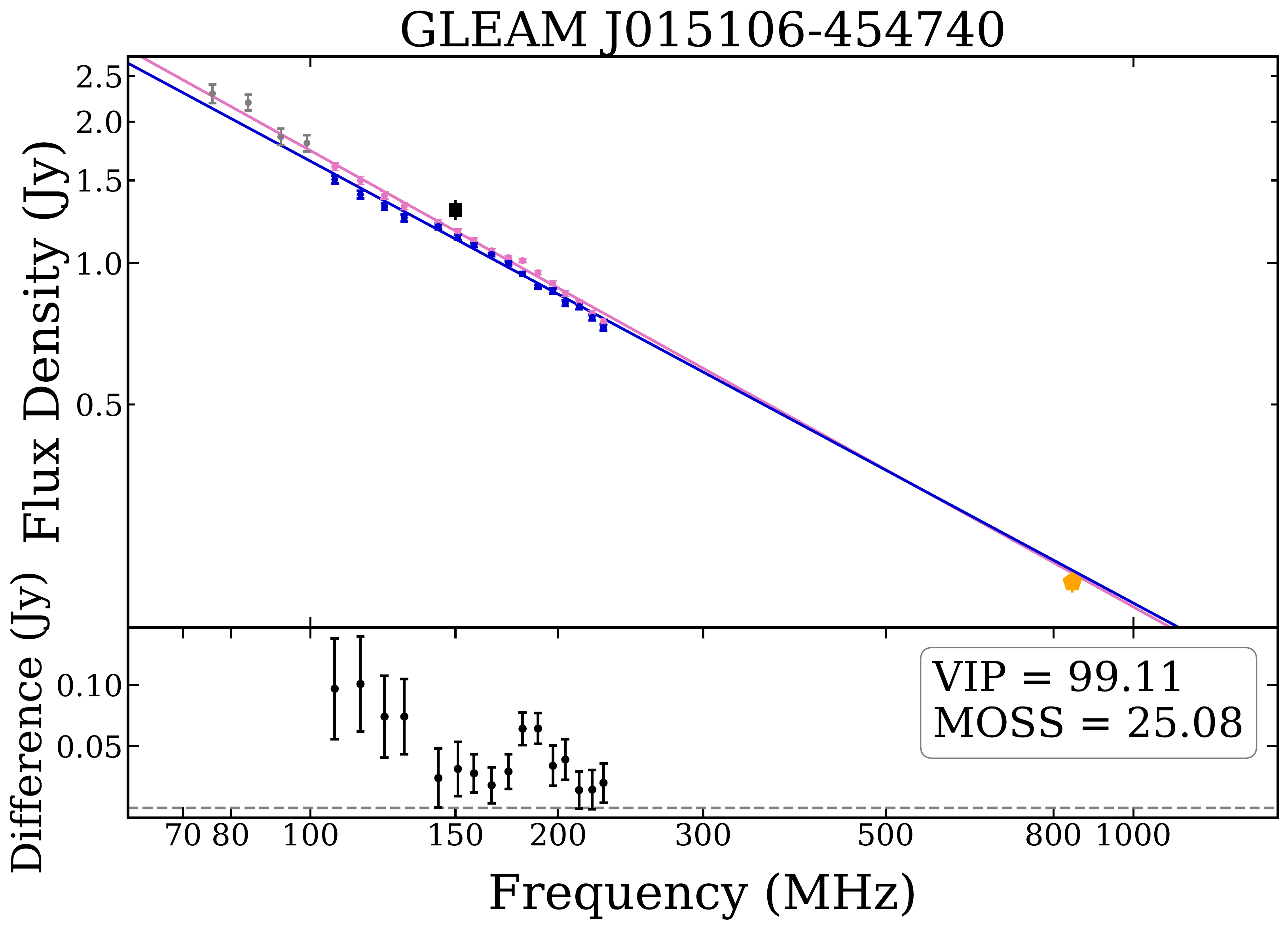} &
\includegraphics[scale=0.15]{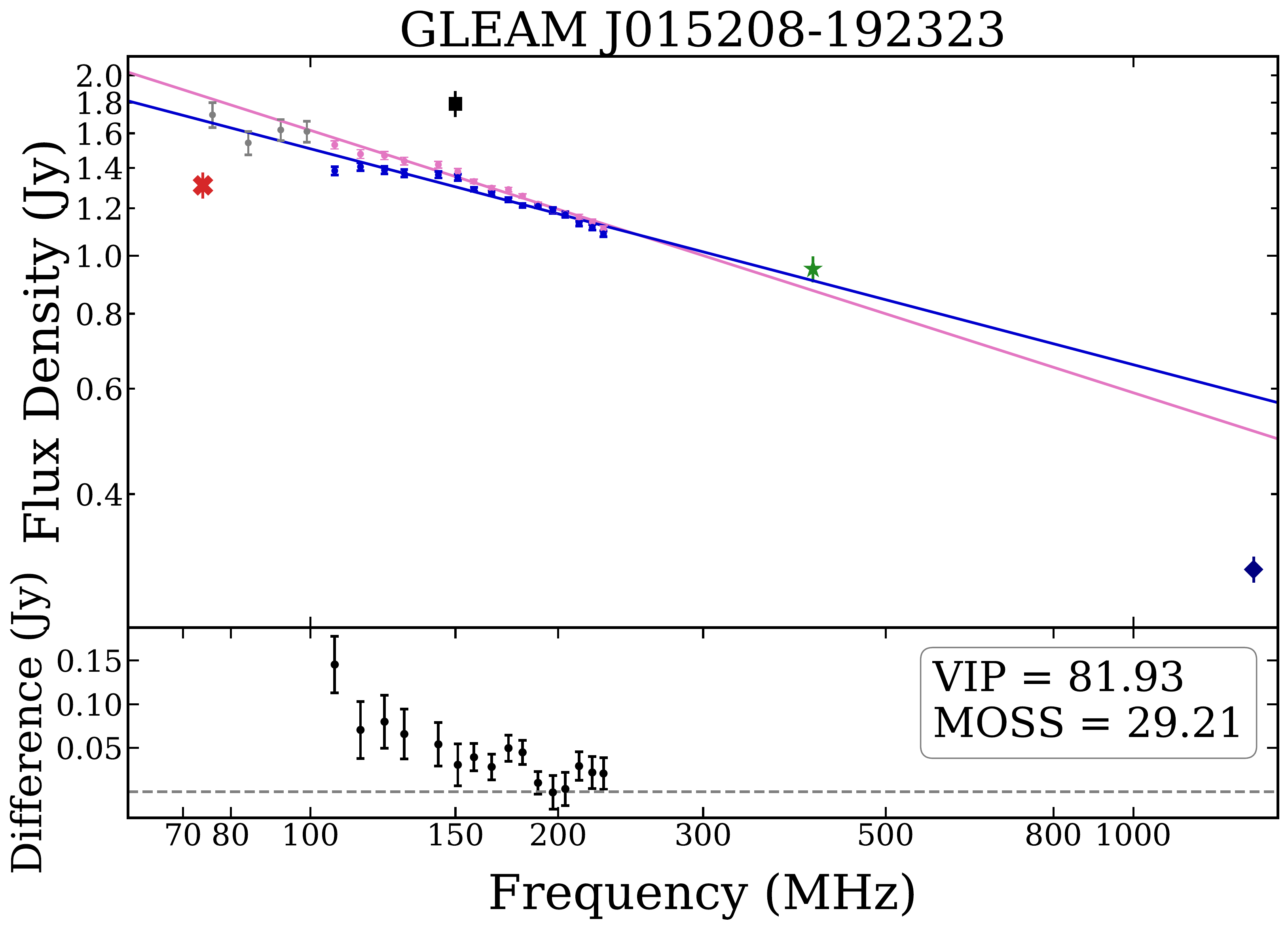} &
\includegraphics[scale=0.15]{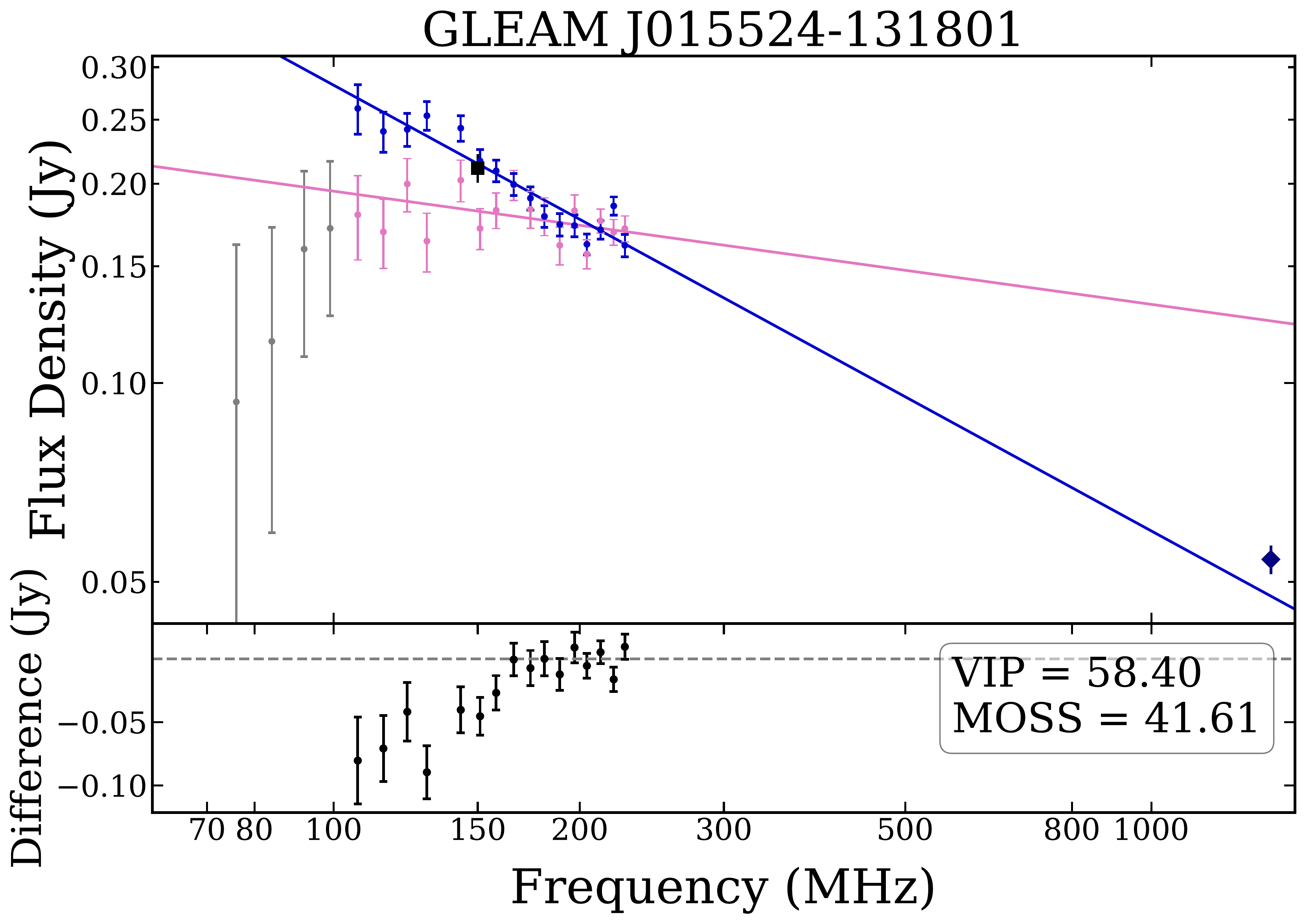} \\
\includegraphics[scale=0.15]{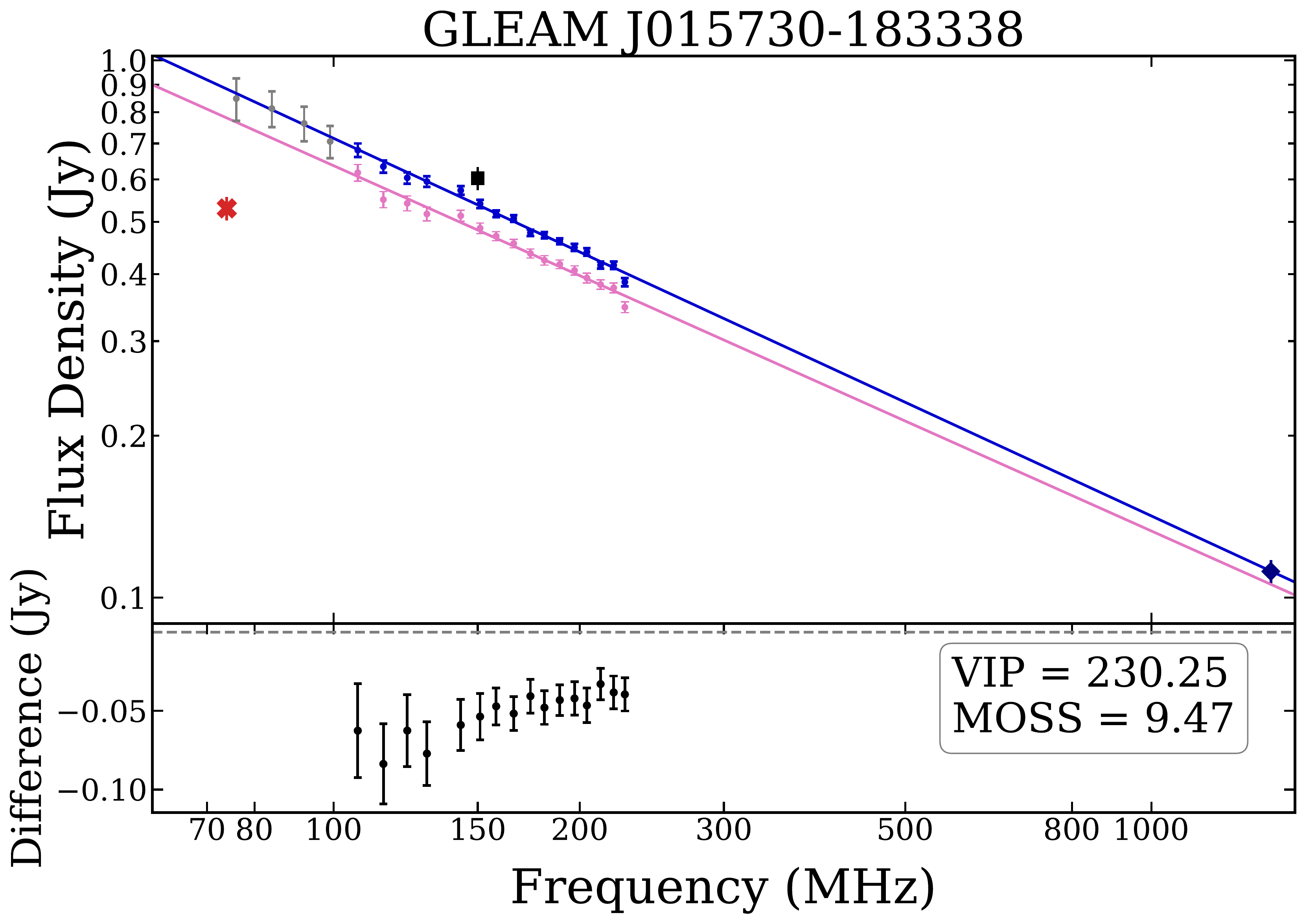} &
\includegraphics[scale=0.15]{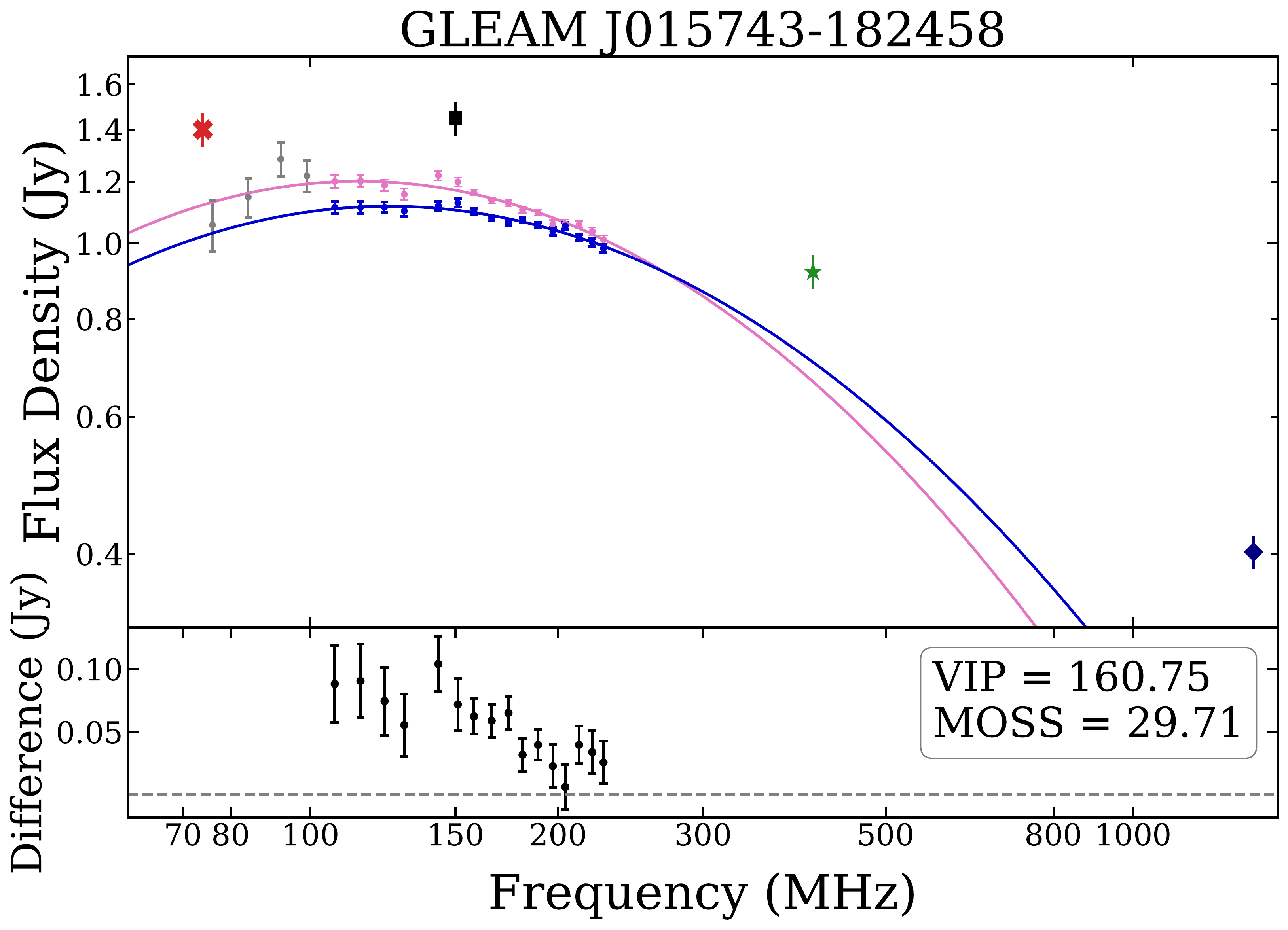} &
\includegraphics[scale=0.15]{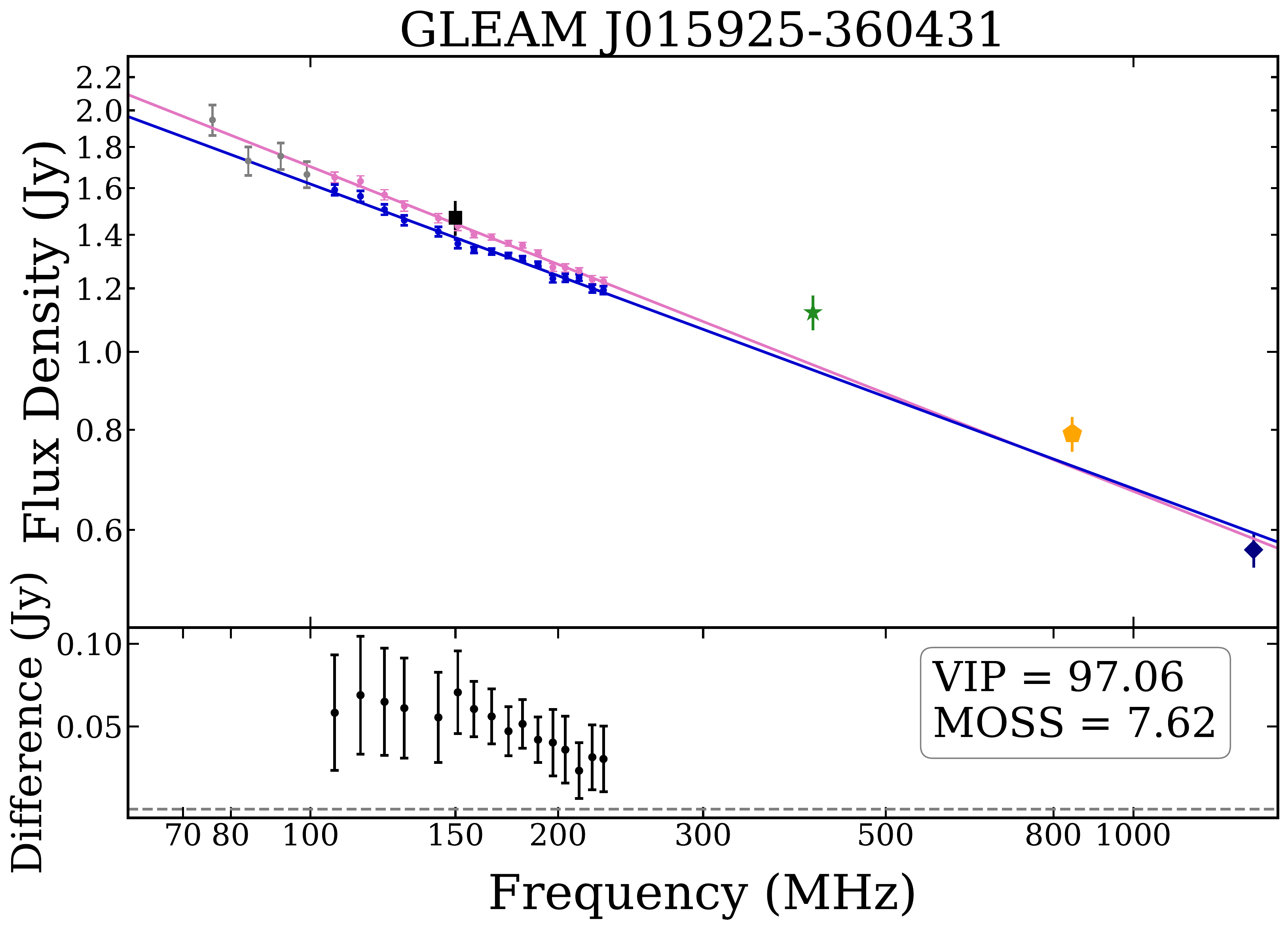} \\
\label{app:fig:pg4}
\end{array}$
\caption{(continued) SEDs for all sources classified as variable according to the VIP. For each source the points represent the following data: GLEAM low frequency (72--100\,MHz) (grey circles), Year 1 (pink circles), Year 2 (blue circles), VLSSr (red cross), TGSS (black square), MRC (green star), SUMSS (yellow pentagon), and NVSS (navy diamond). The models for each year are determined by their classification; a source classified with a peak within the observed band was modelled by a quadratic according to Equation~\ref{eq:quadratic}, remaining sources were modelled by a power-law according to Equation~\ref{eq:plaw}.}
\label{app:fig:pg4}
\end{center}
\end{figure*}
\setcounter{figure}{0}
\begin{figure*}
\begin{center}$
\begin{array}{cccccc}
\includegraphics[scale=0.15]{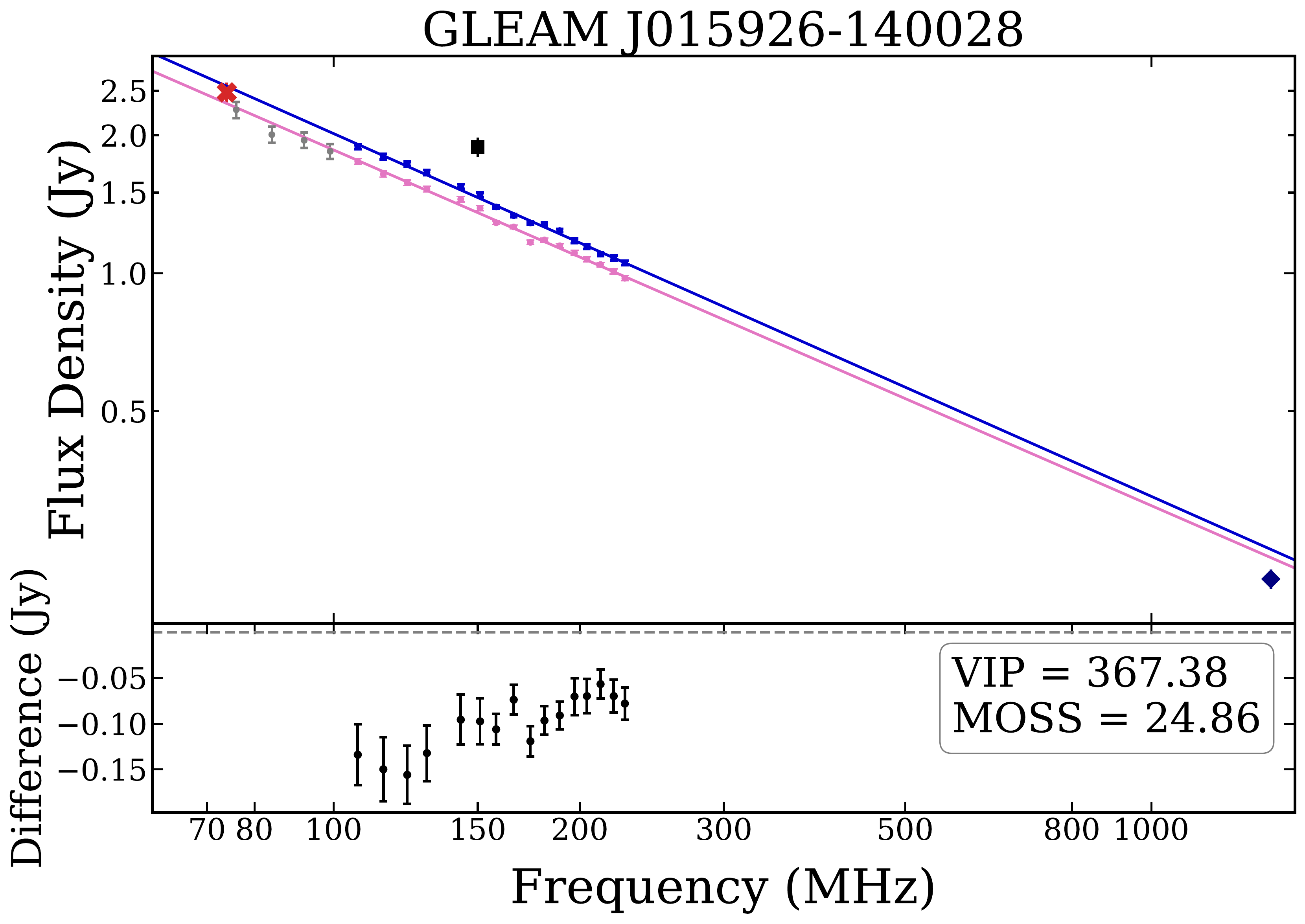} &
\includegraphics[scale=0.15]{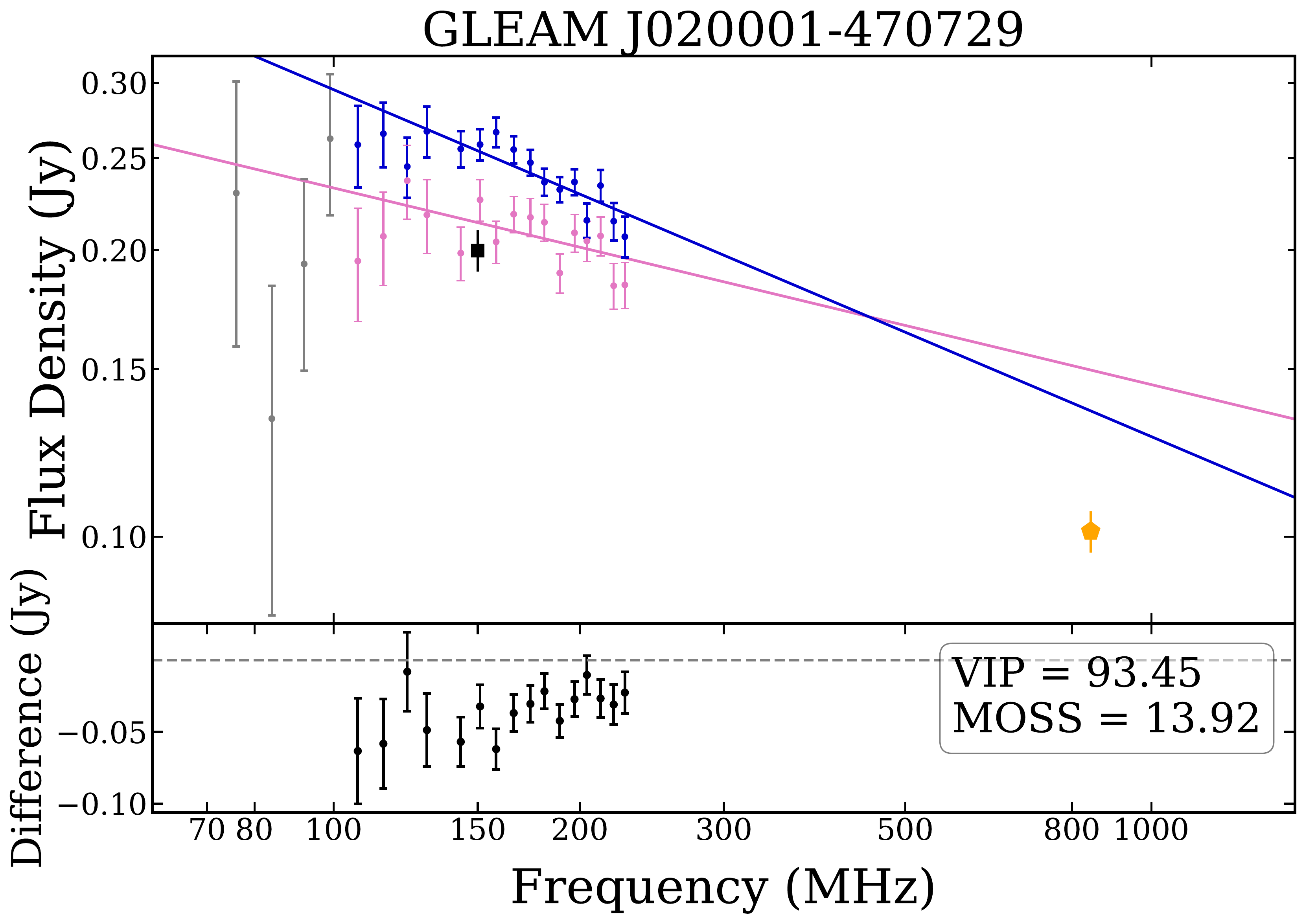} &
\includegraphics[scale=0.15]{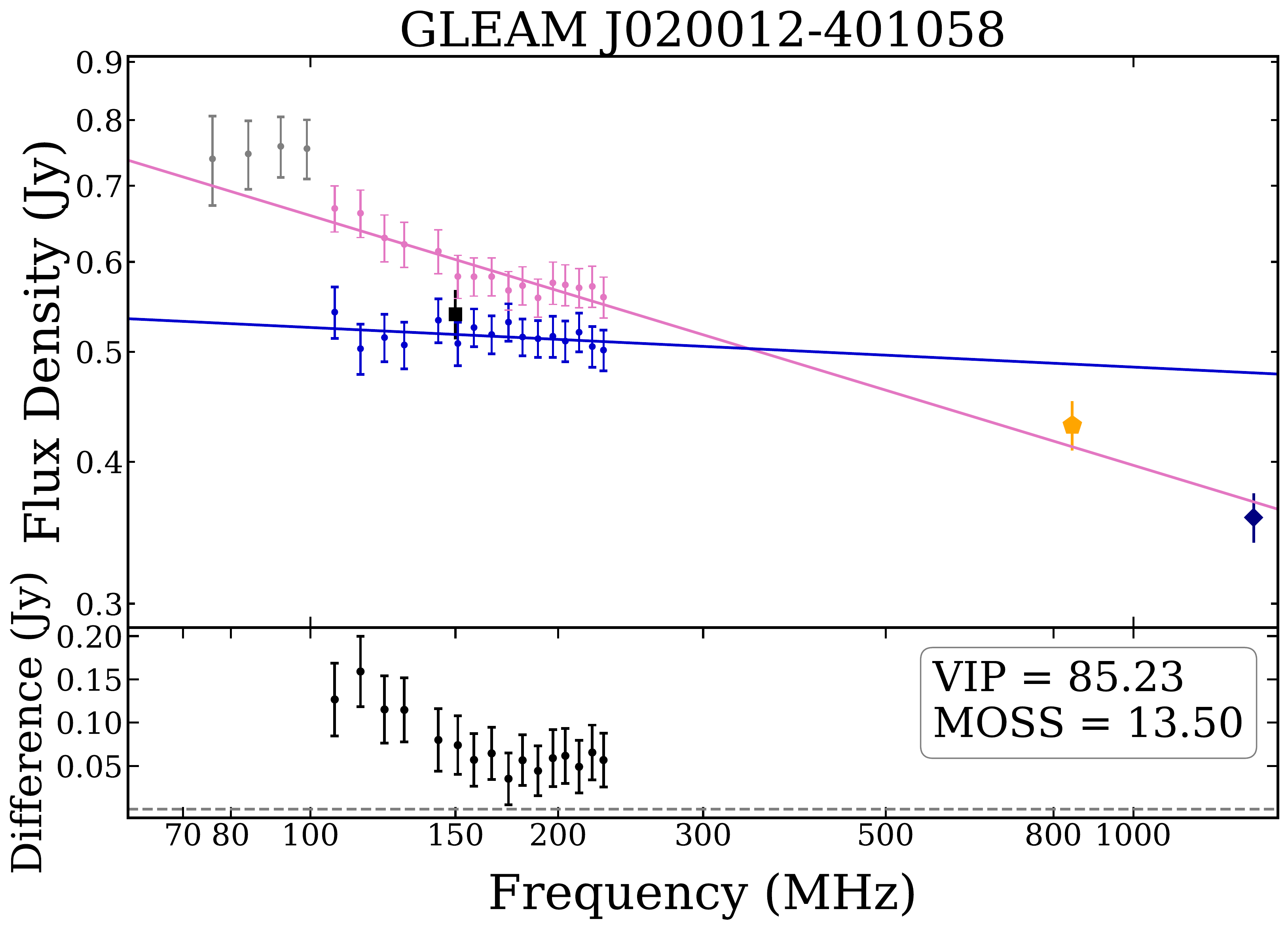} \\
\includegraphics[scale=0.15]{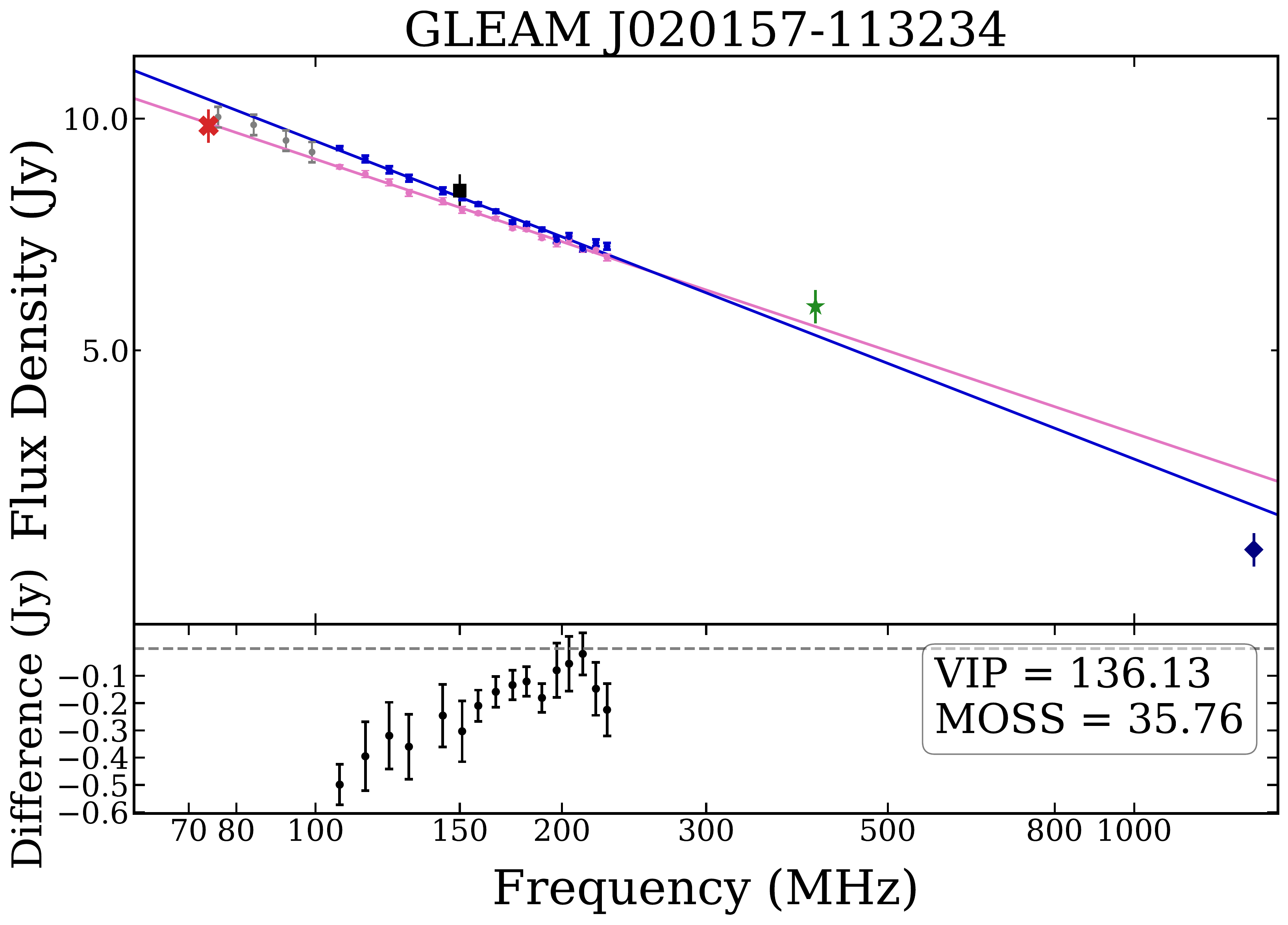} &
\includegraphics[scale=0.15]{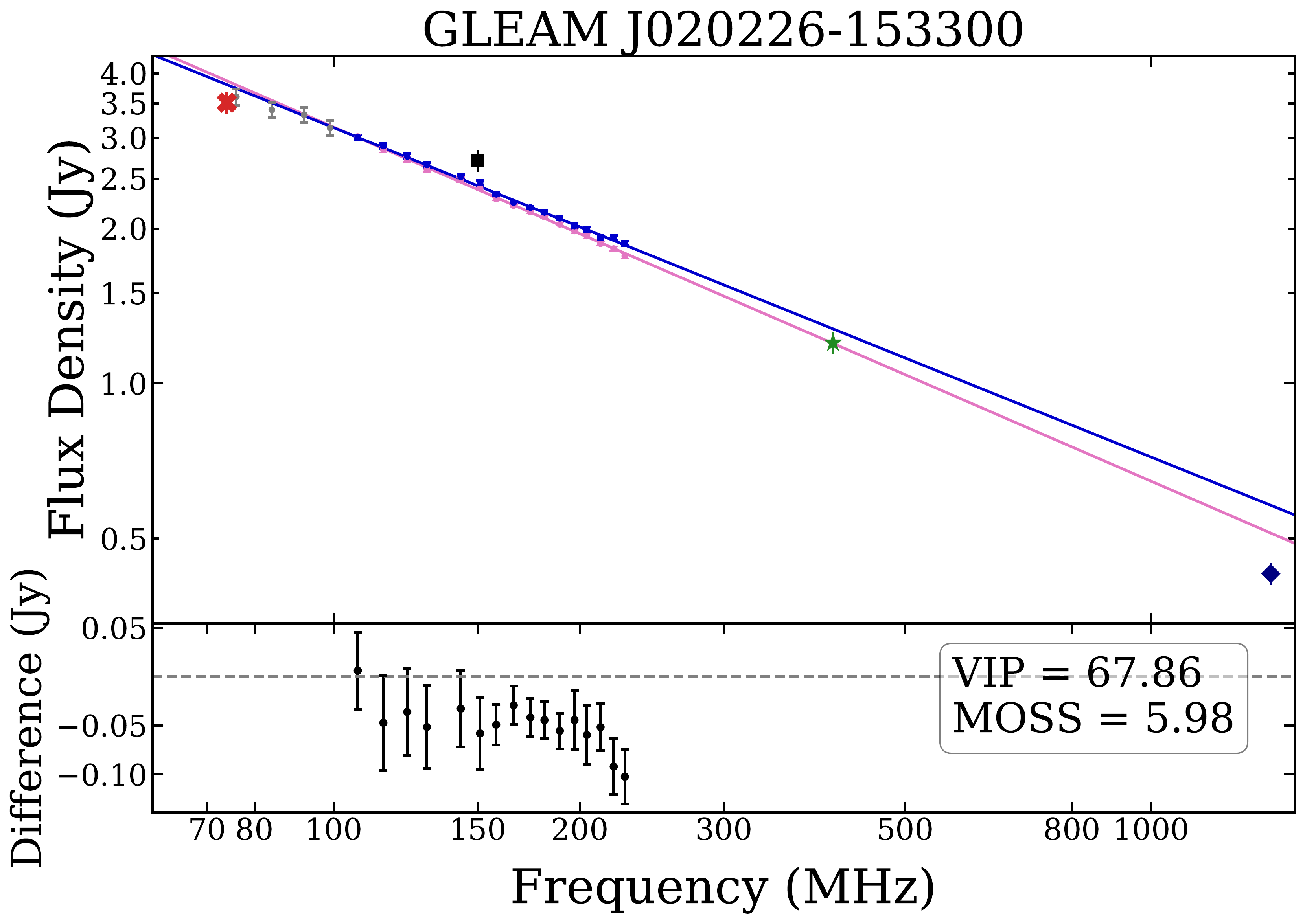} &
\includegraphics[scale=0.15]{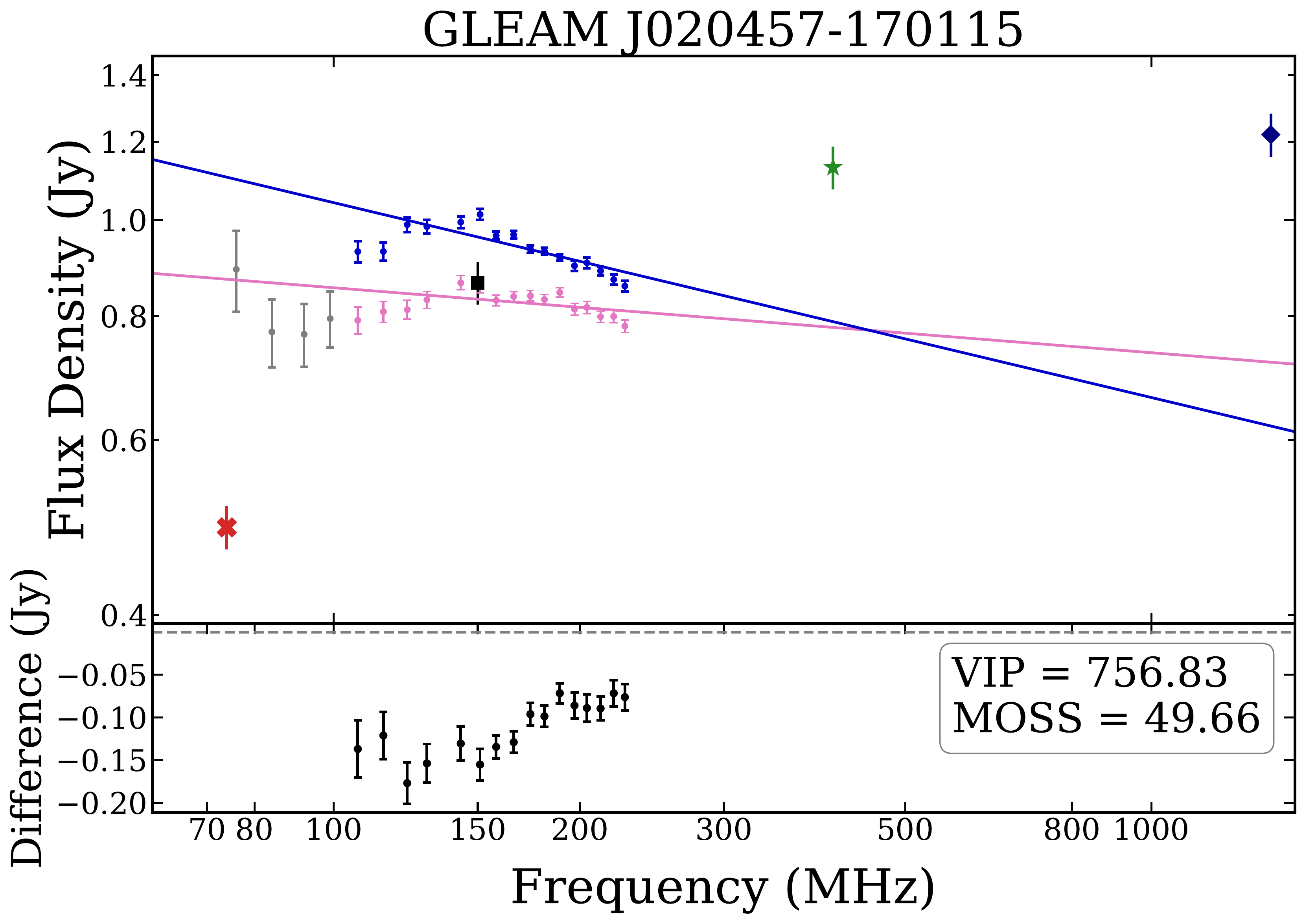} \\
\includegraphics[scale=0.15]{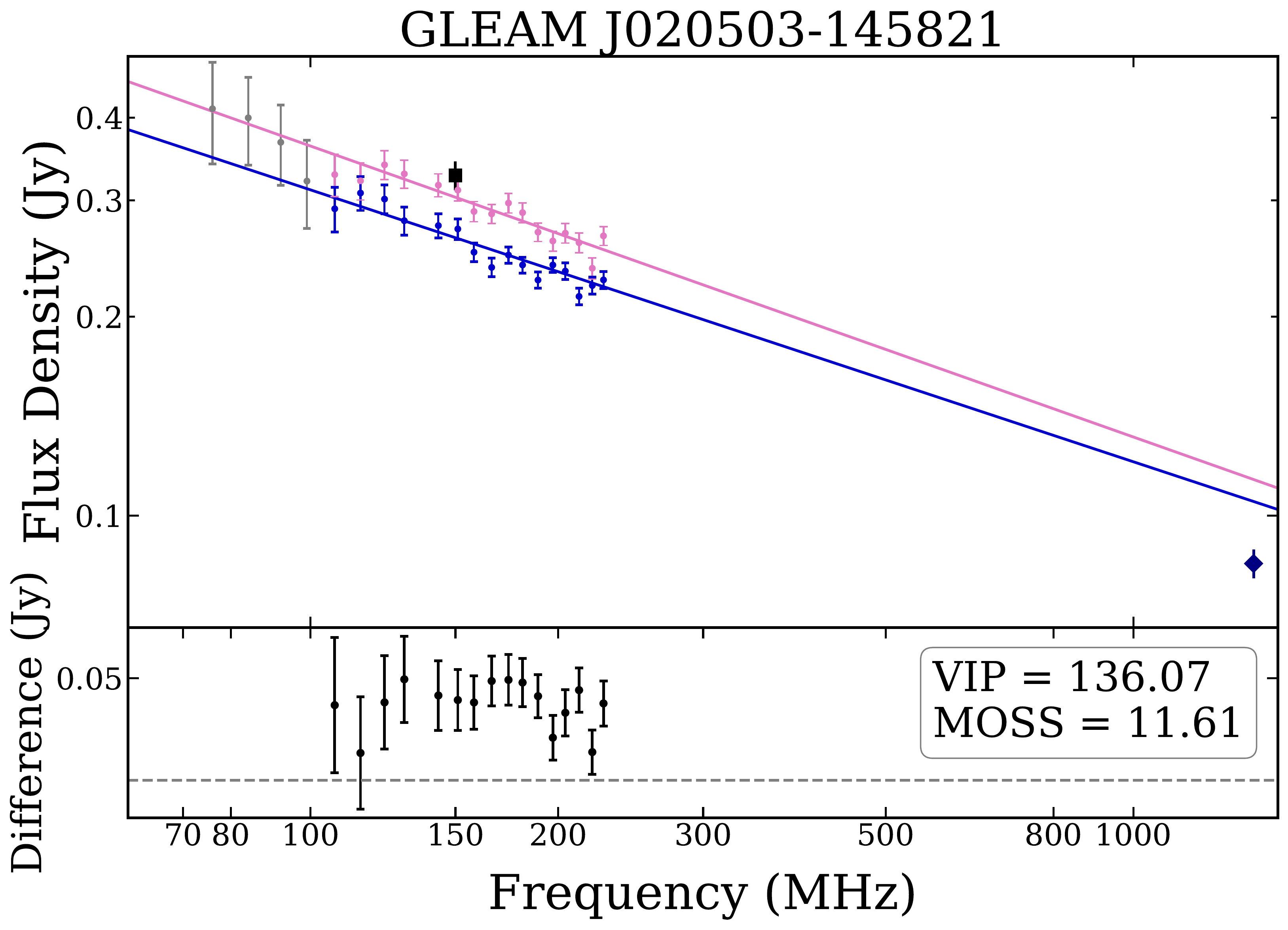} &
\includegraphics[scale=0.15]{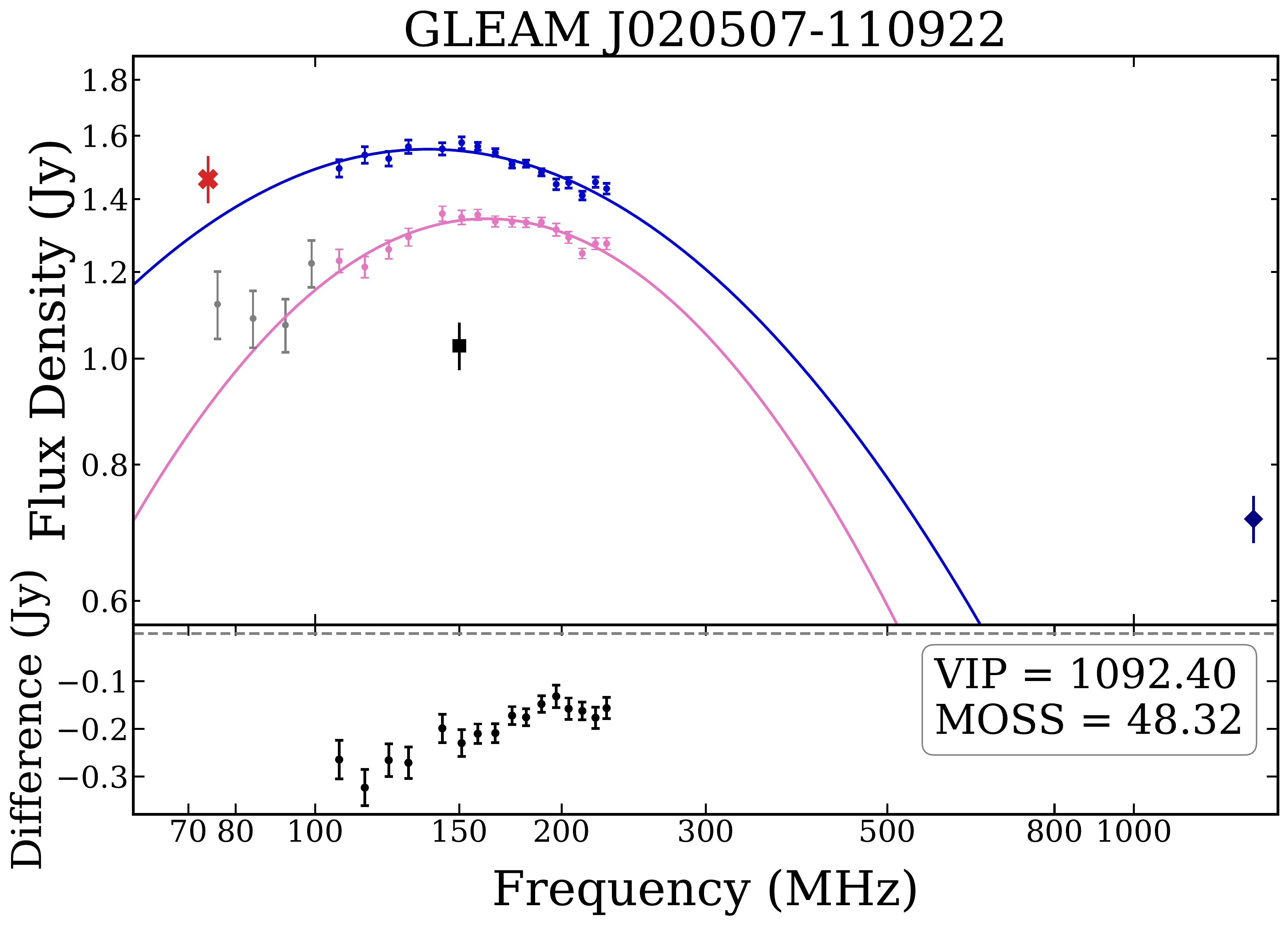} &
\includegraphics[scale=0.15]{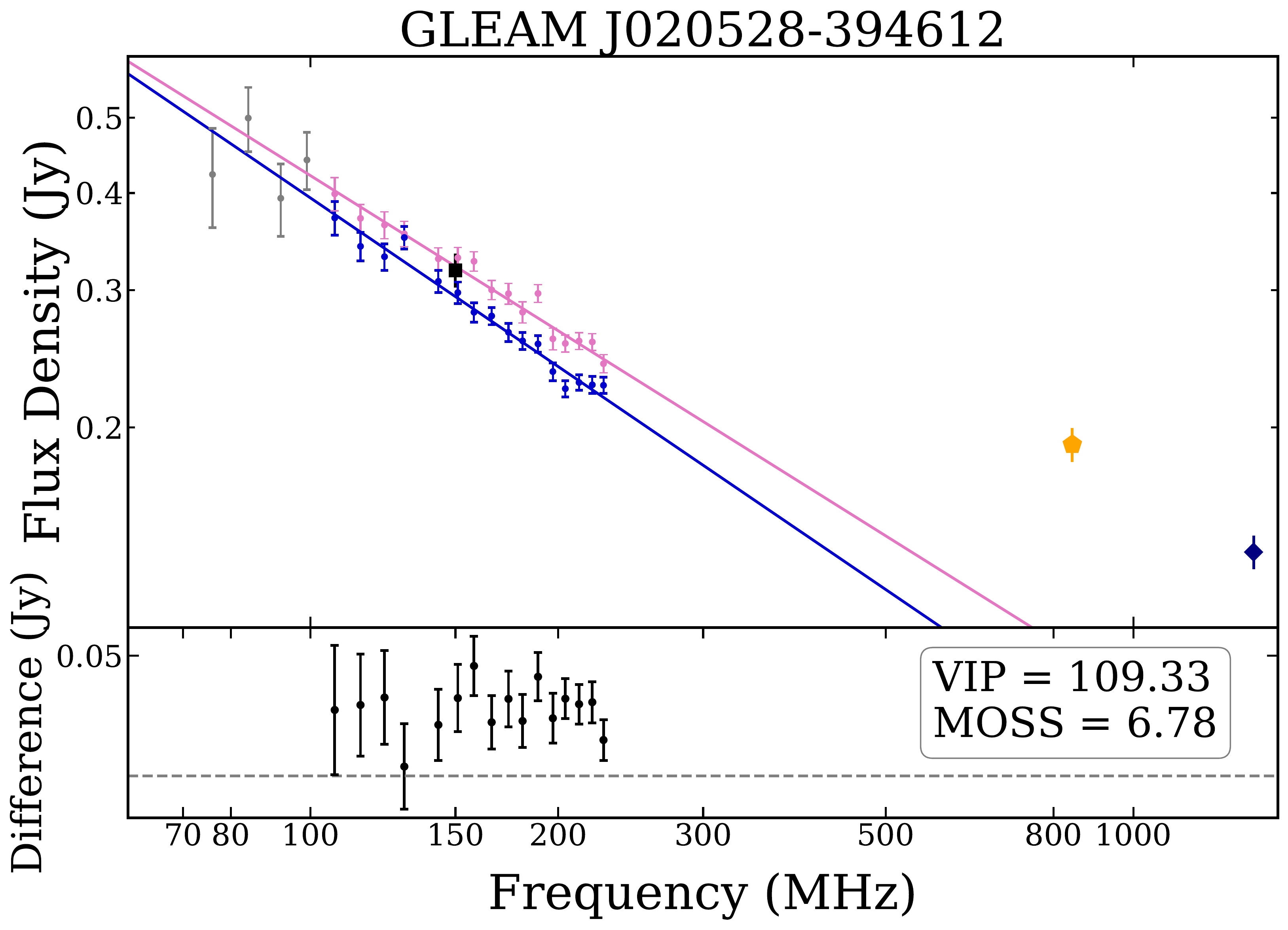} \\
\includegraphics[scale=0.15]{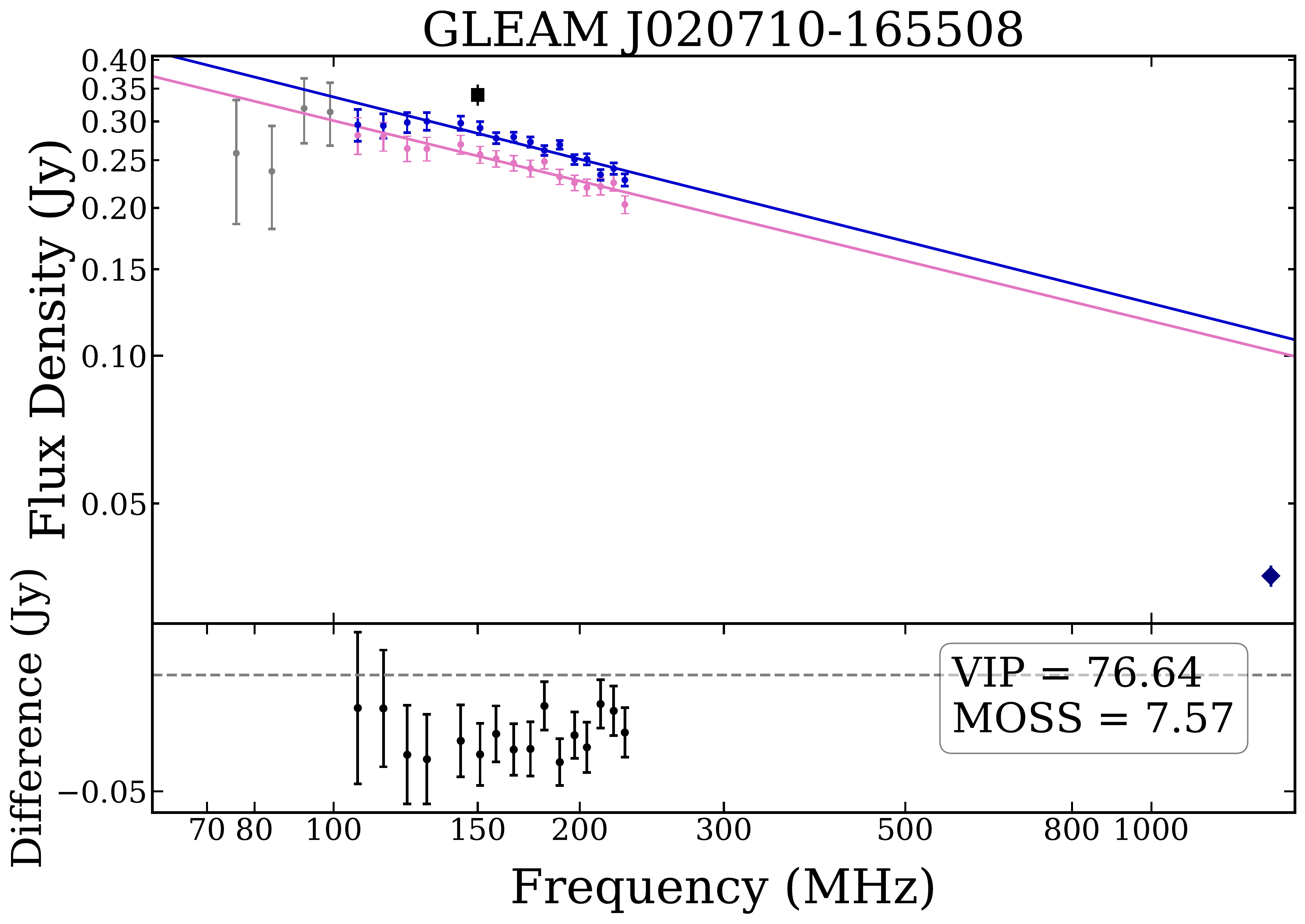} &
\includegraphics[scale=0.15]{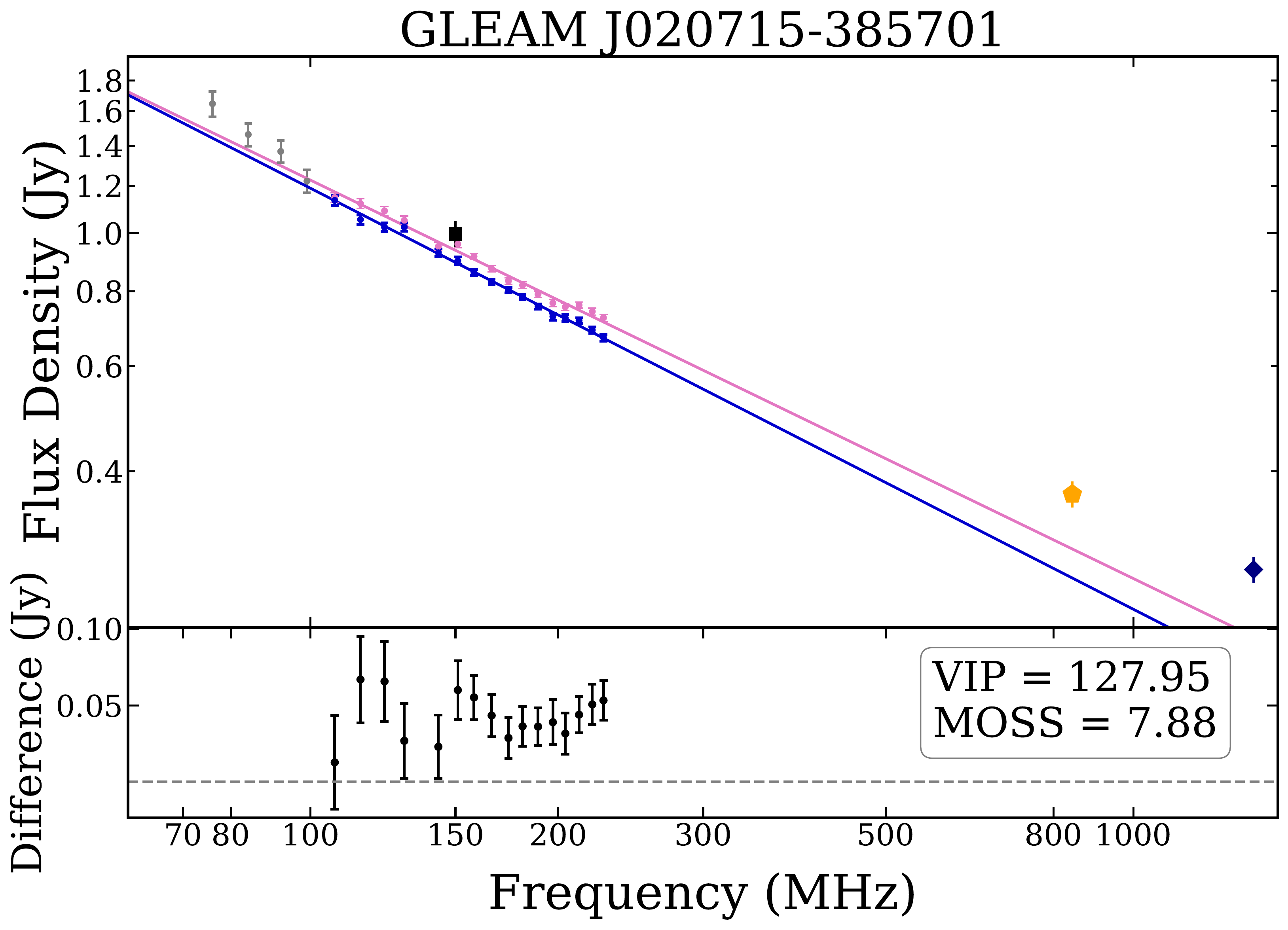} &
\includegraphics[scale=0.15]{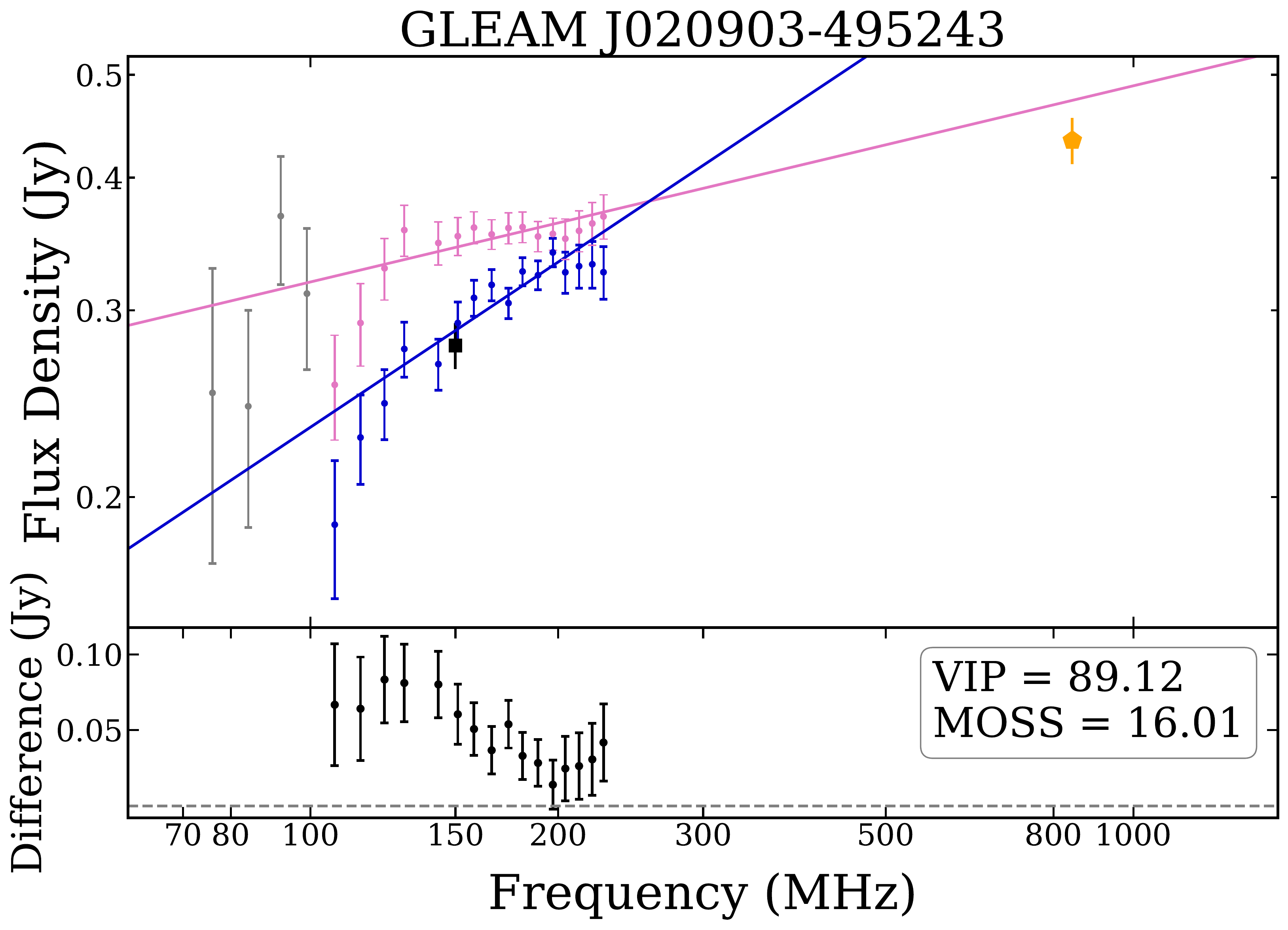} \\
\includegraphics[scale=0.15]{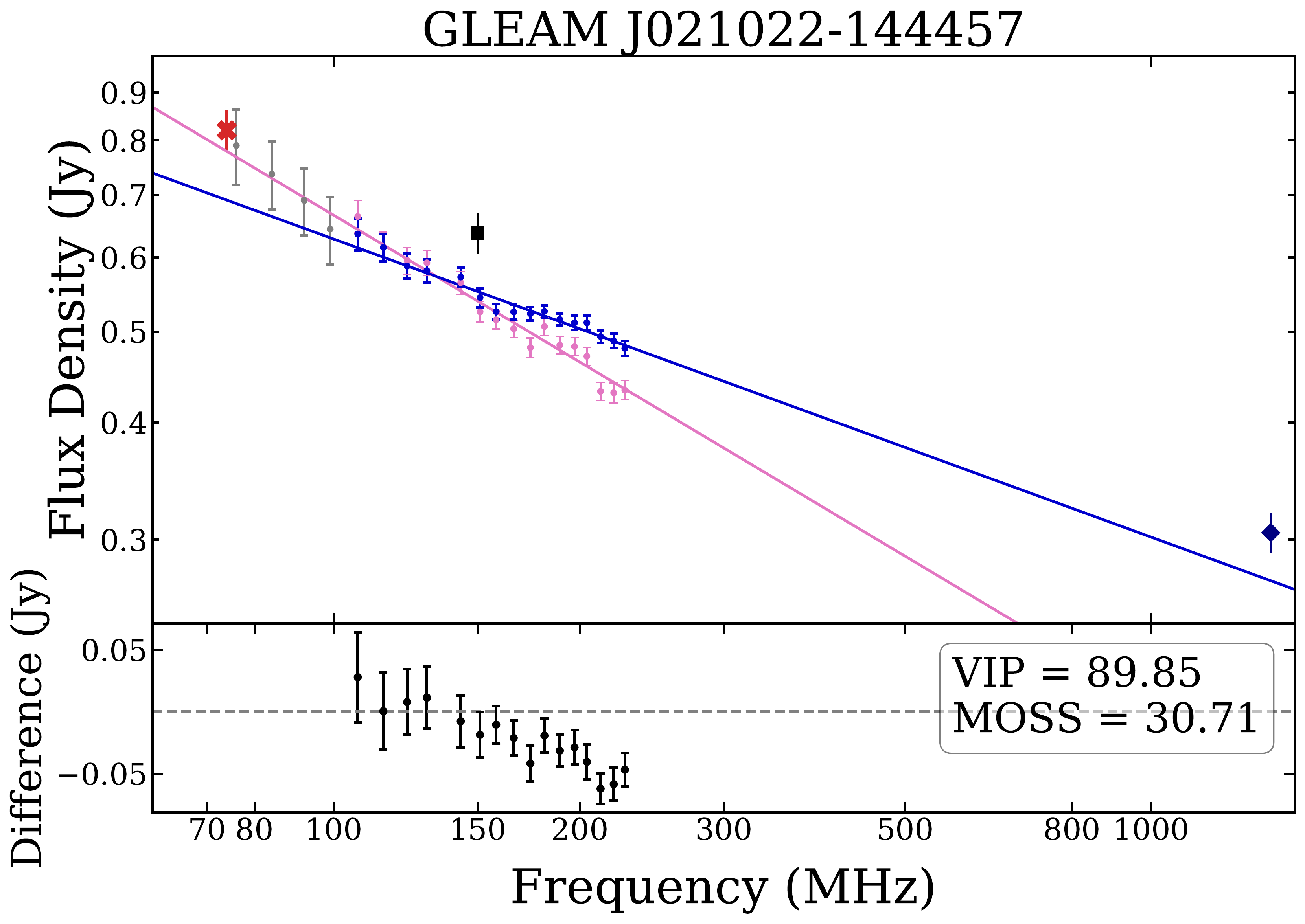} &
\includegraphics[scale=0.15]{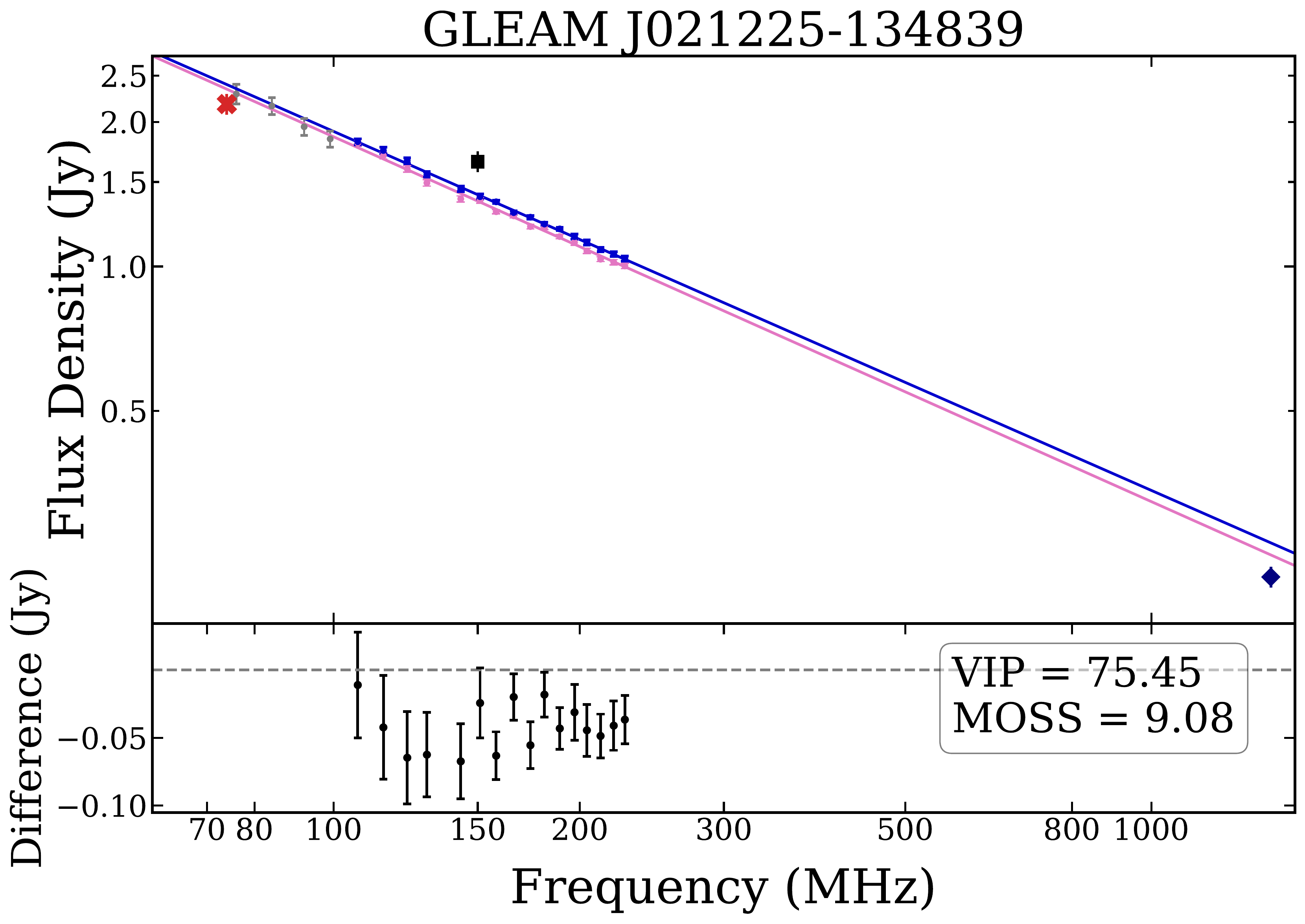} &
\includegraphics[scale=0.15]{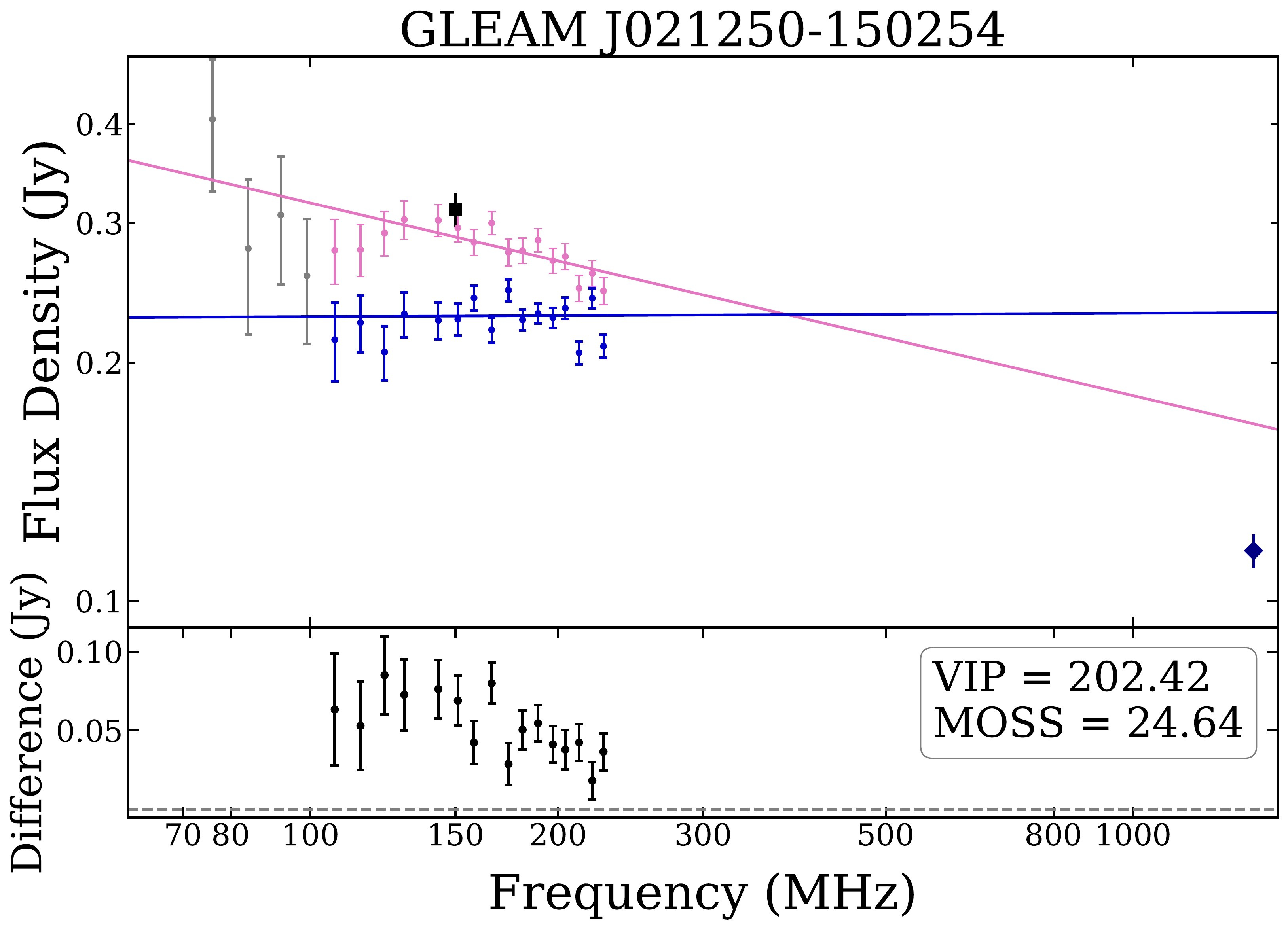} \\
\includegraphics[scale=0.15]{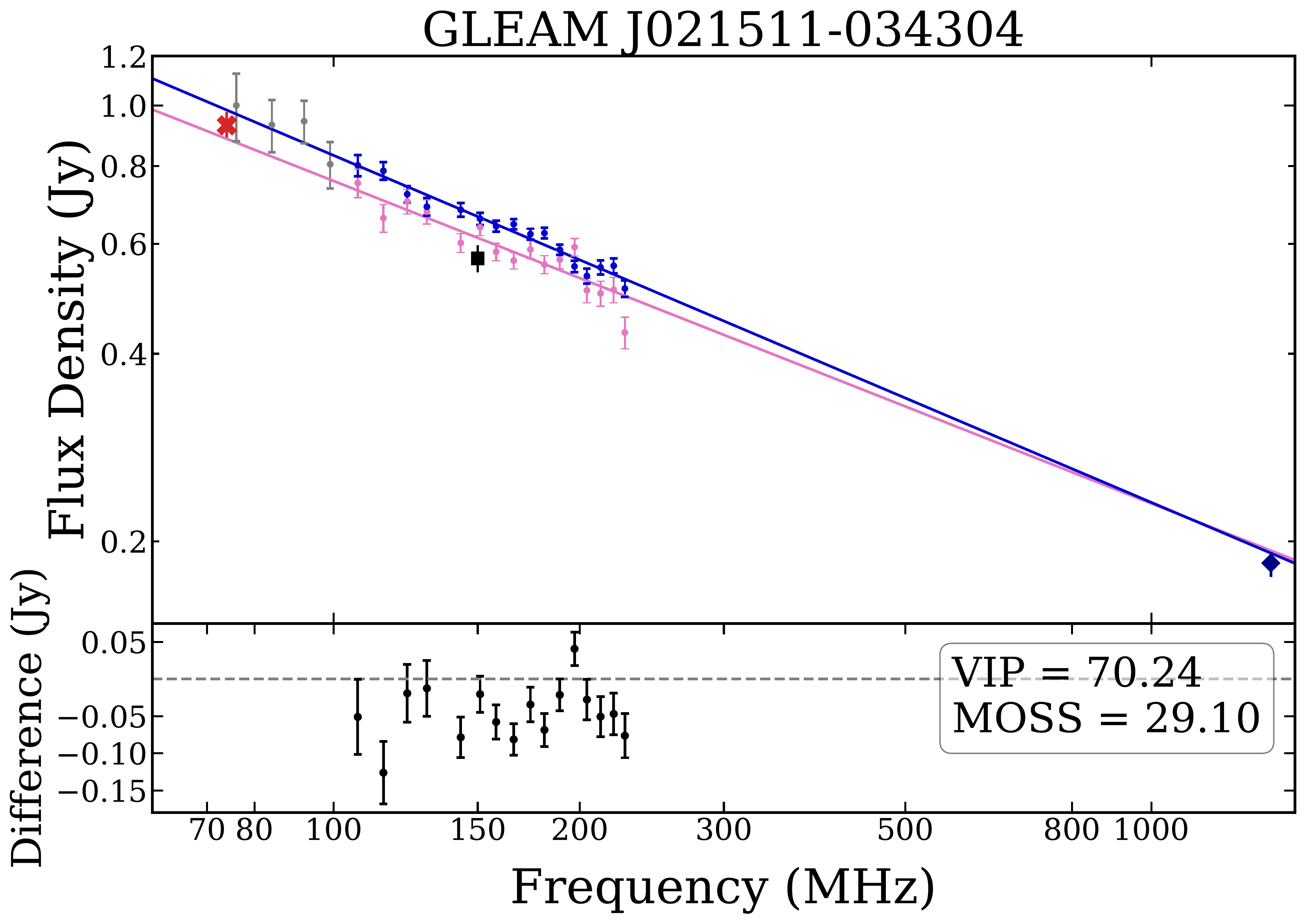} &
\includegraphics[scale=0.15]{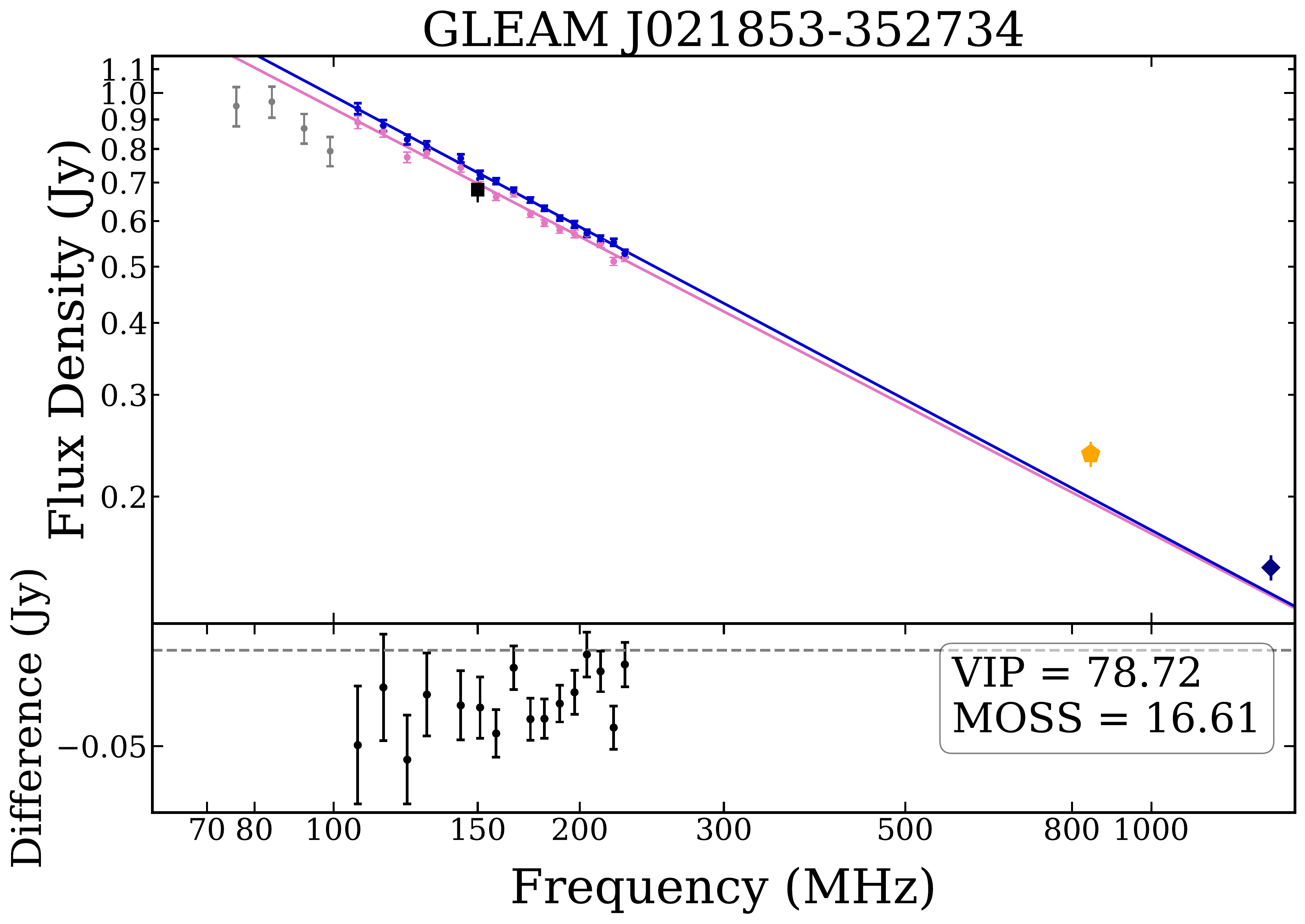} &
\includegraphics[scale=0.15]{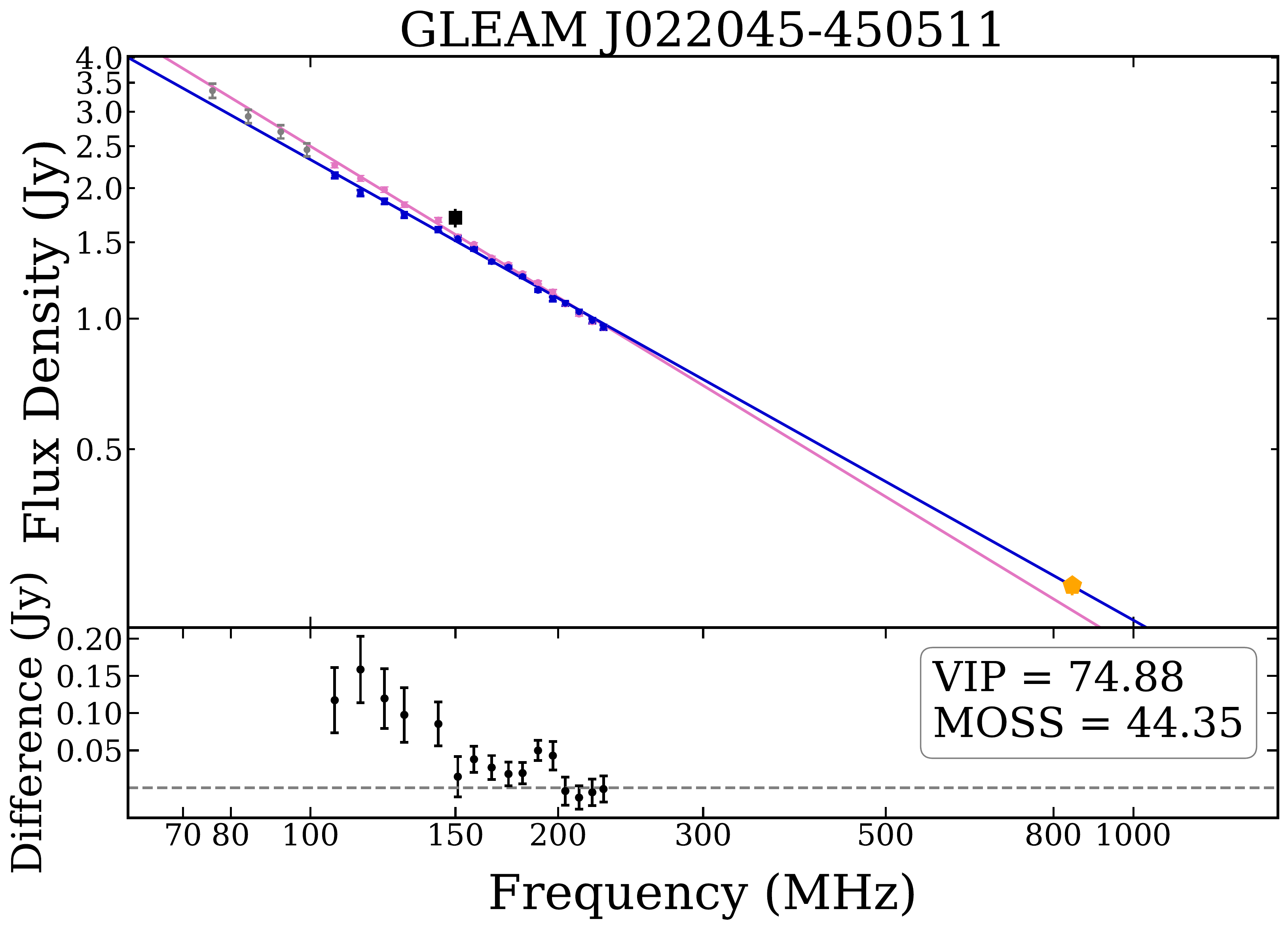} \\
\end{array}$
\caption{(continued) SEDs for all sources classified as variable according to the VIP. For each source the points represent the following data: GLEAM low frequency (72--100\,MHz) (grey circles), Year 1 (pink circles), Year 2 (blue circles), VLSSr (red cross), TGSS (black square), MRC (green star), SUMSS (yellow pentagon), and NVSS (navy diamond). The models for each year are determined by their classification; a source classified with a peak within the observed band was modelled by a quadratic according to Equation~\ref{eq:quadratic}, remaining sources were modelled by a power-law according to Equation~\ref{eq:plaw}.}
\label{app:fig:pg5}
\end{center}
\end{figure*}
\setcounter{figure}{0}
\begin{figure*}
\begin{center}$
\begin{array}{cccccc}
\includegraphics[scale=0.15]{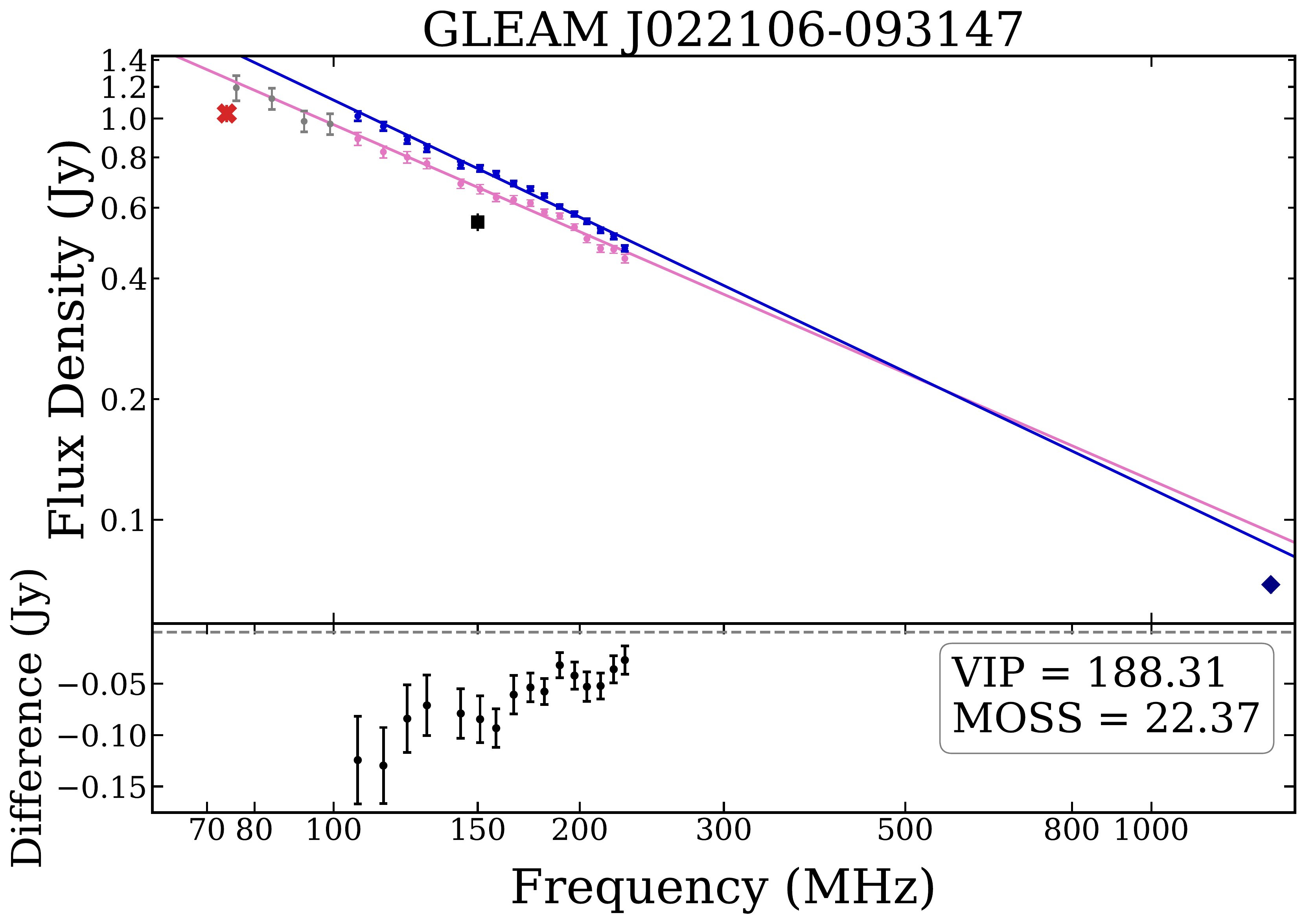} &
\includegraphics[scale=0.15]{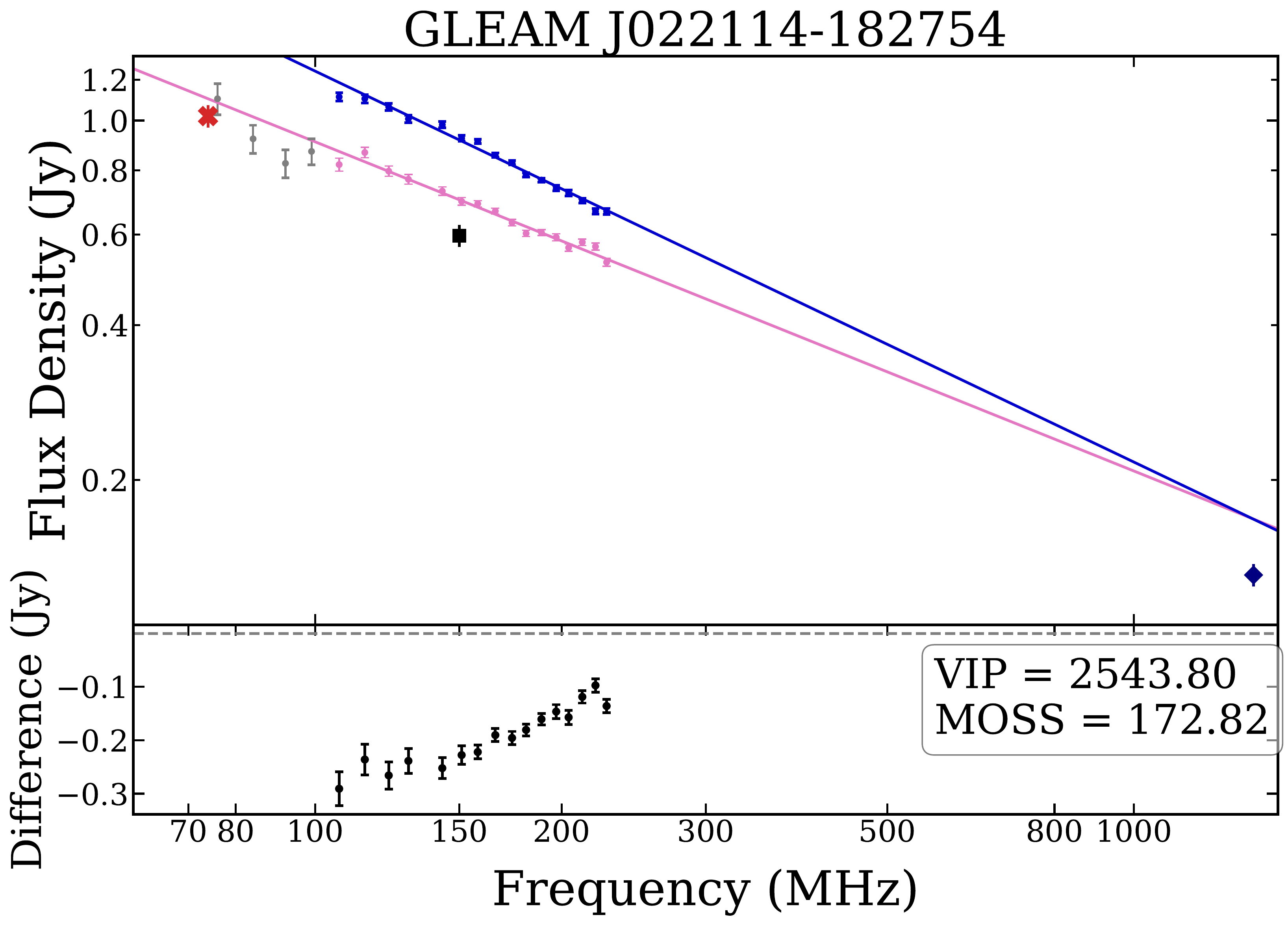} &
\includegraphics[scale=0.15]{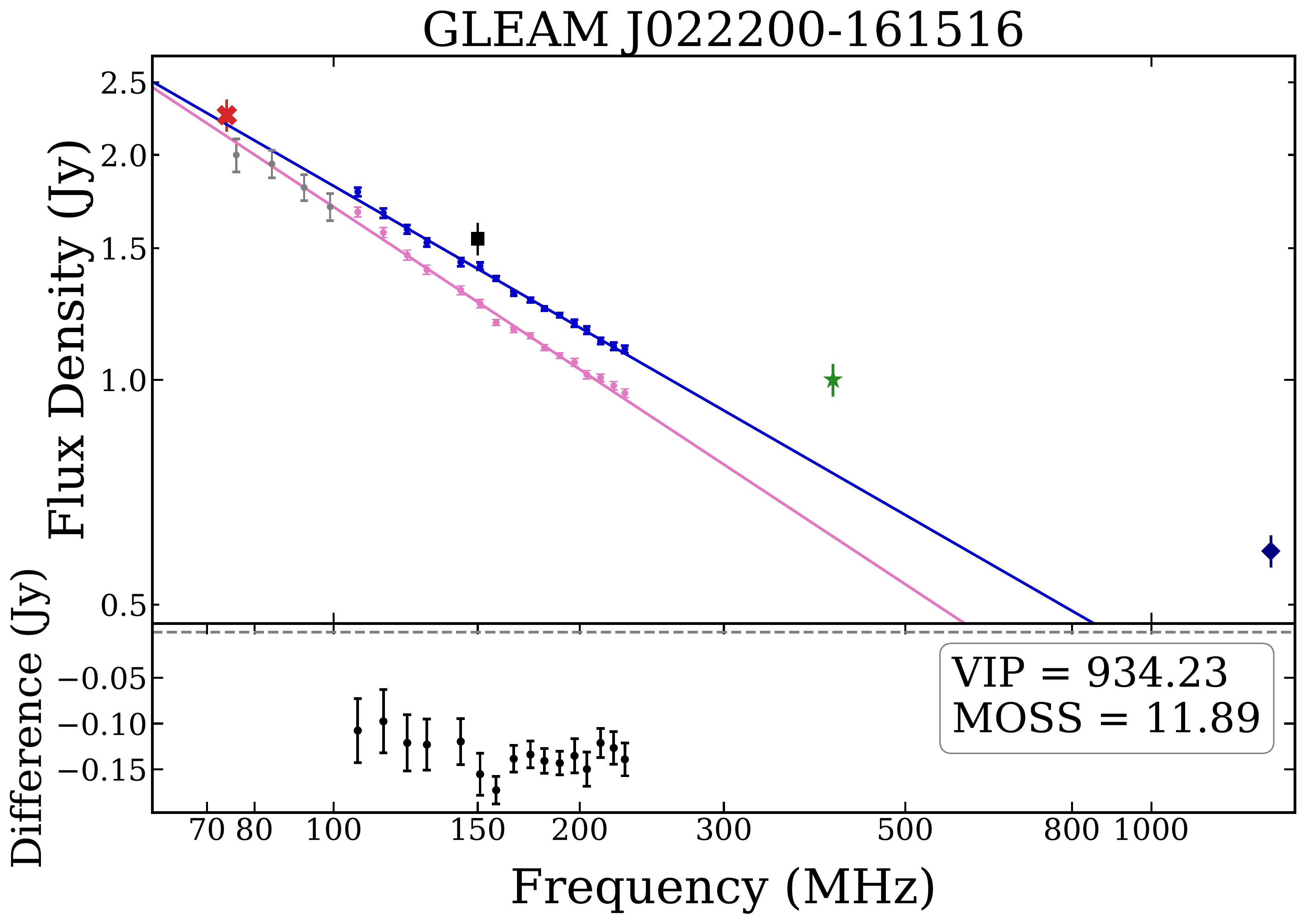} \\
\includegraphics[scale=0.15]{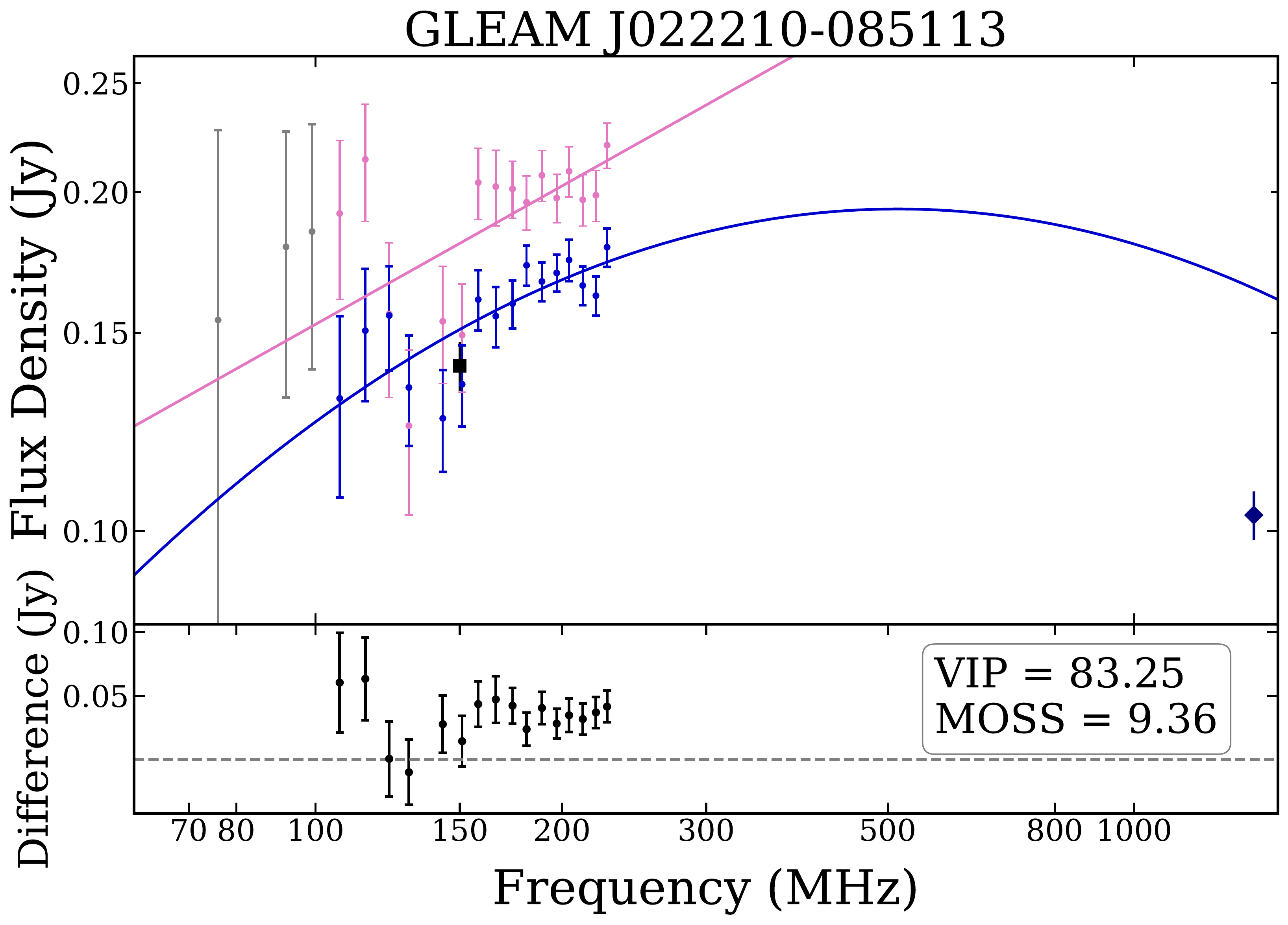} &
\includegraphics[scale=0.15]{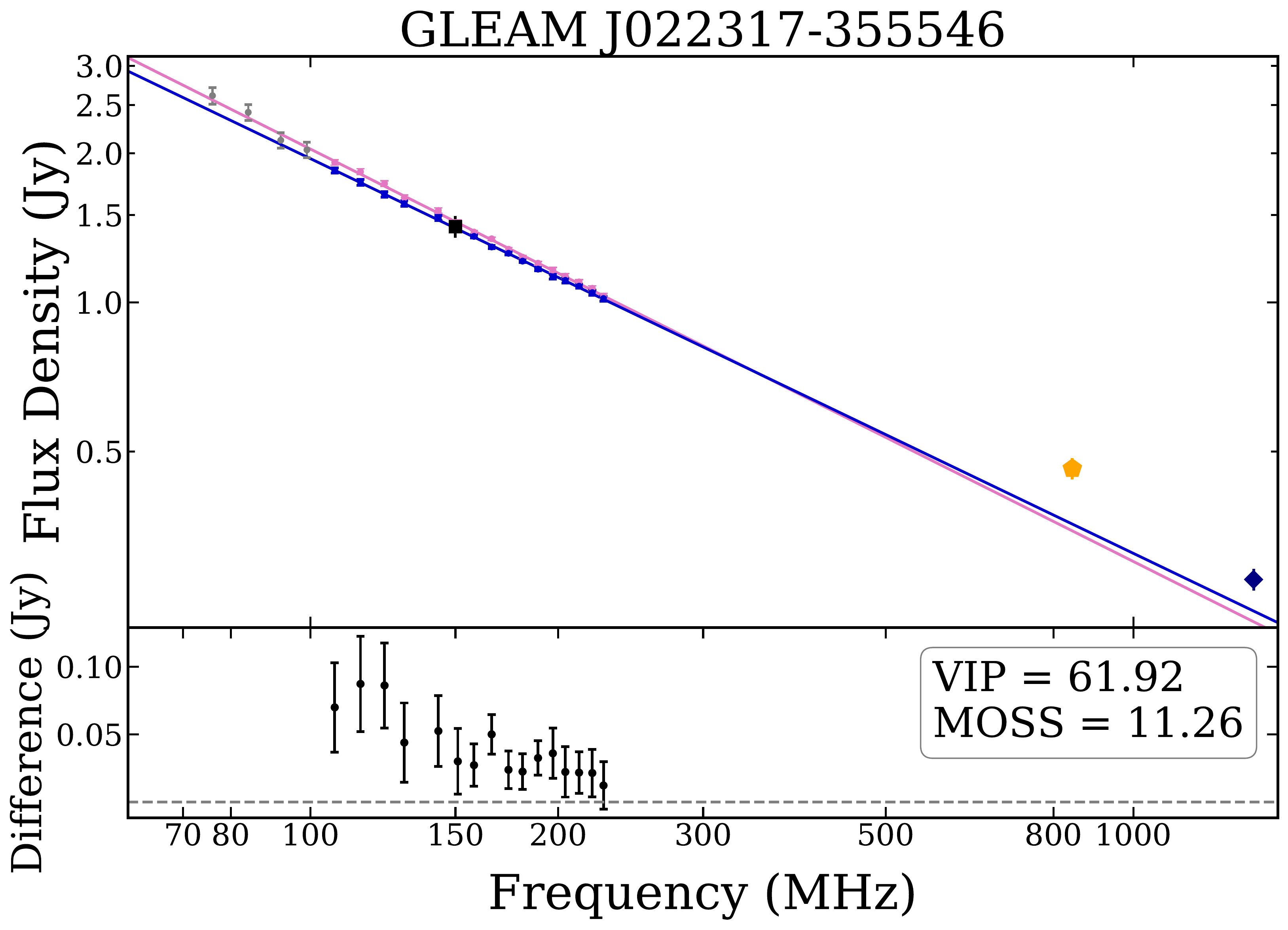} &
\includegraphics[scale=0.15]{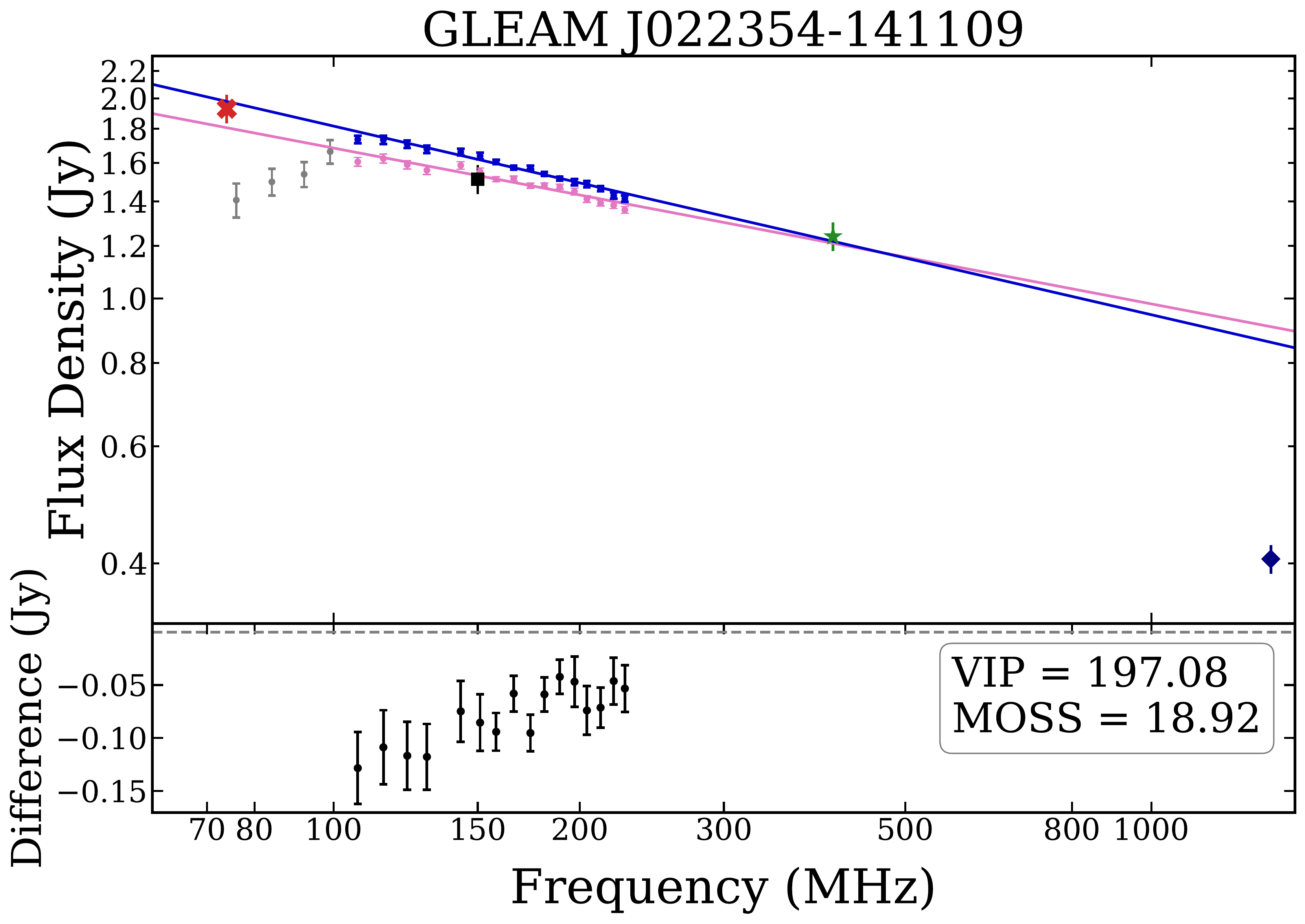} \\
\includegraphics[scale=0.15]{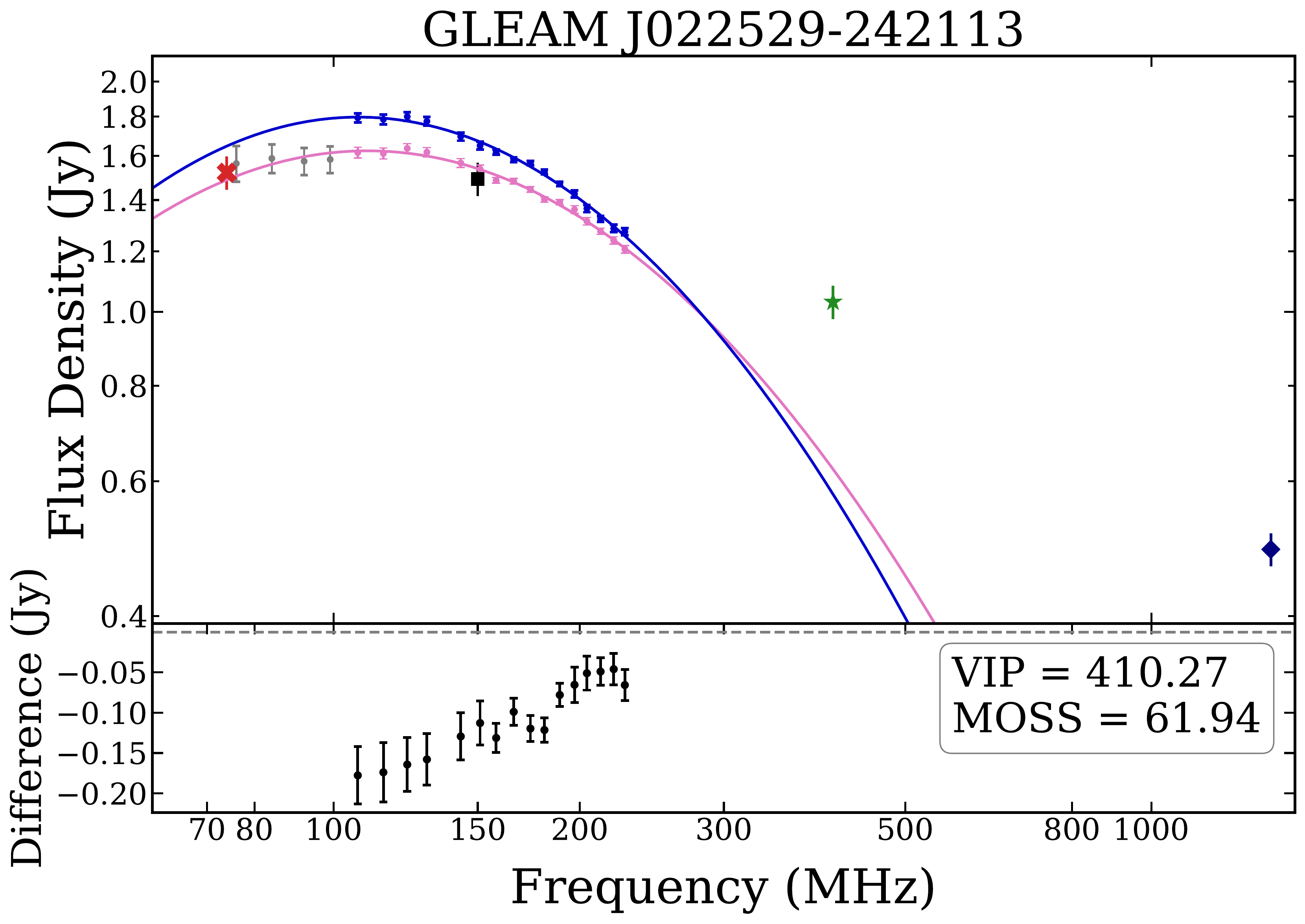} &
\includegraphics[scale=0.15]{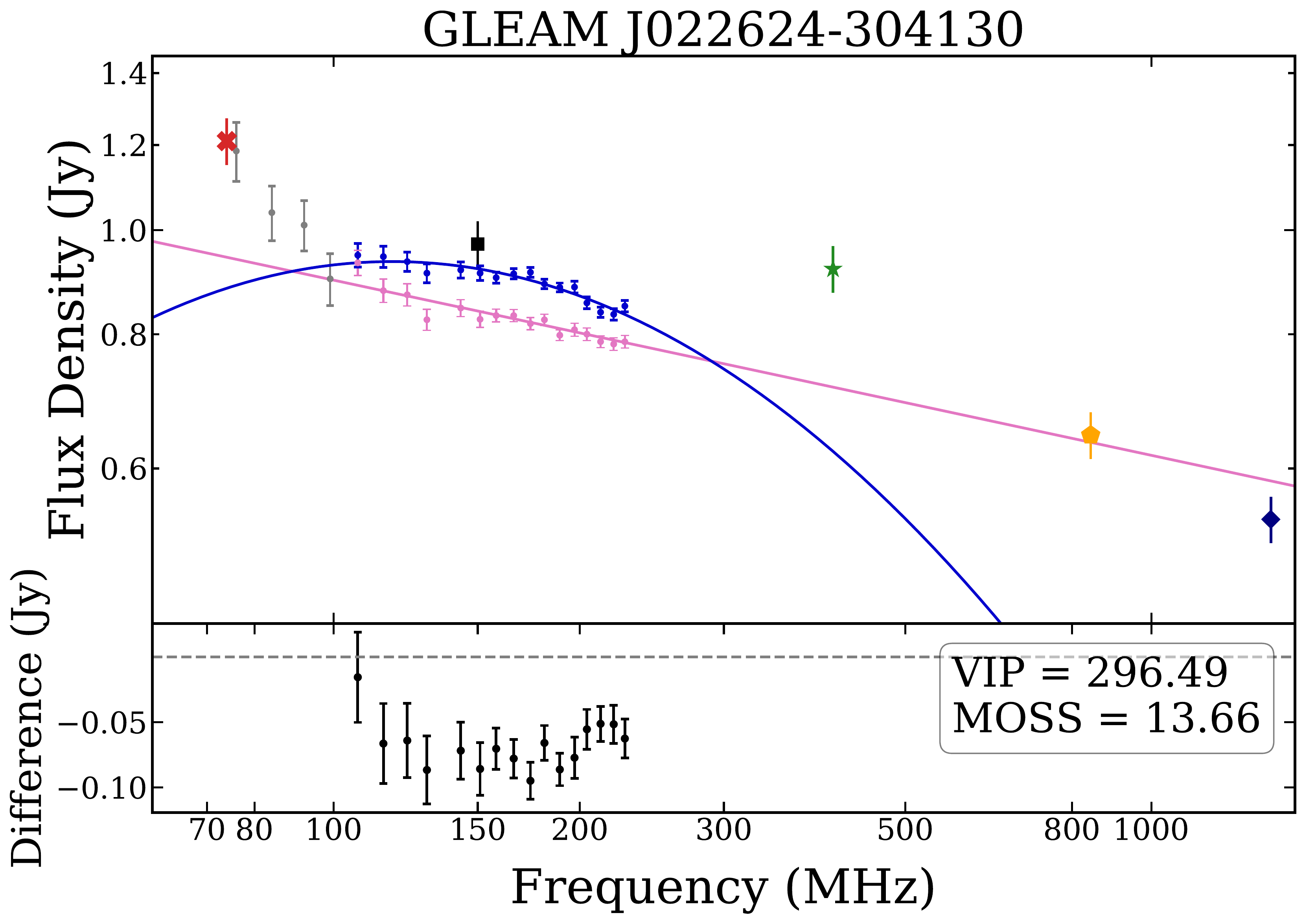} &
\includegraphics[scale=0.15]{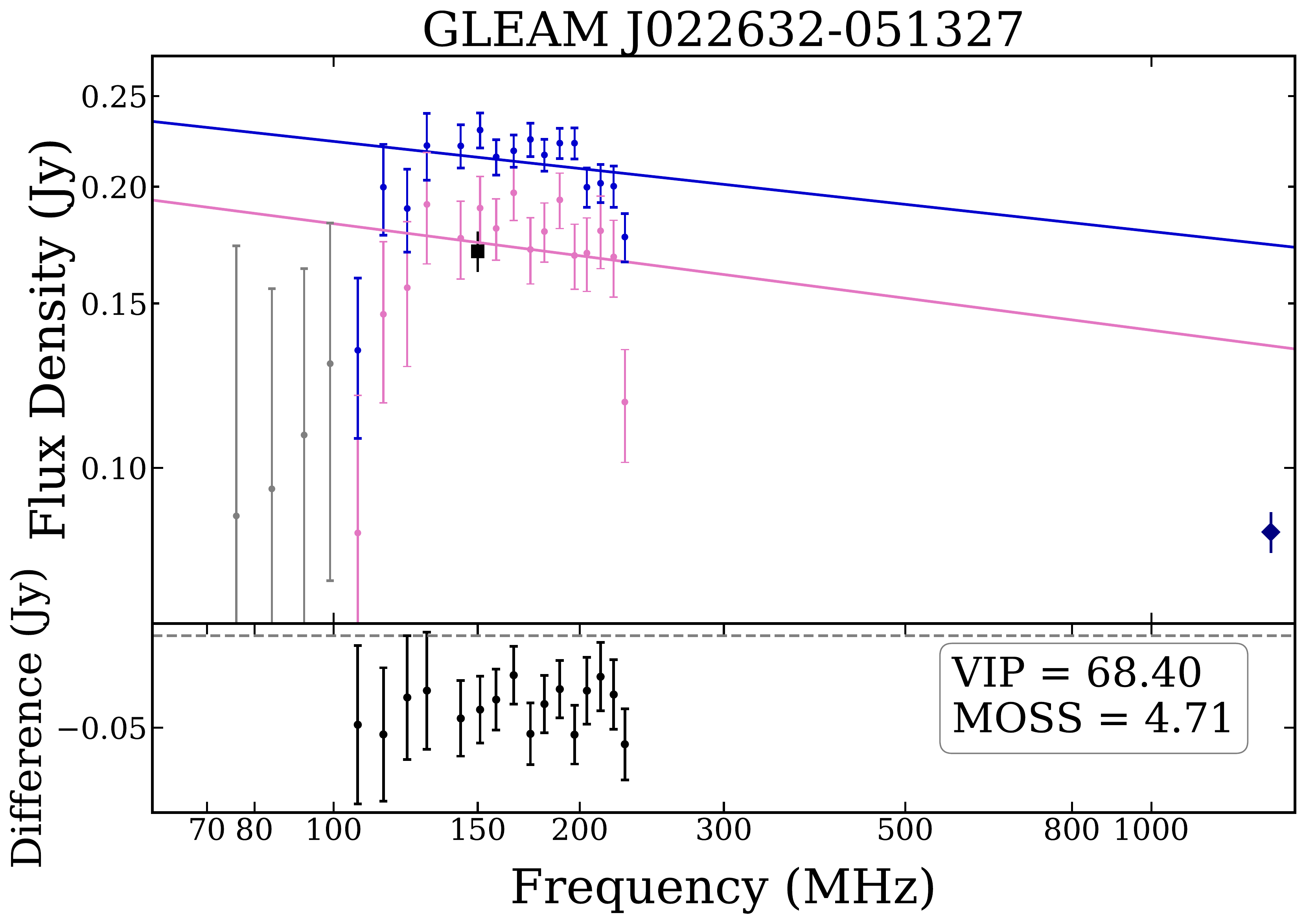} \\
\includegraphics[scale=0.15]{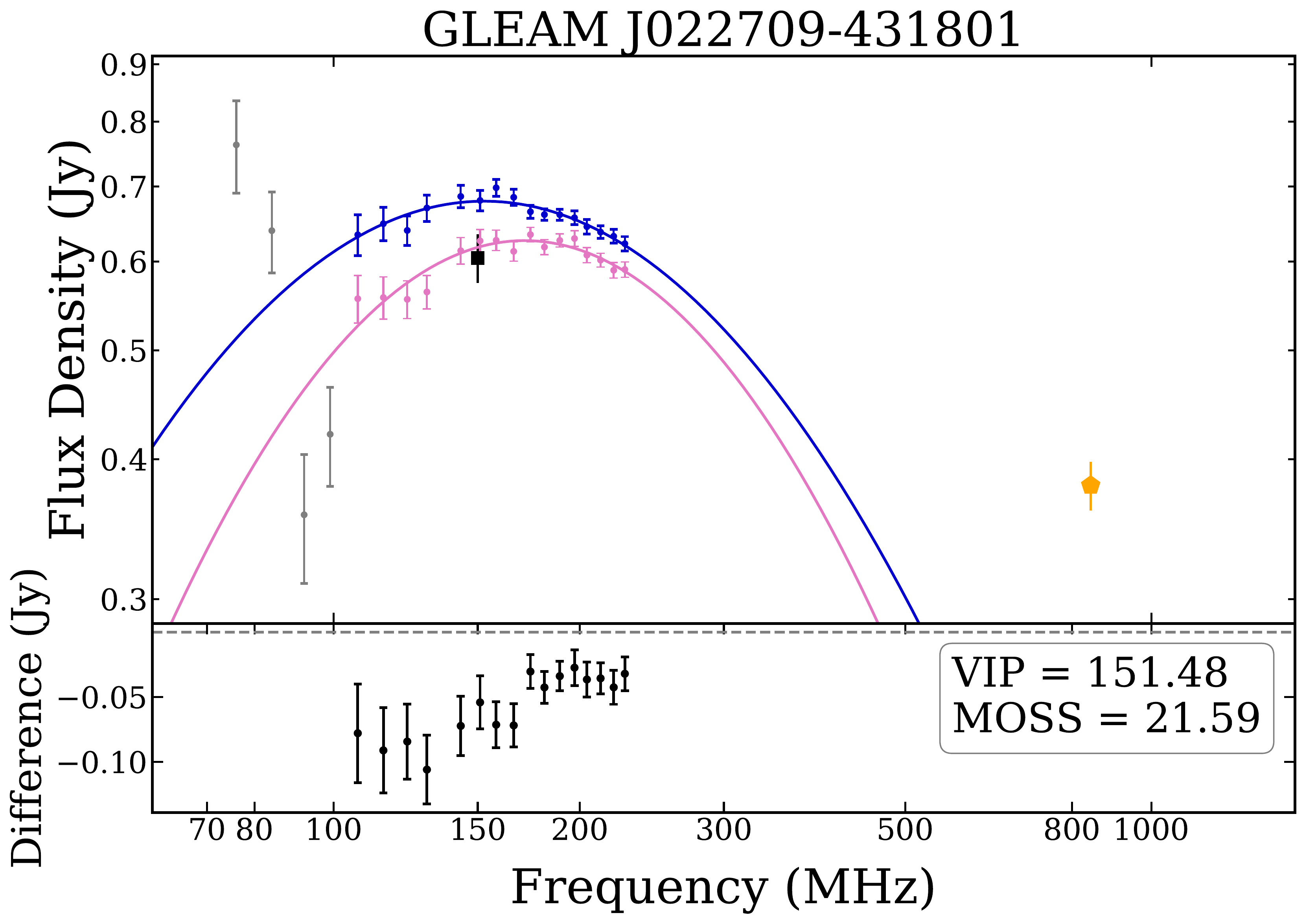} &
\includegraphics[scale=0.15]{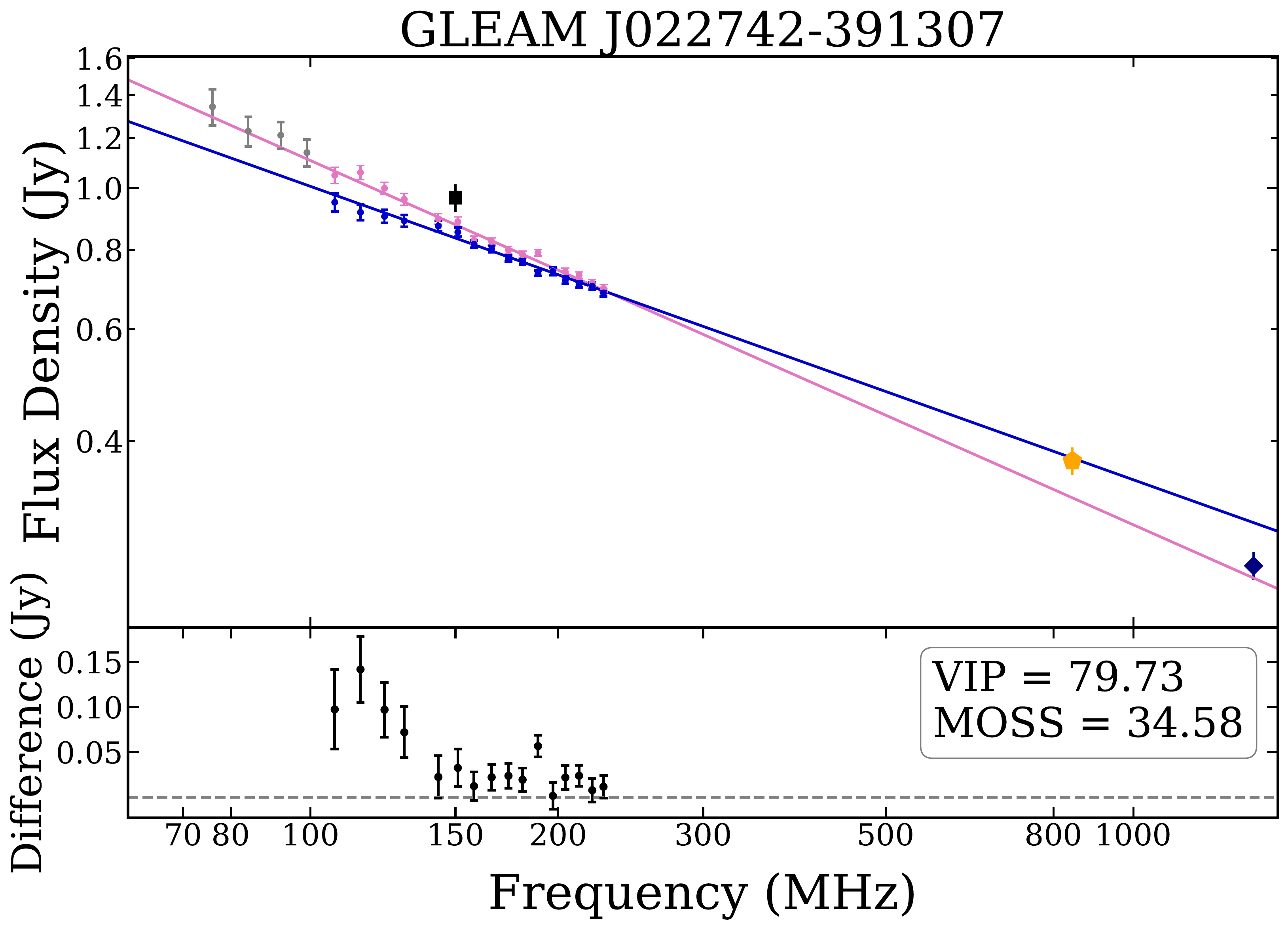} &
\includegraphics[scale=0.15]{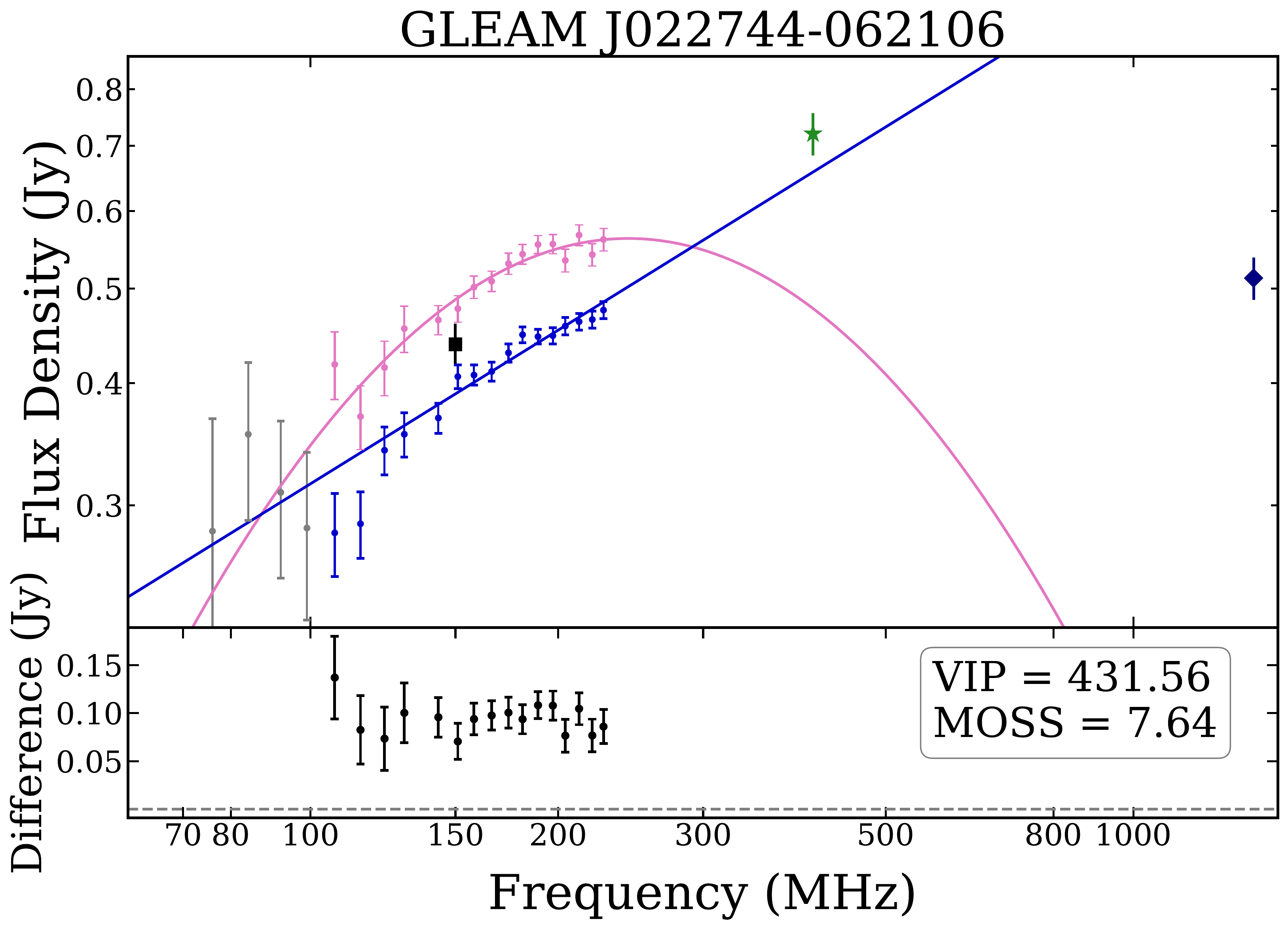} \\
\includegraphics[scale=0.15]{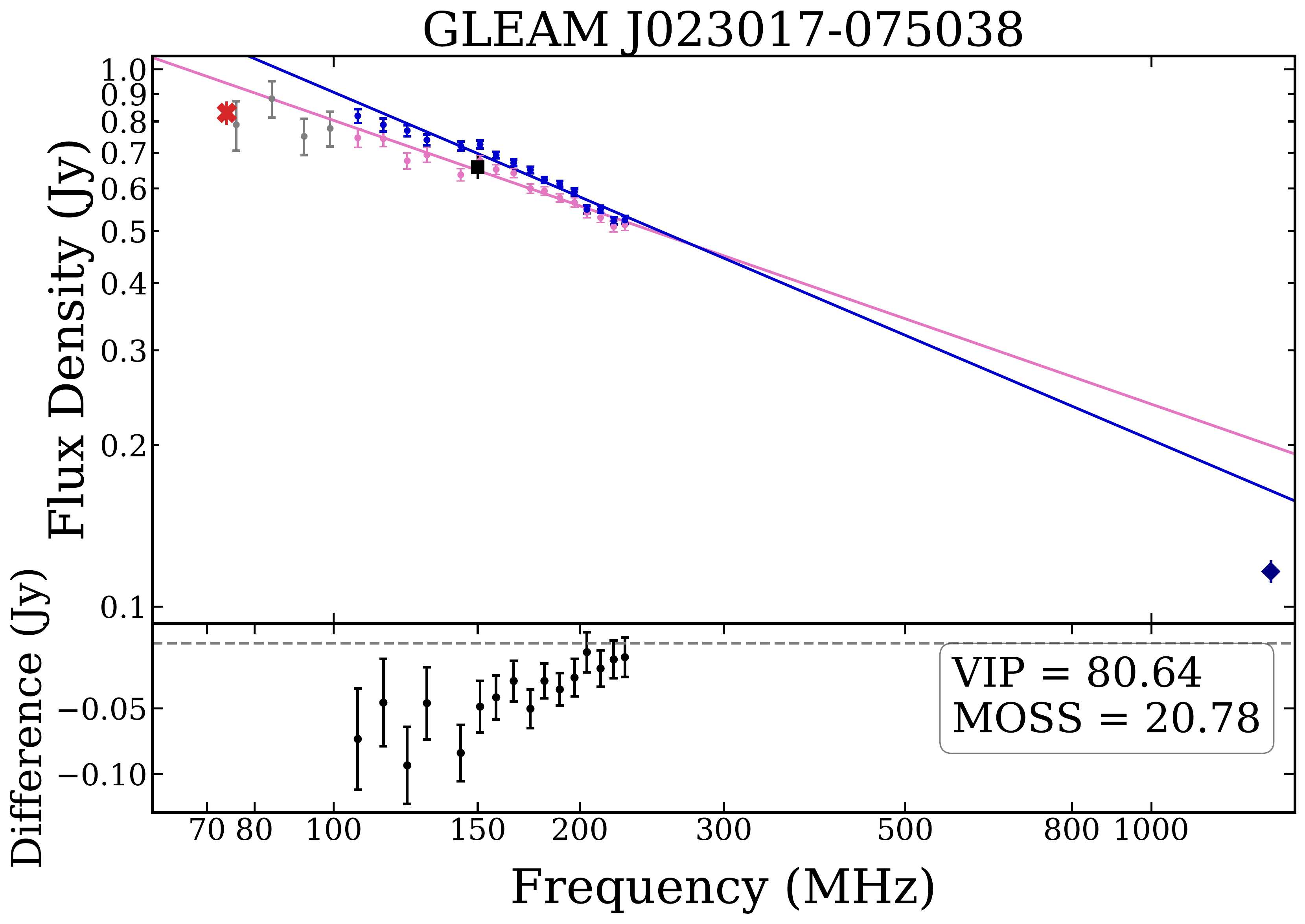} &
\includegraphics[scale=0.15]{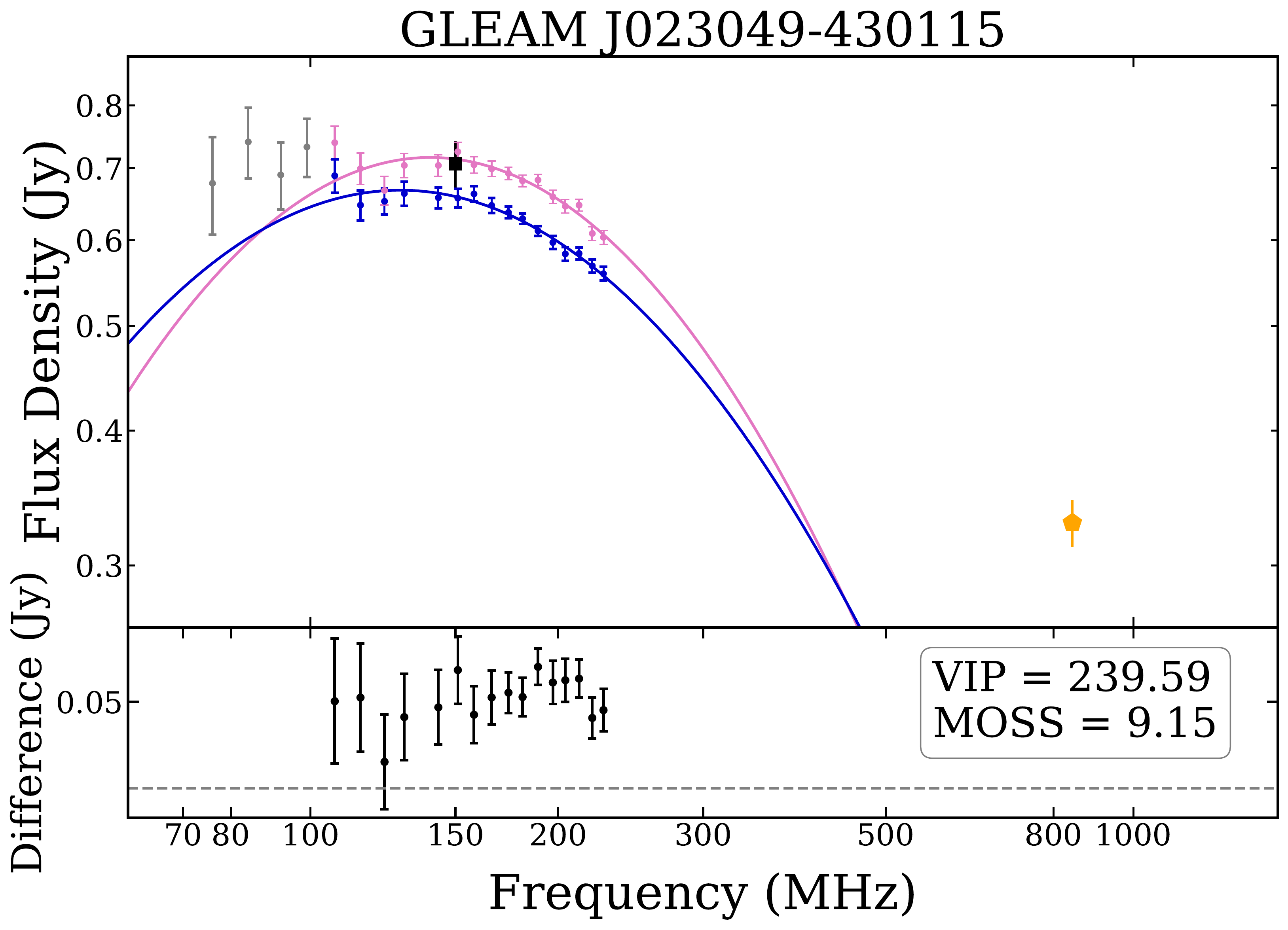} &
\includegraphics[scale=0.15]{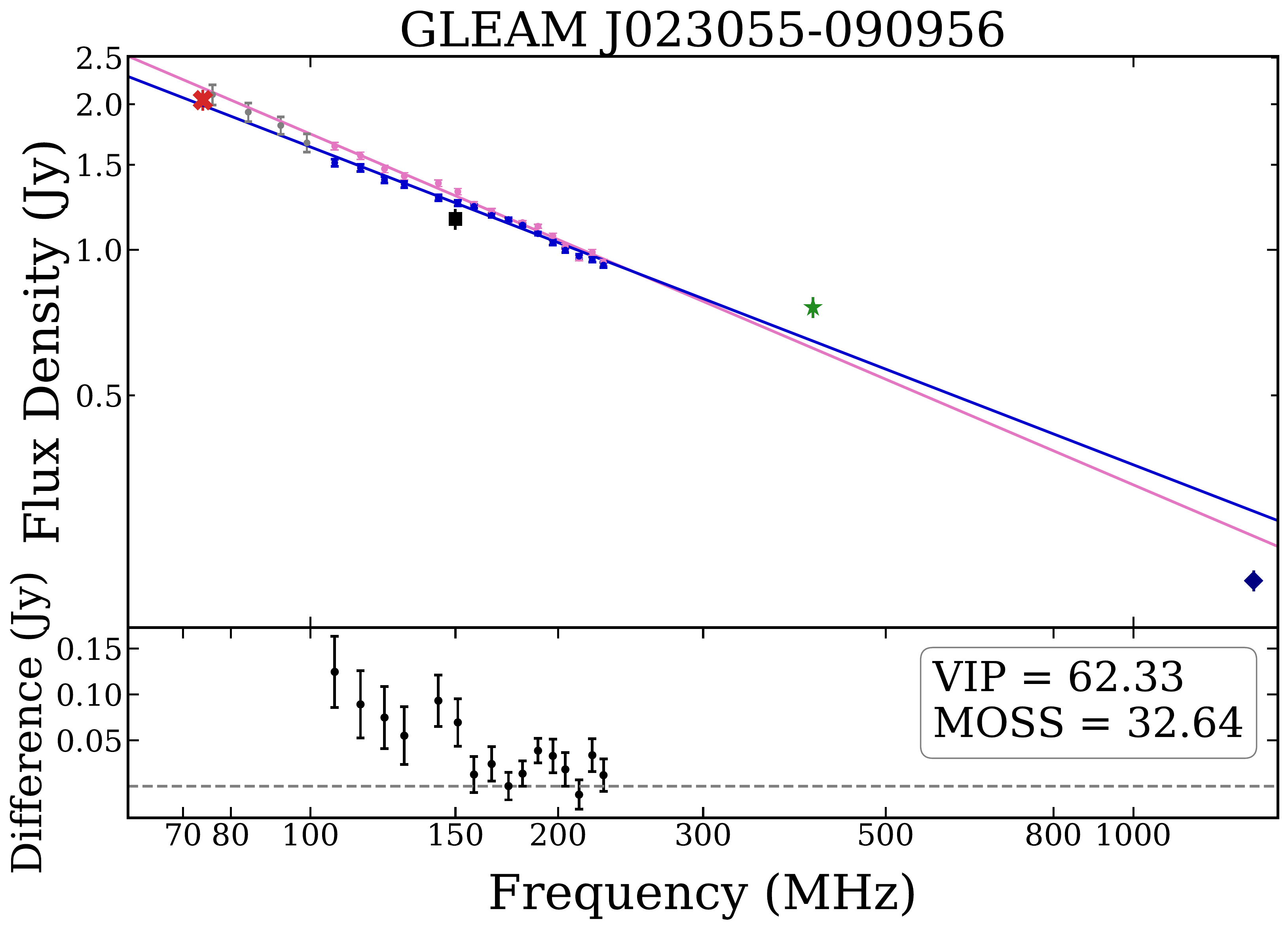} \\
\includegraphics[scale=0.15]{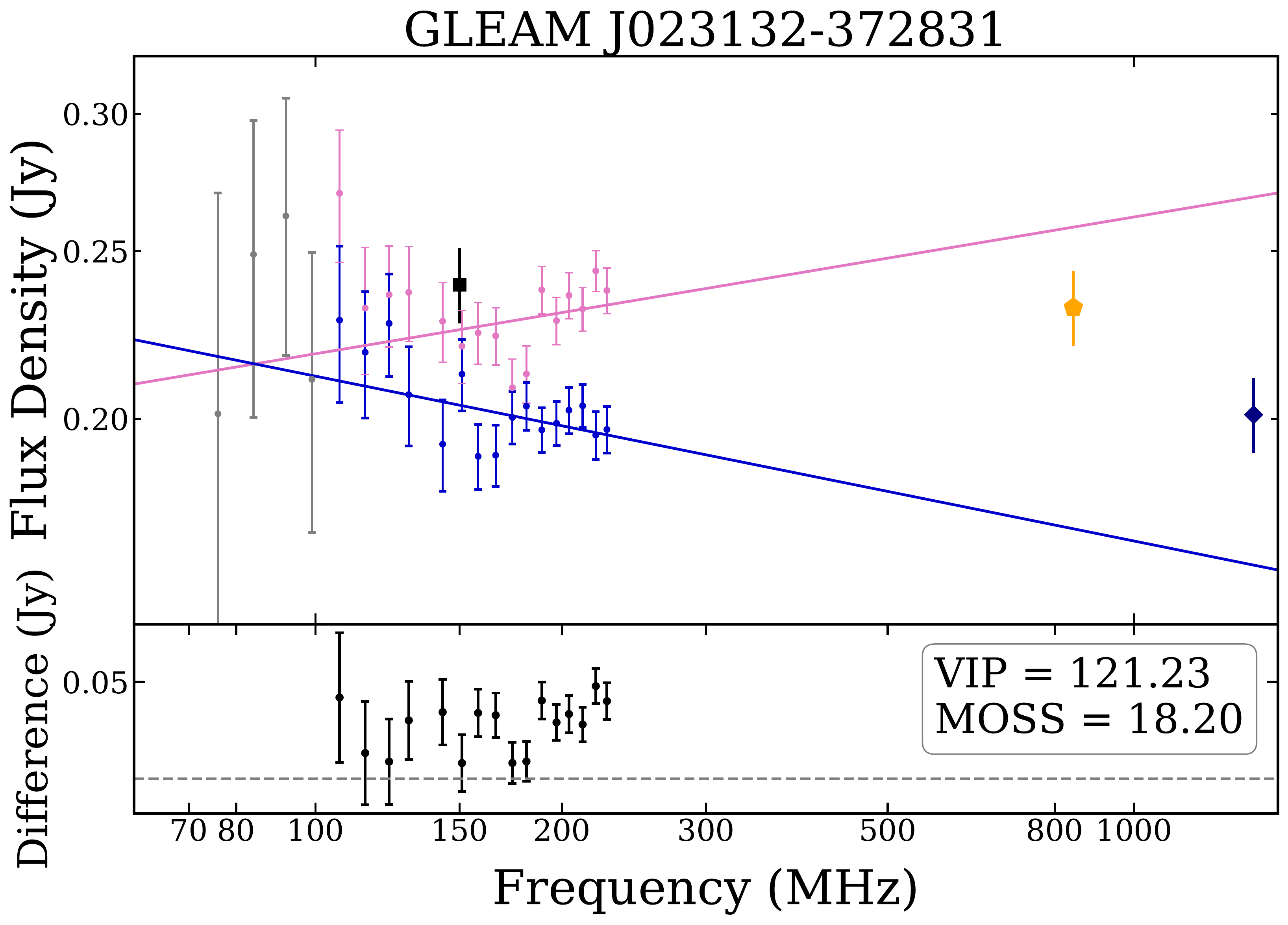} &
\includegraphics[scale=0.15]{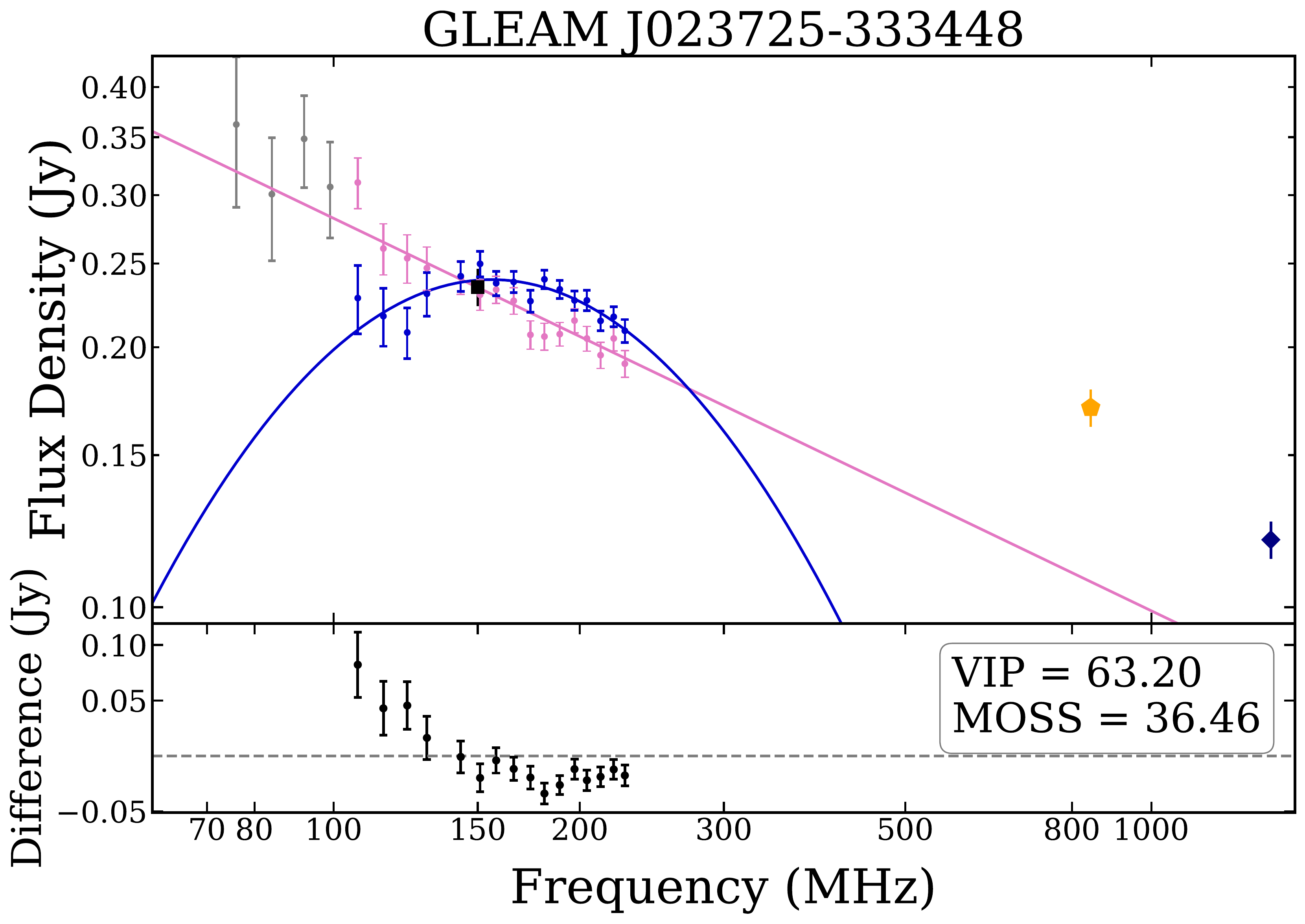} &
\includegraphics[scale=0.15]{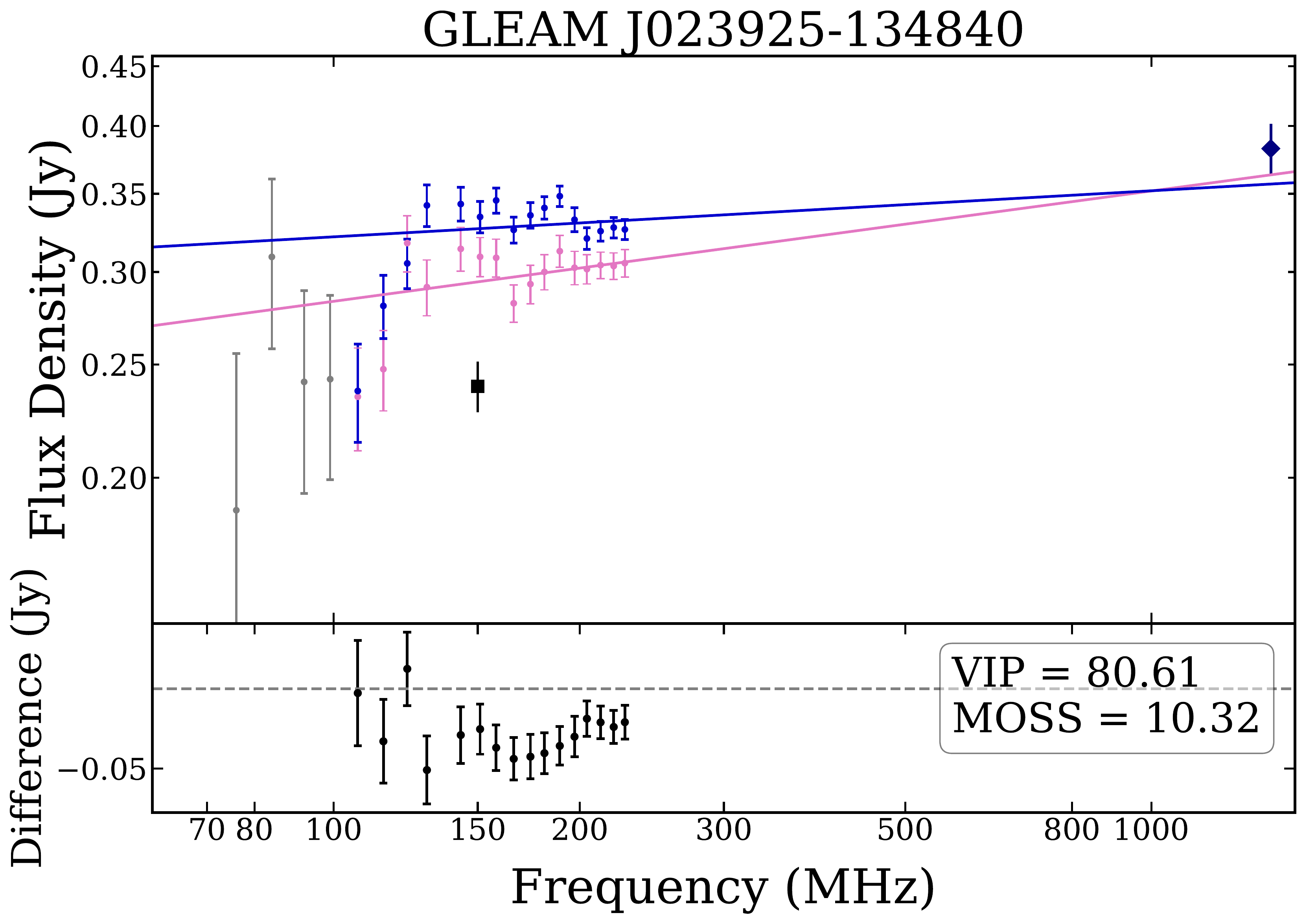} \\
\end{array}$
\caption{(continued) SEDs for all sources classified as variable according to the VIP. For each source the points represent the following data: GLEAM low frequency (72--100\,MHz) (grey circles), Year 1 (pink circles), Year 2 (blue circles), VLSSr (red cross), TGSS (black square), MRC (green star), SUMSS (yellow pentagon), and NVSS (navy diamond). The models for each year are determined by their classification; a source classified with a peak within the observed band was modelled by a quadratic according to Equation~\ref{eq:quadratic}, remaining sources were modelled by a power-law according to Equation~\ref{eq:plaw}.}
\label{app:fig:pg6}
\end{center}
\end{figure*}
\setcounter{figure}{0}
\begin{figure*}
\begin{center}$
\begin{array}{cccccc}
\includegraphics[scale=0.15]{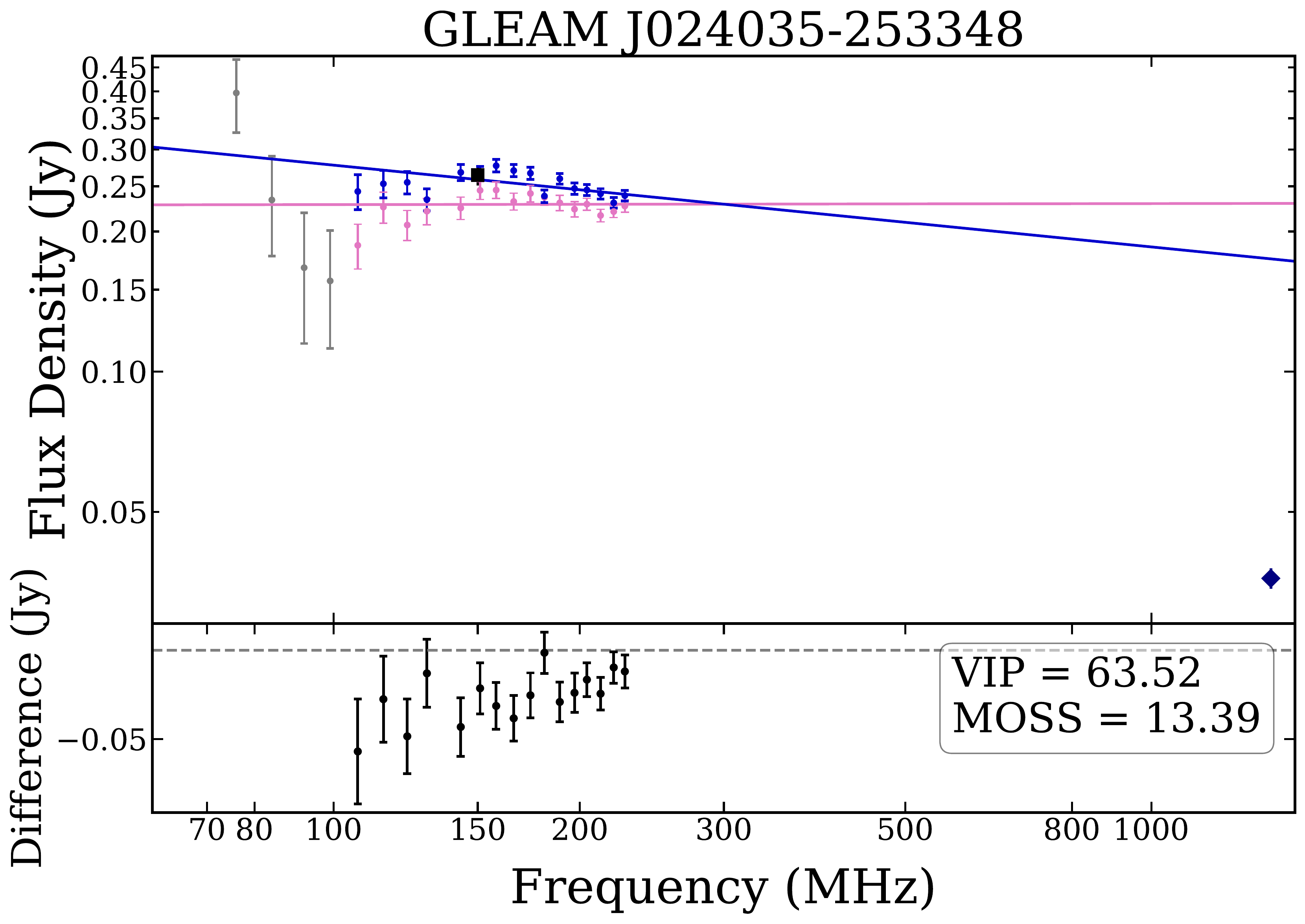} &
\includegraphics[scale=0.15]{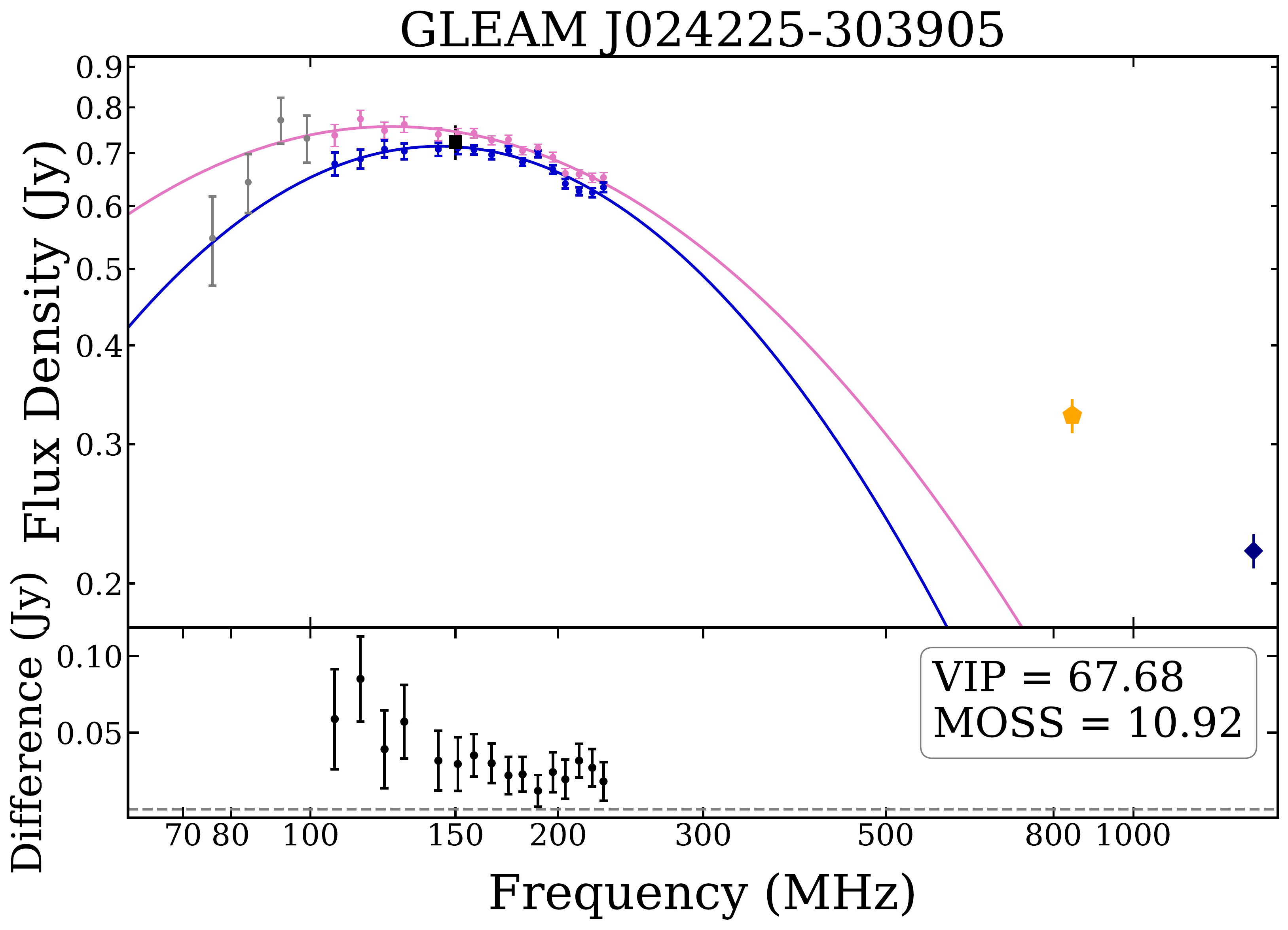} &
\includegraphics[scale=0.15]{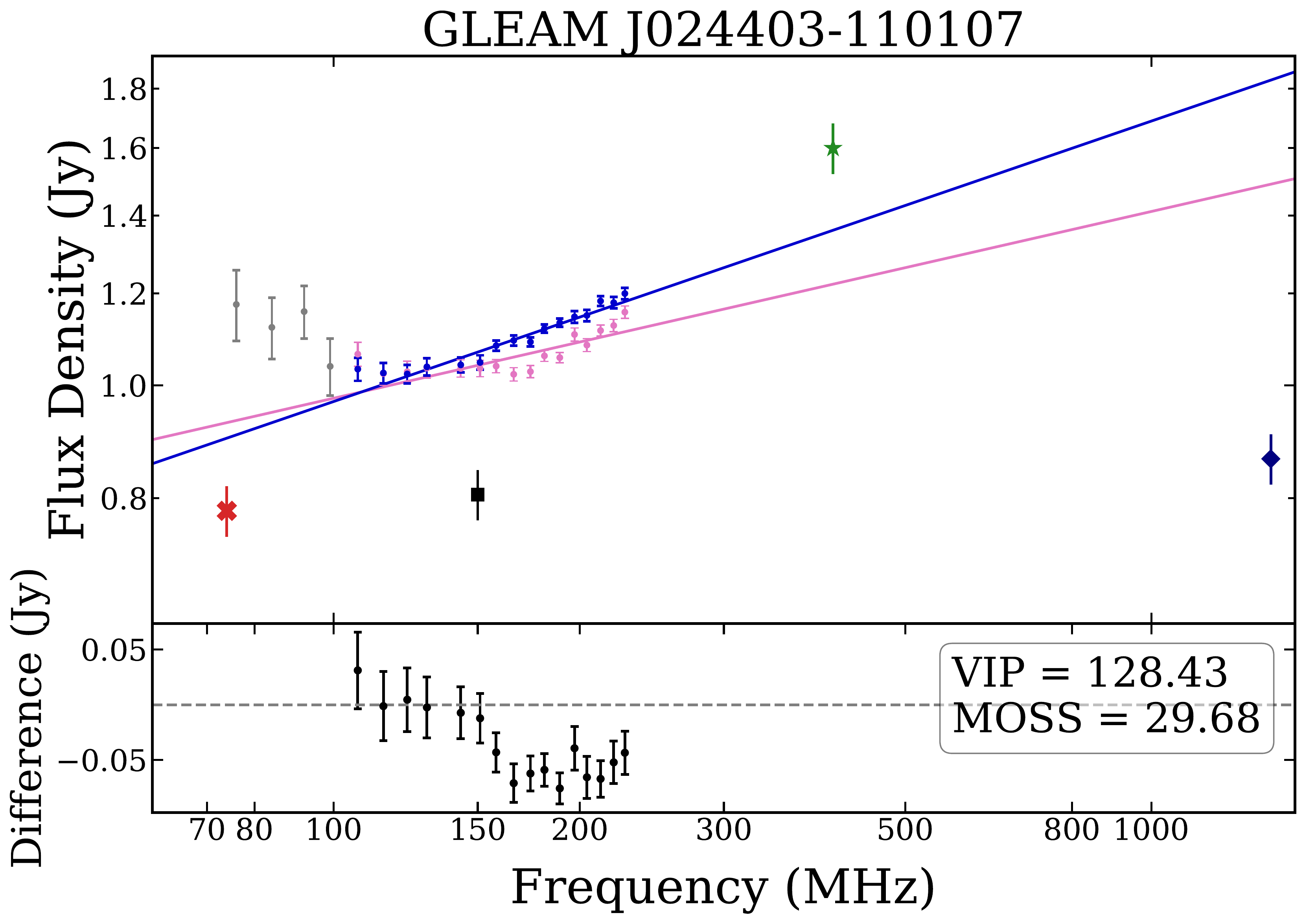} \\
\includegraphics[scale=0.15]{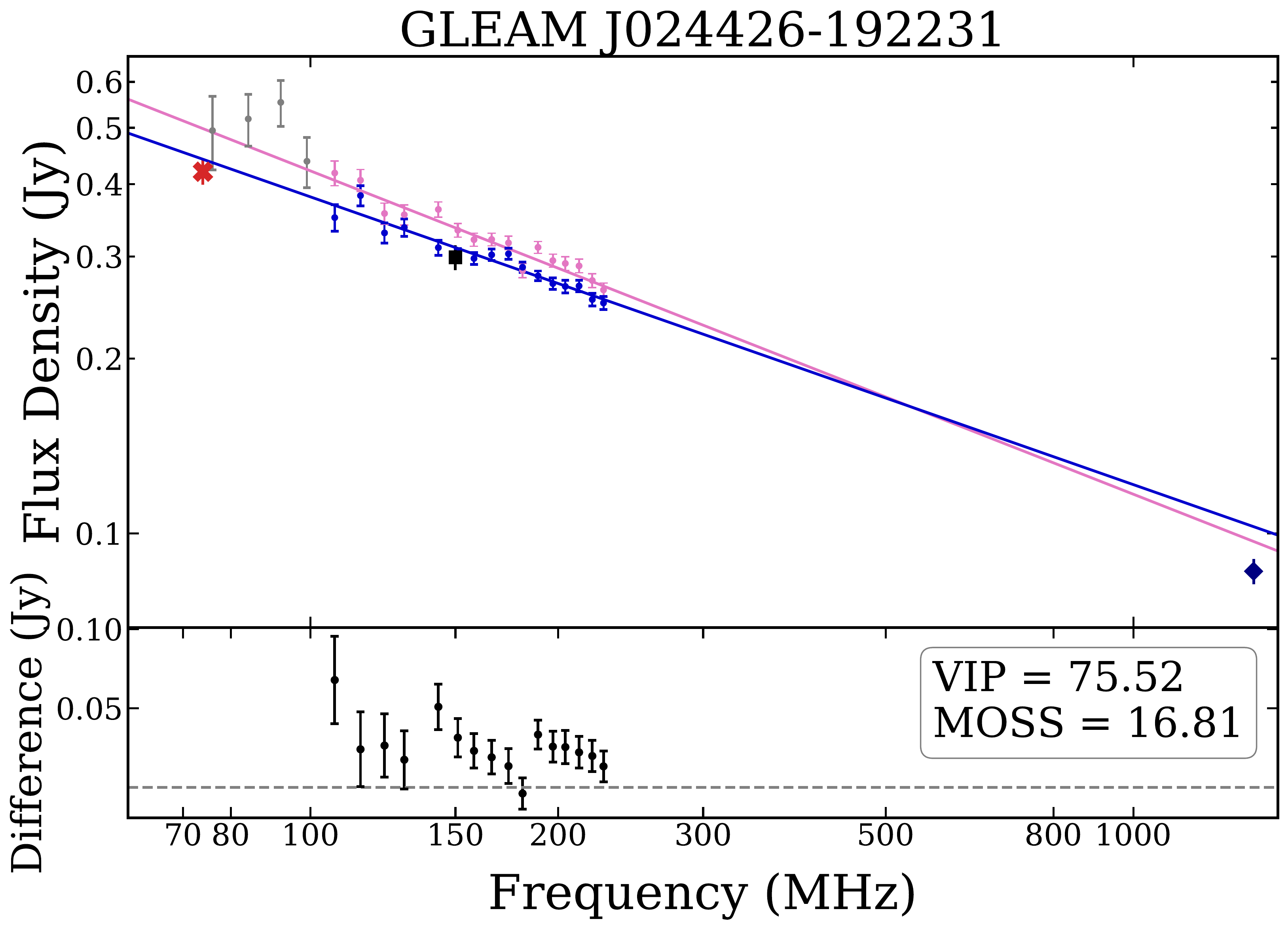} &
\includegraphics[scale=0.15]{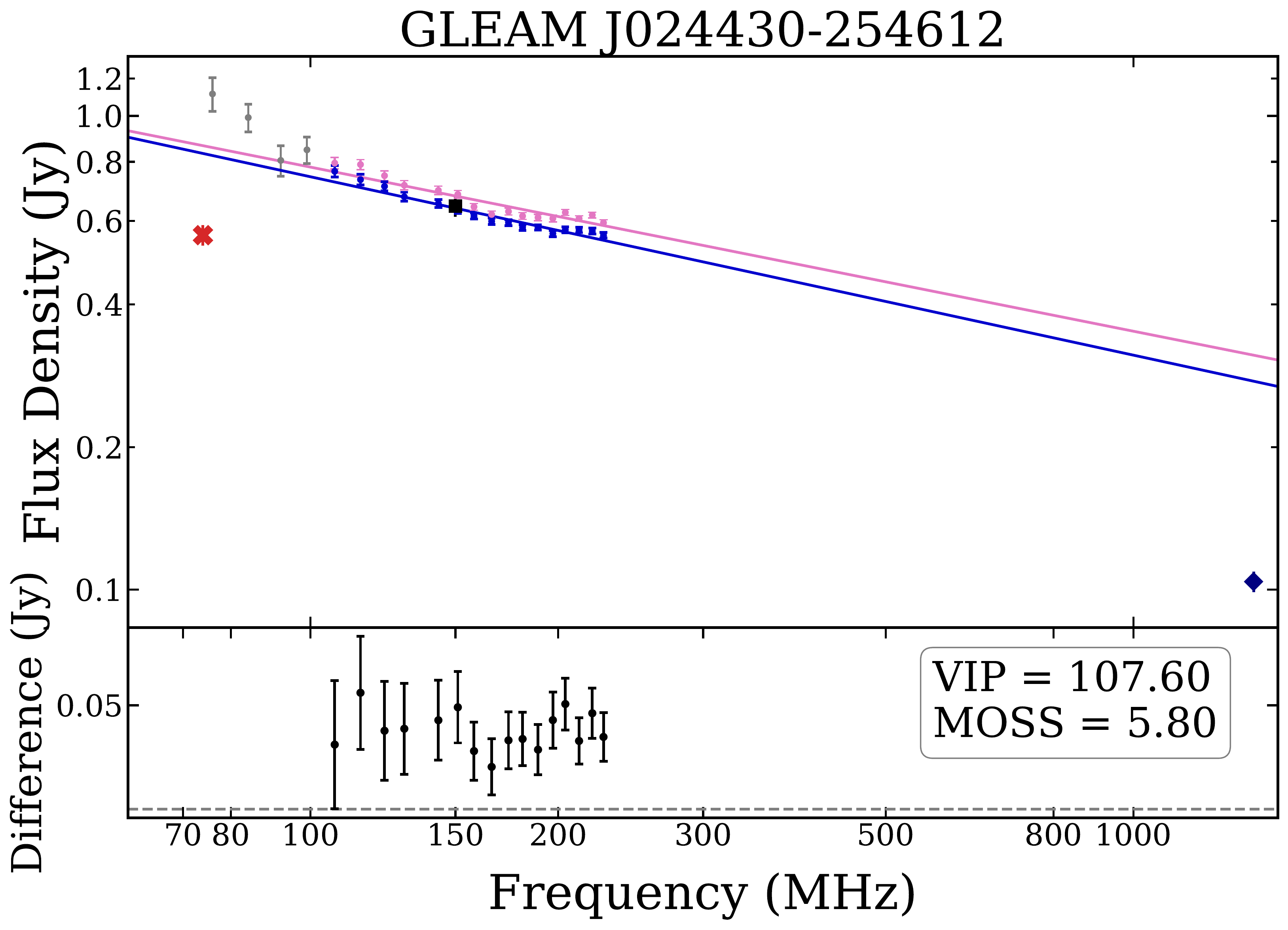} &
\includegraphics[scale=0.15]{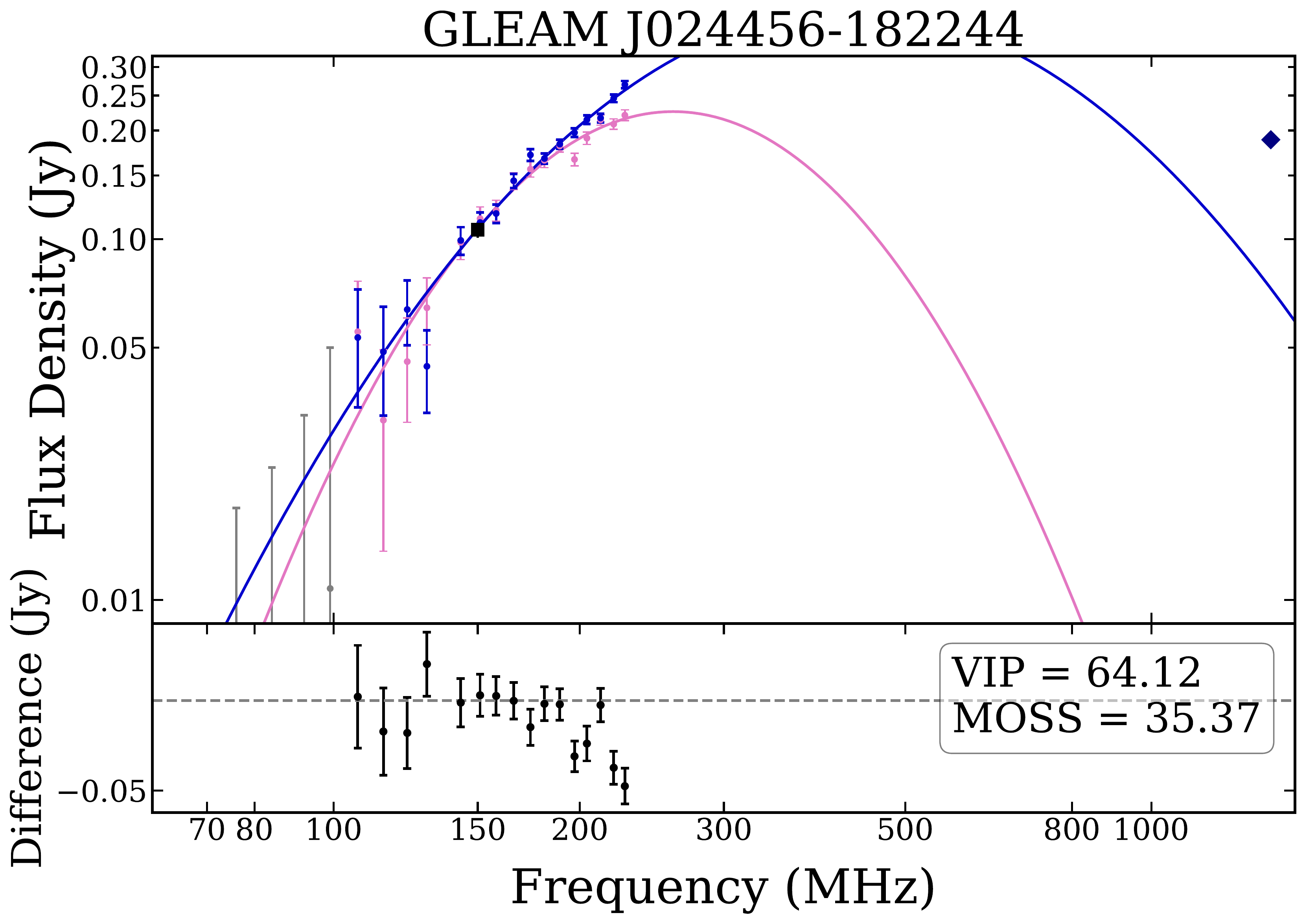} \\
\includegraphics[scale=0.15]{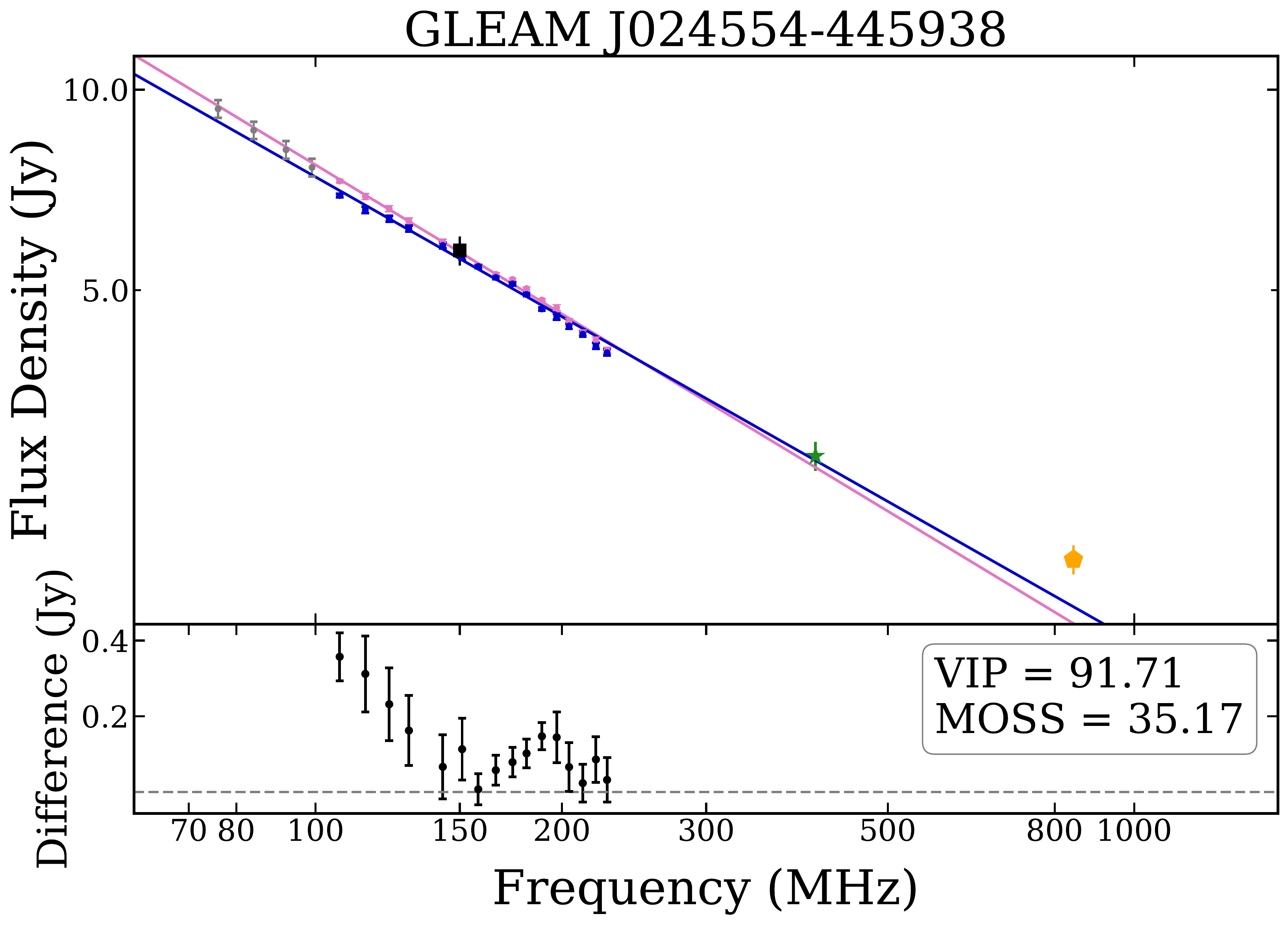} &
\includegraphics[scale=0.15]{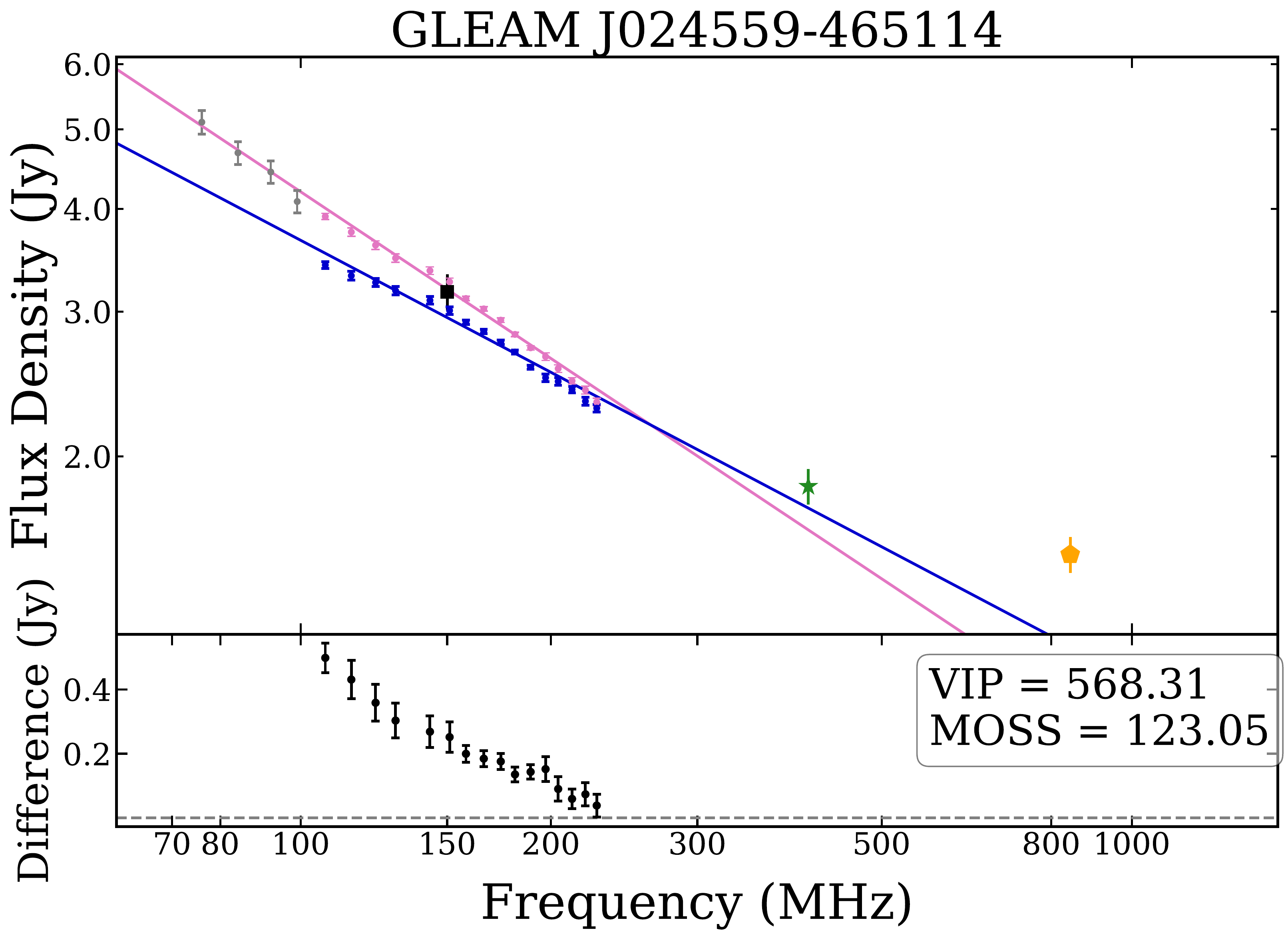} &
\includegraphics[scale=0.15]{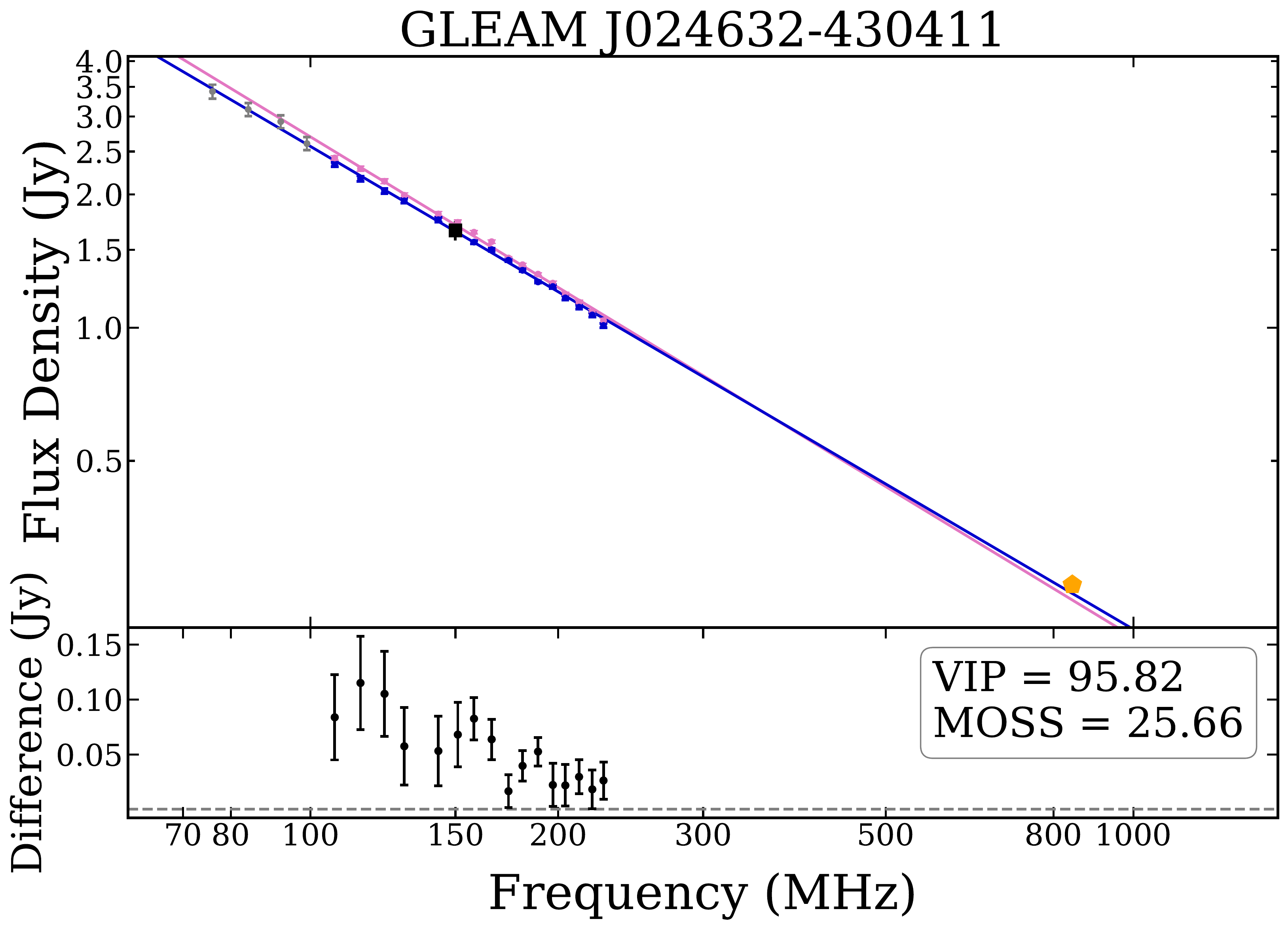} \\
\includegraphics[scale=0.15]{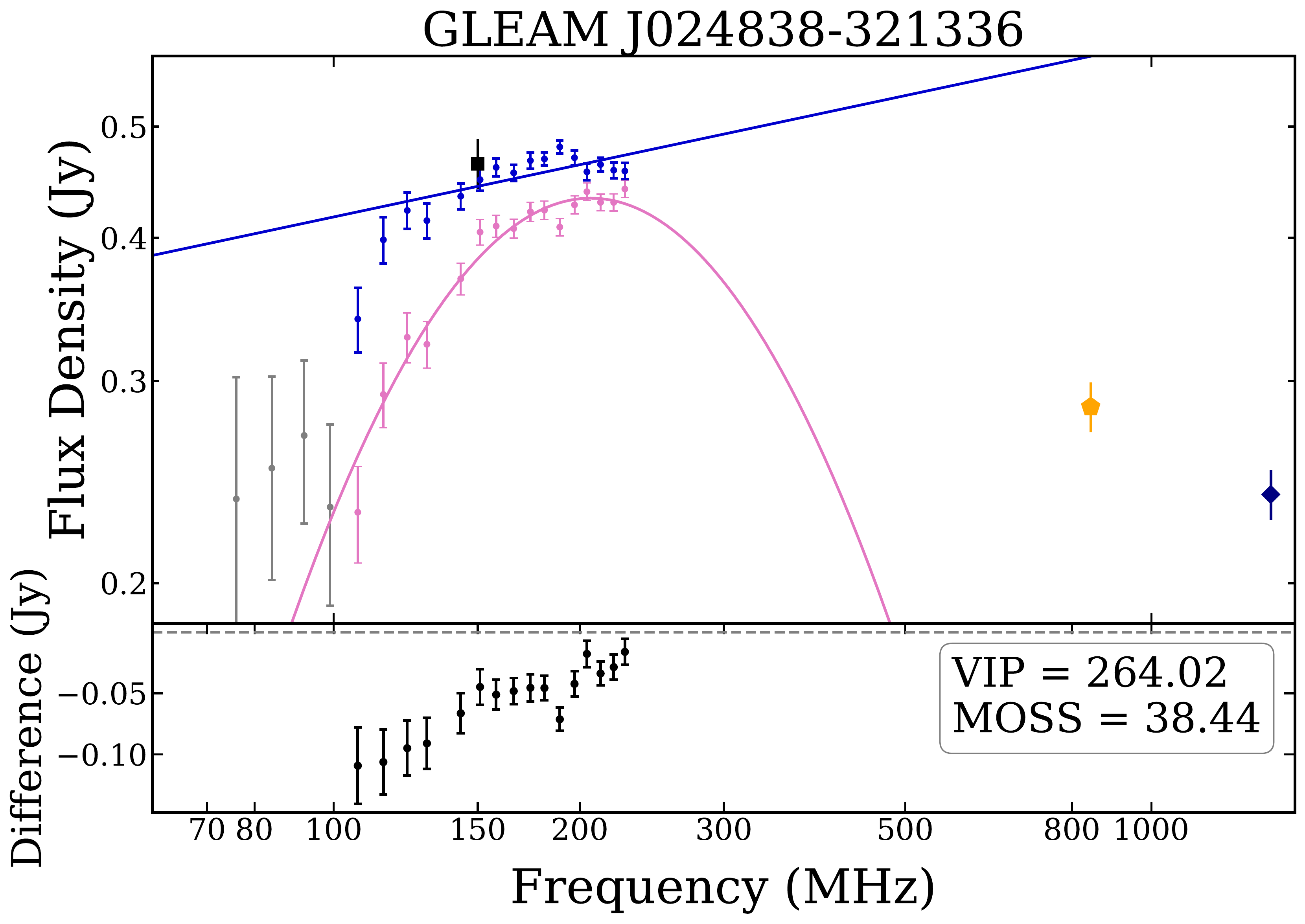} &
\includegraphics[scale=0.15]{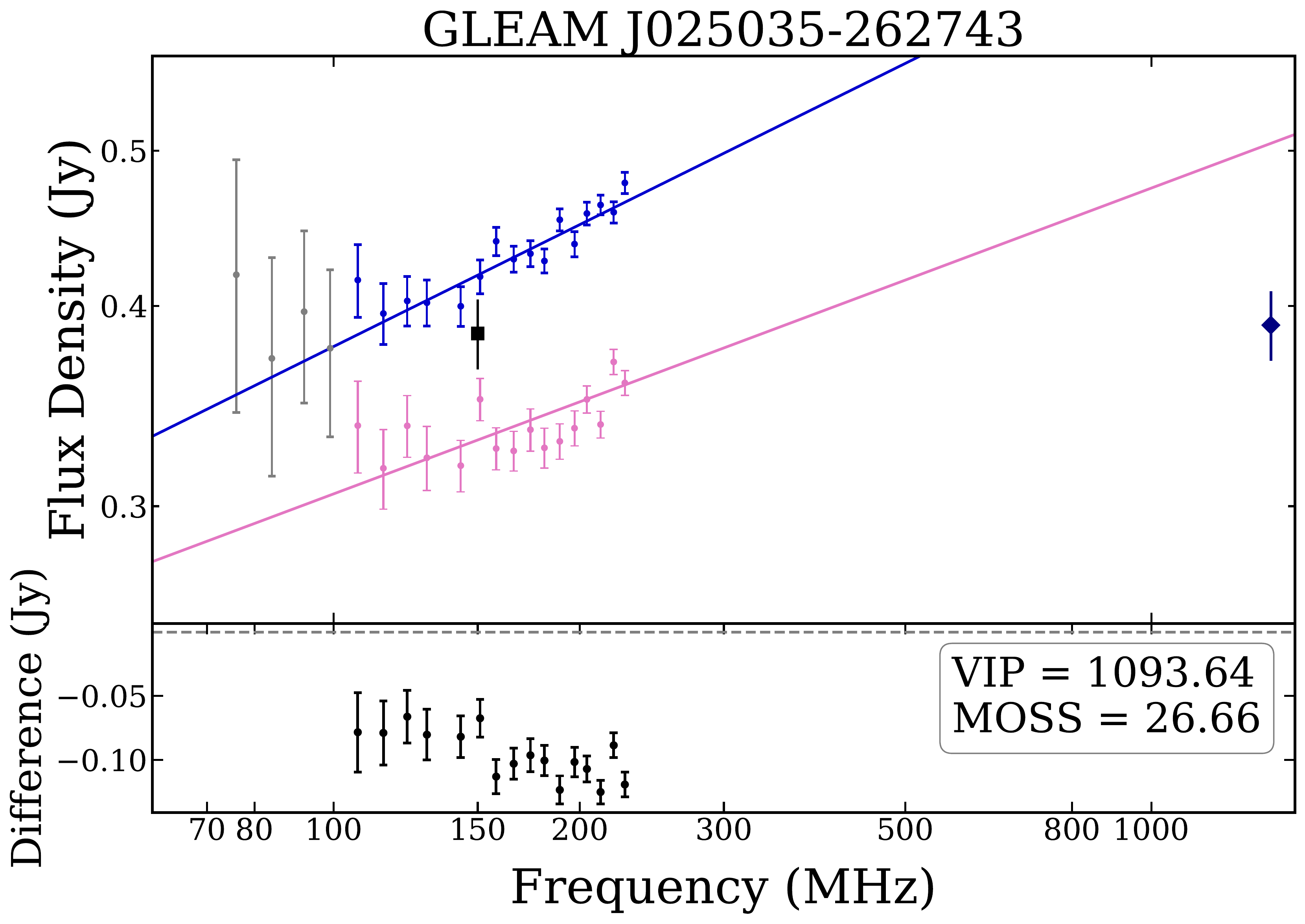} &
\includegraphics[scale=0.15]{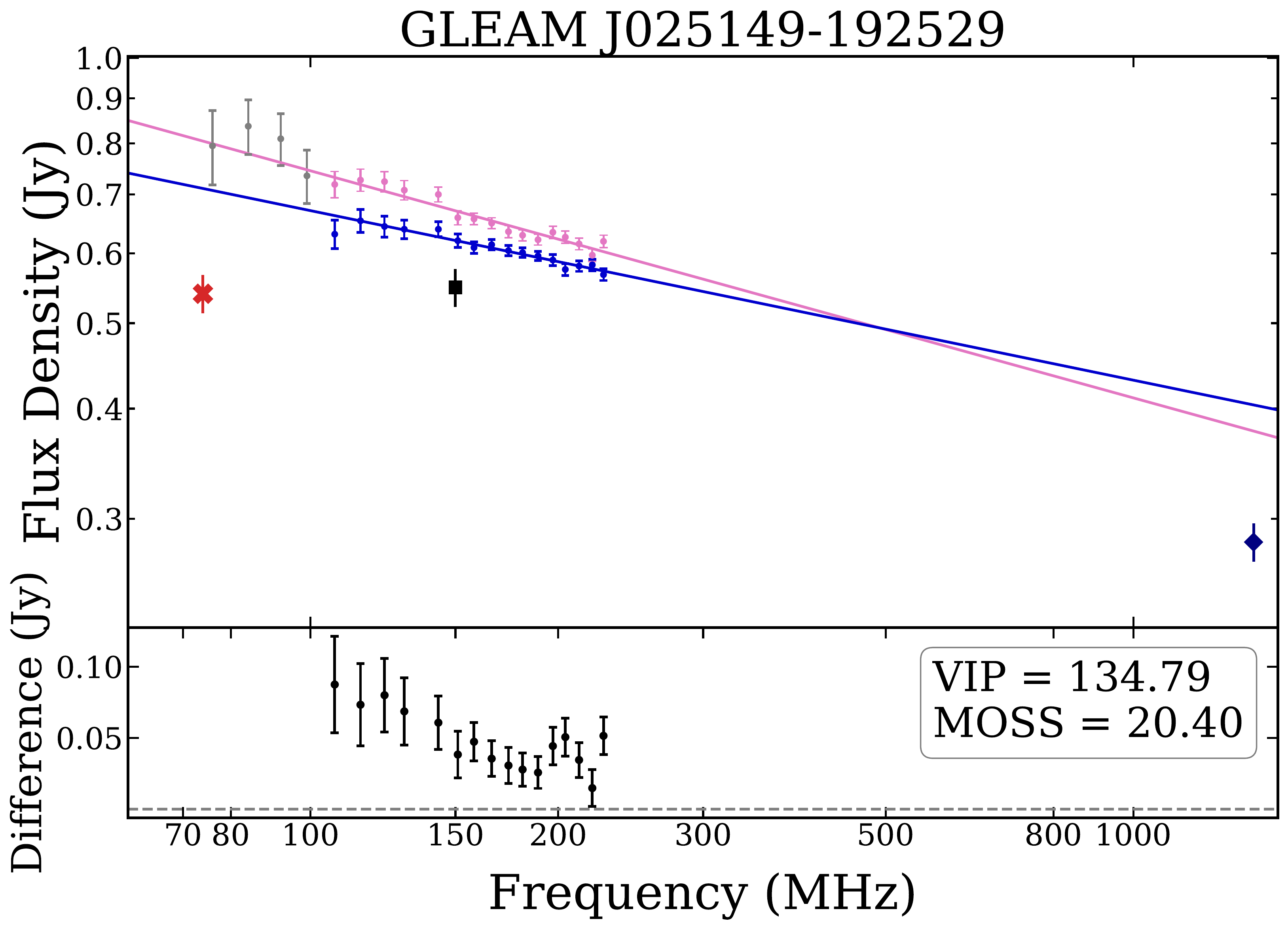} \\
\includegraphics[scale=0.15]{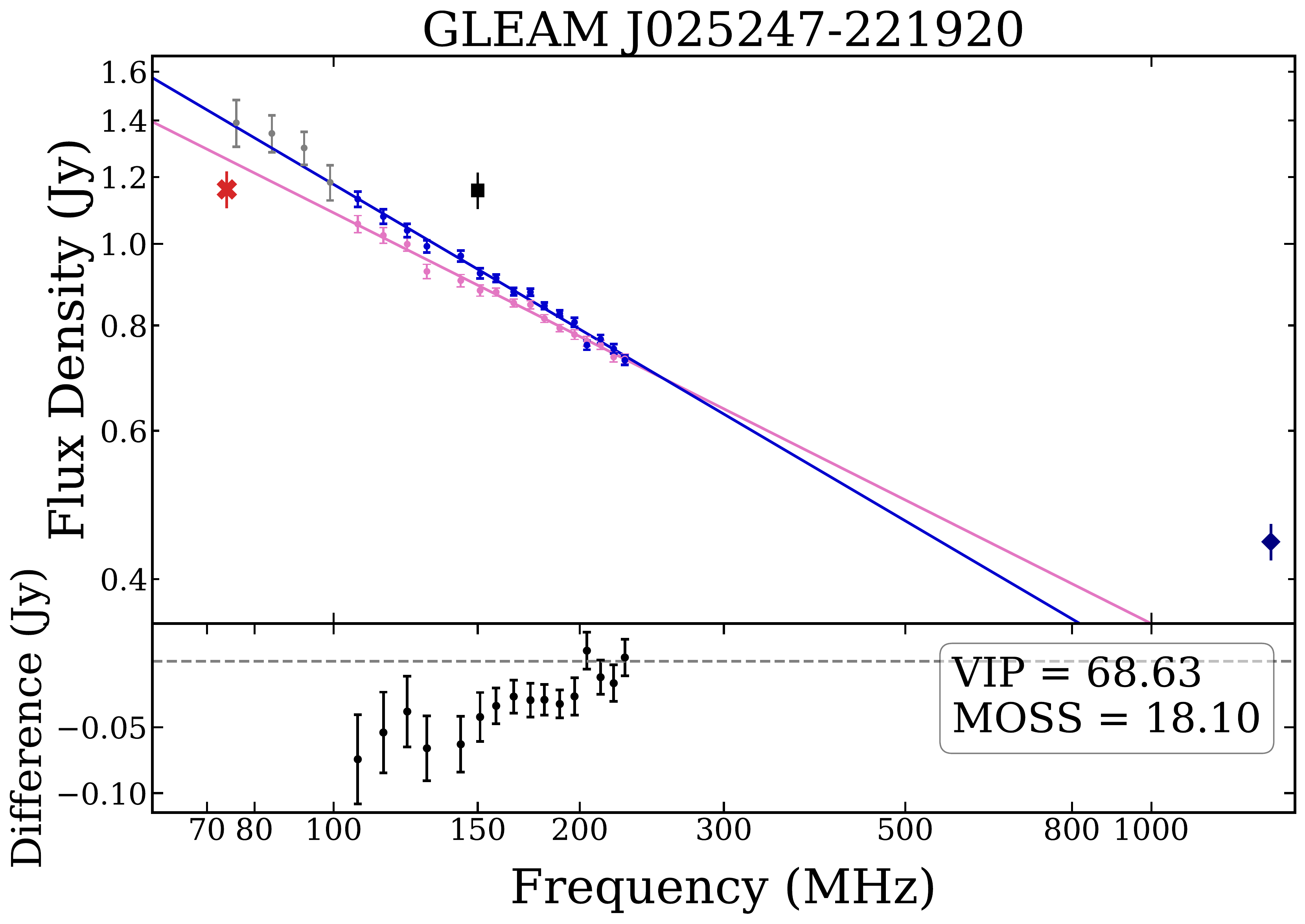} &
\includegraphics[scale=0.15]{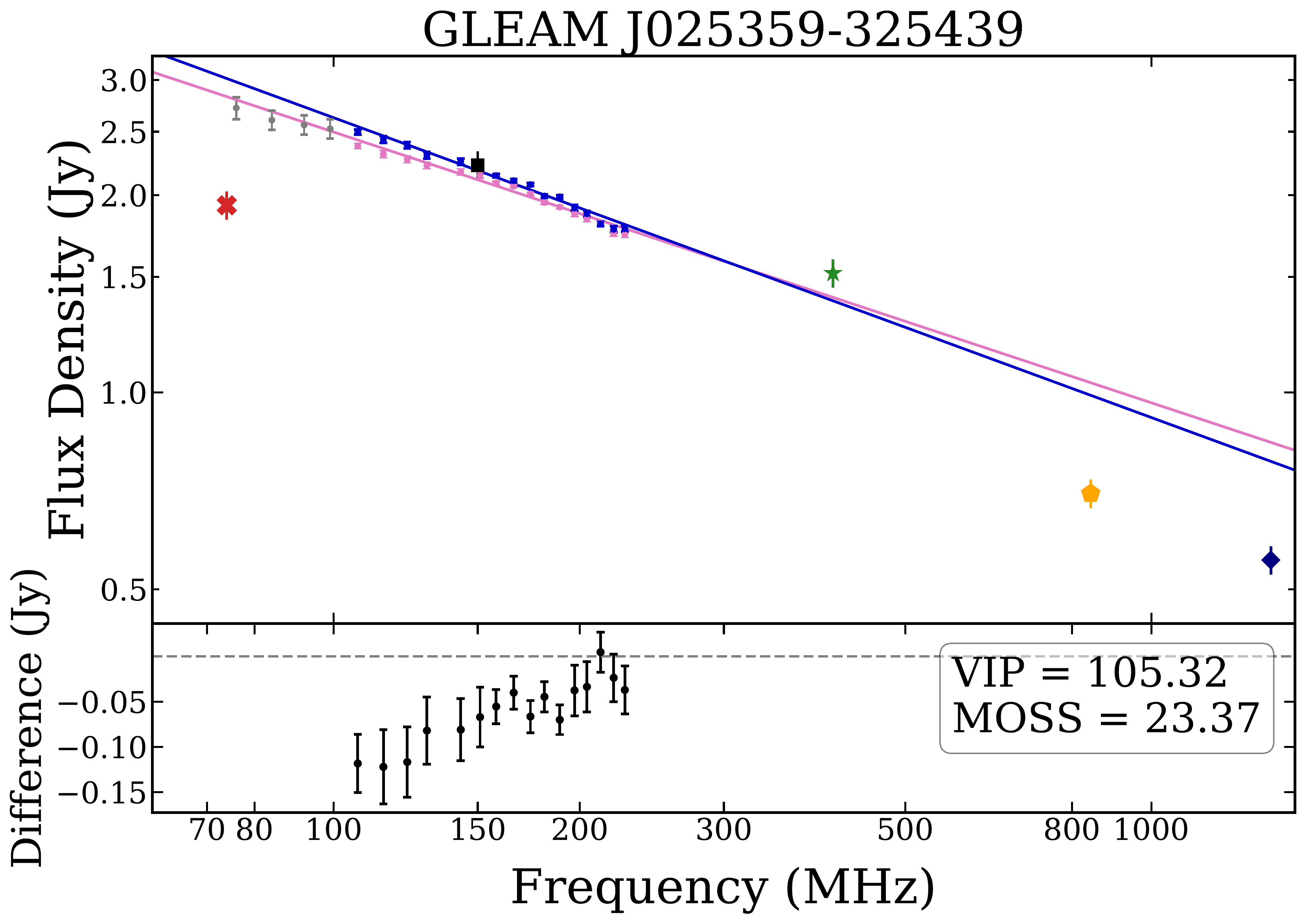} &
\includegraphics[scale=0.15]{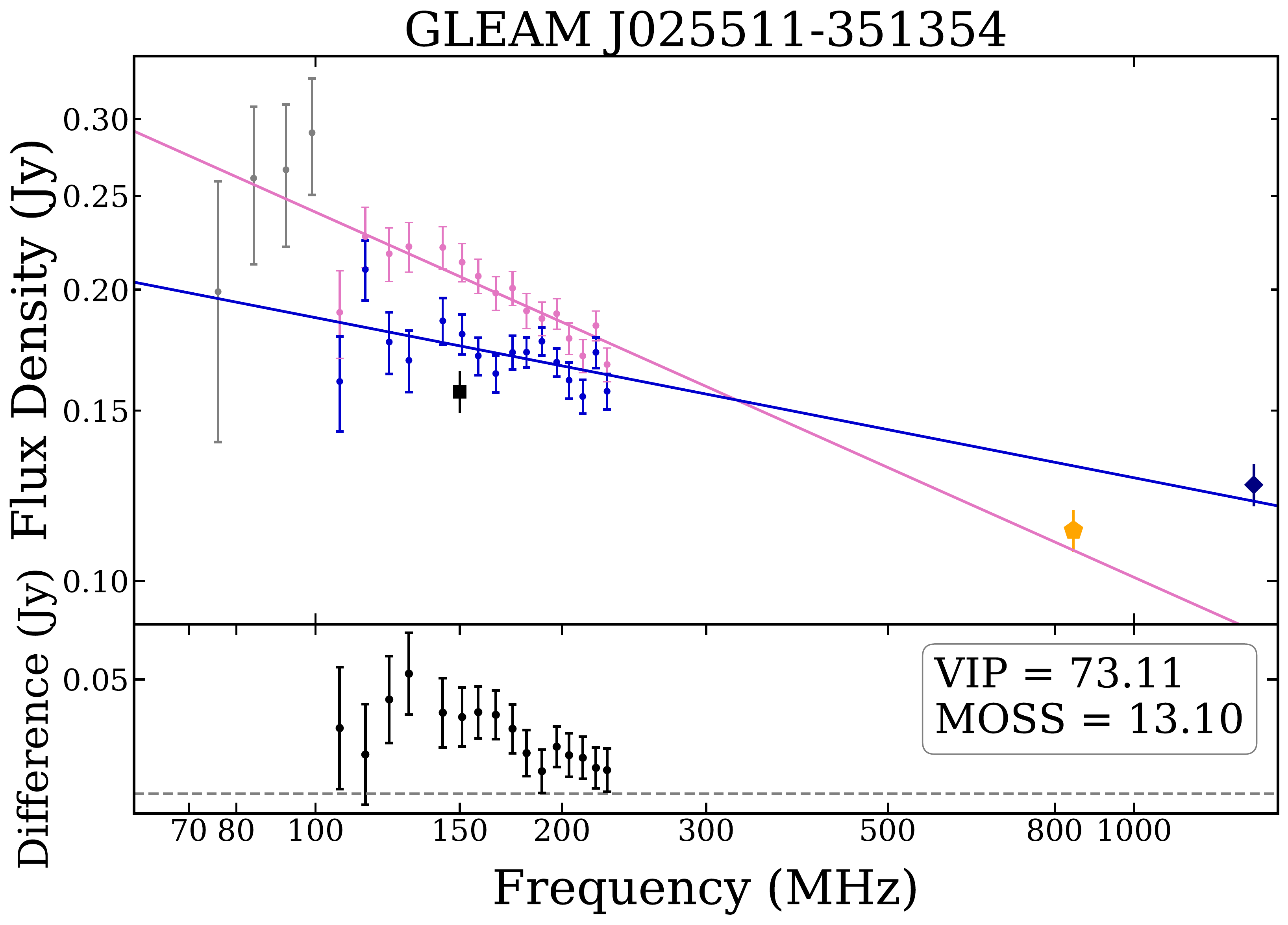} \\
\includegraphics[scale=0.15]{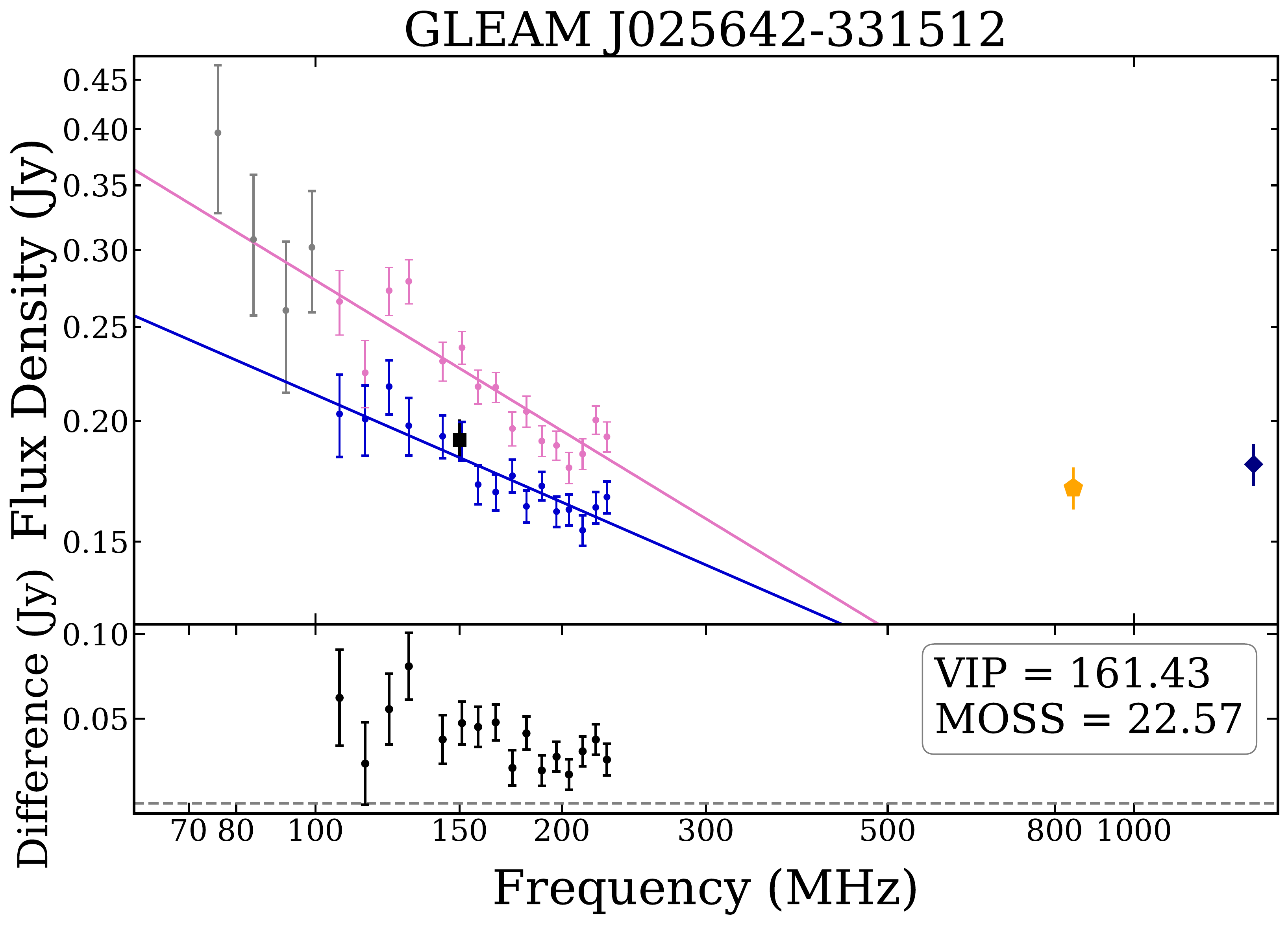} &
\includegraphics[scale=0.15]{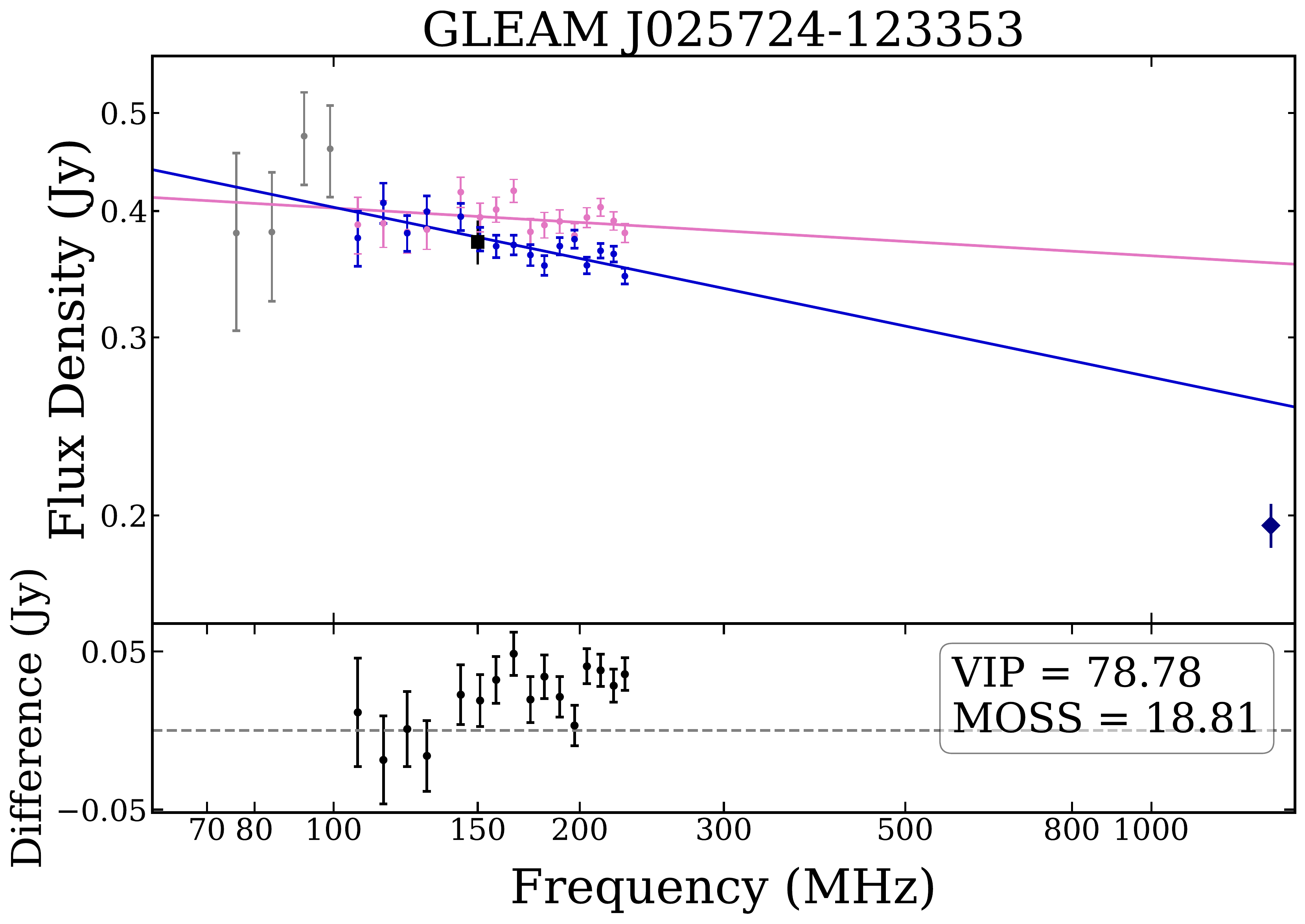} &
\includegraphics[scale=0.15]{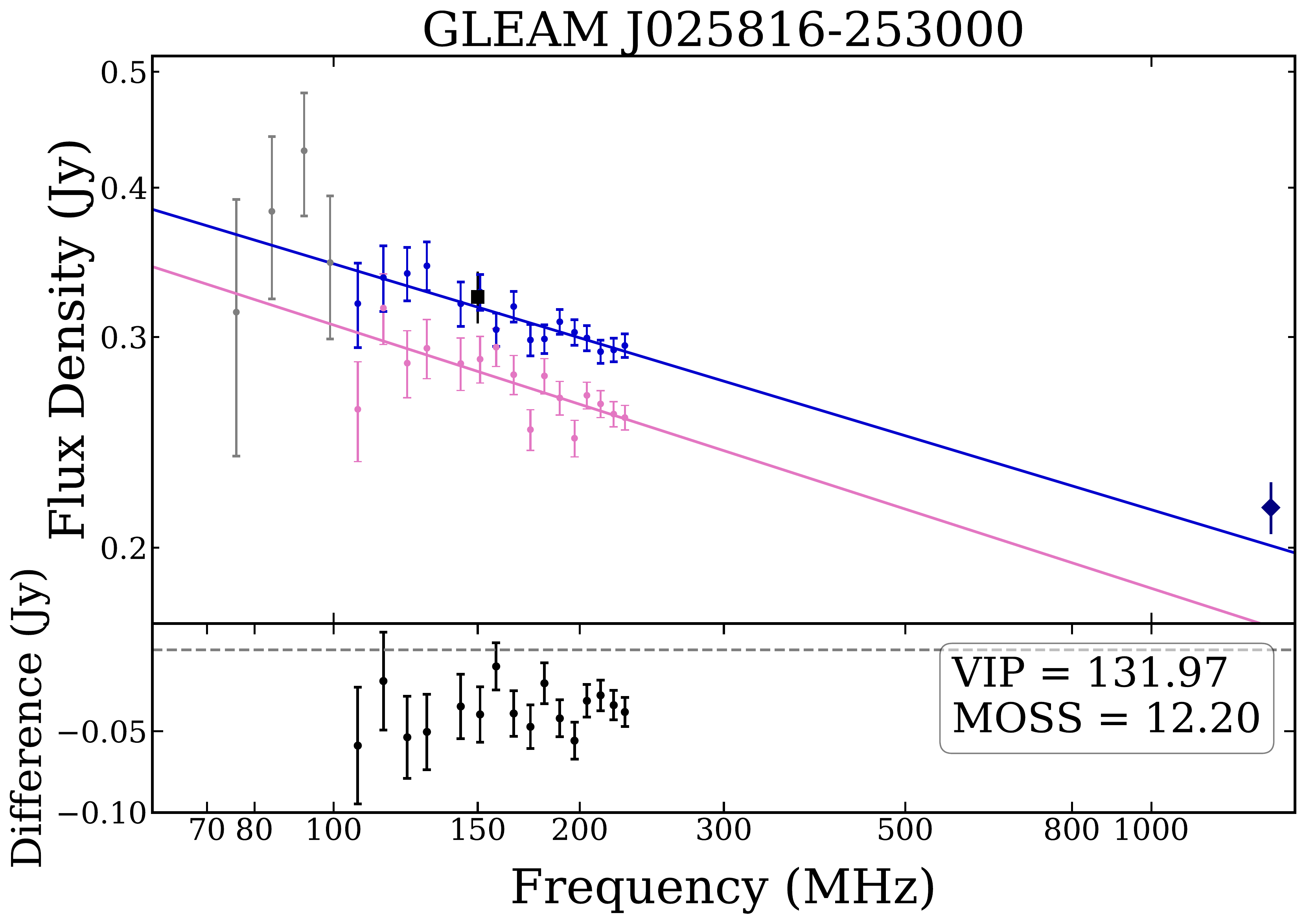} \\
\end{array}$
\caption{(continued) SEDs for all sources classified as variable according to the VIP. For each source the points represent the following data: GLEAM low frequency (72--100\,MHz) (grey circles), Year 1 (pink circles), Year 2 (blue circles), VLSSr (red cross), TGSS (black square), MRC (green star), SUMSS (yellow pentagon), and NVSS (navy diamond). The models for each year are determined by their classification; a source classified with a peak within the observed band was modelled by a quadratic according to Equation~\ref{eq:quadratic}, remaining sources were modelled by a power-law according to Equation~\ref{eq:plaw}.}
\label{app:fig:pg7}
\end{center}
\end{figure*}
\setcounter{figure}{0}
\begin{figure*}
\begin{center}$
\begin{array}{cccccc}
\includegraphics[scale=0.15]{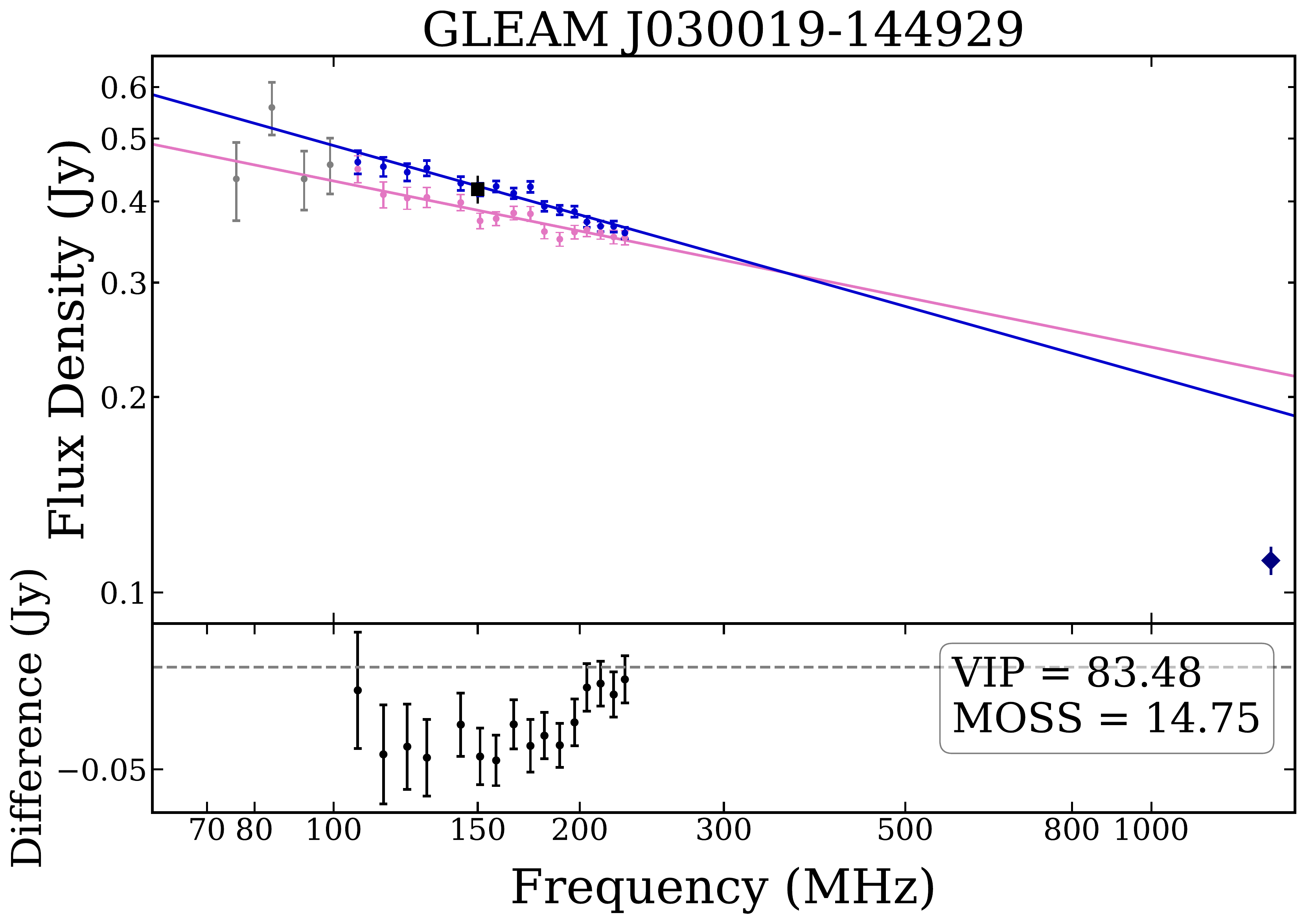} &
\includegraphics[scale=0.15]{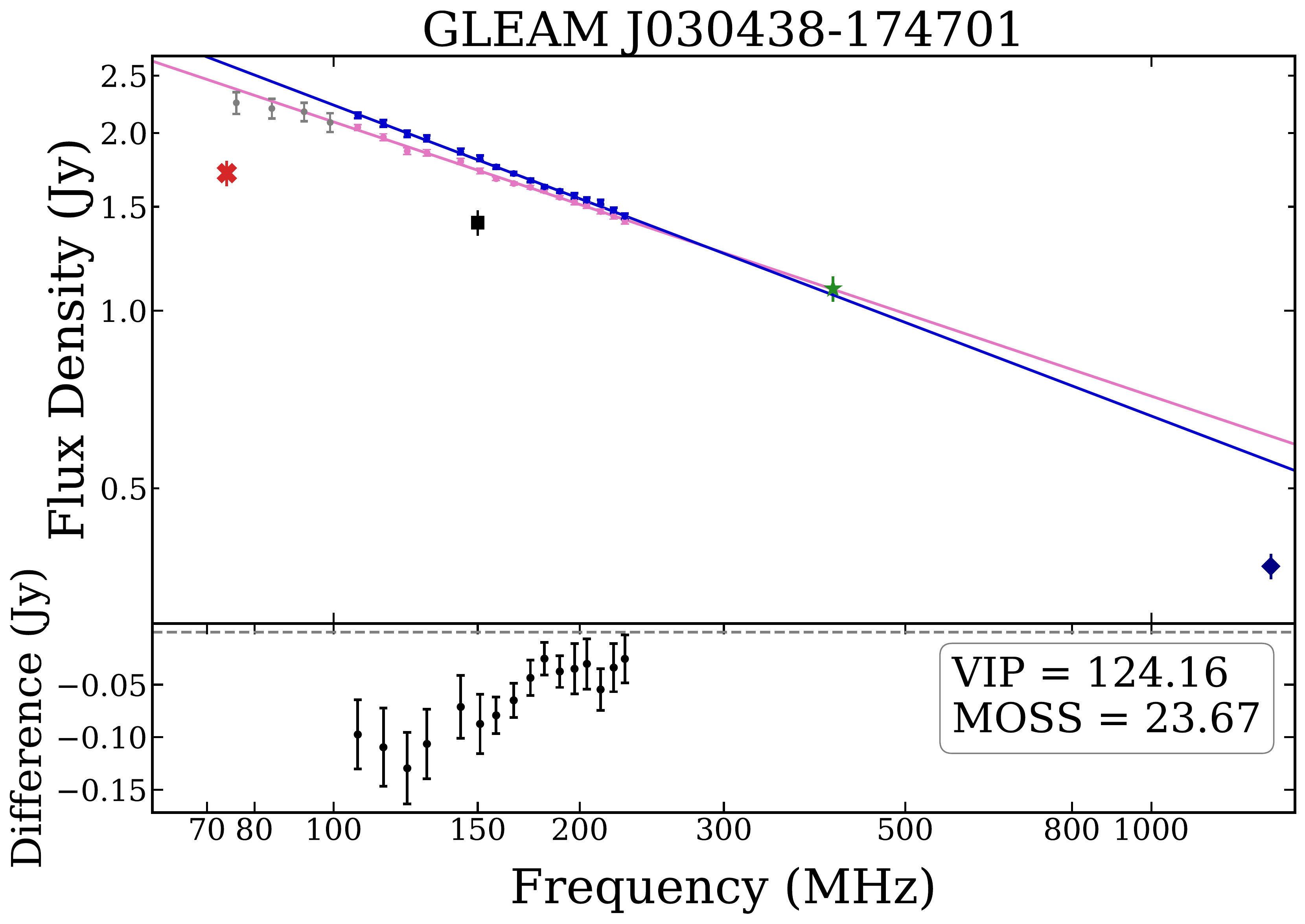} &
\includegraphics[scale=0.15]{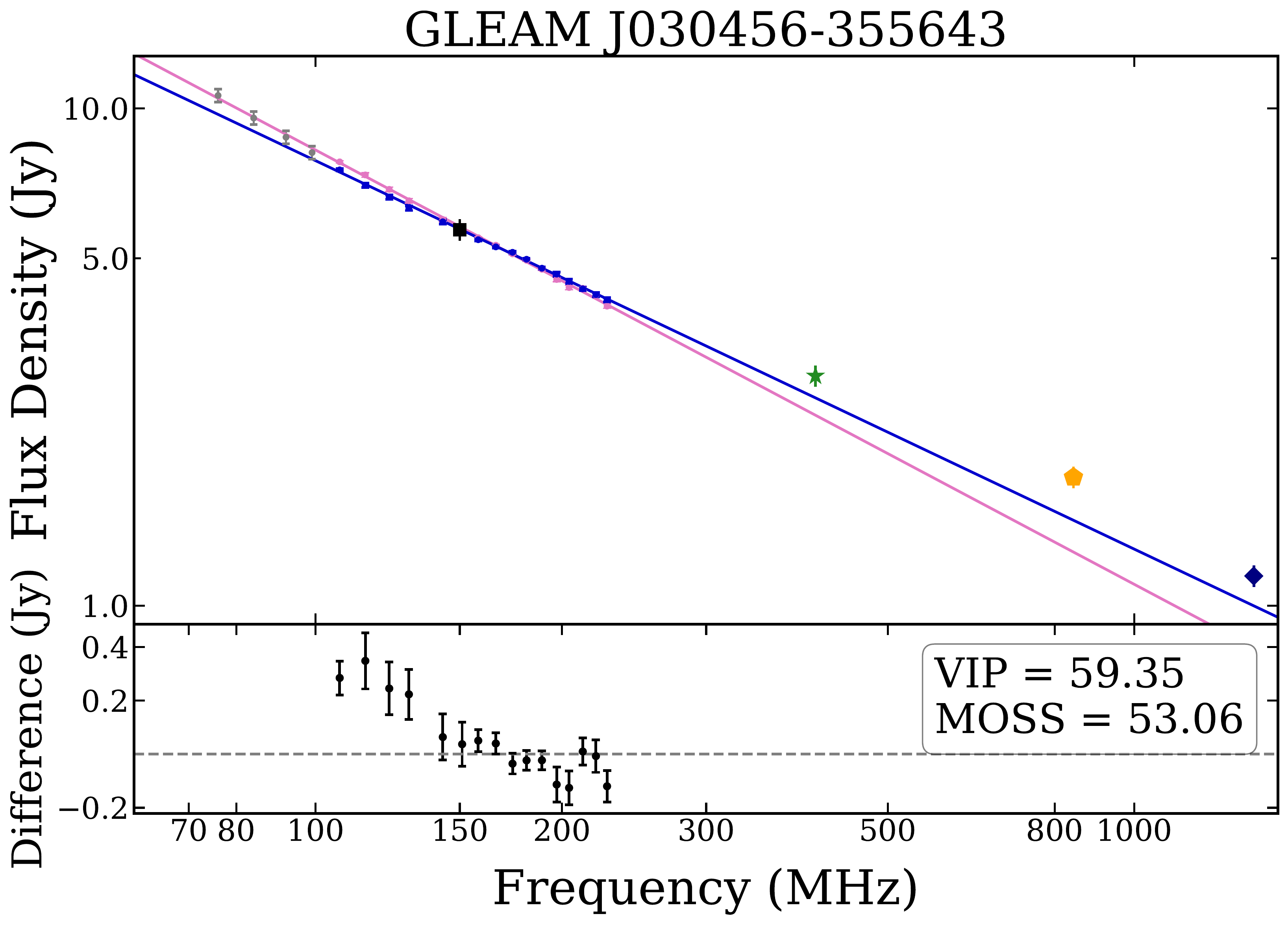} \\
\includegraphics[scale=0.15]{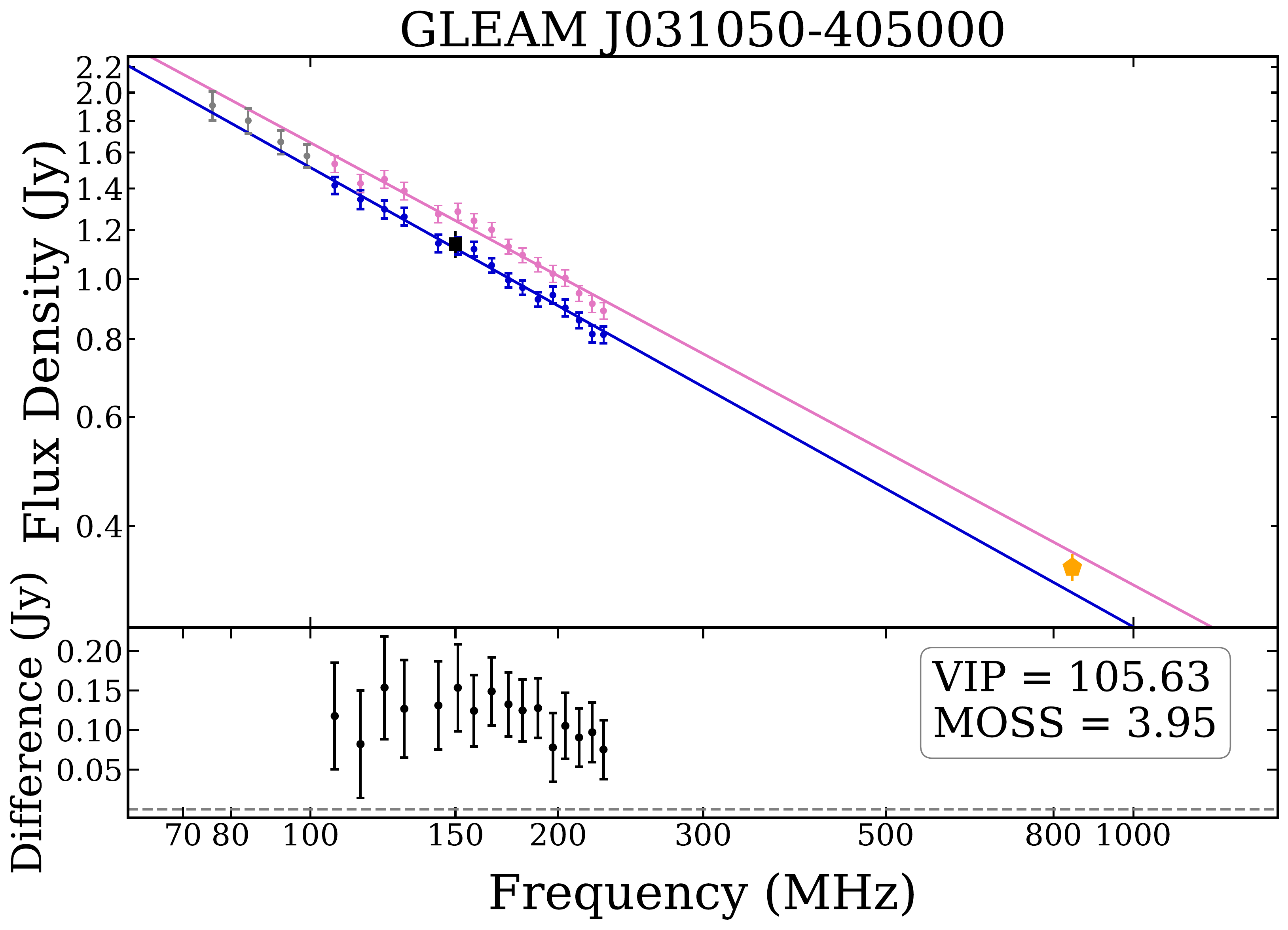} &
\includegraphics[scale=0.15]{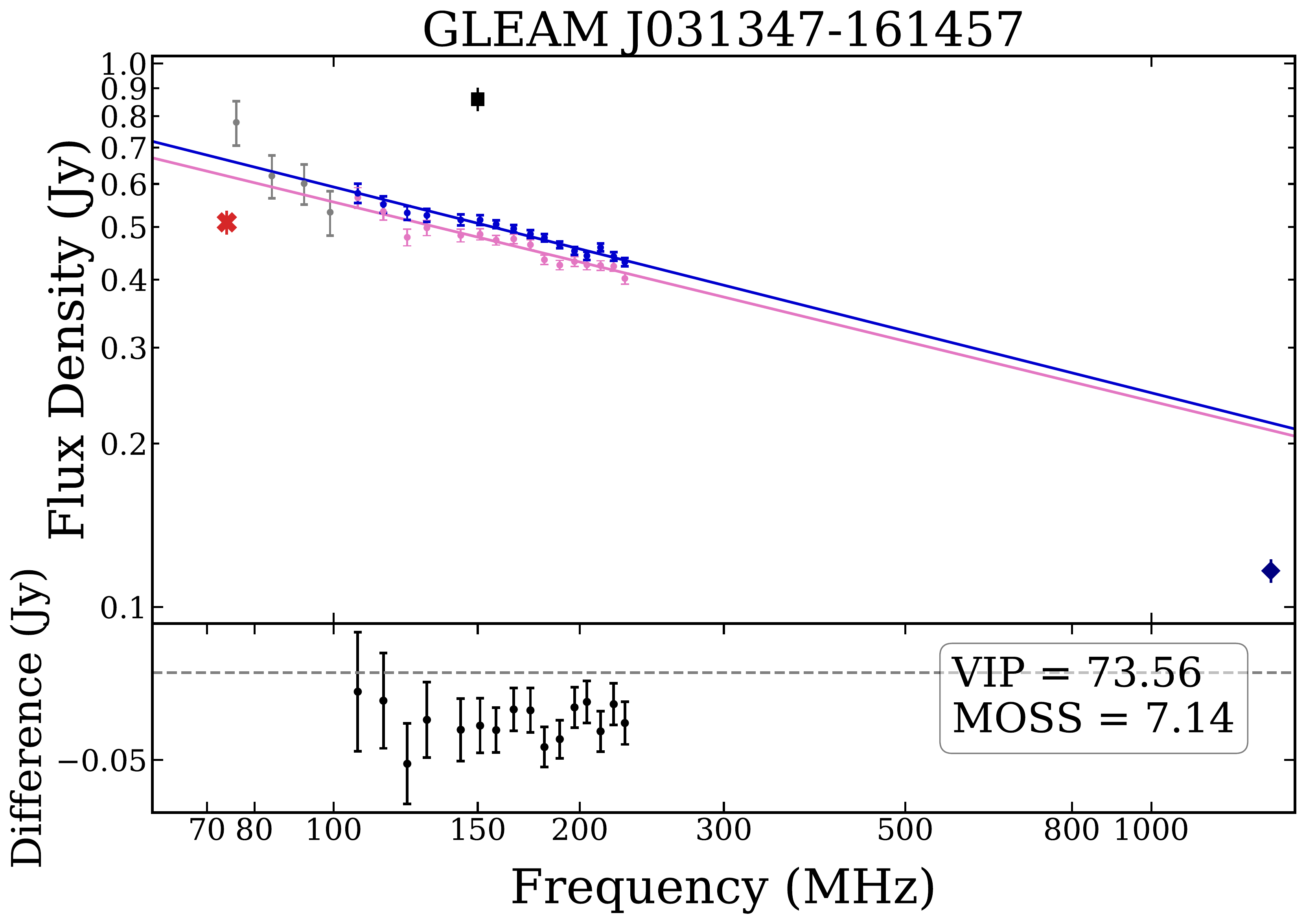} &
\includegraphics[scale=0.15]{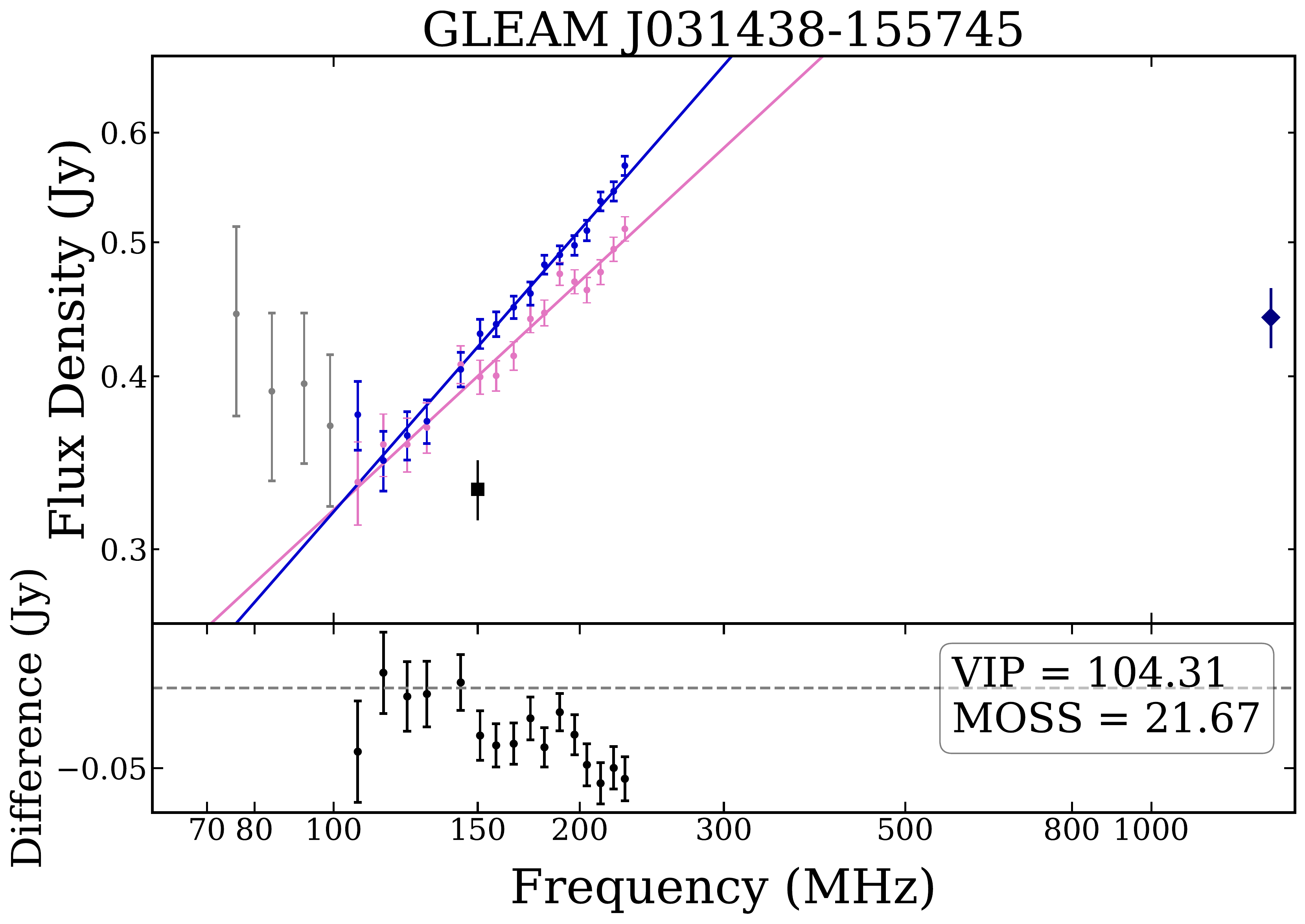} \\
\includegraphics[scale=0.15]{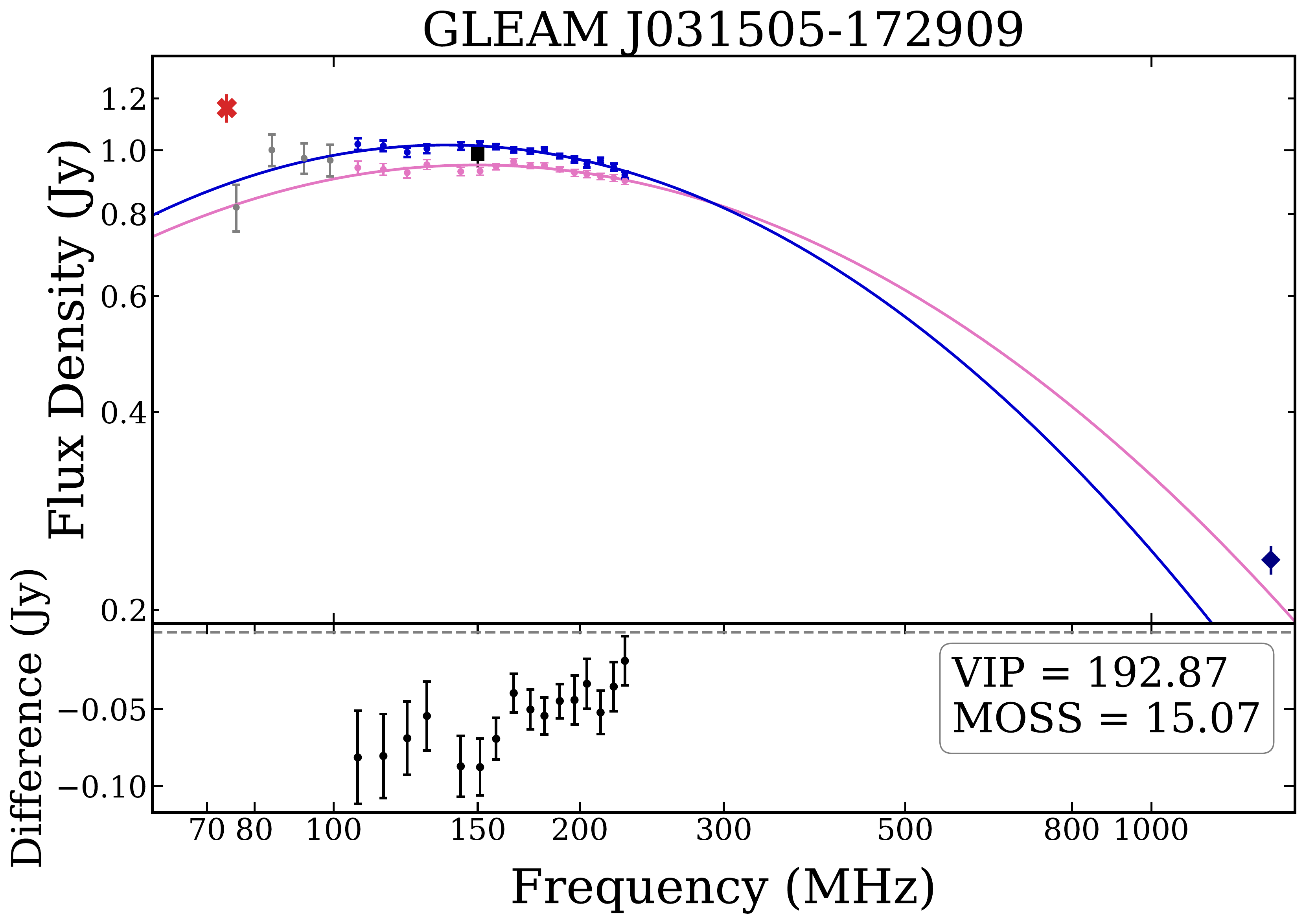} &
\includegraphics[scale=0.15]{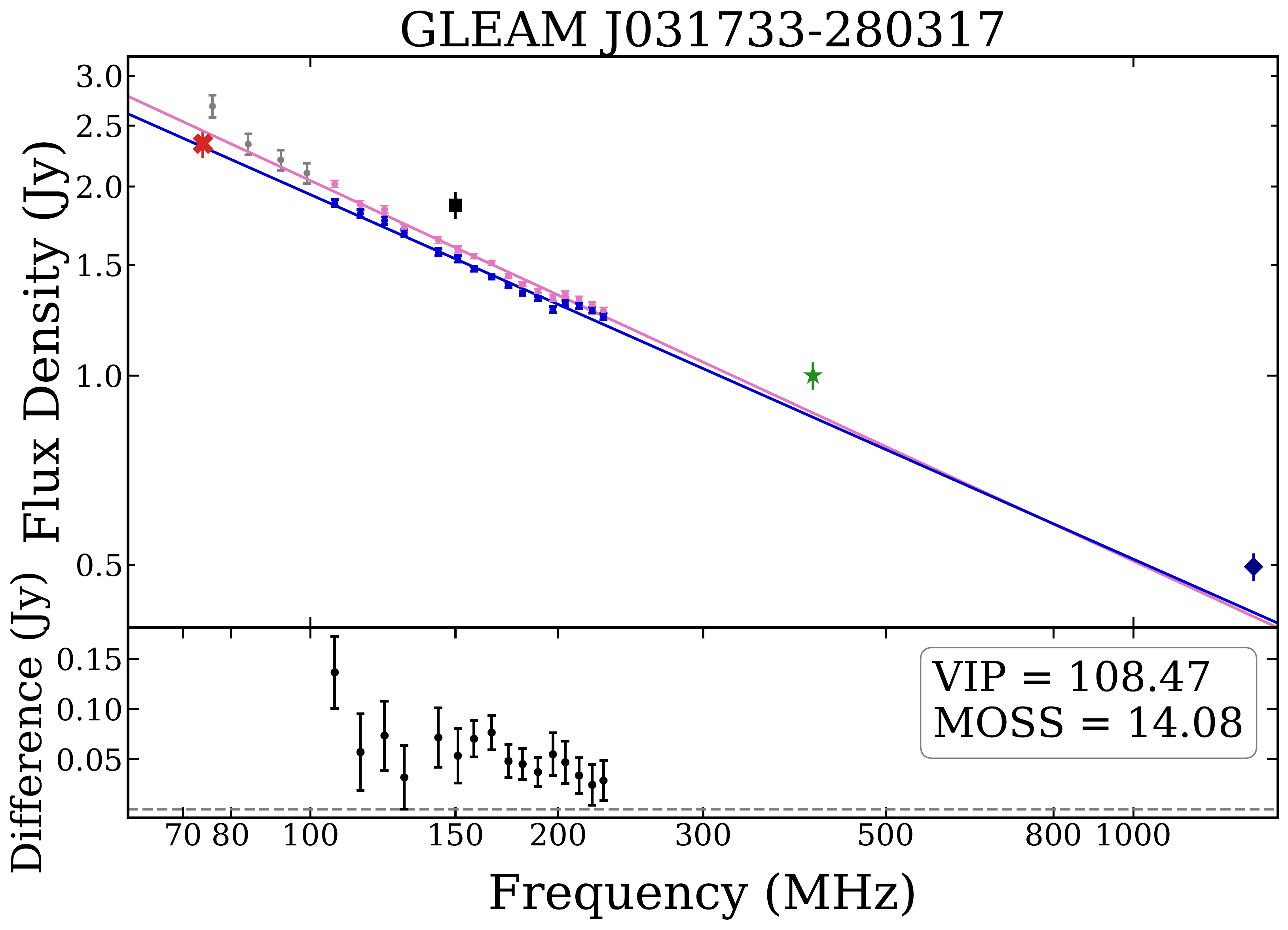} &
\includegraphics[scale=0.15]{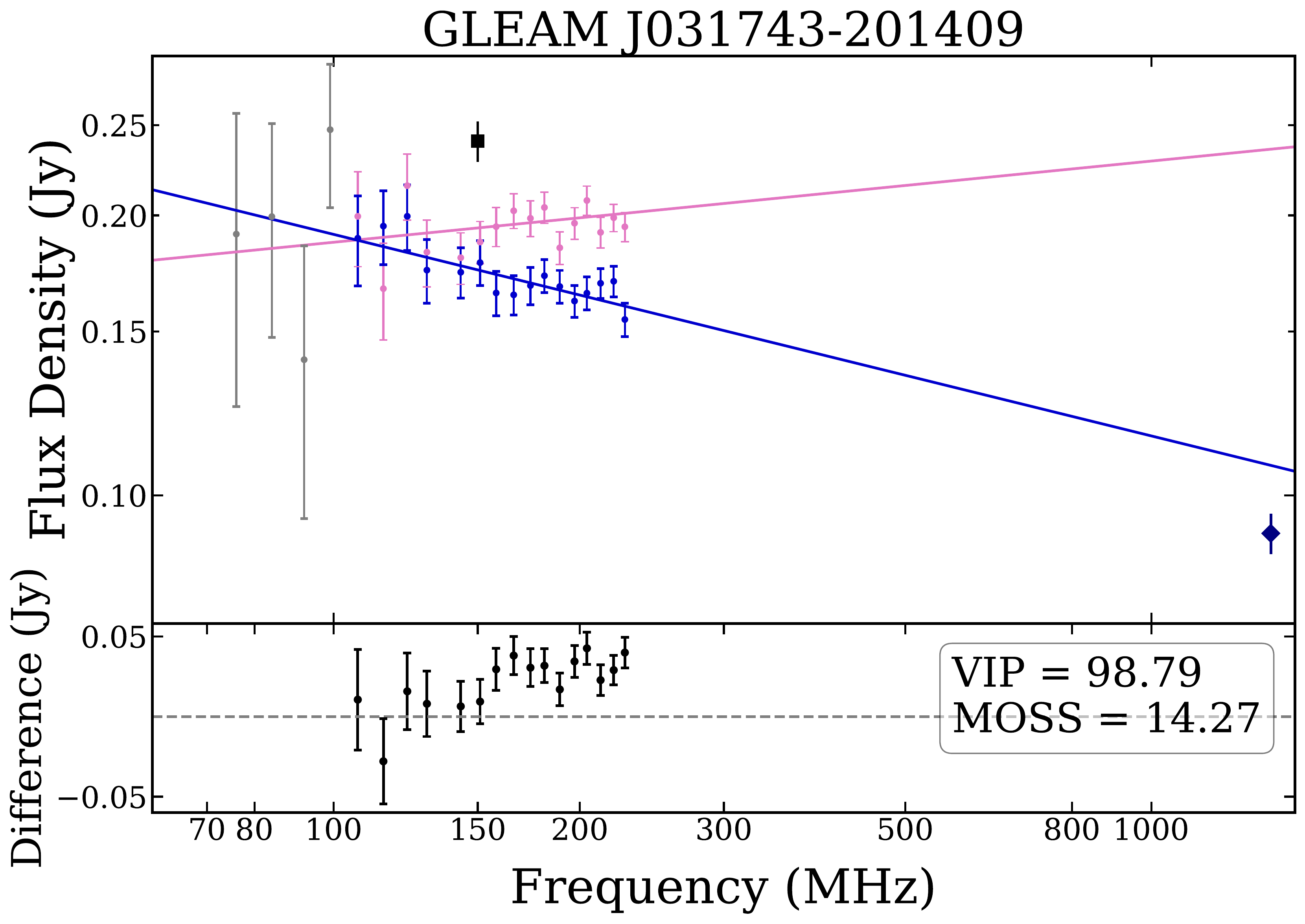} \\
\includegraphics[scale=0.15]{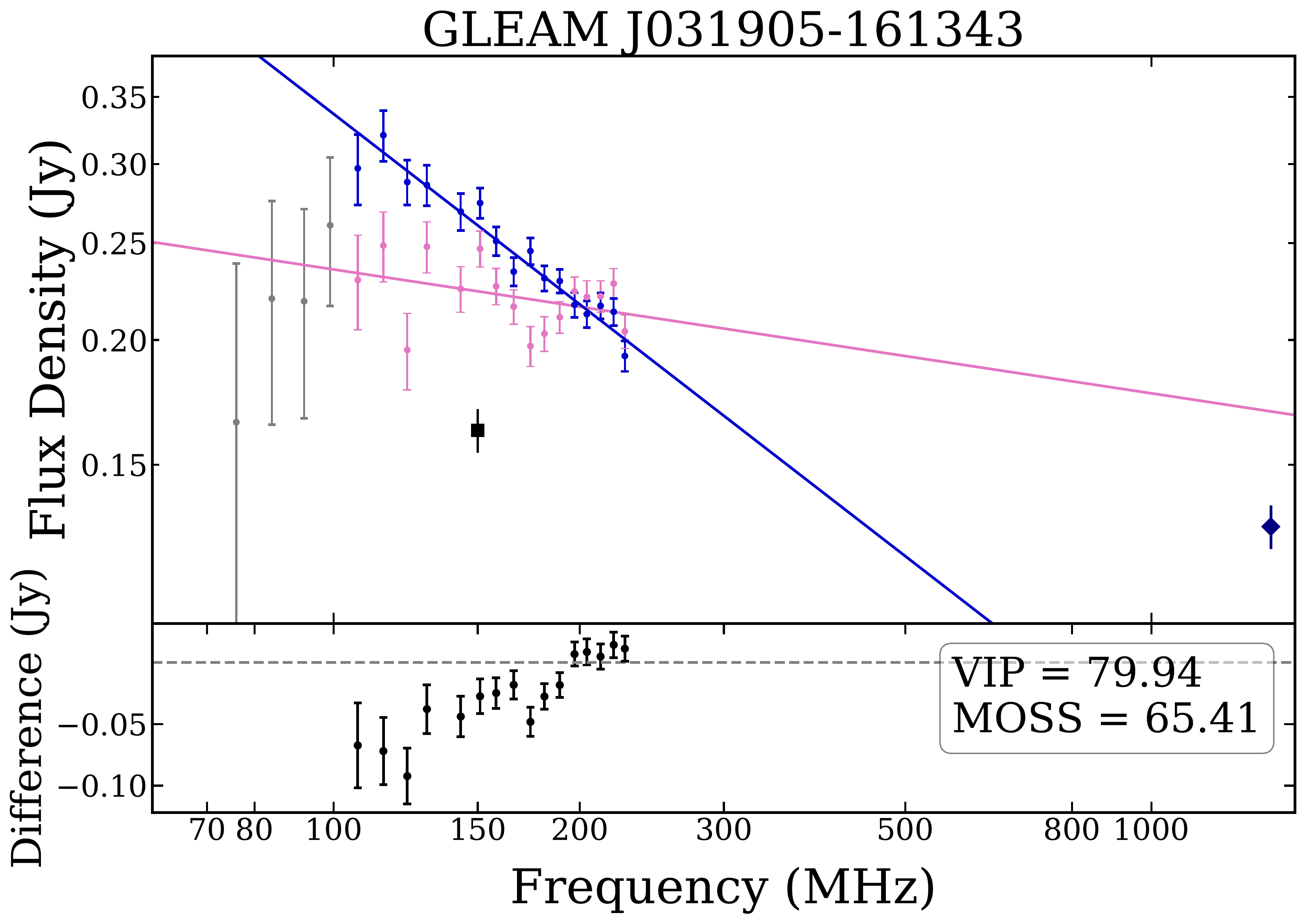} &
\includegraphics[scale=0.15]{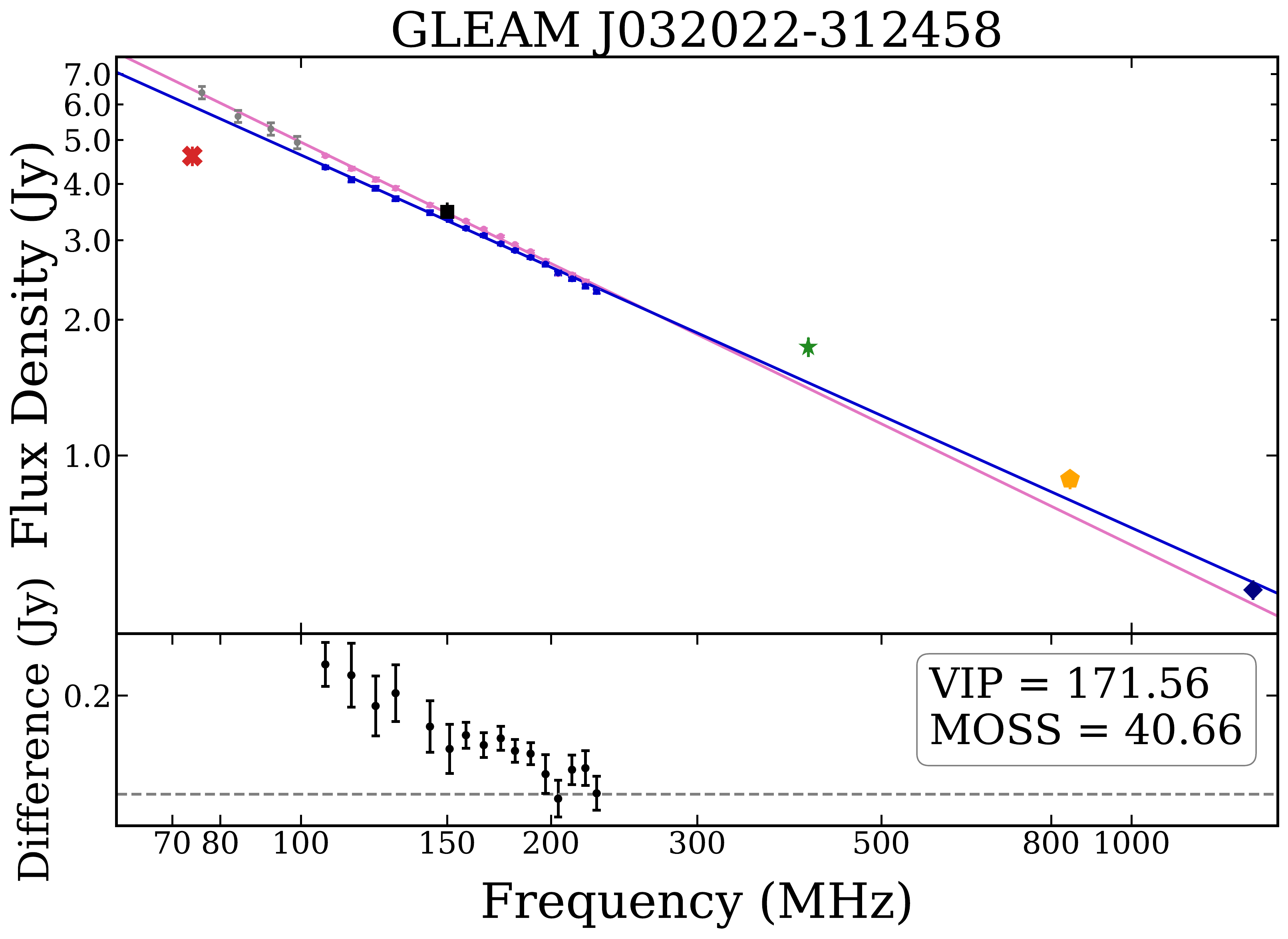} &
\includegraphics[scale=0.15]{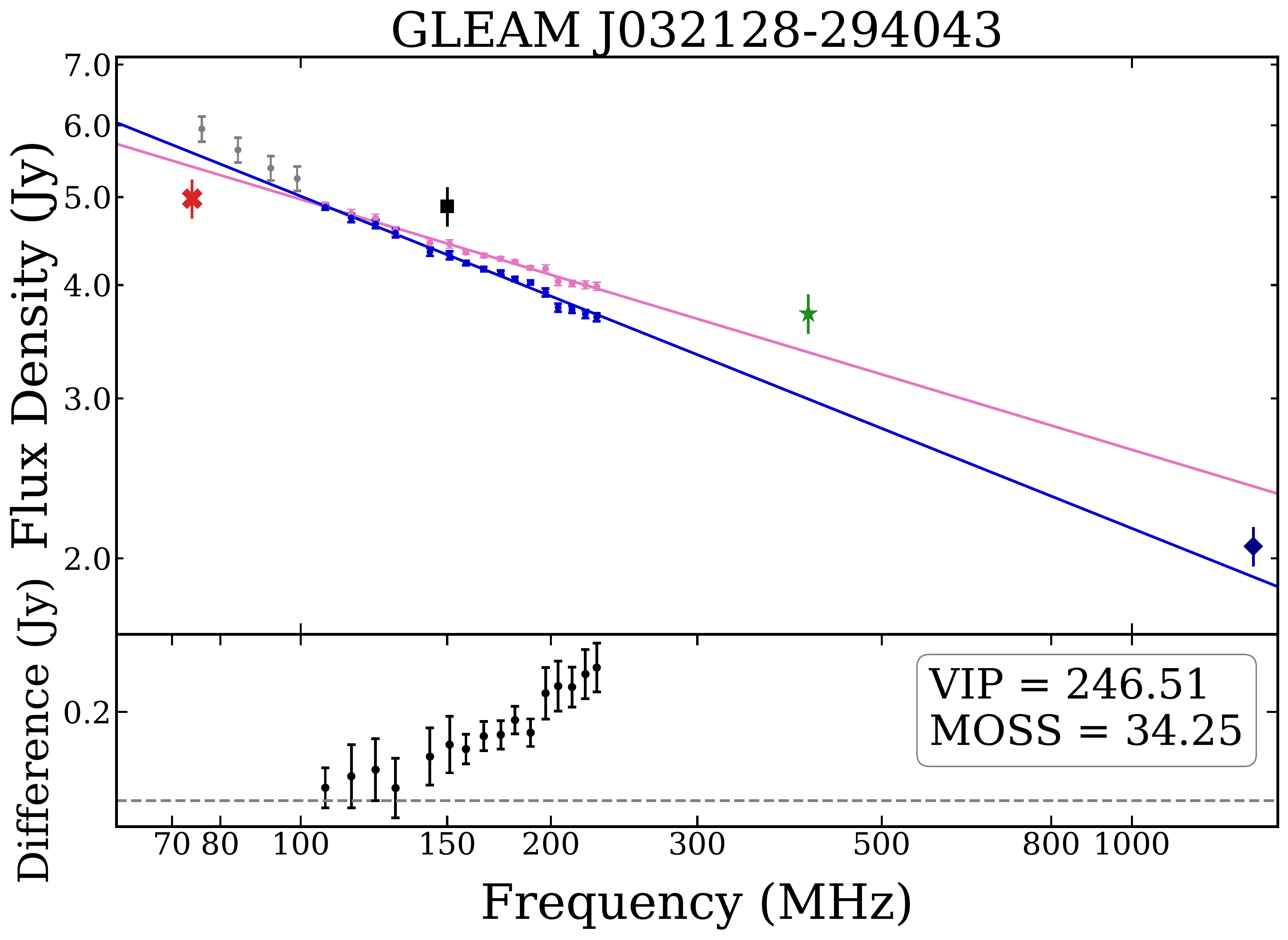} \\
\includegraphics[scale=0.15]{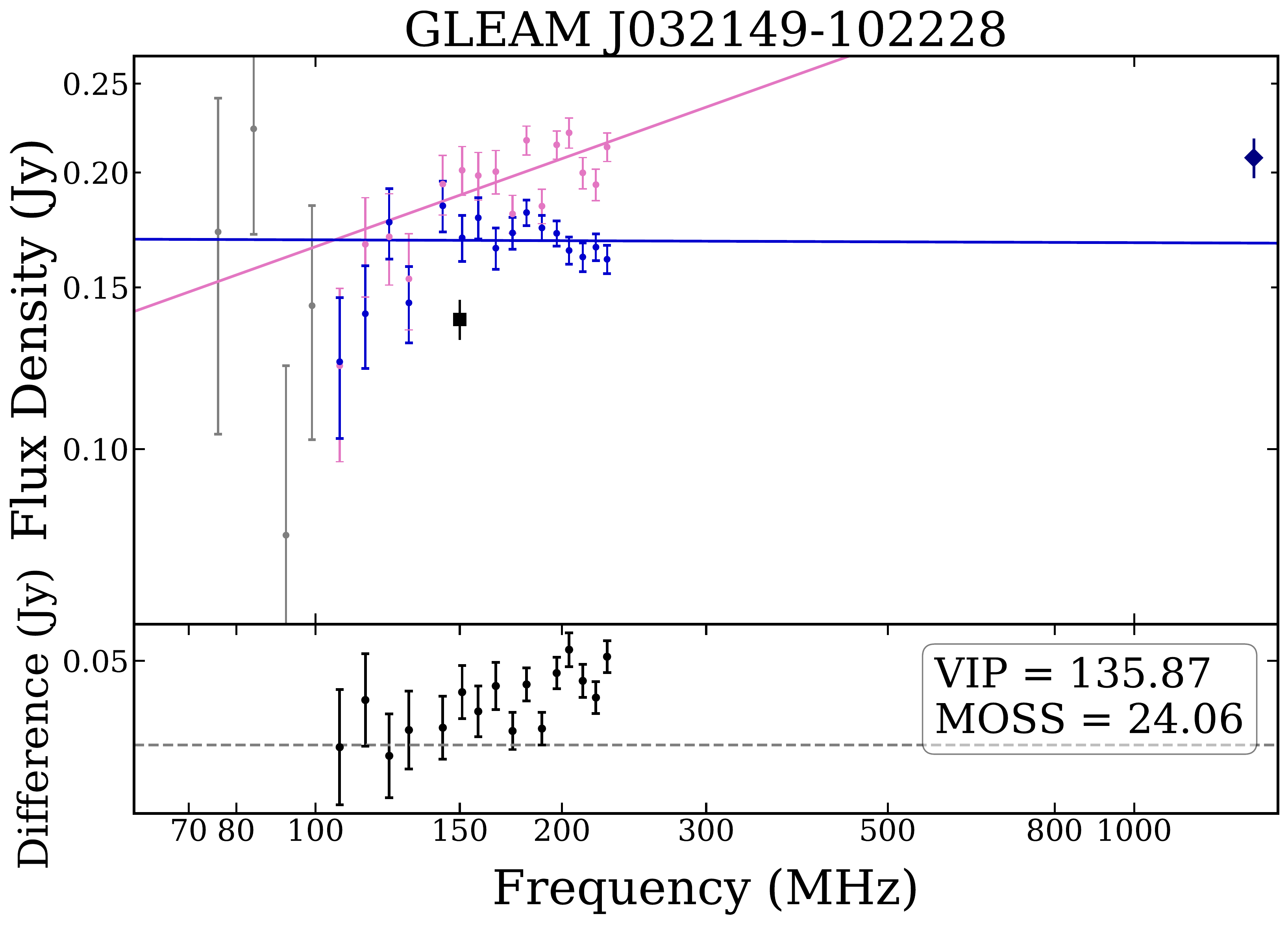} &
\includegraphics[scale=0.15]{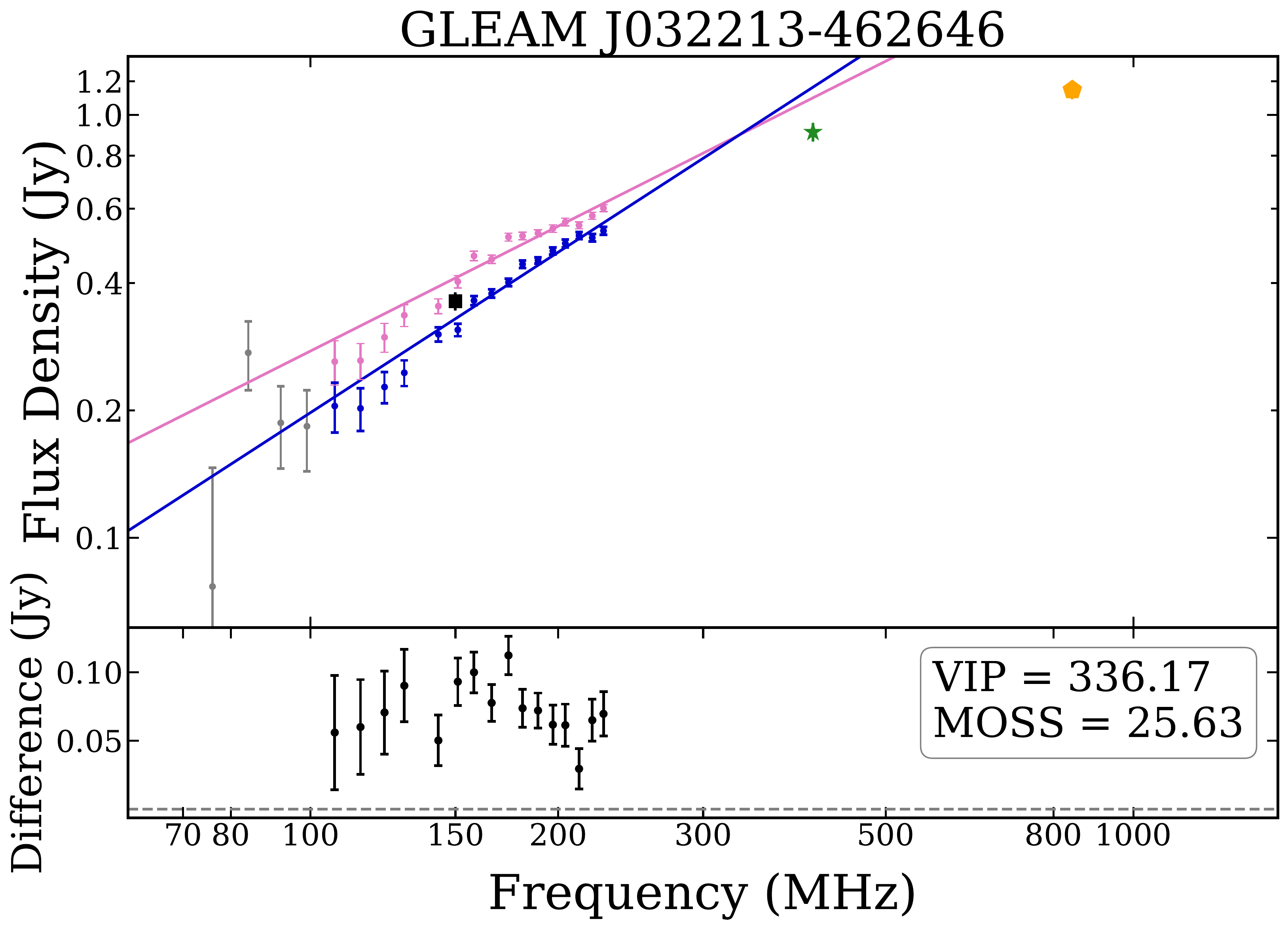} &
\includegraphics[scale=0.15]{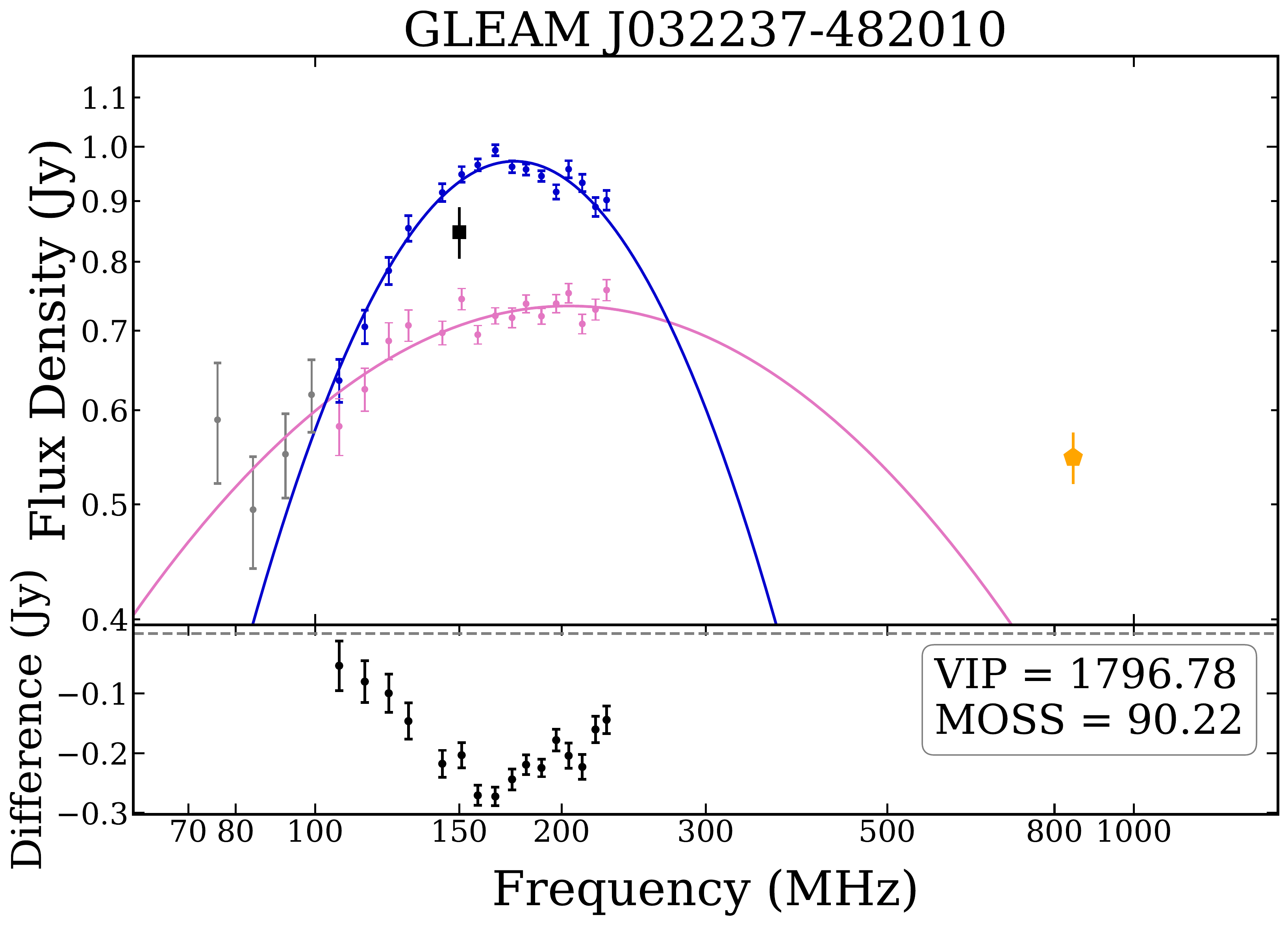} \\
\includegraphics[scale=0.15]{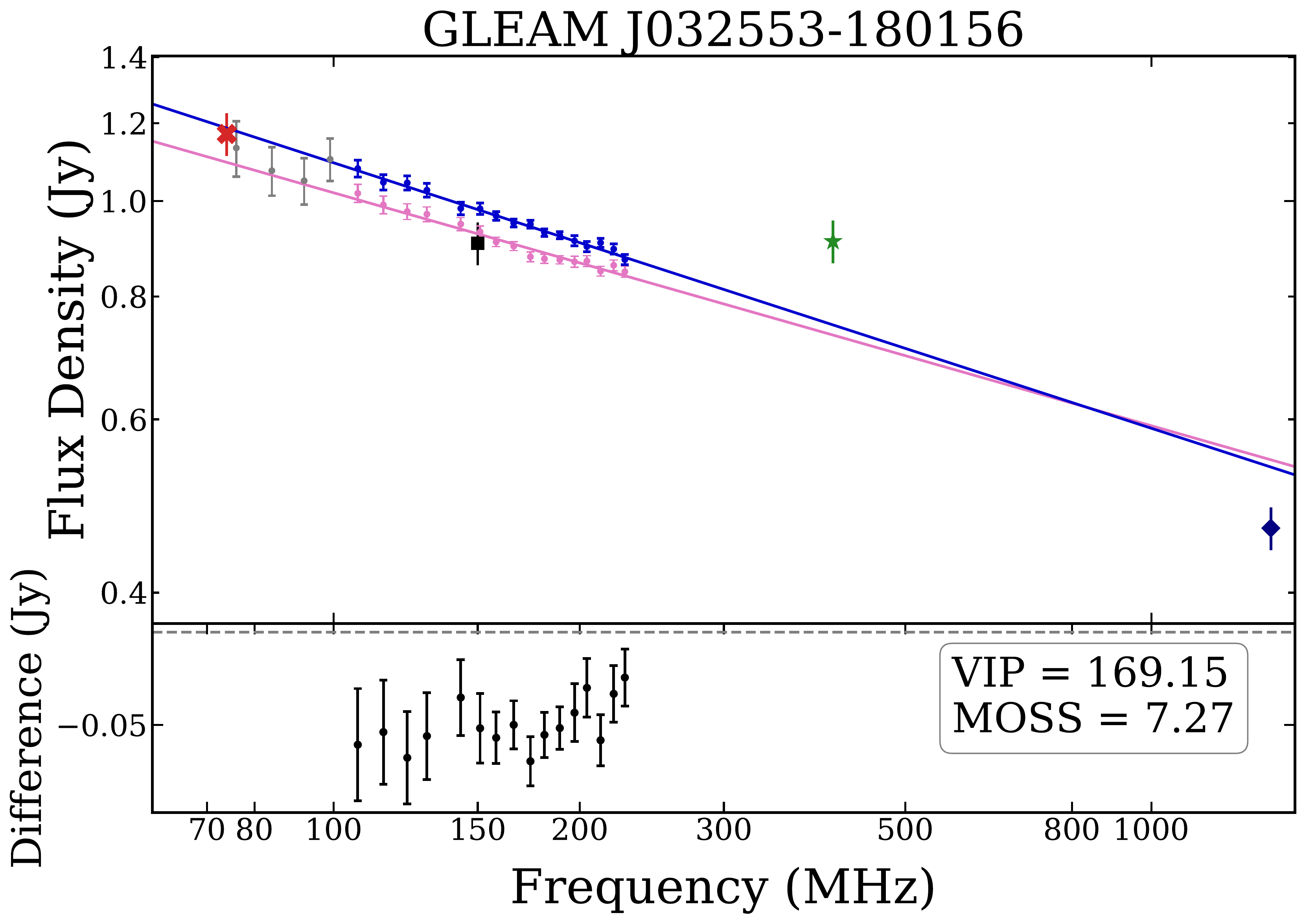} &
\includegraphics[scale=0.15]{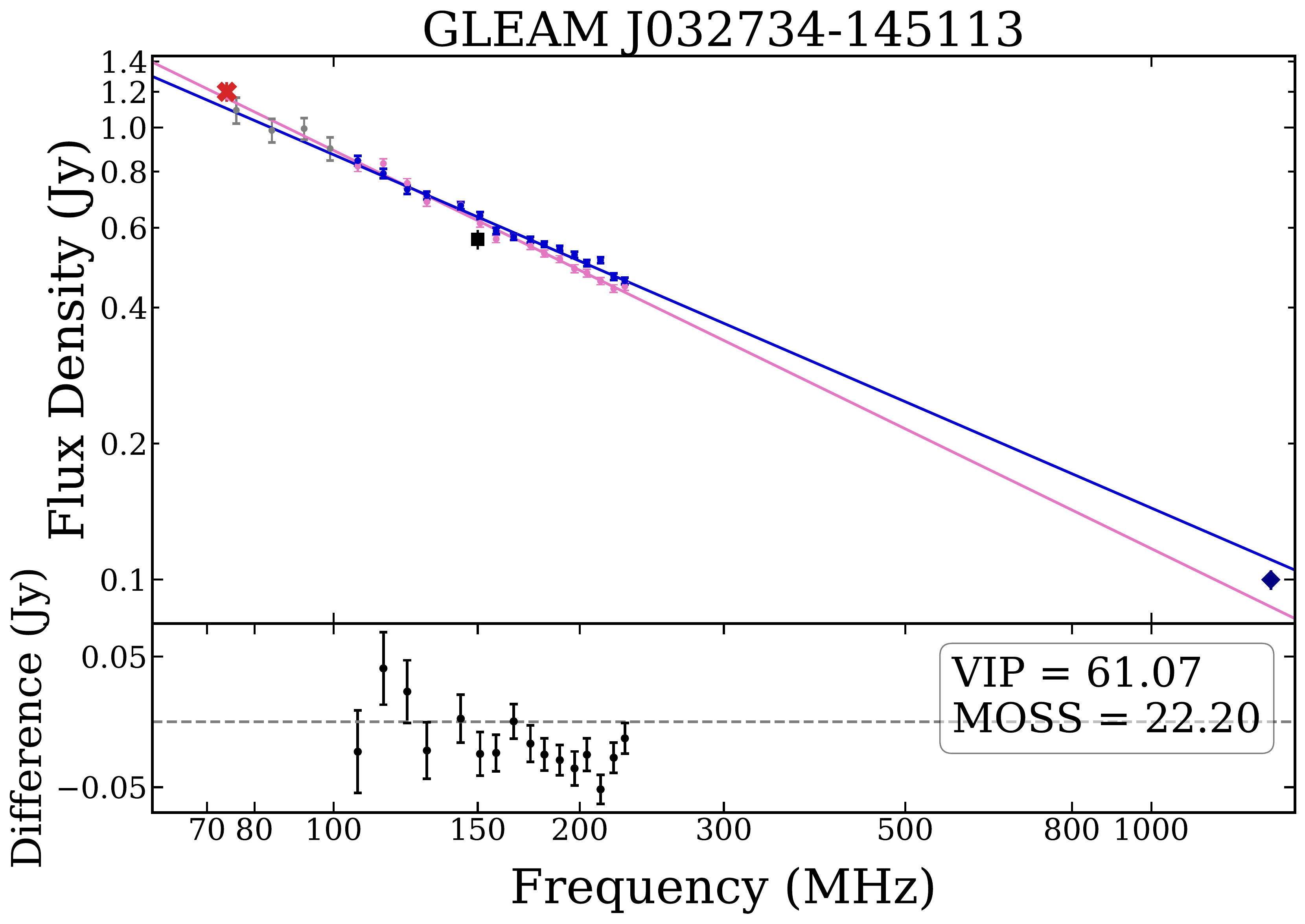} &
\includegraphics[scale=0.15]{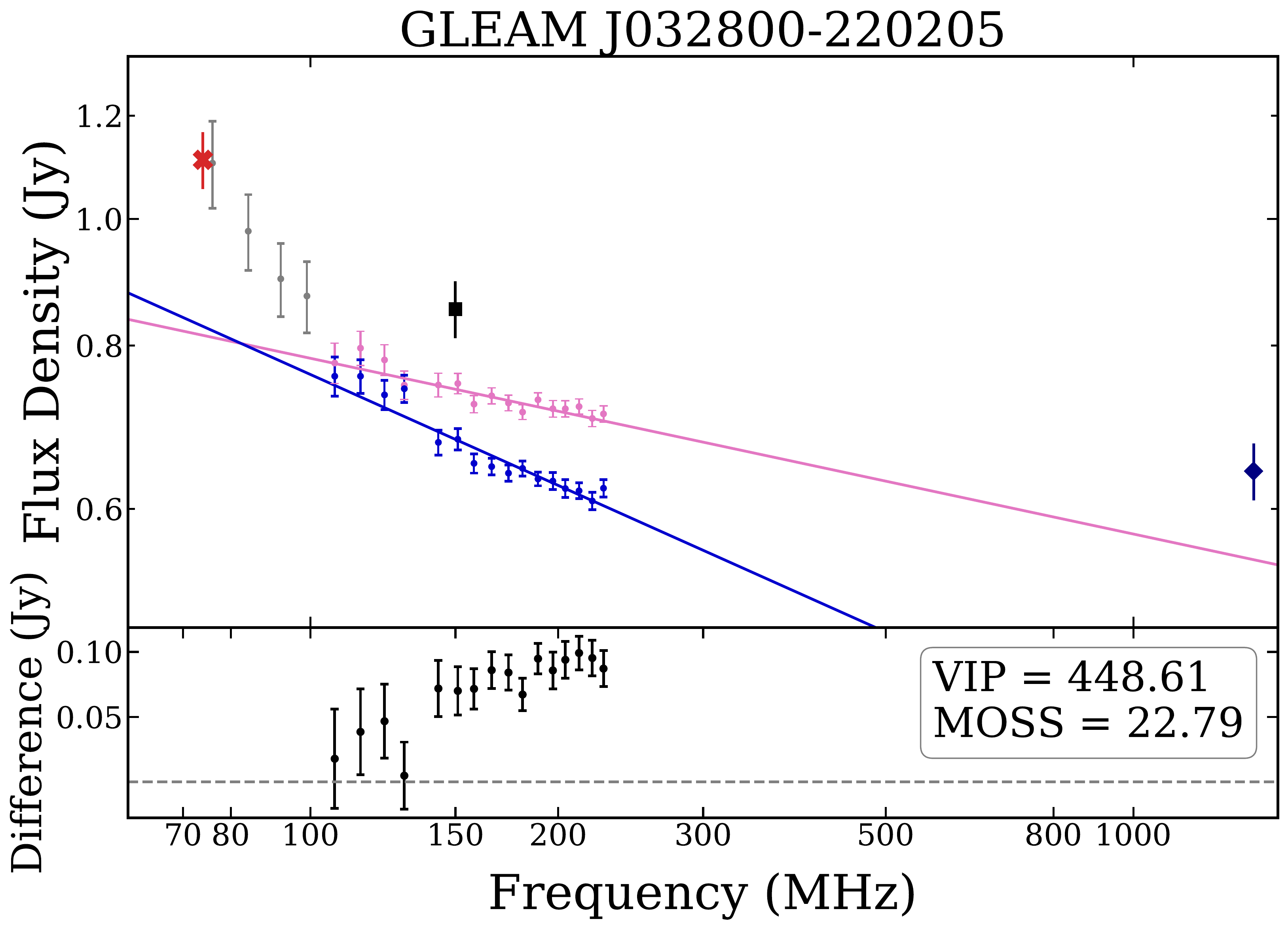} \\
\end{array}$
\caption{(continued) SEDs for all sources classified as variable according to the VIP. For each source the points represent the following data: GLEAM low frequency (72--100\,MHz) (grey circles), Year 1 (pink circles), Year 2 (blue circles), VLSSr (red cross), TGSS (black square), MRC (green star), SUMSS (yellow pentagon), and NVSS (navy diamond). The models for each year are determined by their classification; a source classified with a peak within the observed band was modelled by a quadratic according to Equation~\ref{eq:quadratic}, remaining sources were modelled by a power-law according to Equation~\ref{eq:plaw}.}
\label{app:fig:pg8}
\end{center}
\end{figure*}
\setcounter{figure}{0}
\begin{figure*}
\begin{center}$
\begin{array}{cccccc}
\includegraphics[scale=0.15]{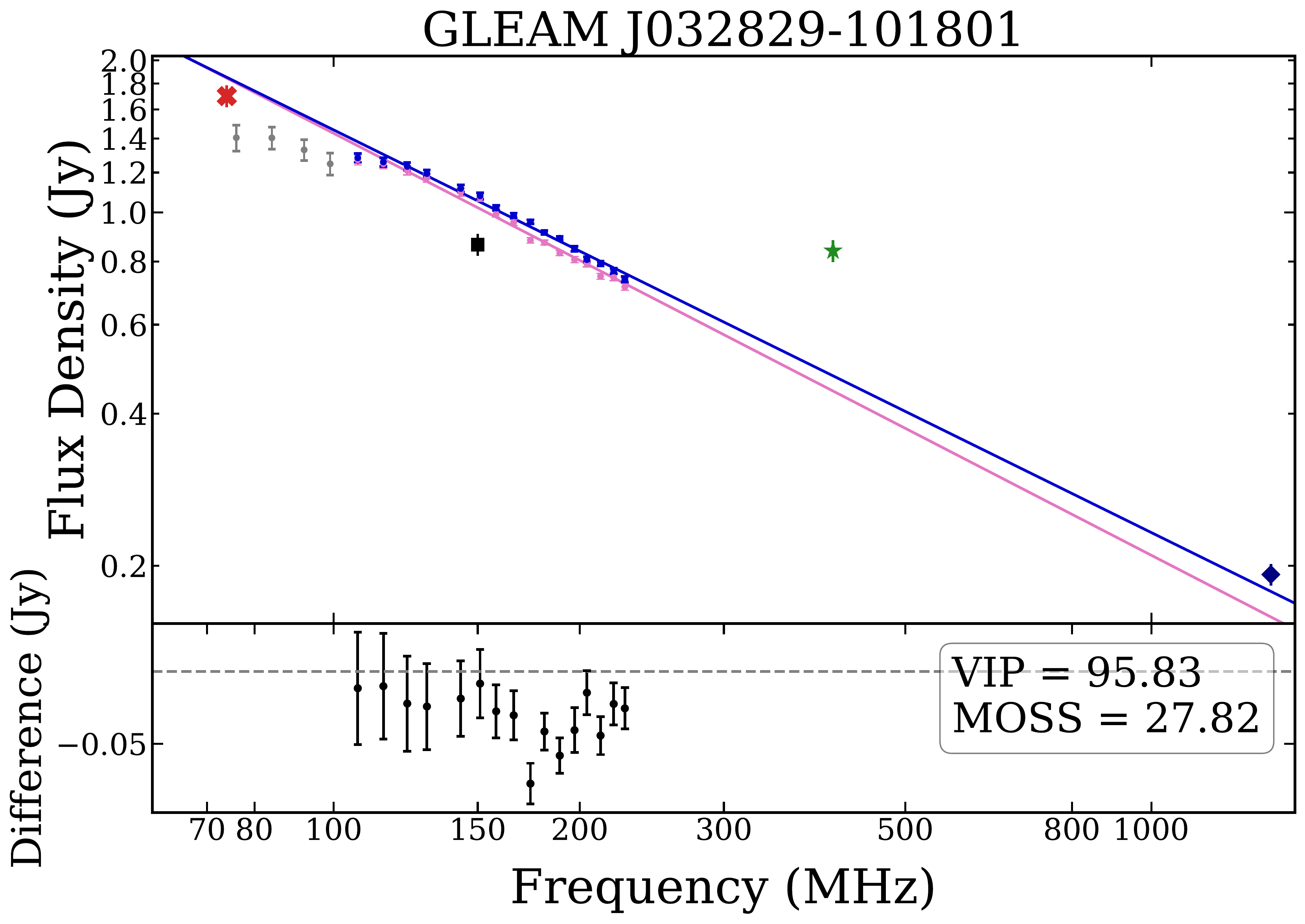} &
\includegraphics[scale=0.15]{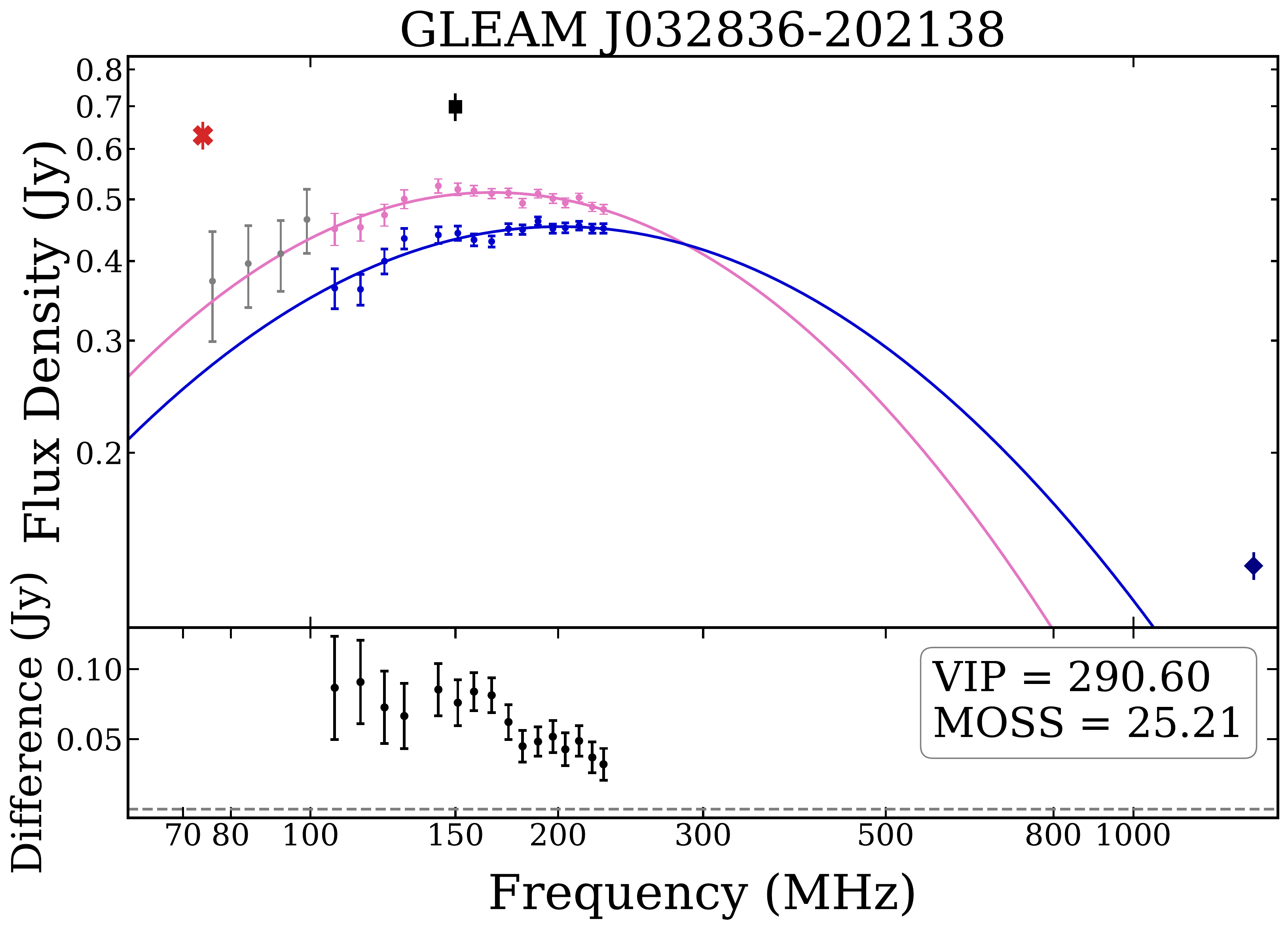} &
\includegraphics[scale=0.15]{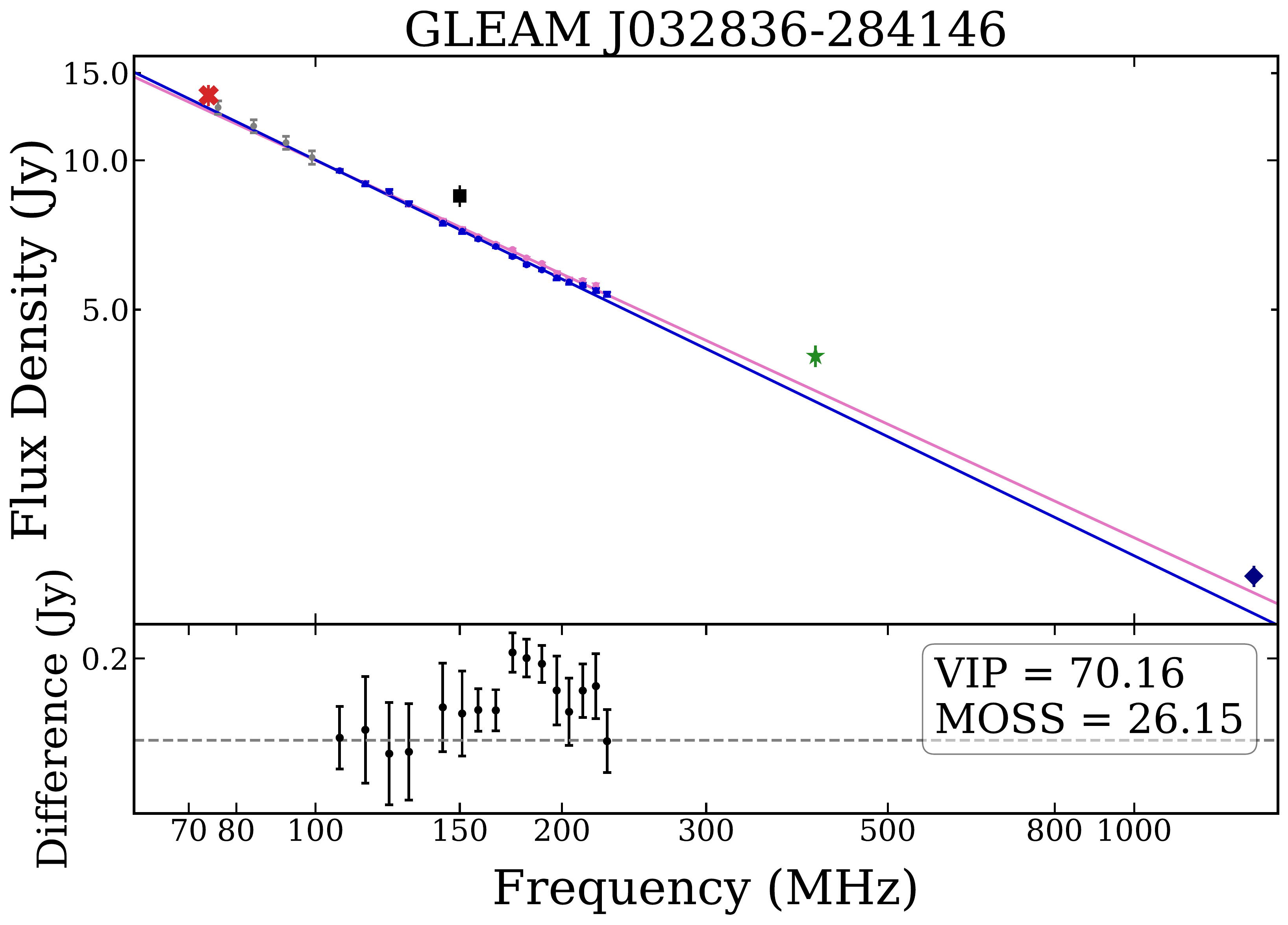} \\
\includegraphics[scale=0.15]{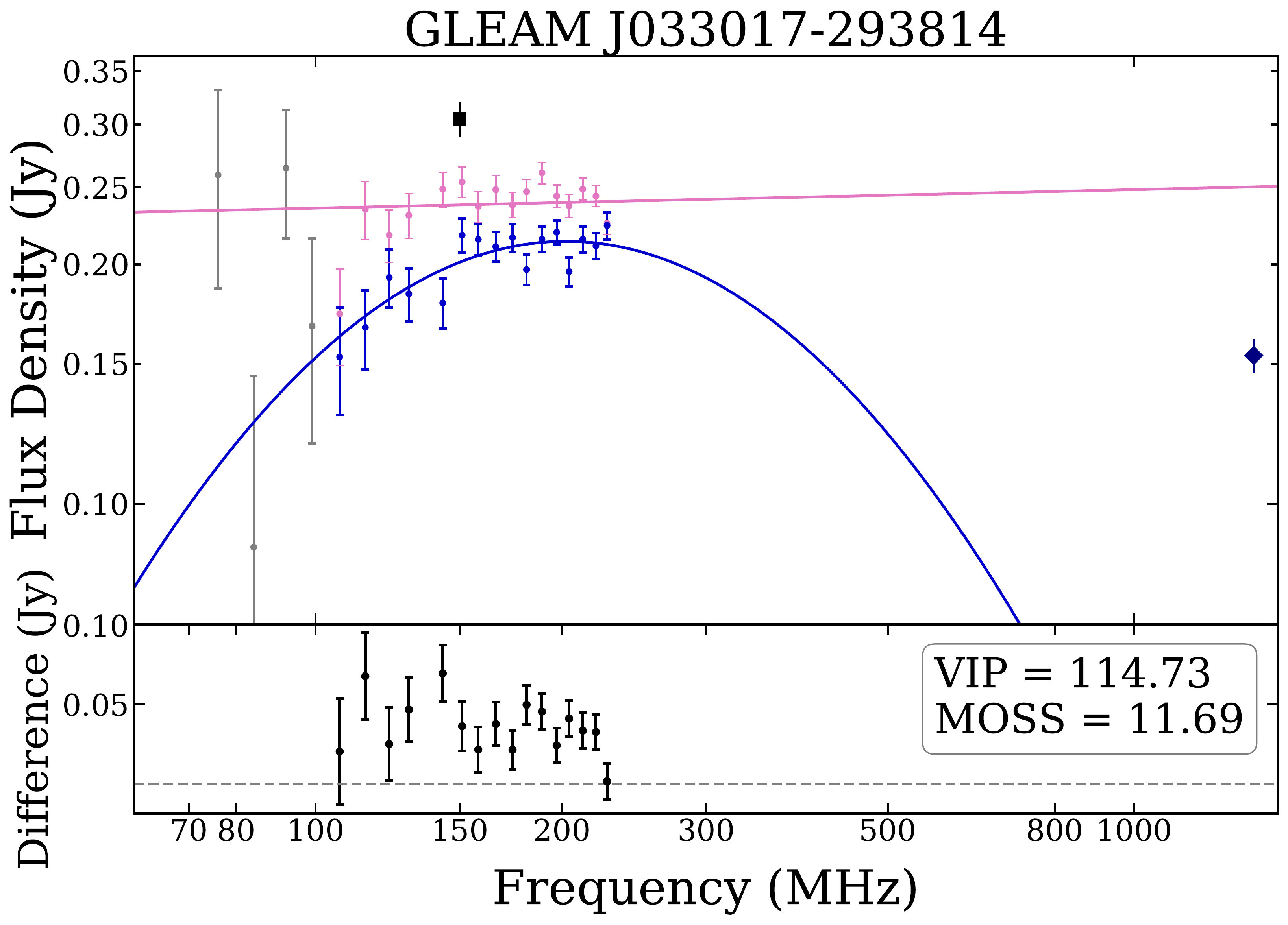} &
\includegraphics[scale=0.15]{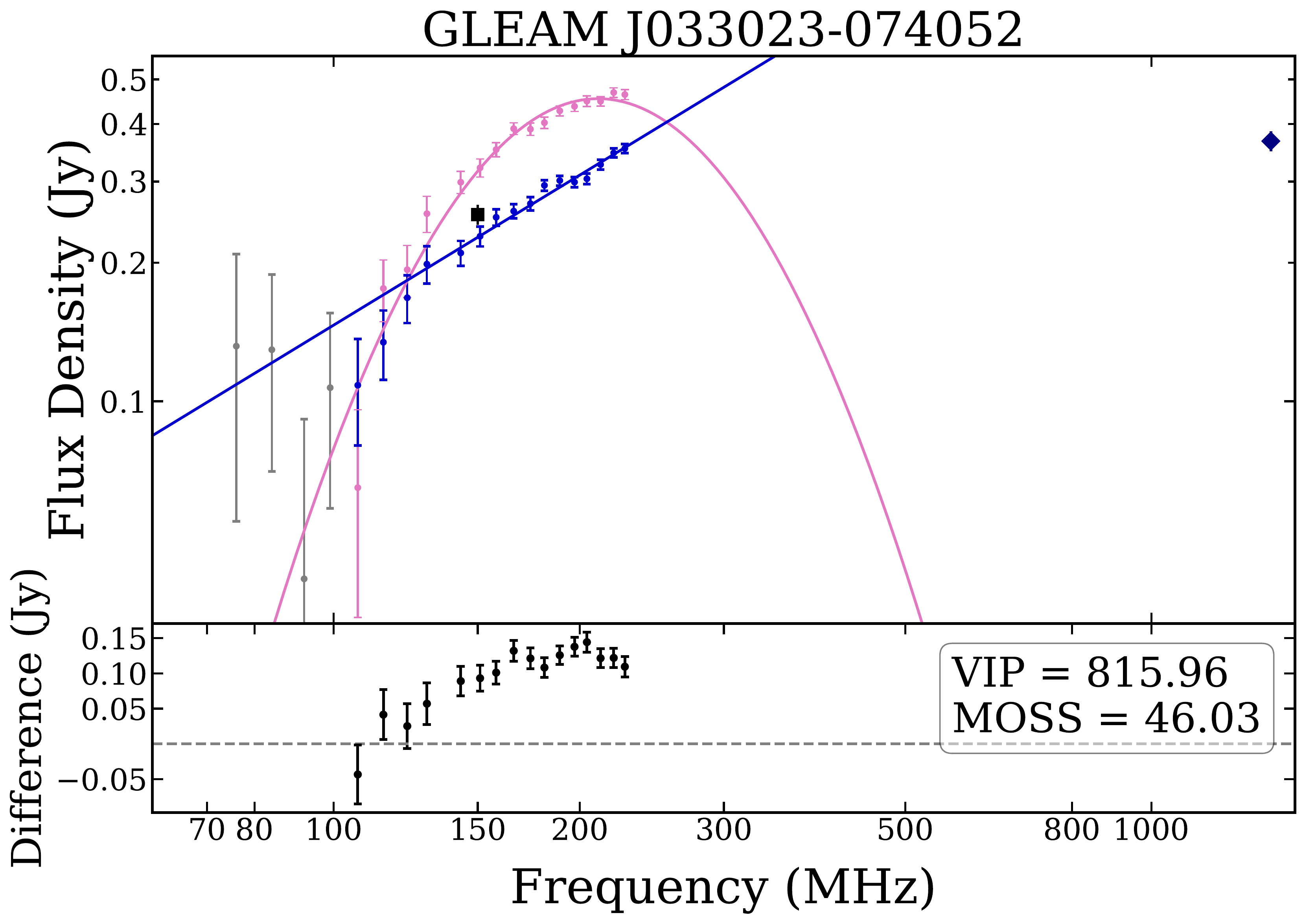} &
\includegraphics[scale=0.15]{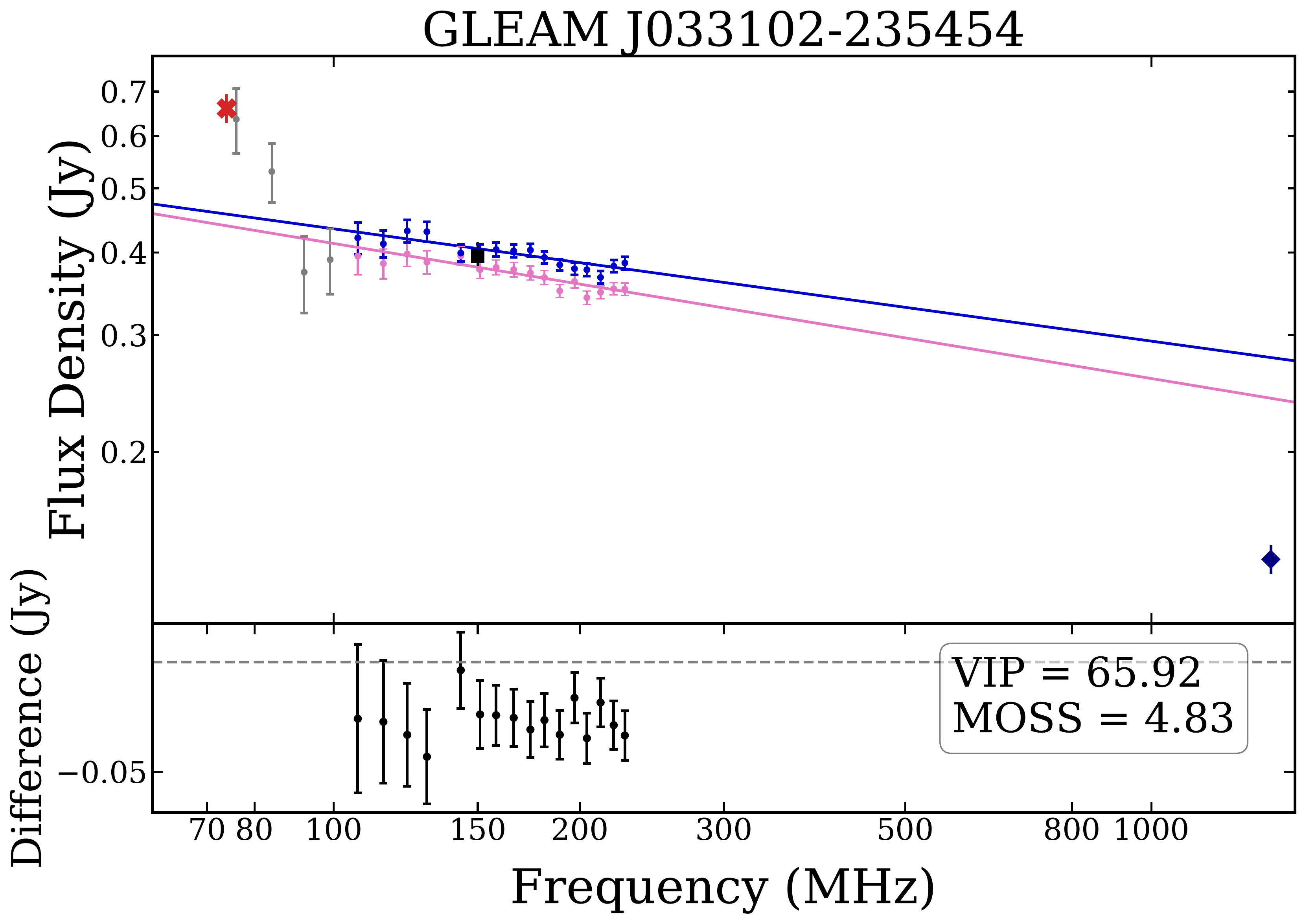} \\
\includegraphics[scale=0.15]{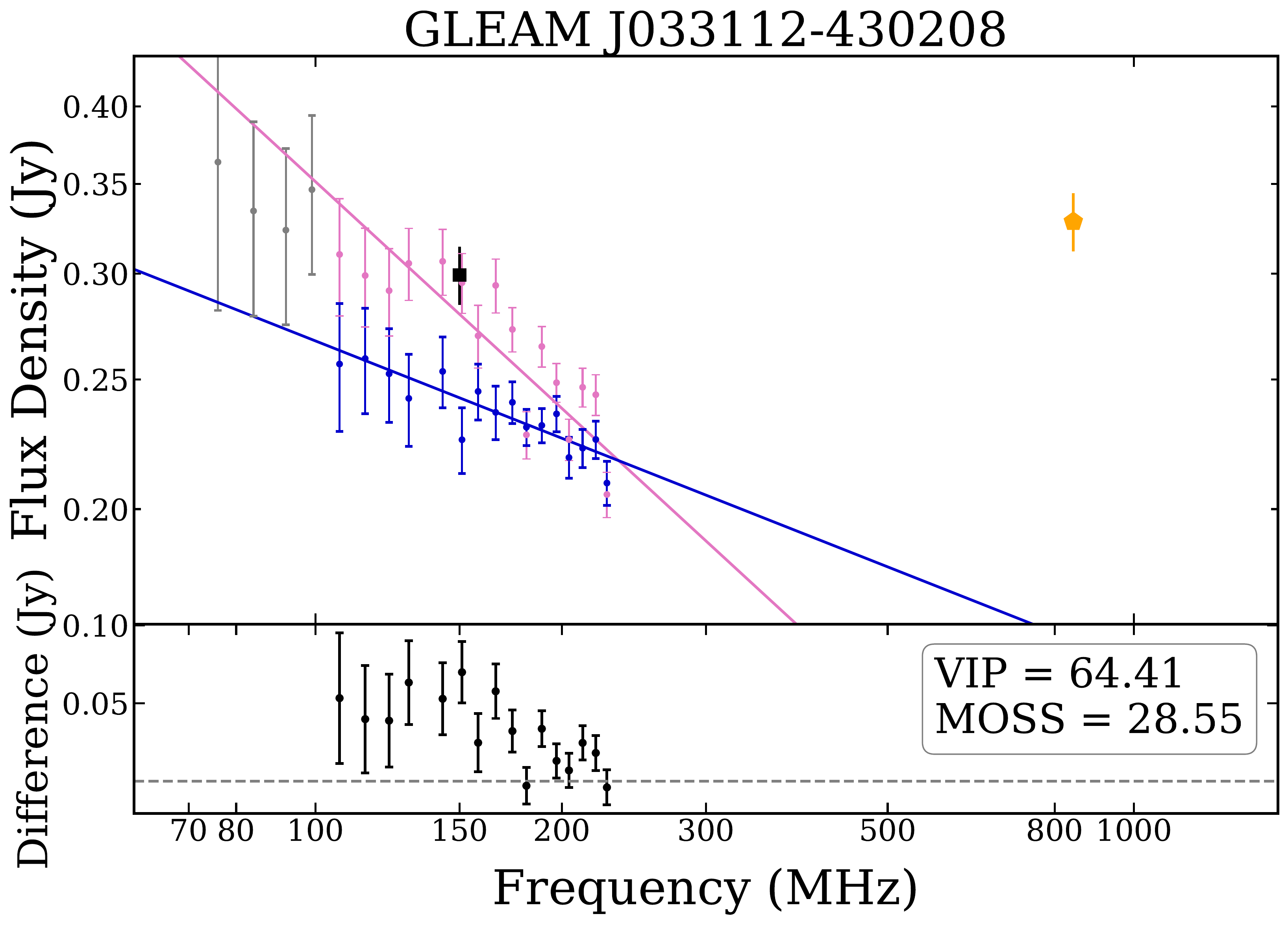} &
\includegraphics[scale=0.15]{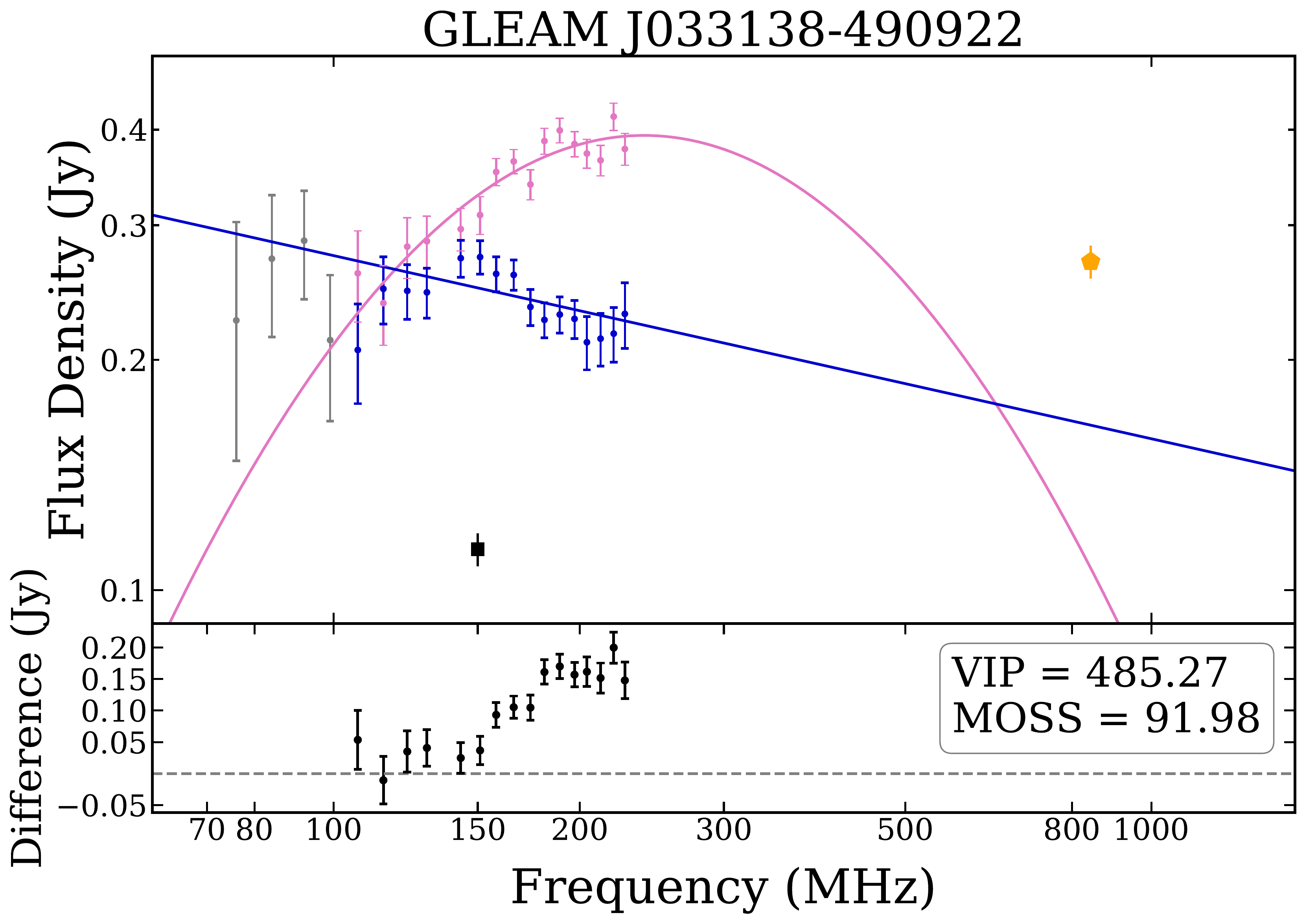} &
\includegraphics[scale=0.15]{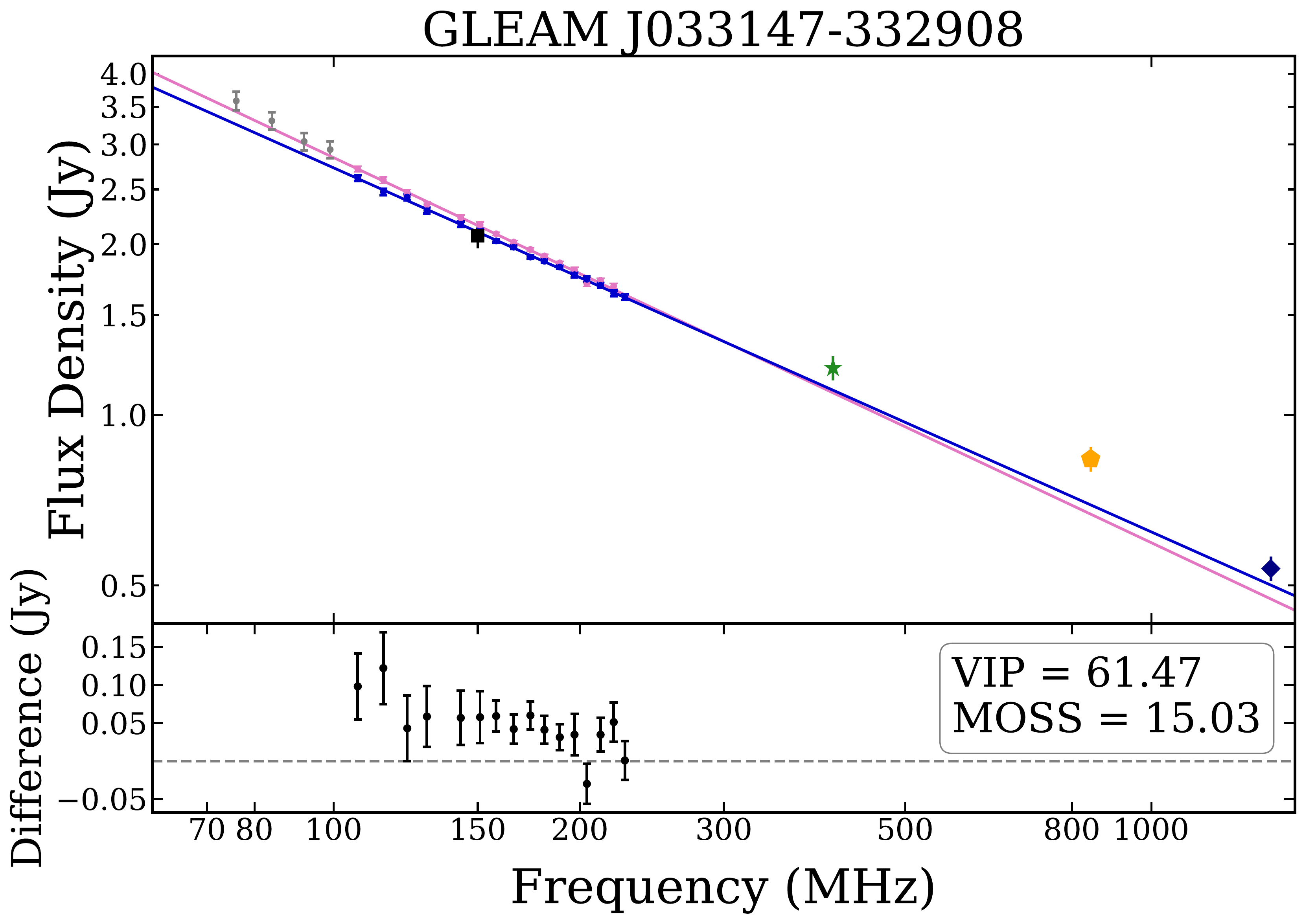} \\
\includegraphics[scale=0.15]{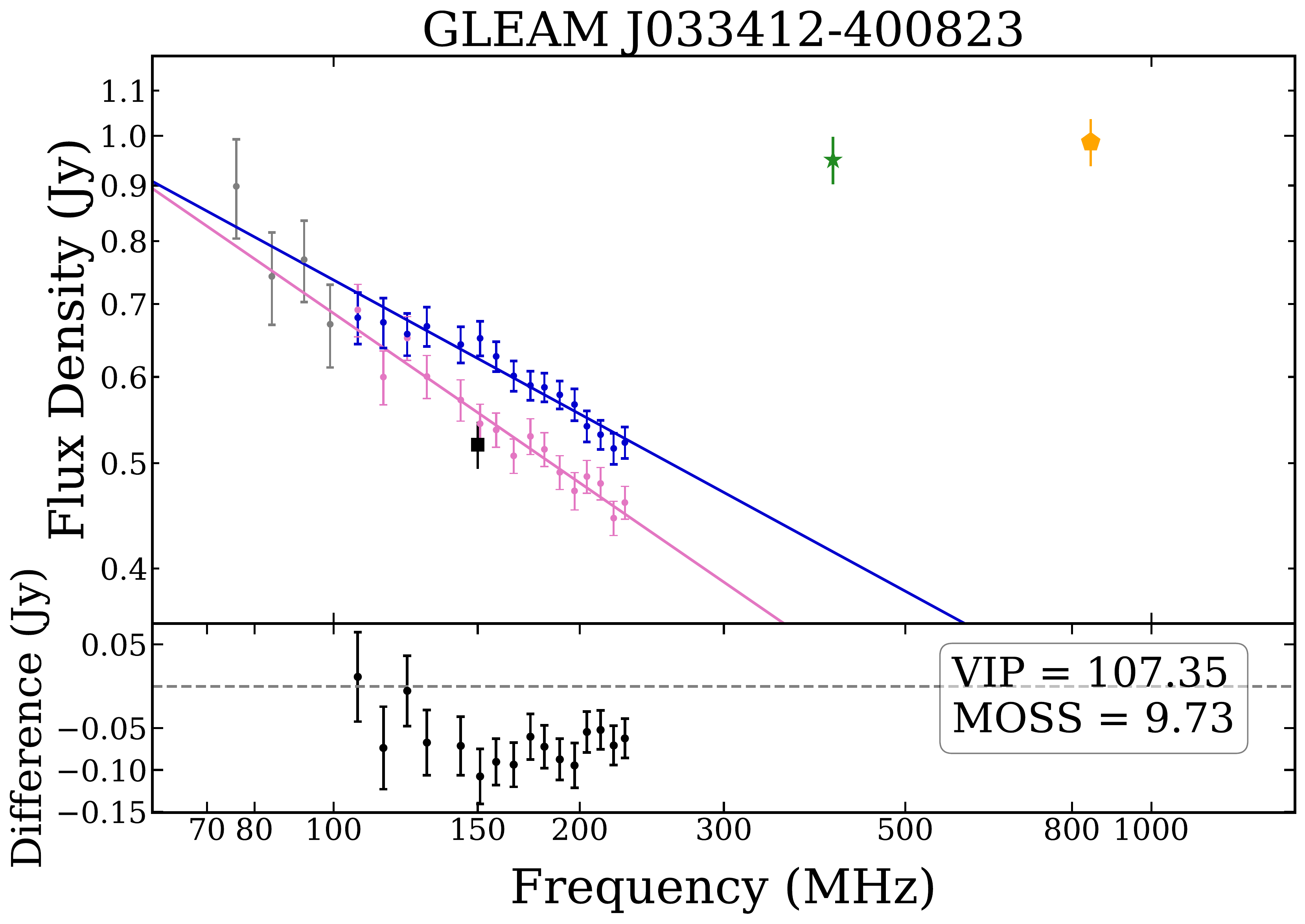} &
\includegraphics[scale=0.15]{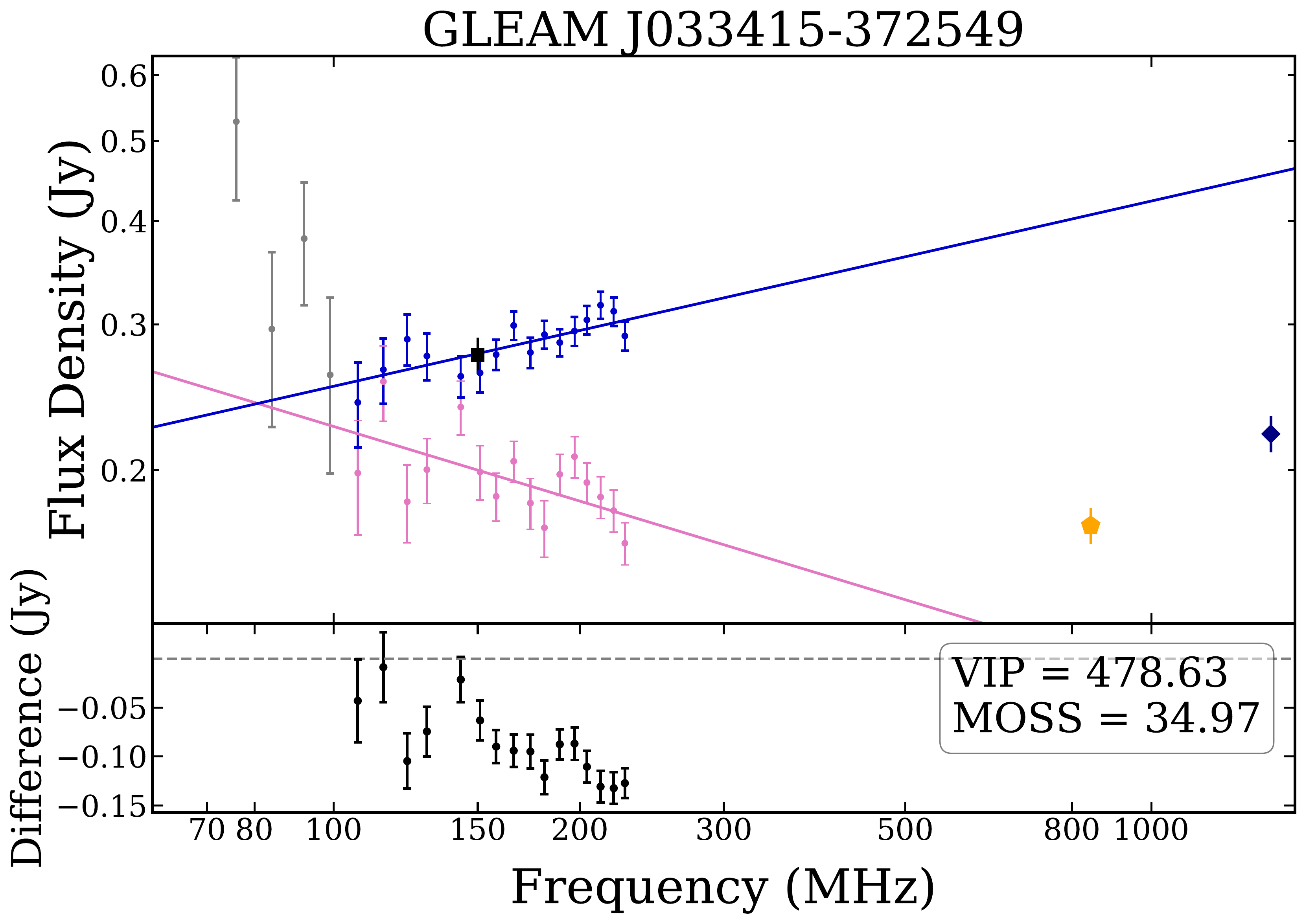} &
\includegraphics[scale=0.15]{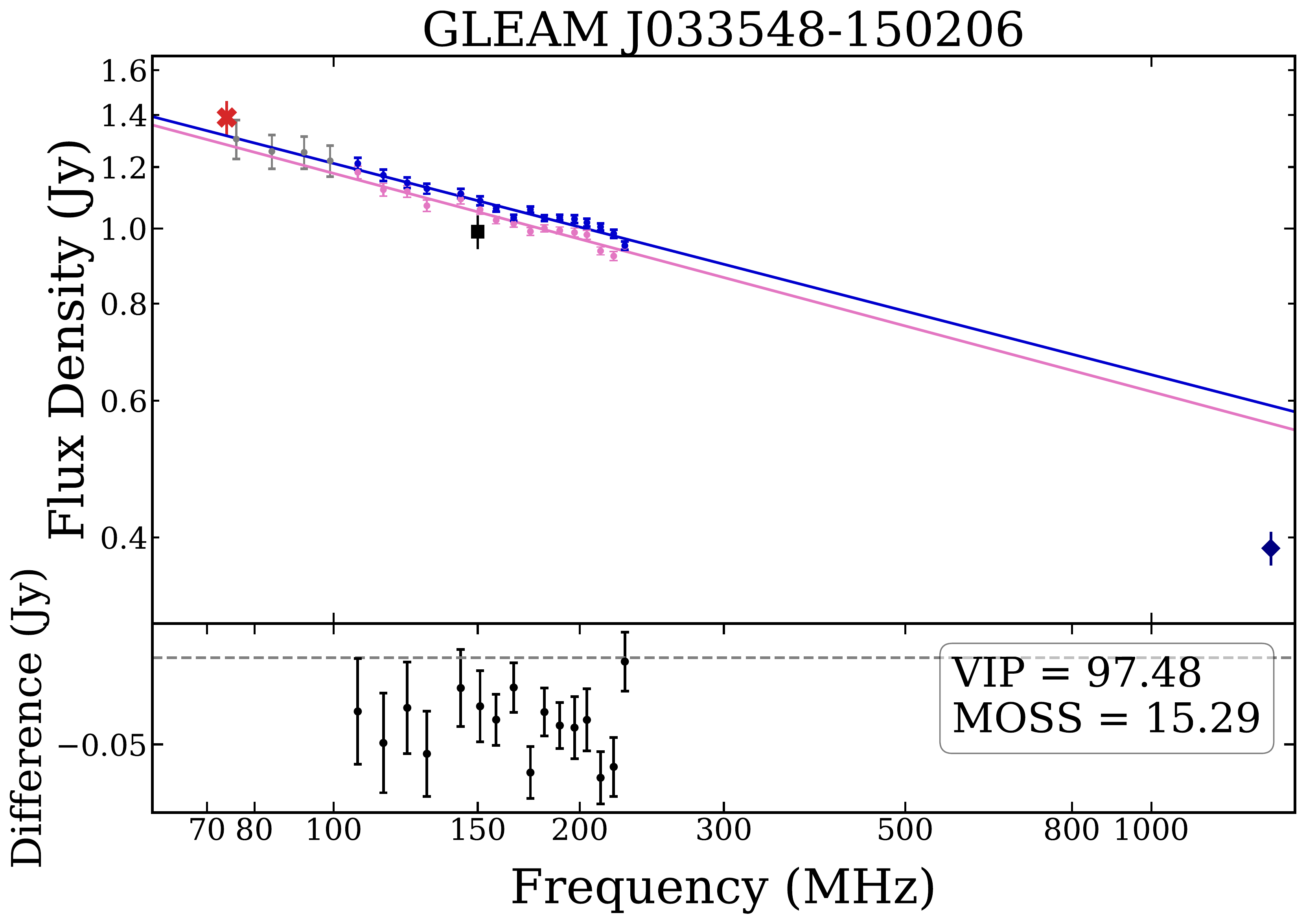} \\
\includegraphics[scale=0.15]{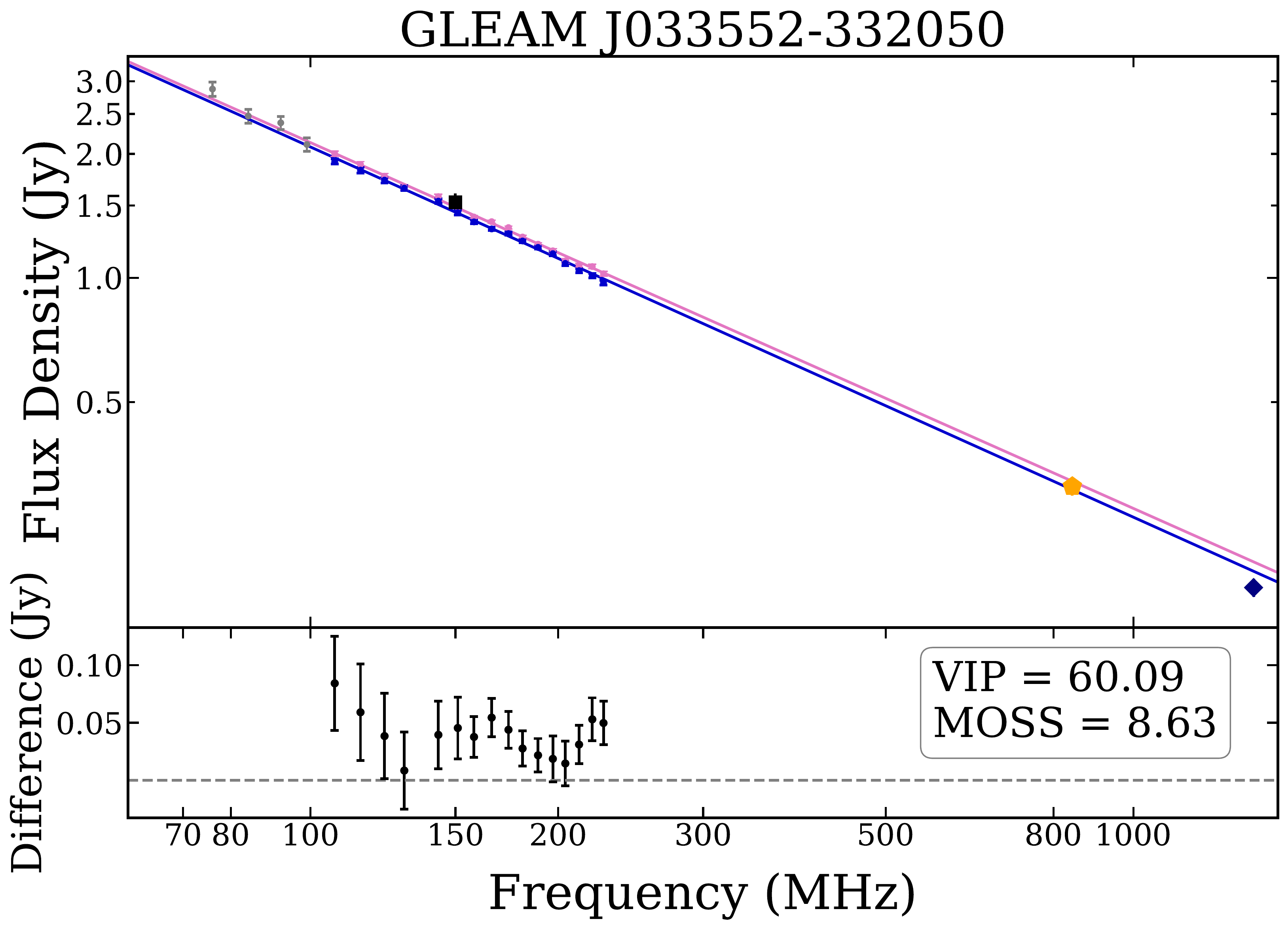} &
\includegraphics[scale=0.15]{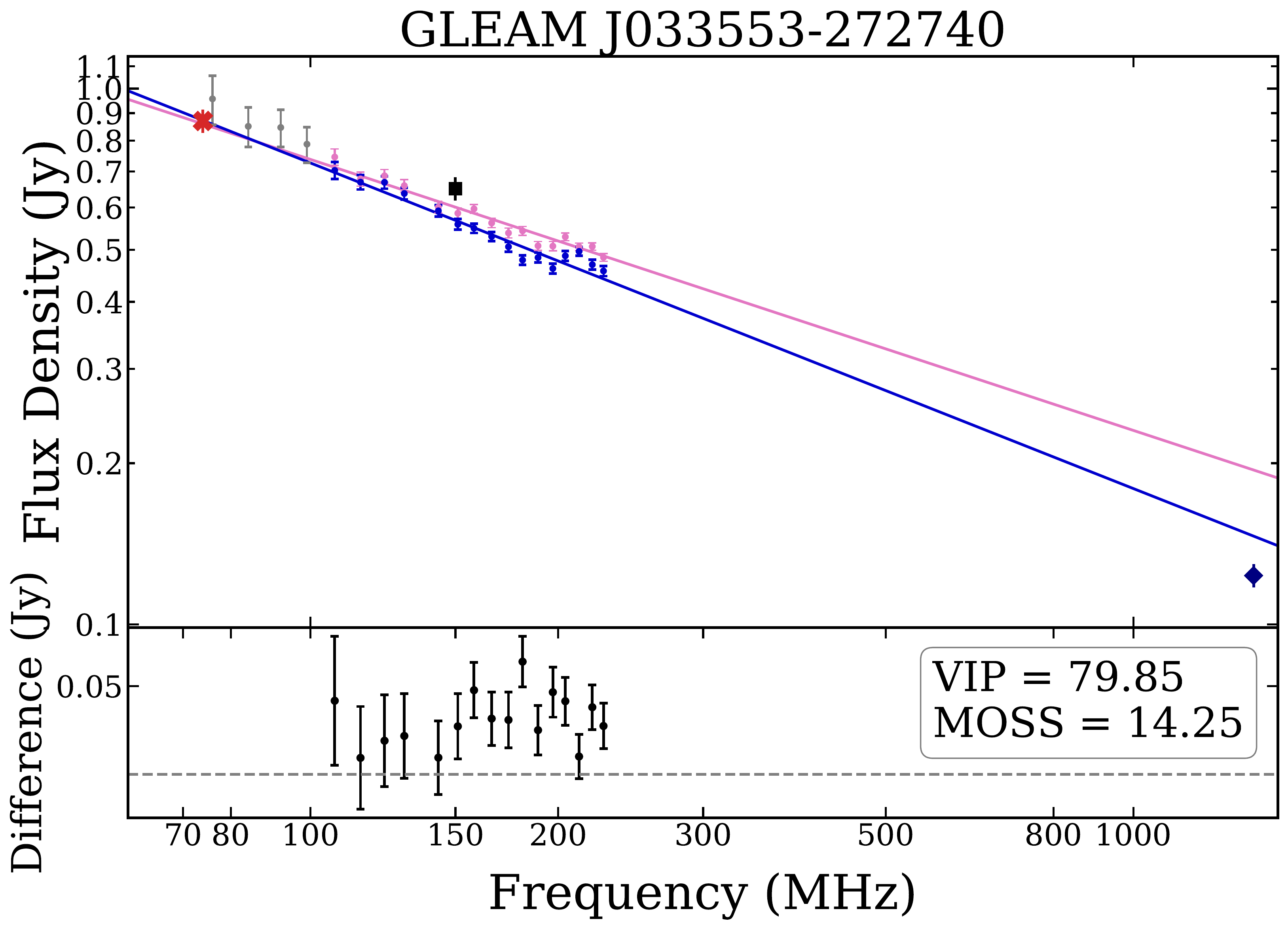} &
\includegraphics[scale=0.15]{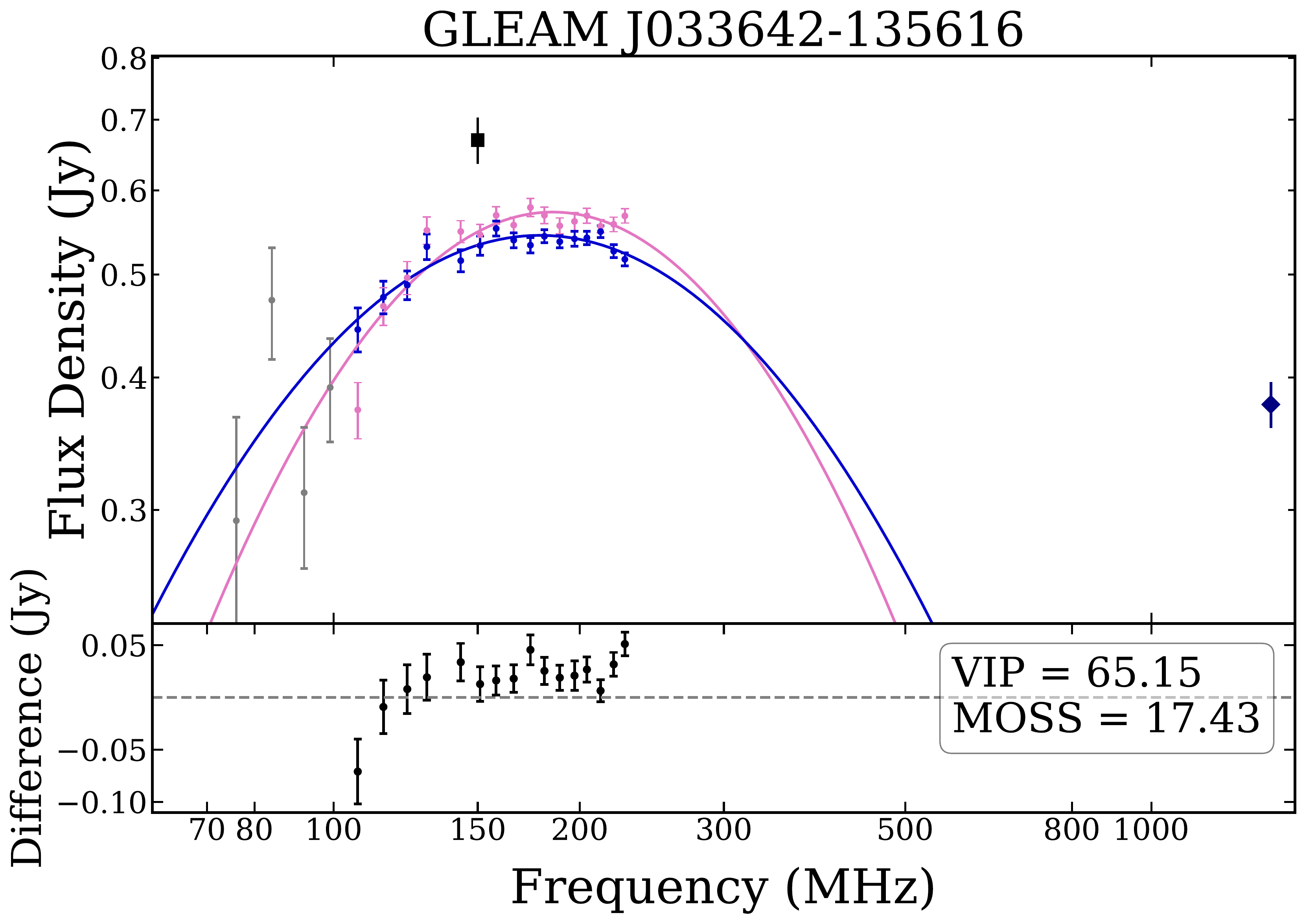} \\
\includegraphics[scale=0.15]{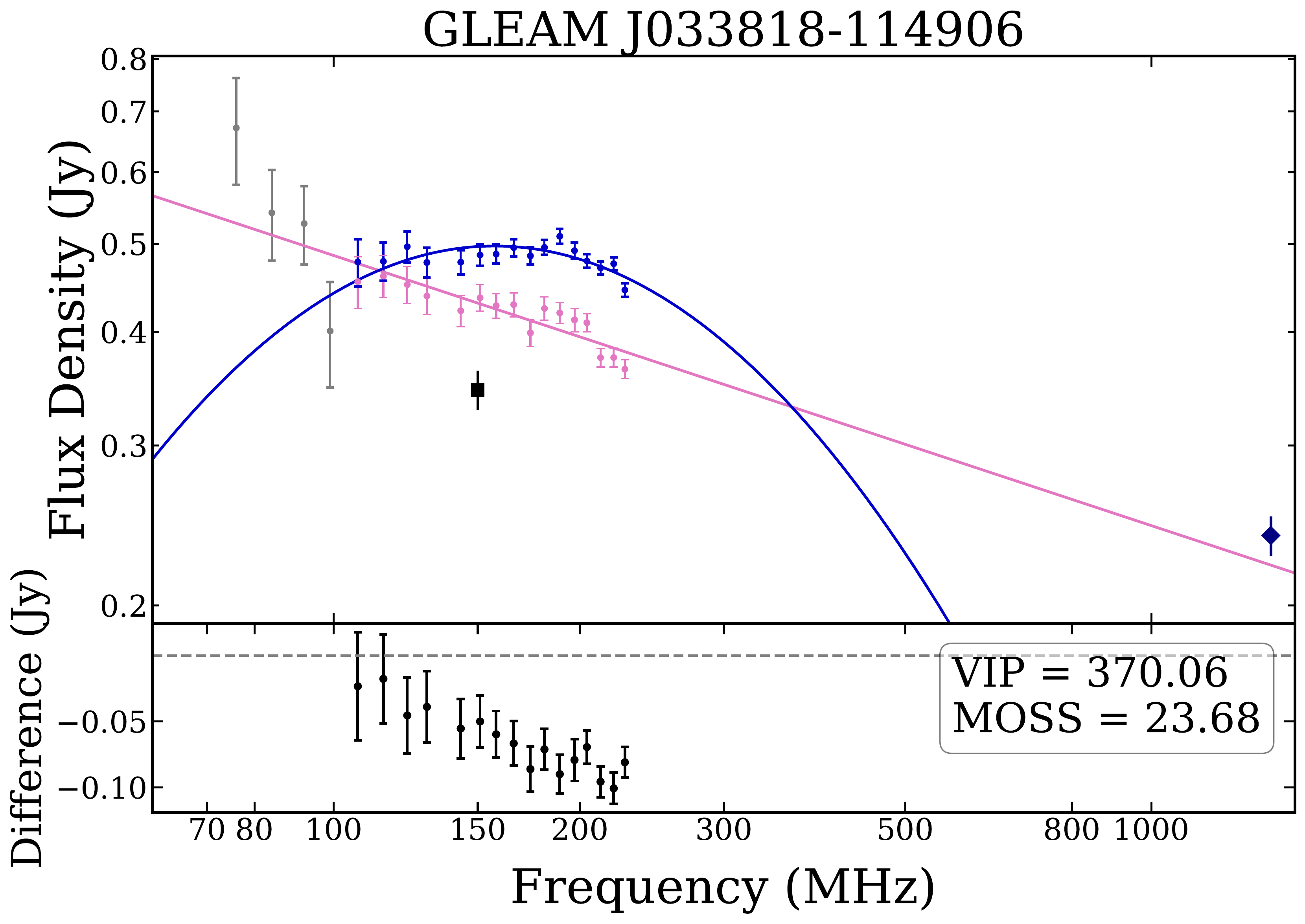} &
\includegraphics[scale=0.15]{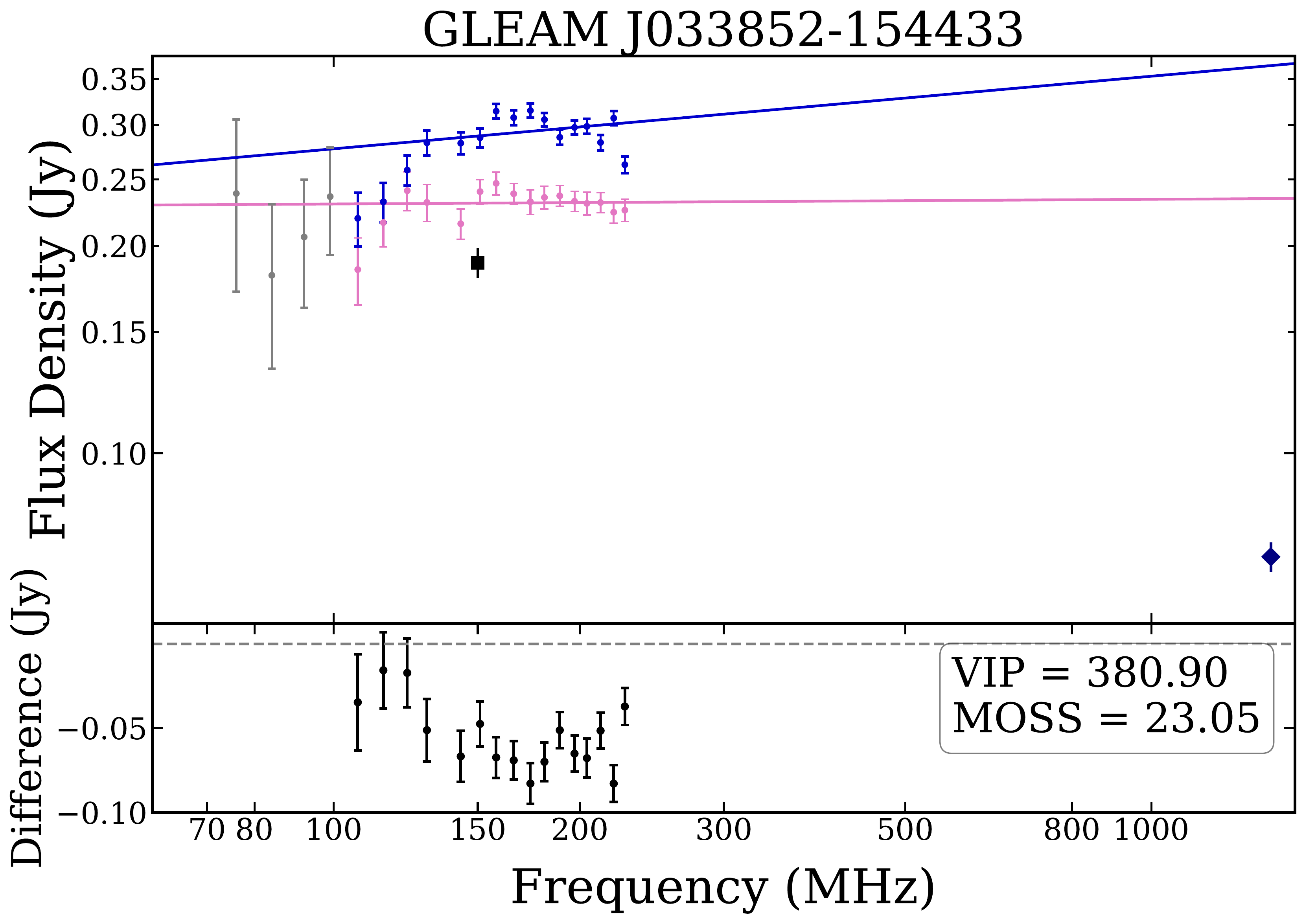} &
\includegraphics[scale=0.15]{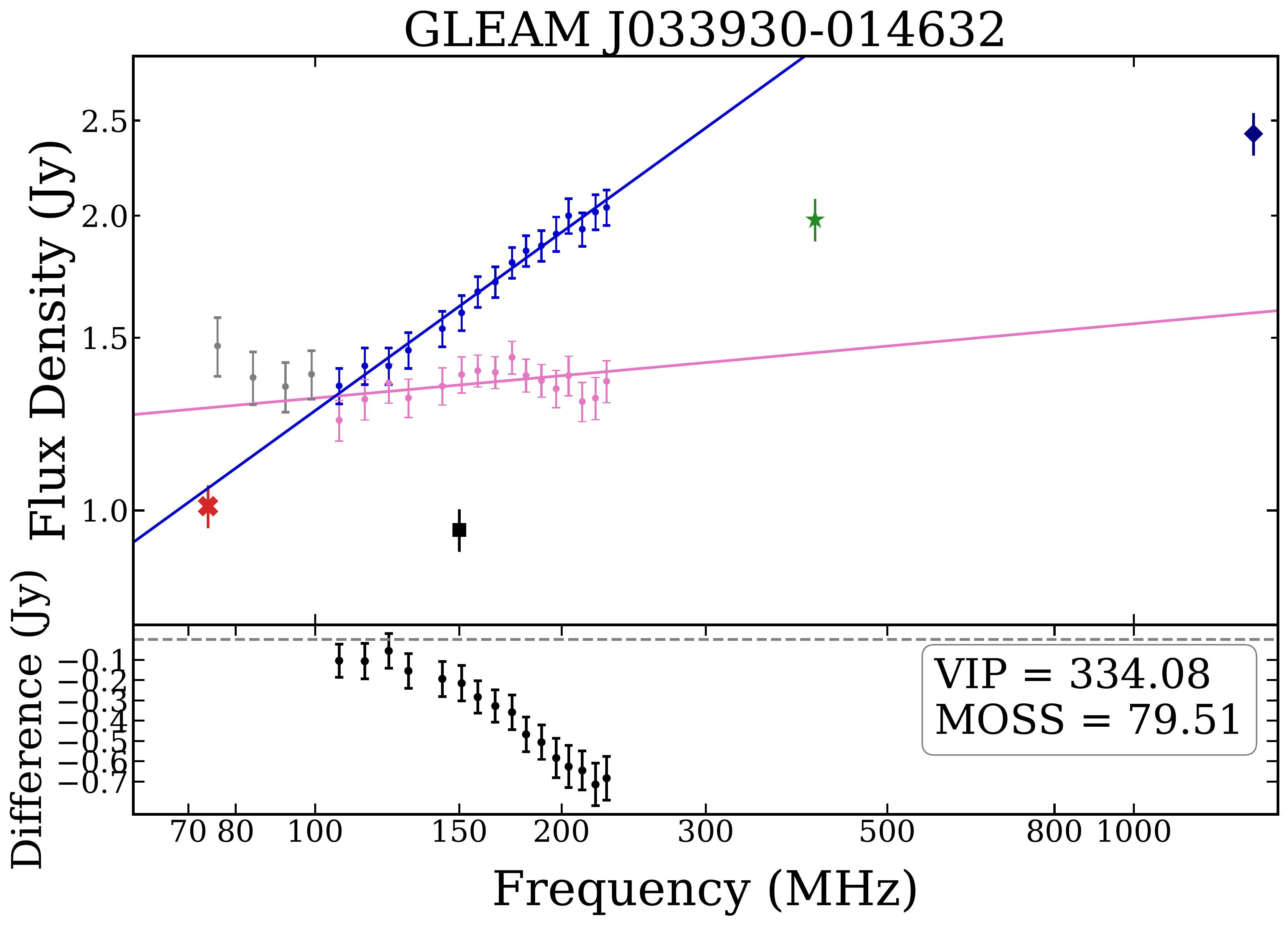} \\
\end{array}$
\caption{(continued) SEDs for all sources classified as variable according to the VIP. For each source the points represent the following data: GLEAM low frequency (72--100\,MHz) (grey circles), Year 1 (pink circles), Year 2 (blue circles), VLSSr (red cross), TGSS (black square), MRC (green star), SUMSS (yellow pentagon), and NVSS (navy diamond). The models for each year are determined by their classification; a source classified with a peak within the observed band was modelled by a quadratic according to Equation~\ref{eq:quadratic}, remaining sources were modelled by a power-law according to Equation~\ref{eq:plaw}.}
\label{app:fig:pg9}
\end{center}
\end{figure*}
\setcounter{figure}{0}
\begin{figure*}
\begin{center}$
\begin{array}{cccccc}
\includegraphics[scale=0.15]{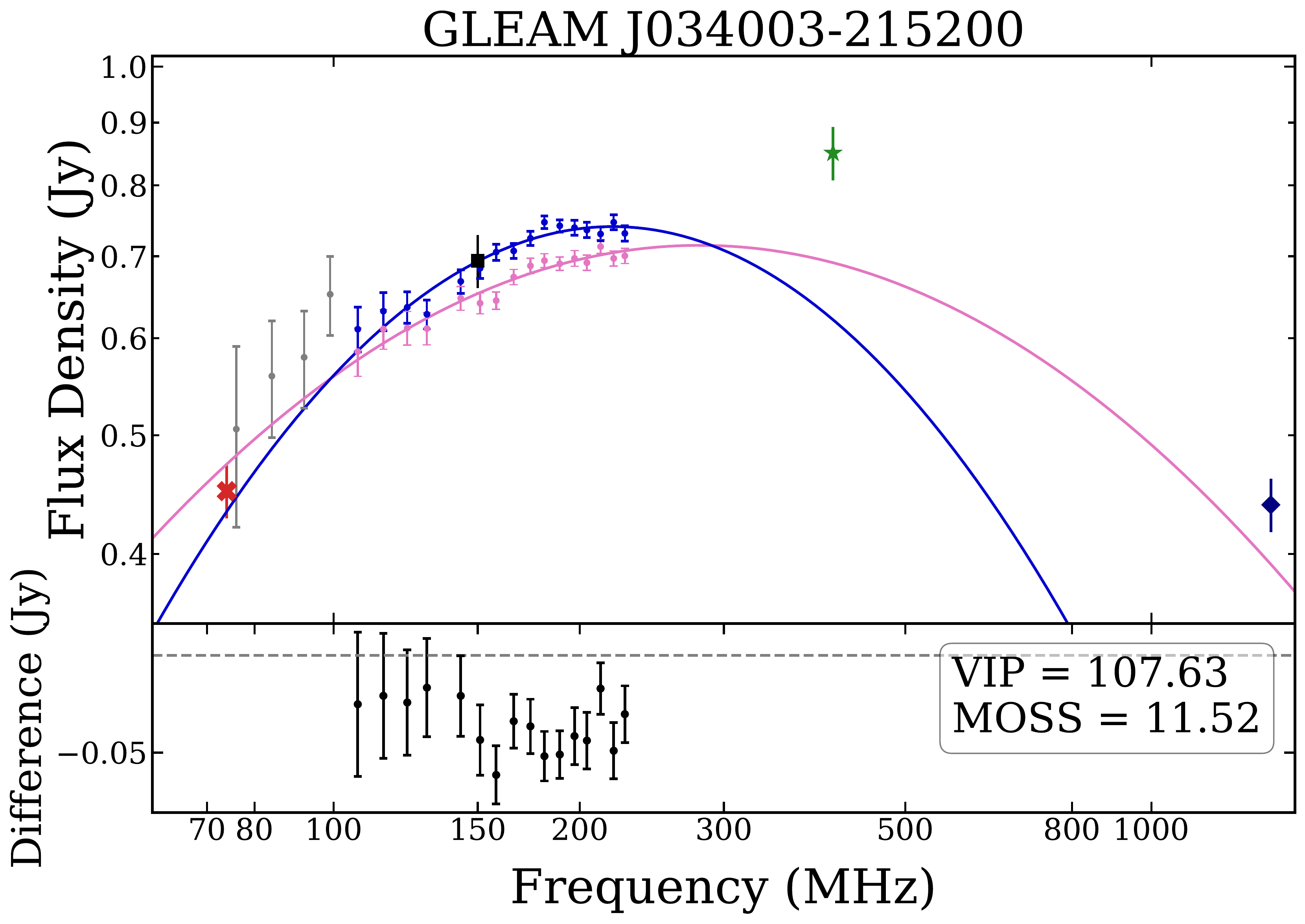} &
\includegraphics[scale=0.15]{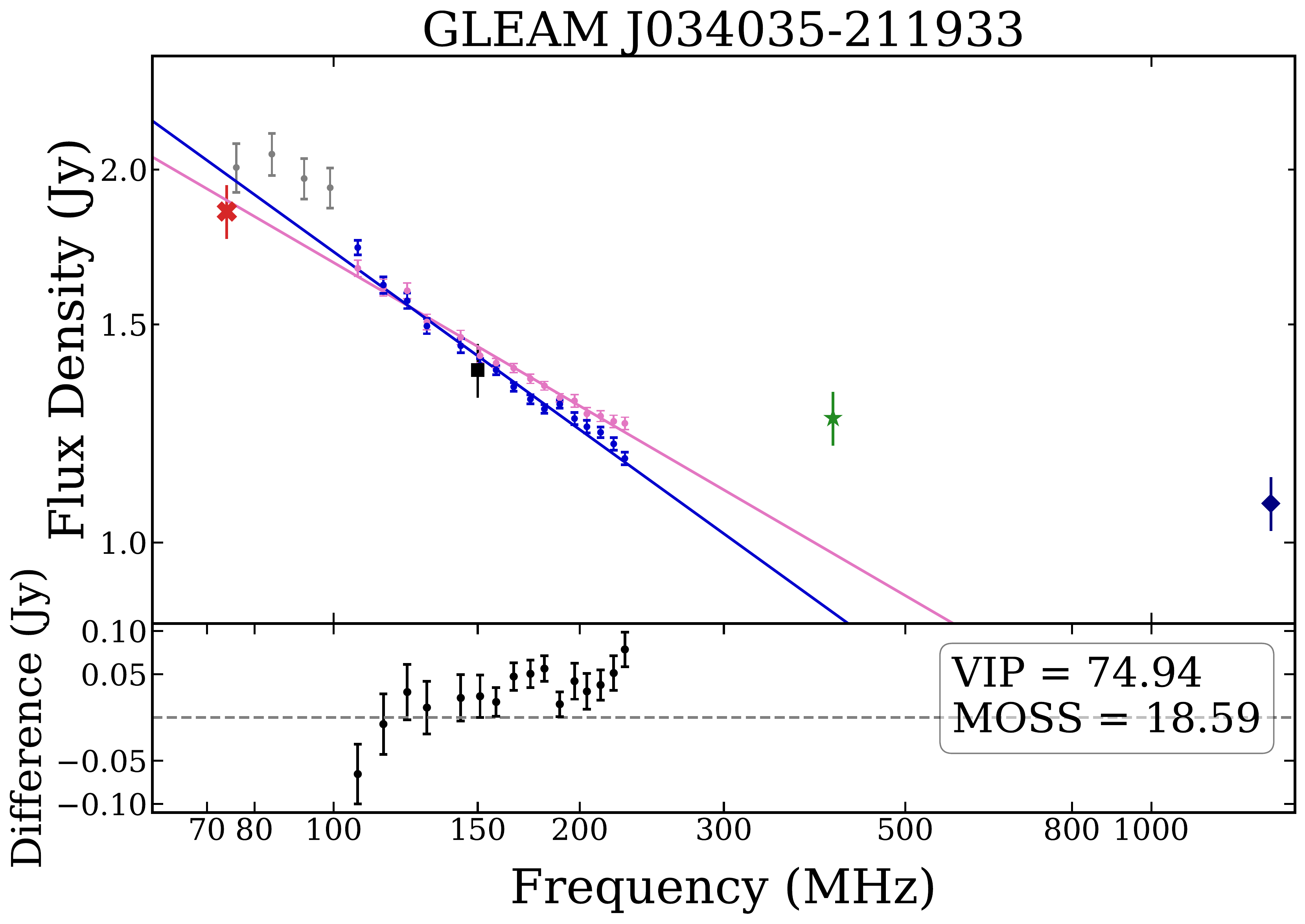} &
\includegraphics[scale=0.15]{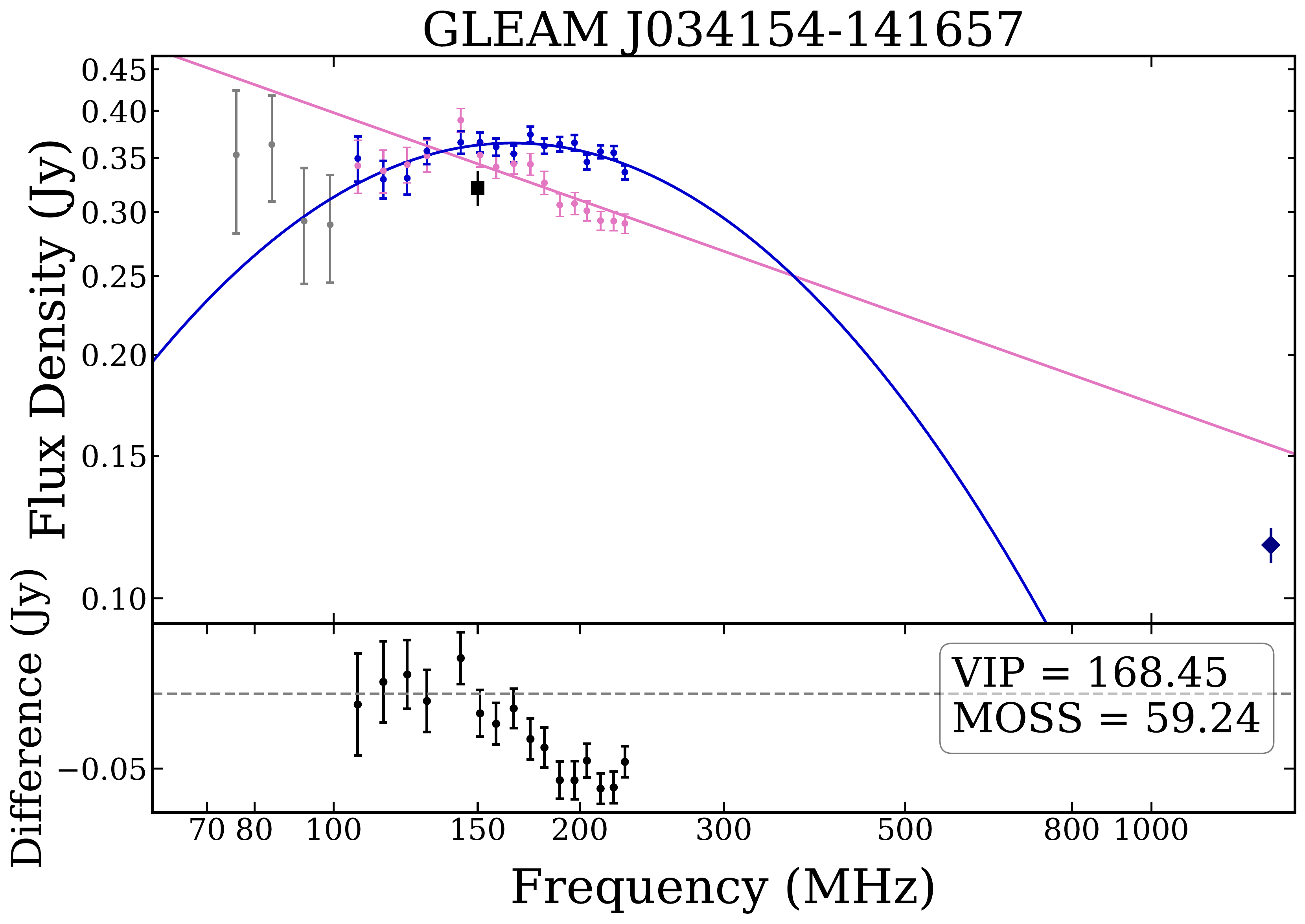} \\
\includegraphics[scale=0.15]{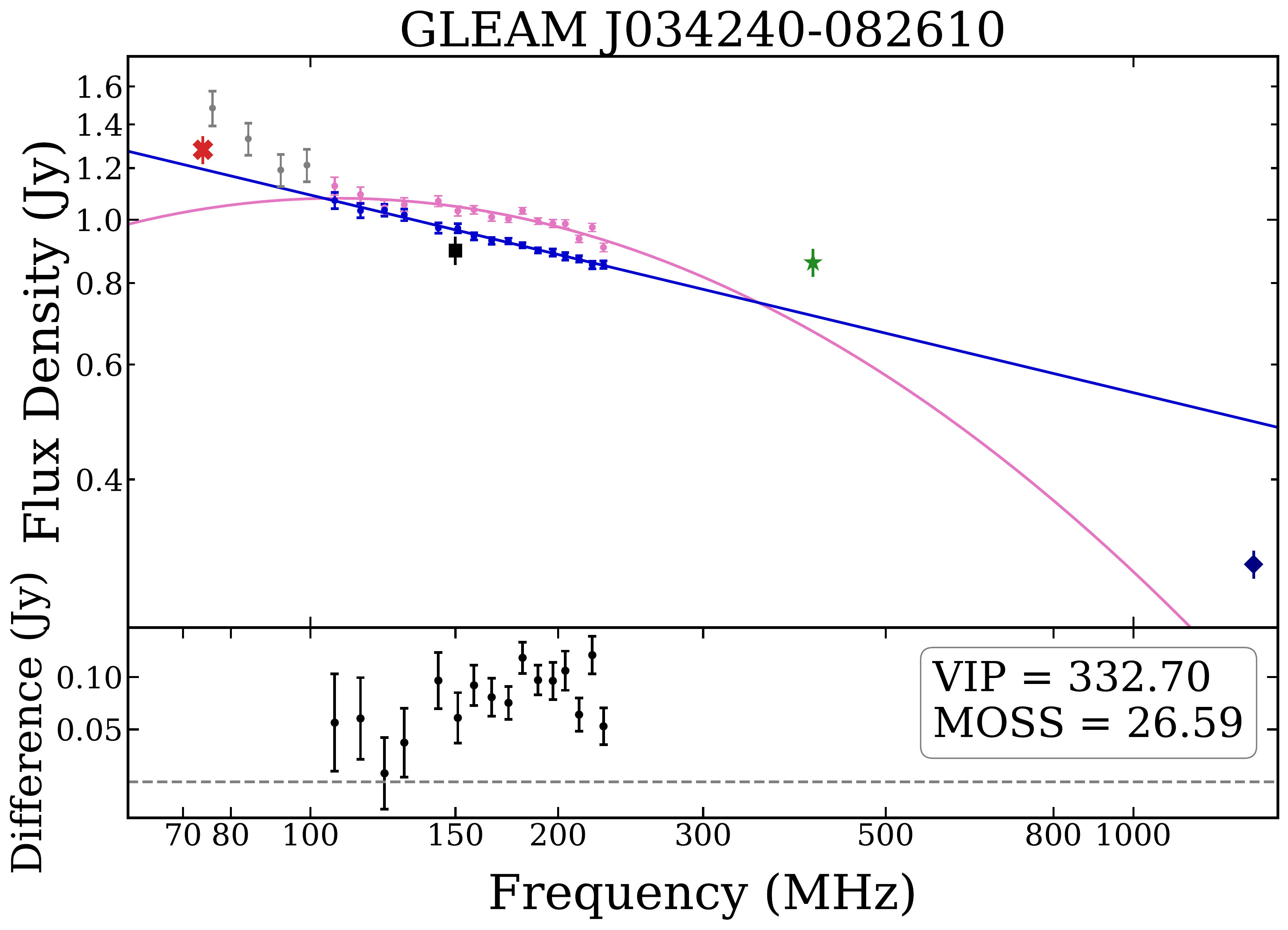} &
\includegraphics[scale=0.15]{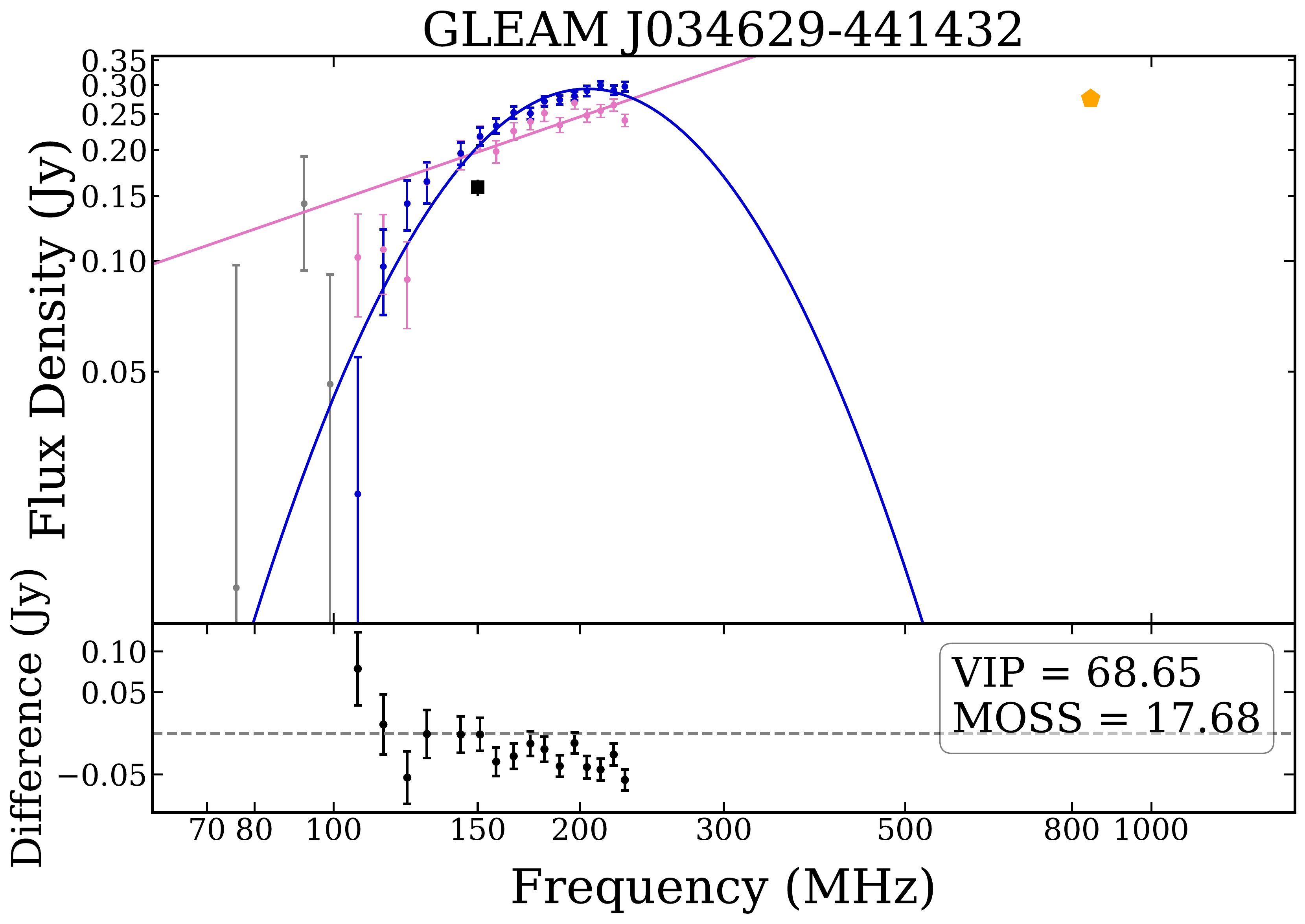} &
\includegraphics[scale=0.15]{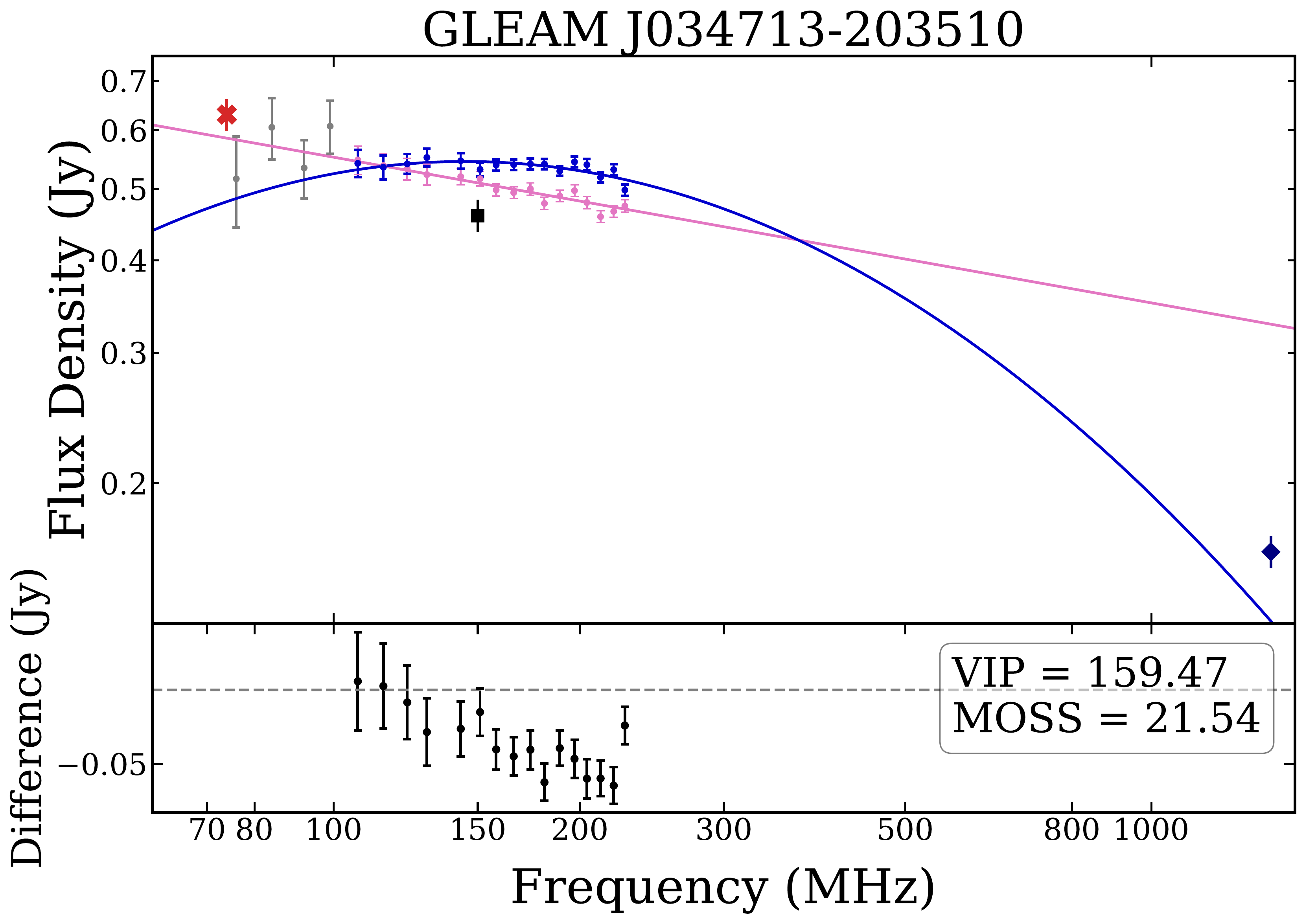} \\
\includegraphics[scale=0.15]{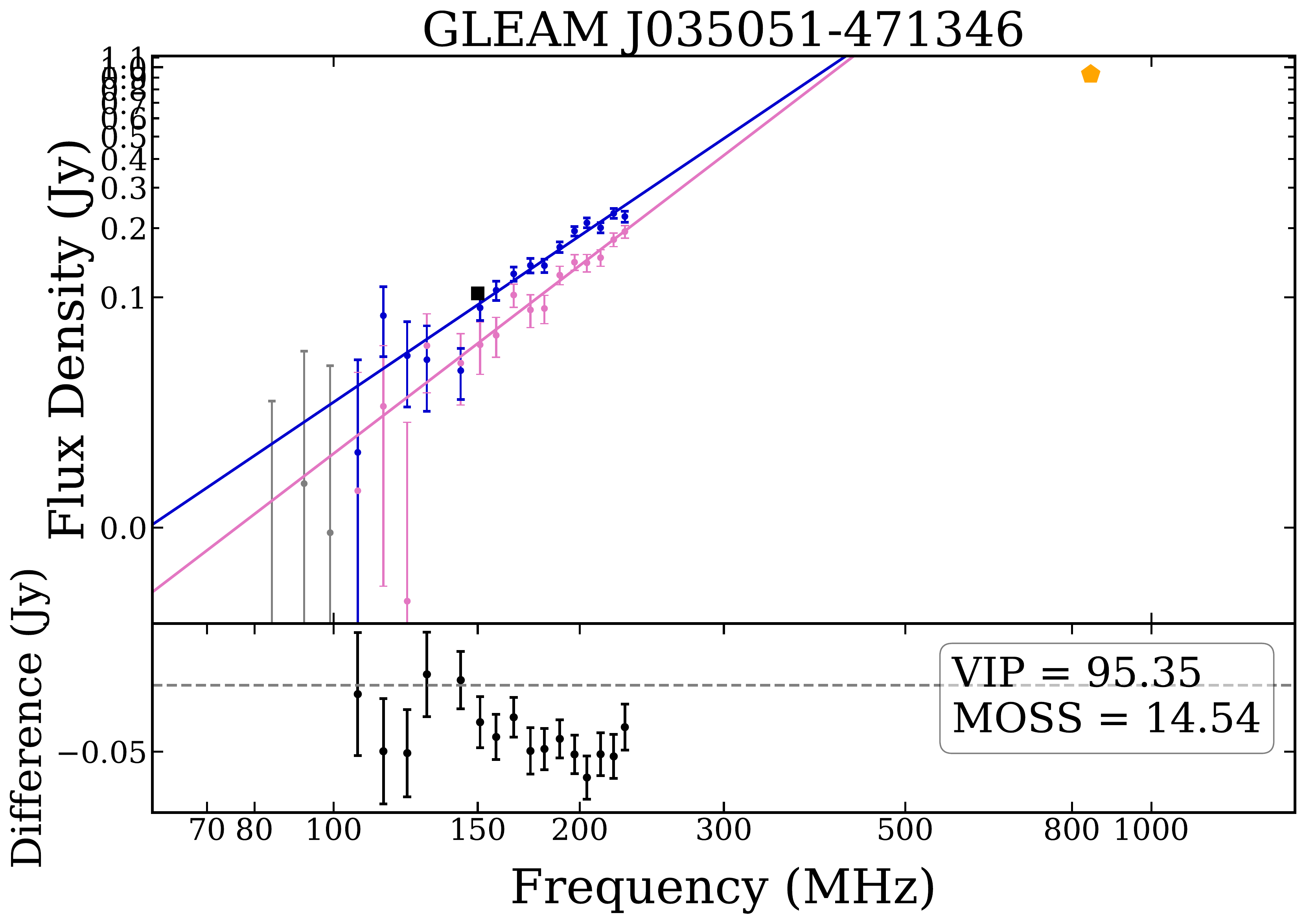} &
\includegraphics[scale=0.15]{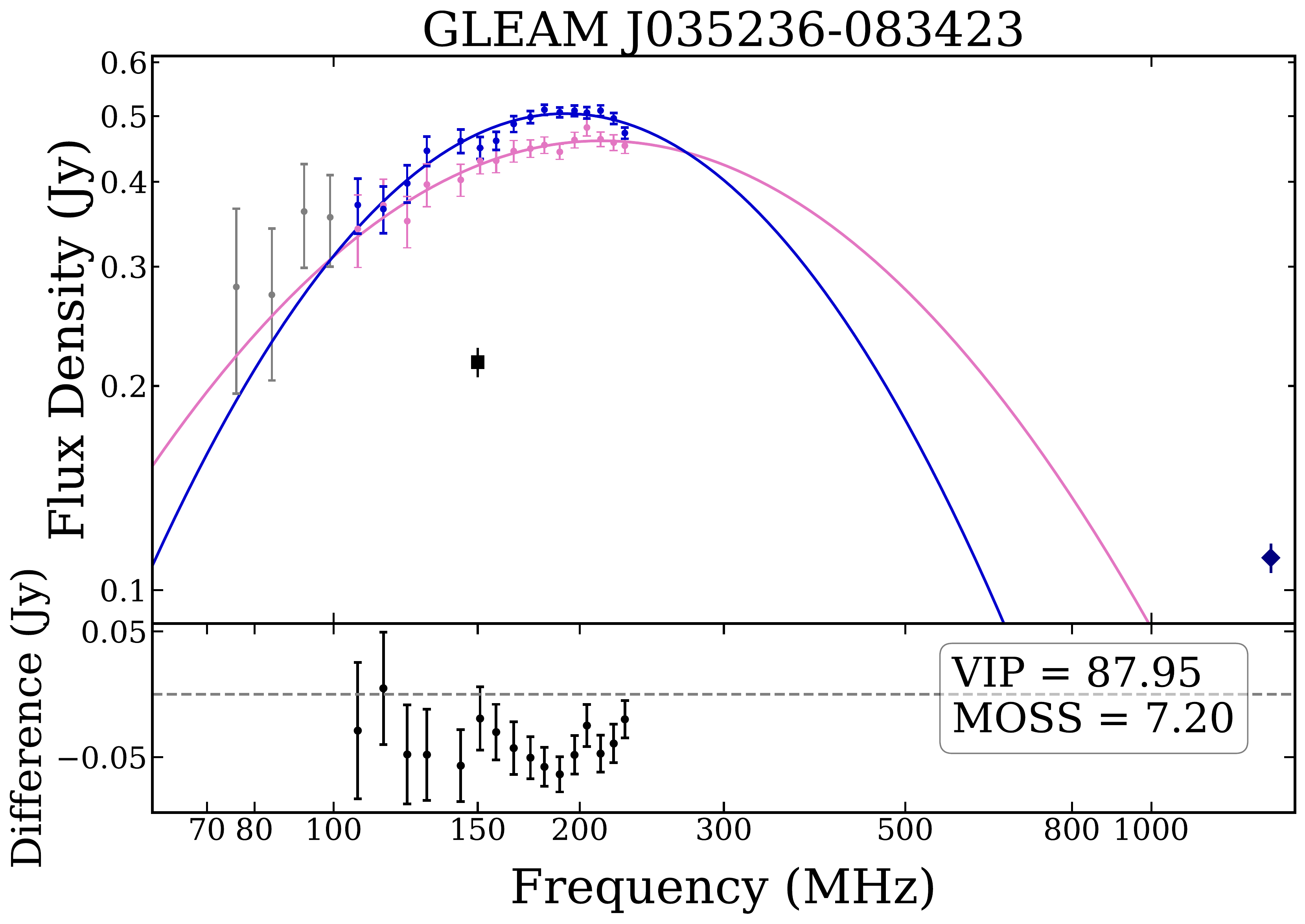} &
\includegraphics[scale=0.15]{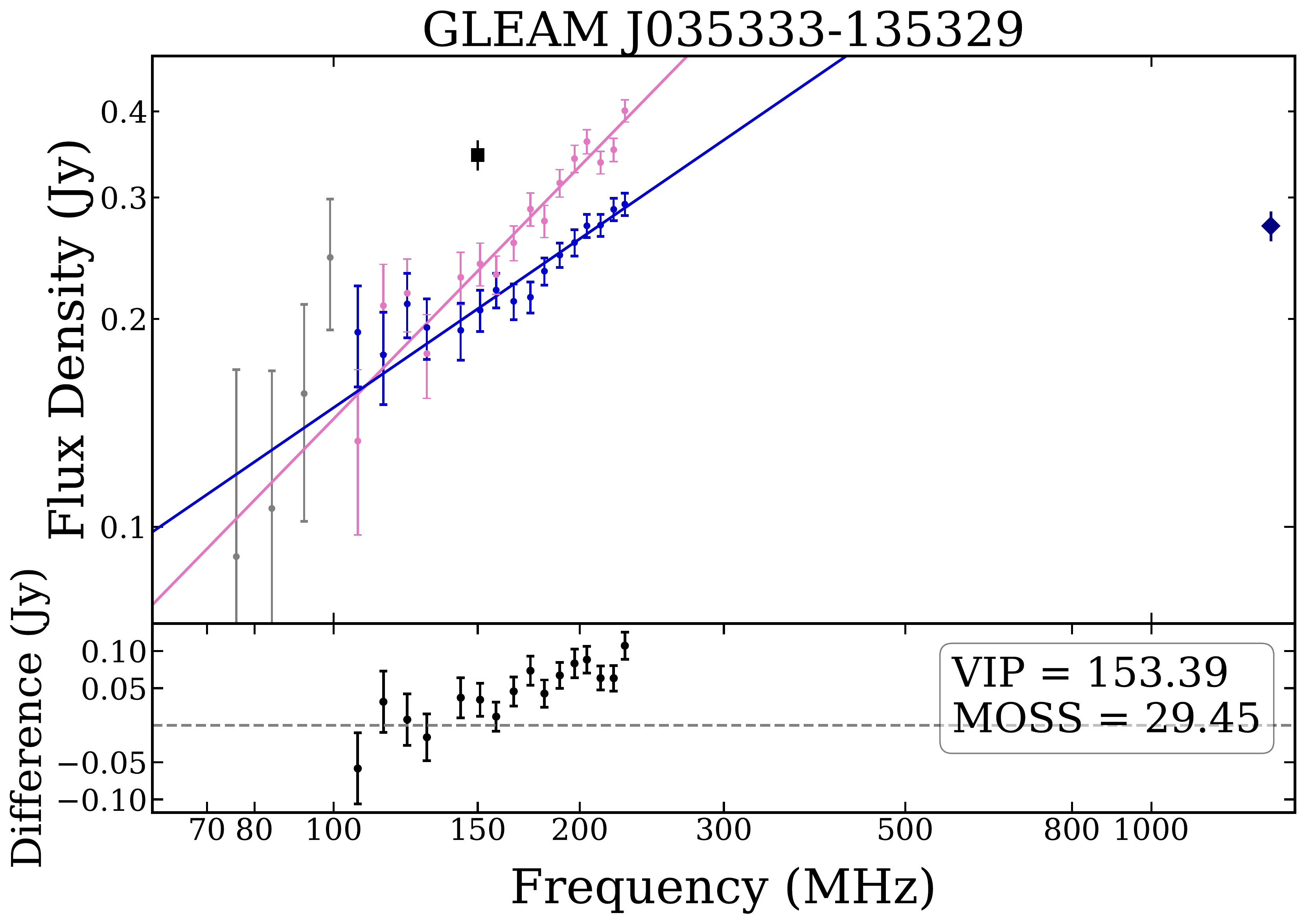} \\
\includegraphics[scale=0.15]{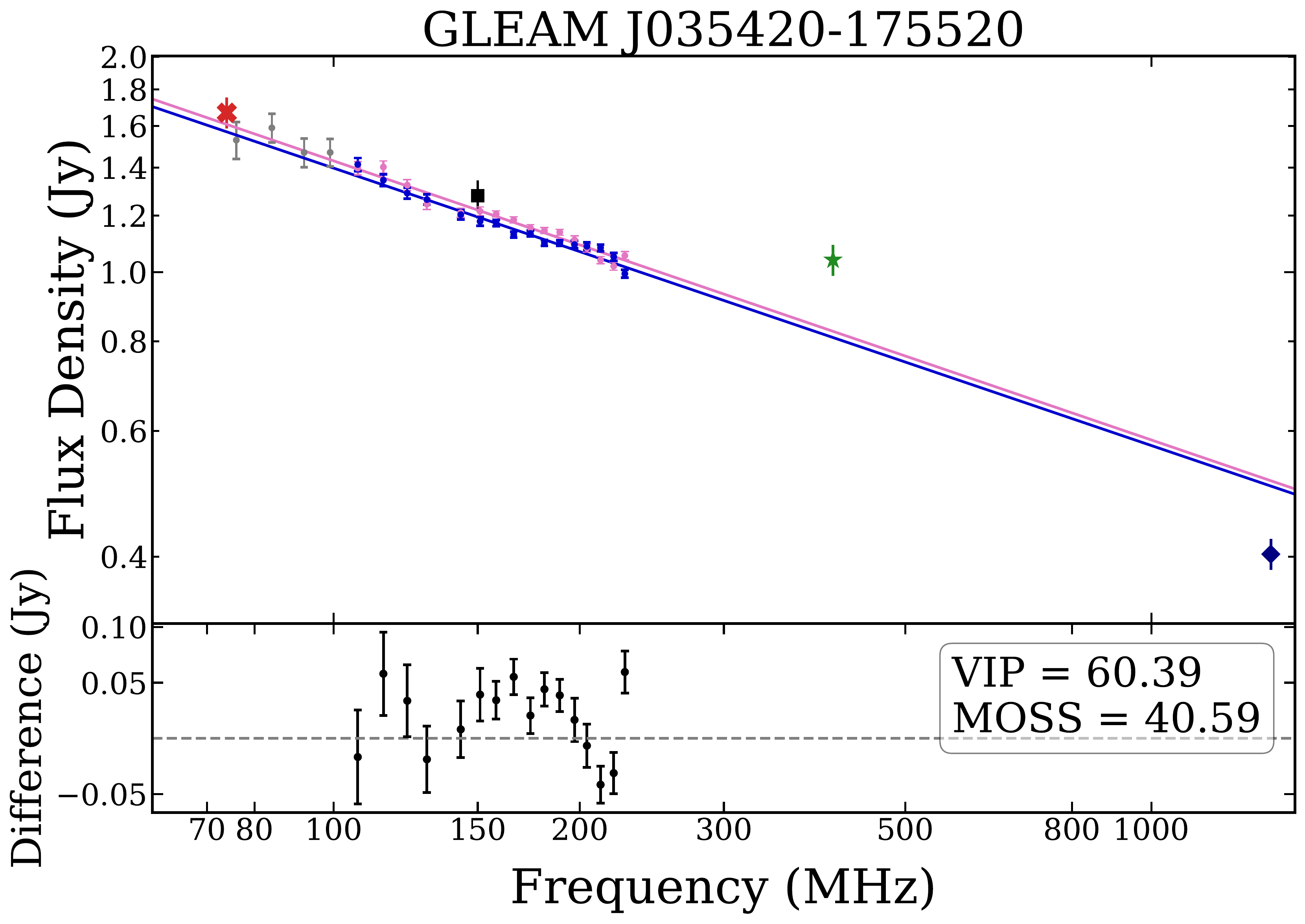} &
\includegraphics[scale=0.15]{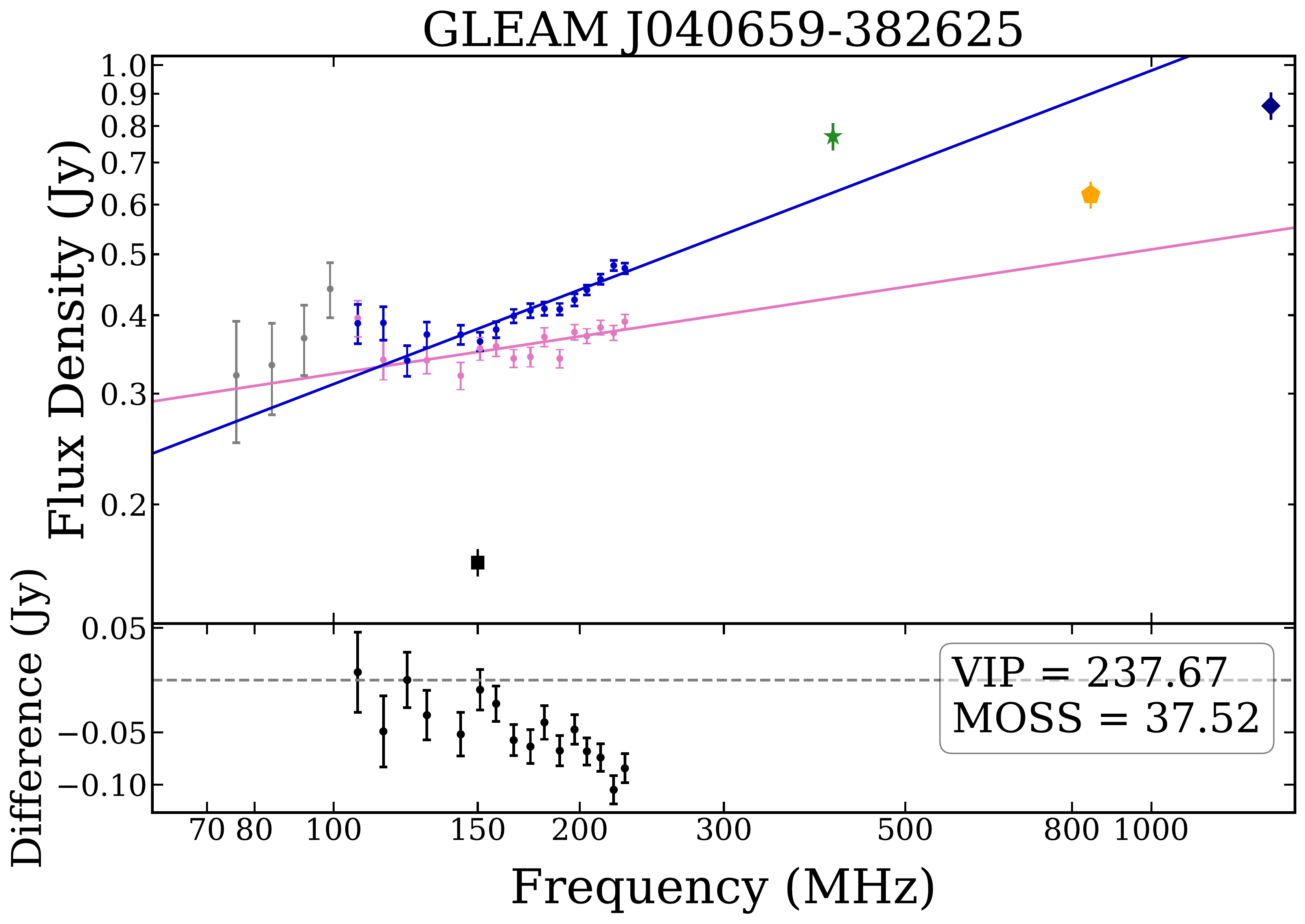} &
\includegraphics[scale=0.15]{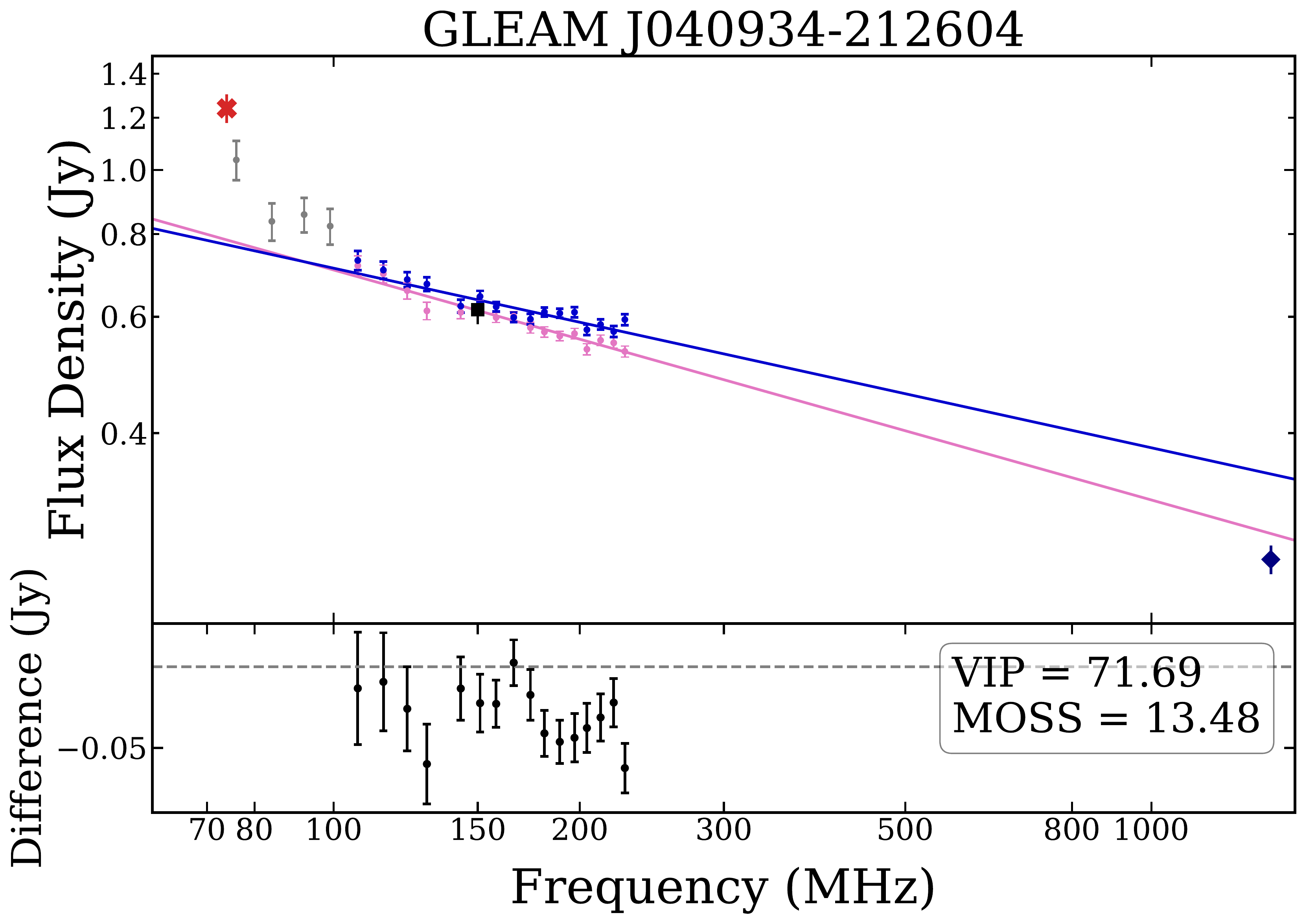} \\
\includegraphics[scale=0.15]{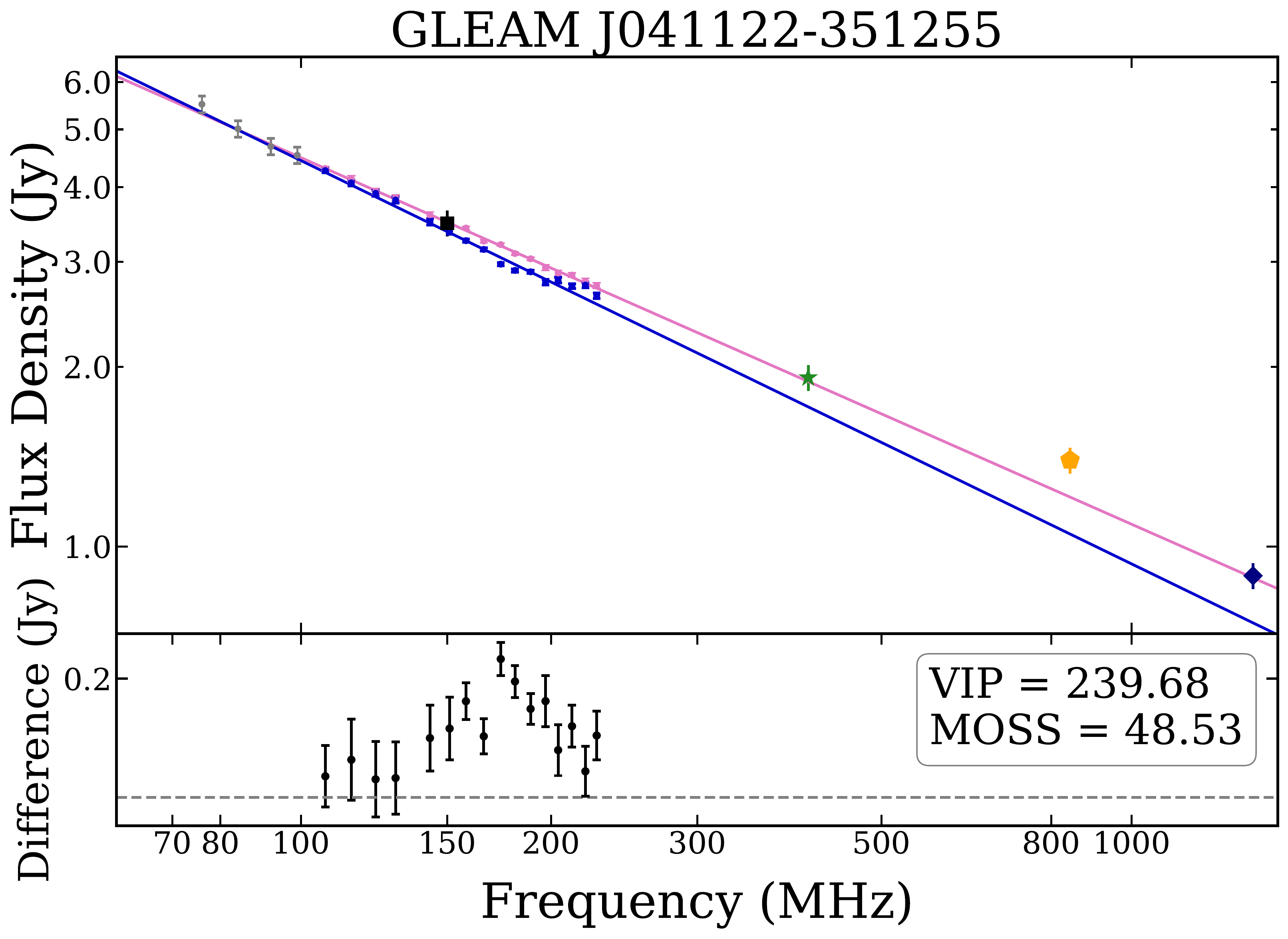} &
\includegraphics[scale=0.15]{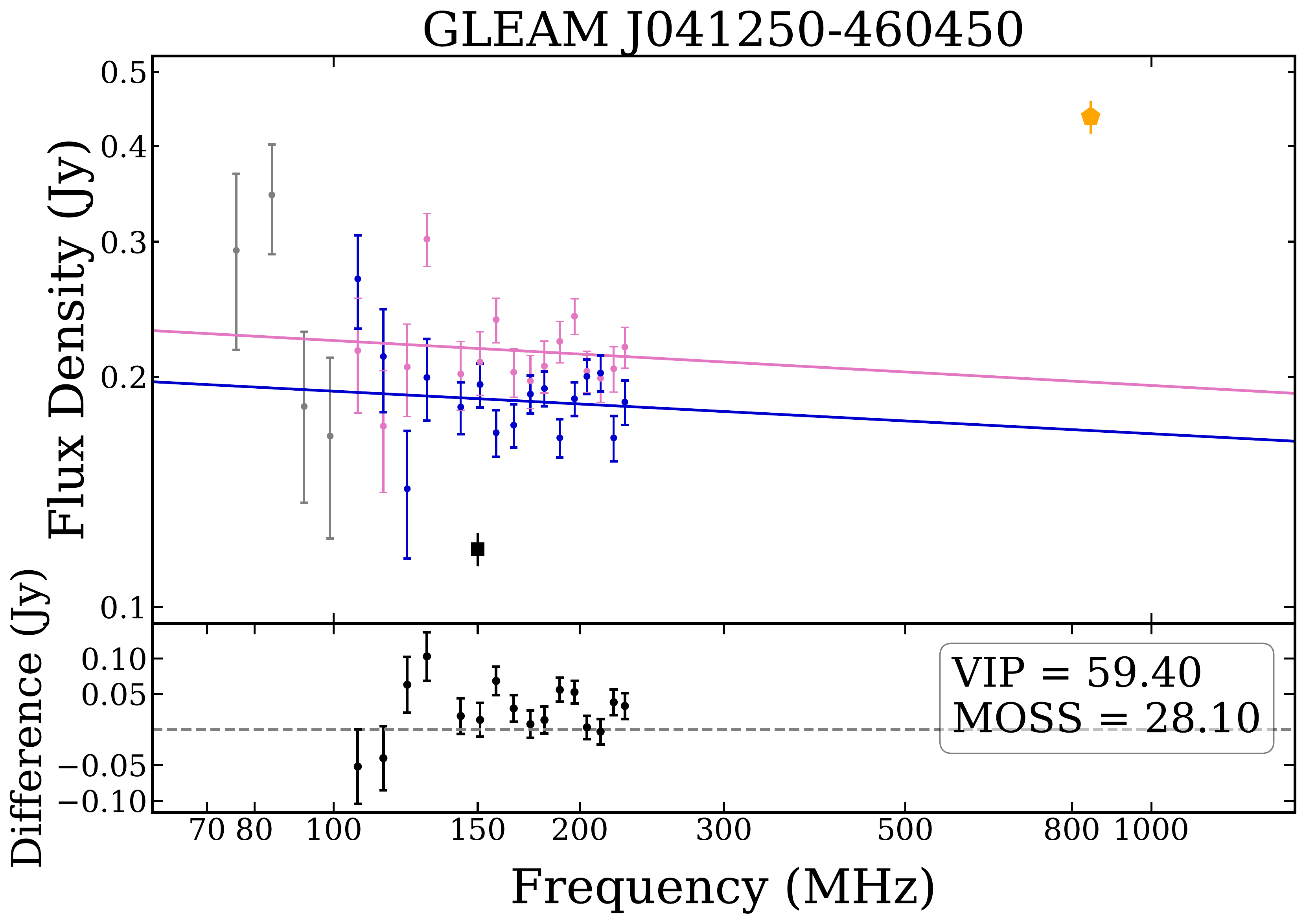} &
\includegraphics[scale=0.15]{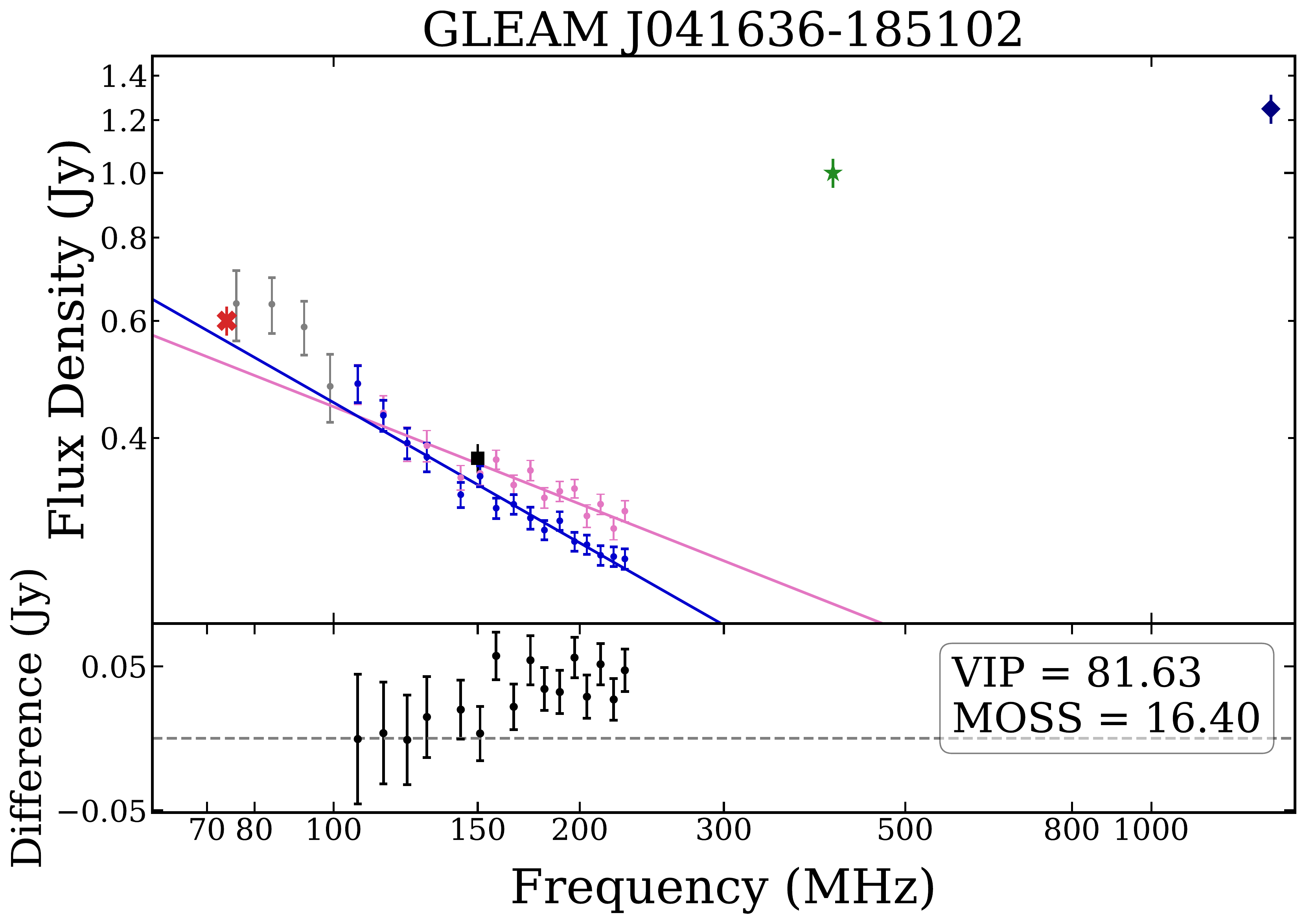} \\
\includegraphics[scale=0.15]{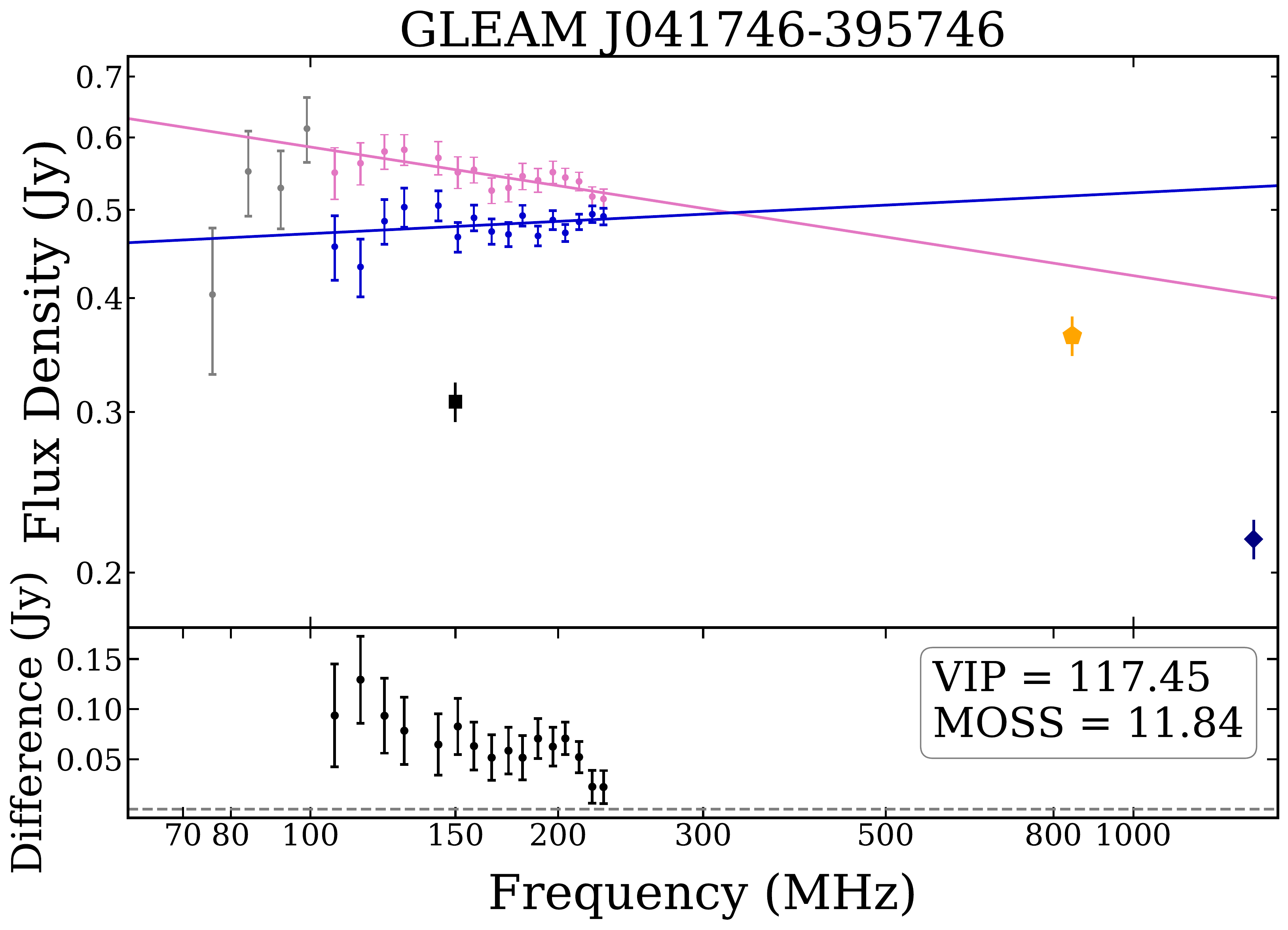} &
\includegraphics[scale=0.15]{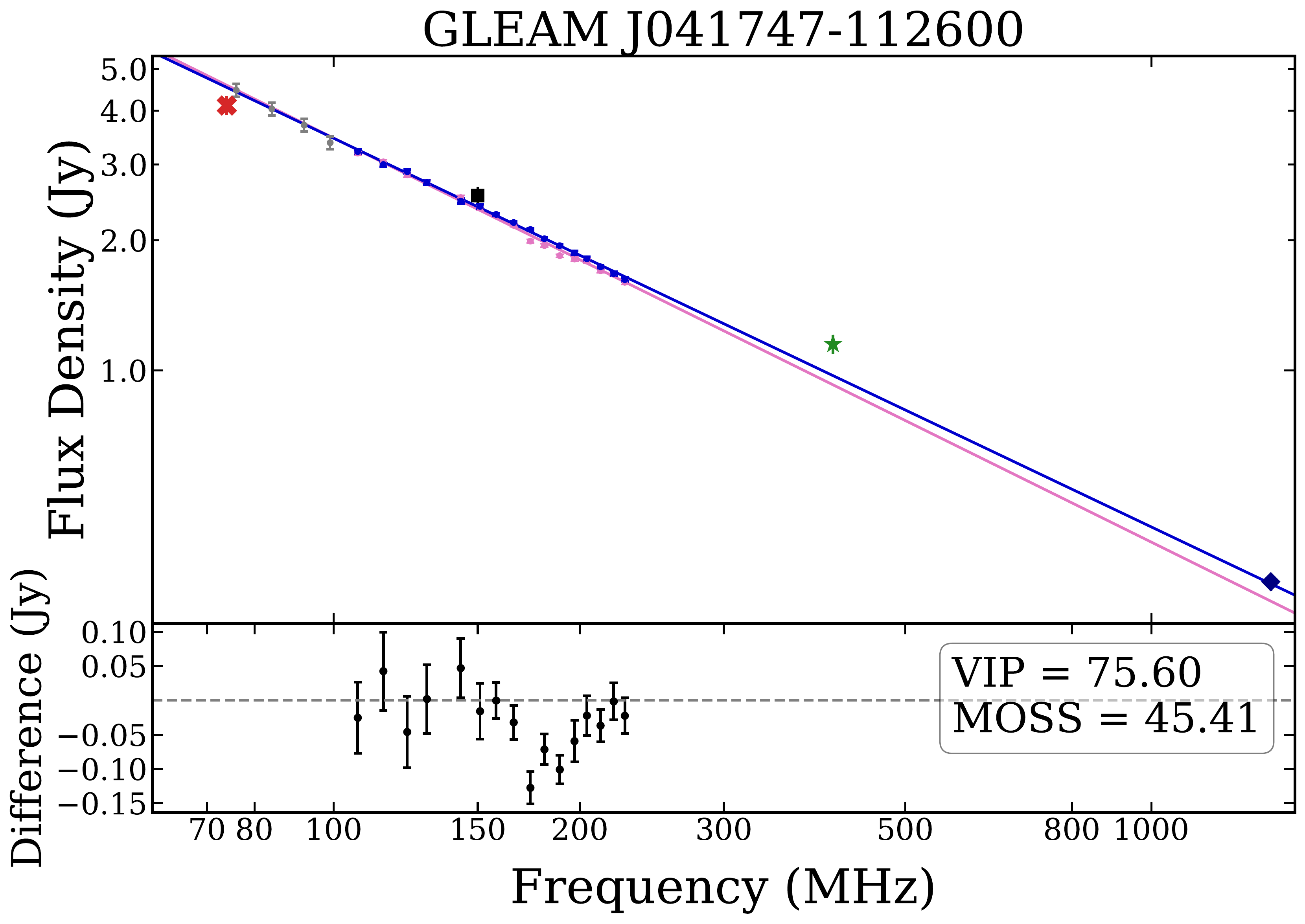} &
\includegraphics[scale=0.15]{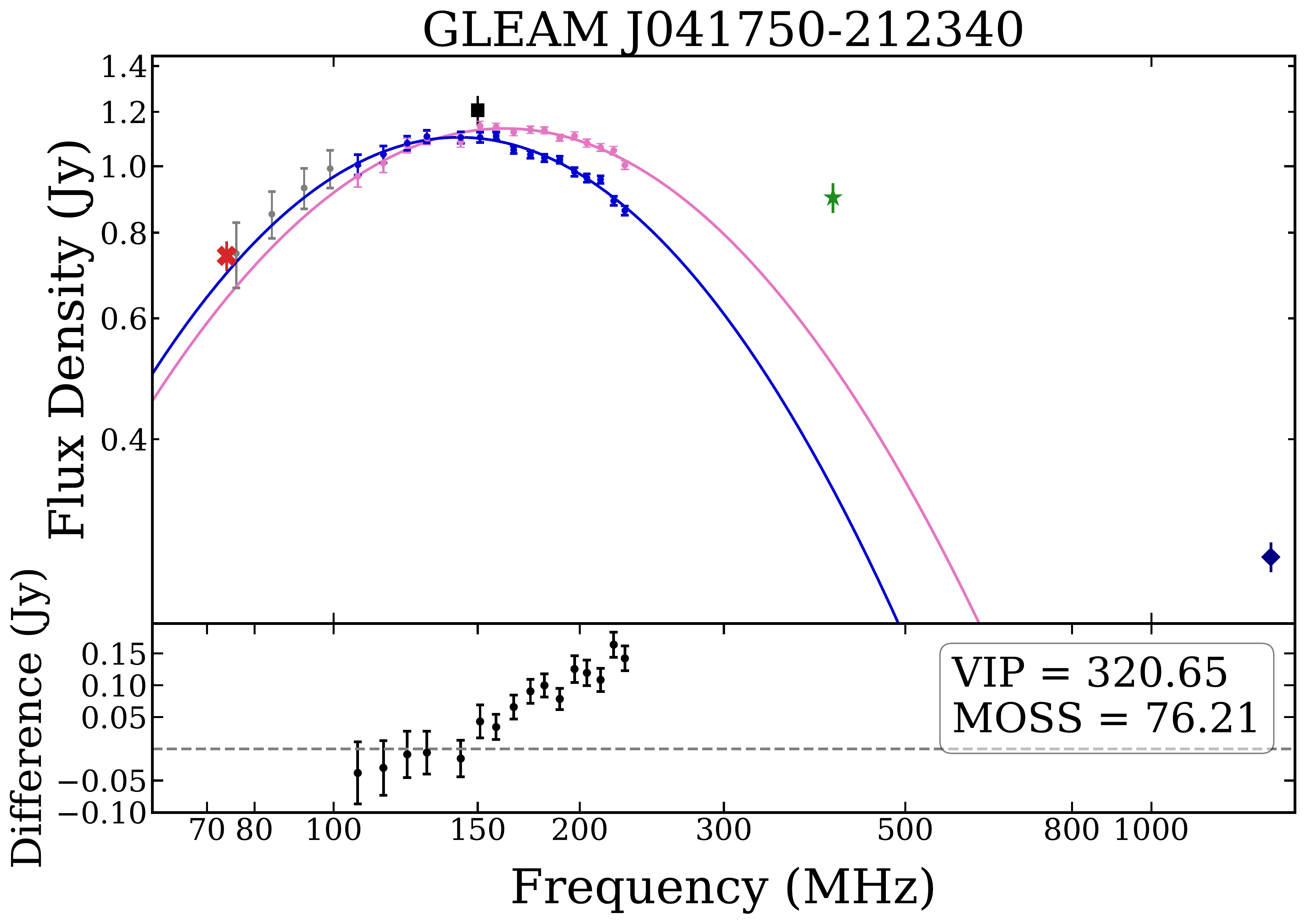} \\
\end{array}$
\caption{(continued) SEDs for all sources classified as variable according to the VIP. For each source the points represent the following data: GLEAM low frequency (72--100\,MHz) (grey circles), Year 1 (pink circles), Year 2 (blue circles), VLSSr (red cross), TGSS (black square), MRC (green star), SUMSS (yellow pentagon), and NVSS (navy diamond). The models for each year are determined by their classification; a source classified with a peak within the observed band was modelled by a quadratic according to Equation~\ref{eq:quadratic}, remaining sources were modelled by a power-law according to Equation~\ref{eq:plaw}.}
\label{app:fig:pg10}
\end{center}
\end{figure*}
\setcounter{figure}{0}
\begin{figure*}
\begin{center}$
\begin{array}{cccccc}
\includegraphics[scale=0.15]{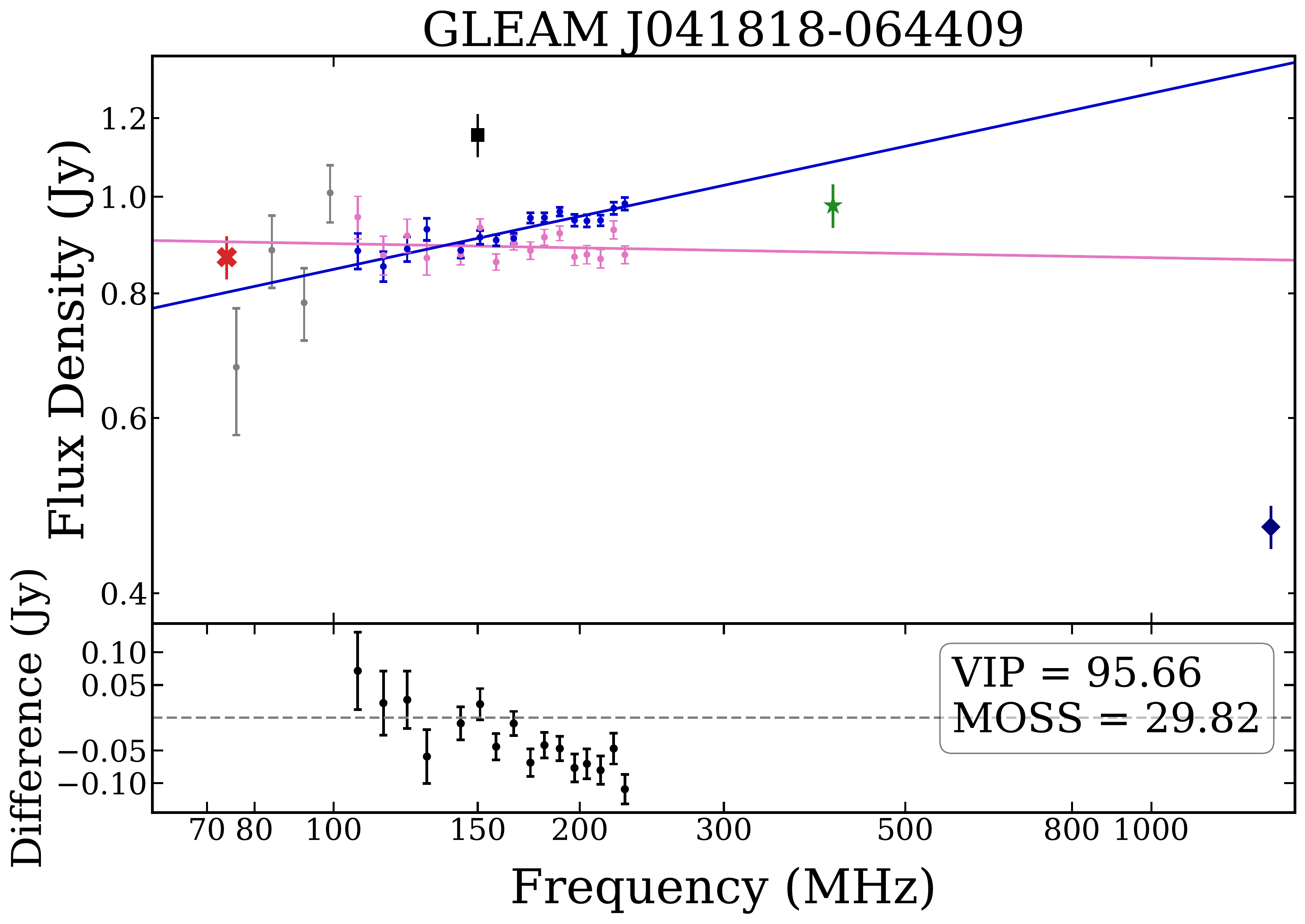} &
\includegraphics[scale=0.15]{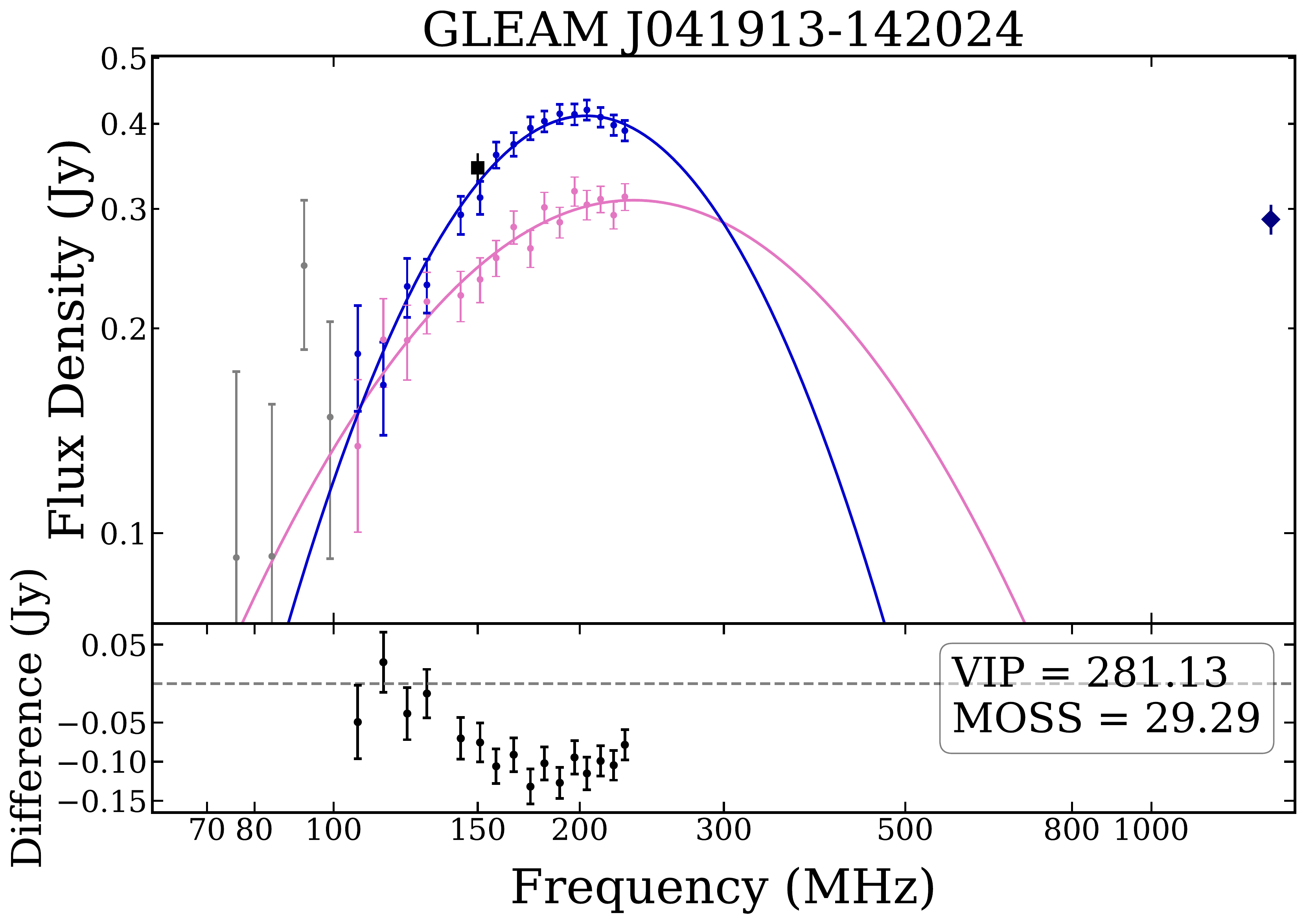} &
\includegraphics[scale=0.15]{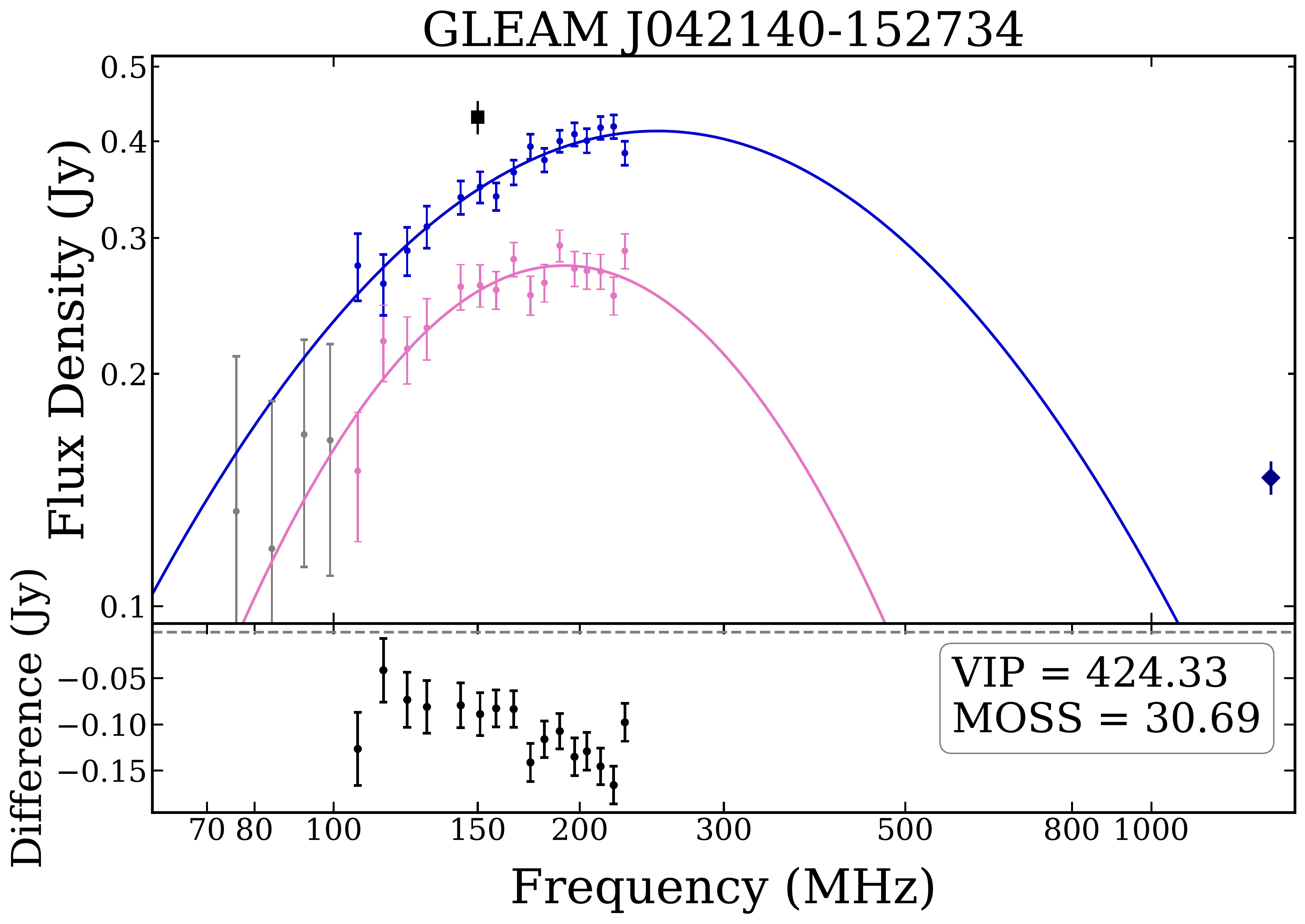} \\
\includegraphics[scale=0.15]{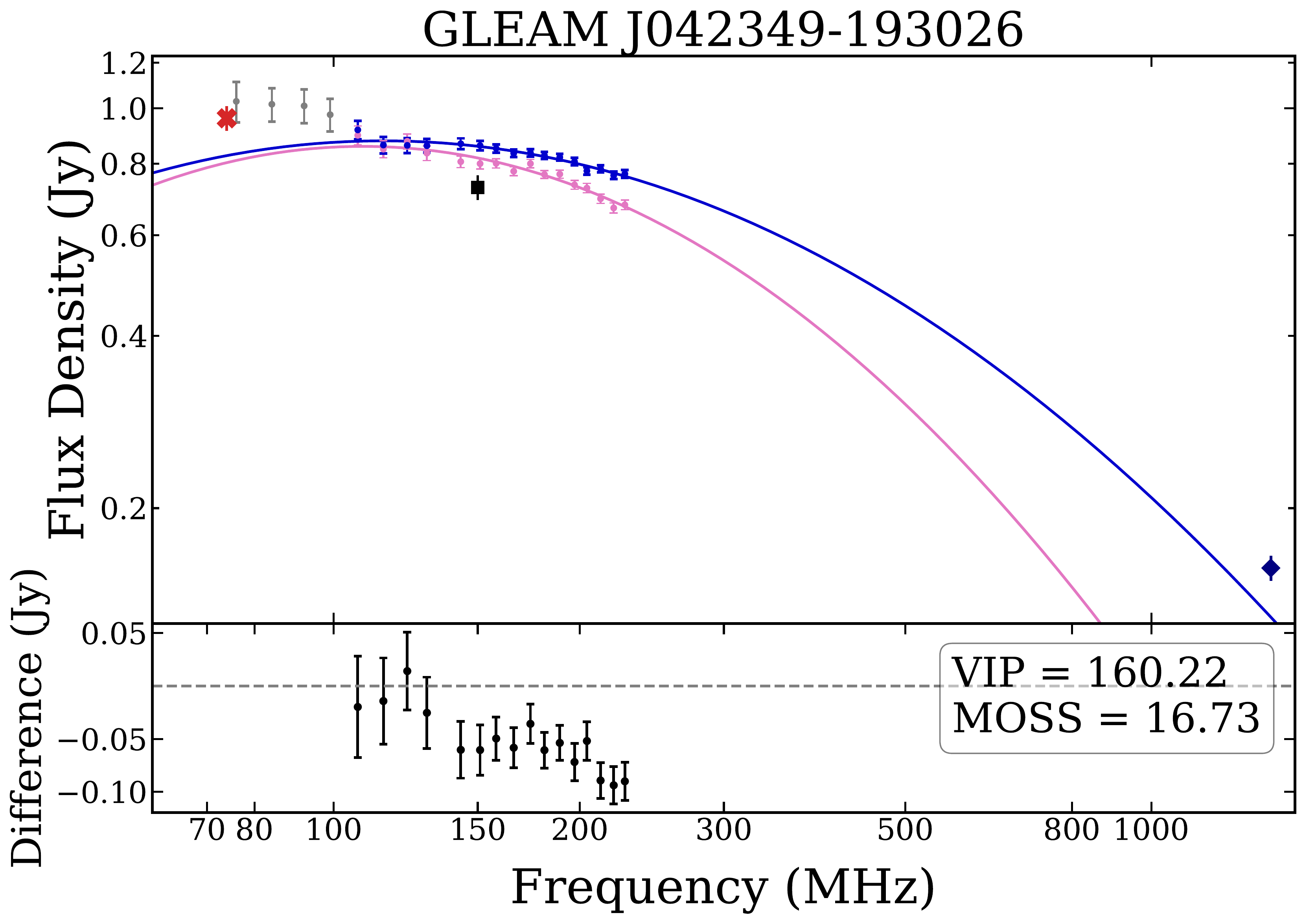} &
\includegraphics[scale=0.15]{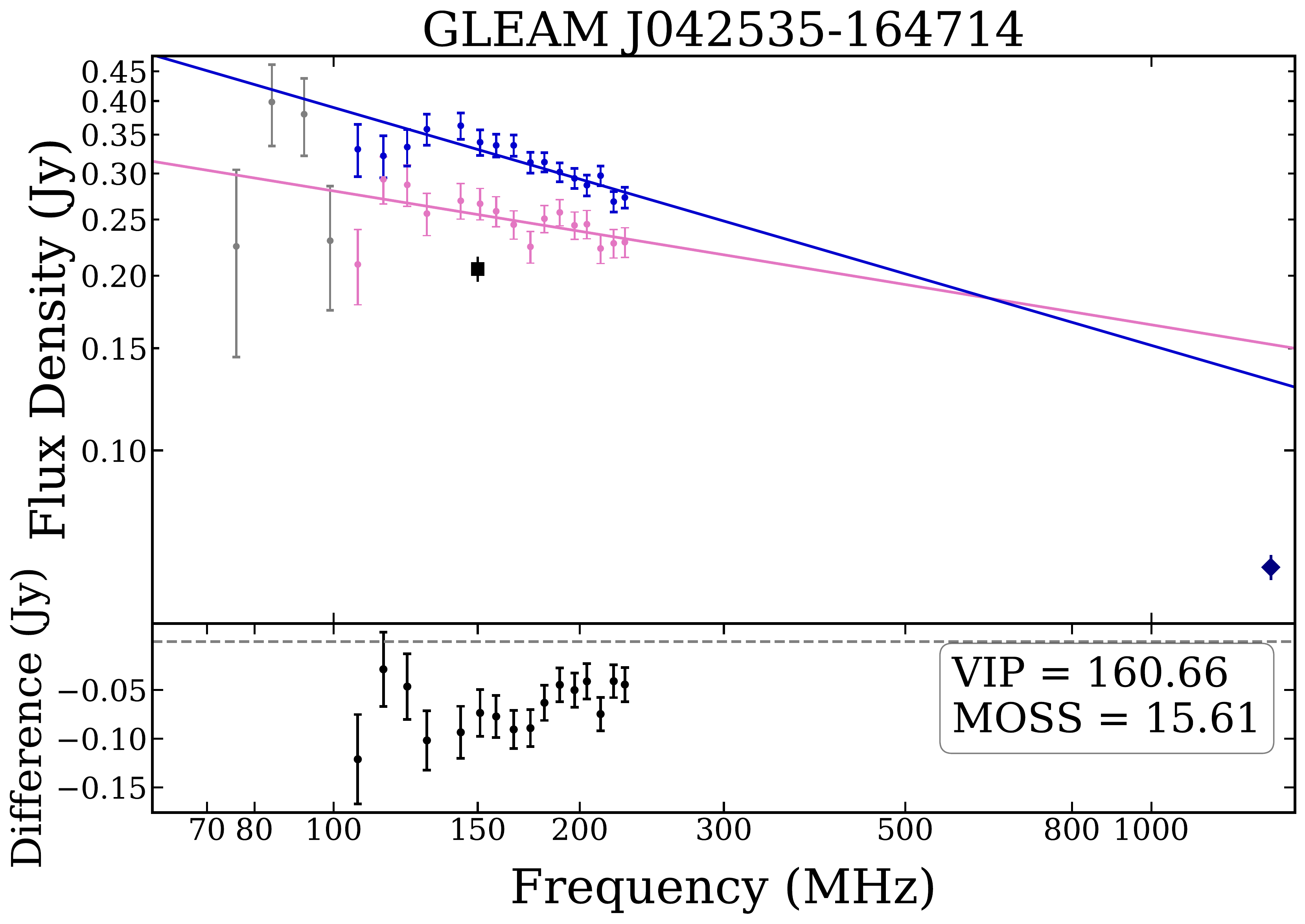} &
\includegraphics[scale=0.15]{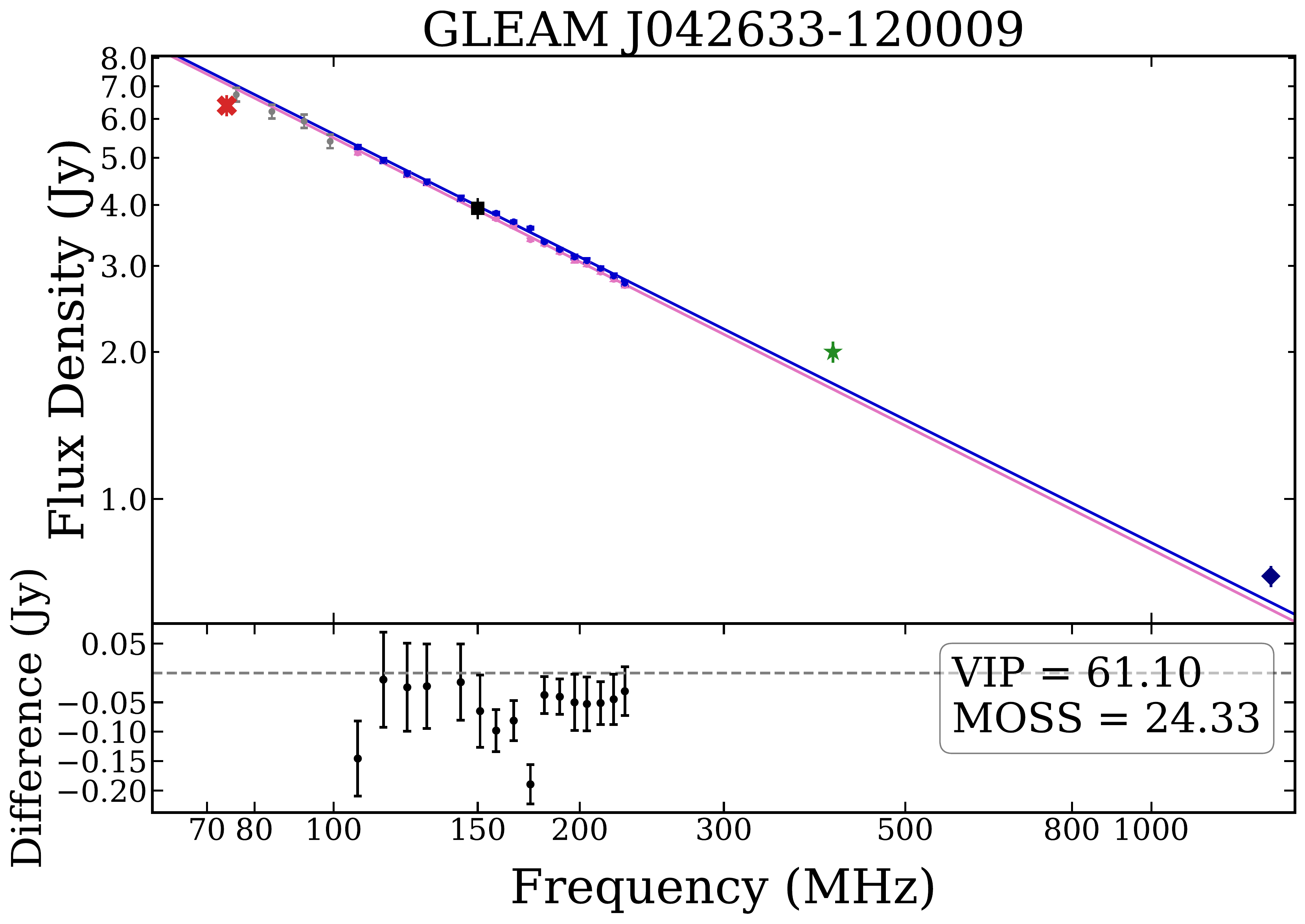} \\
\includegraphics[scale=0.15]{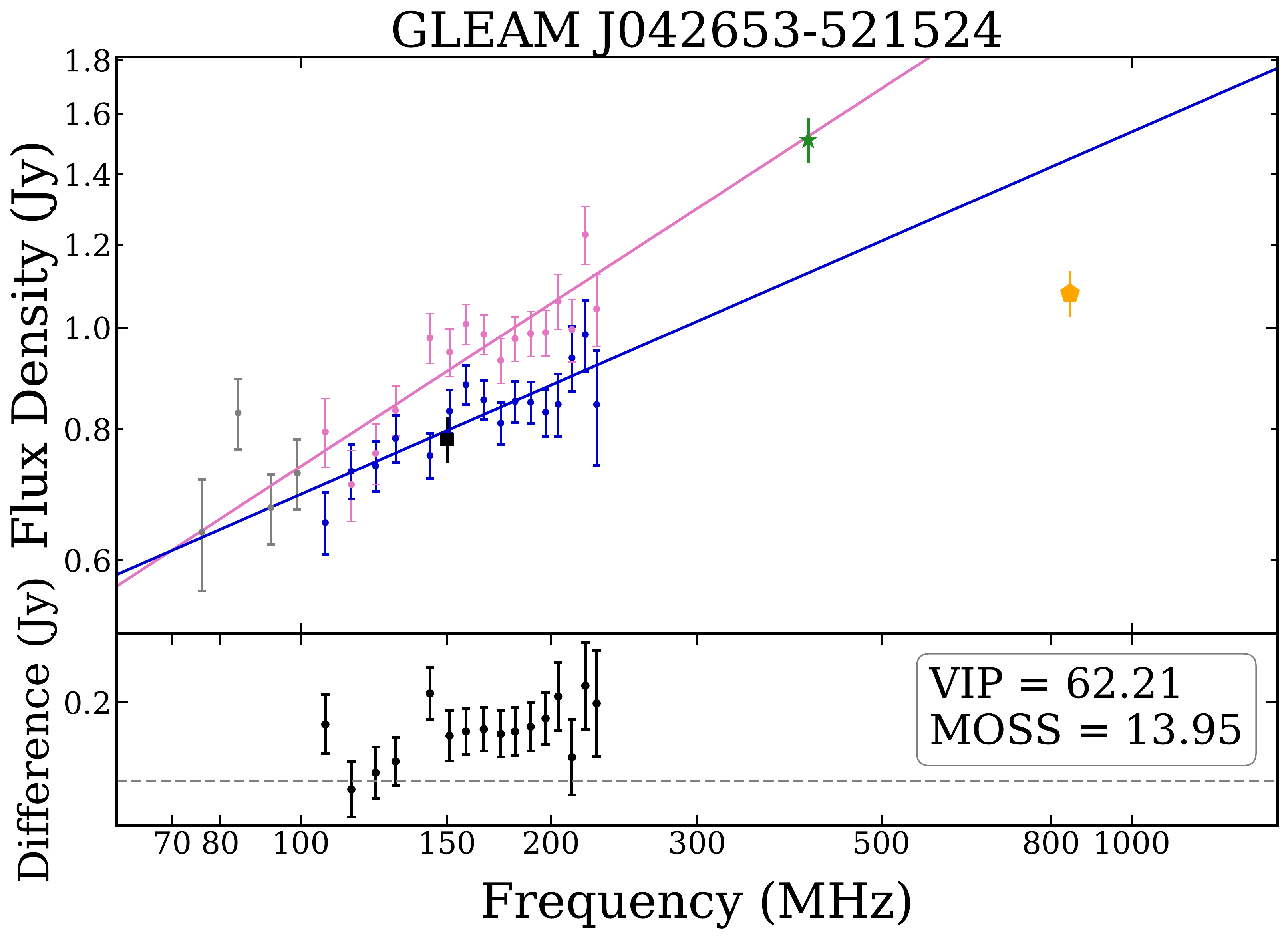} &
\includegraphics[scale=0.15]{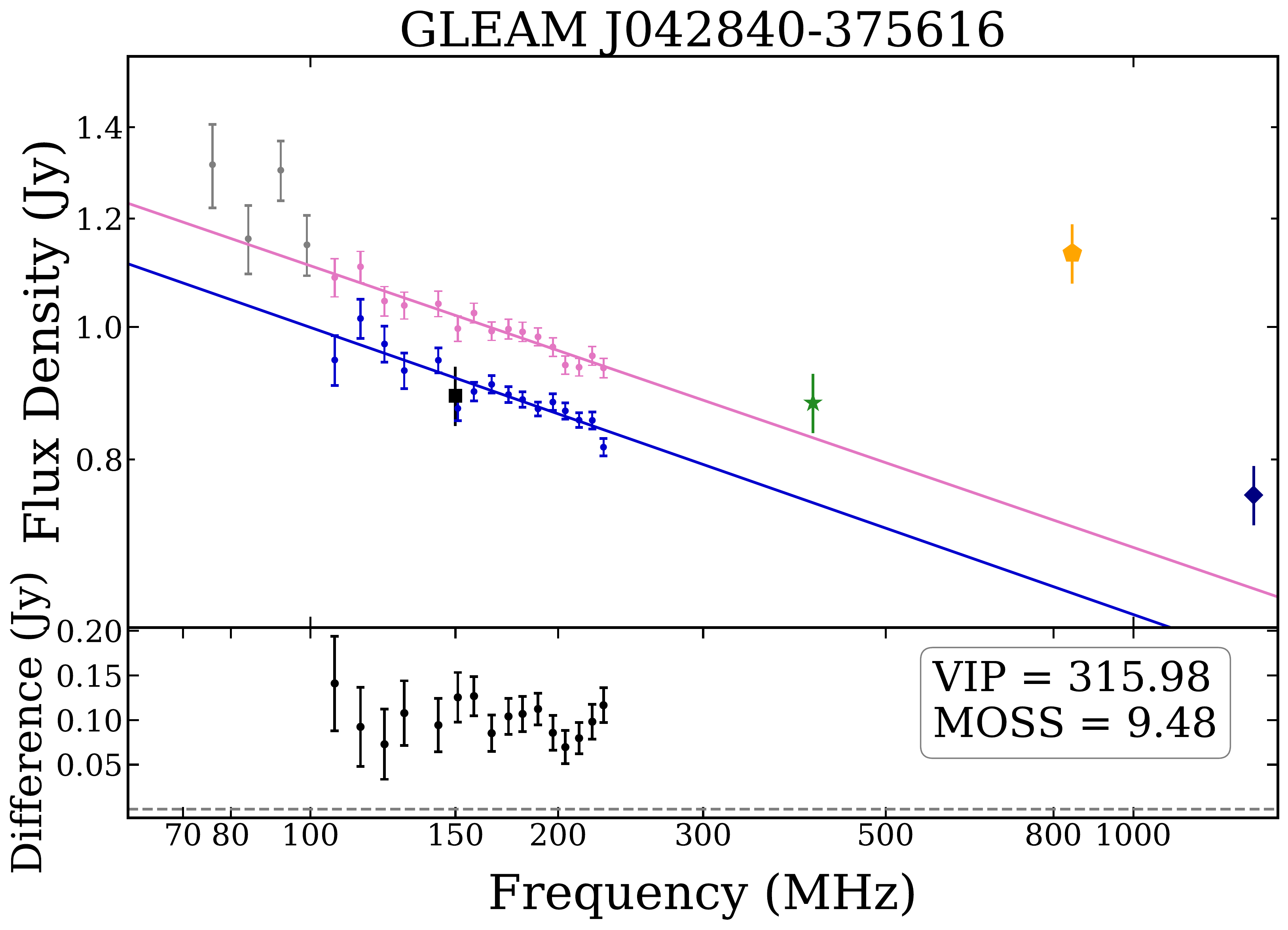} &
\includegraphics[scale=0.15]{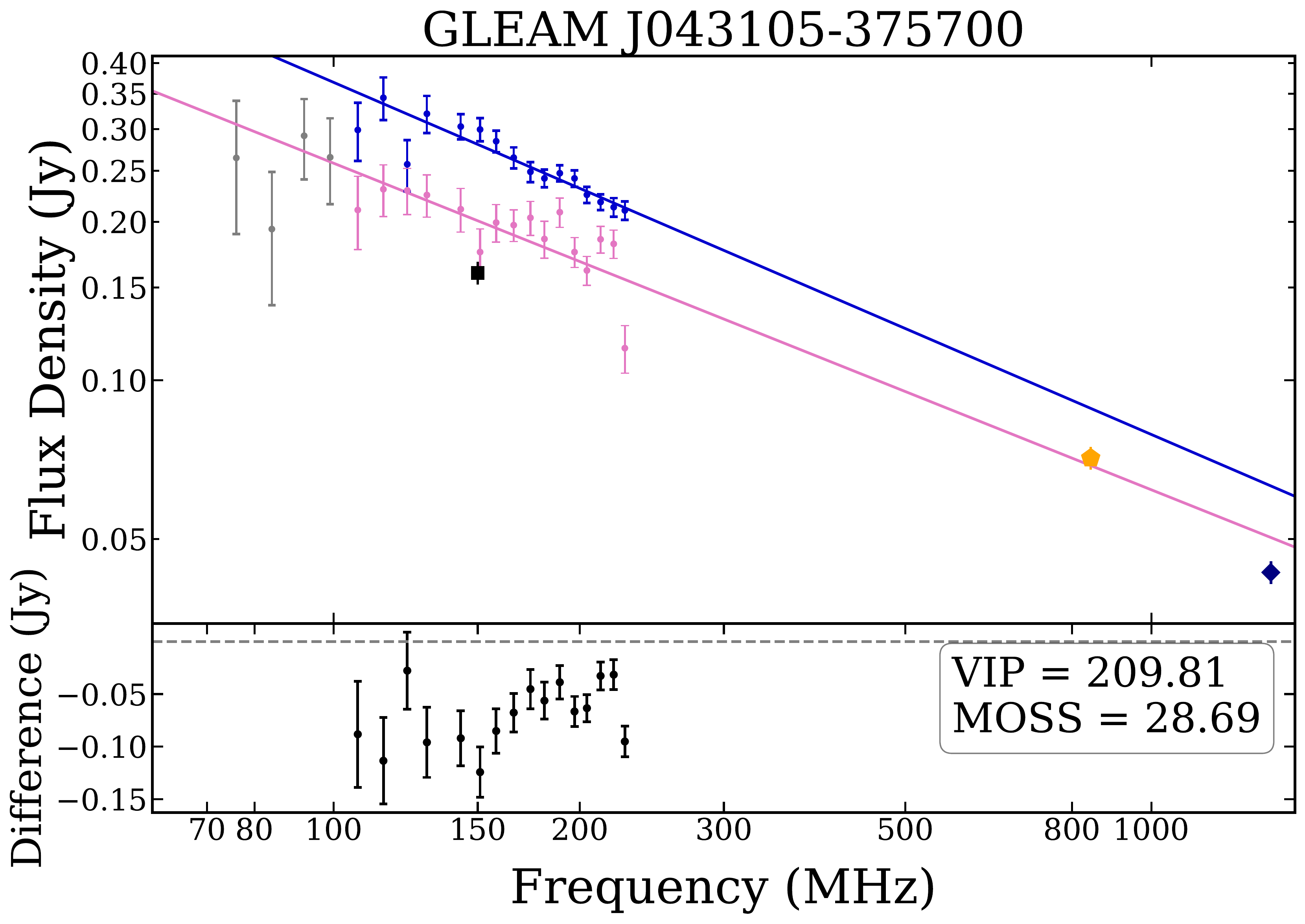} \\
\includegraphics[scale=0.15]{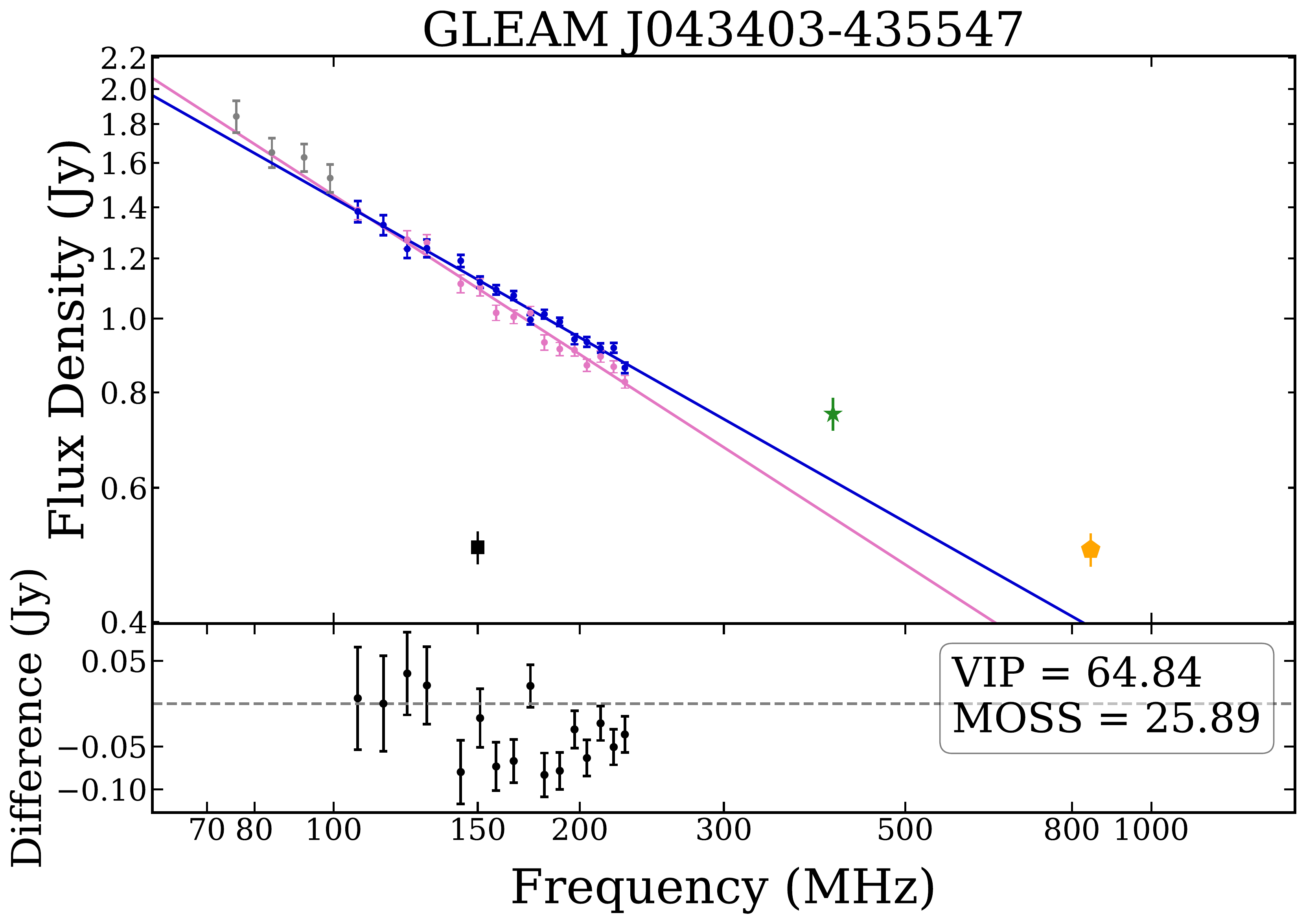} &
\includegraphics[scale=0.15]{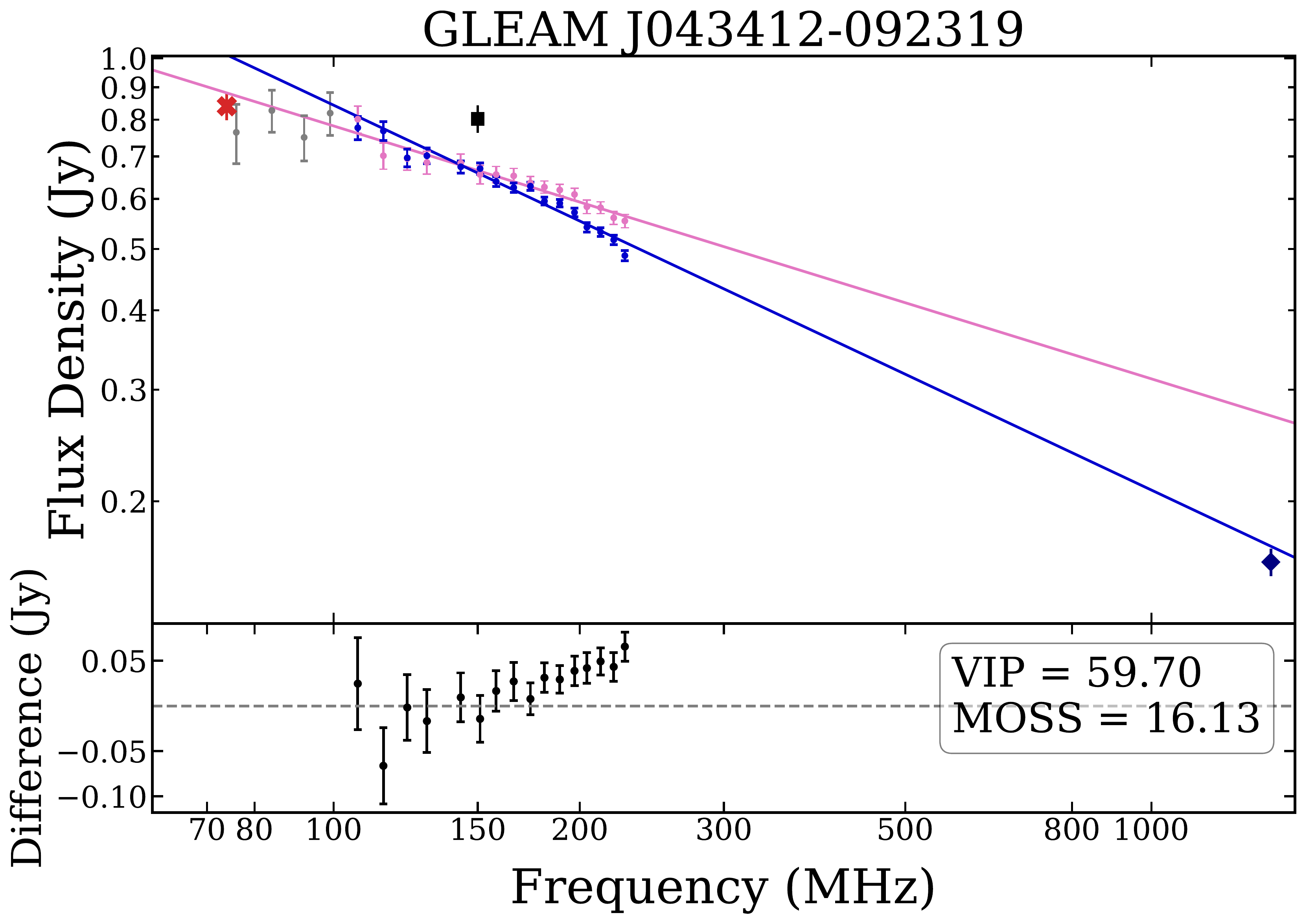} &
\includegraphics[scale=0.15]{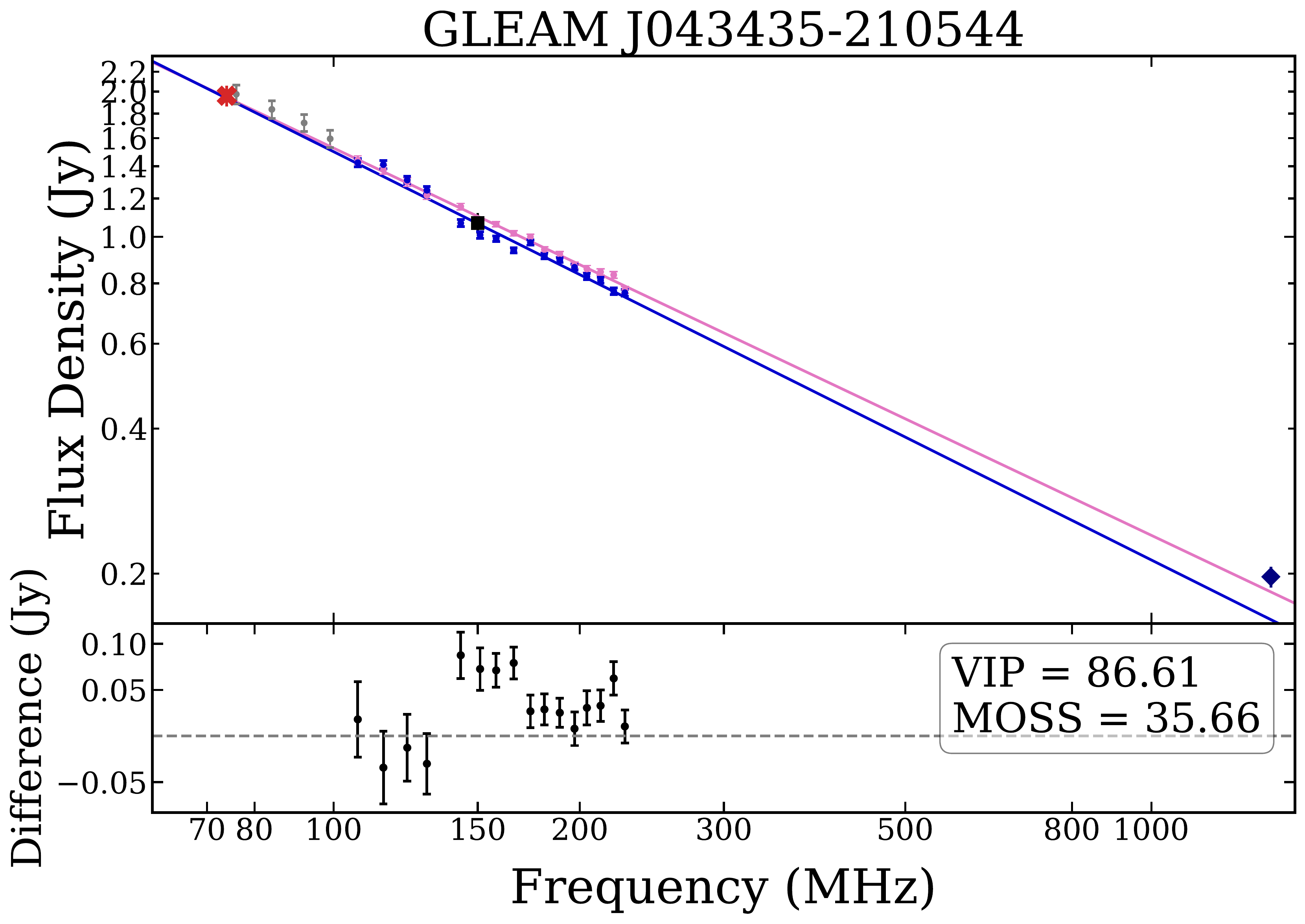} \\
\includegraphics[scale=0.15]{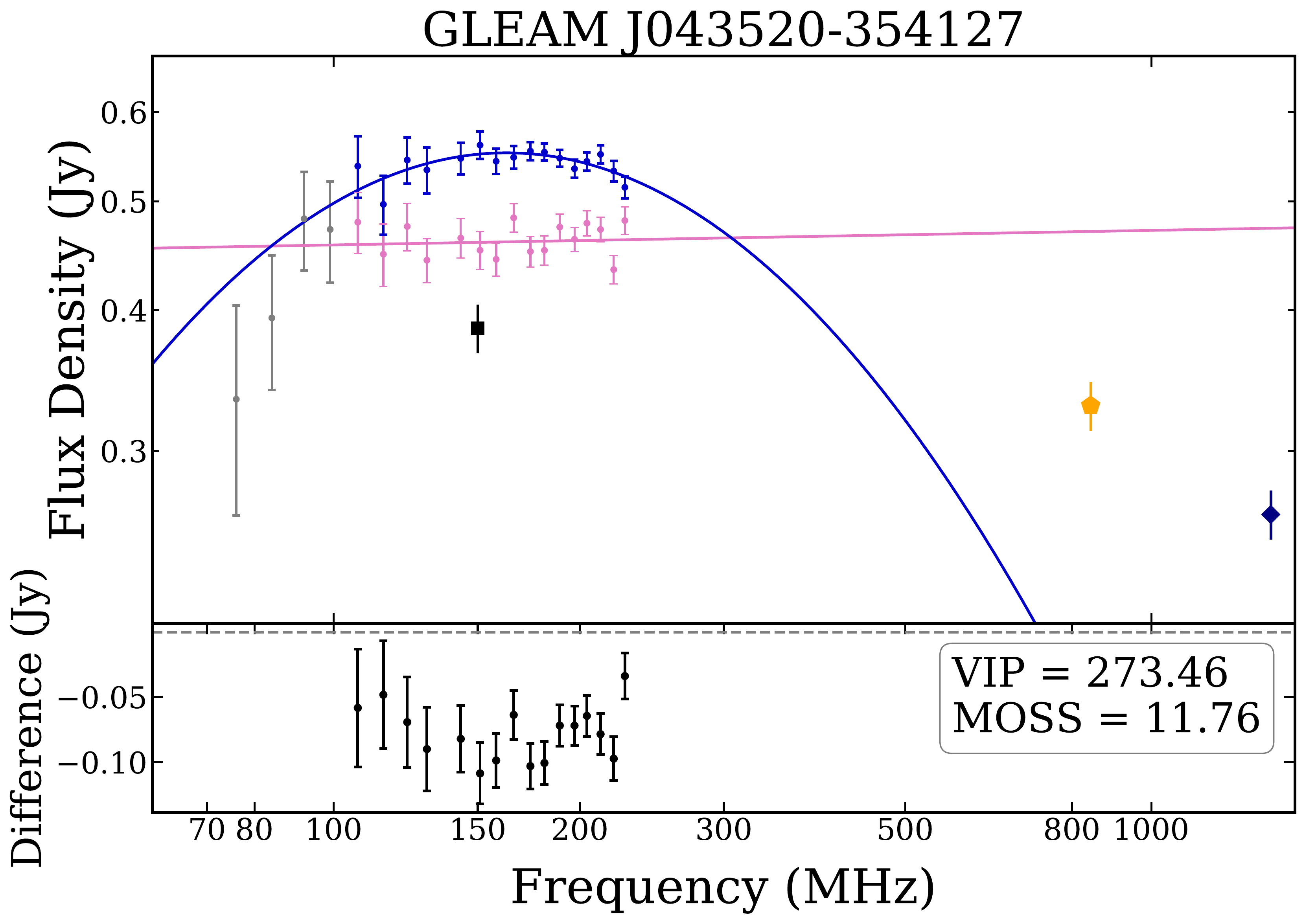} &
\includegraphics[scale=0.15]{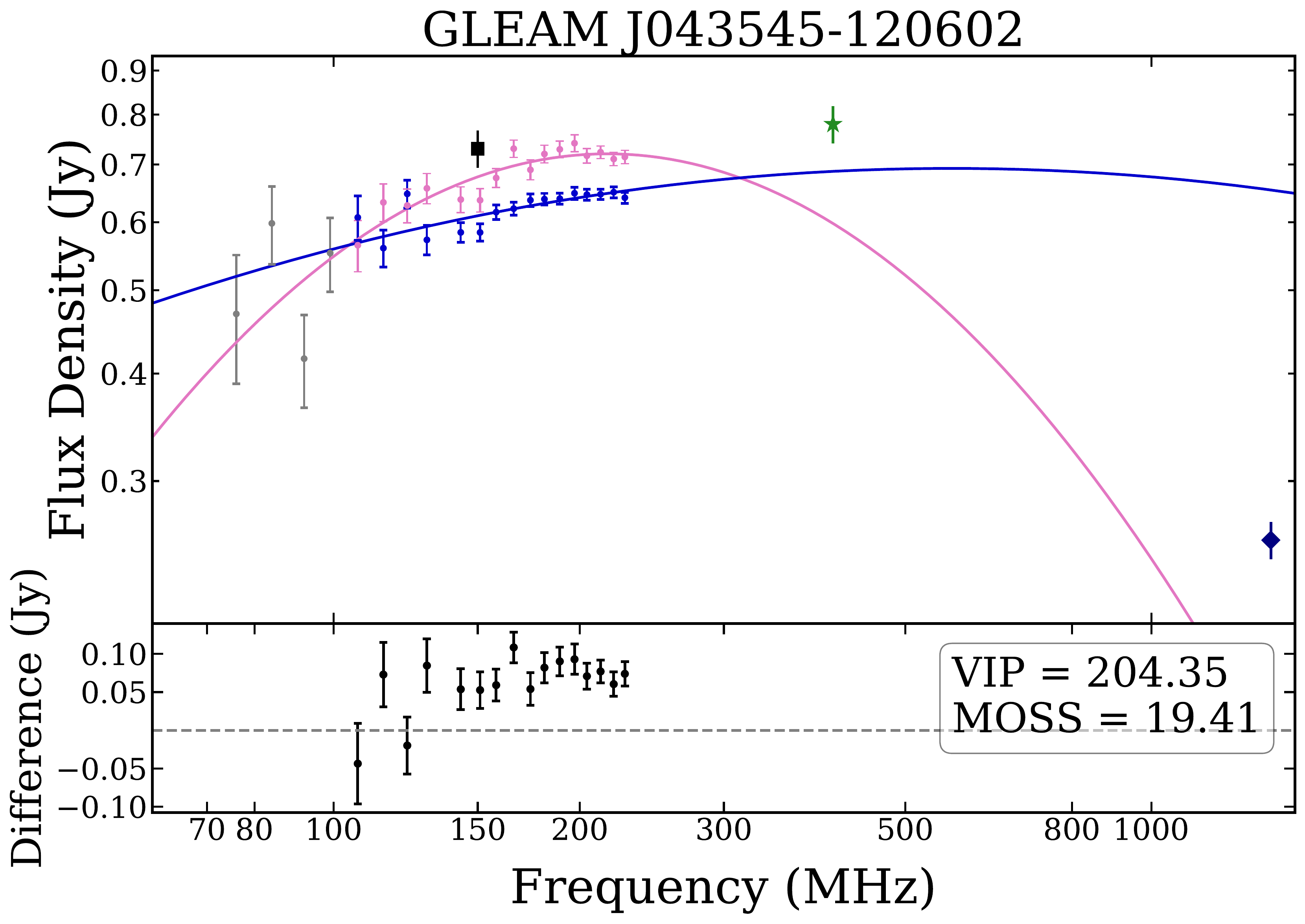} &
\includegraphics[scale=0.15]{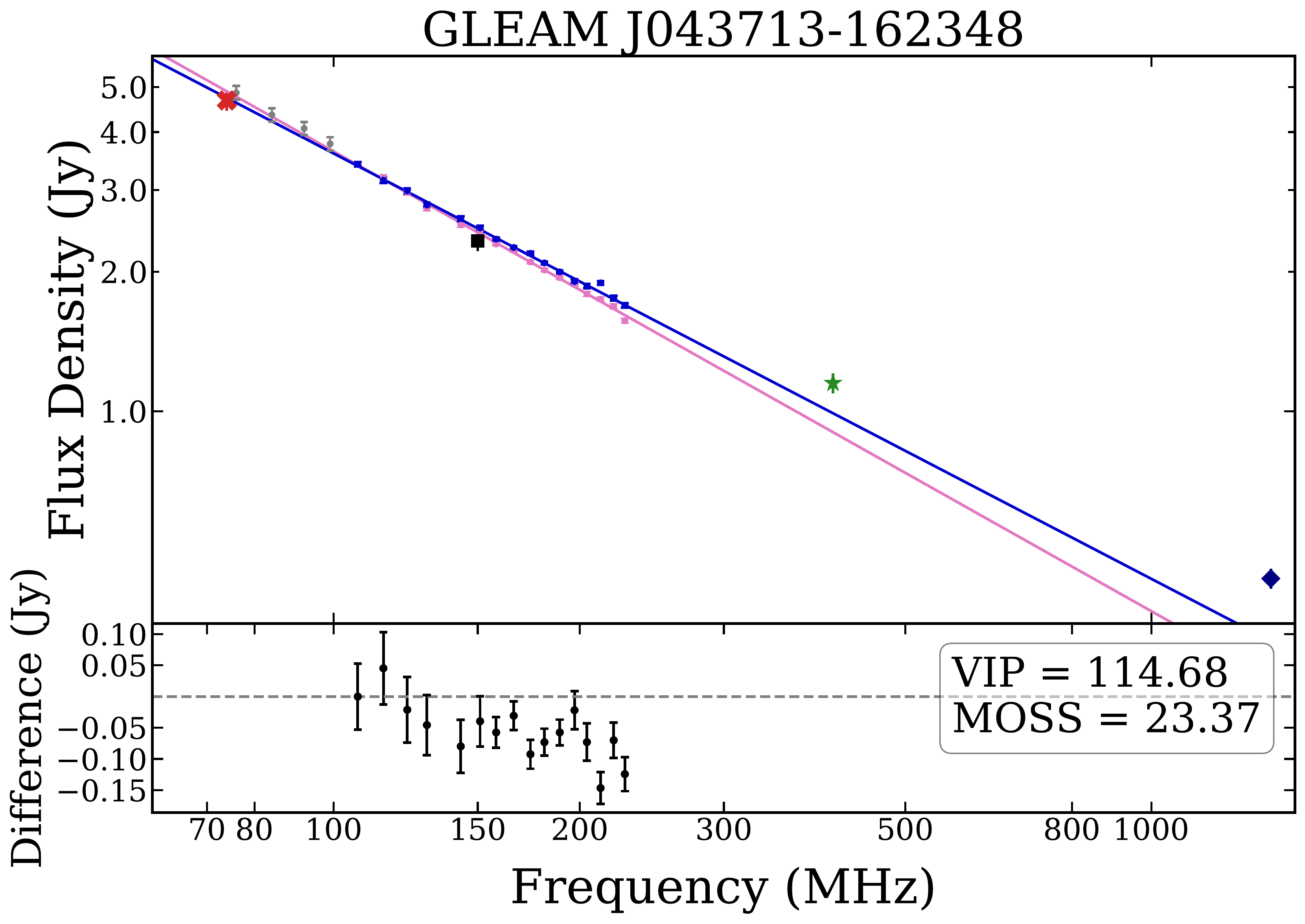} \\
\includegraphics[scale=0.15]{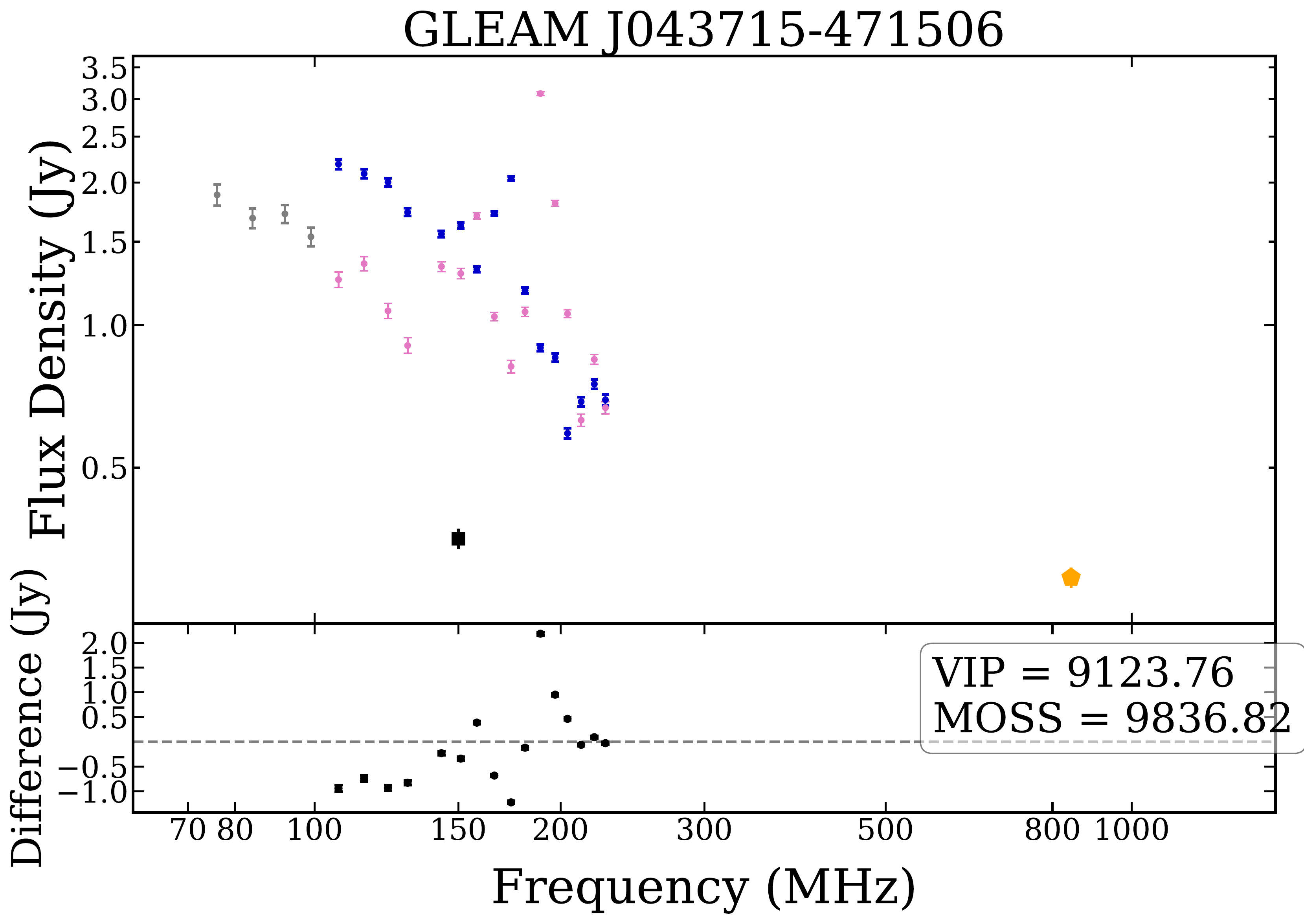} &
\includegraphics[scale=0.15]{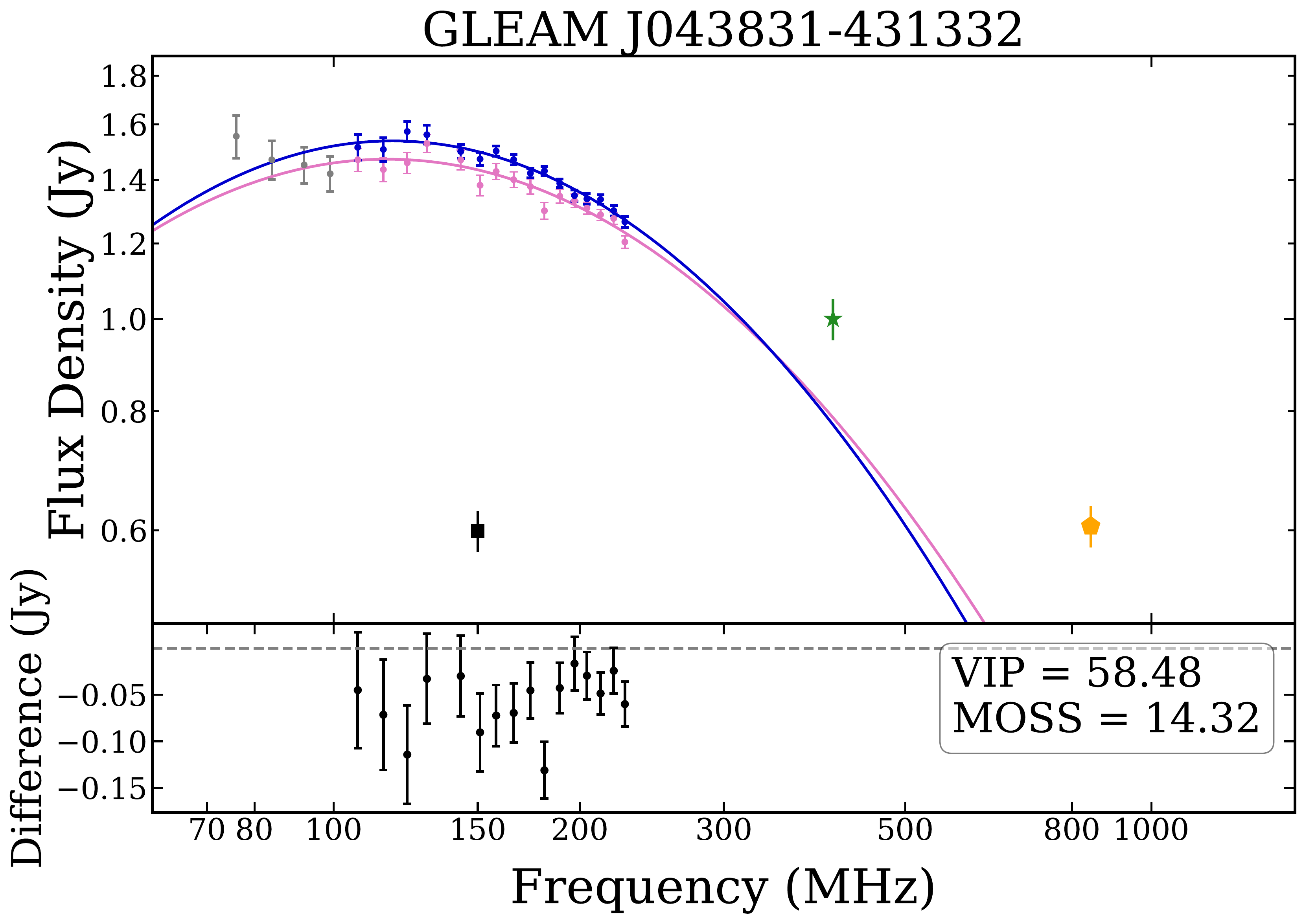} &
\includegraphics[scale=0.15]{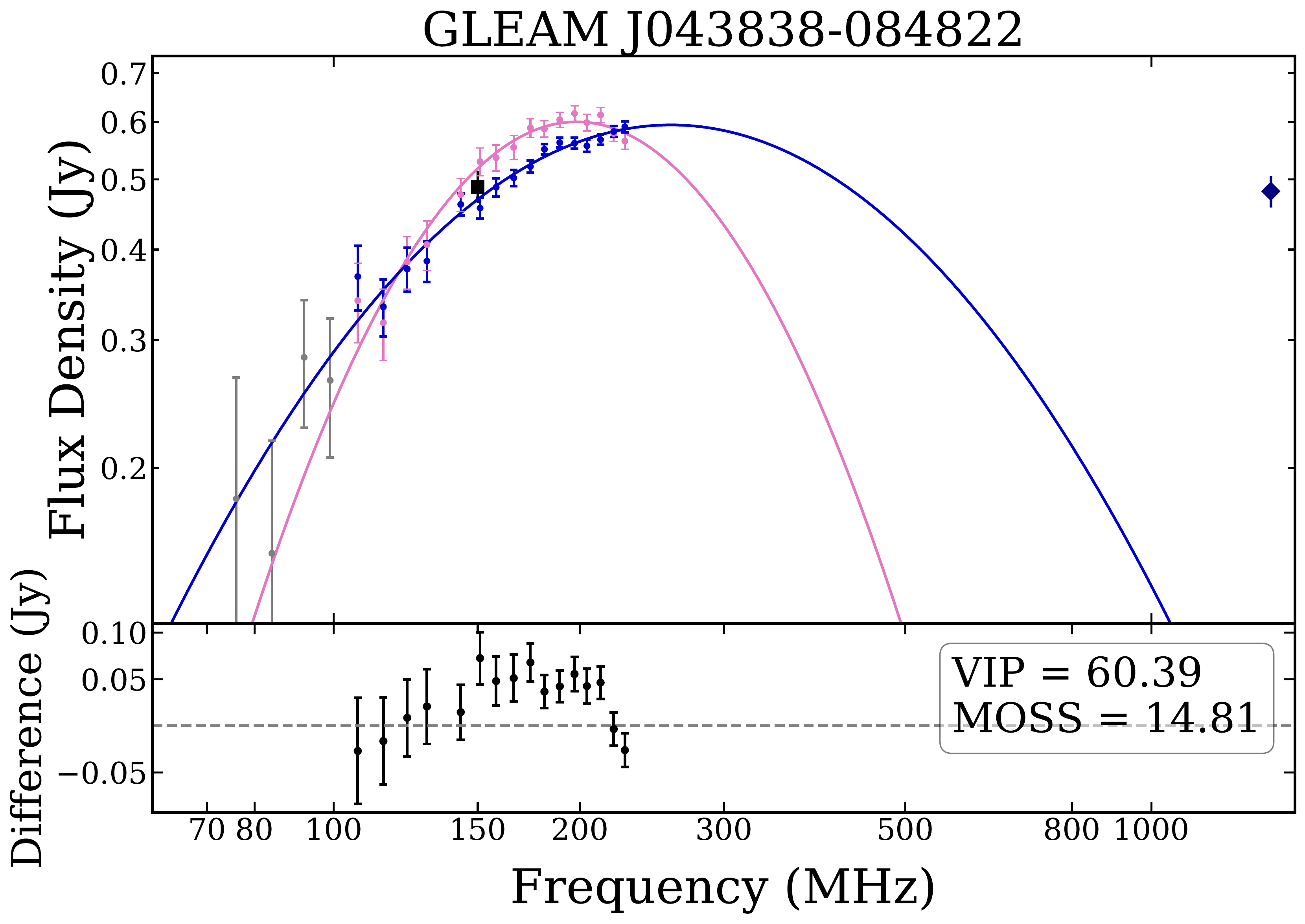} \\
\end{array}$
\caption{(continued) SEDs for all sources classified as variable according to the VIP. For each source the points represent the following data: GLEAM low frequency (72--100\,MHz) (grey circles), Year 1 (pink circles), Year 2 (blue circles), VLSSr (red cross), TGSS (black square), MRC (green star), SUMSS (yellow pentagon), and NVSS (navy diamond). The models for each year are determined by their classification; a source classified with a peak within the observed band was modelled by a quadratic according to Equation~\ref{eq:quadratic}, remaining sources were modelled by a power-law according to Equation~\ref{eq:plaw}.}
\label{app:fig:pg11}
\end{center}
\end{figure*}
\setcounter{figure}{0}
\begin{figure*}
\begin{center}$
\begin{array}{cccccc}
\includegraphics[scale=0.15]{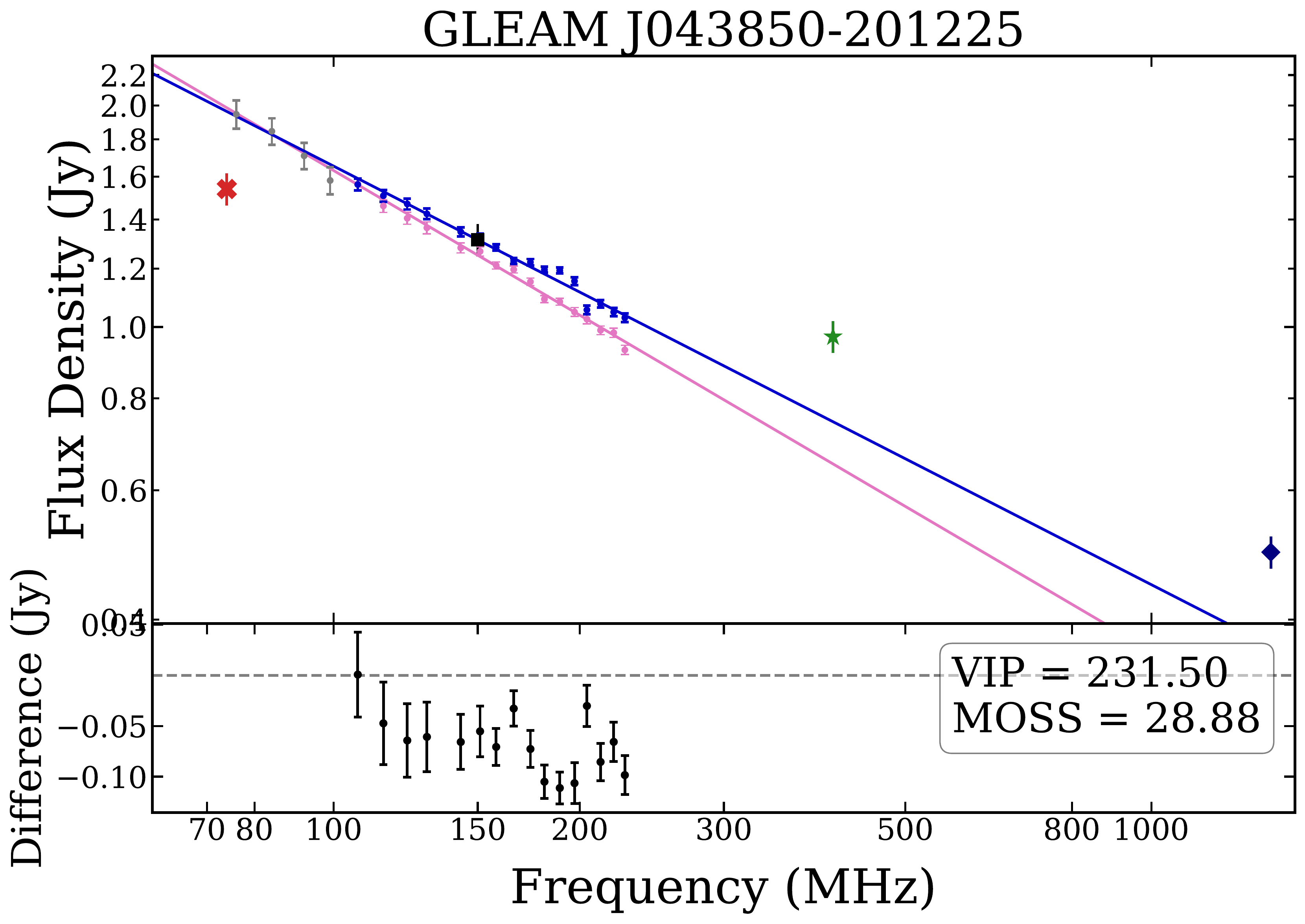} &
\includegraphics[scale=0.15]{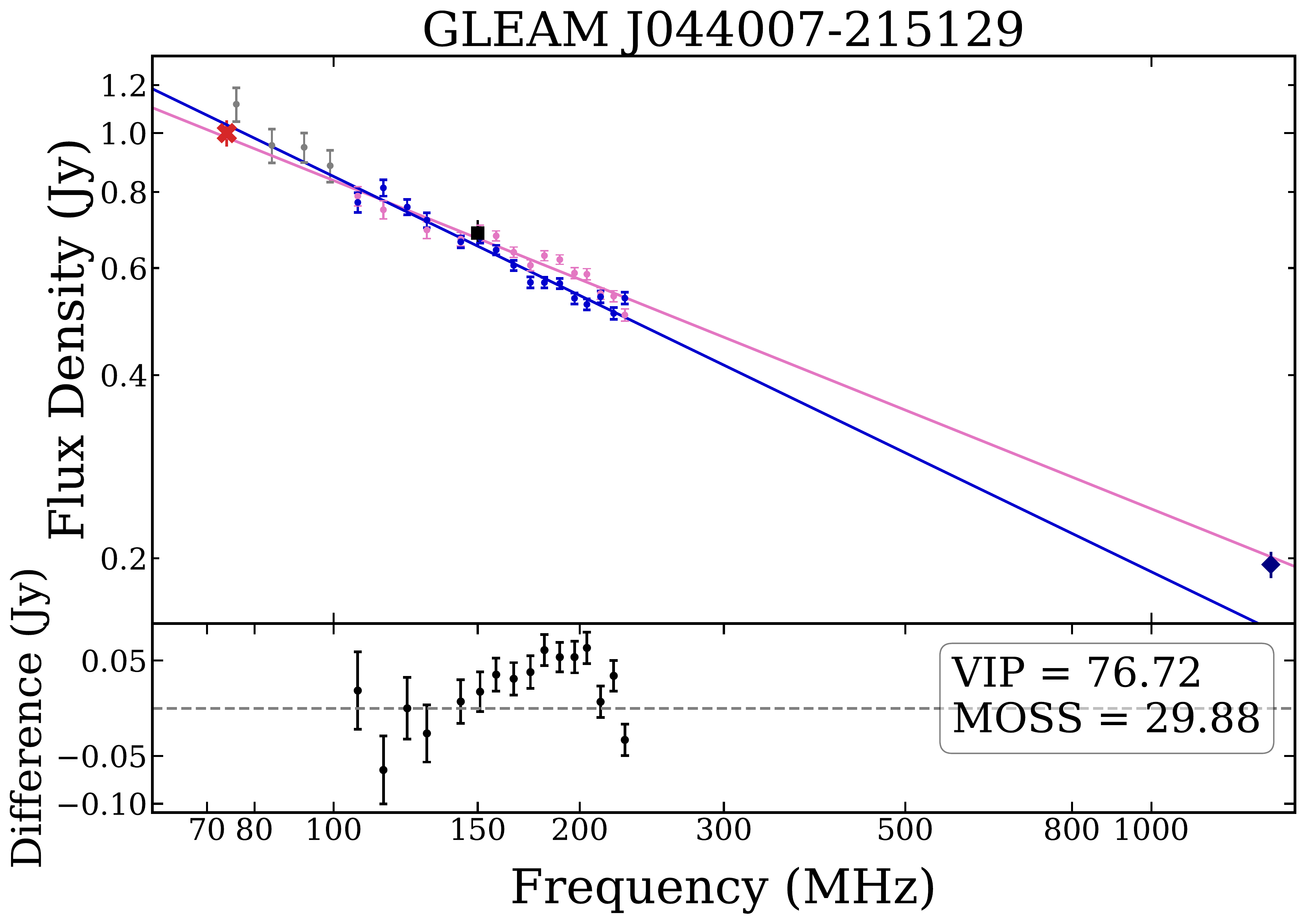} &
\includegraphics[scale=0.15]{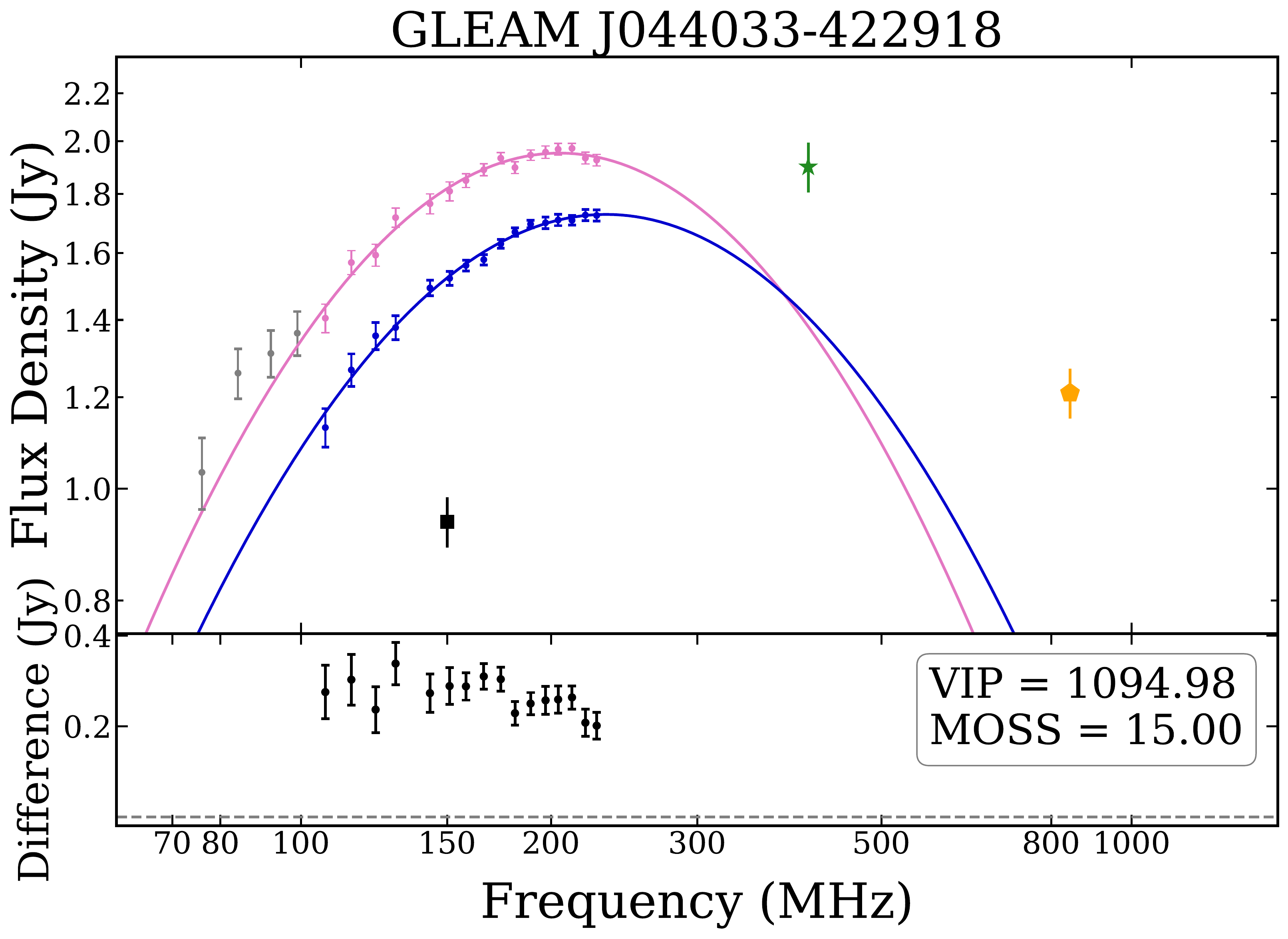} \\
\includegraphics[scale=0.15]{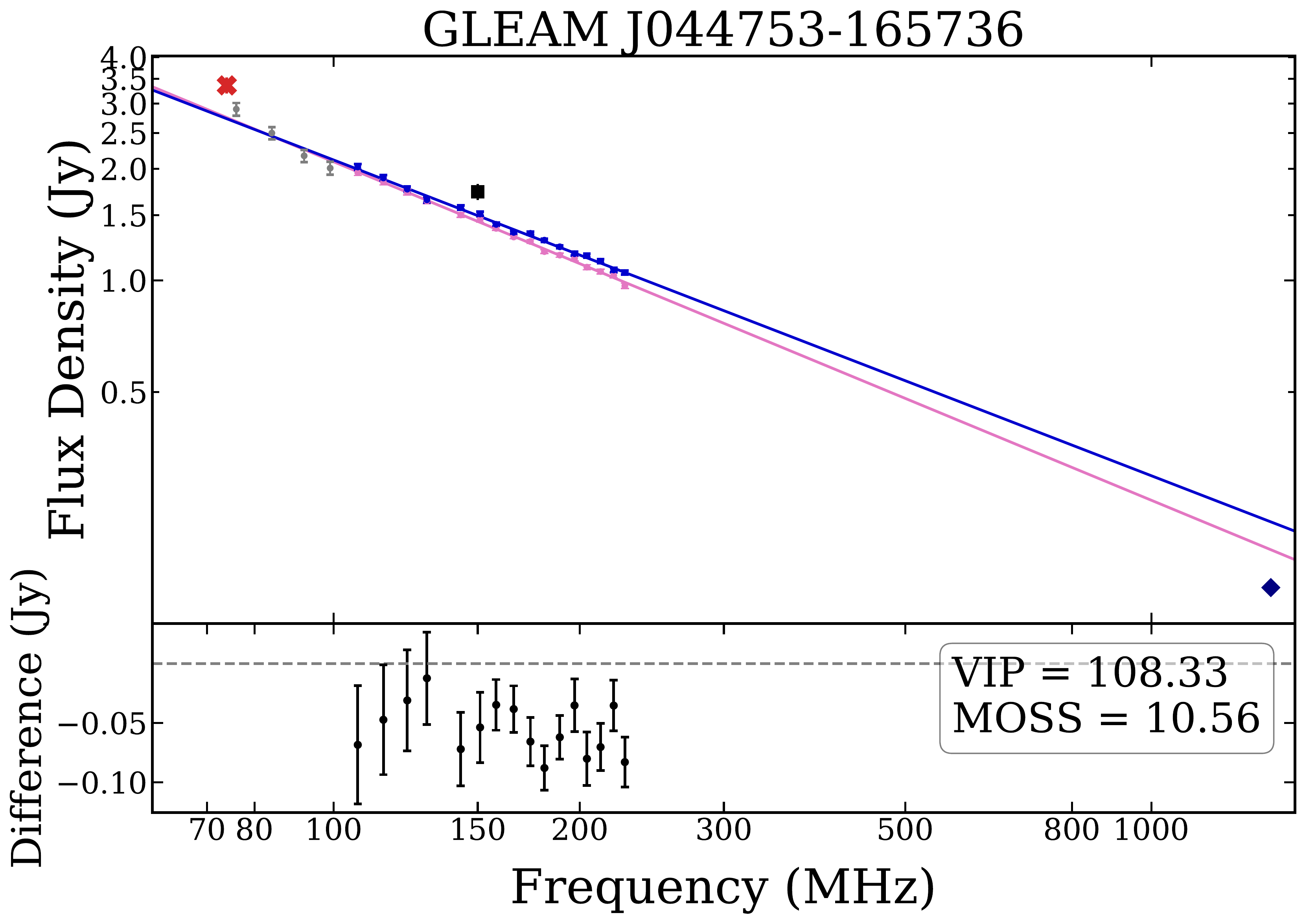} &
\includegraphics[scale=0.15]{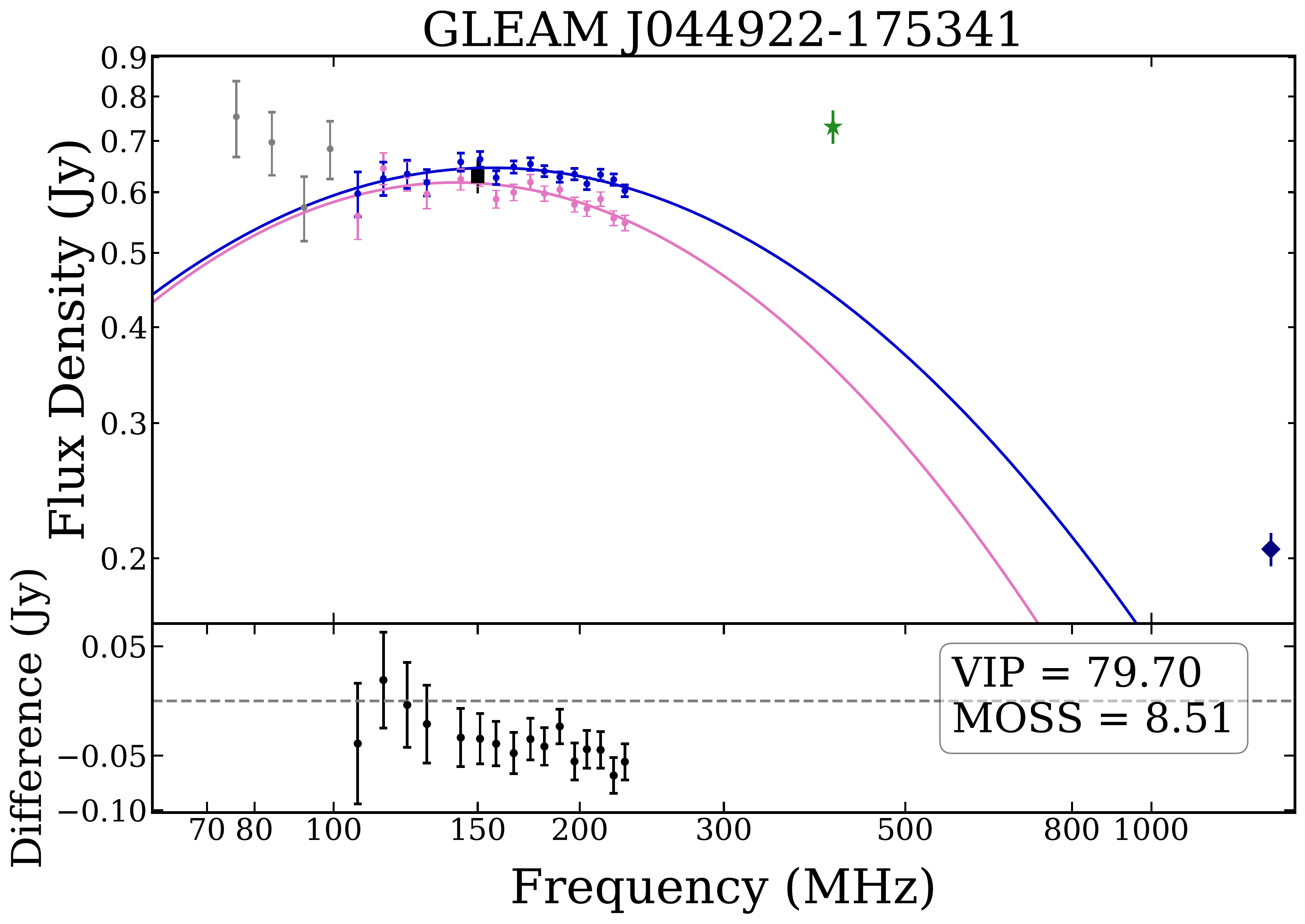} &
\includegraphics[scale=0.15]{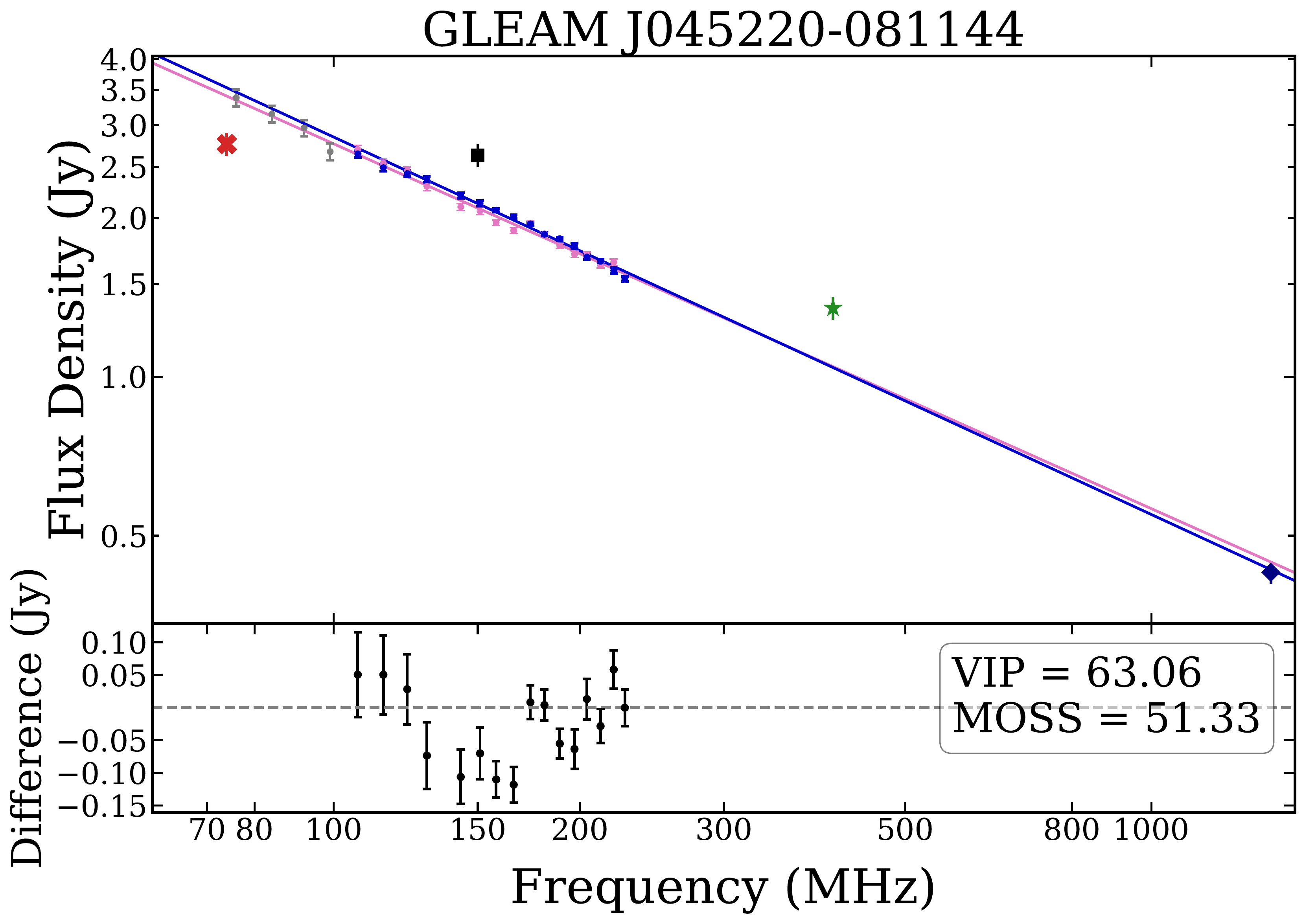} \\
\includegraphics[scale=0.15]{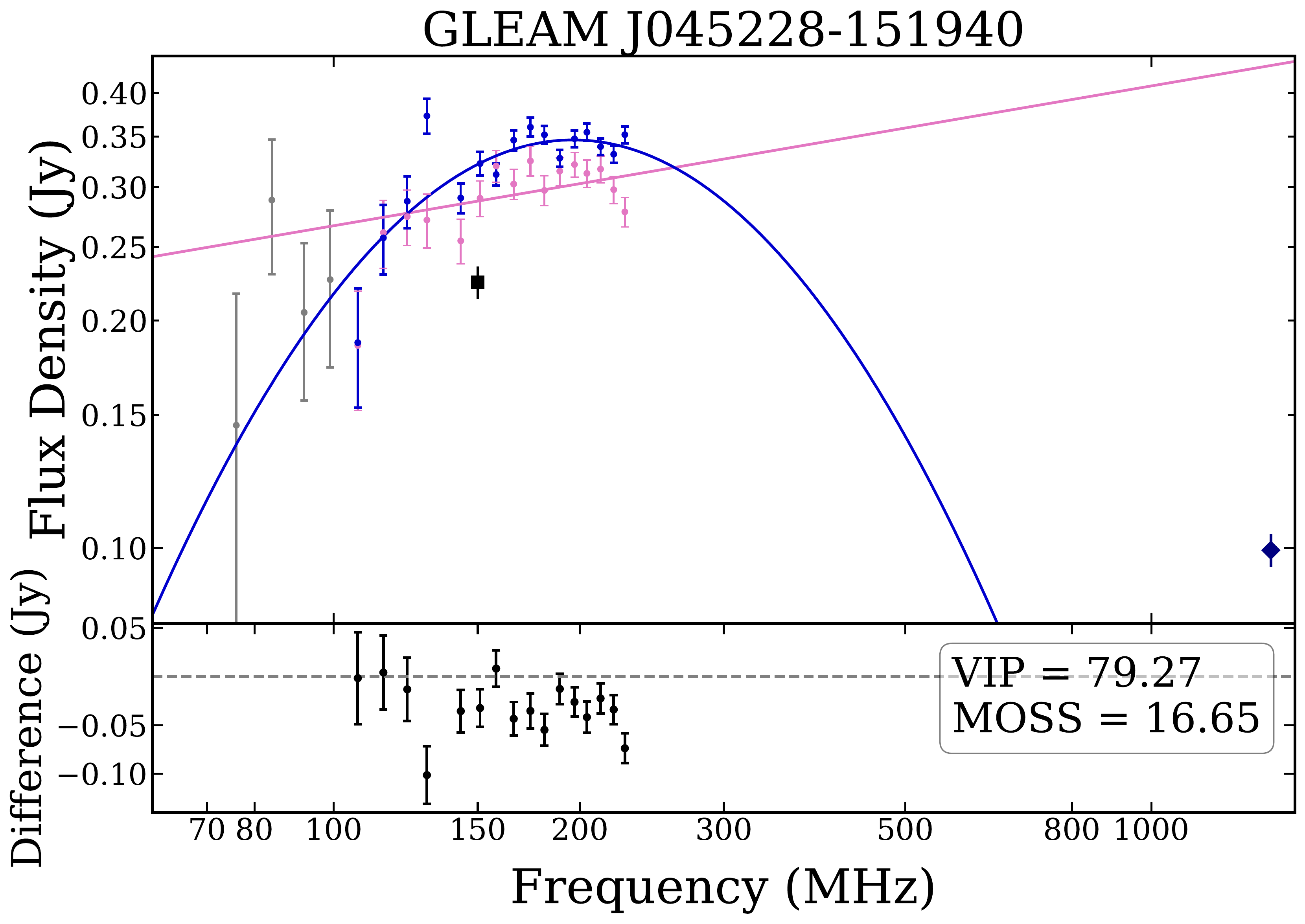} &
\includegraphics[scale=0.15]{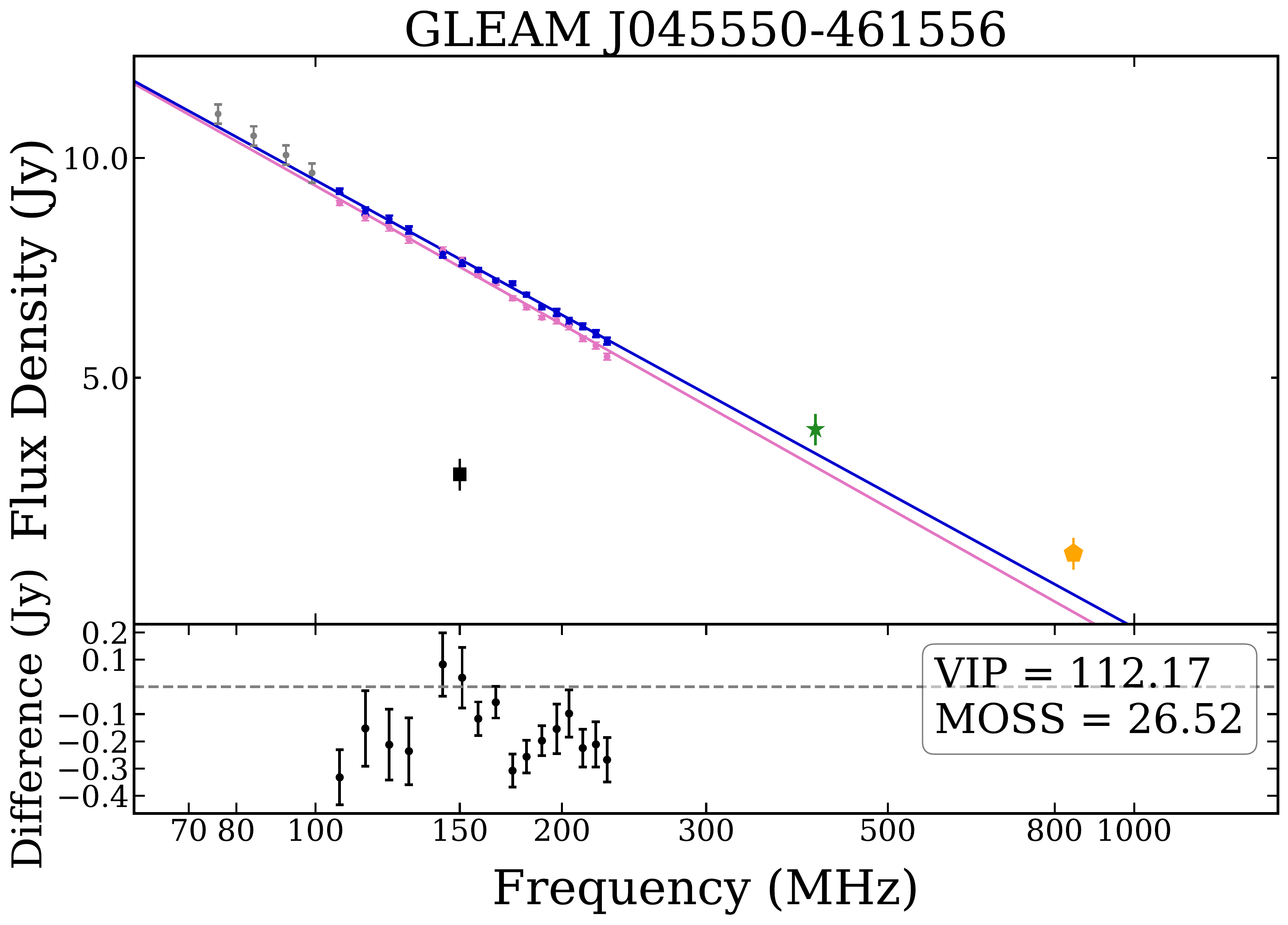} &
\includegraphics[scale=0.15]{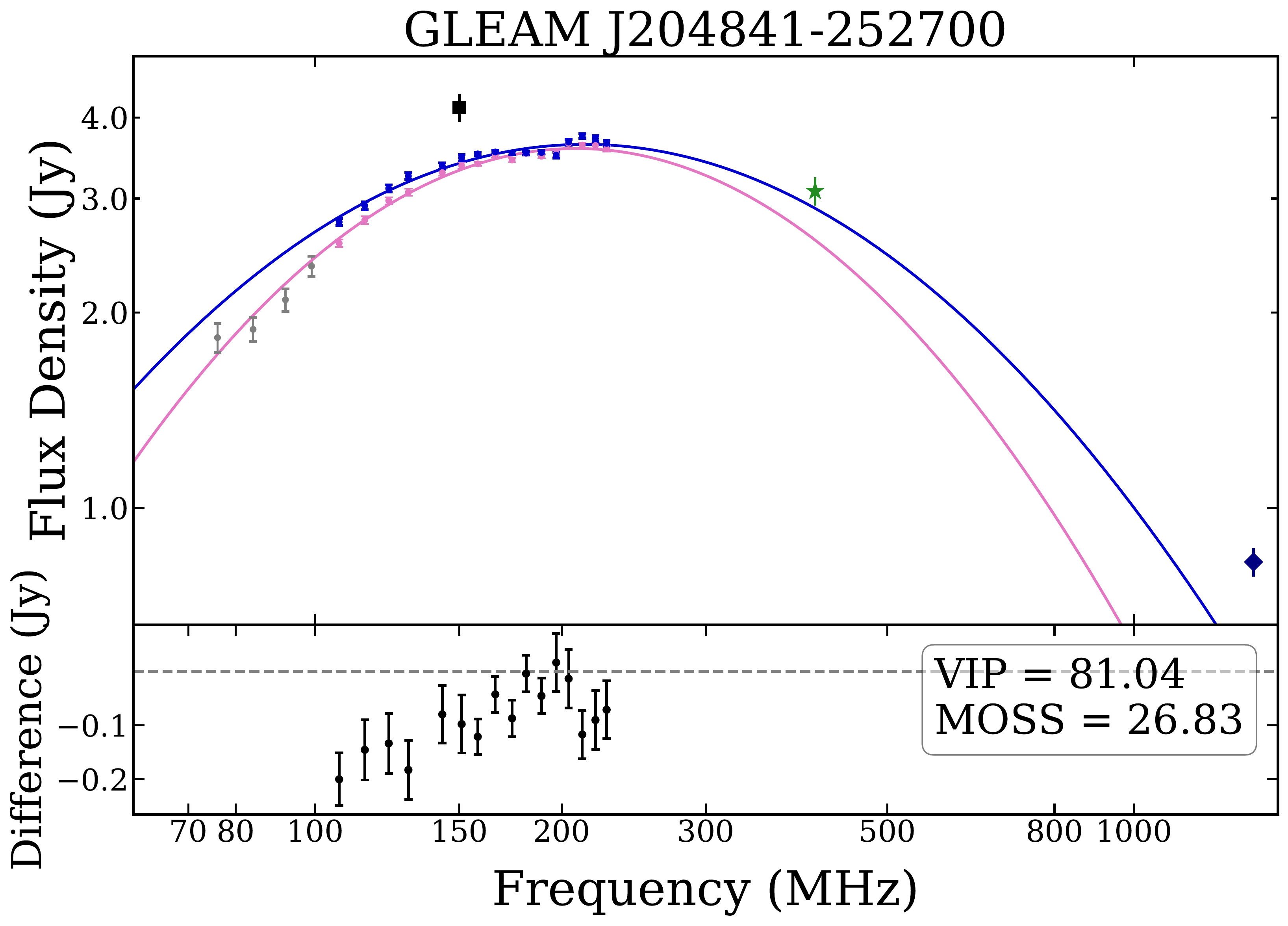} \\
\includegraphics[scale=0.15]{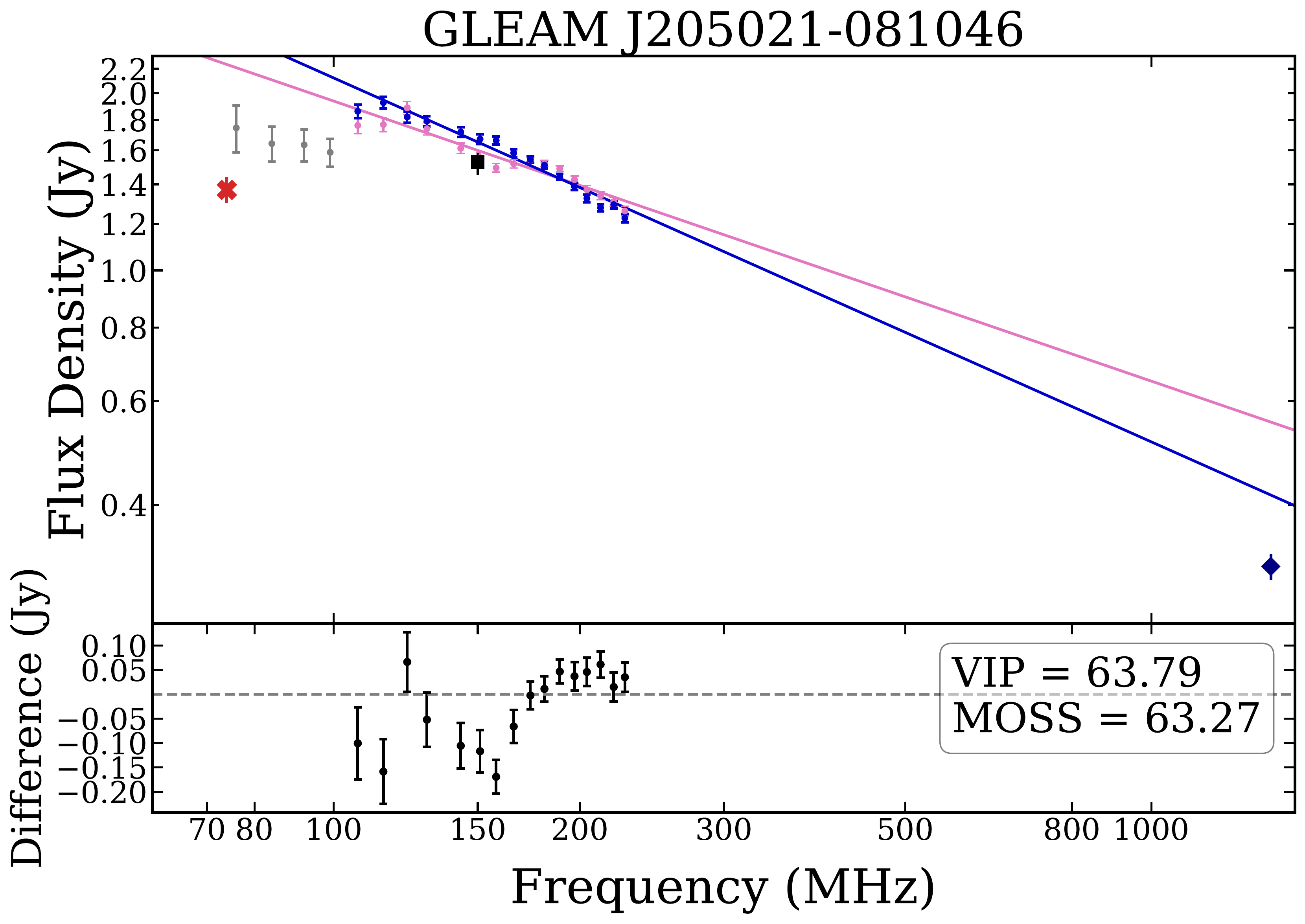} &
\includegraphics[scale=0.15]{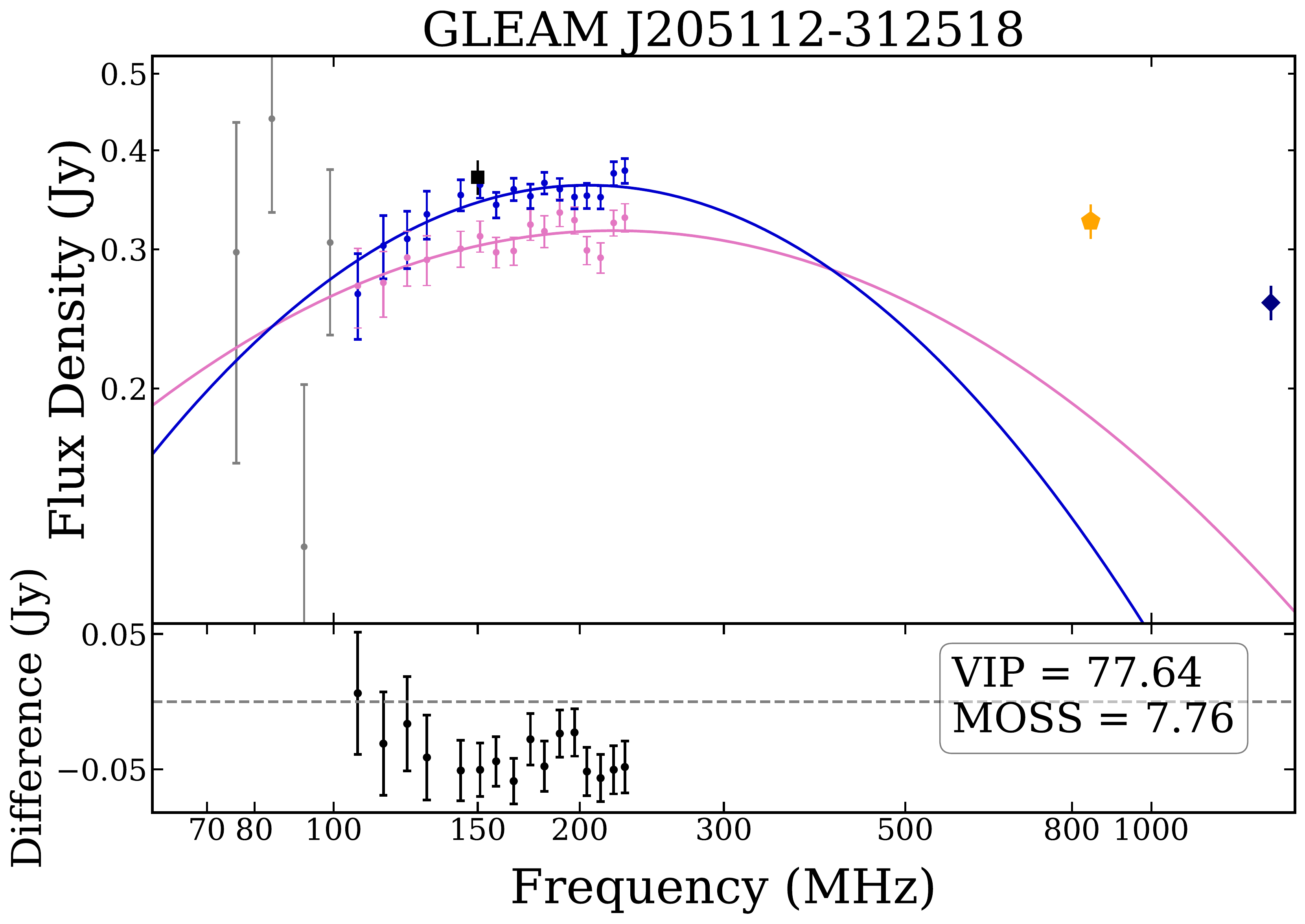} &
\includegraphics[scale=0.15]{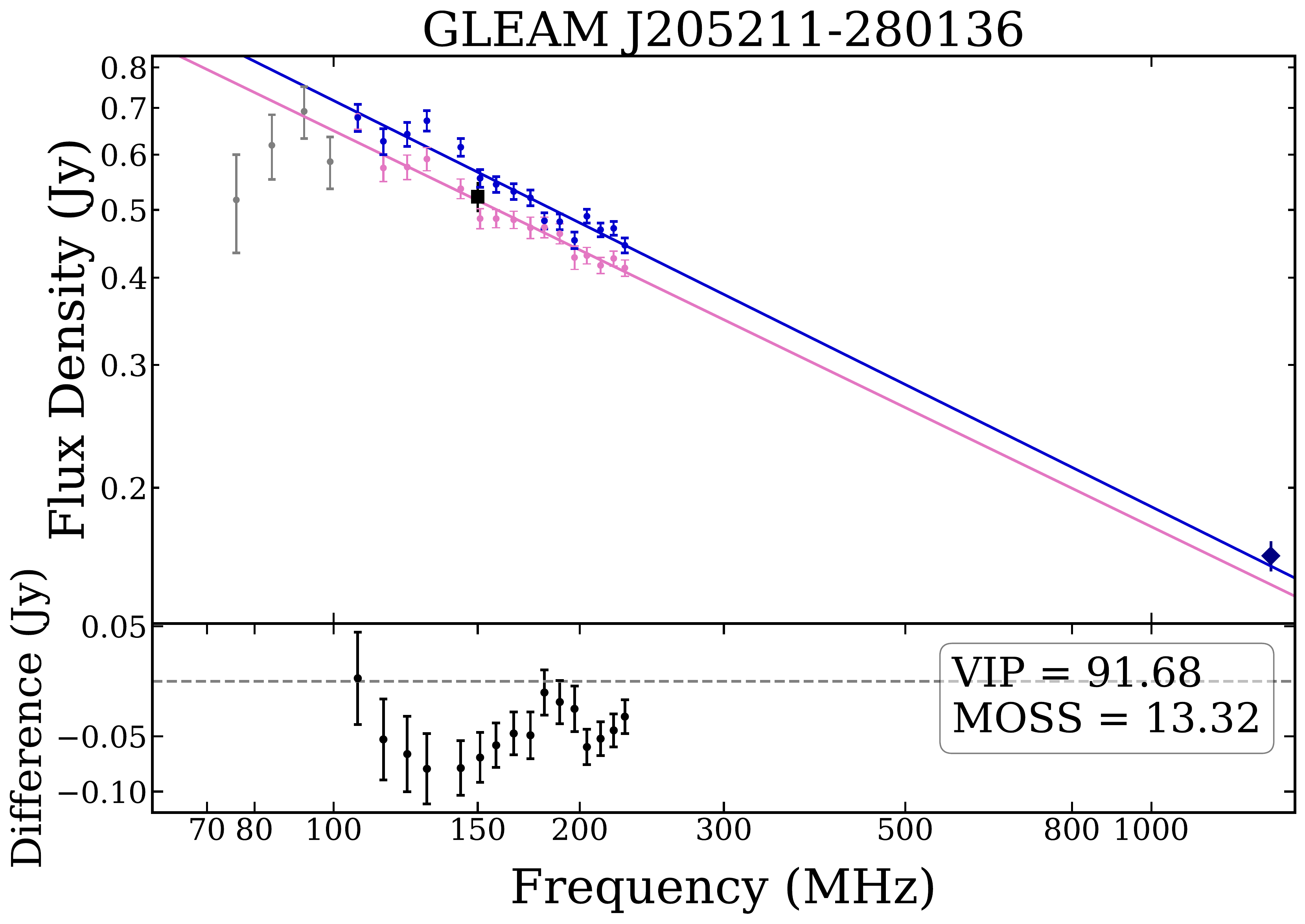} \\
\includegraphics[scale=0.15]{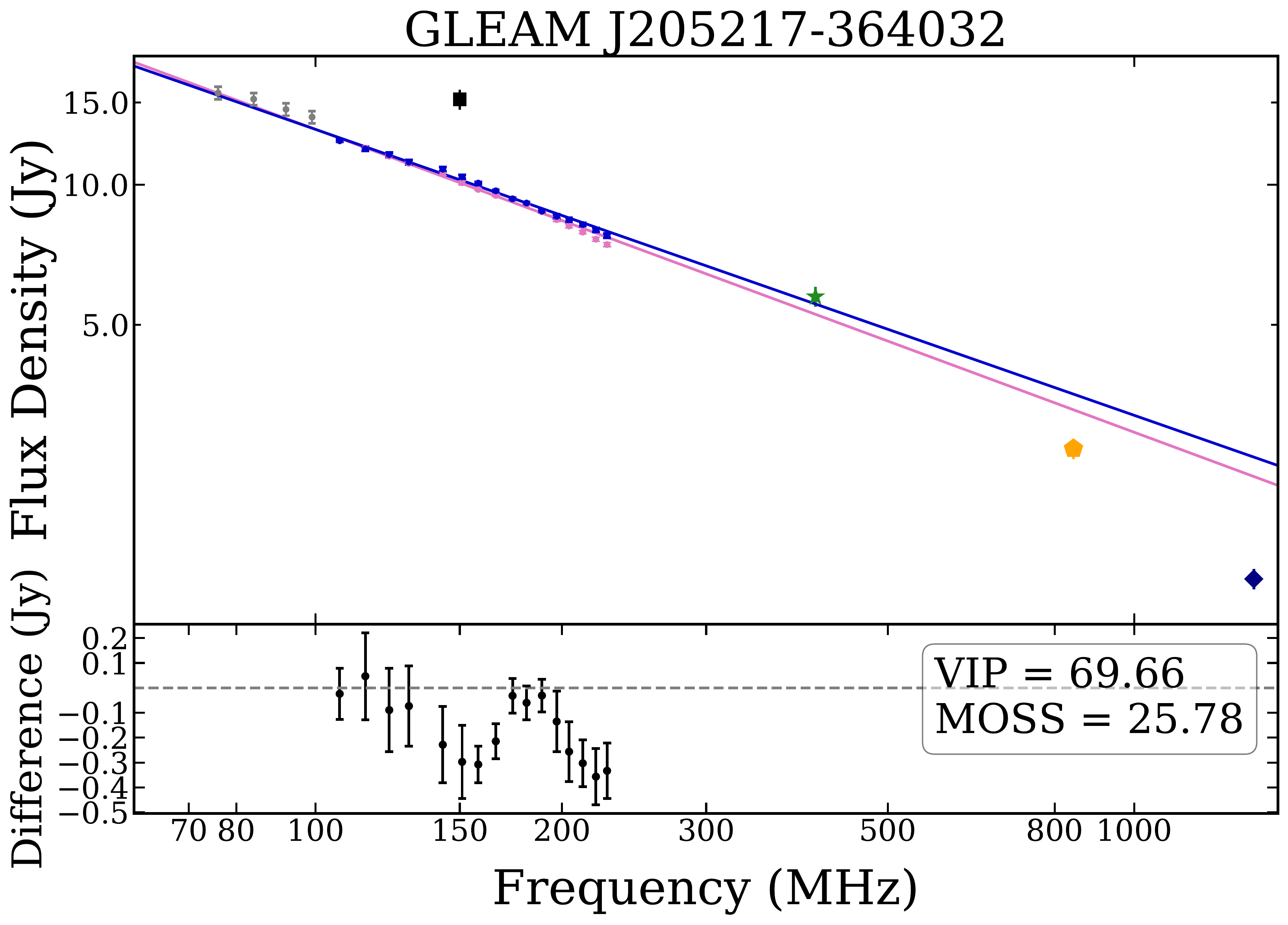} &
\includegraphics[scale=0.15]{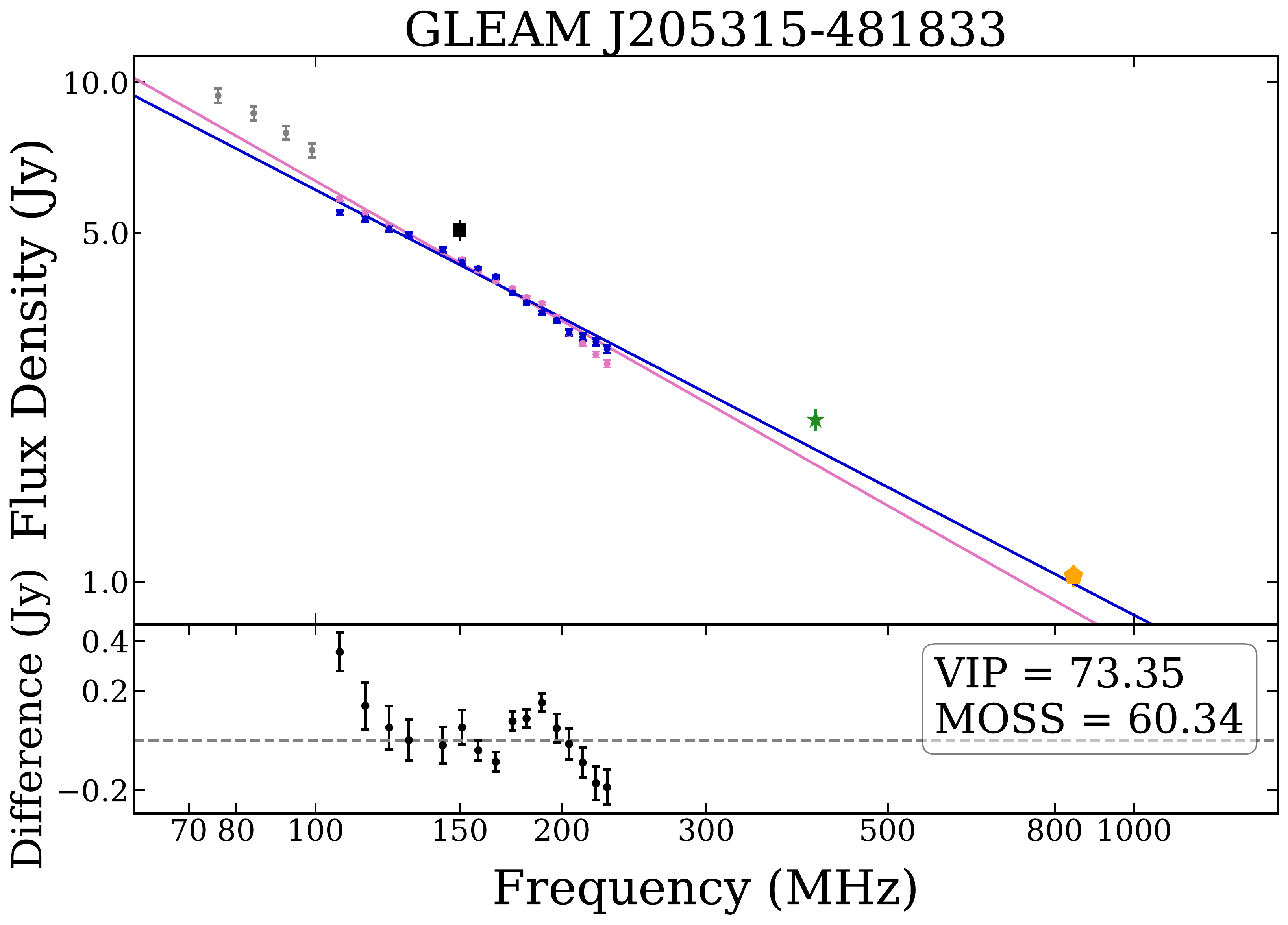} &
\includegraphics[scale=0.15]{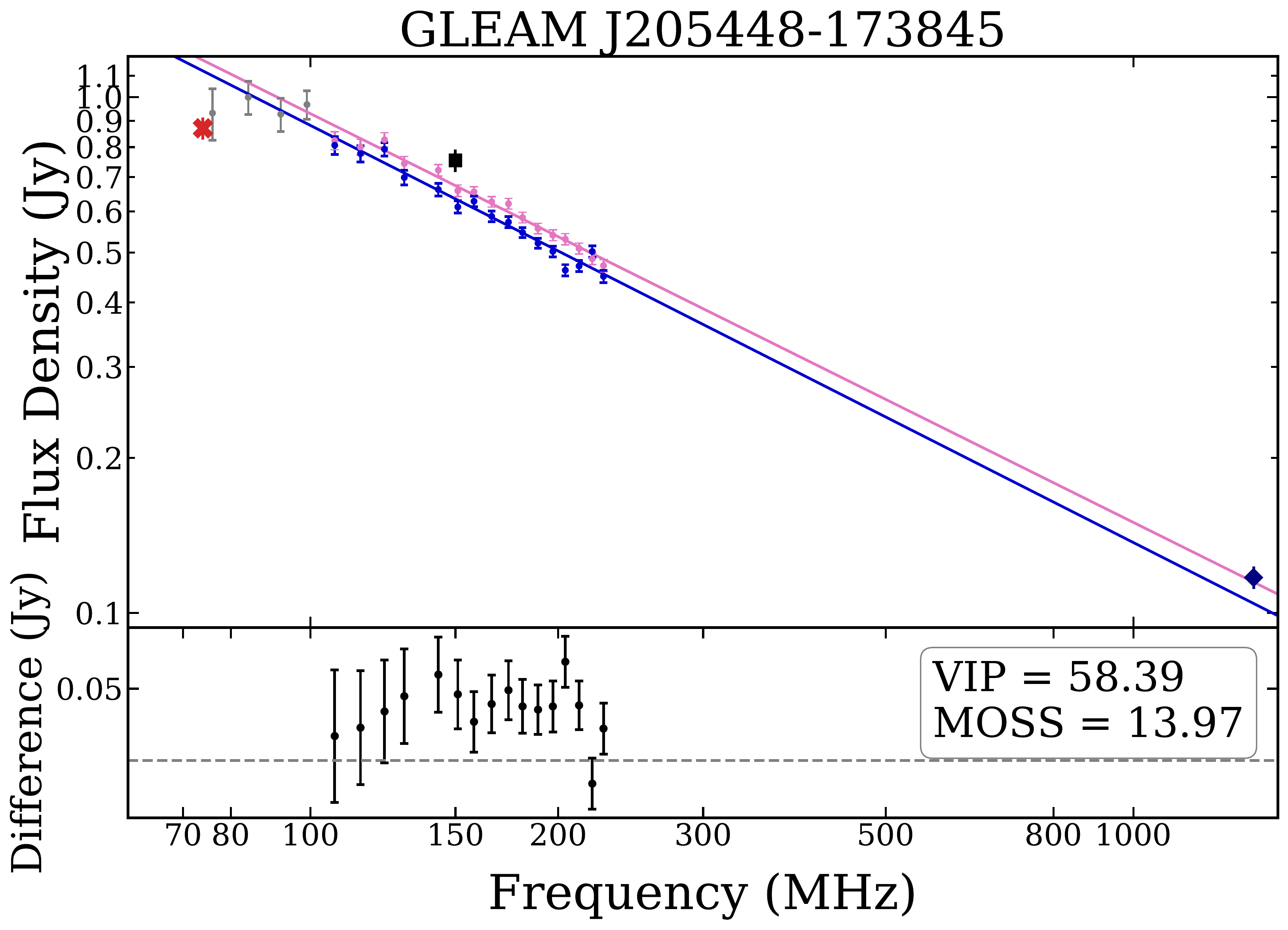} \\
\includegraphics[scale=0.15]{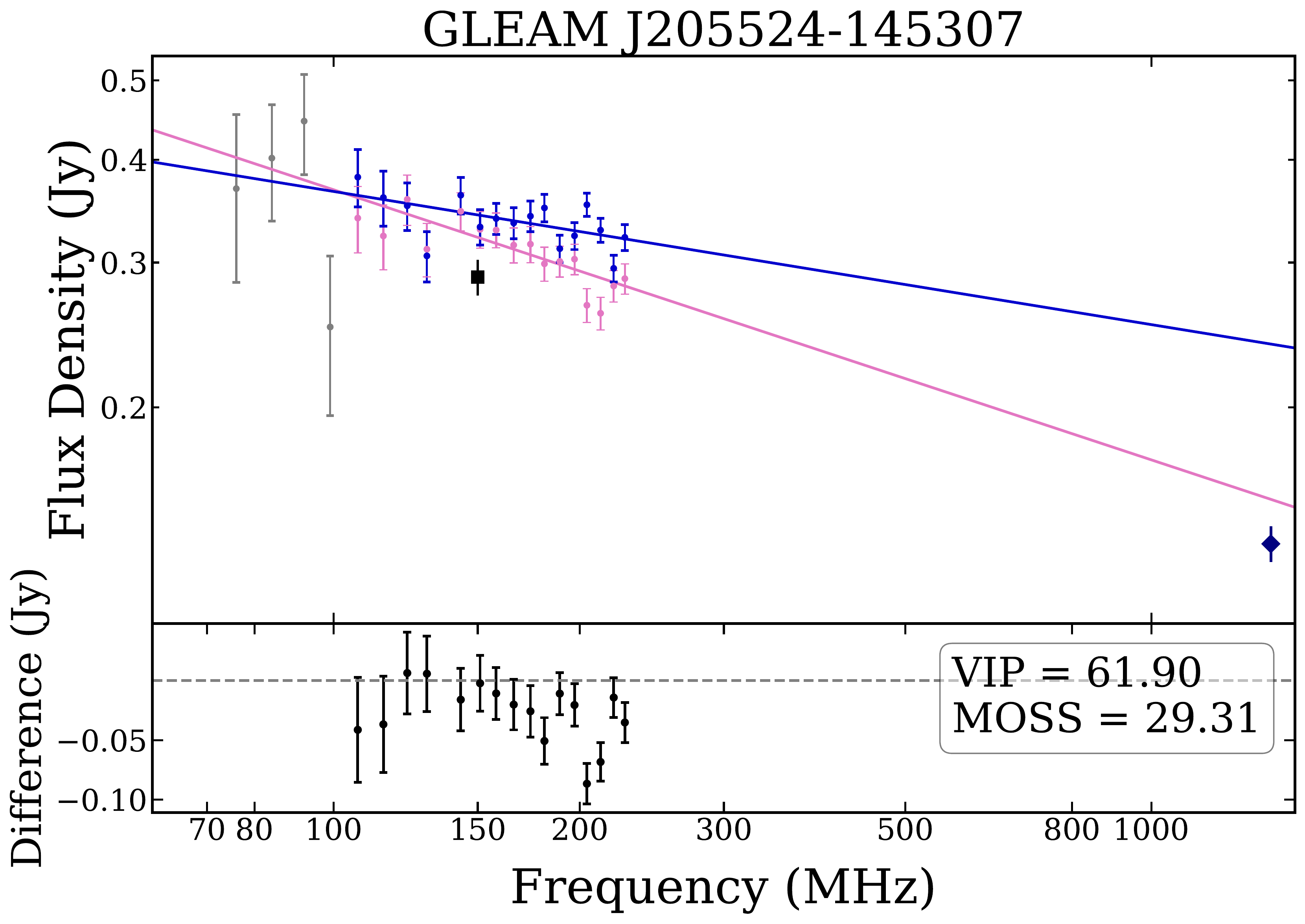} &
\includegraphics[scale=0.15]{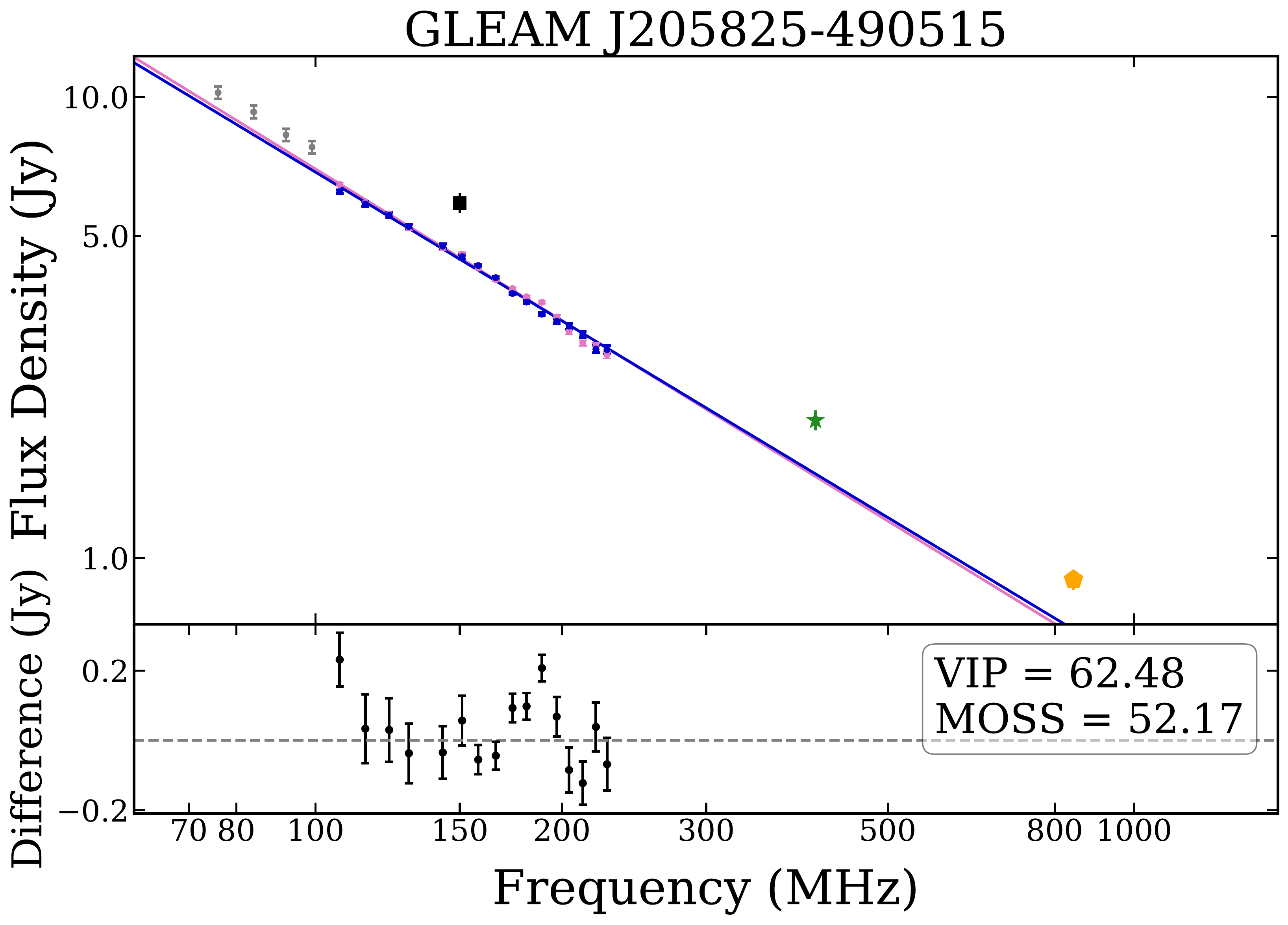} &
\includegraphics[scale=0.15]{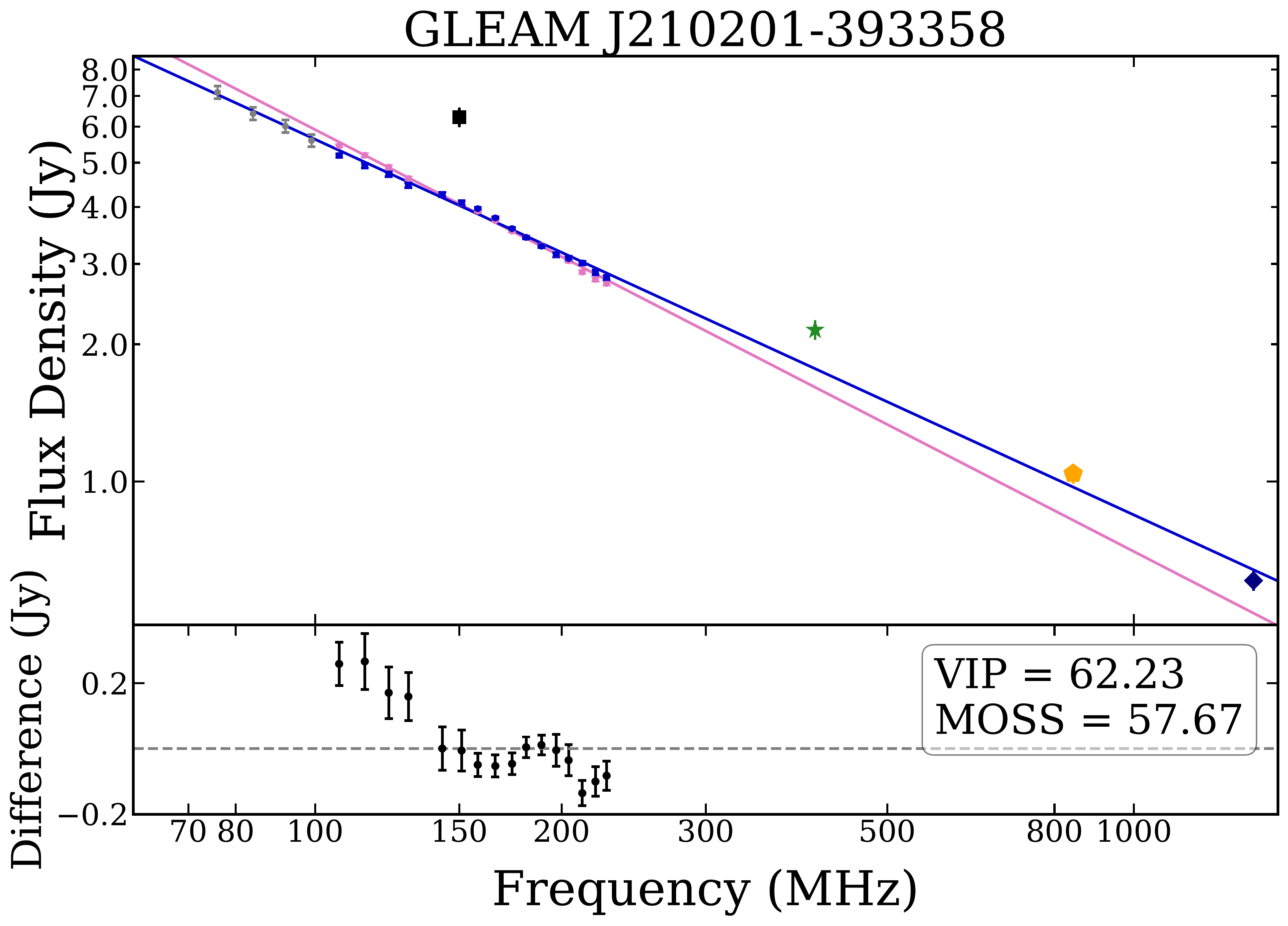} \\
\end{array}$
\caption{(continued) SEDs for all sources classified as variable according to the VIP. For each source the points represent the following data: GLEAM low frequency (72--100\,MHz) (grey circles), Year 1 (pink circles), Year 2 (blue circles), VLSSr (red cross), TGSS (black square), MRC (green star), SUMSS (yellow pentagon), and NVSS (navy diamond). The models for each year are determined by their classification; a source classified with a peak within the observed band was modelled by a quadratic according to Equation~\ref{eq:quadratic}, remaining sources were modelled by a power-law according to Equation~\ref{eq:plaw}.}
\label{app:fig:pg12}
\end{center}
\end{figure*}
\setcounter{figure}{0}
\begin{figure*}
\begin{center}$
\begin{array}{cccccc}
\includegraphics[scale=0.15]{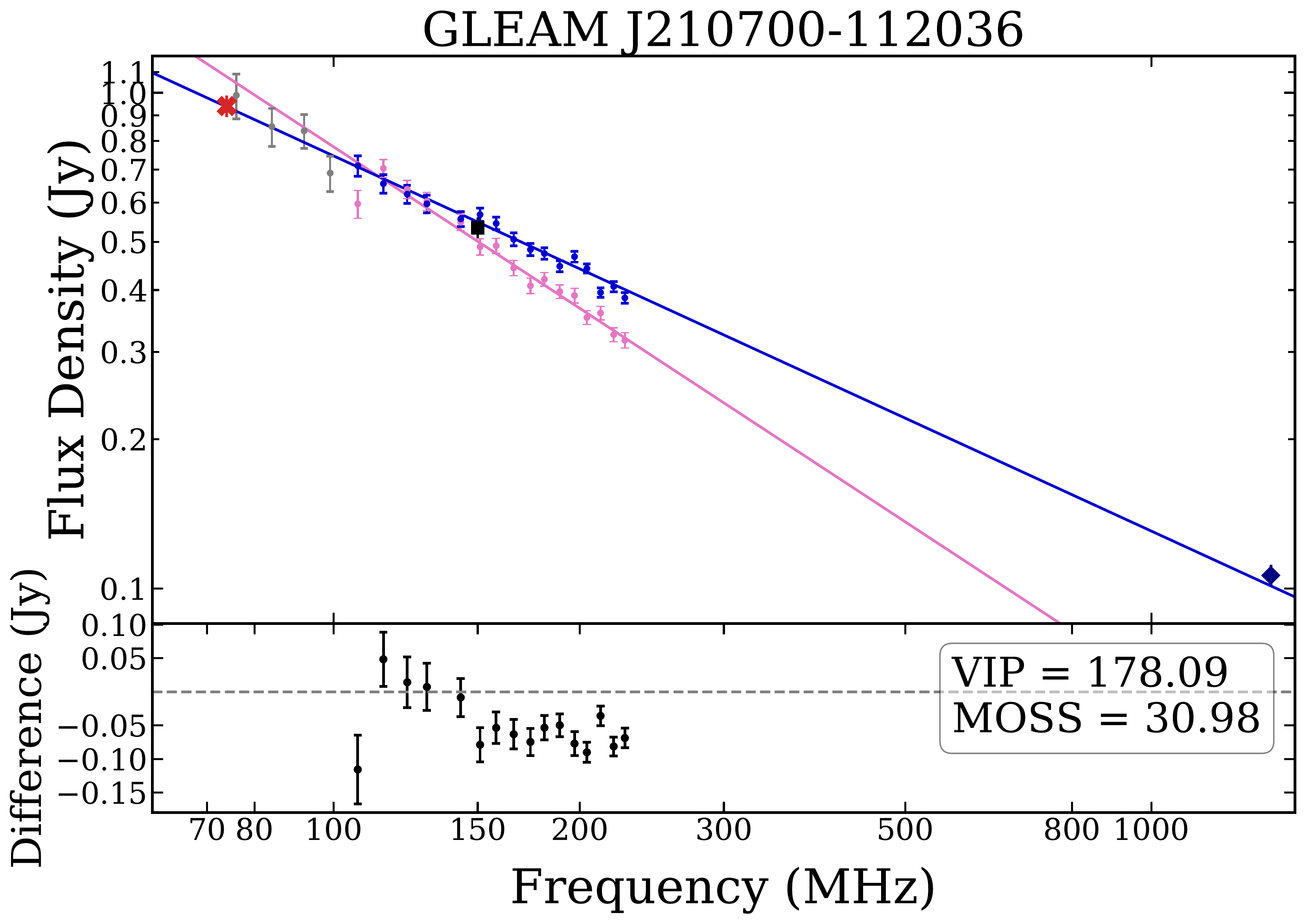} &
\includegraphics[scale=0.15]{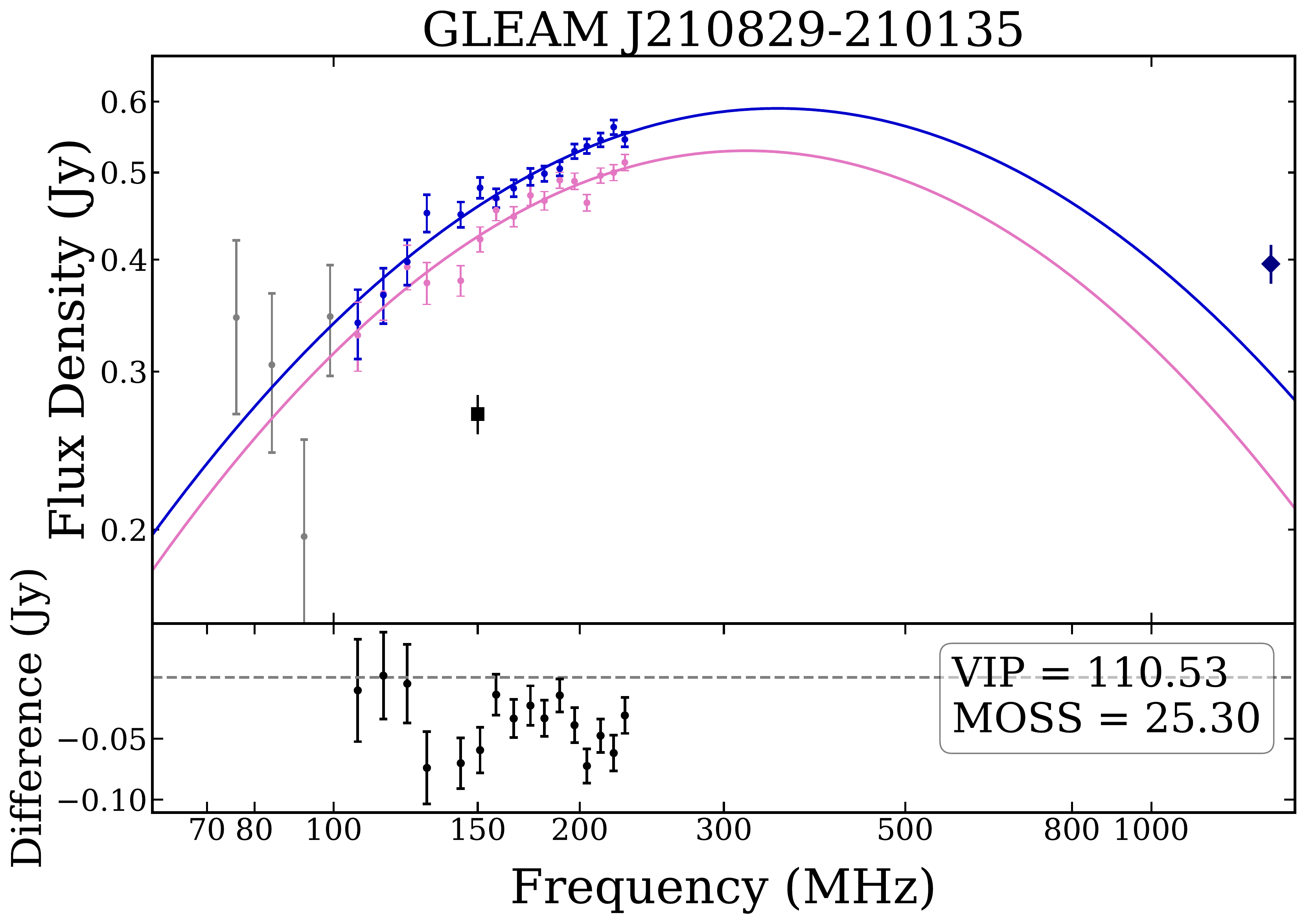} &
\includegraphics[scale=0.15]{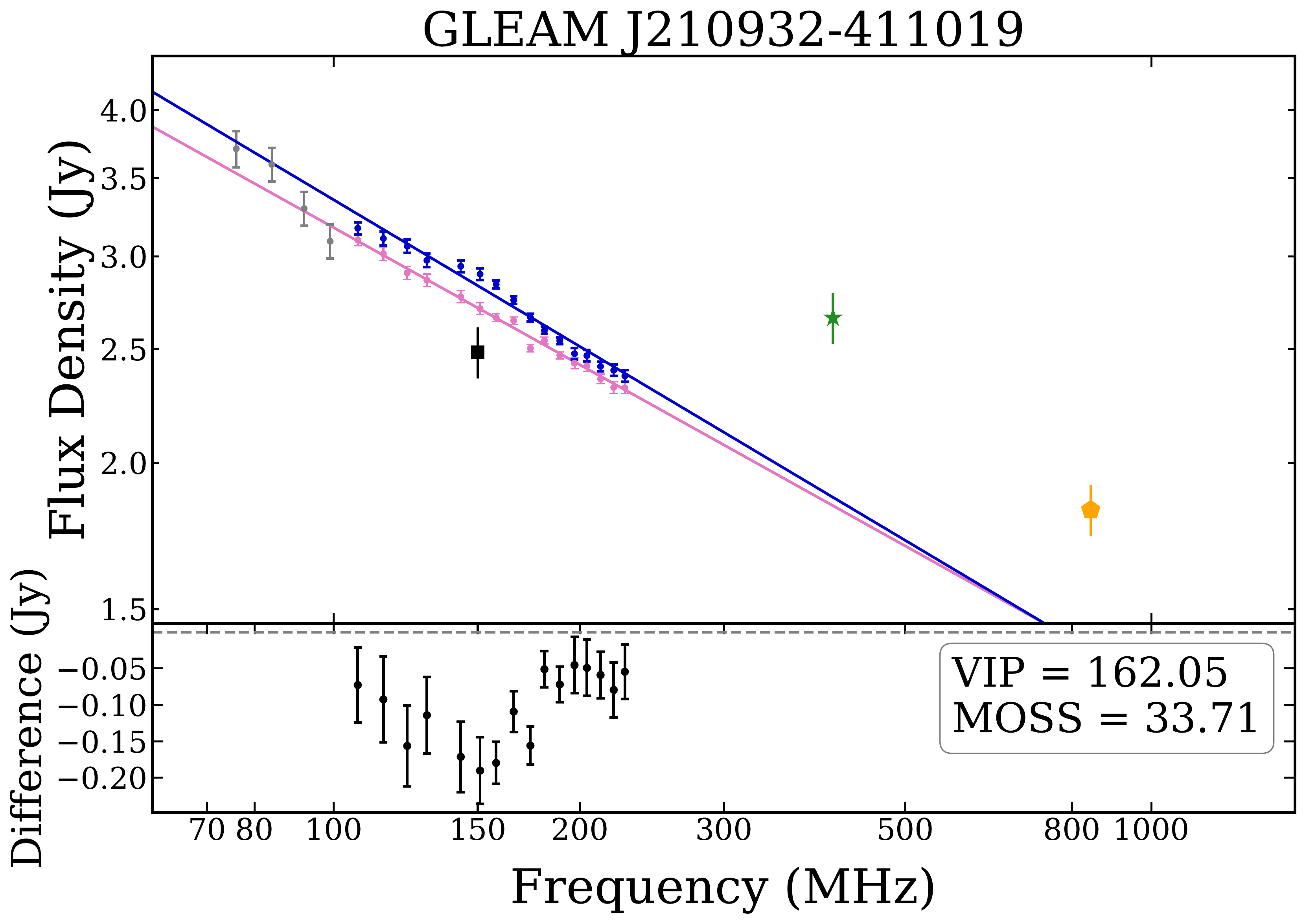} \\
\includegraphics[scale=0.15]{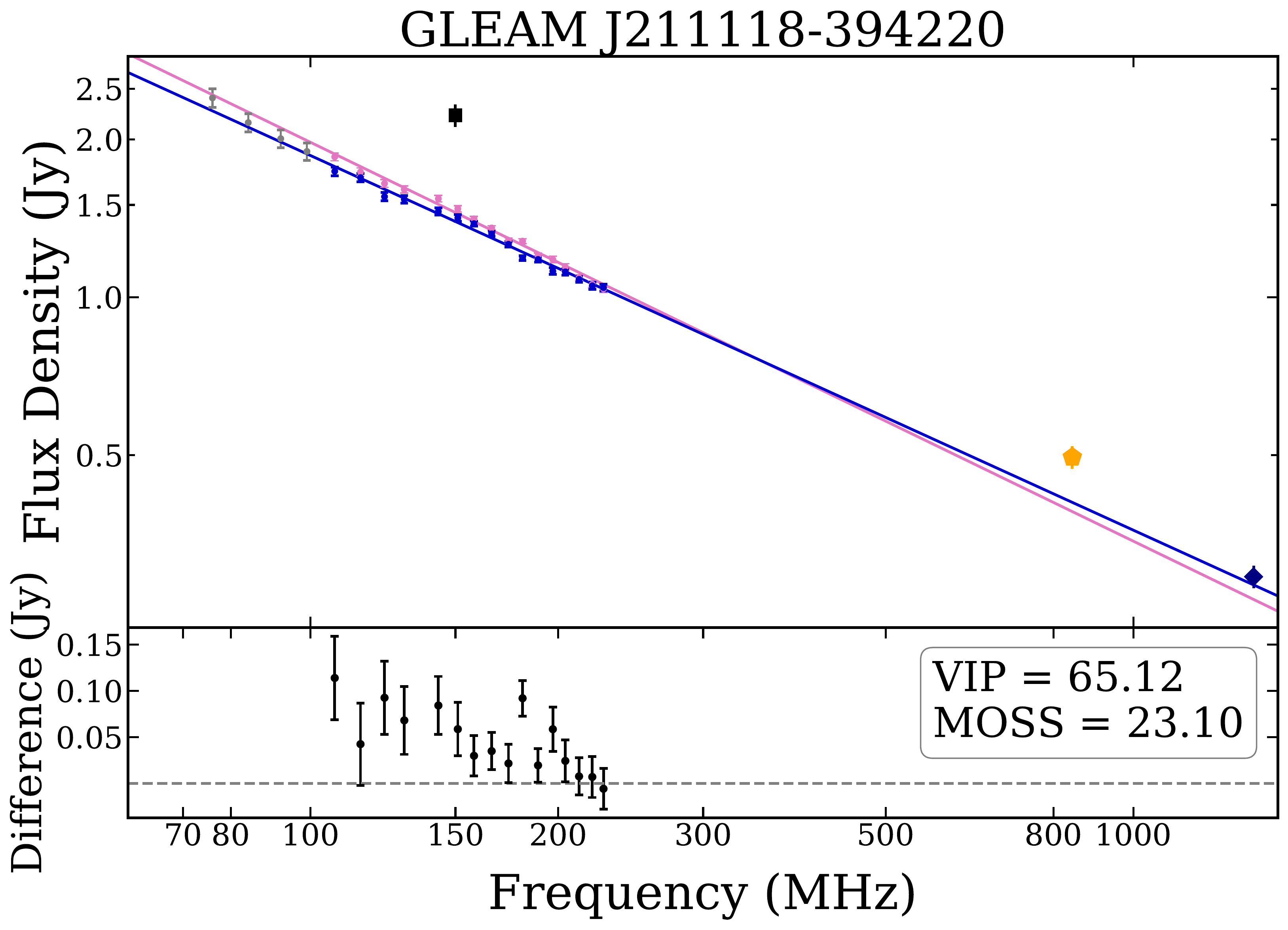} &
\includegraphics[scale=0.15]{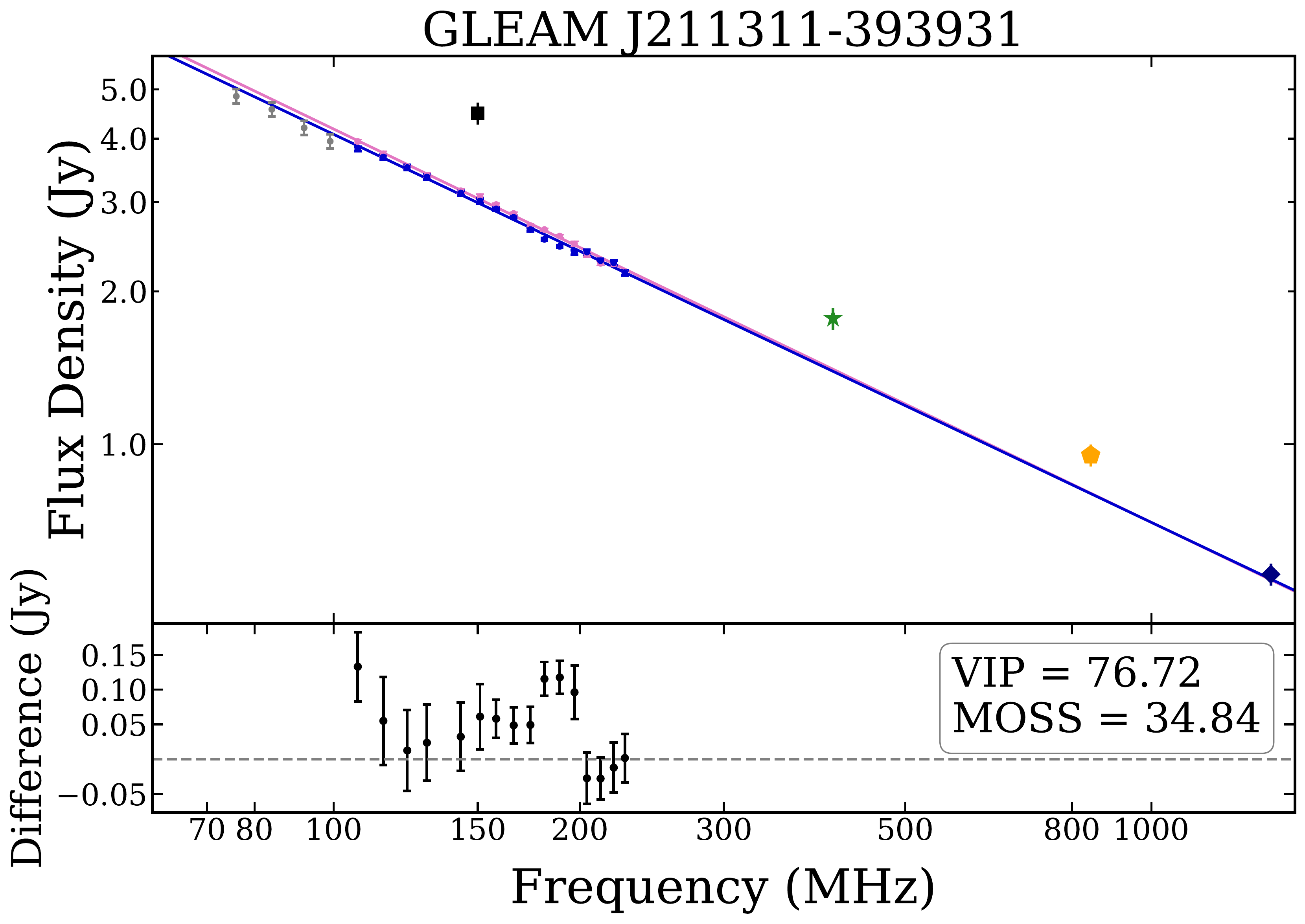} &
\includegraphics[scale=0.15]{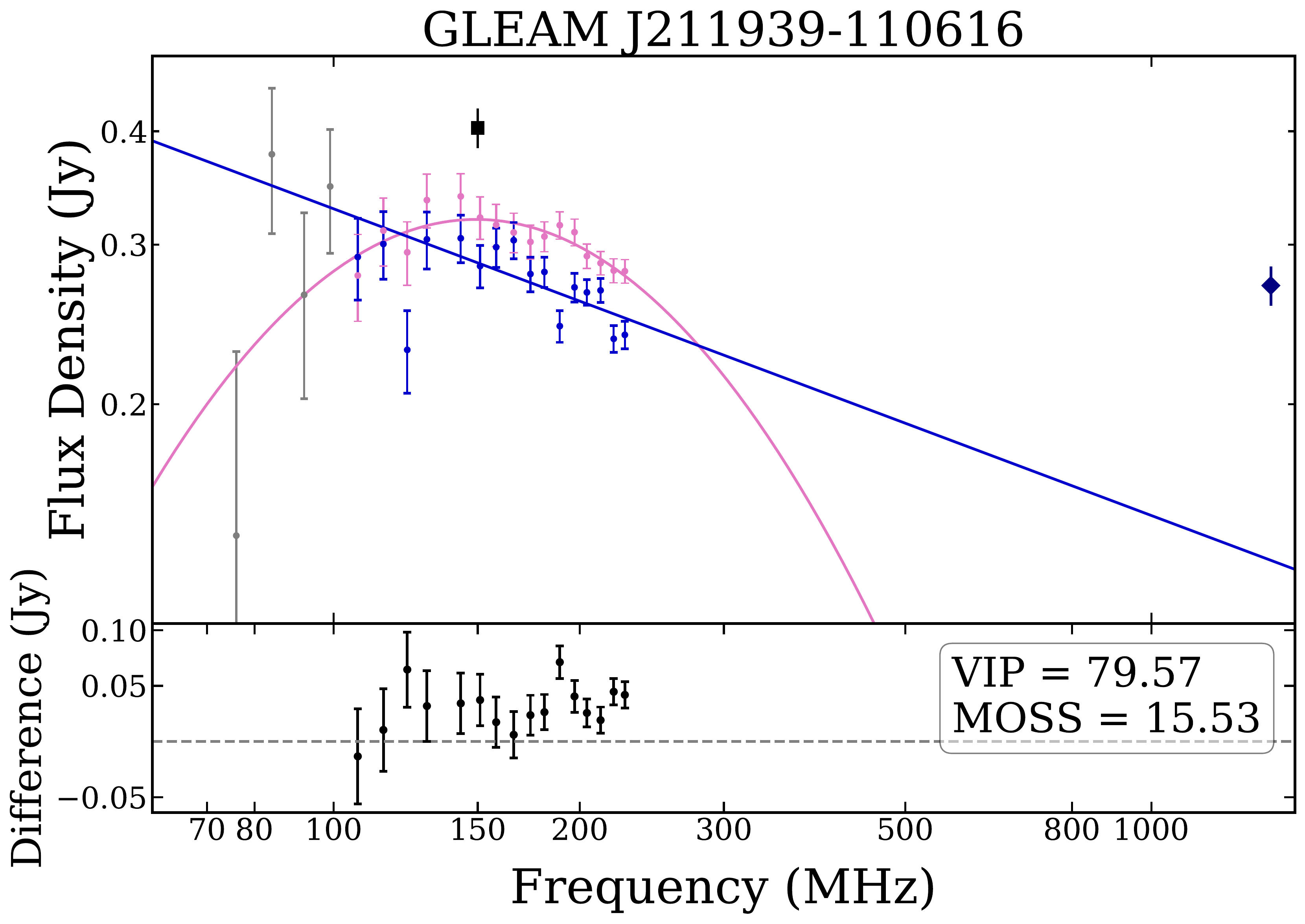} \\
\includegraphics[scale=0.15]{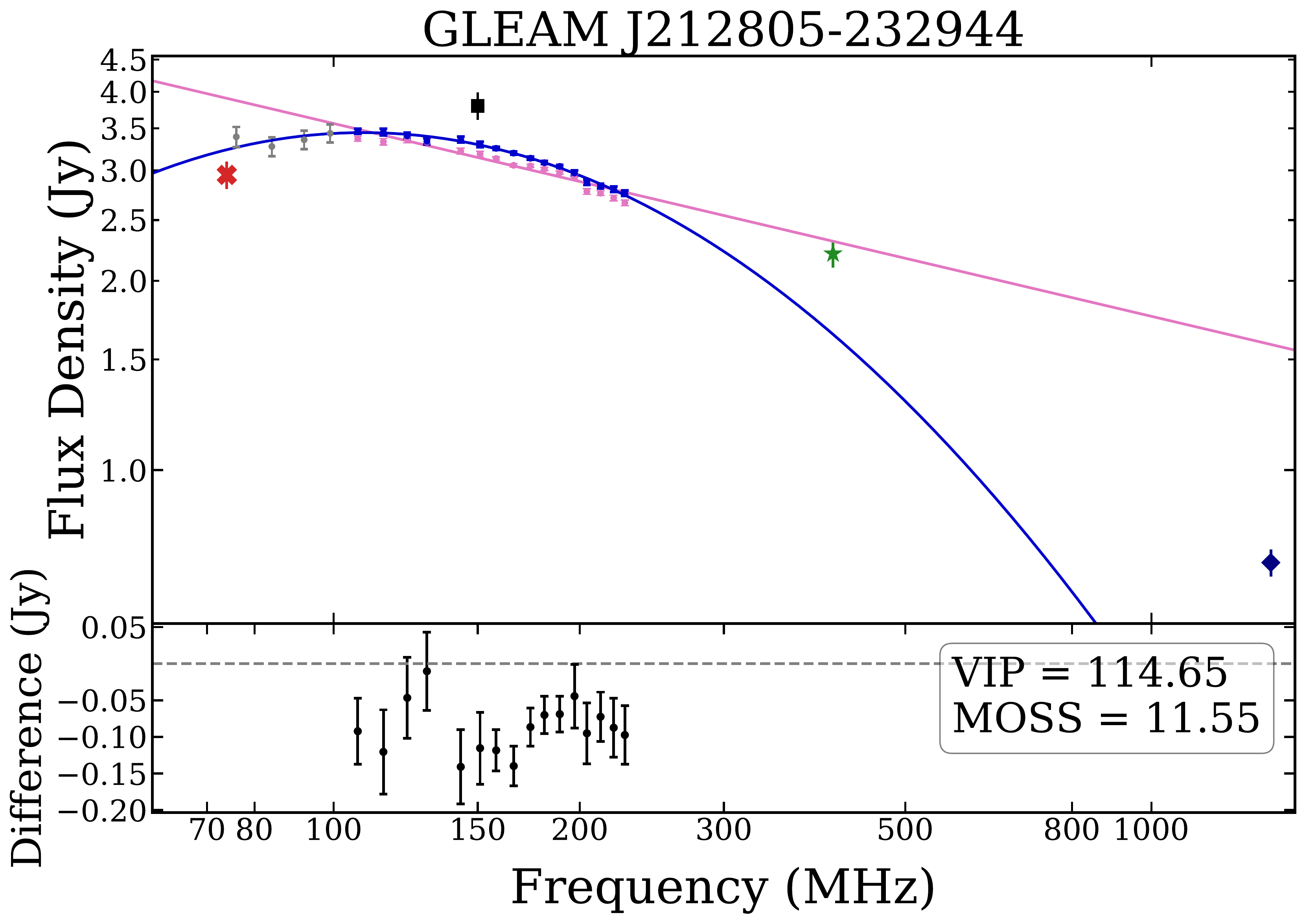} &
\includegraphics[scale=0.15]{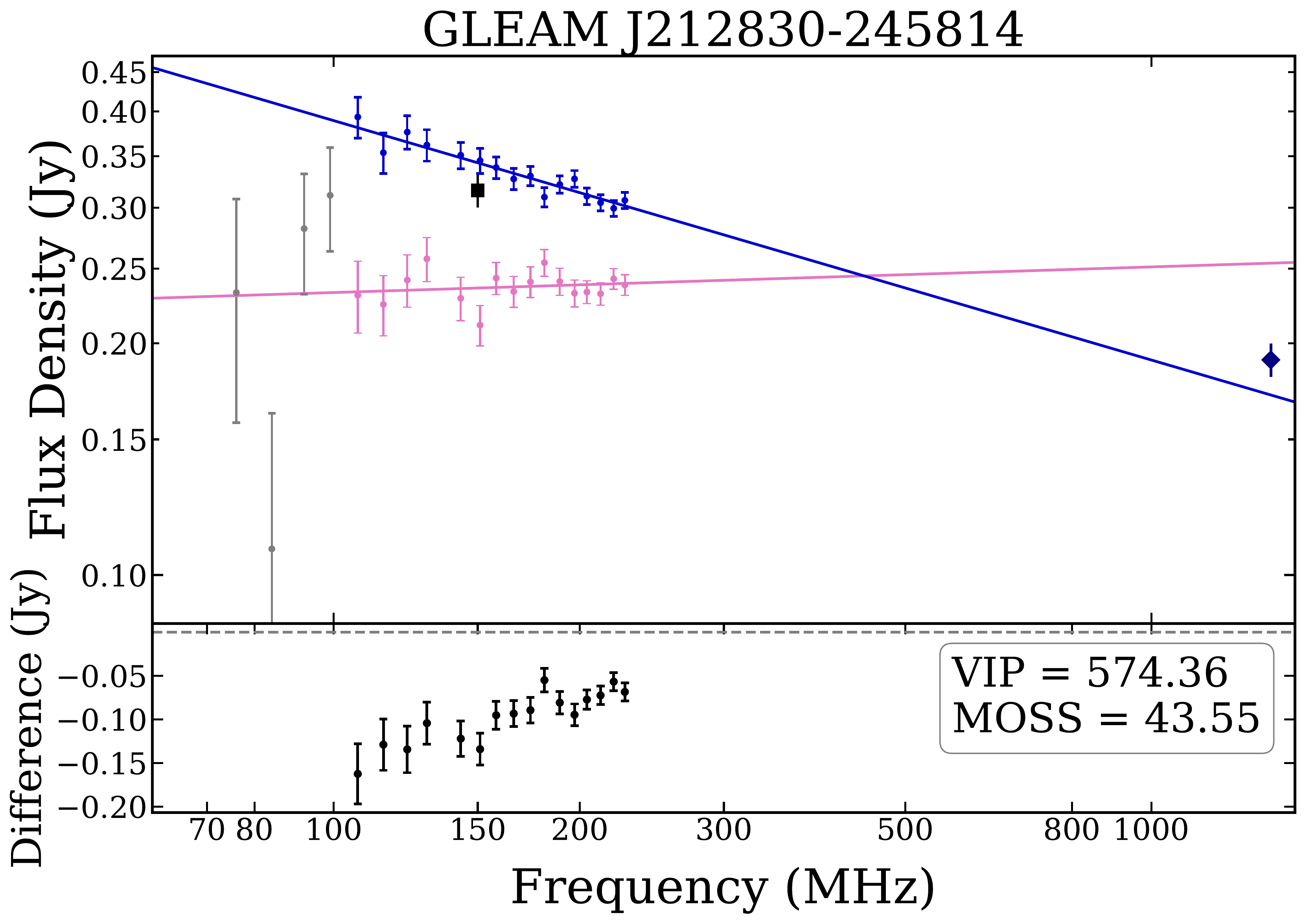} &
\includegraphics[scale=0.15]{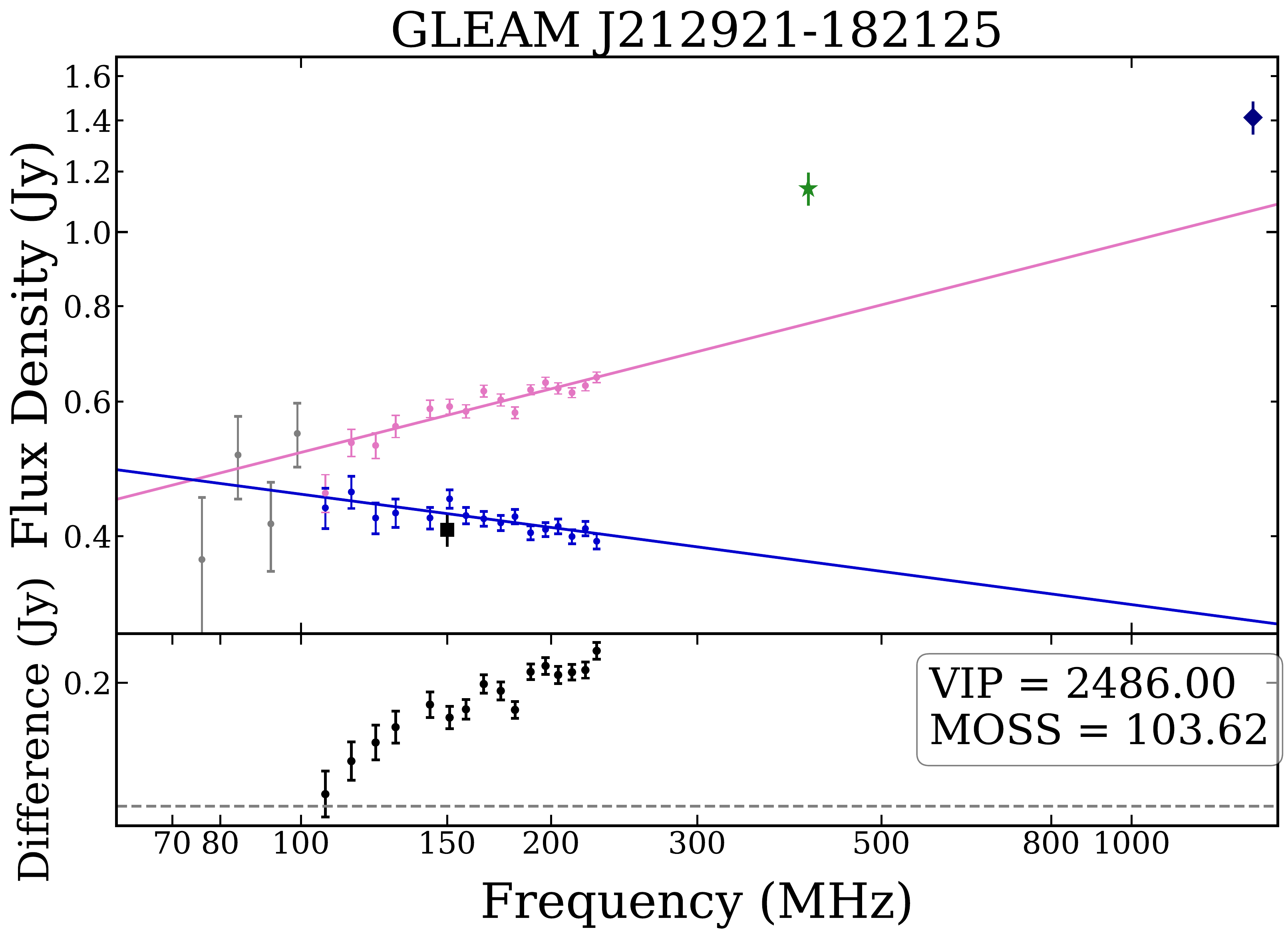} \\
\includegraphics[scale=0.15]{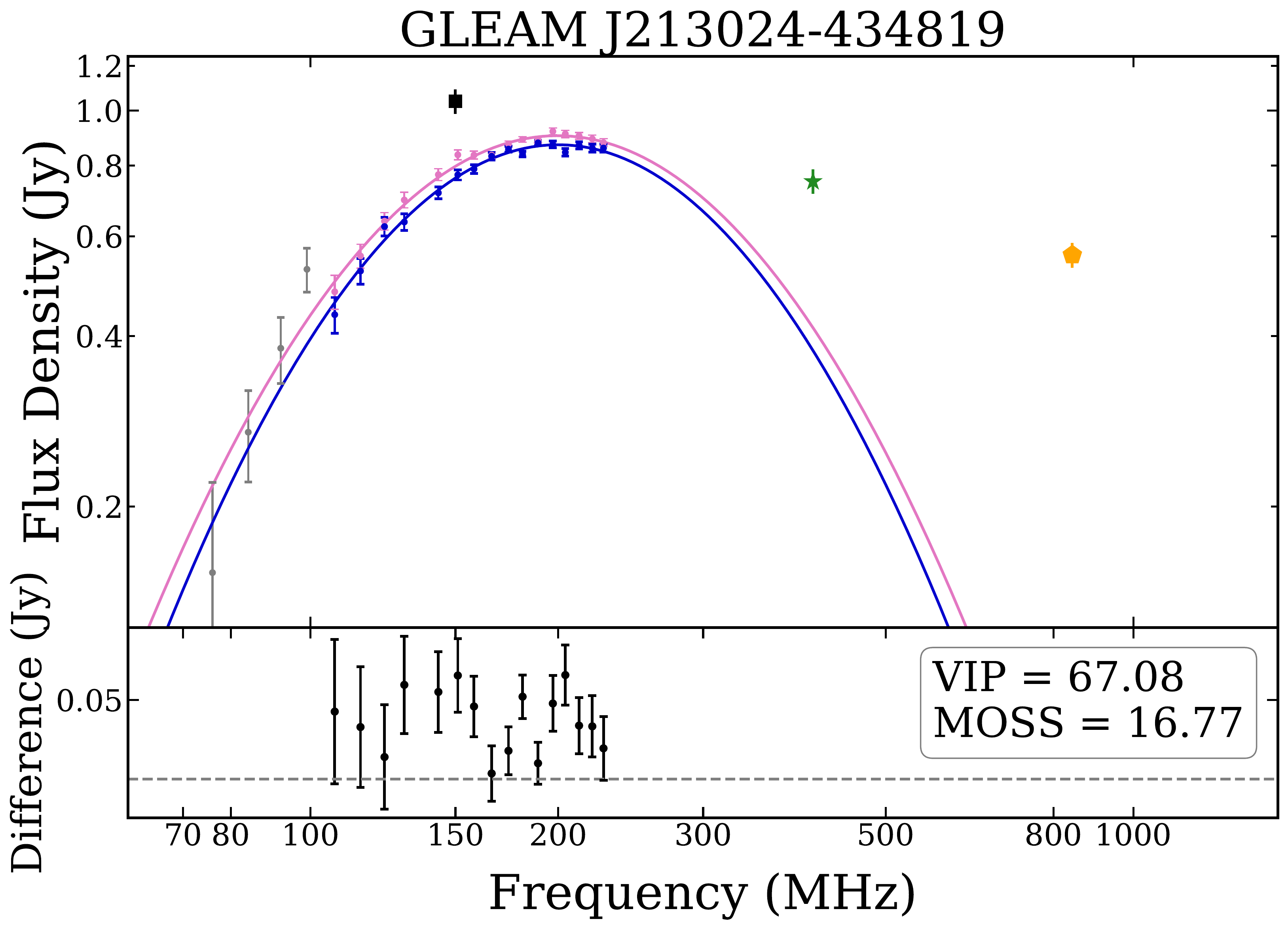} &
\includegraphics[scale=0.15]{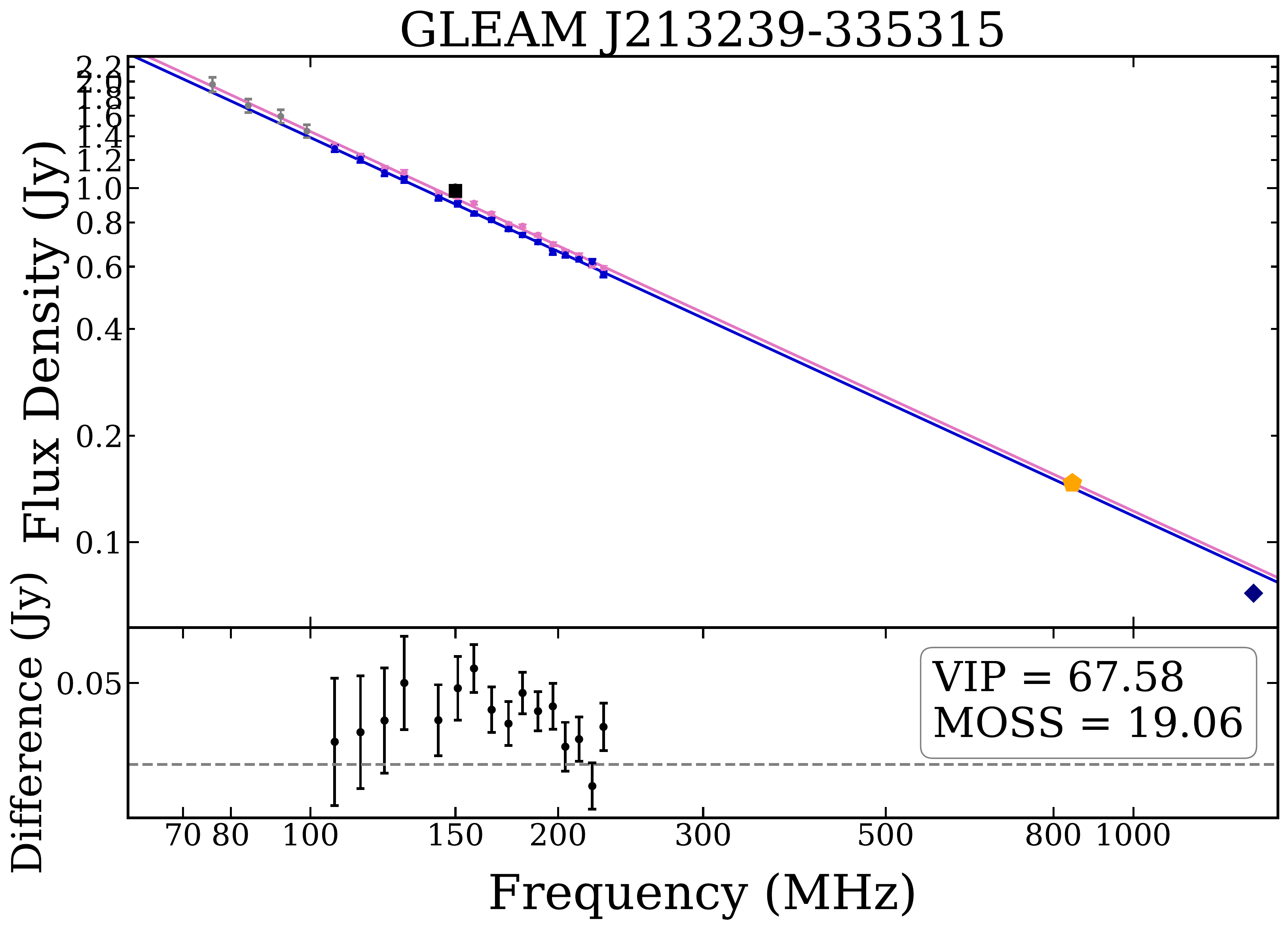} &
\includegraphics[scale=0.15]{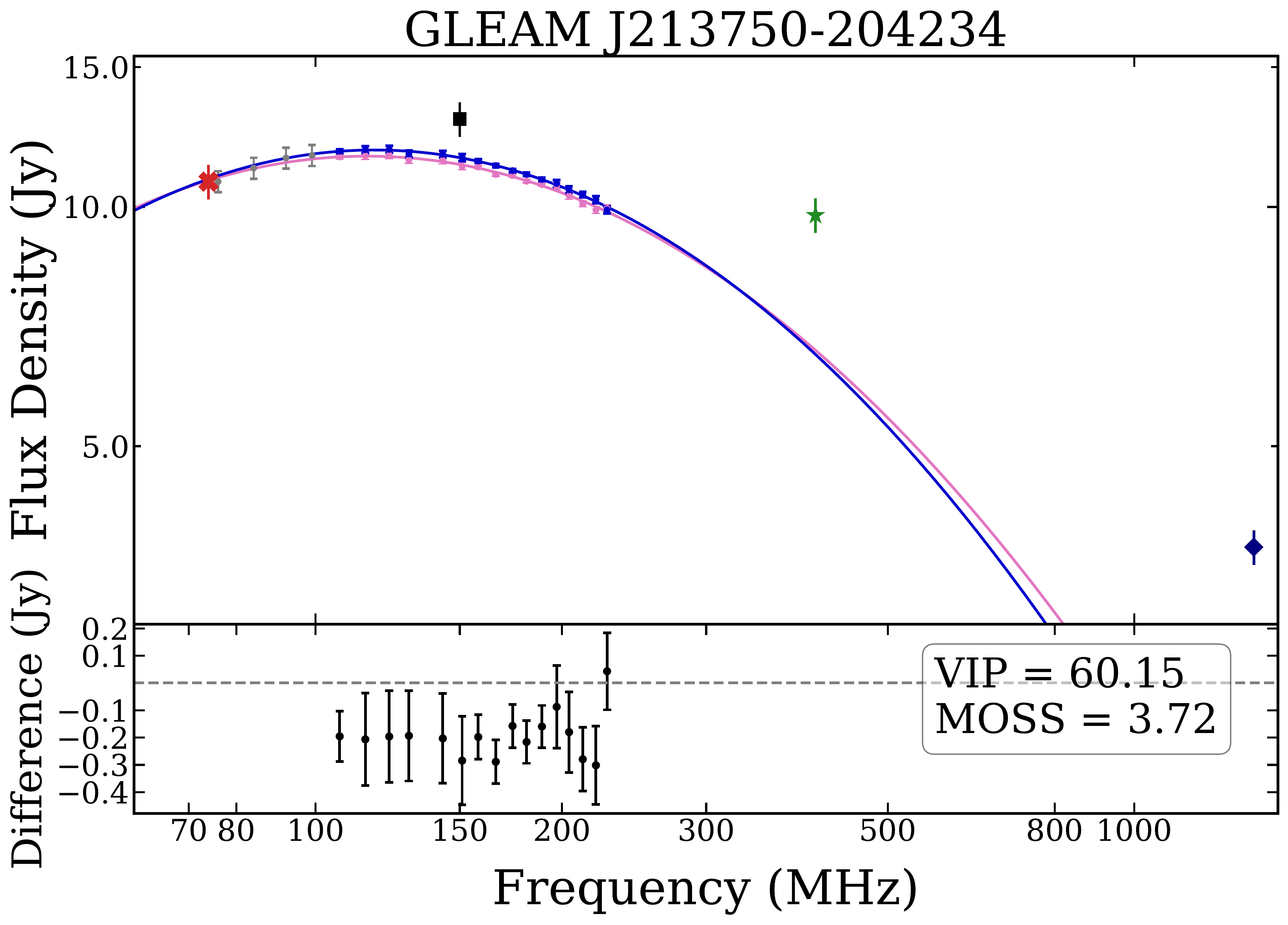} \\
\includegraphics[scale=0.15]{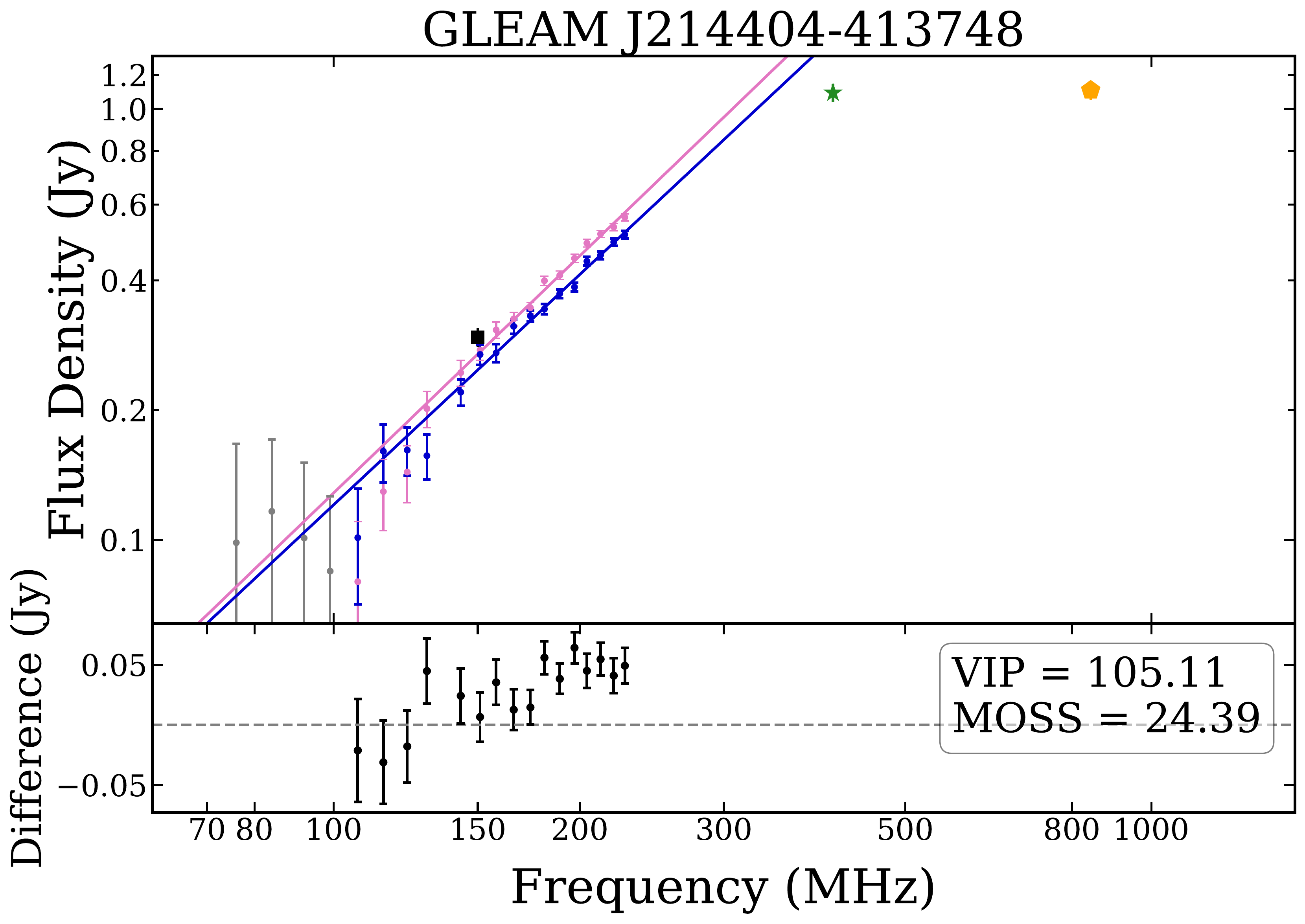} &
\includegraphics[scale=0.15]{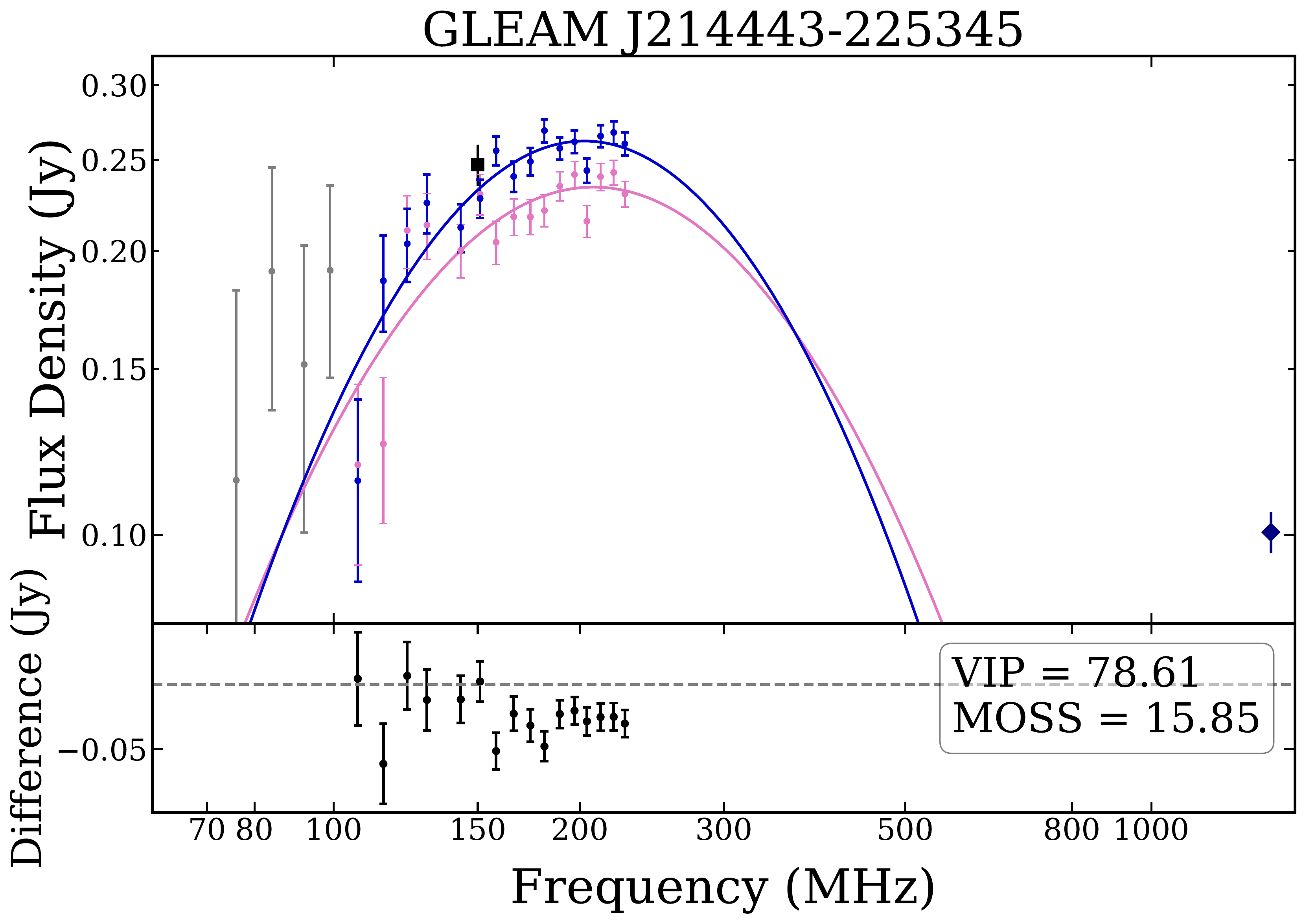} &
\includegraphics[scale=0.15]{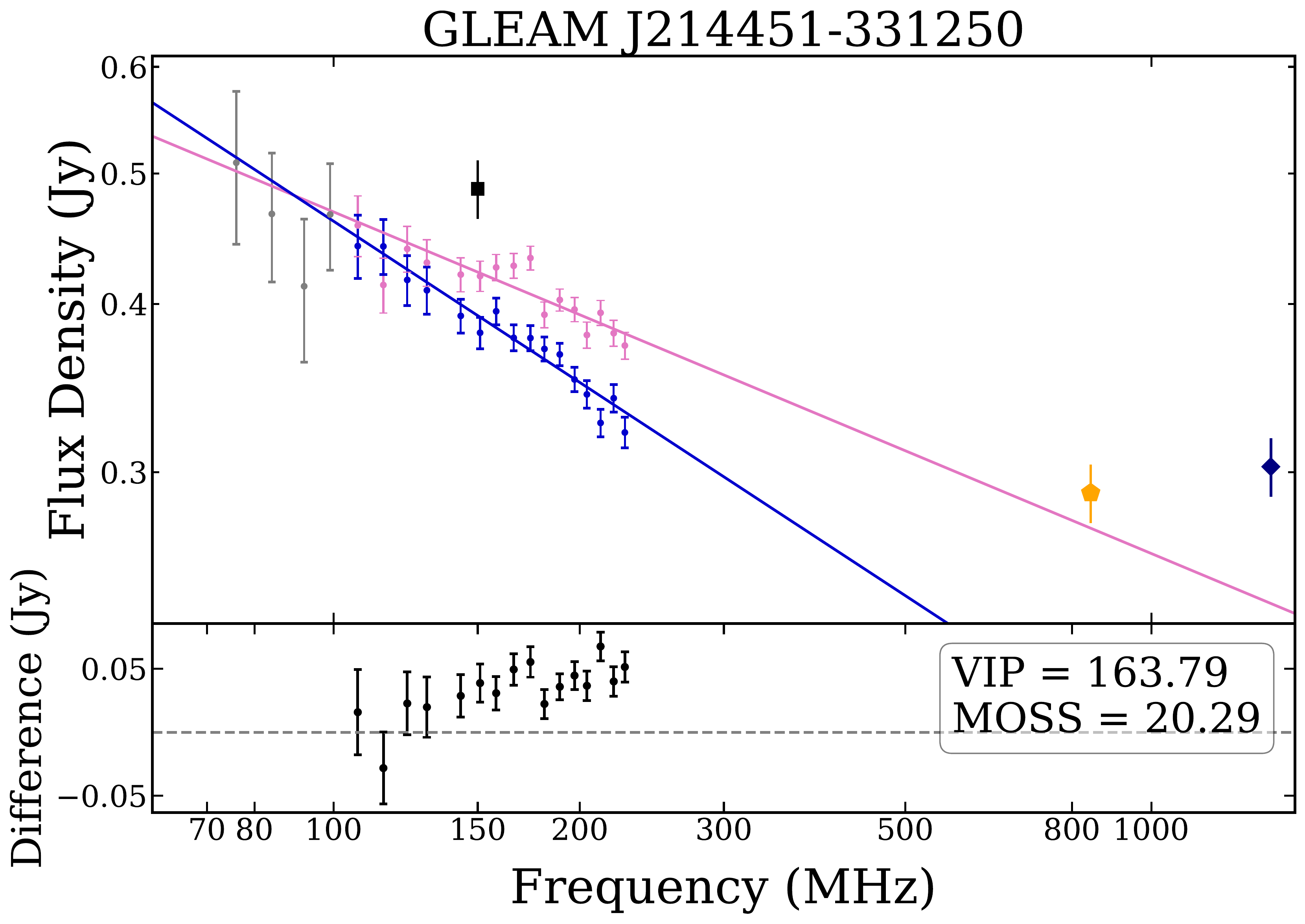} \\
\includegraphics[scale=0.15]{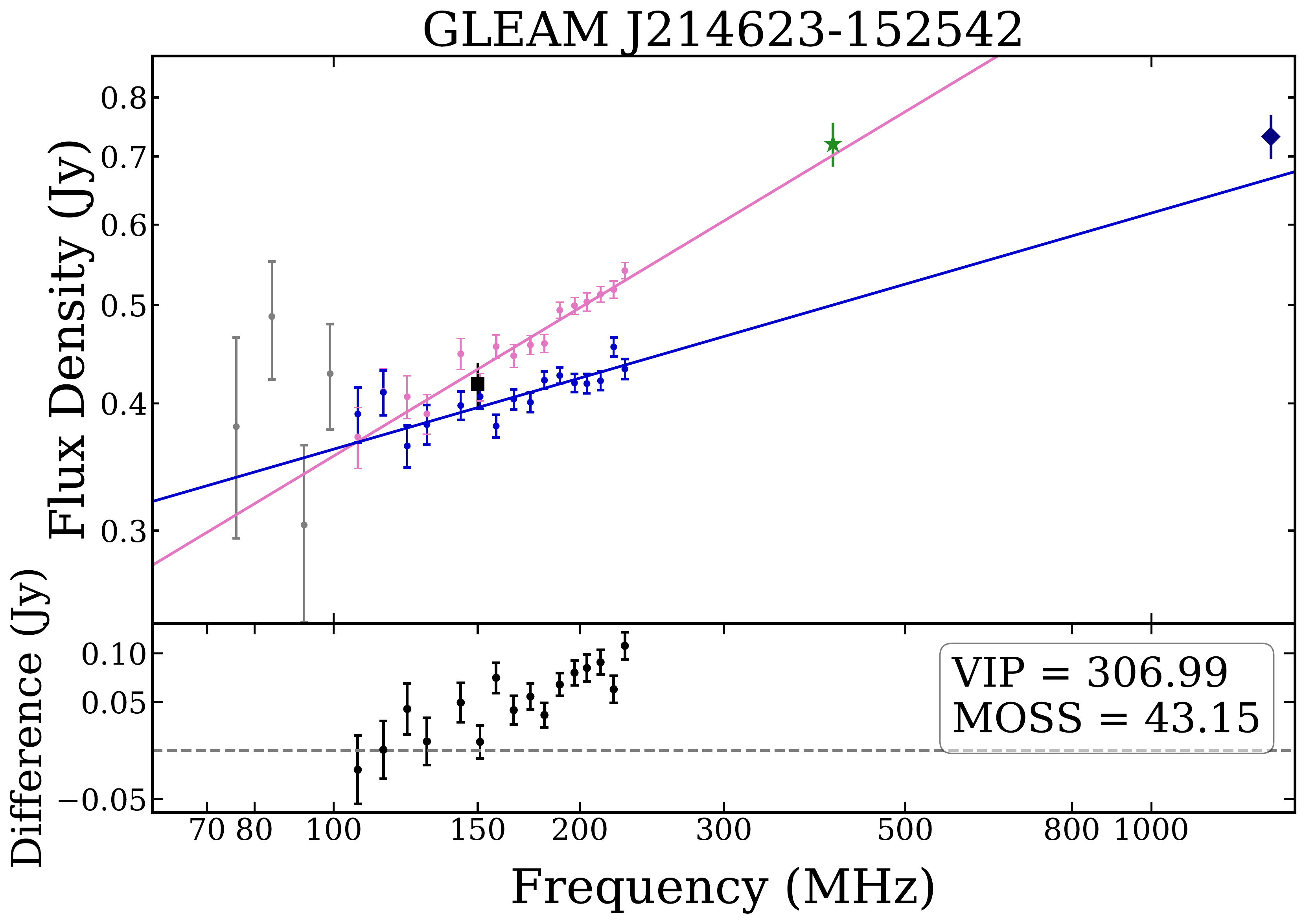} &
\includegraphics[scale=0.15]{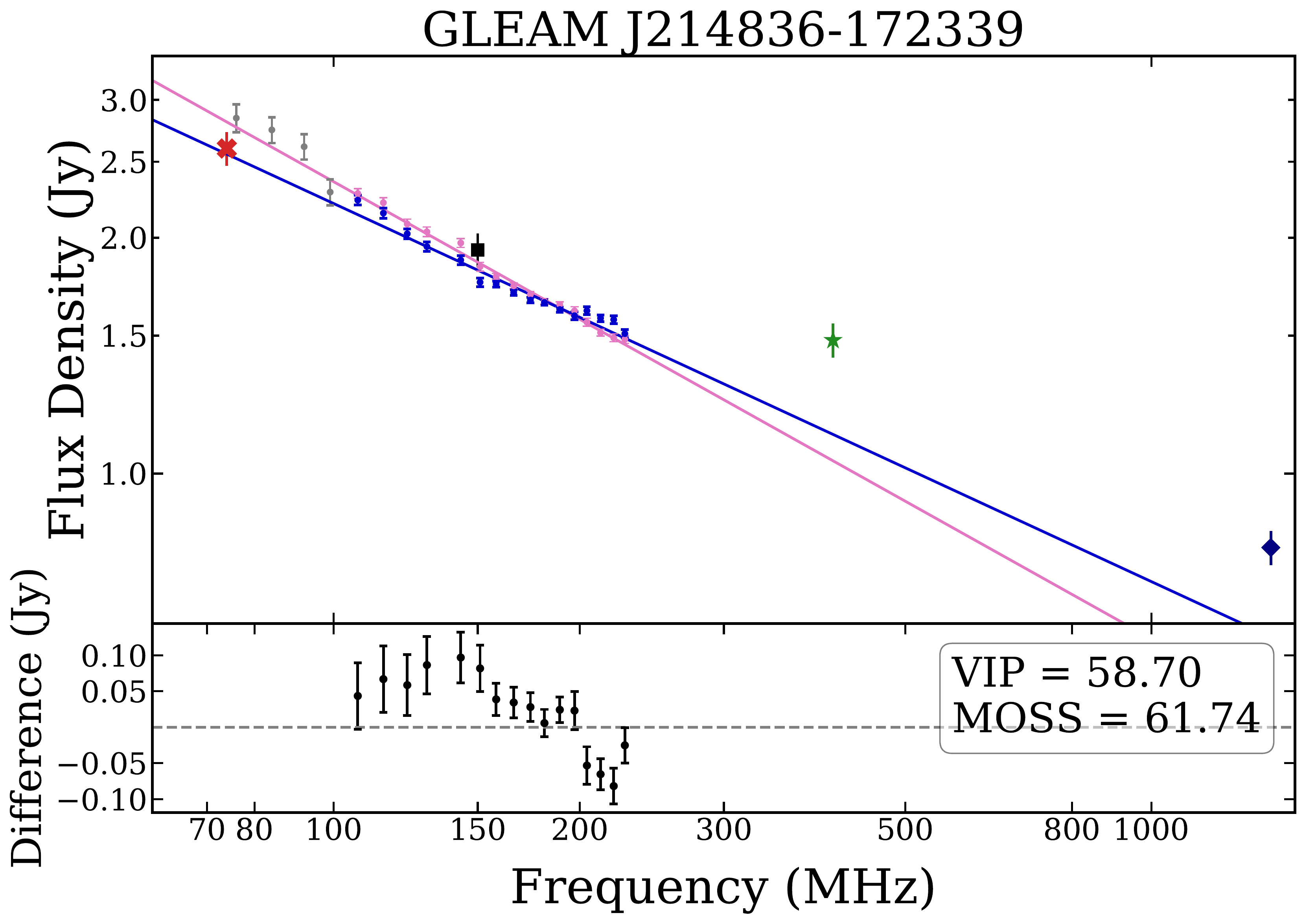} &
\includegraphics[scale=0.15]{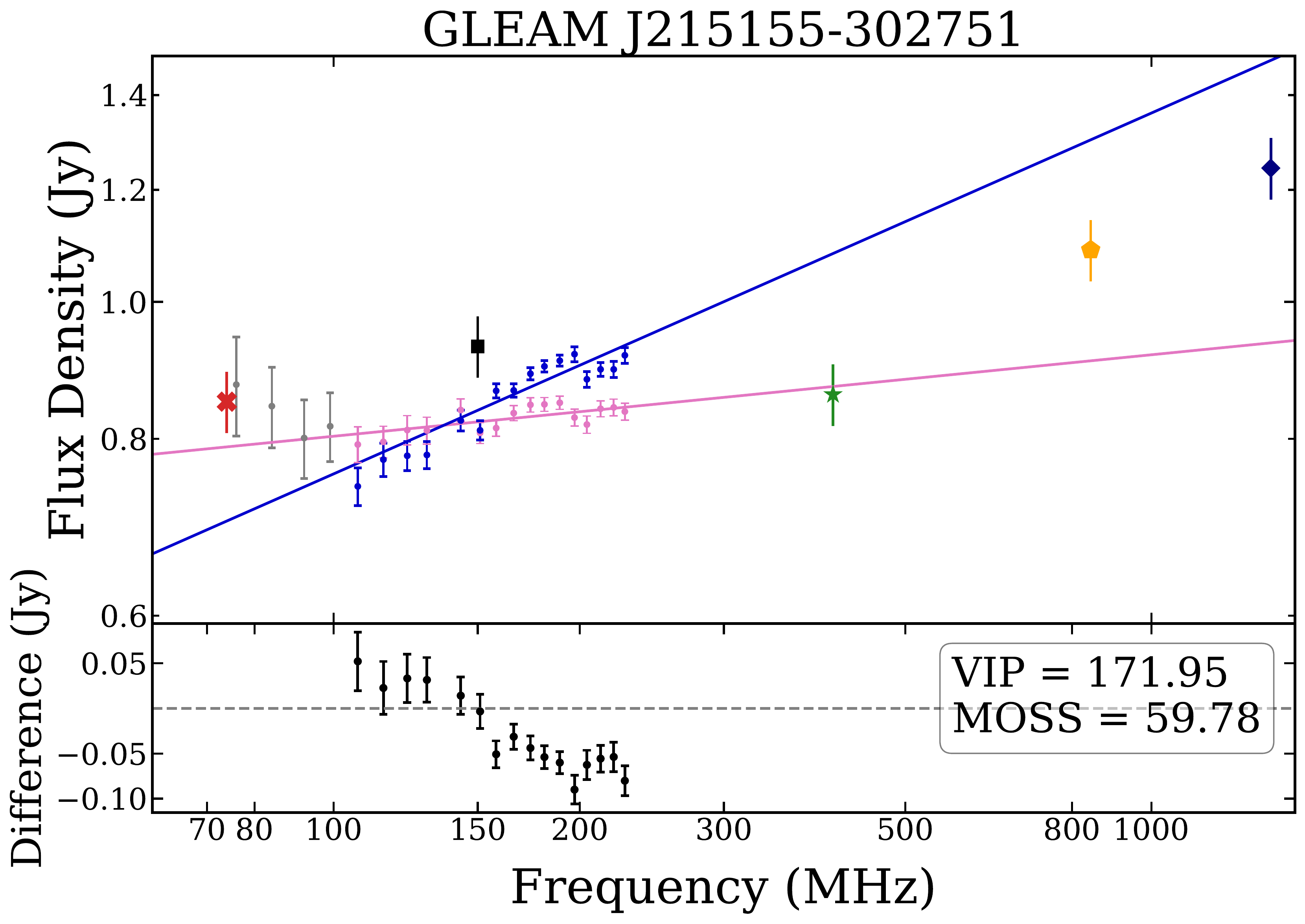} \\
\end{array}$
\caption{(continued) SEDs for all sources classified as variable according to the VIP. For each source the points represent the following data: GLEAM low frequency (72--100\,MHz) (grey circles), Year 1 (pink circles), Year 2 (blue circles), VLSSr (red cross), TGSS (black square), MRC (green star), SUMSS (yellow pentagon), and NVSS (navy diamond). The models for each year are determined by their classification; a source classified with a peak within the observed band was modelled by a quadratic according to Equation~\ref{eq:quadratic}, remaining sources were modelled by a power-law according to Equation~\ref{eq:plaw}.}
\label{app:fig:pg13}
\end{center}
\end{figure*}
\setcounter{figure}{0}
\begin{figure*}
\begin{center}$
\begin{array}{cccccc}
\includegraphics[scale=0.15]{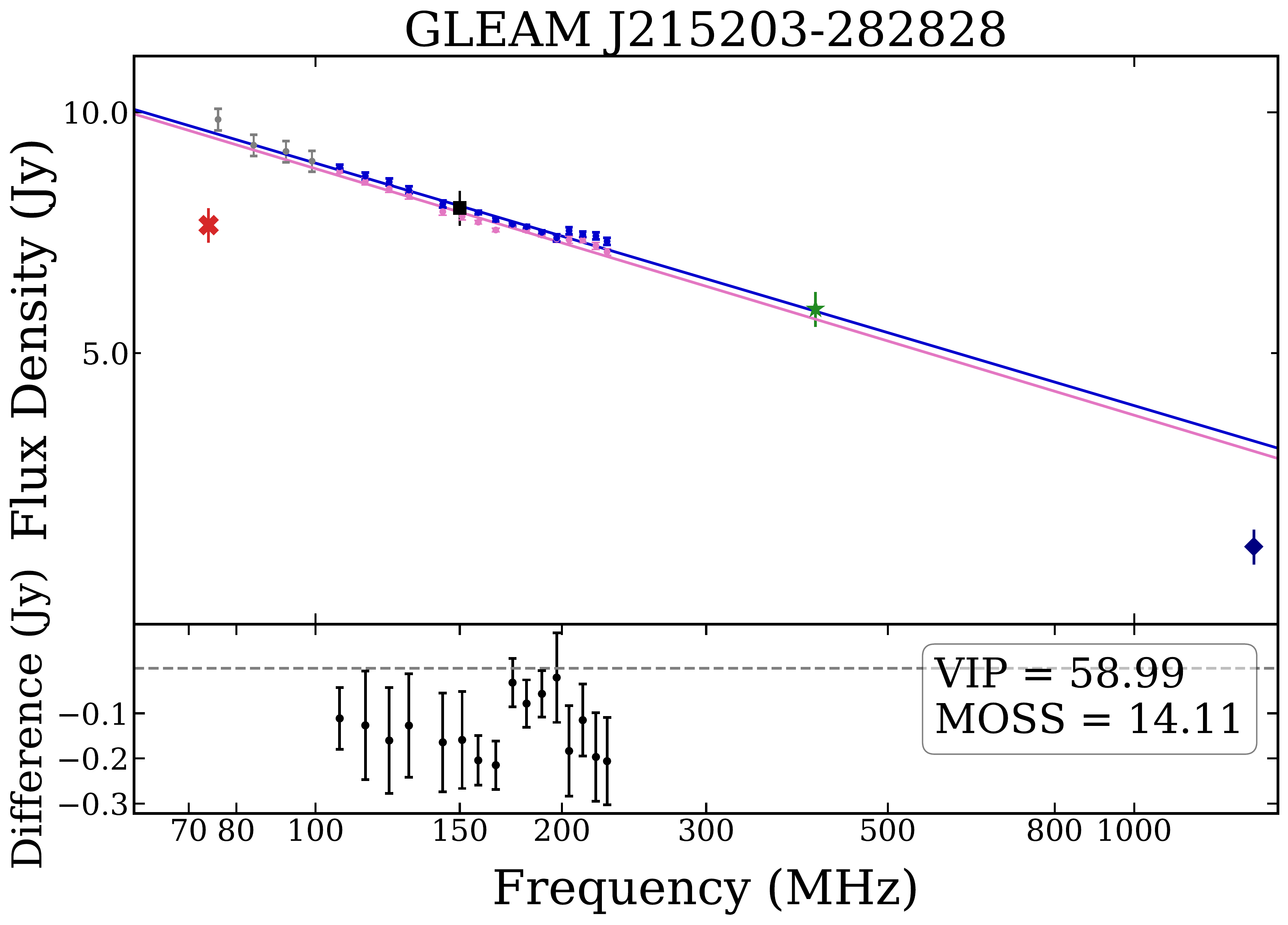} &
\includegraphics[scale=0.15]{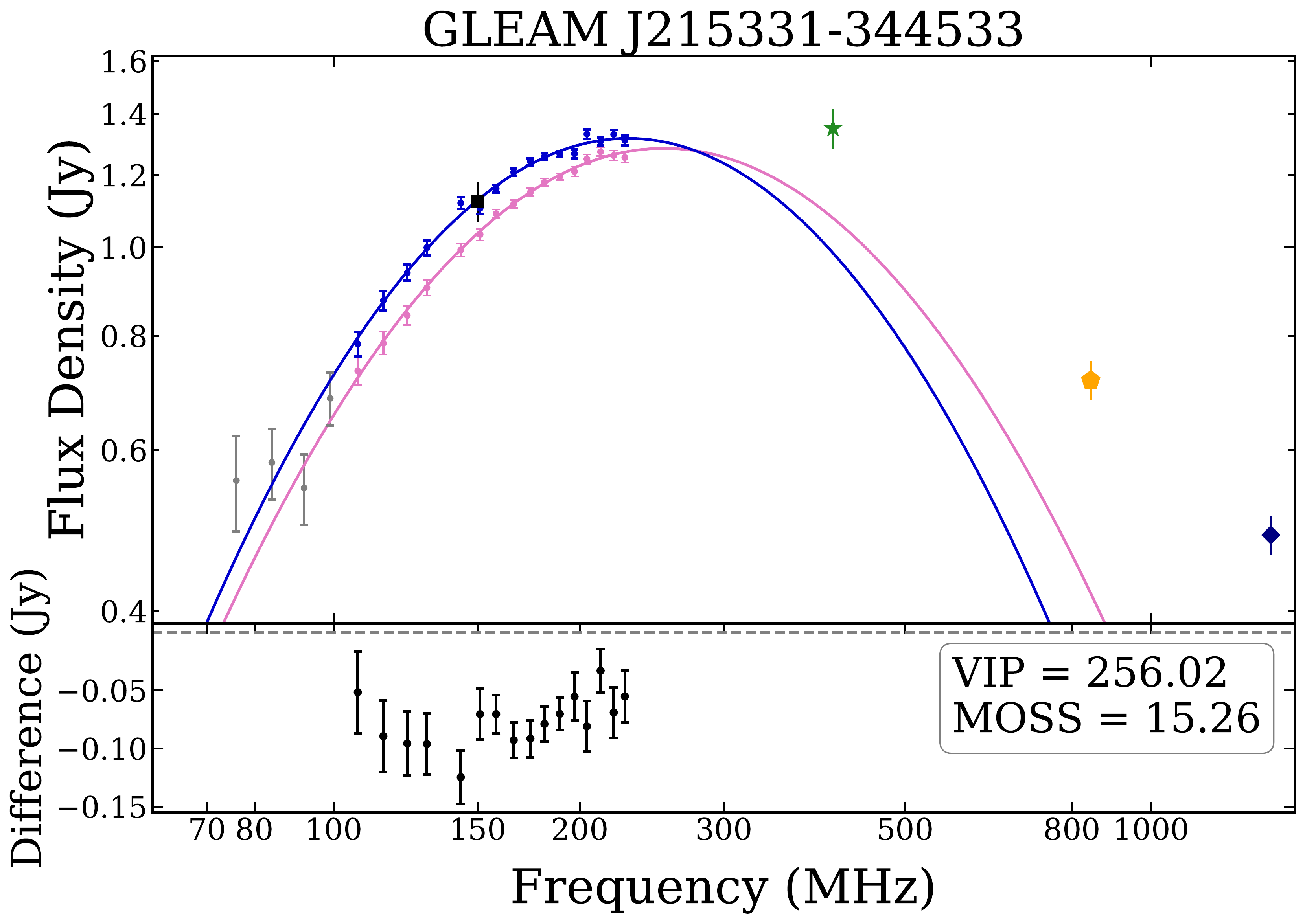} &
\includegraphics[scale=0.15]{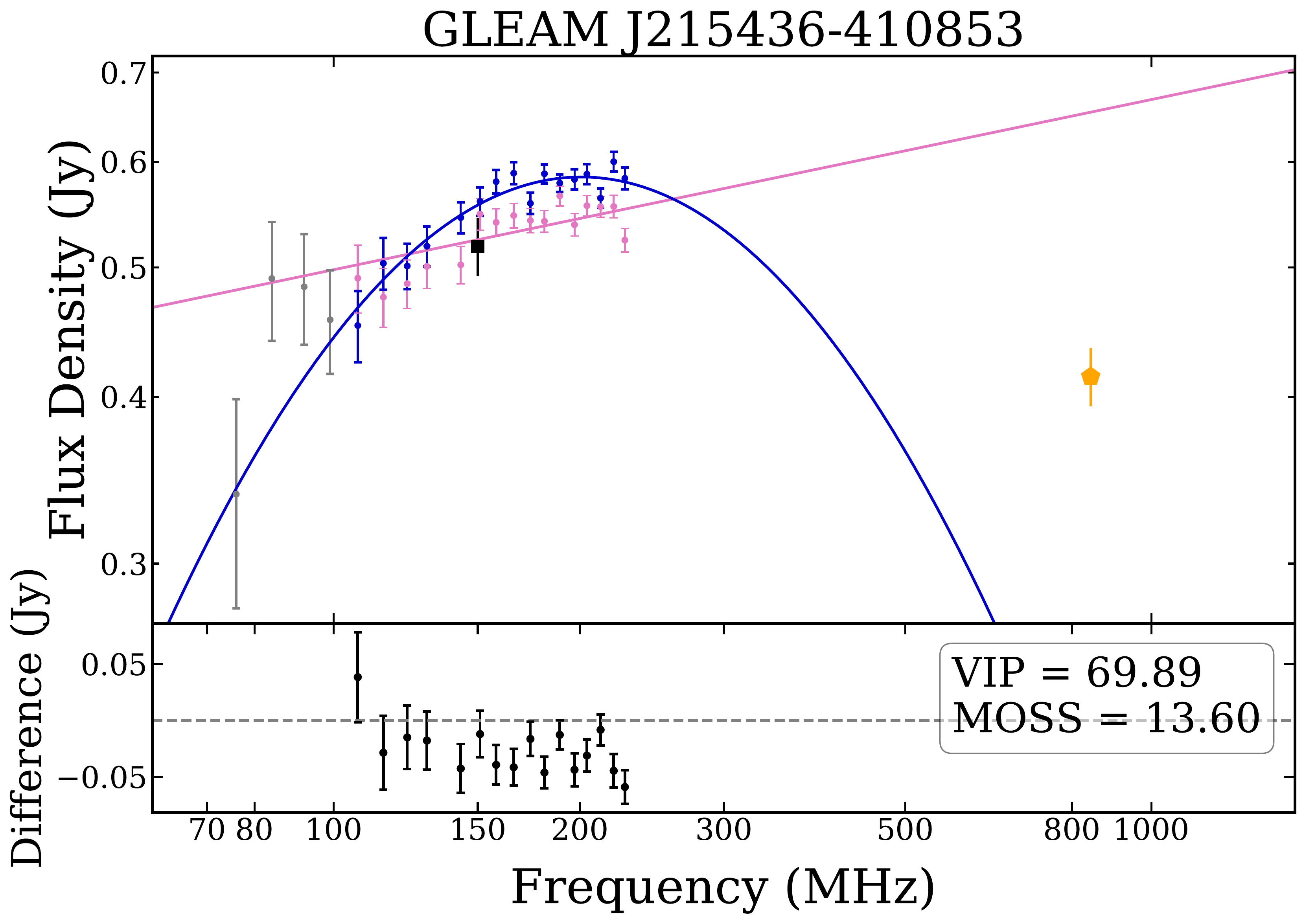} \\
\includegraphics[scale=0.15]{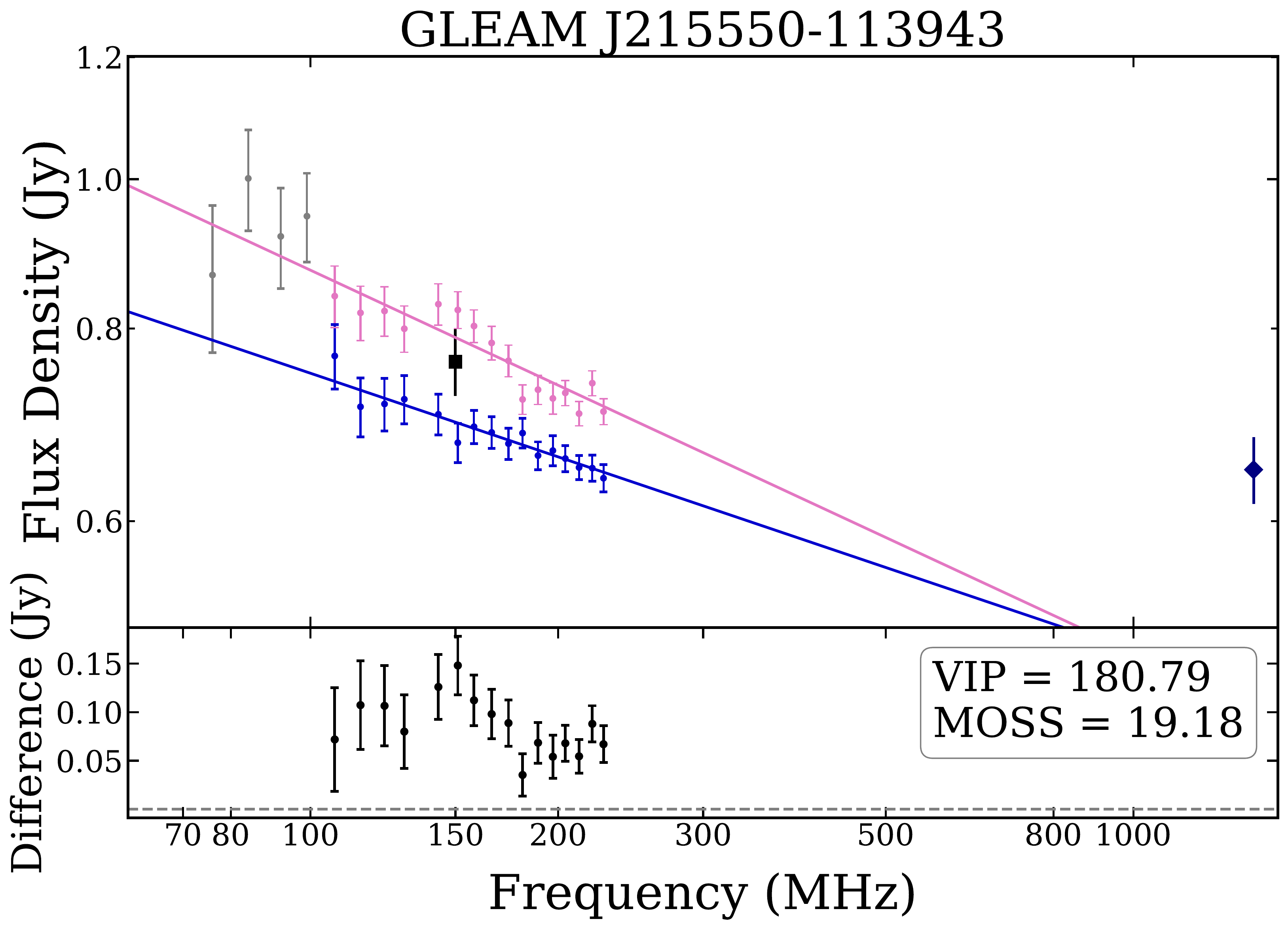} &
\includegraphics[scale=0.15]{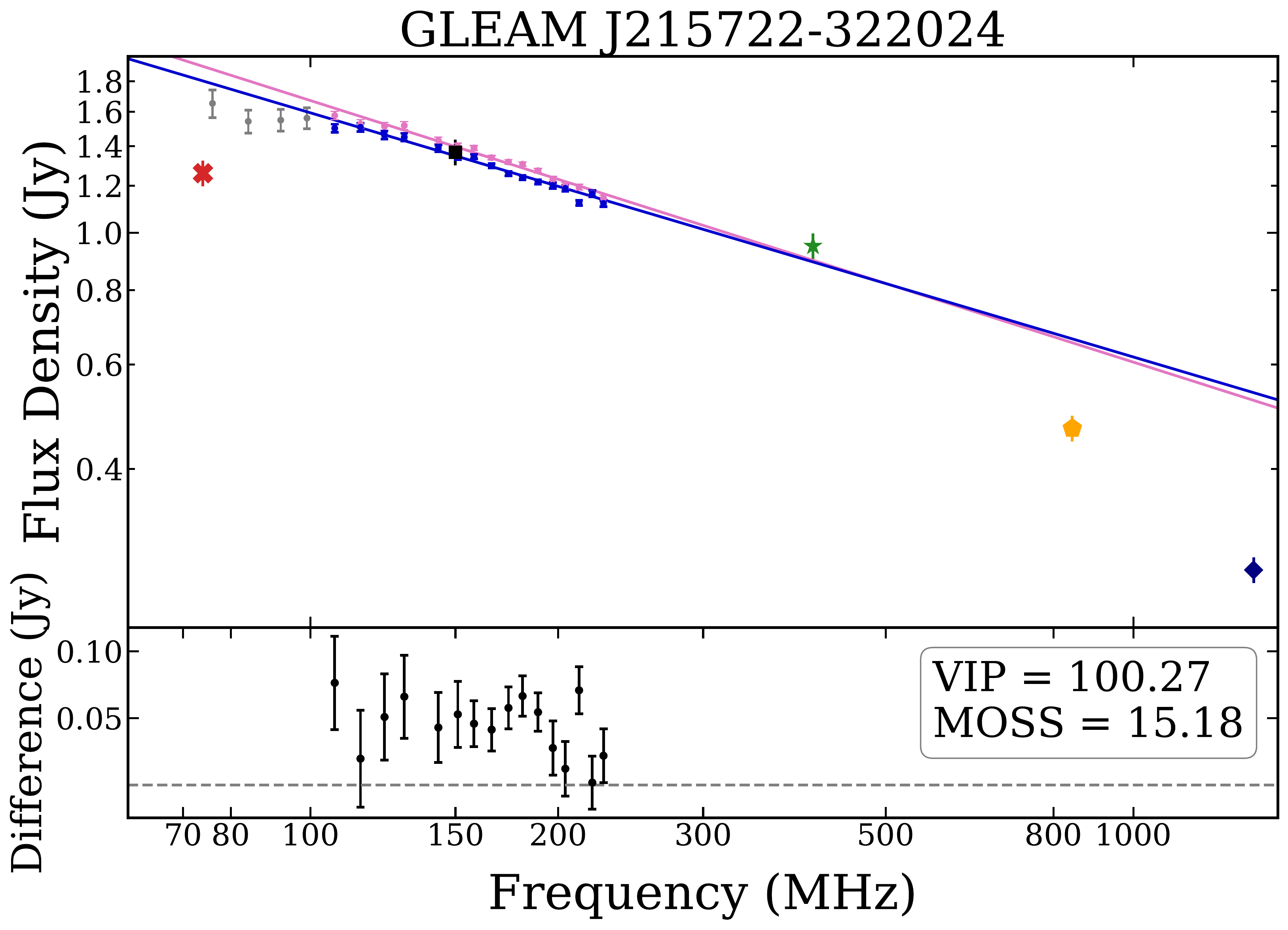} &
\includegraphics[scale=0.15]{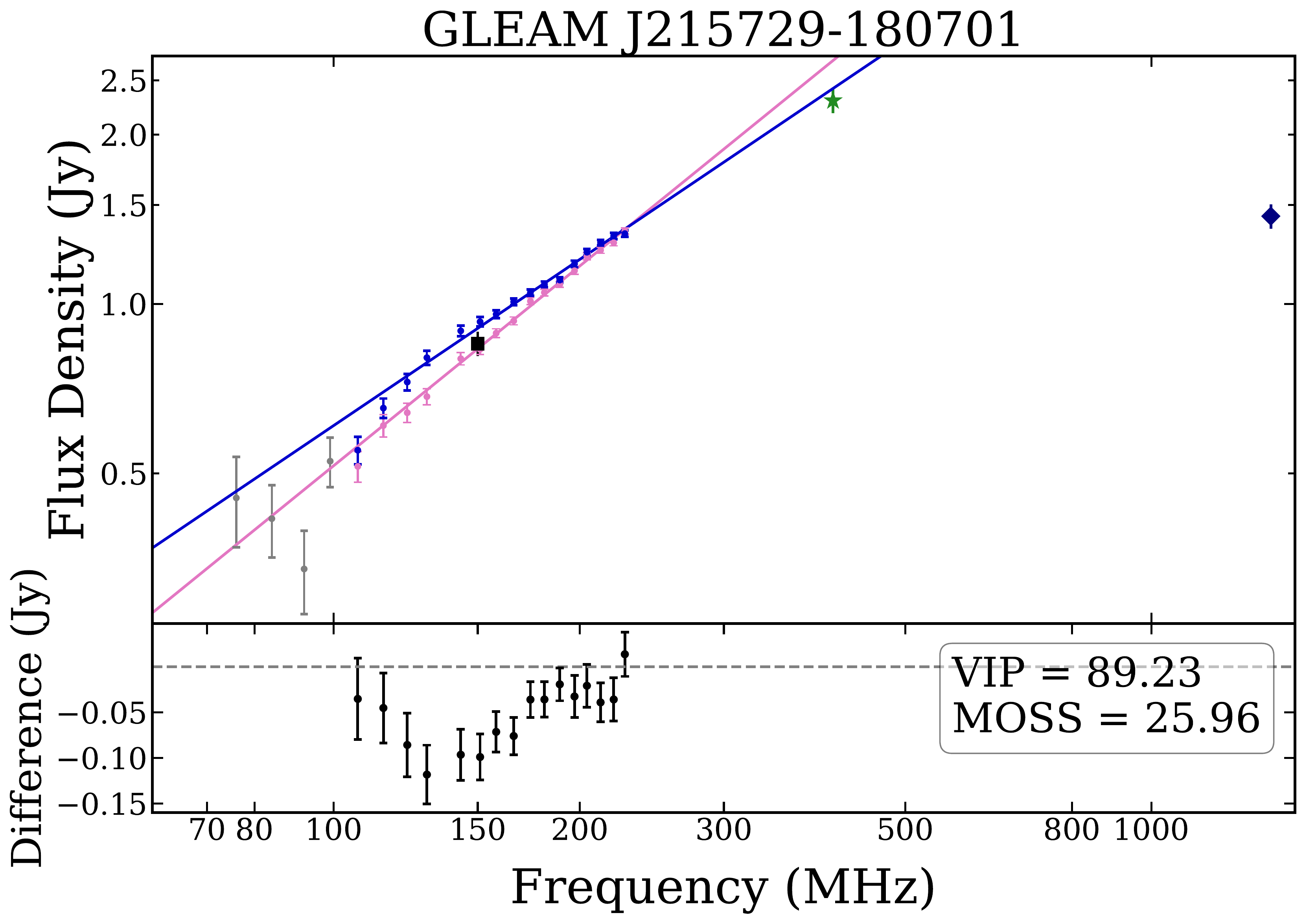} \\
\includegraphics[scale=0.15]{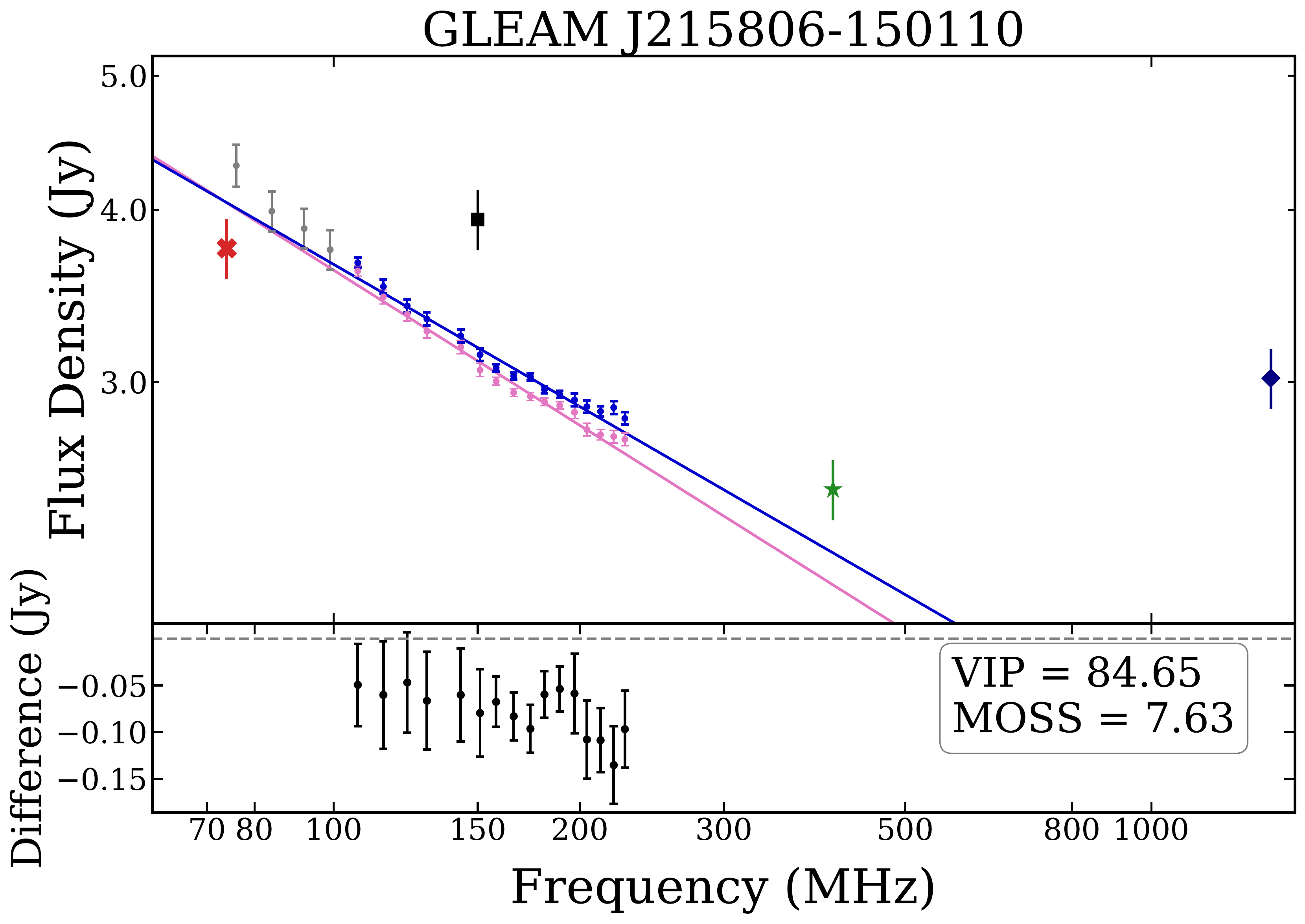} &
\includegraphics[scale=0.15]{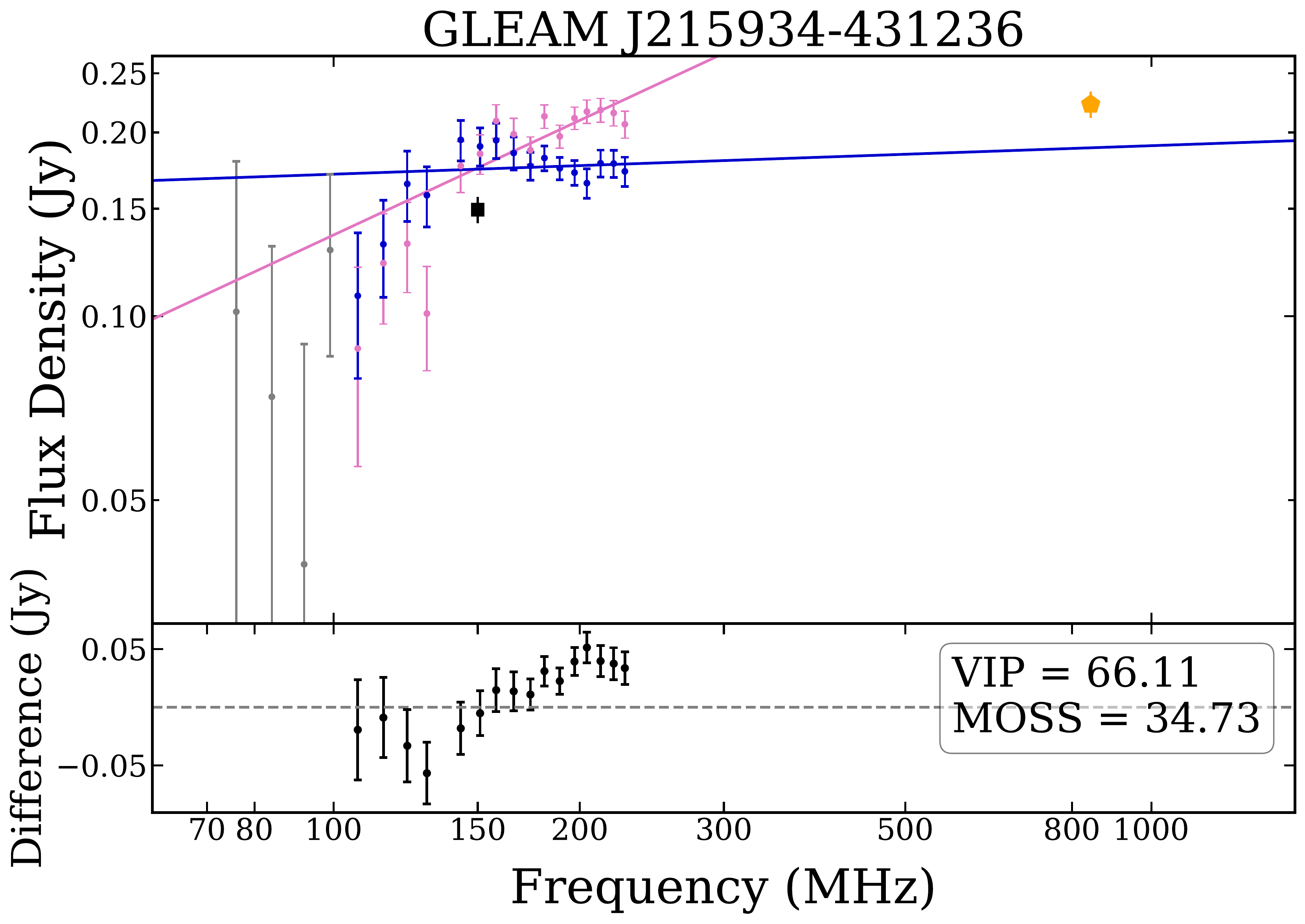} &
\includegraphics[scale=0.15]{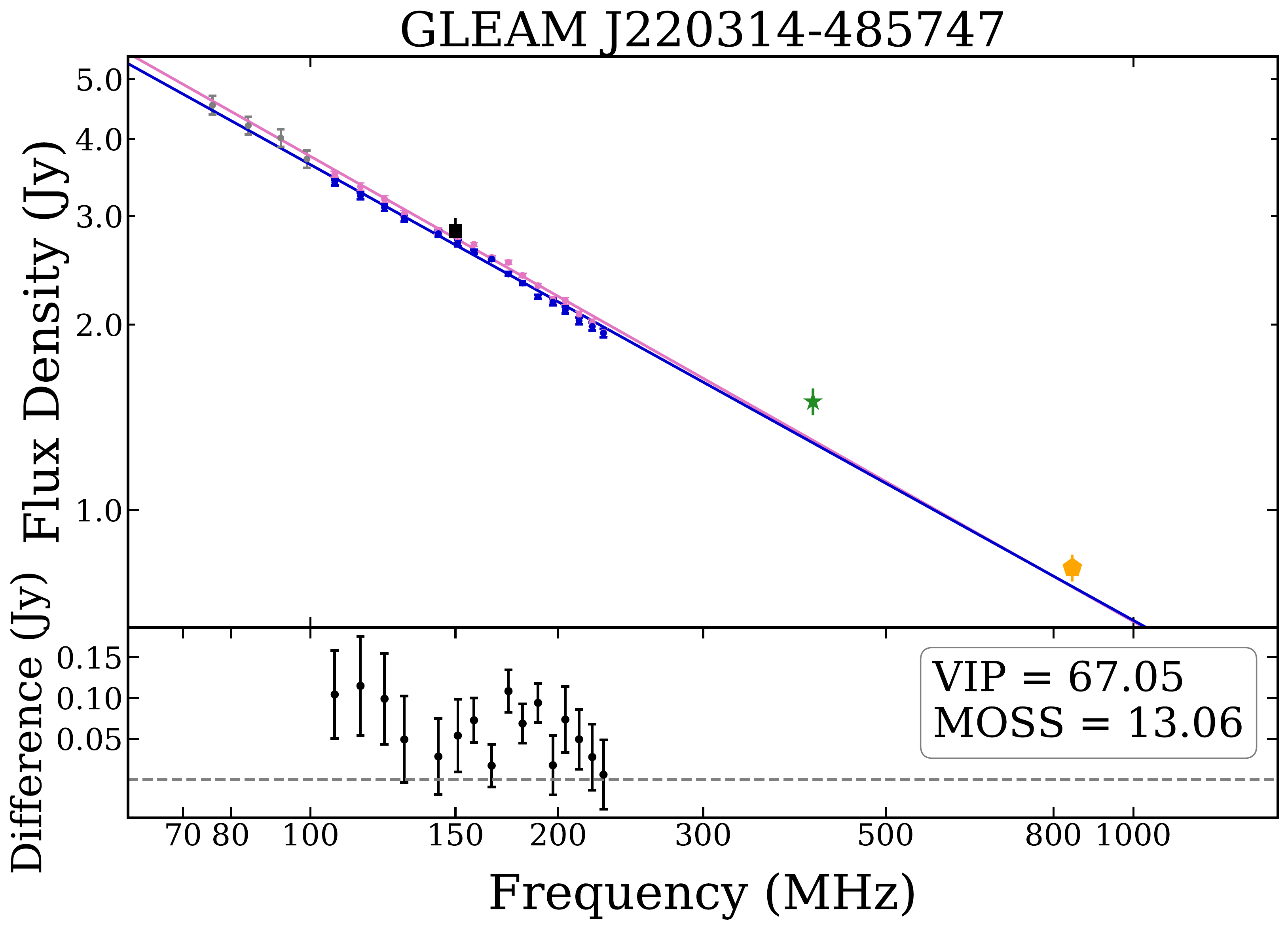} \\
\includegraphics[scale=0.15]{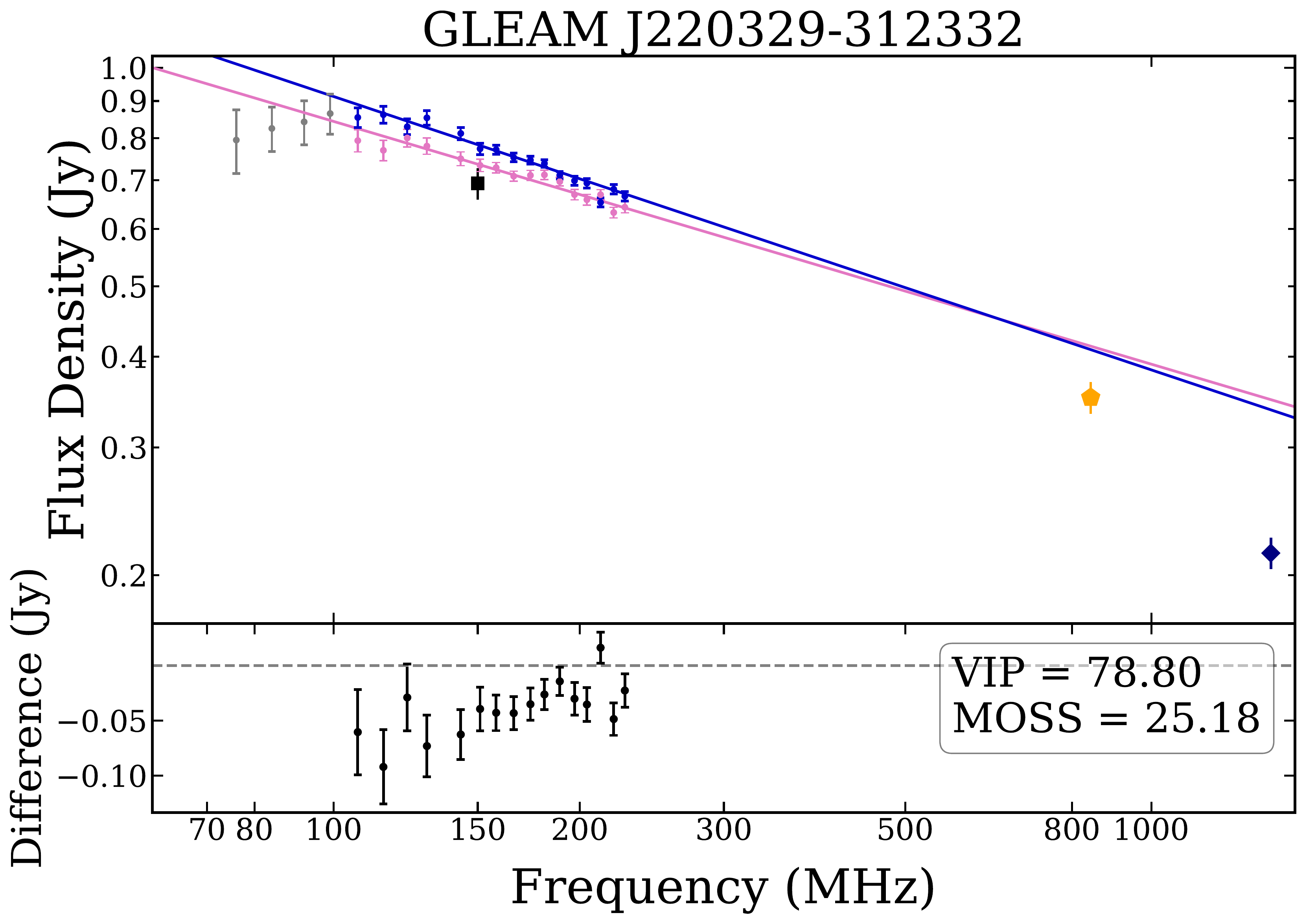} &
\includegraphics[scale=0.15]{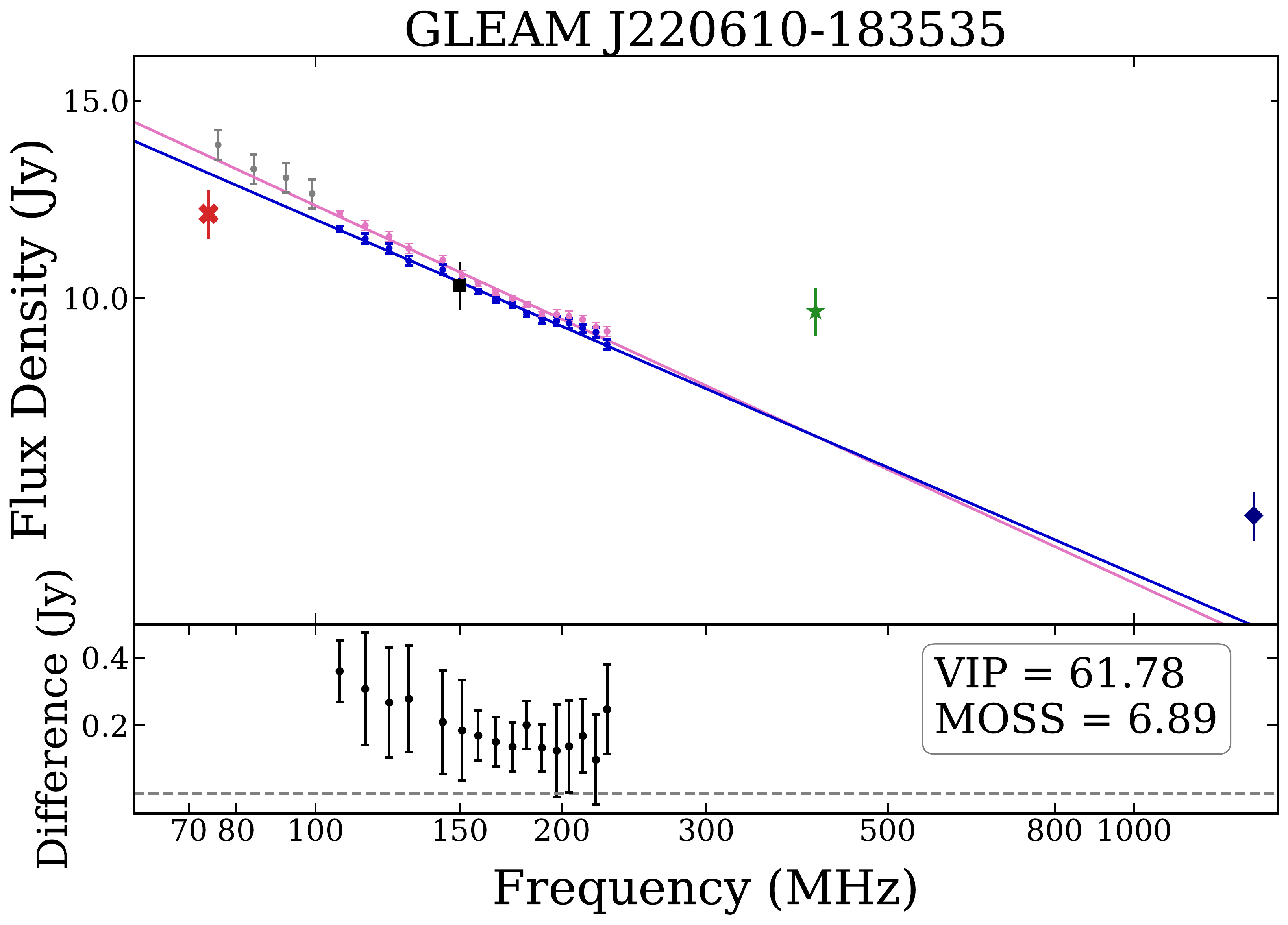} &
\includegraphics[scale=0.15]{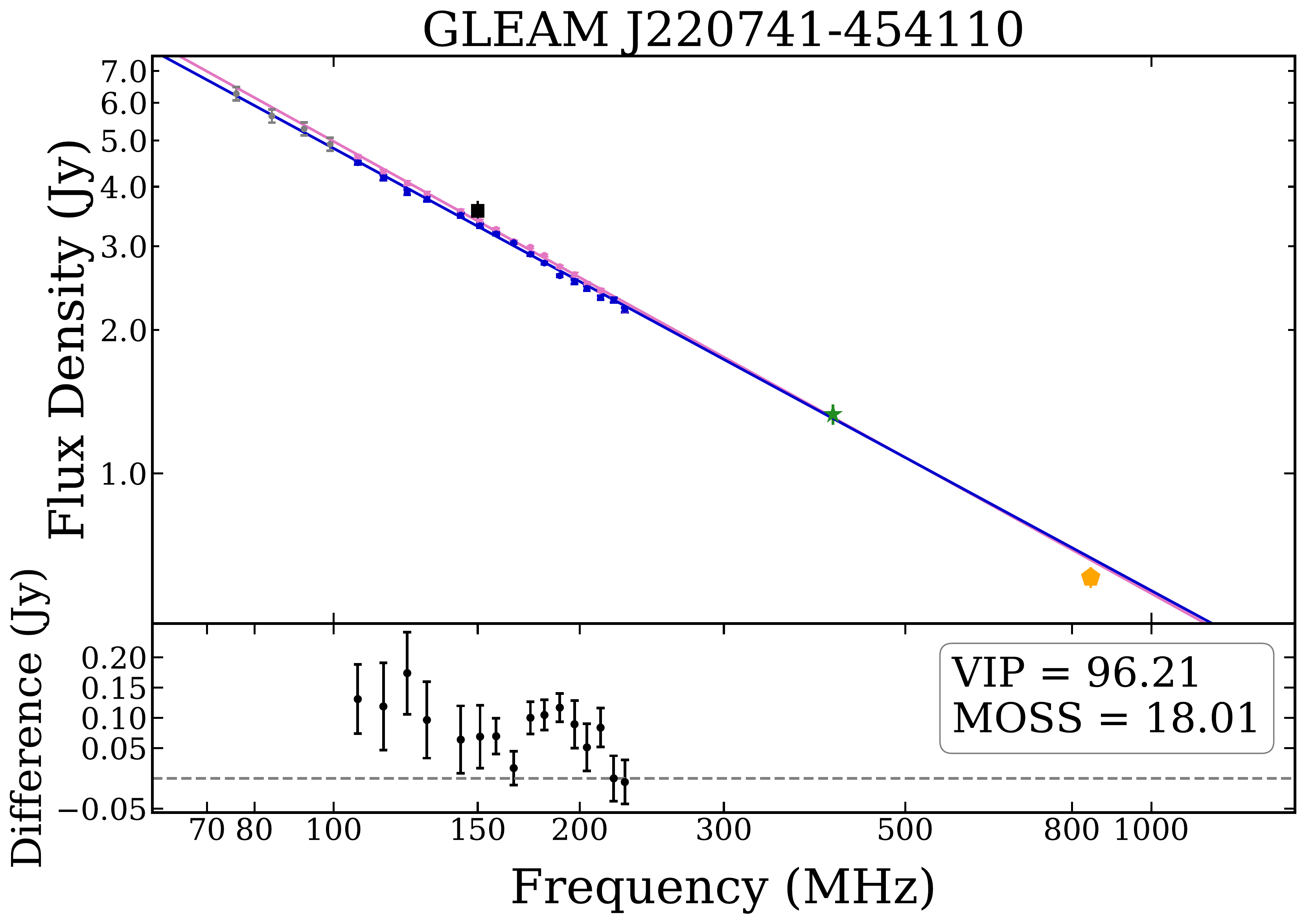} \\
\includegraphics[scale=0.15]{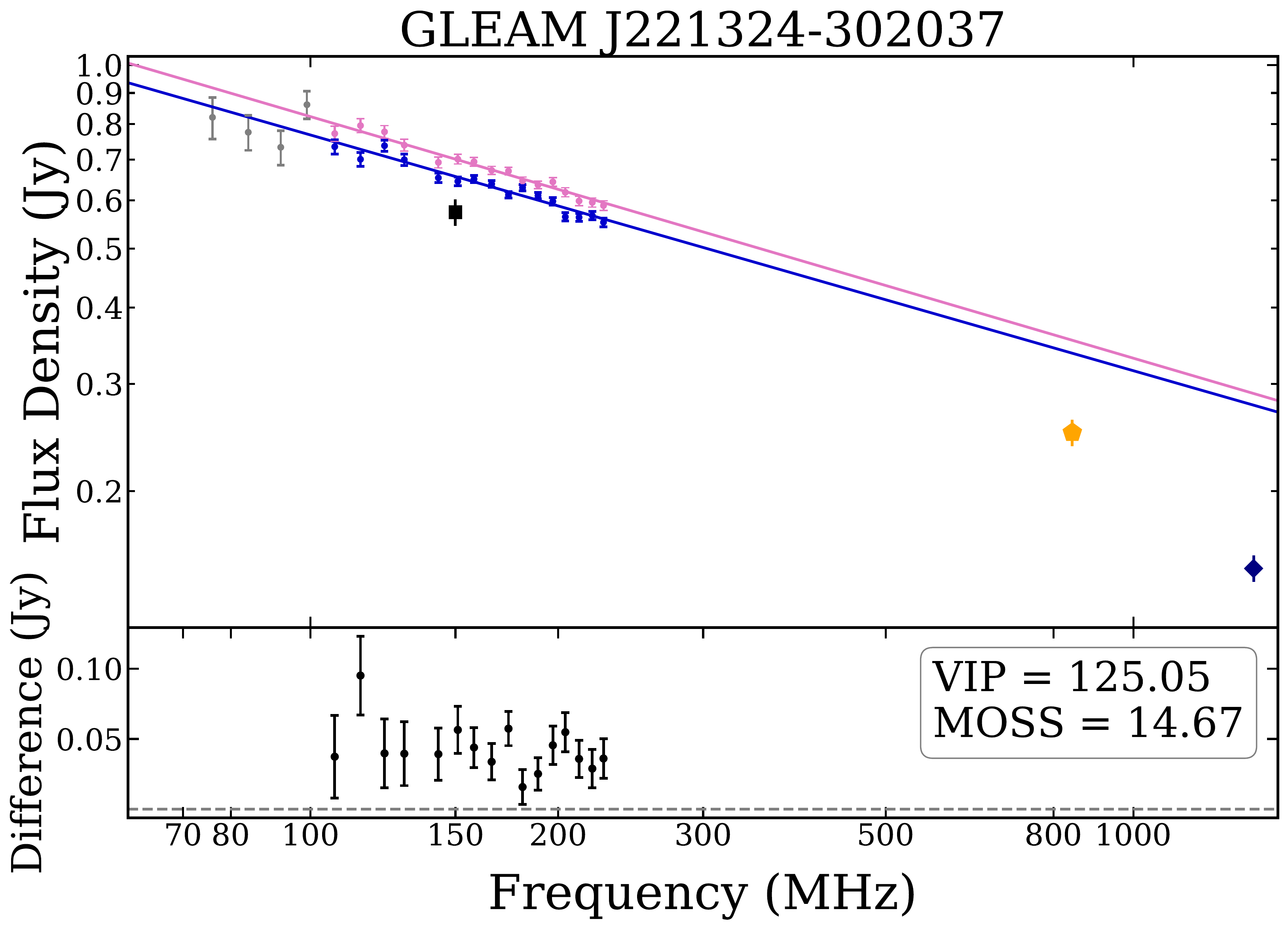} &
\includegraphics[scale=0.15]{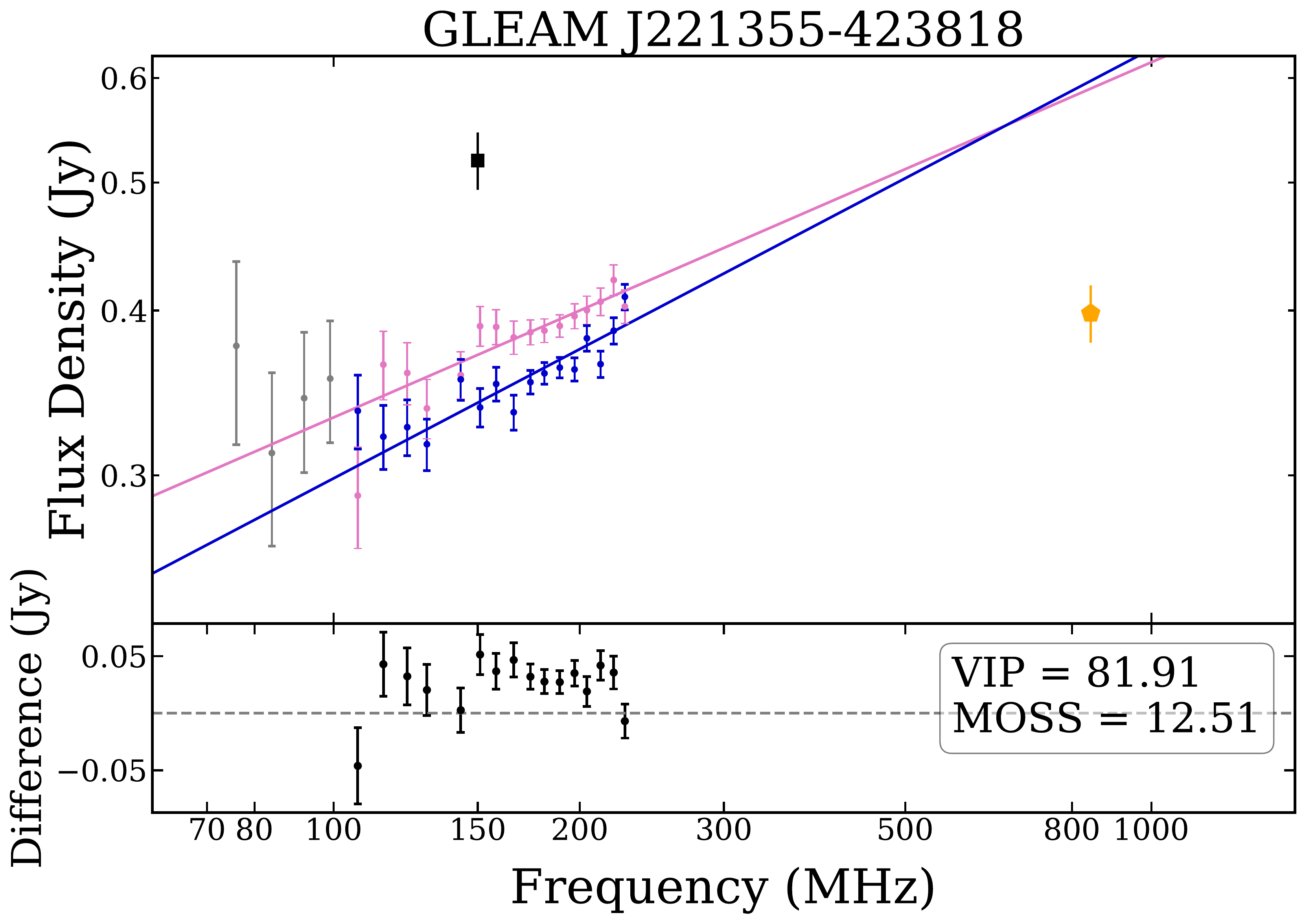} &
\includegraphics[scale=0.15]{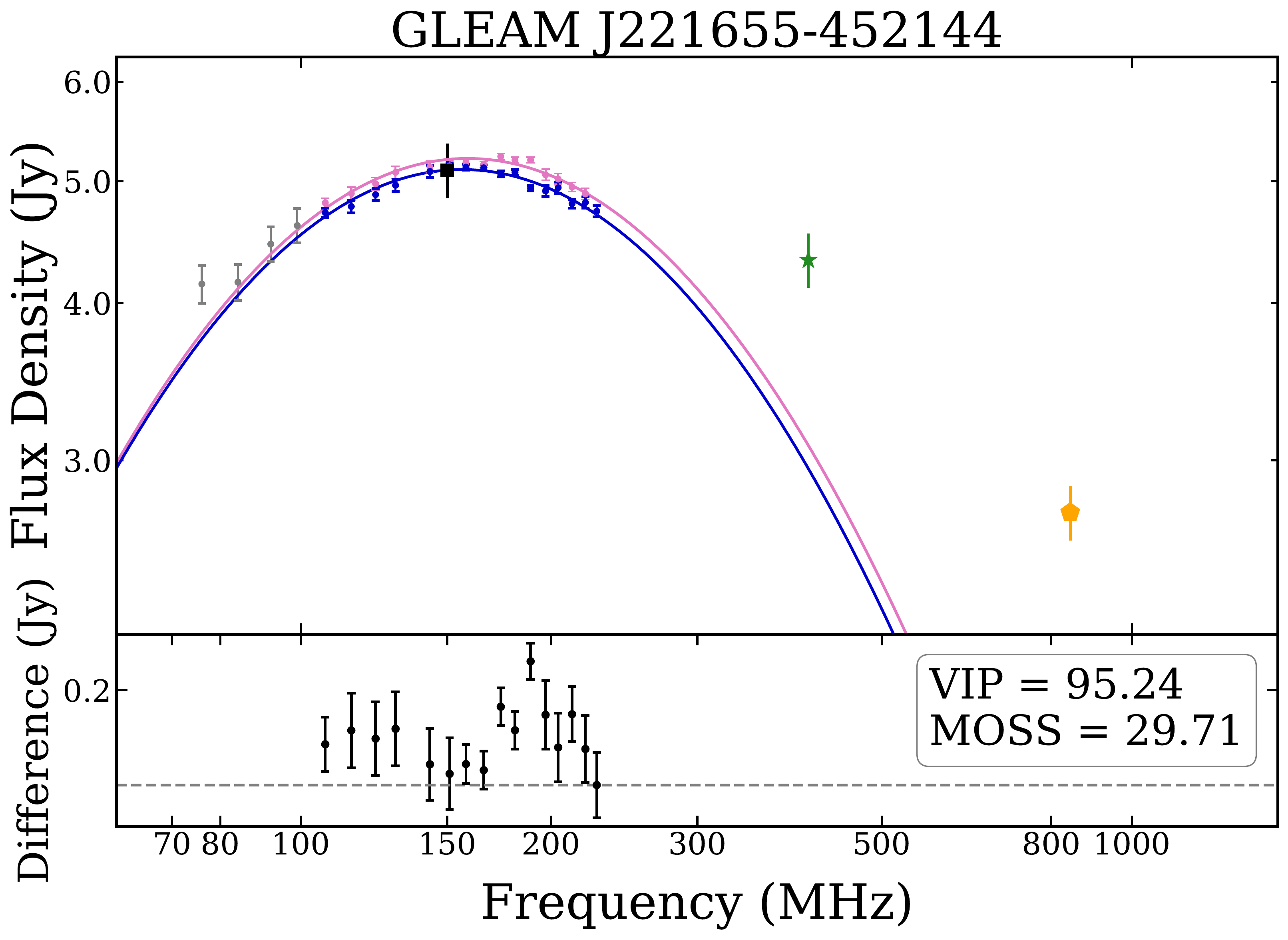} \\
\includegraphics[scale=0.15]{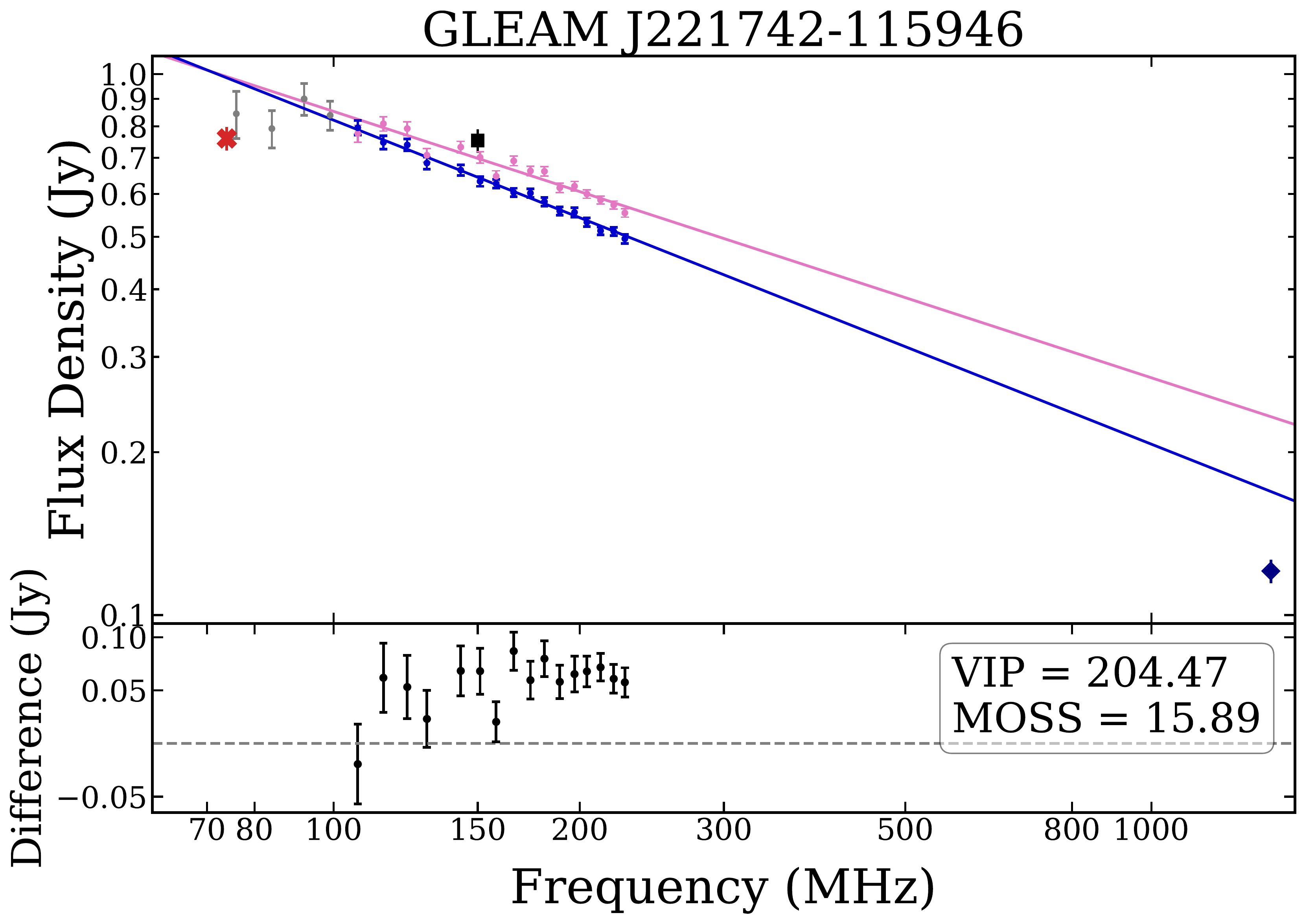} &
\includegraphics[scale=0.15]{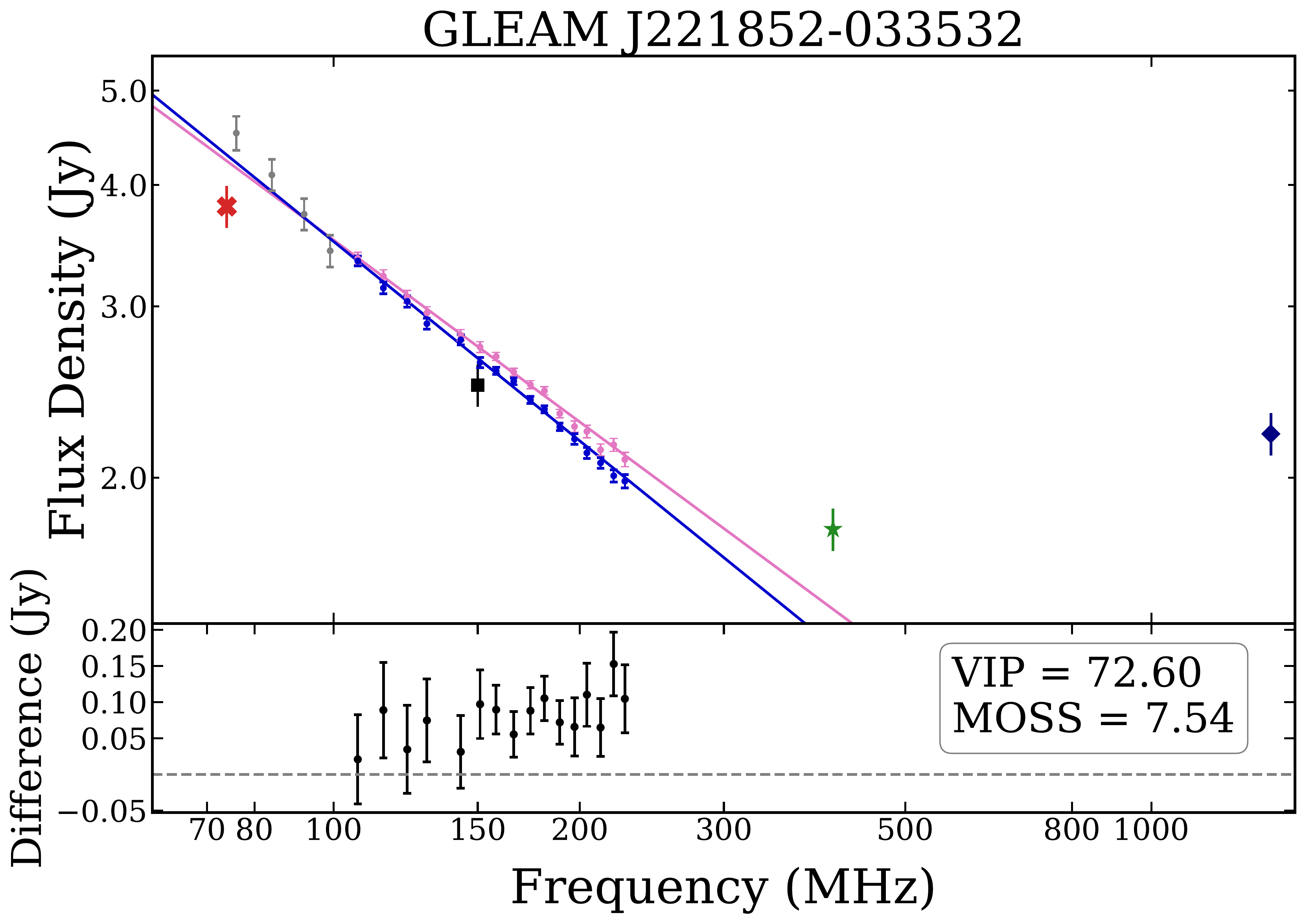} &
\includegraphics[scale=0.15]{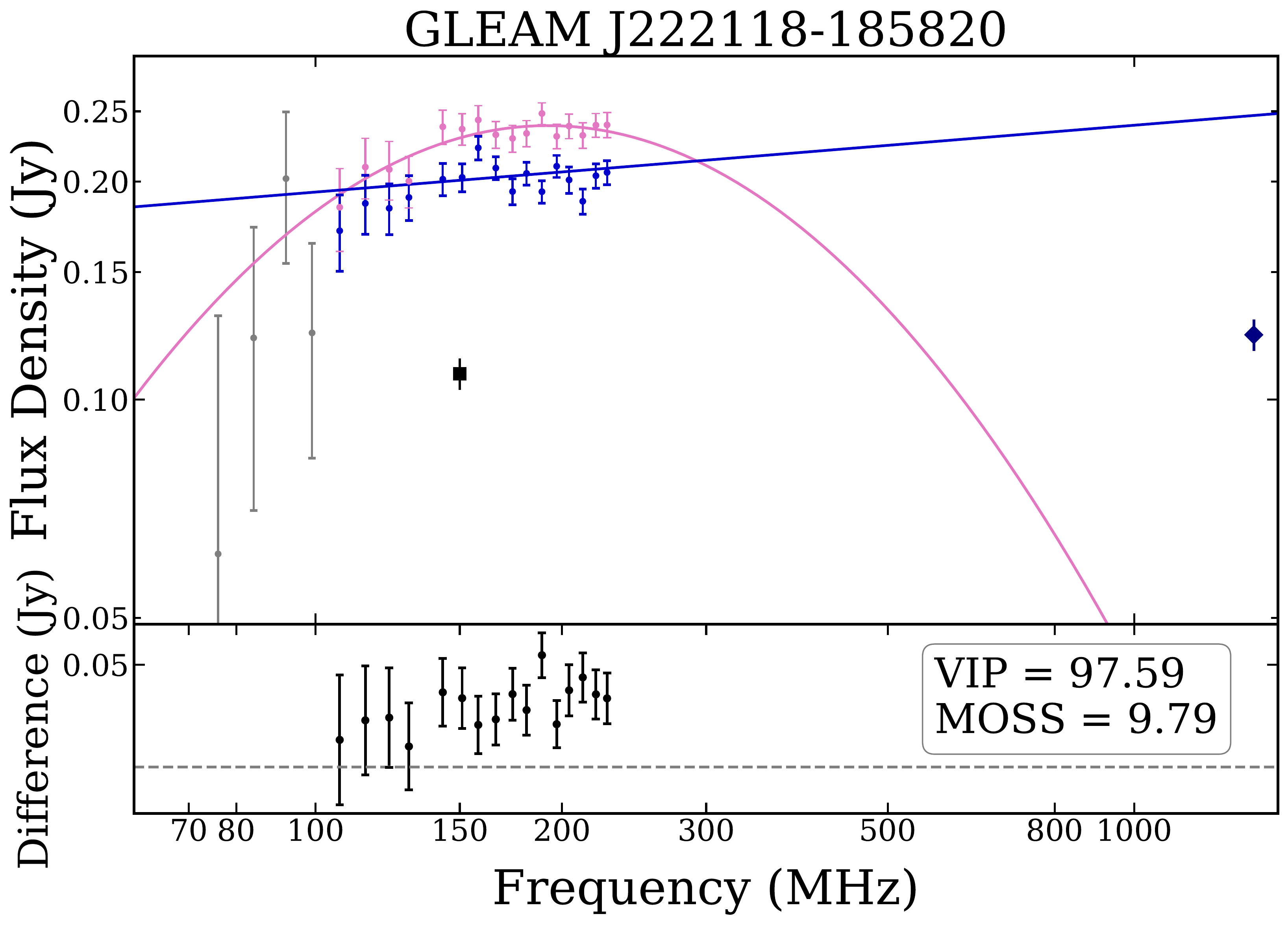} \\
\end{array}$
\caption{(continued) SEDs for all sources classified as variable according to the VIP. For each source the points represent the following data: GLEAM low frequency (72--100\,MHz) (grey circles), Year 1 (pink circles), Year 2 (blue circles), VLSSr (red cross), TGSS (black square), MRC (green star), SUMSS (yellow pentagon), and NVSS (navy diamond). The models for each year are determined by their classification; a source classified with a peak within the observed band was modelled by a quadratic according to Equation~\ref{eq:quadratic}, remaining sources were modelled by a power-law according to Equation~\ref{eq:plaw}.}
\label{app:fig:pg14}
\end{center}
\end{figure*}
\setcounter{figure}{0}
\begin{figure*}
\begin{center}$
\begin{array}{cccccc}
\includegraphics[scale=0.15]{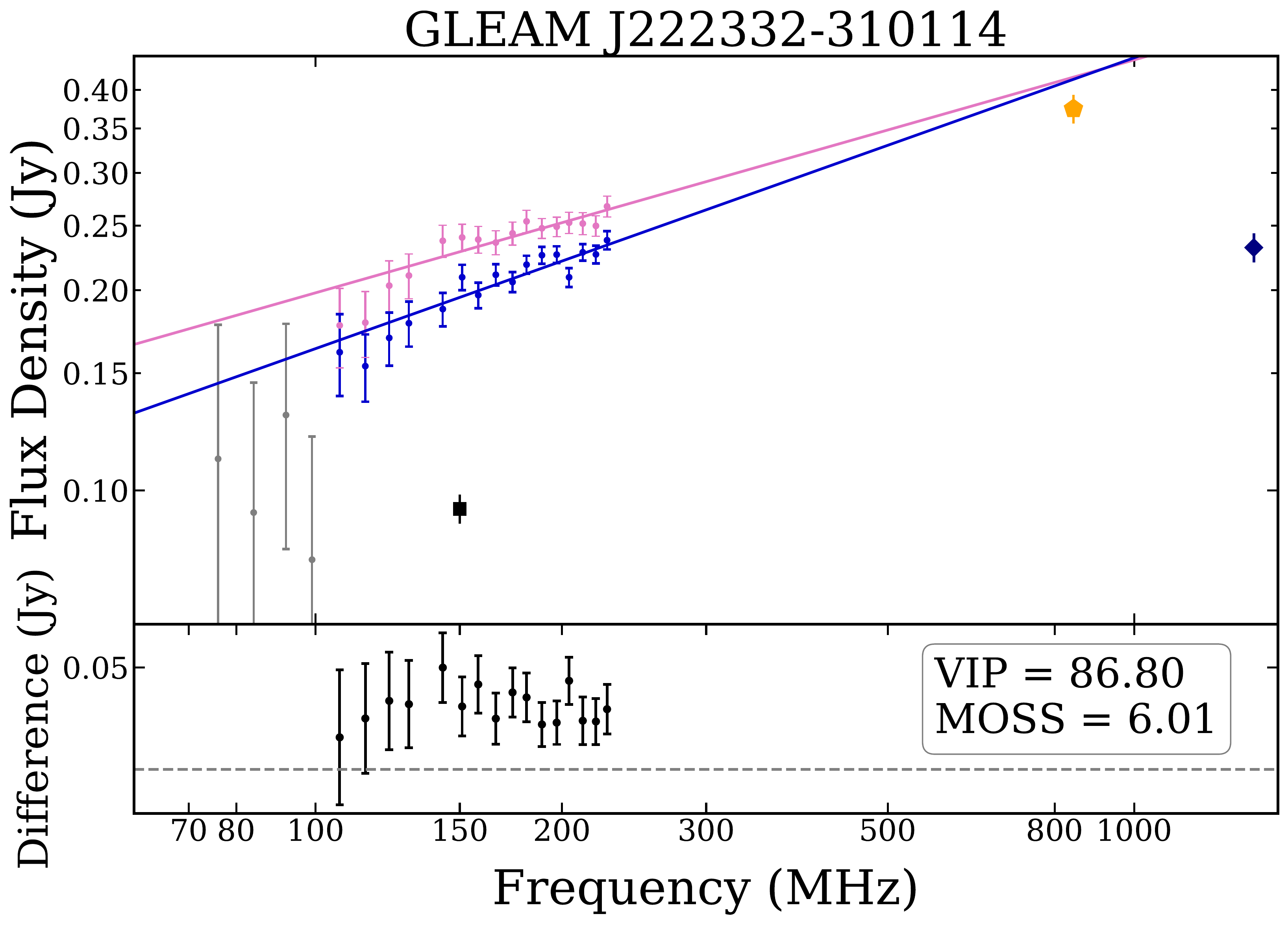} &
\includegraphics[scale=0.15]{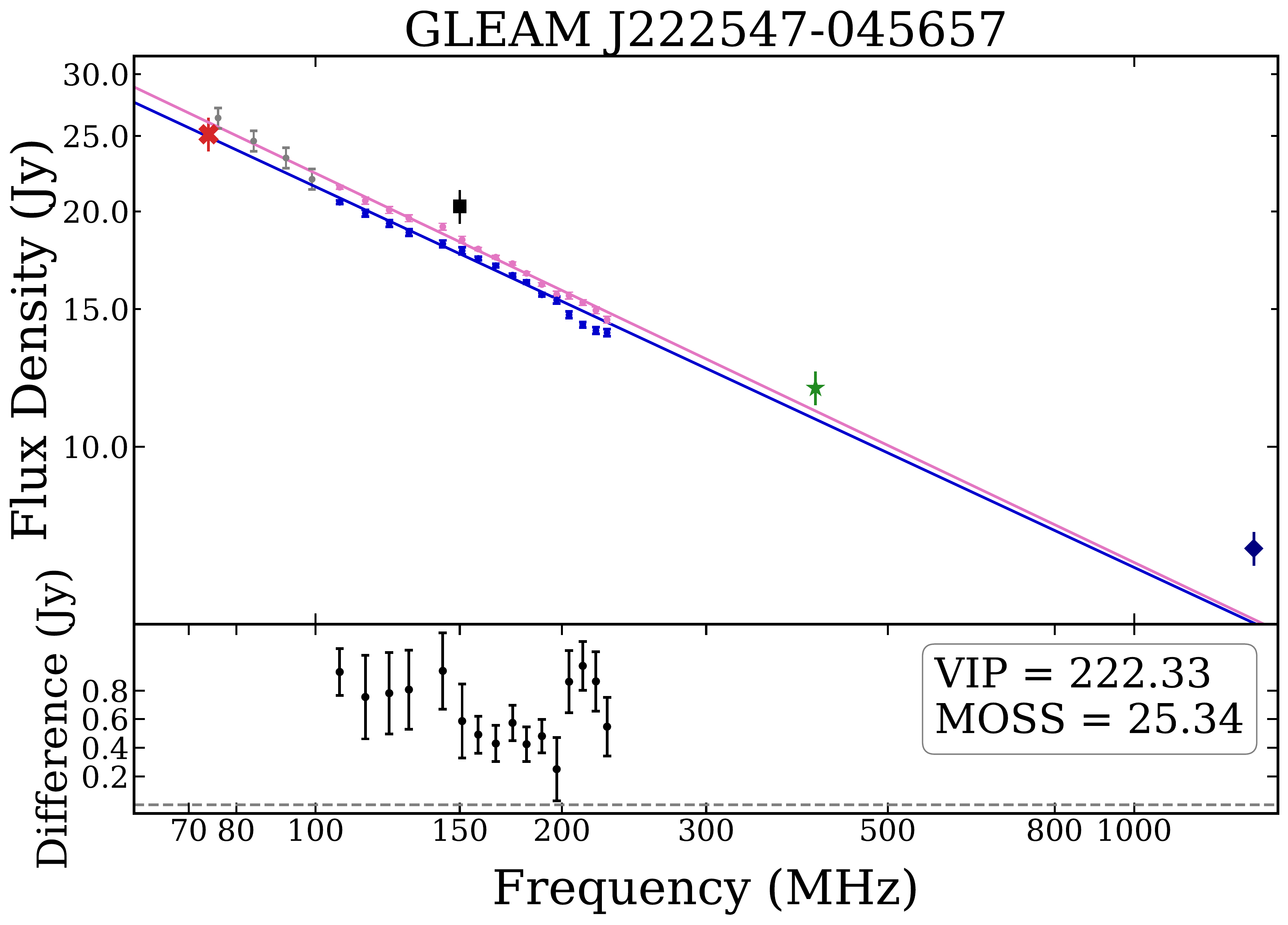} &
\includegraphics[scale=0.15]{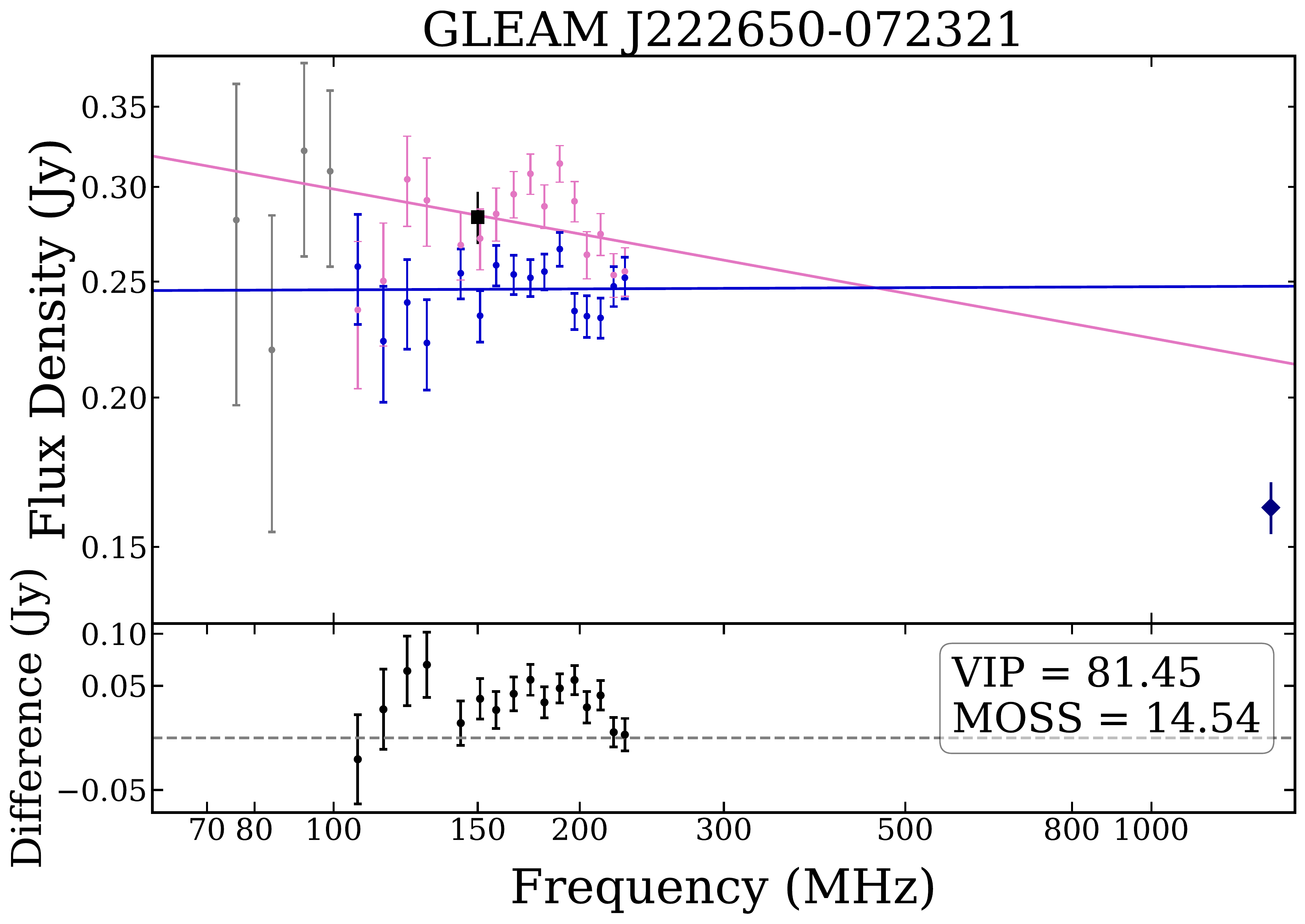} \\
\includegraphics[scale=0.15]{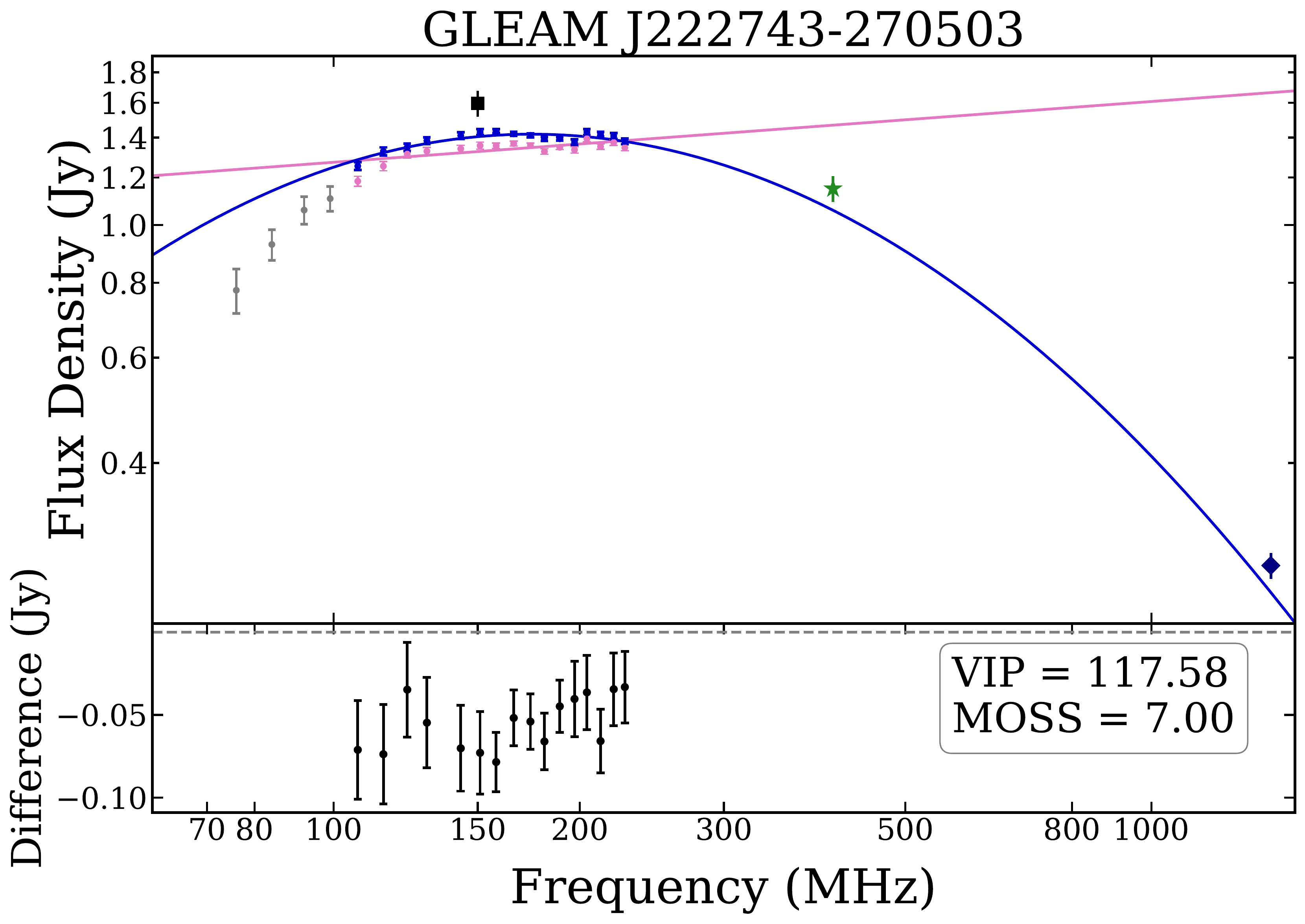} &
\includegraphics[scale=0.15]{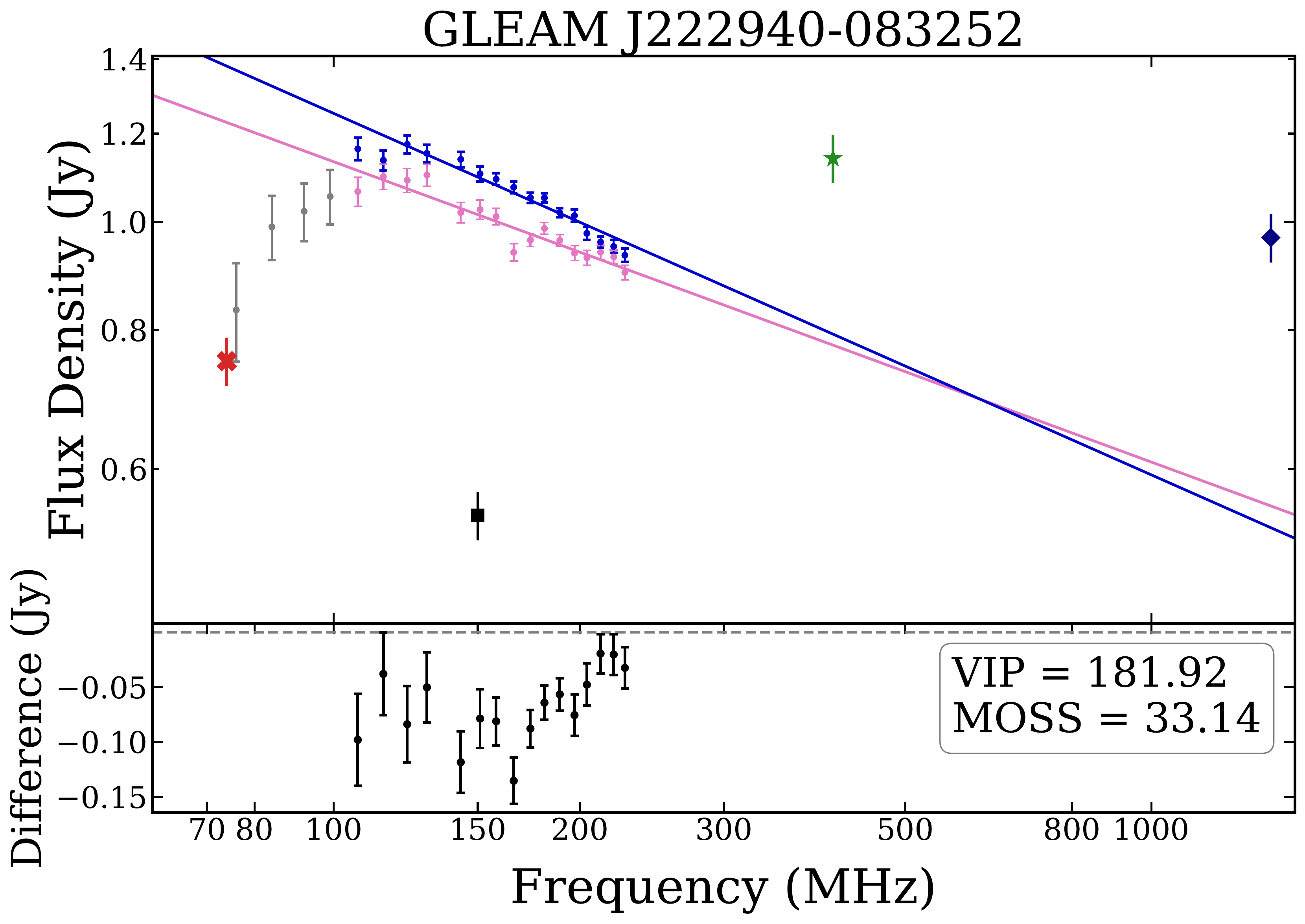} &
\includegraphics[scale=0.15]{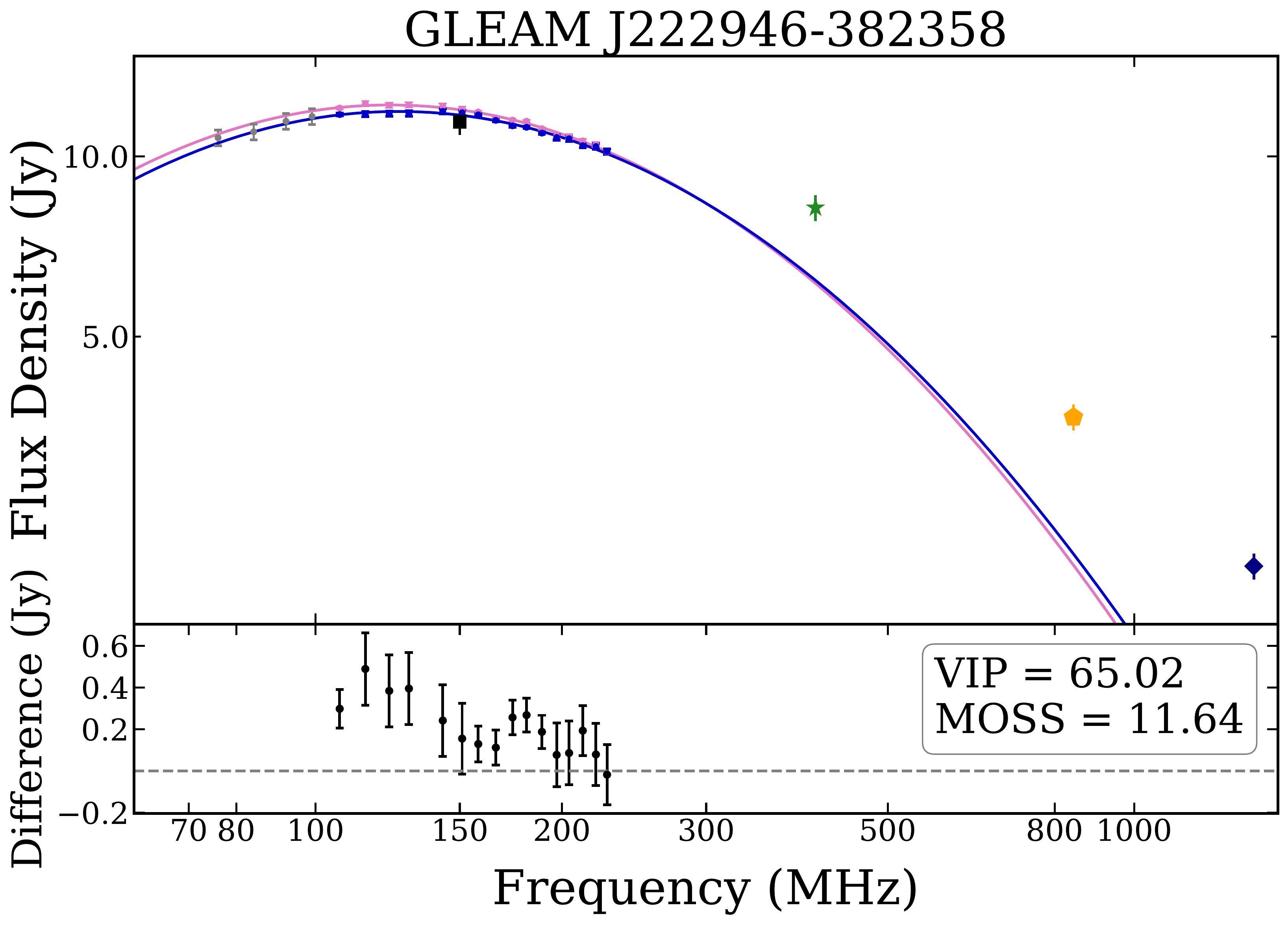} \\
\includegraphics[scale=0.15]{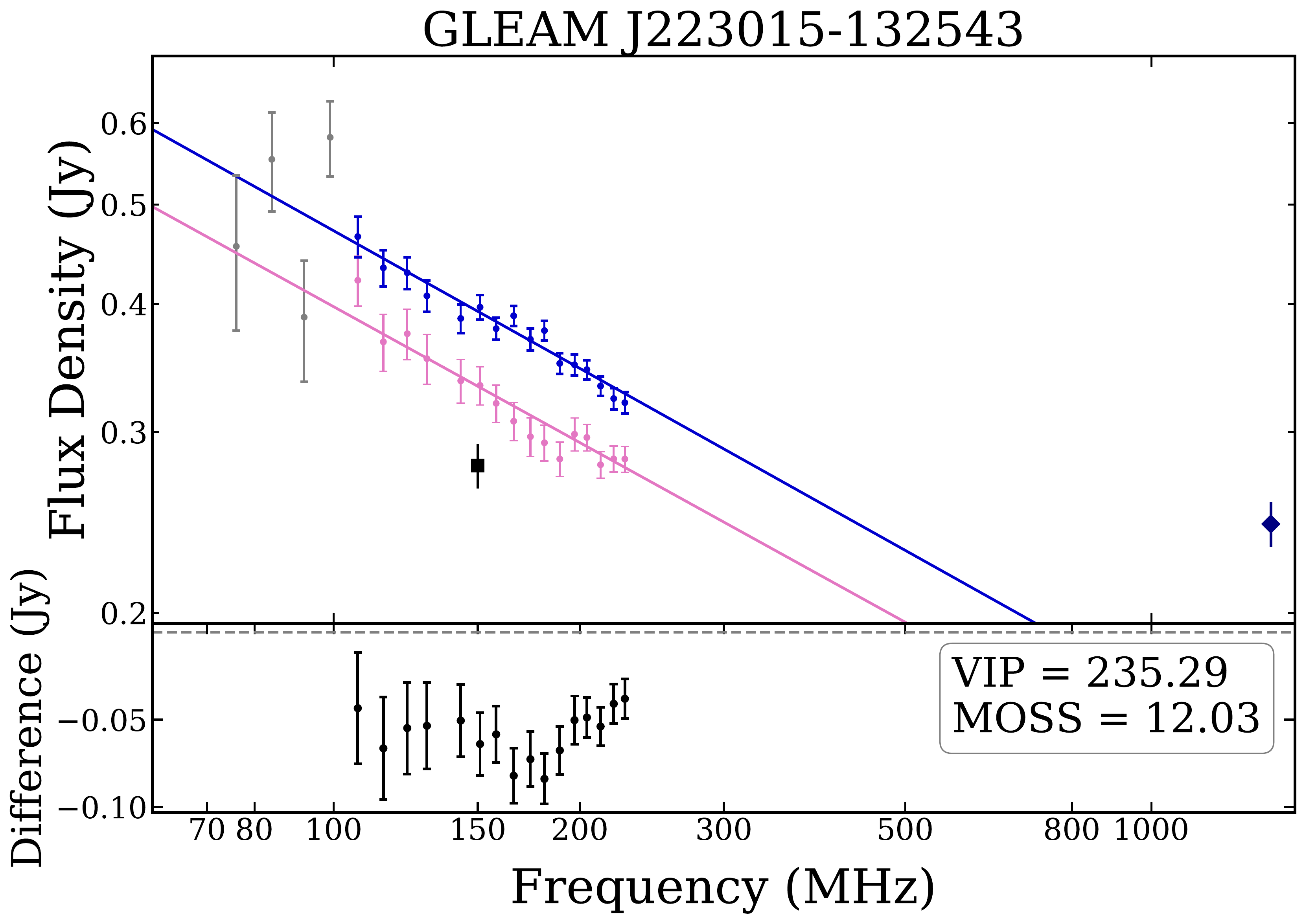} &
\includegraphics[scale=0.15]{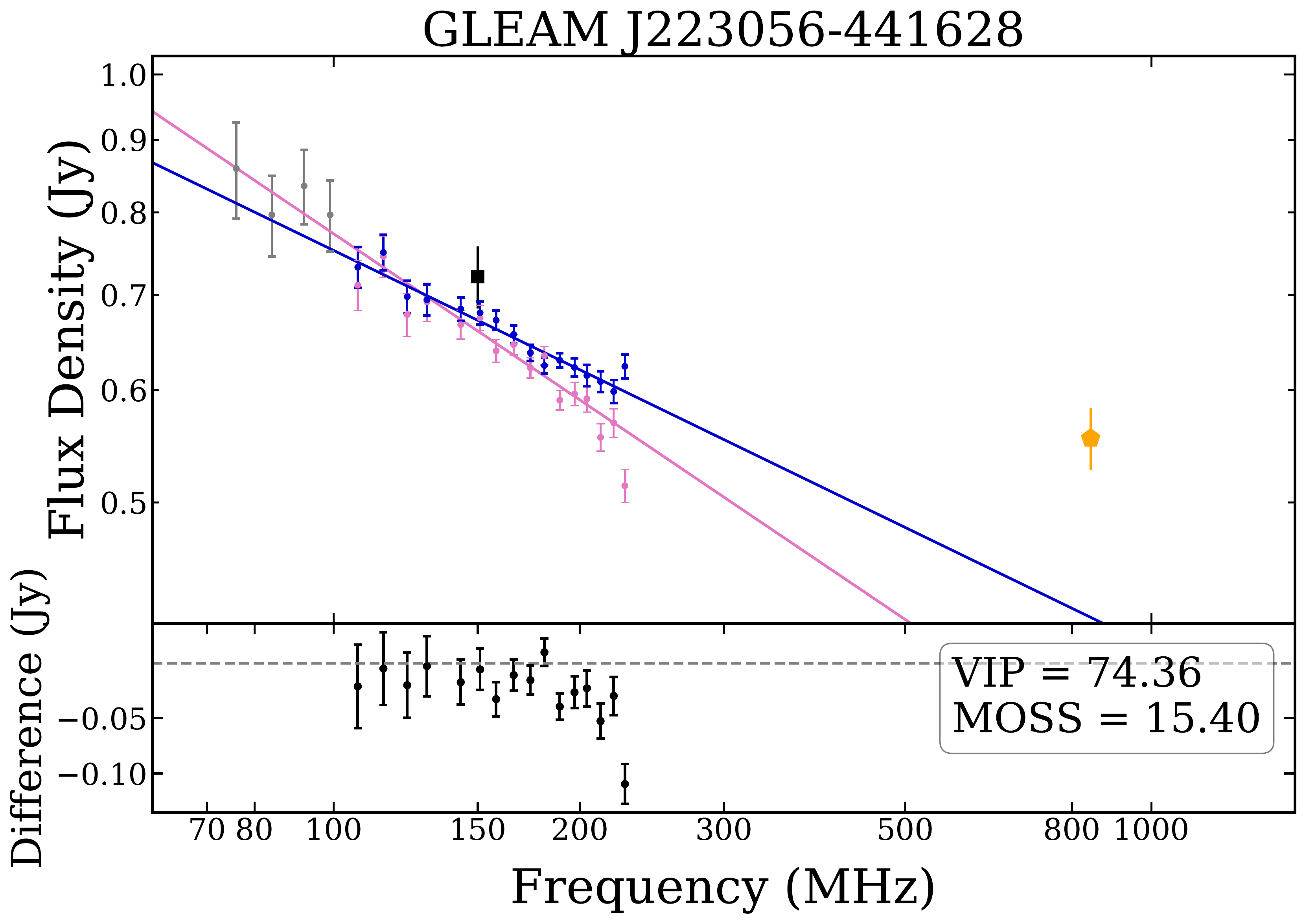} &
\includegraphics[scale=0.15]{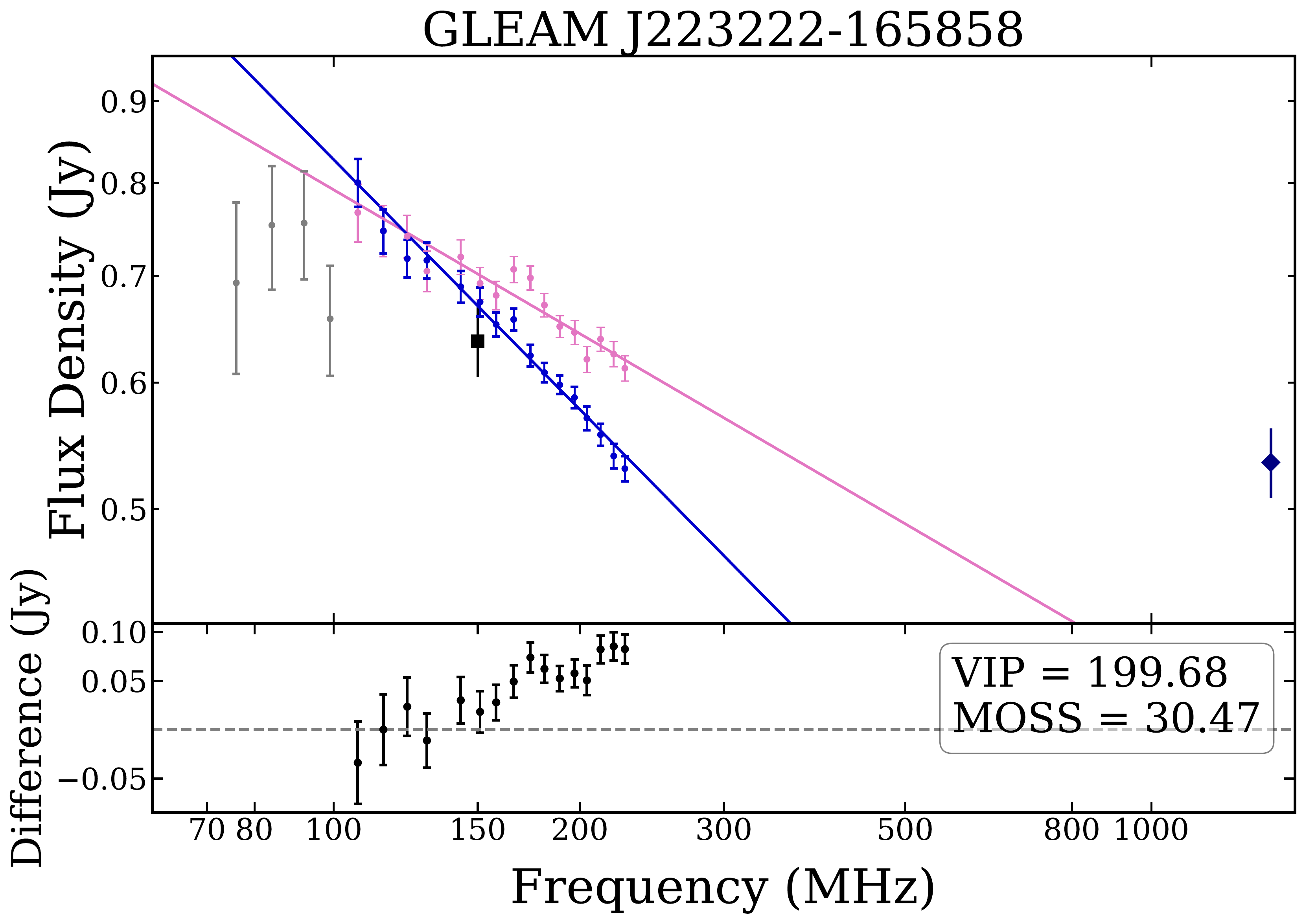} \\
\includegraphics[scale=0.15]{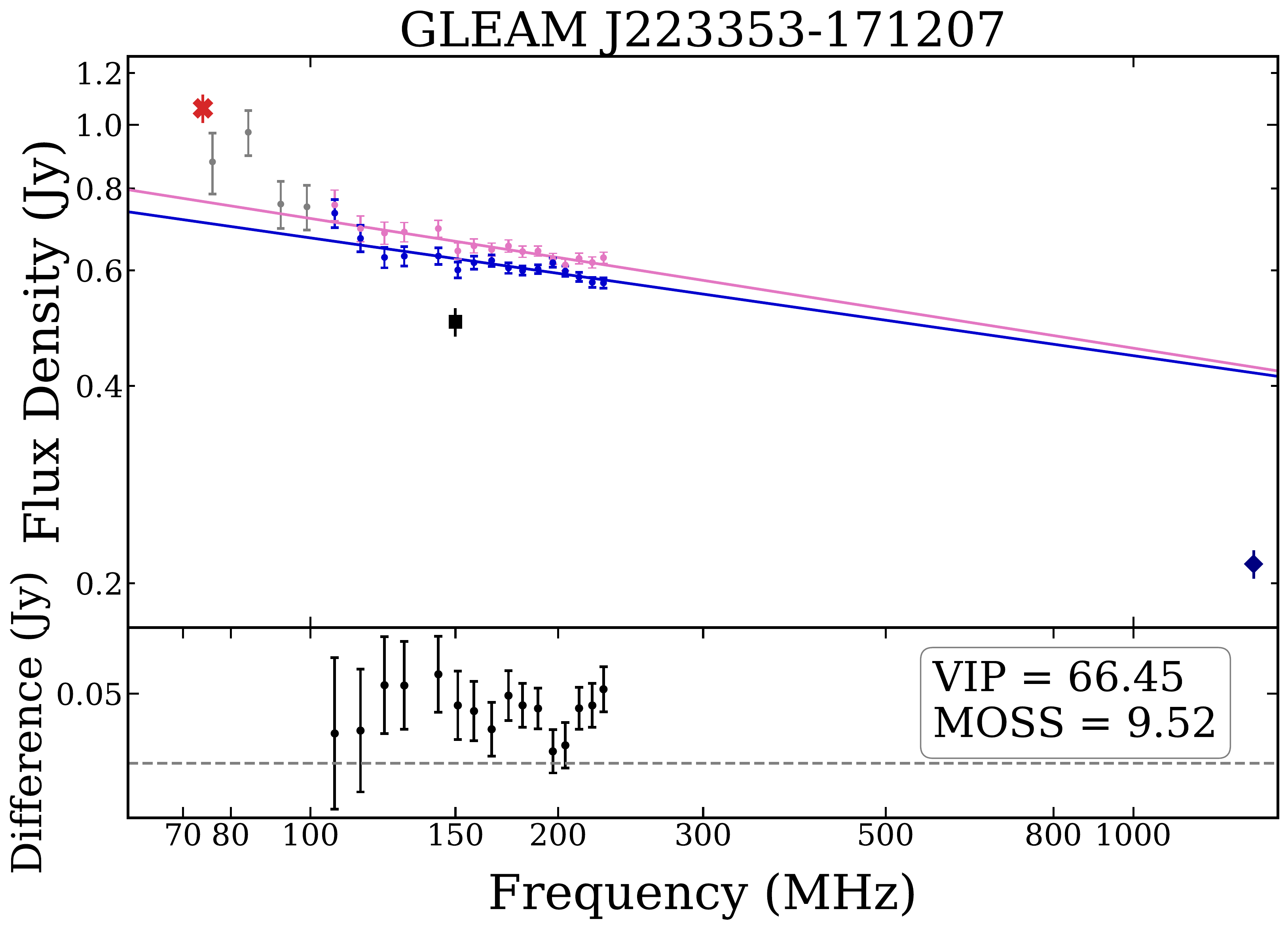} &
\includegraphics[scale=0.15]{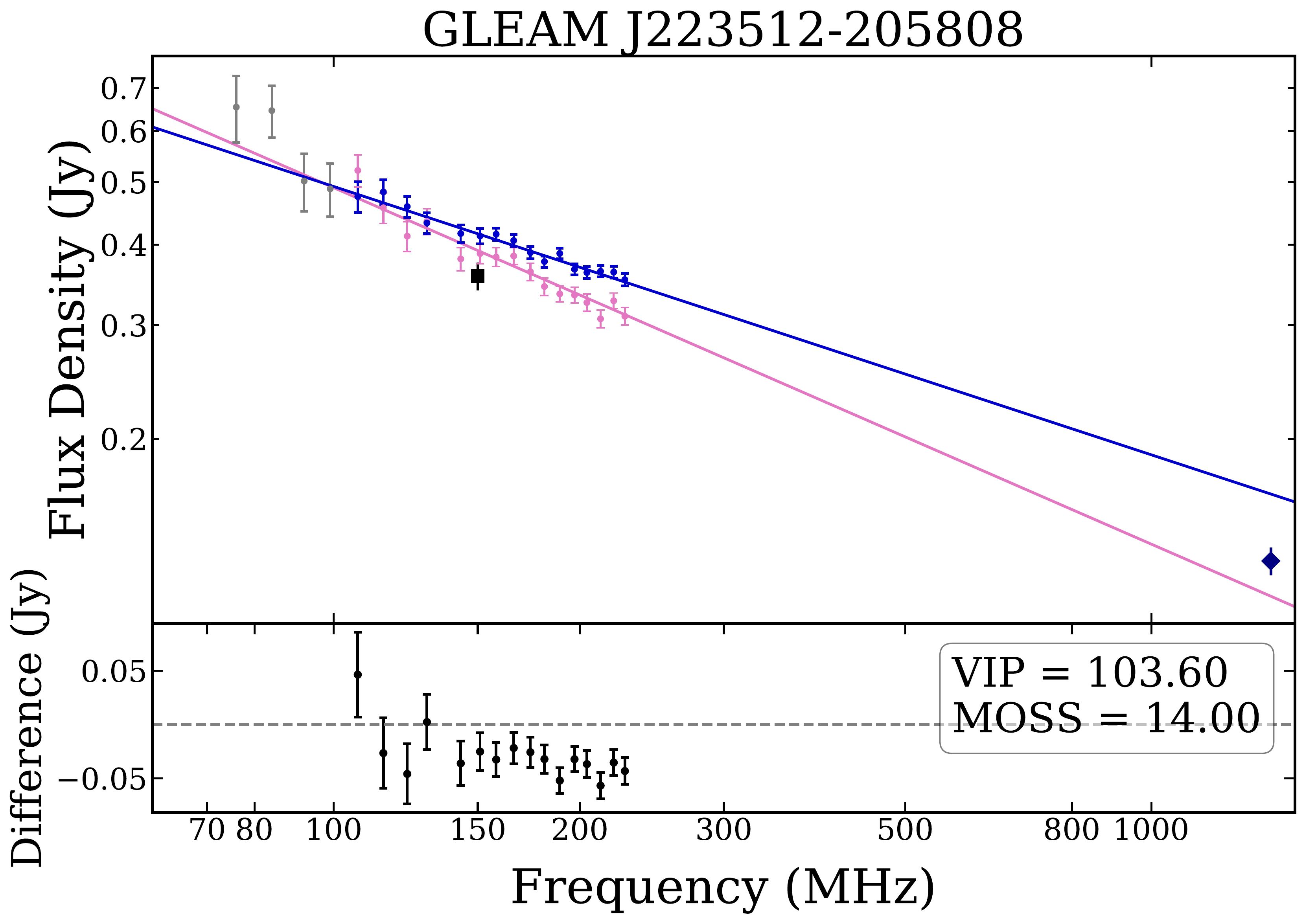} &
\includegraphics[scale=0.15]{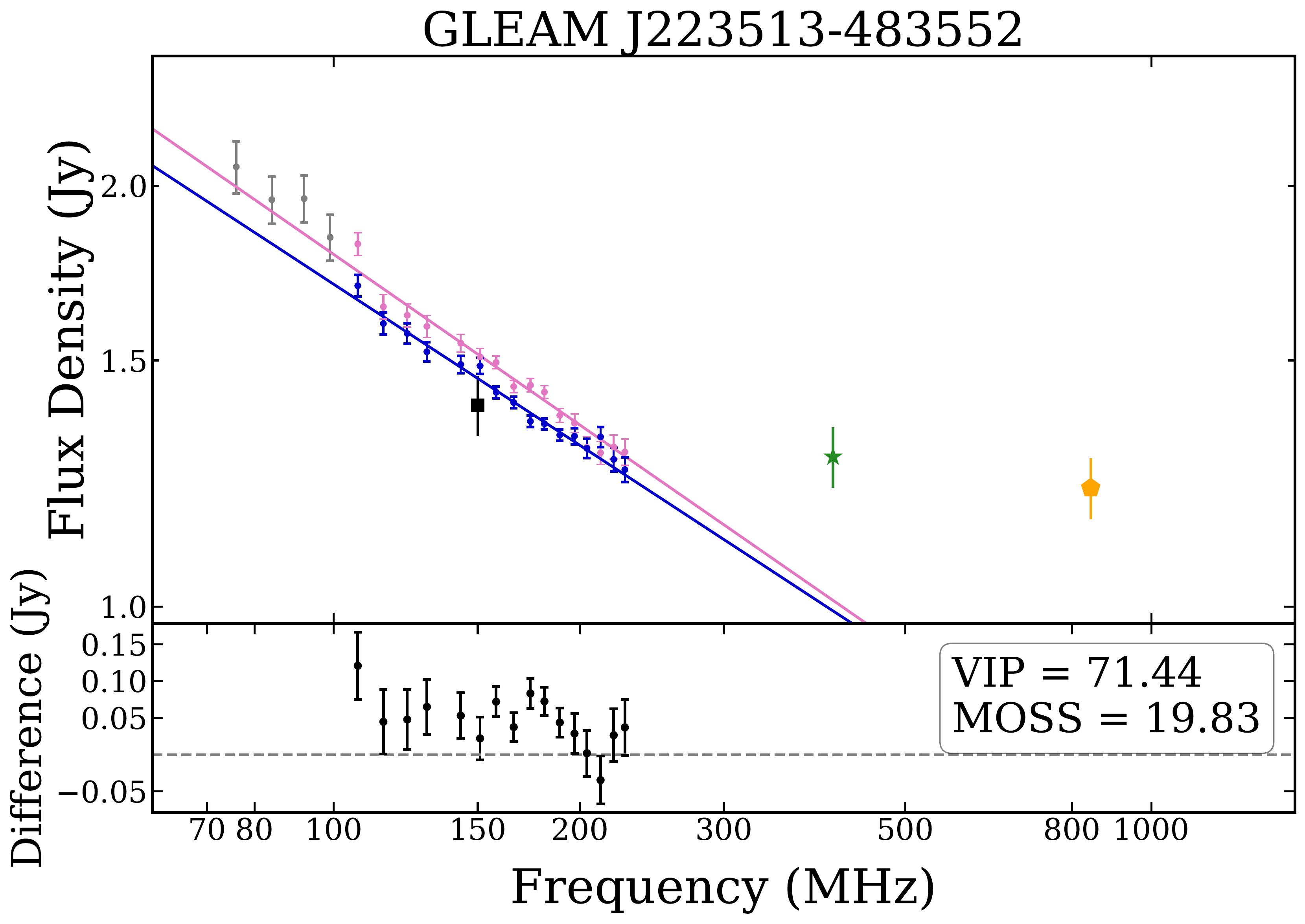} \\
\includegraphics[scale=0.15]{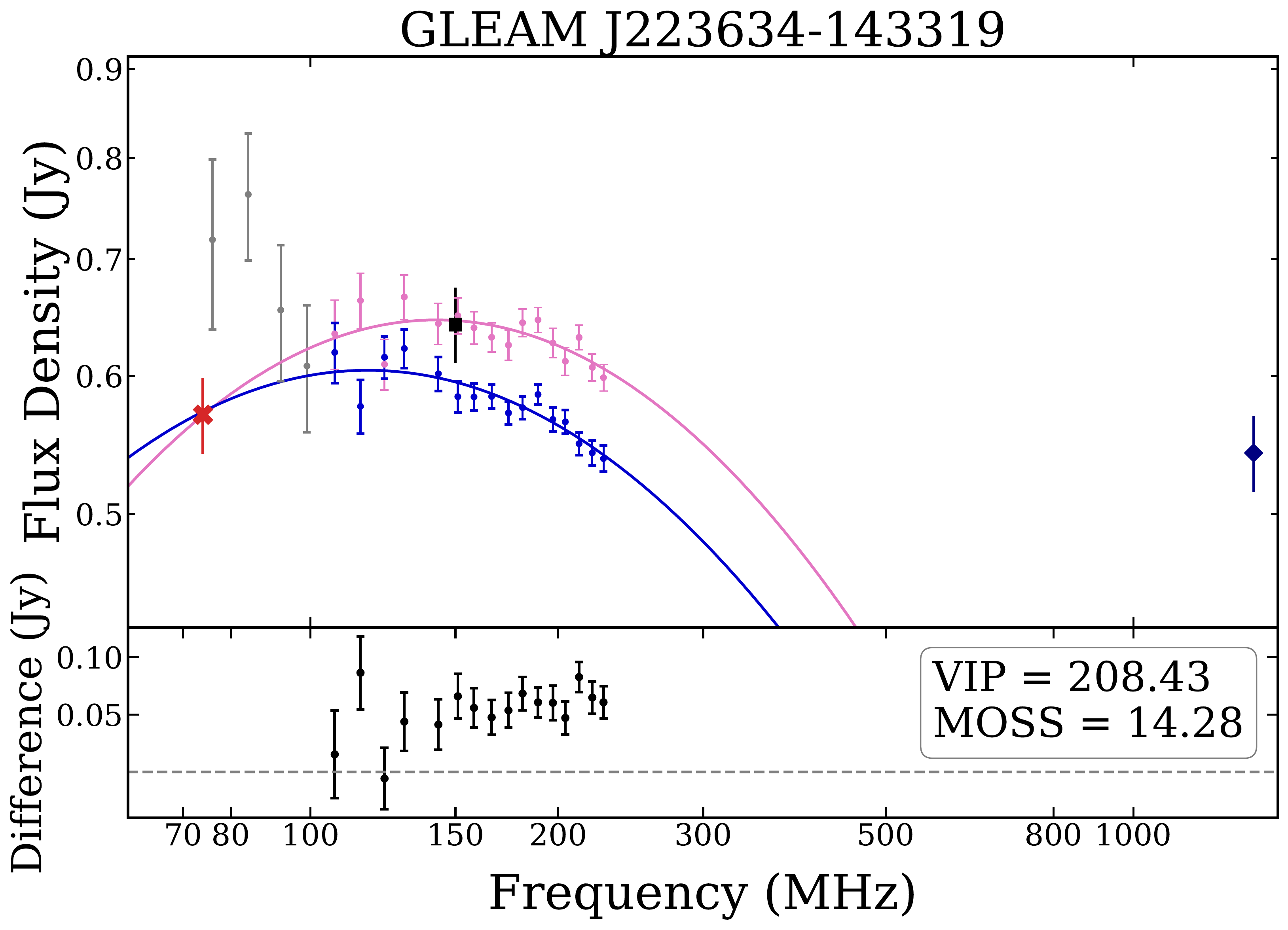} &
\includegraphics[scale=0.15]{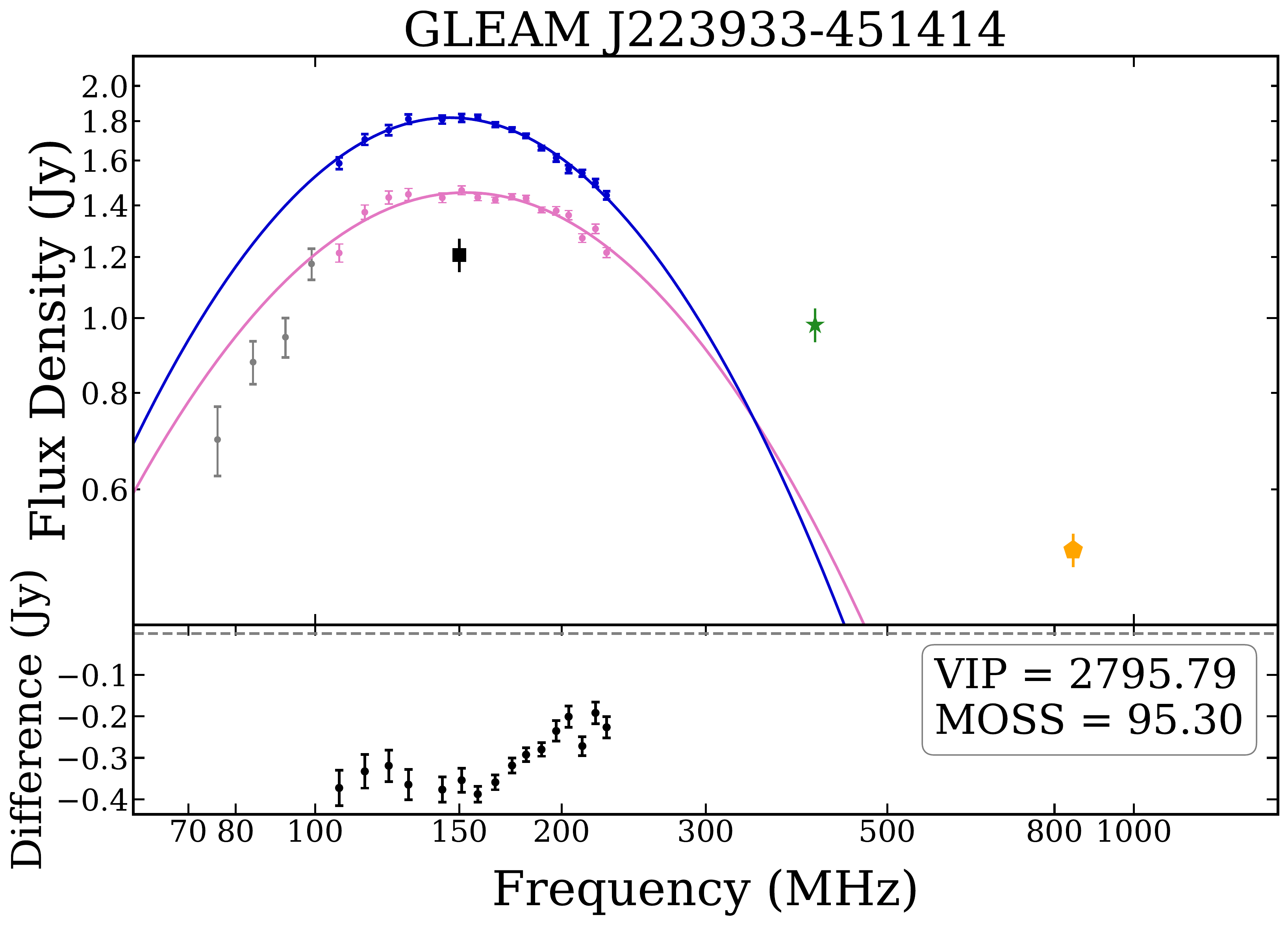} &
\includegraphics[scale=0.15]{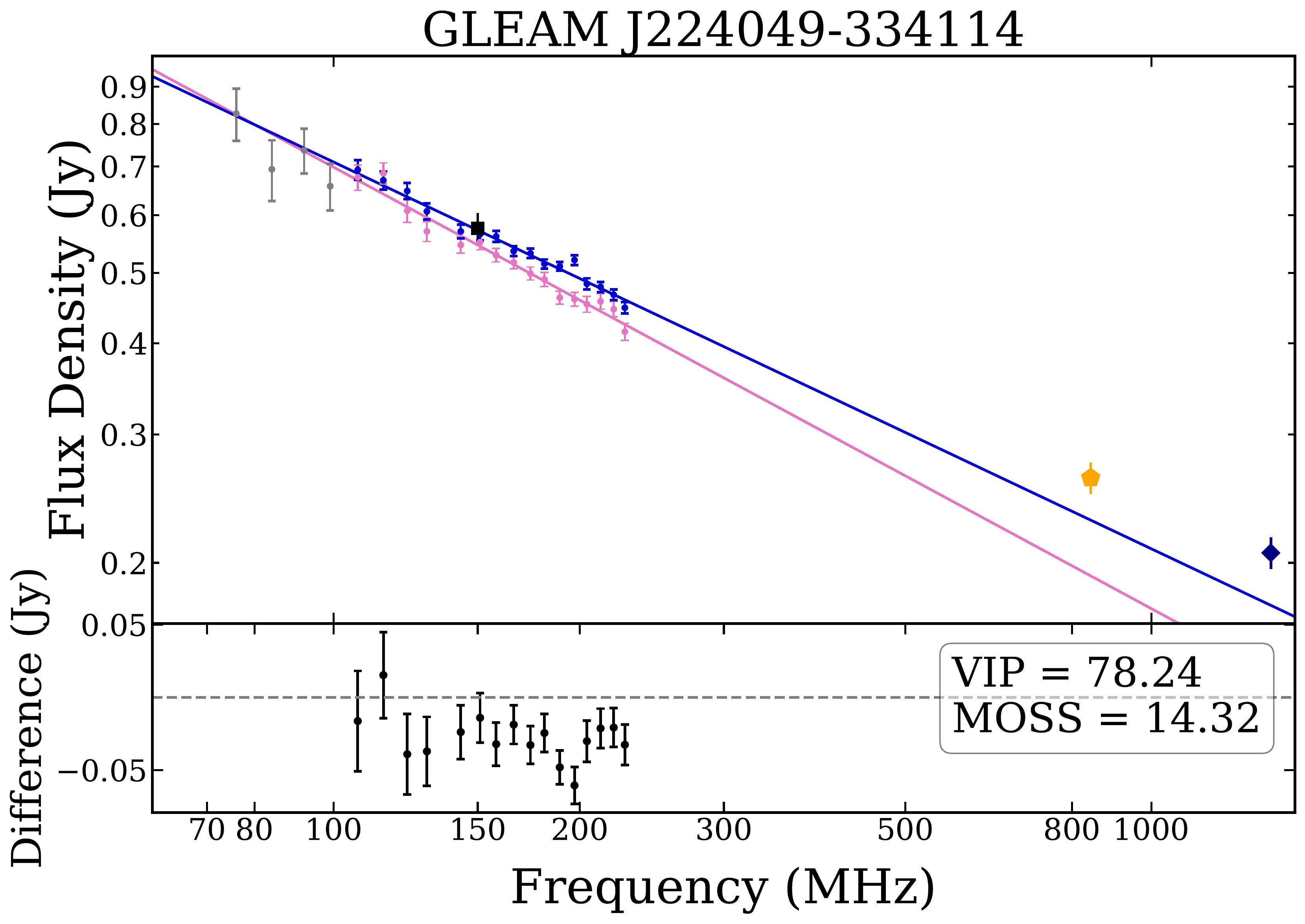} \\
\includegraphics[scale=0.15]{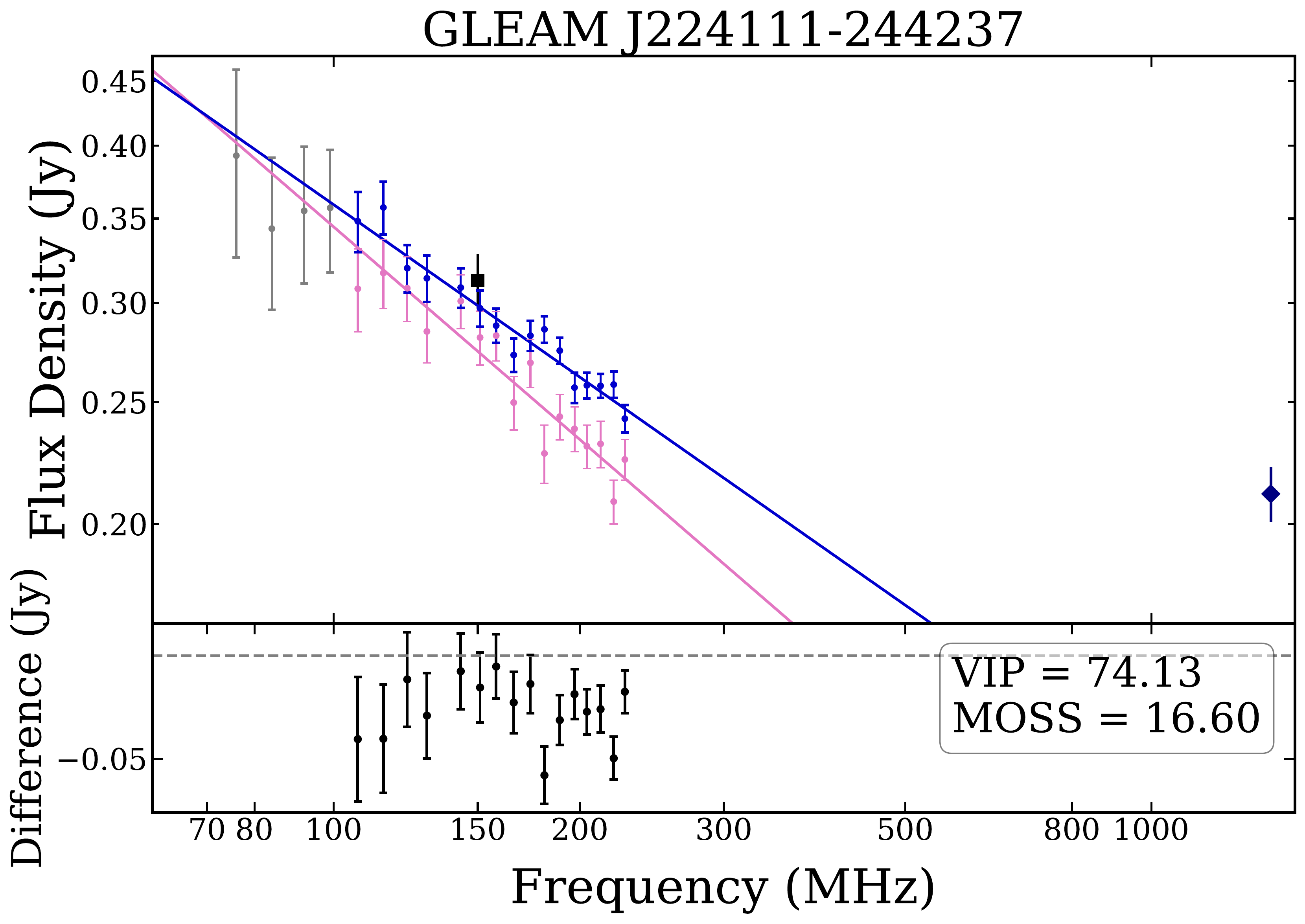} &
\includegraphics[scale=0.15]{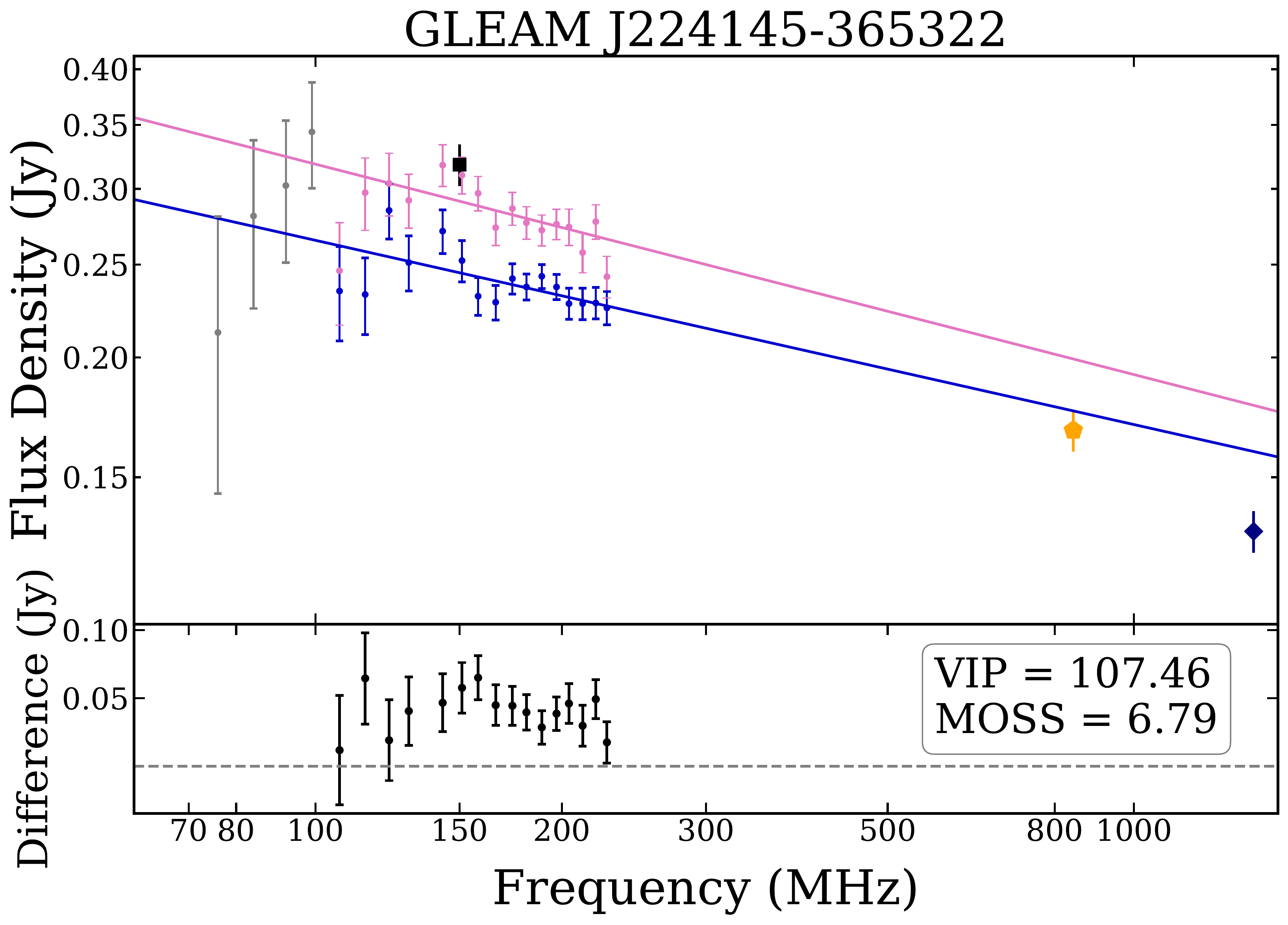} &
\includegraphics[scale=0.15]{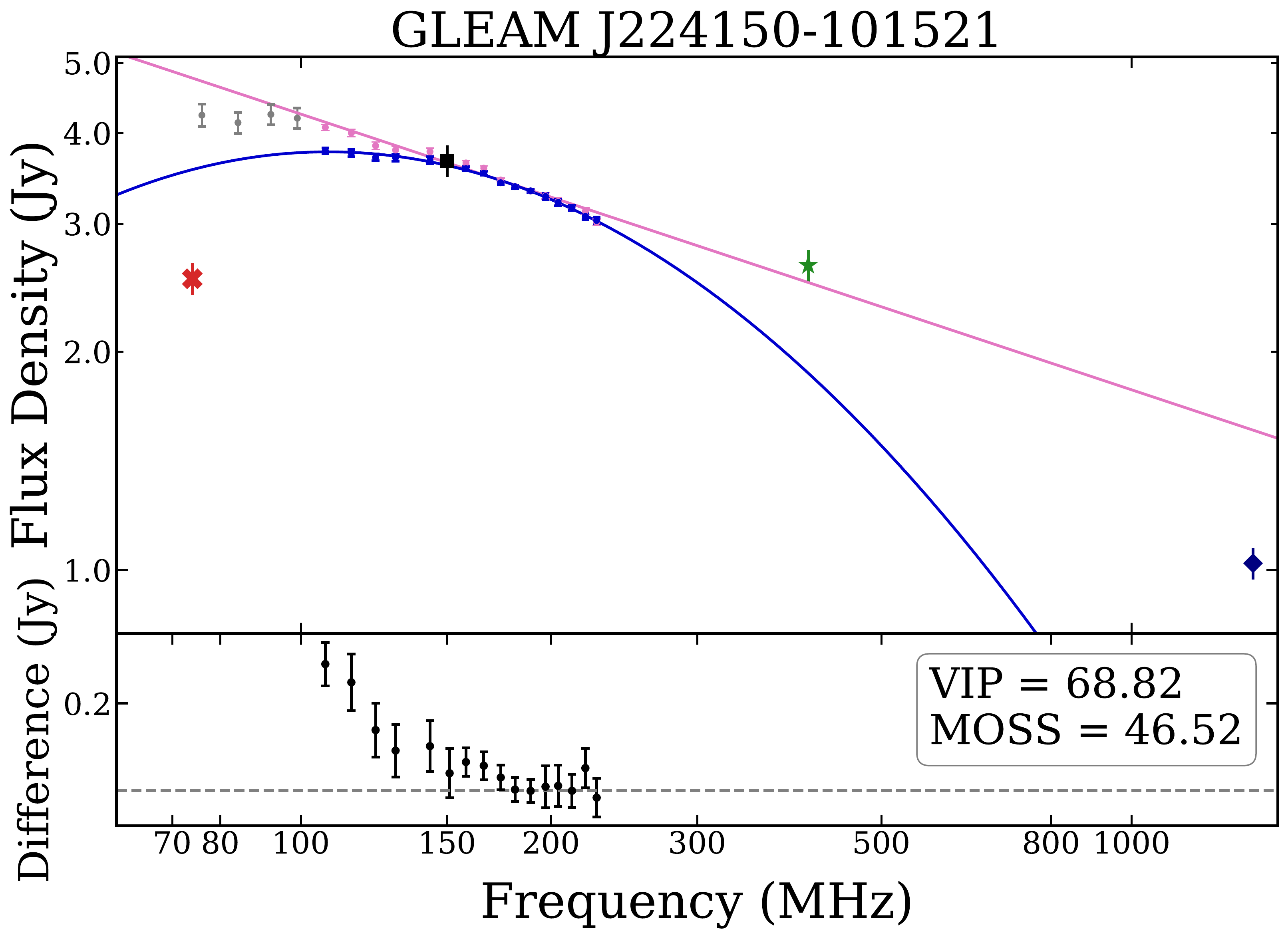} \\
\end{array}$
\caption{(continued) SEDs for all sources classified as variable according to the VIP. For each source the points represent the following data: GLEAM low frequency (72--100\,MHz) (grey circles), Year 1 (pink circles), Year 2 (blue circles), VLSSr (red cross), TGSS (black square), MRC (green star), SUMSS (yellow pentagon), and NVSS (navy diamond). The models for each year are determined by their classification; a source classified with a peak within the observed band was modelled by a quadratic according to Equation~\ref{eq:quadratic}, remaining sources were modelled by a power-law according to Equation~\ref{eq:plaw}.}
\label{app:fig:pg15}
\end{center}
\end{figure*}
\setcounter{figure}{0}
\begin{figure*}
\begin{center}$
\begin{array}{cccccc}
\includegraphics[scale=0.15]{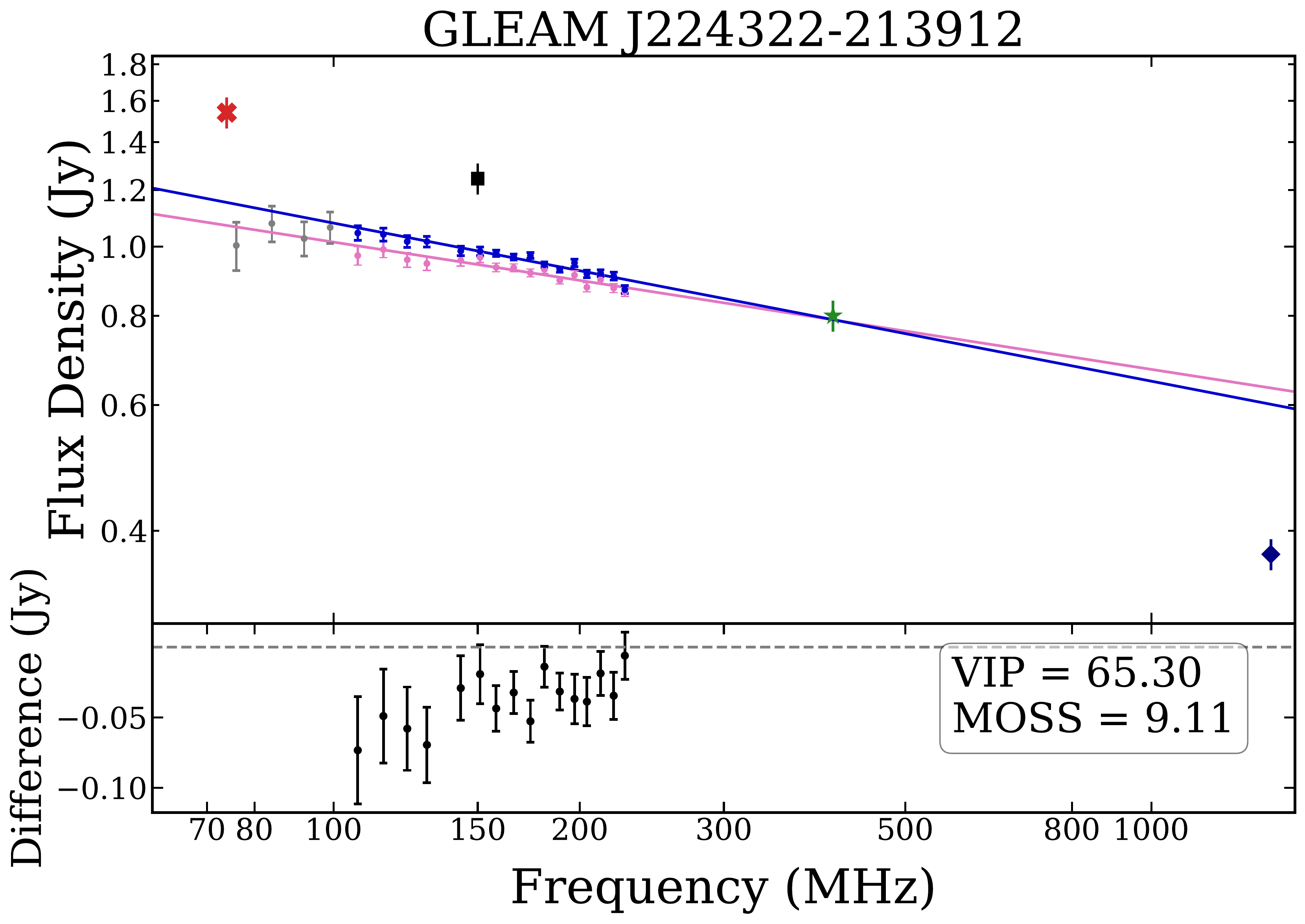} &
\includegraphics[scale=0.15]{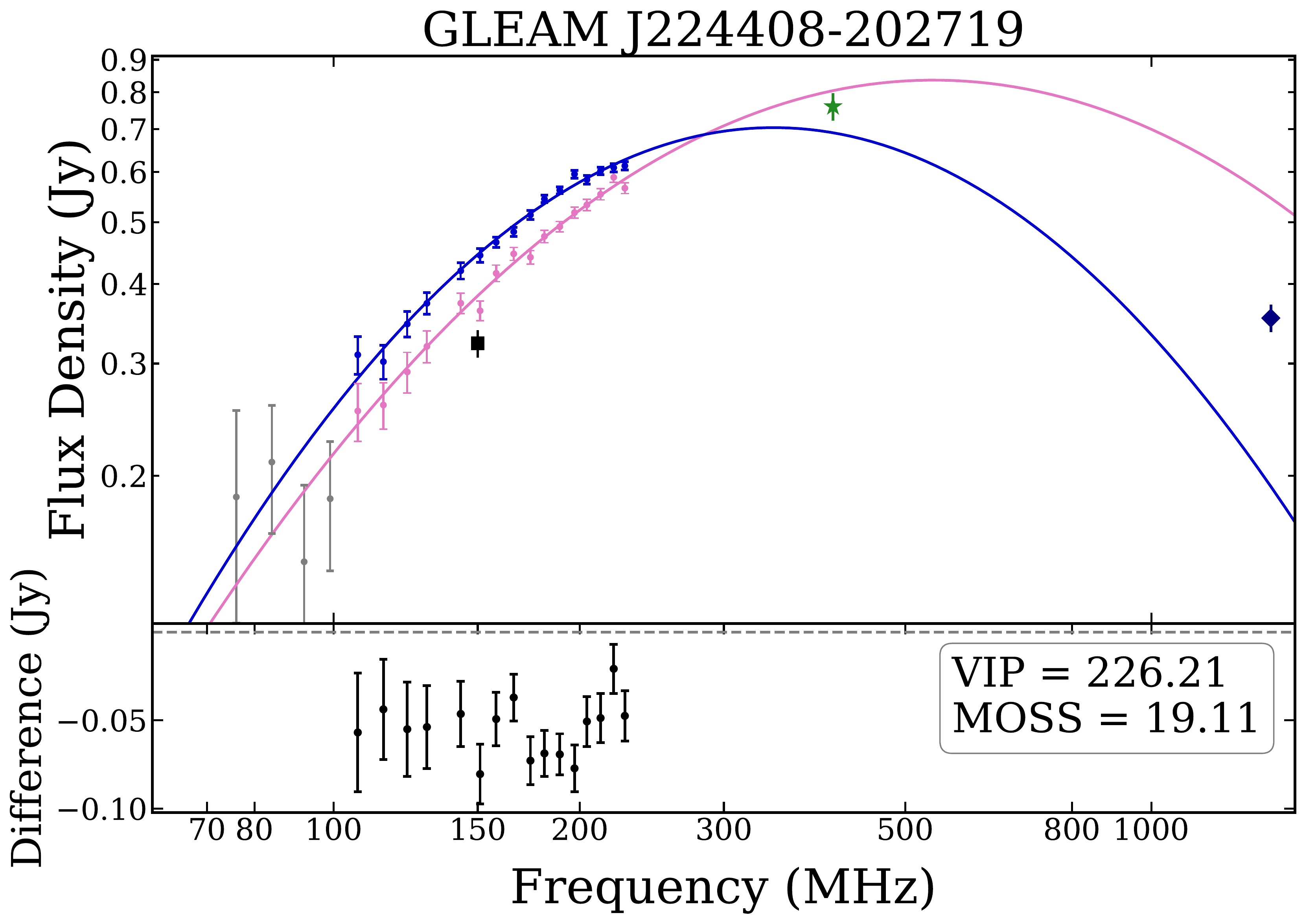} &
\includegraphics[scale=0.15]{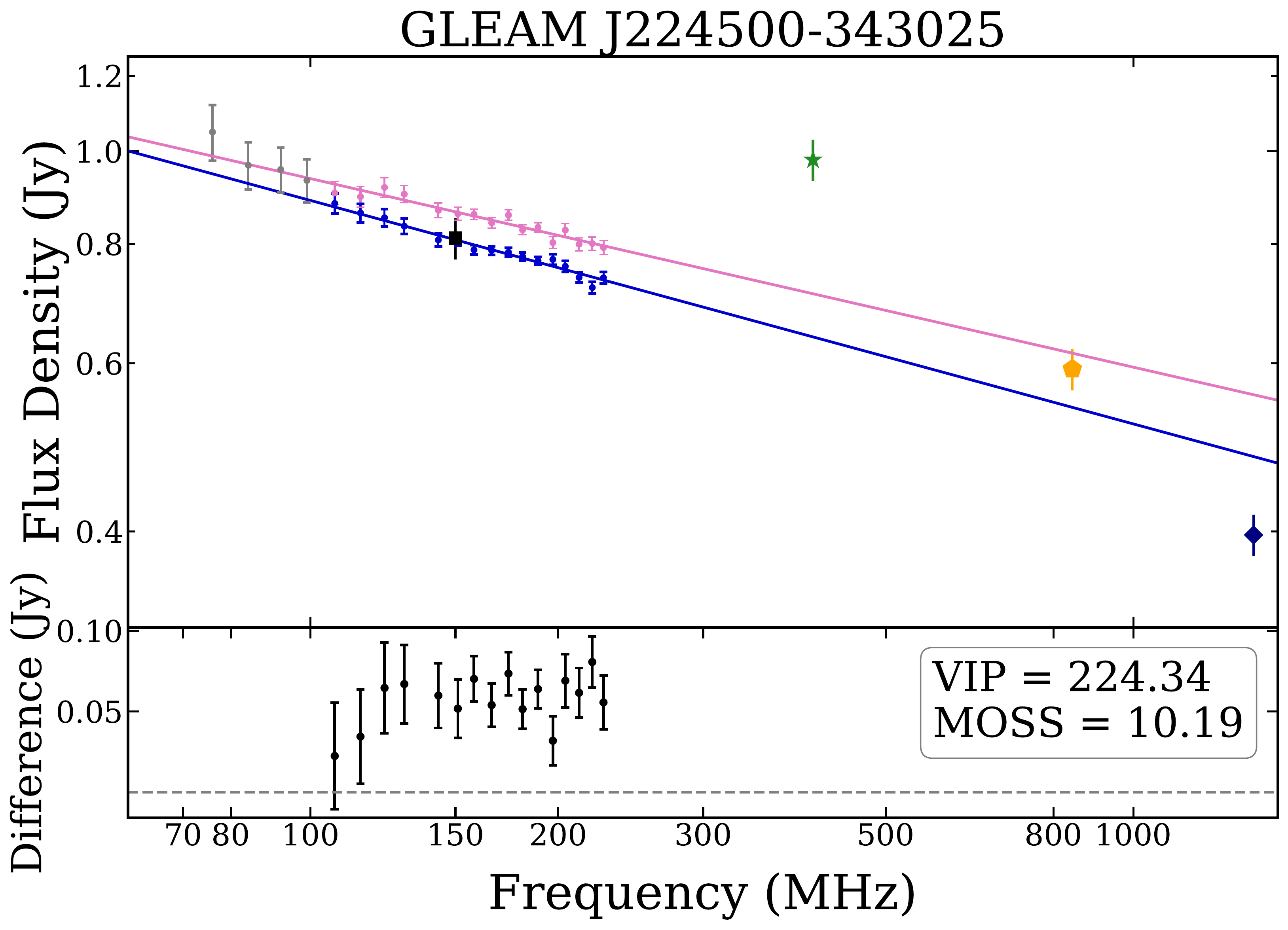} \\
\includegraphics[scale=0.15]{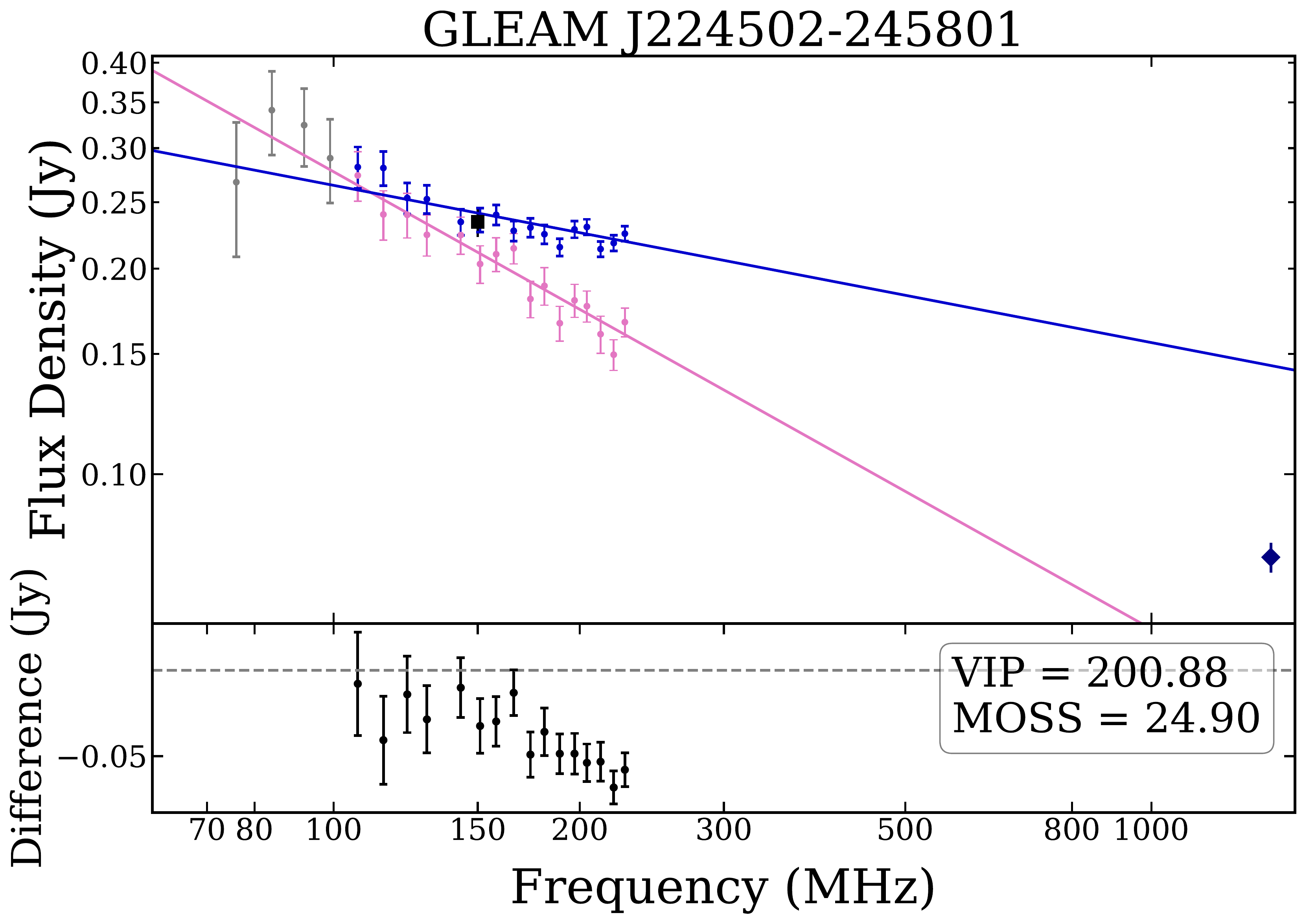} &
\includegraphics[scale=0.15]{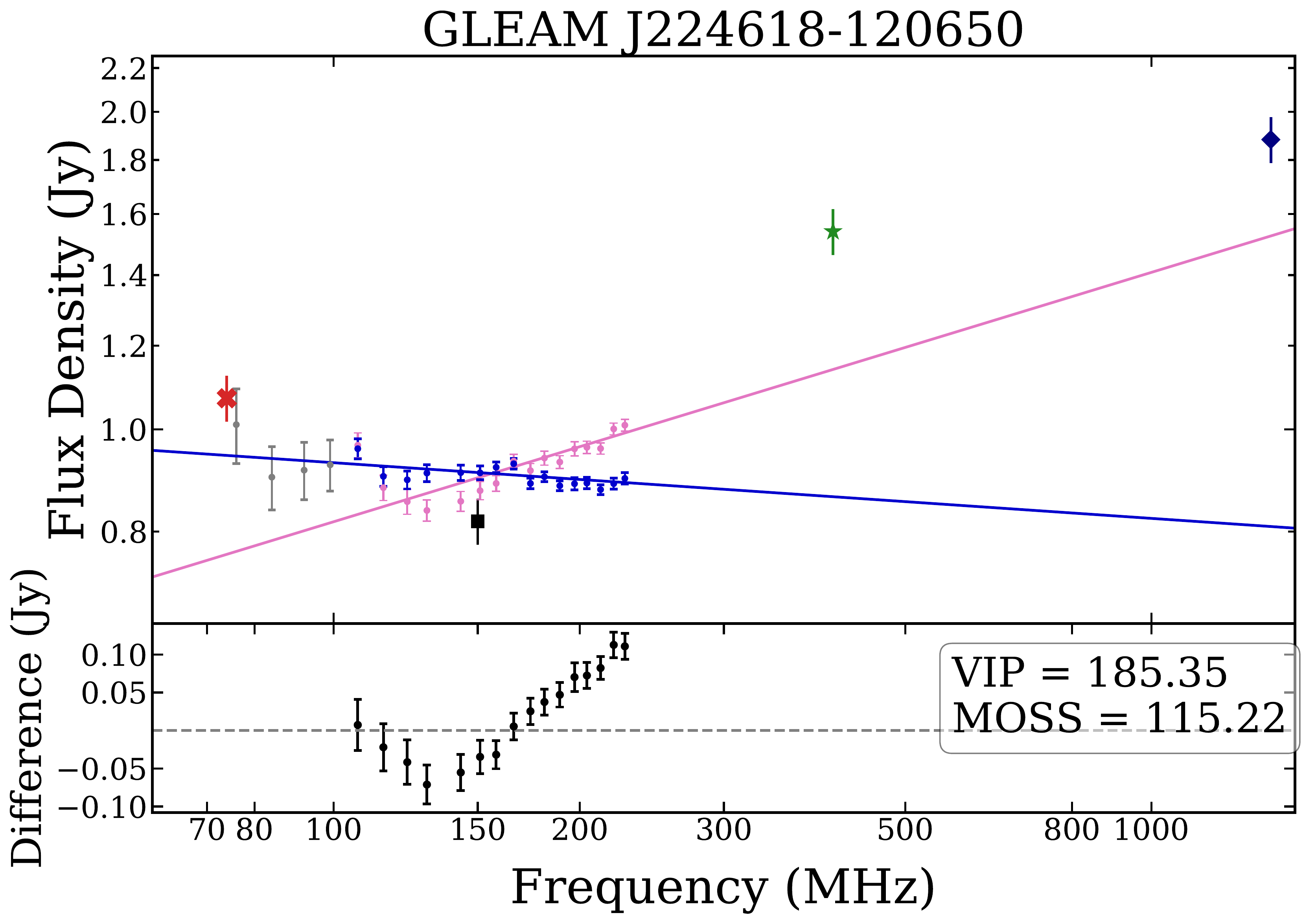} &
\includegraphics[scale=0.15]{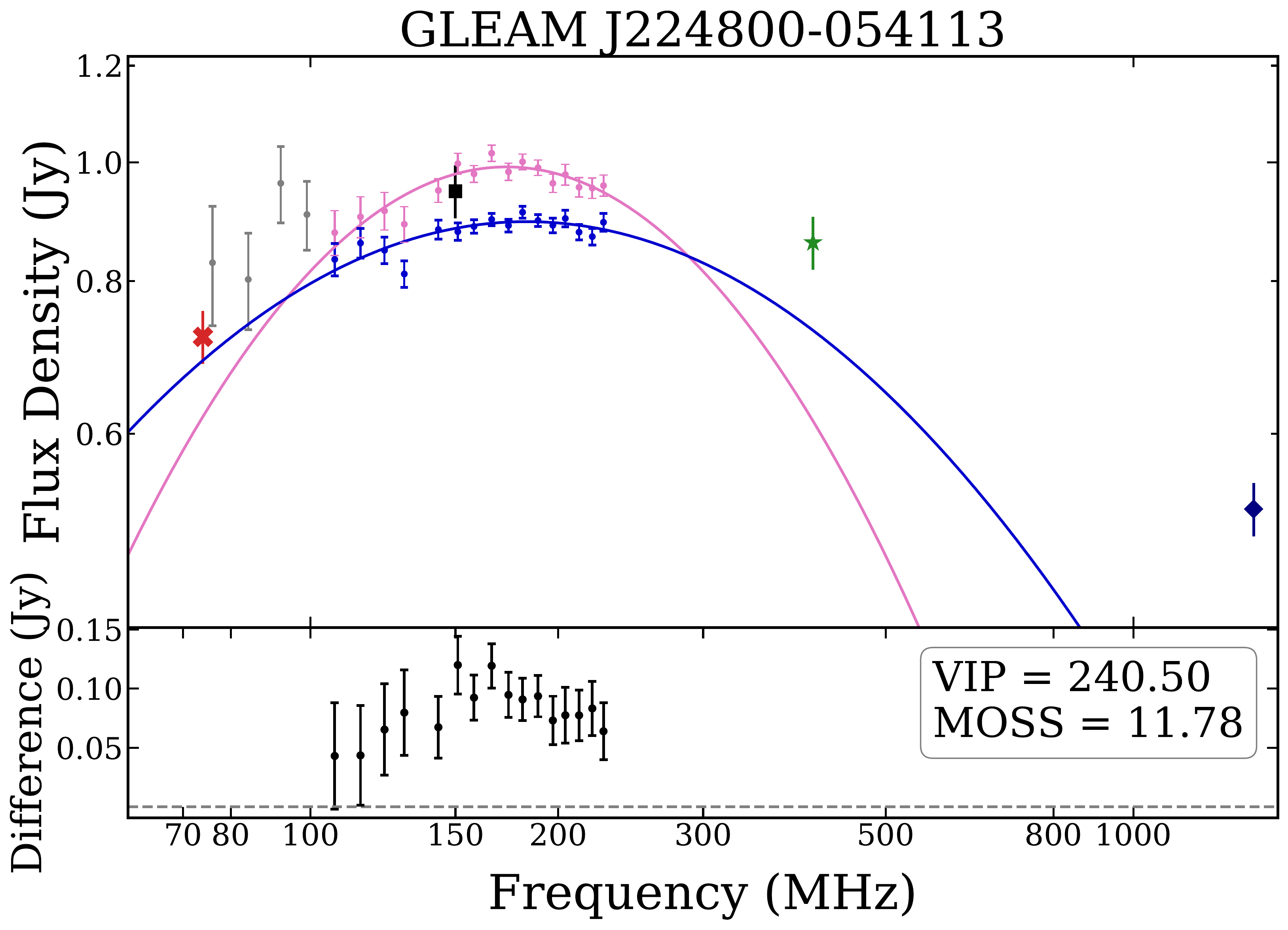} \\
\includegraphics[scale=0.15]{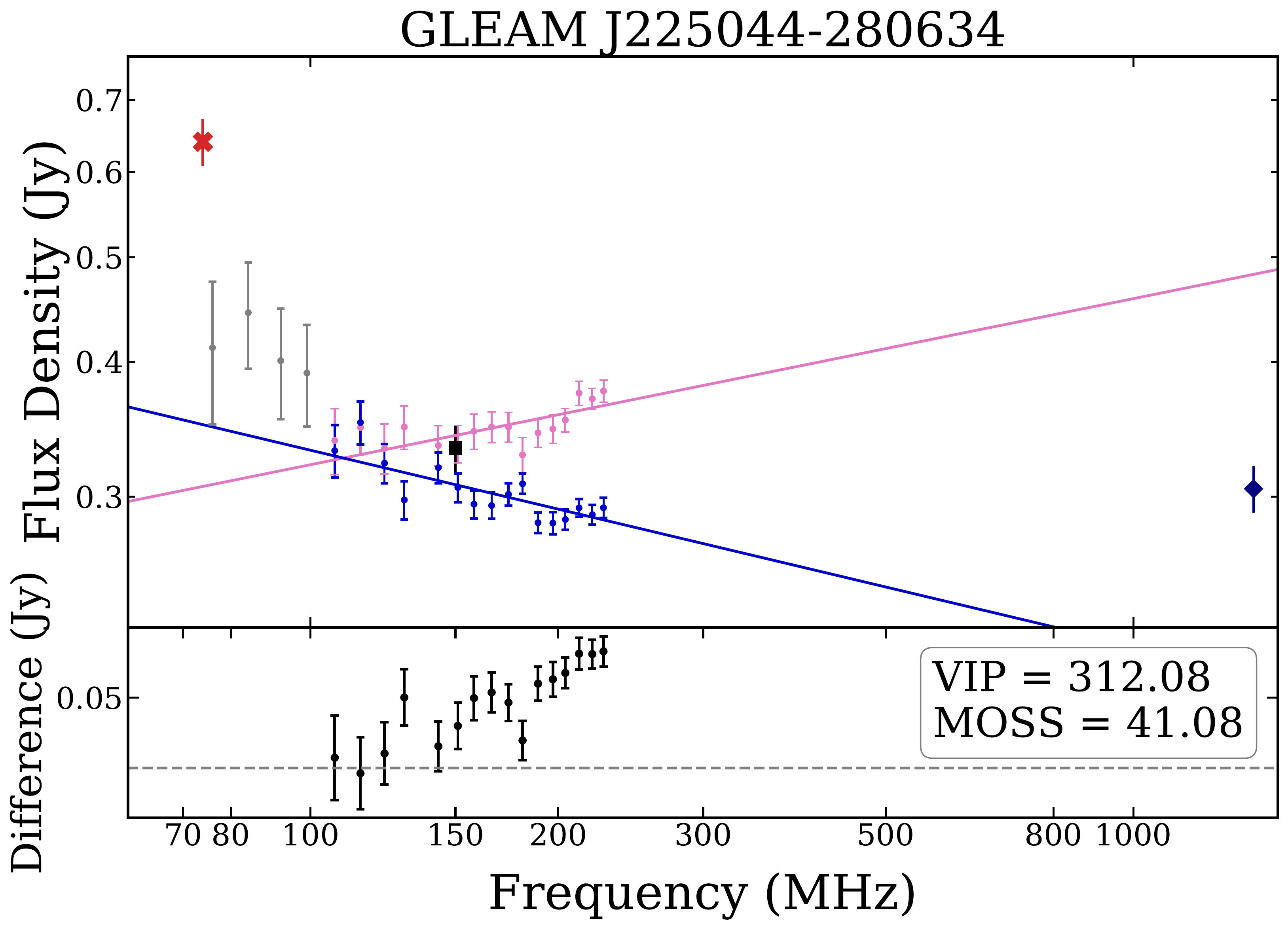} &
\includegraphics[scale=0.15]{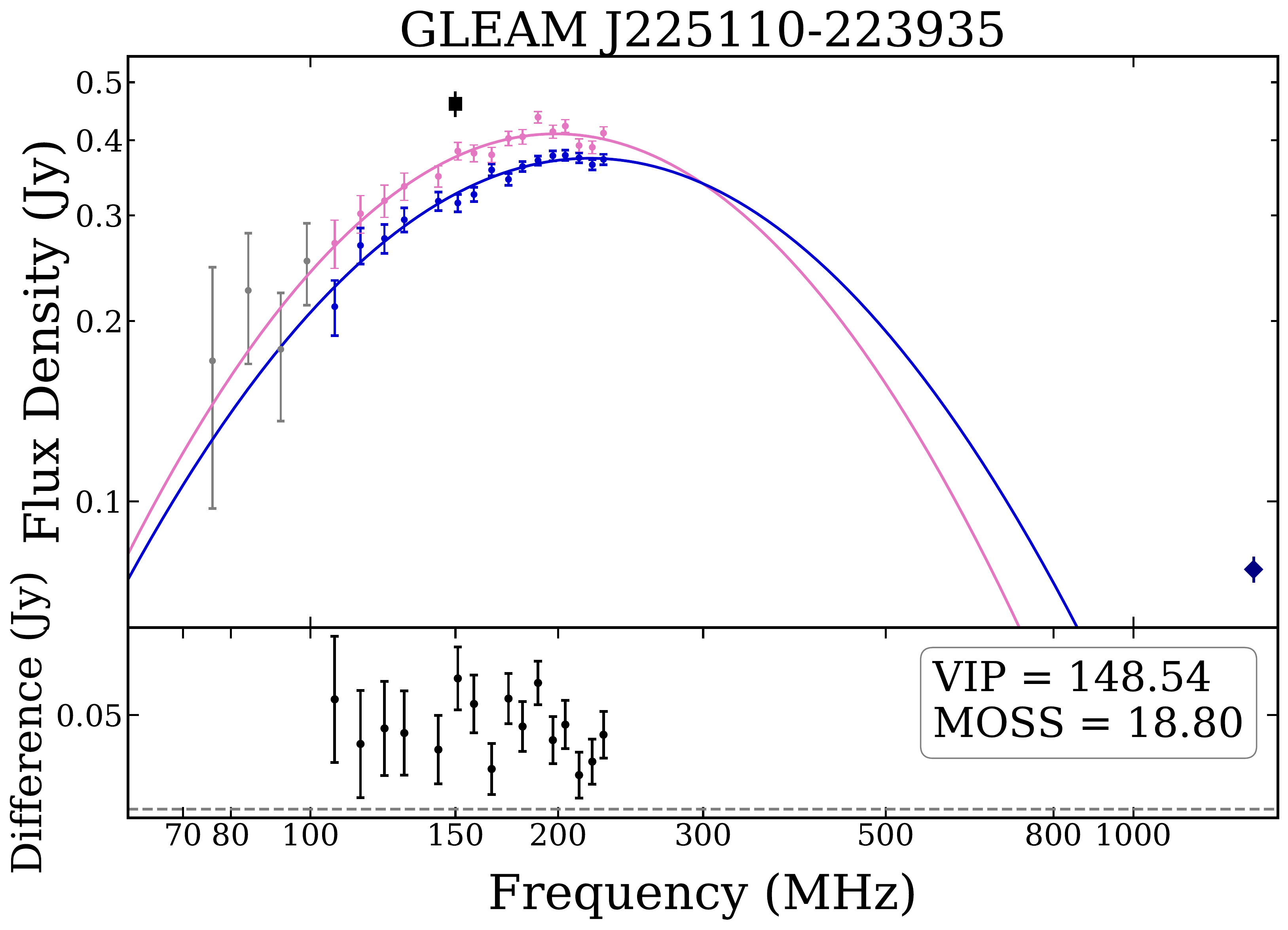} &
\includegraphics[scale=0.15]{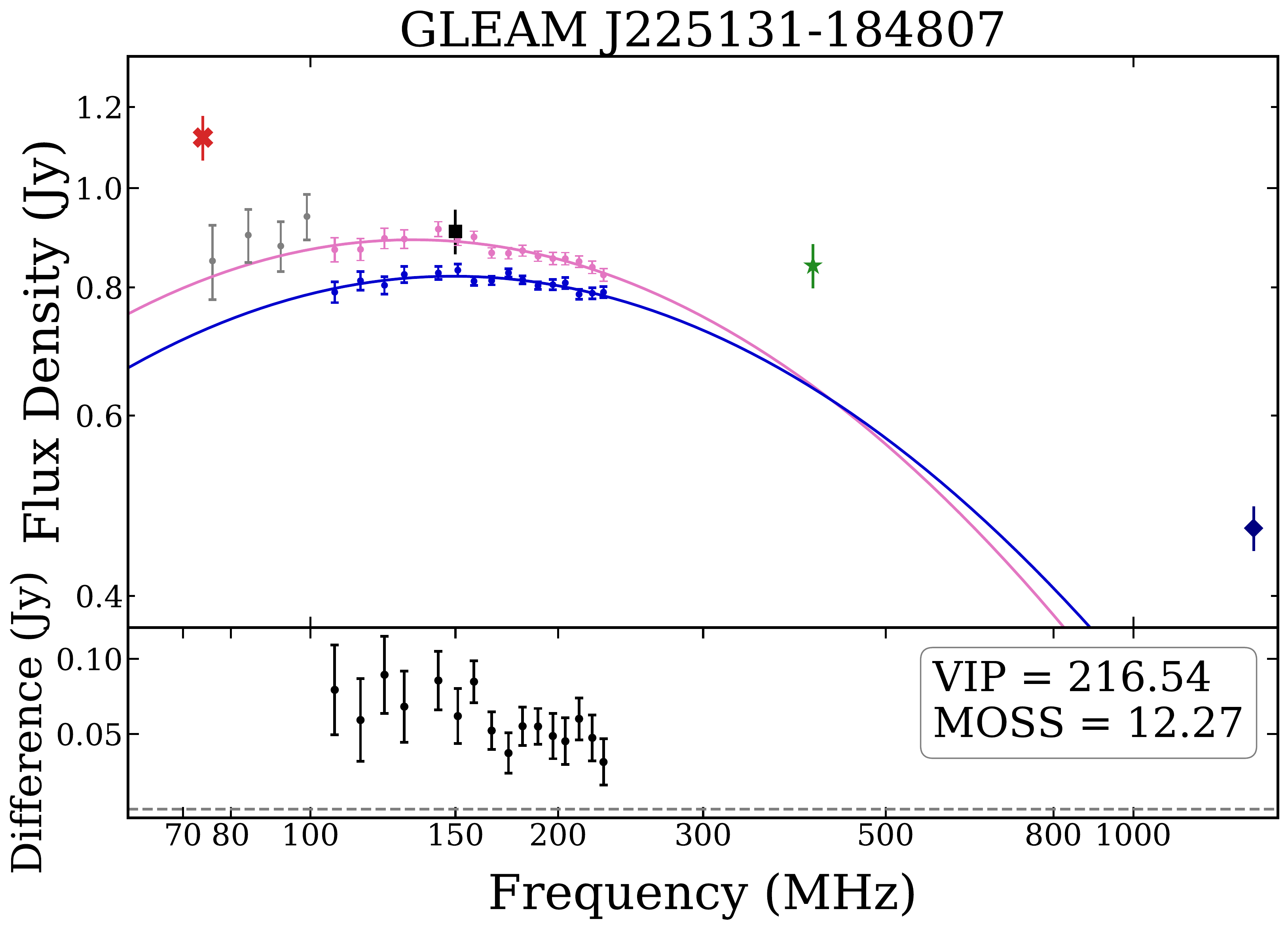} \\
\includegraphics[scale=0.15]{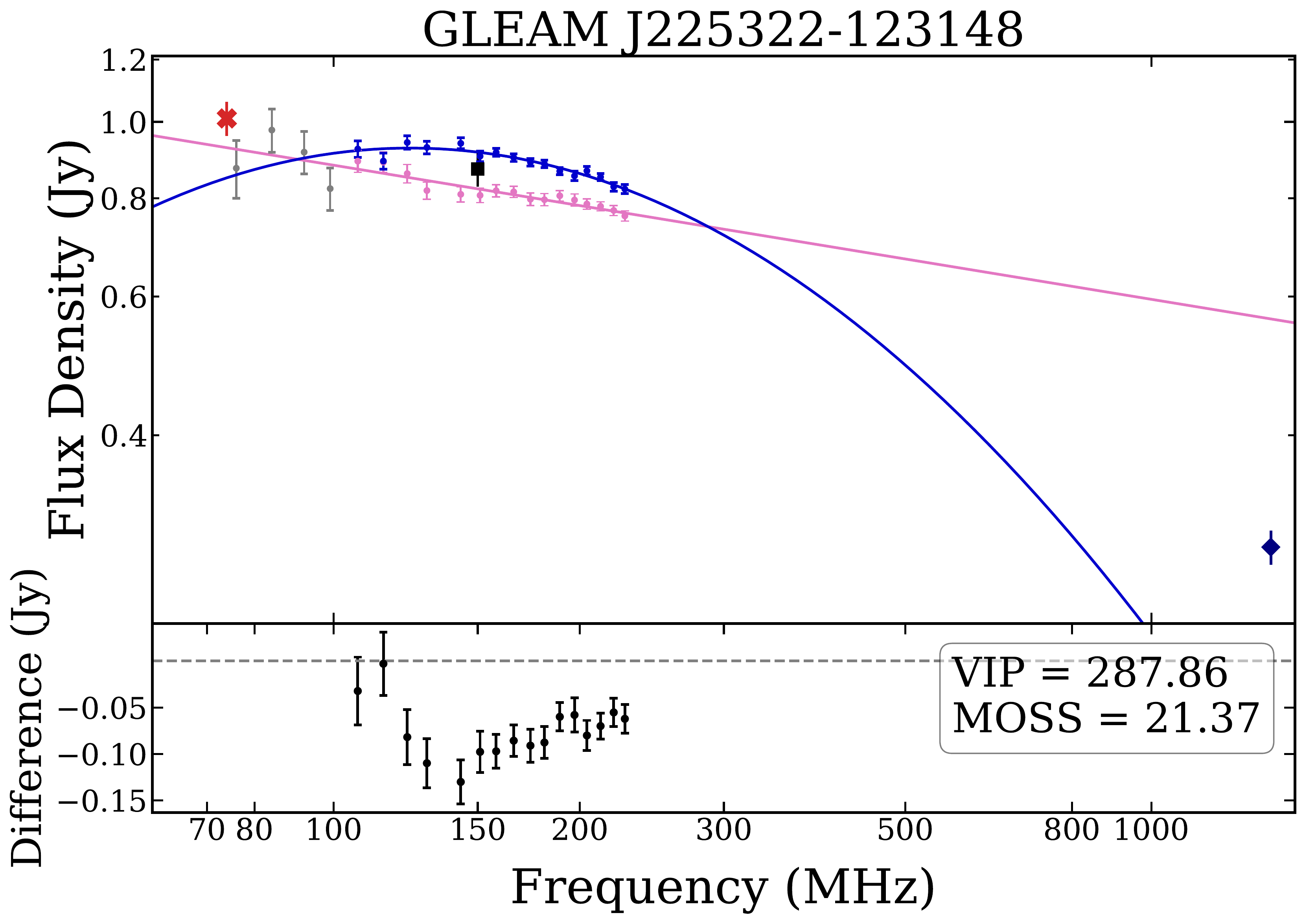} &
\includegraphics[scale=0.15]{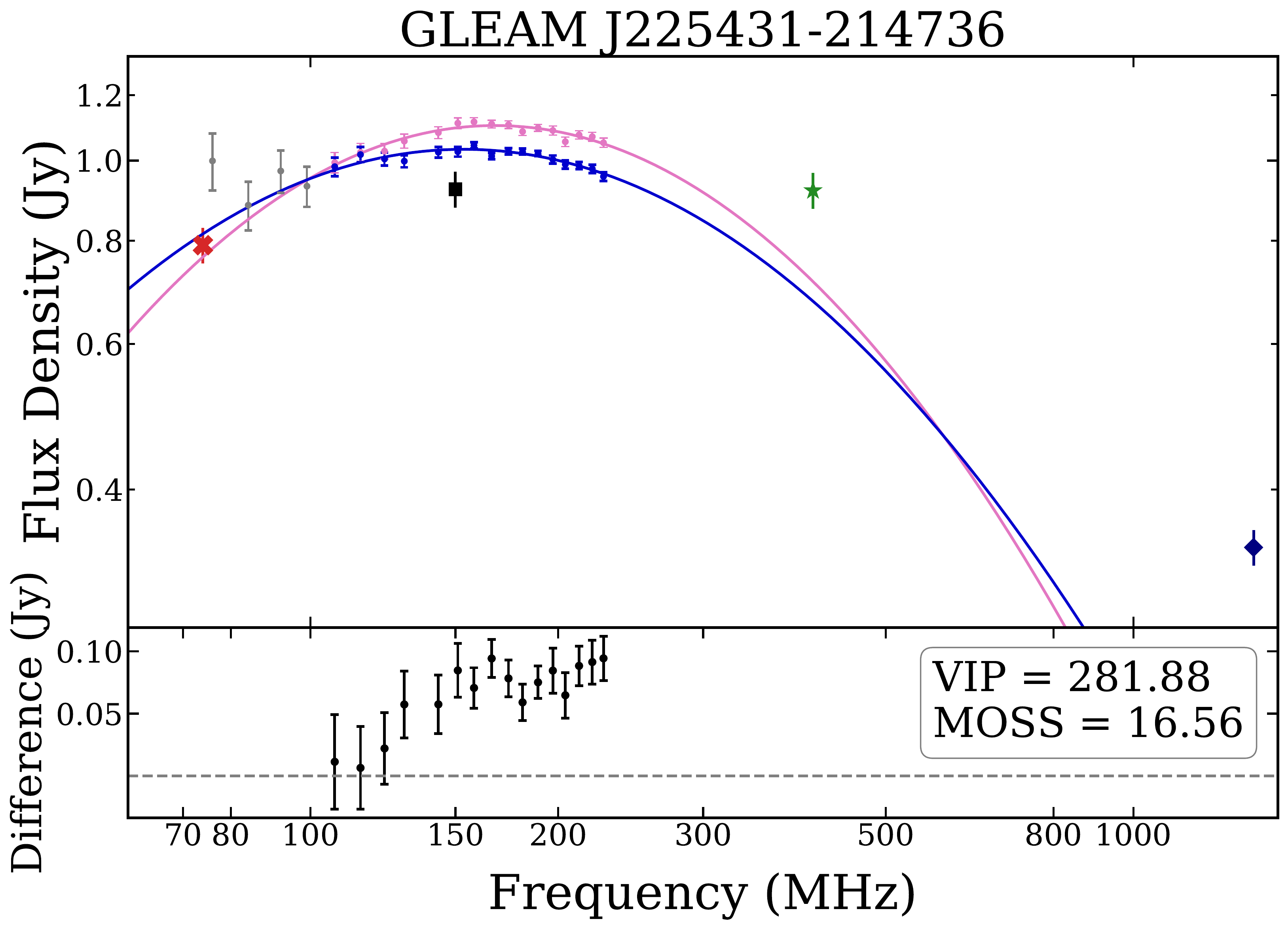} &
\includegraphics[scale=0.15]{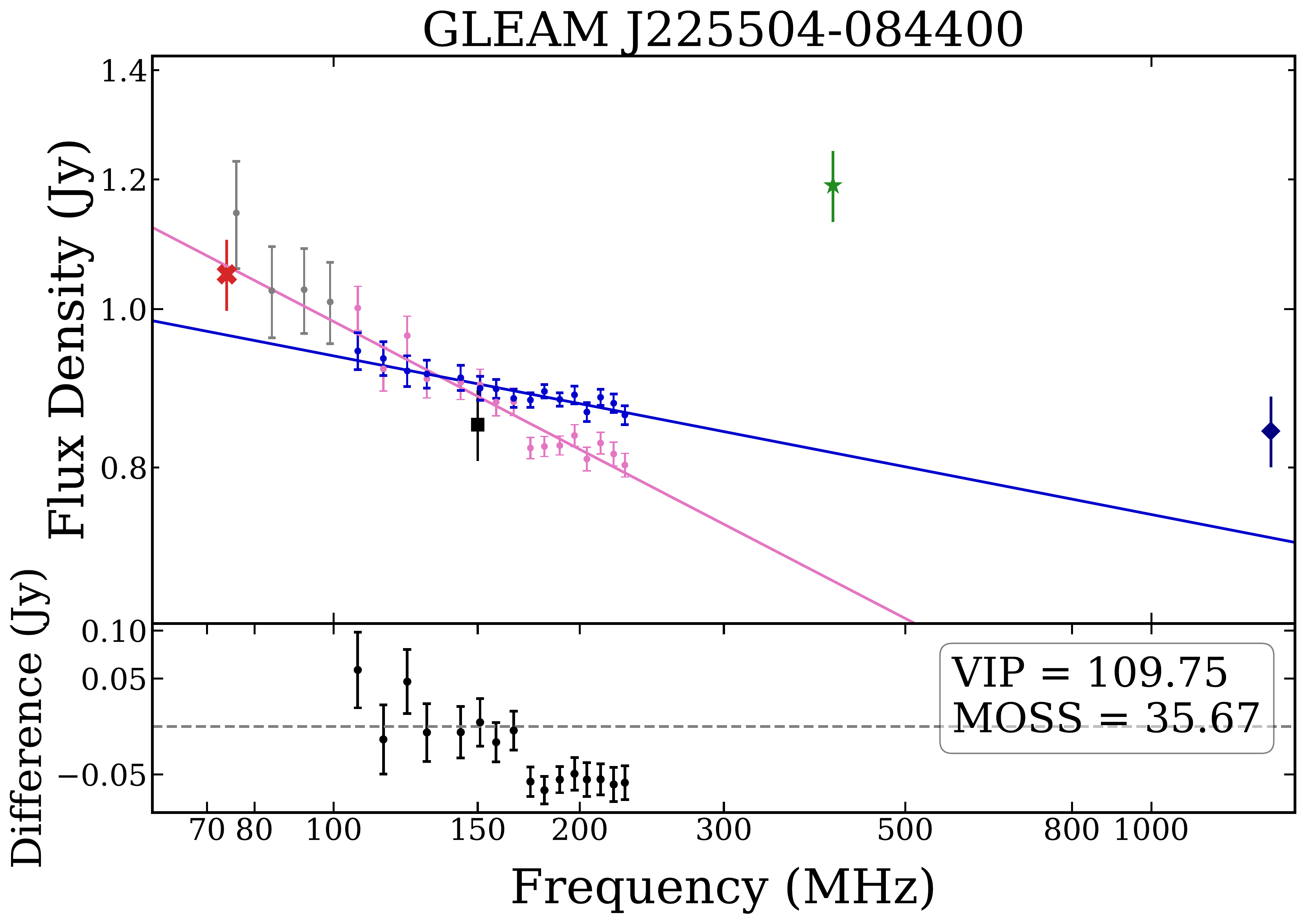} \\
\includegraphics[scale=0.15]{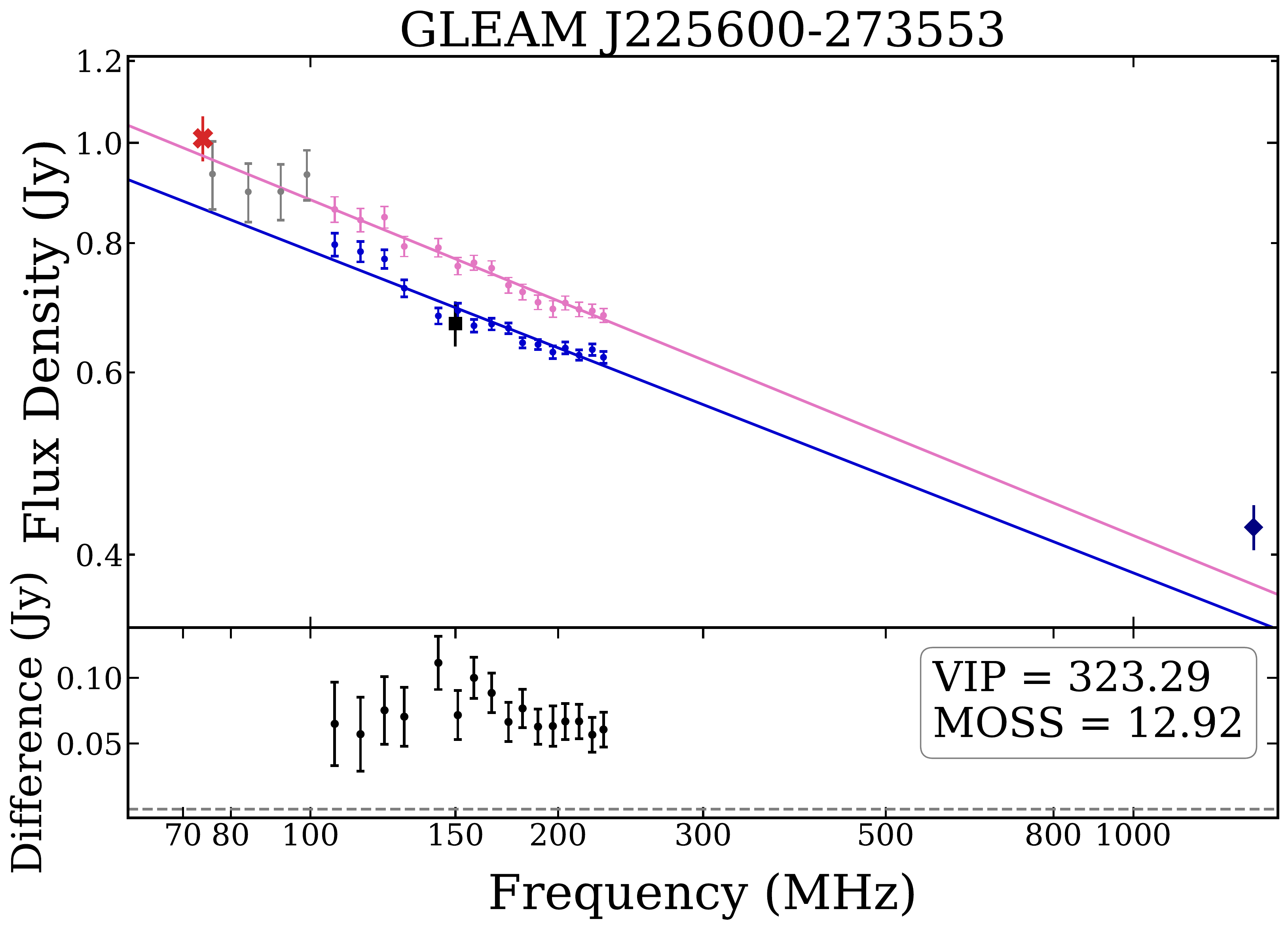} &
\includegraphics[scale=0.15]{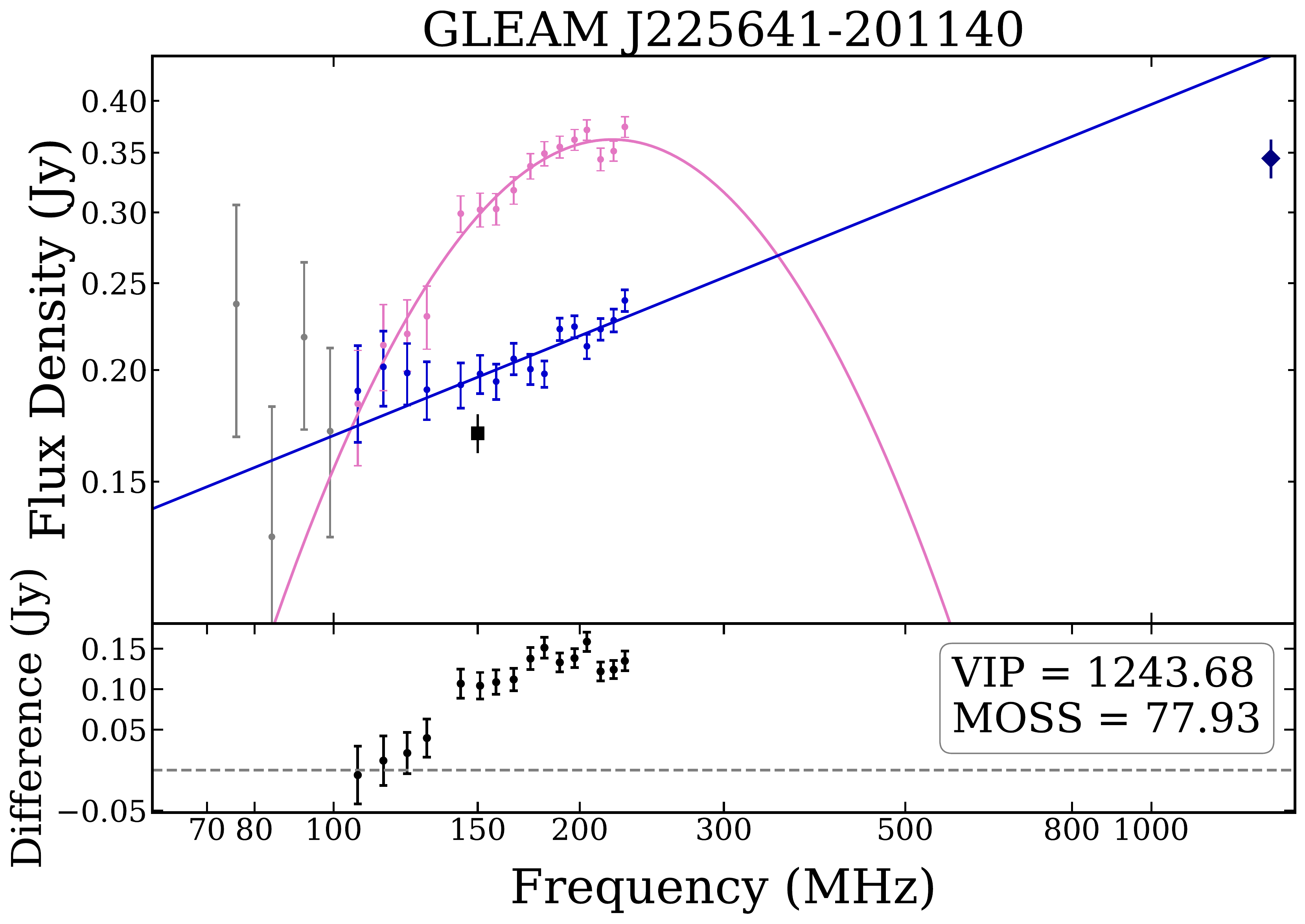} &
\includegraphics[scale=0.15]{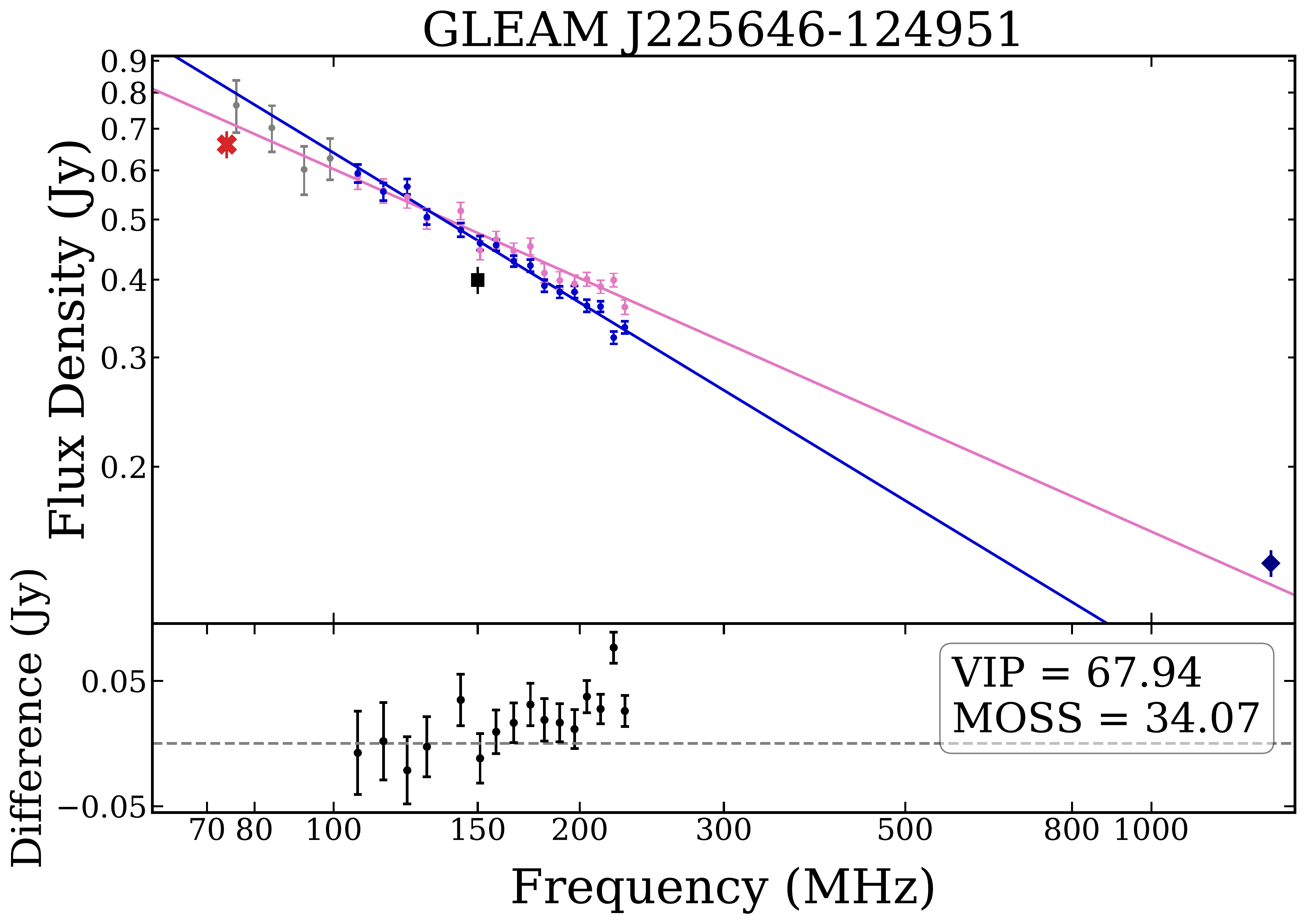} \\
\includegraphics[scale=0.15]{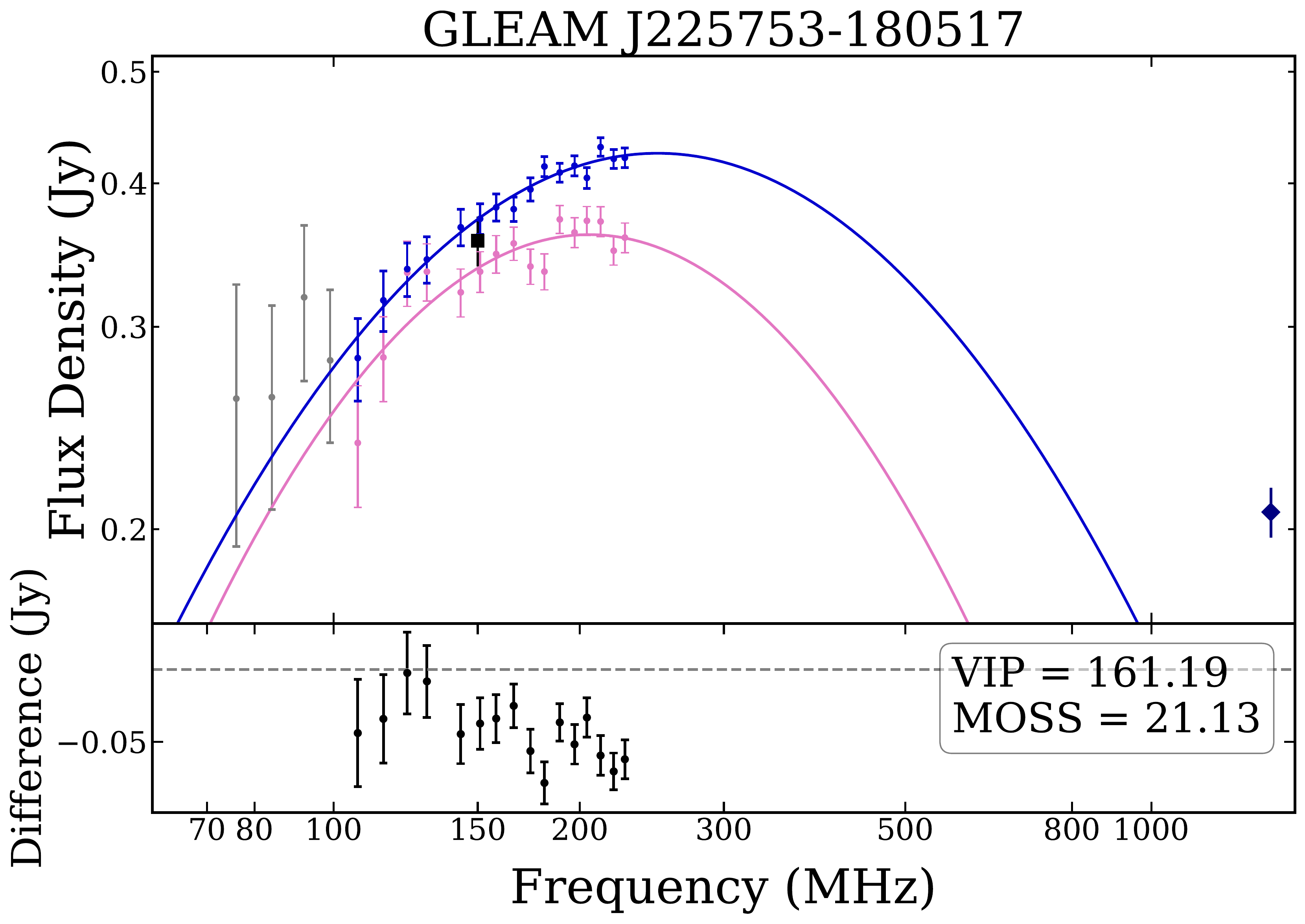} &
\includegraphics[scale=0.15]{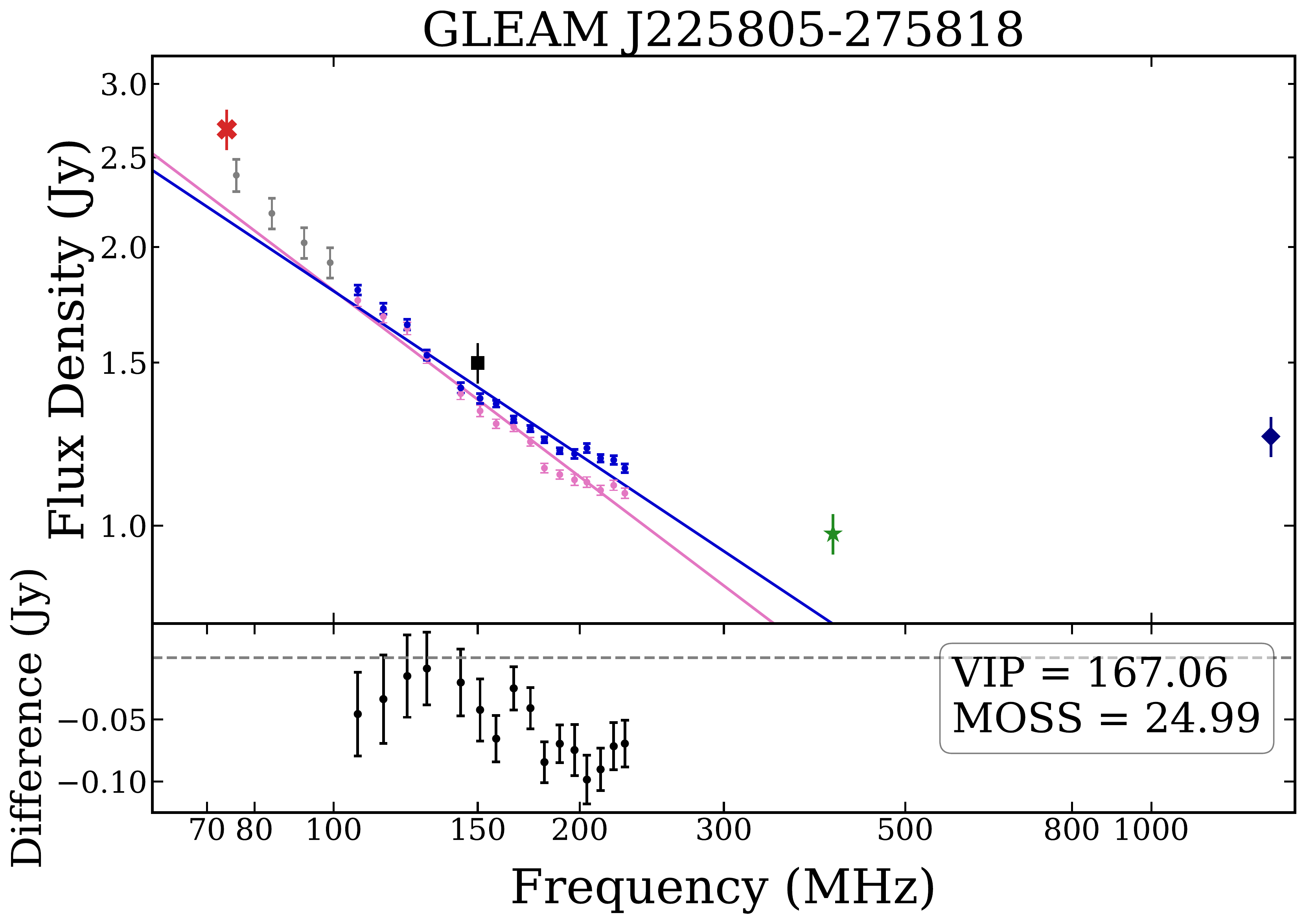} &
\includegraphics[scale=0.15]{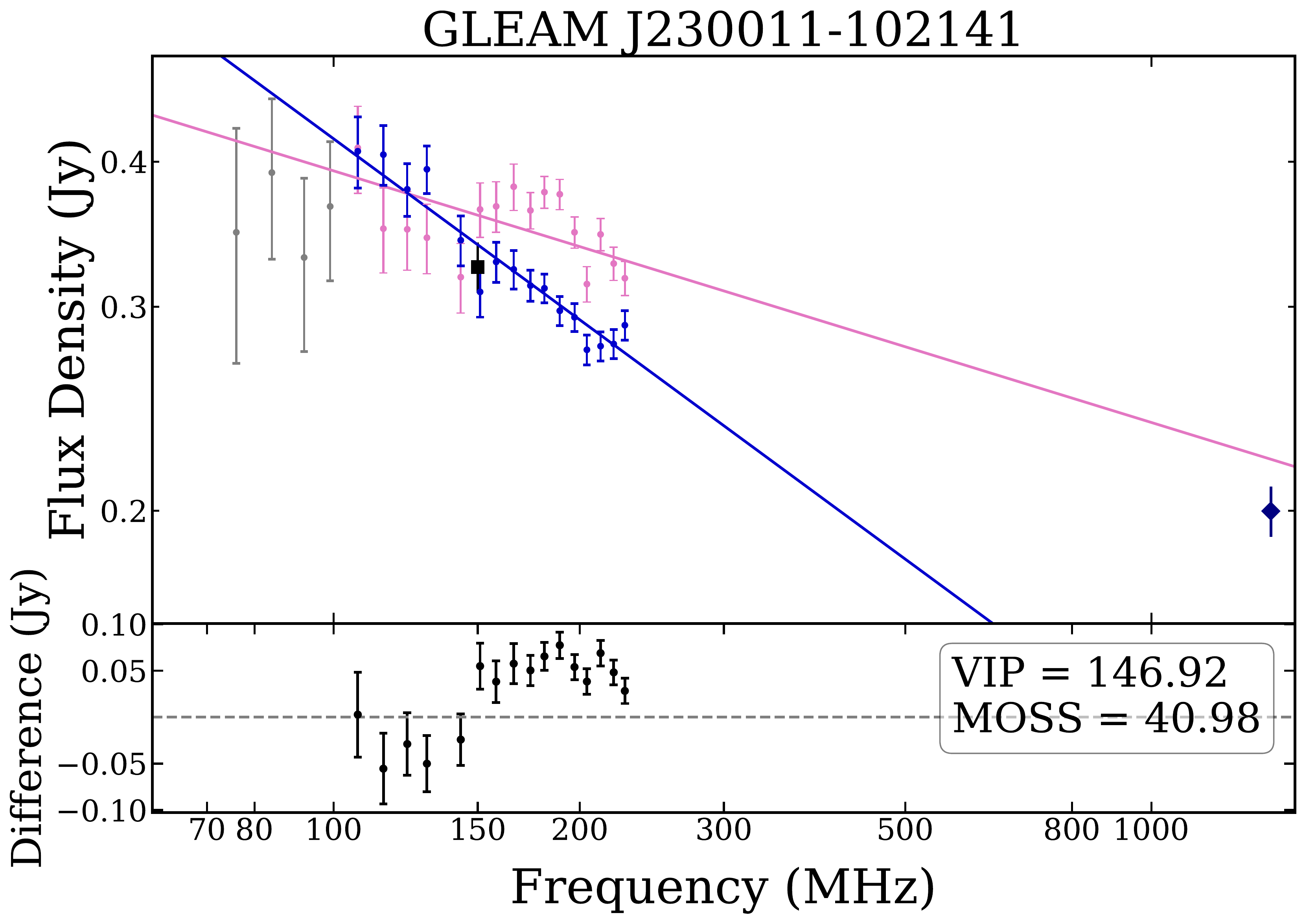} \\
\end{array}$
\caption{(continued) SEDs for all sources classified as variable according to the VIP. For each source the points represent the following data: GLEAM low frequency (72--100\,MHz) (grey circles), Year 1 (pink circles), Year 2 (blue circles), VLSSr (red cross), TGSS (black square), MRC (green star), SUMSS (yellow pentagon), and NVSS (navy diamond). The models for each year are determined by their classification; a source classified with a peak within the observed band was modelled by a quadratic according to Equation~\ref{eq:quadratic}, remaining sources were modelled by a power-law according to Equation~\ref{eq:plaw}.}
\label{app:fig:pg16}
\end{center}
\end{figure*}
\setcounter{figure}{0}
\begin{figure*}
\begin{center}$
\begin{array}{cccccc}
\includegraphics[scale=0.15]{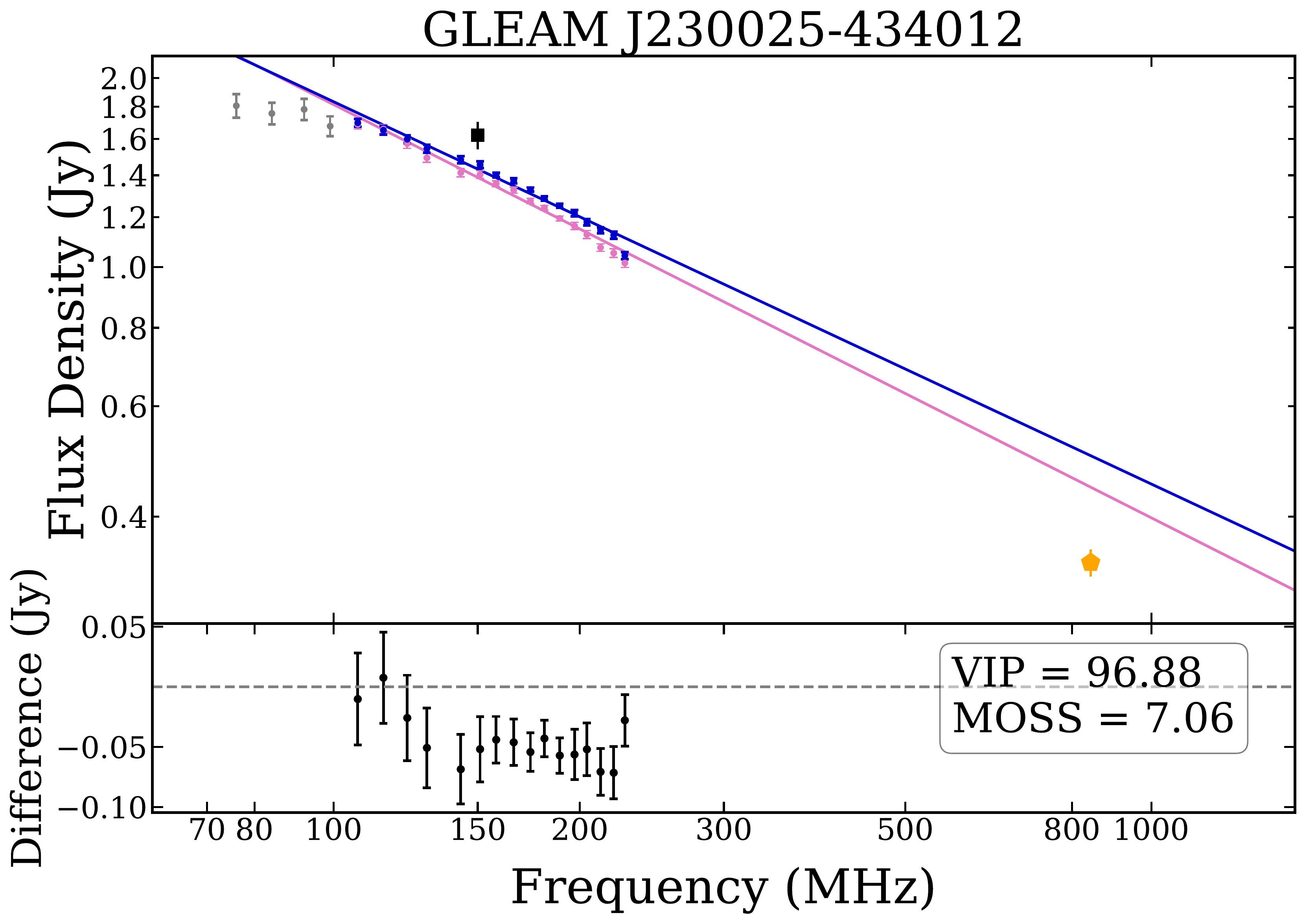} &
\includegraphics[scale=0.15]{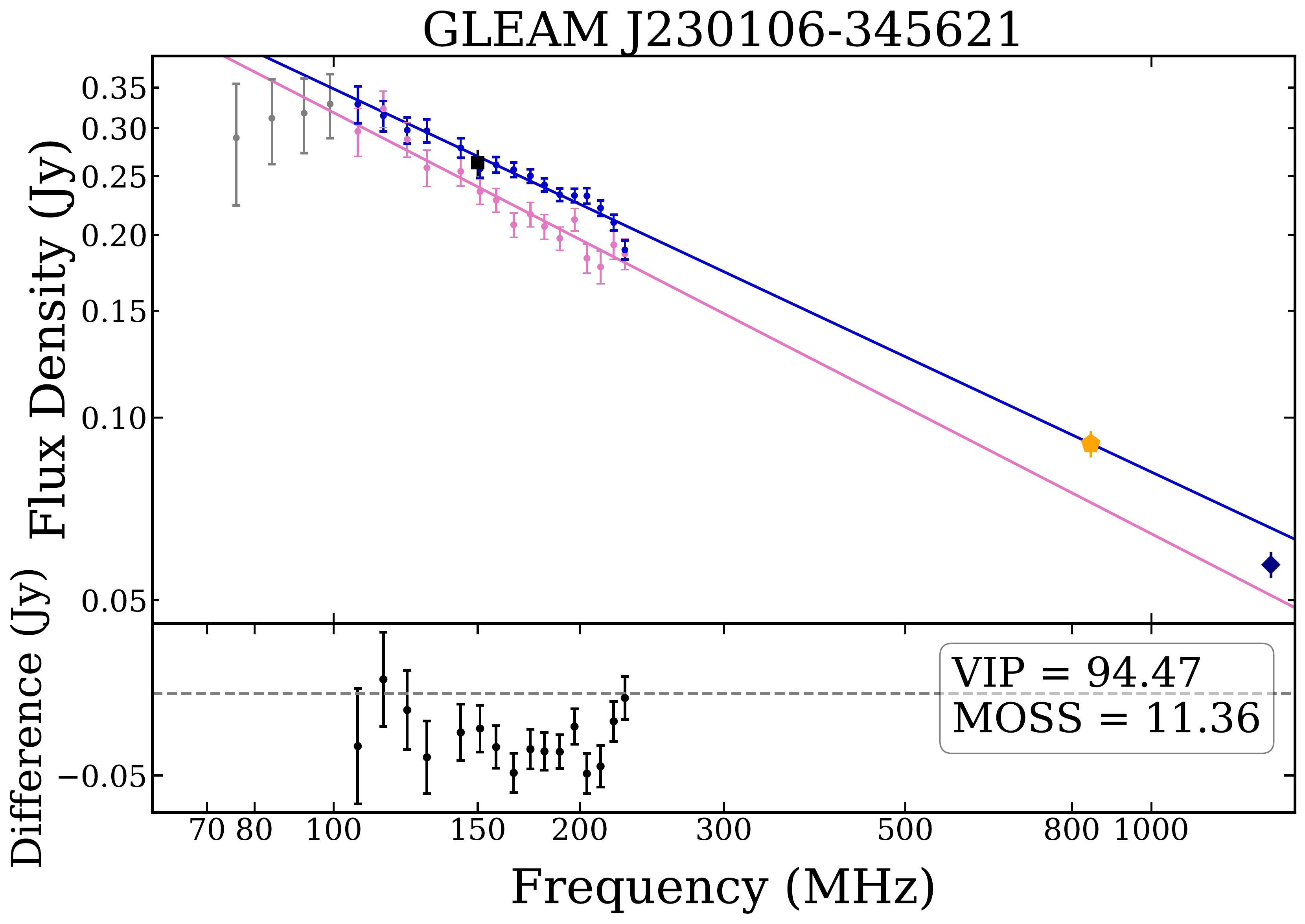} &
\includegraphics[scale=0.15]{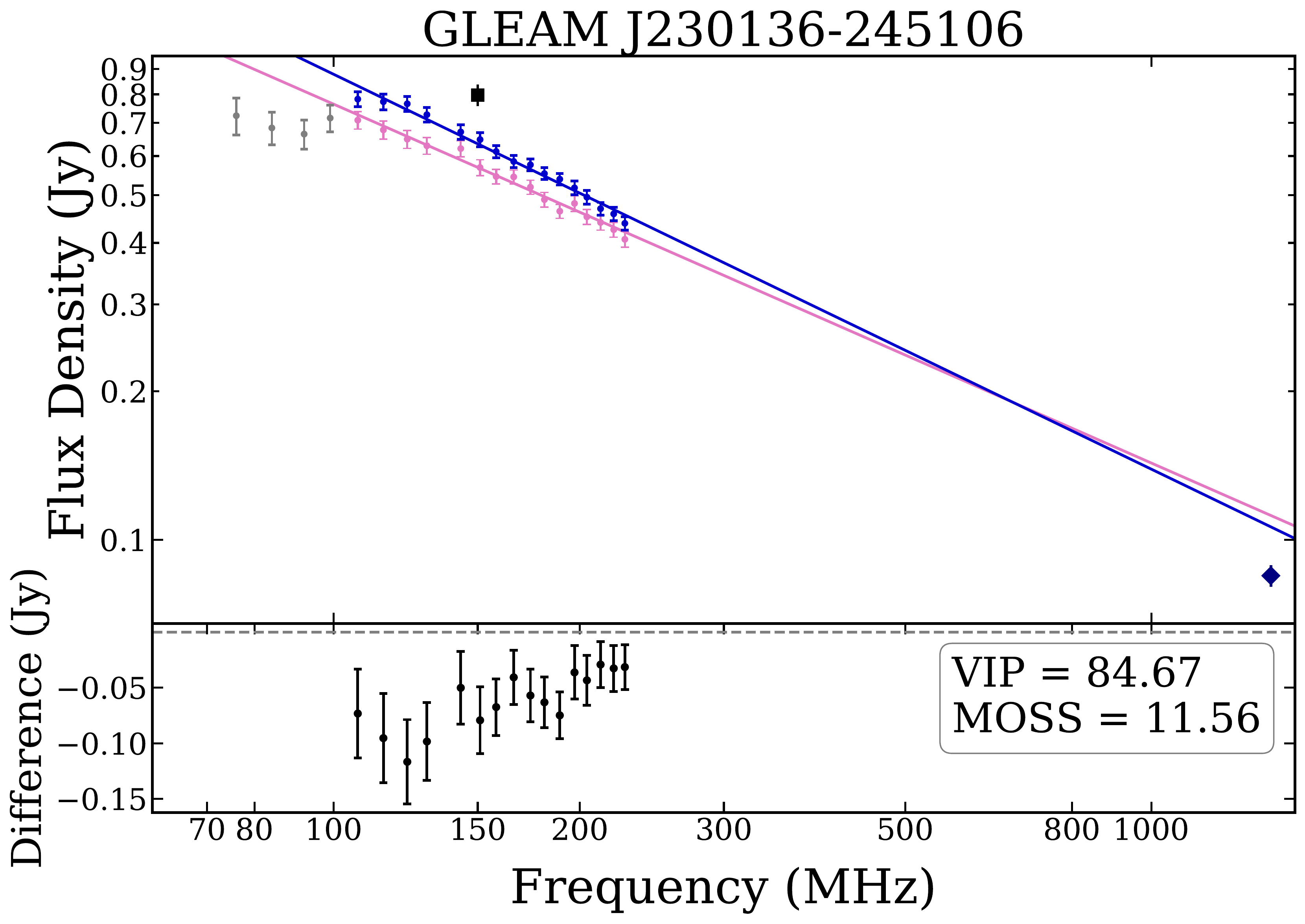} \\
\includegraphics[scale=0.15]{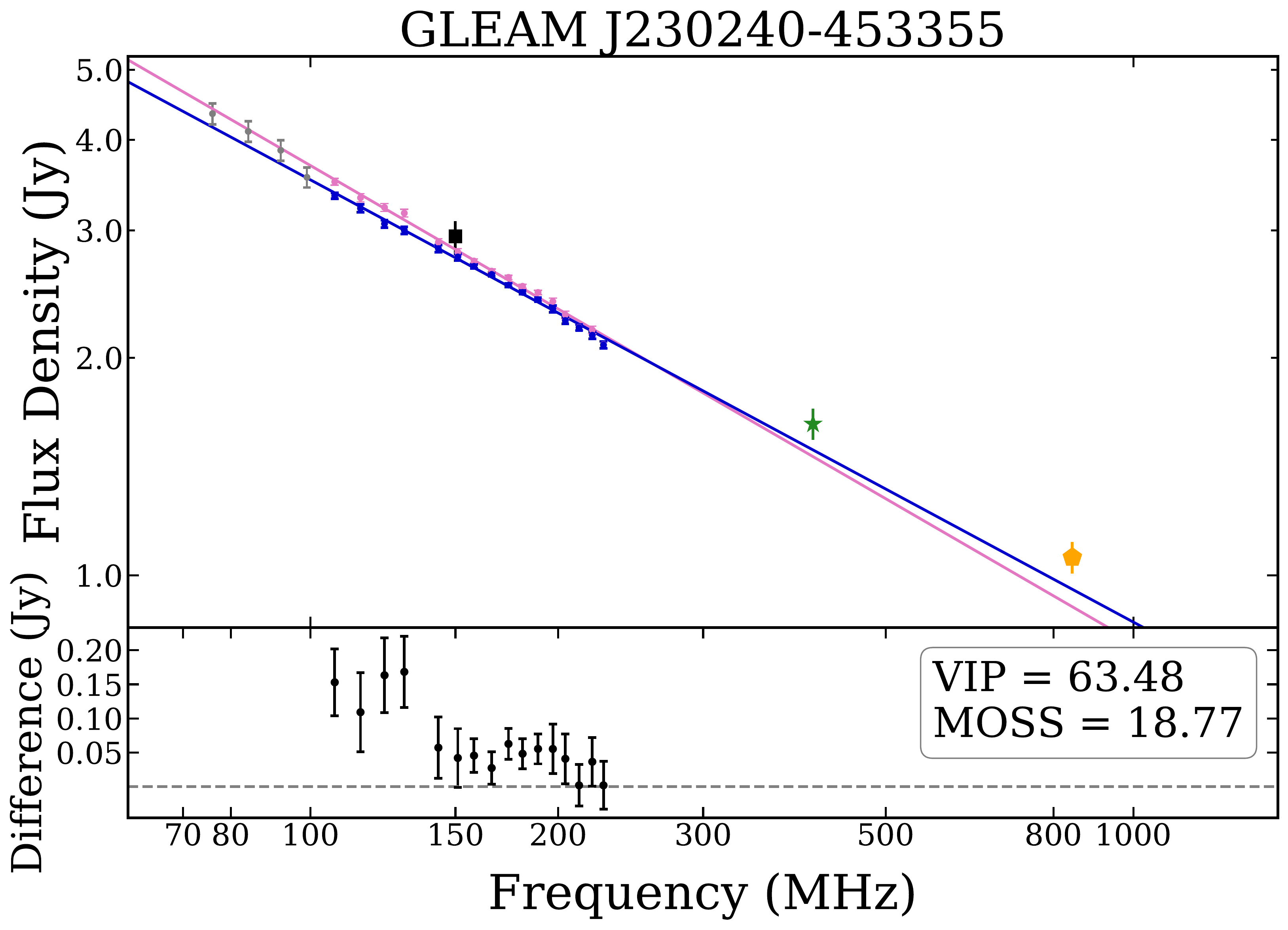} &
\includegraphics[scale=0.15]{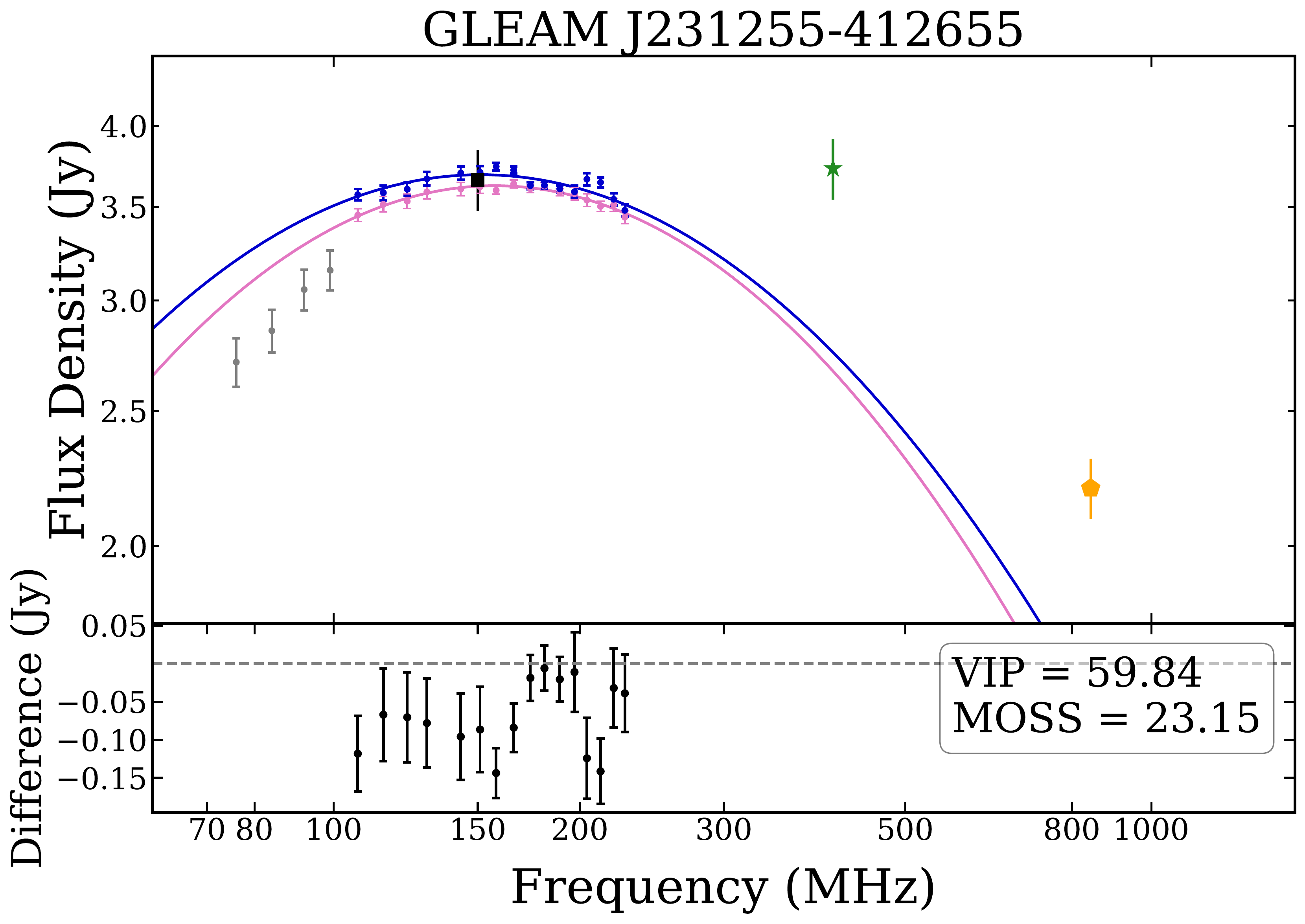} &
\includegraphics[scale=0.15]{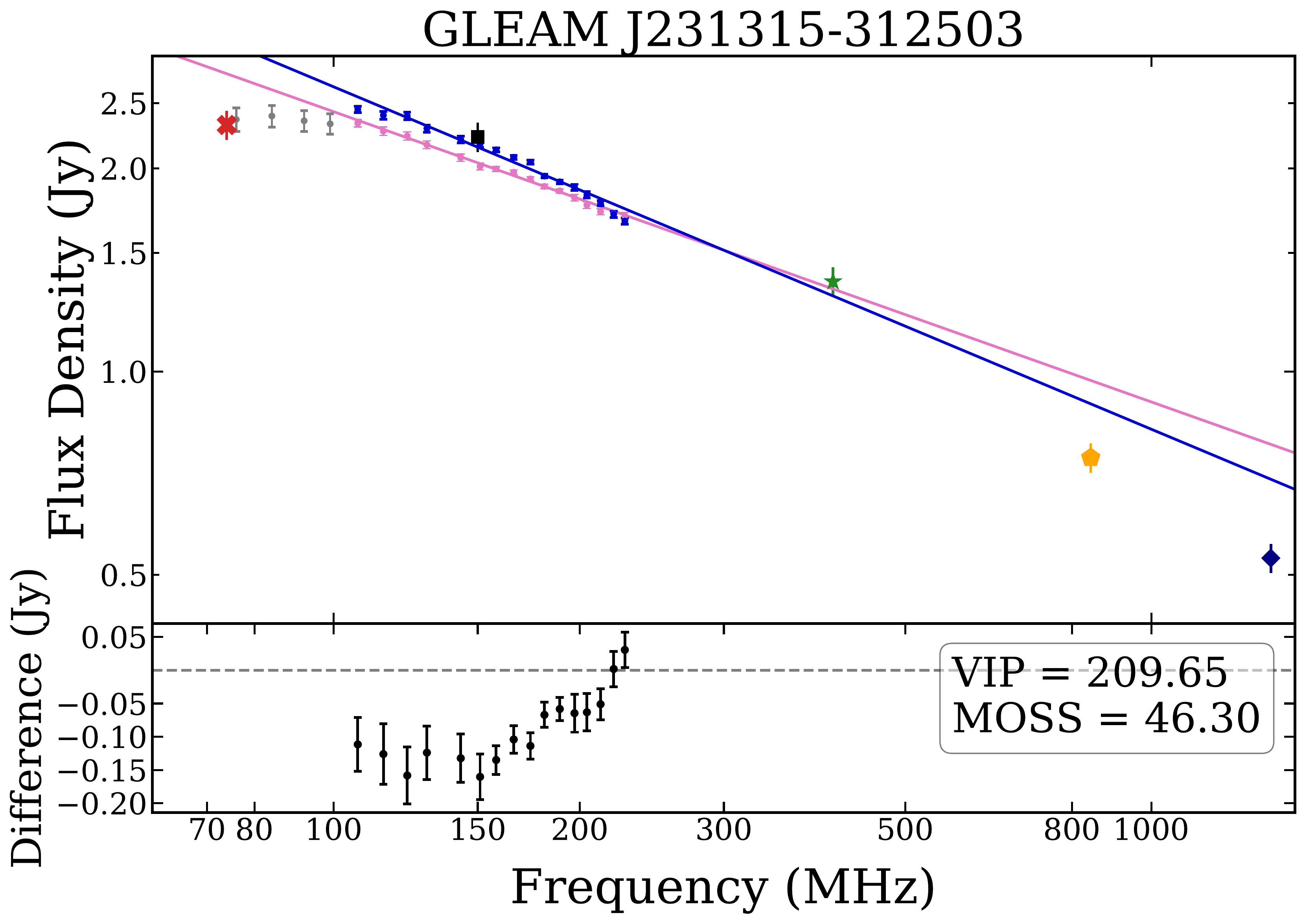} \\
\includegraphics[scale=0.15]{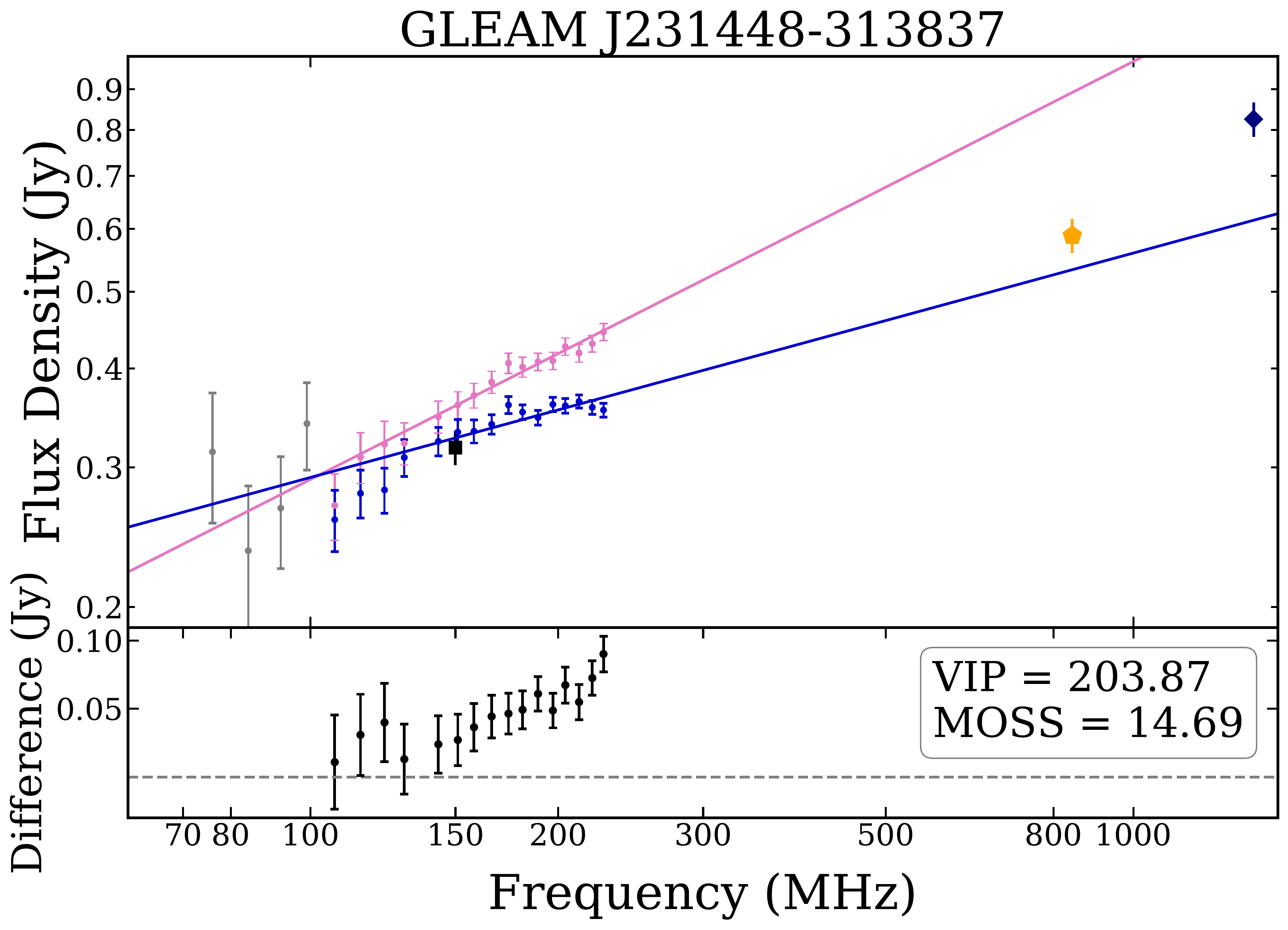} &
\includegraphics[scale=0.15]{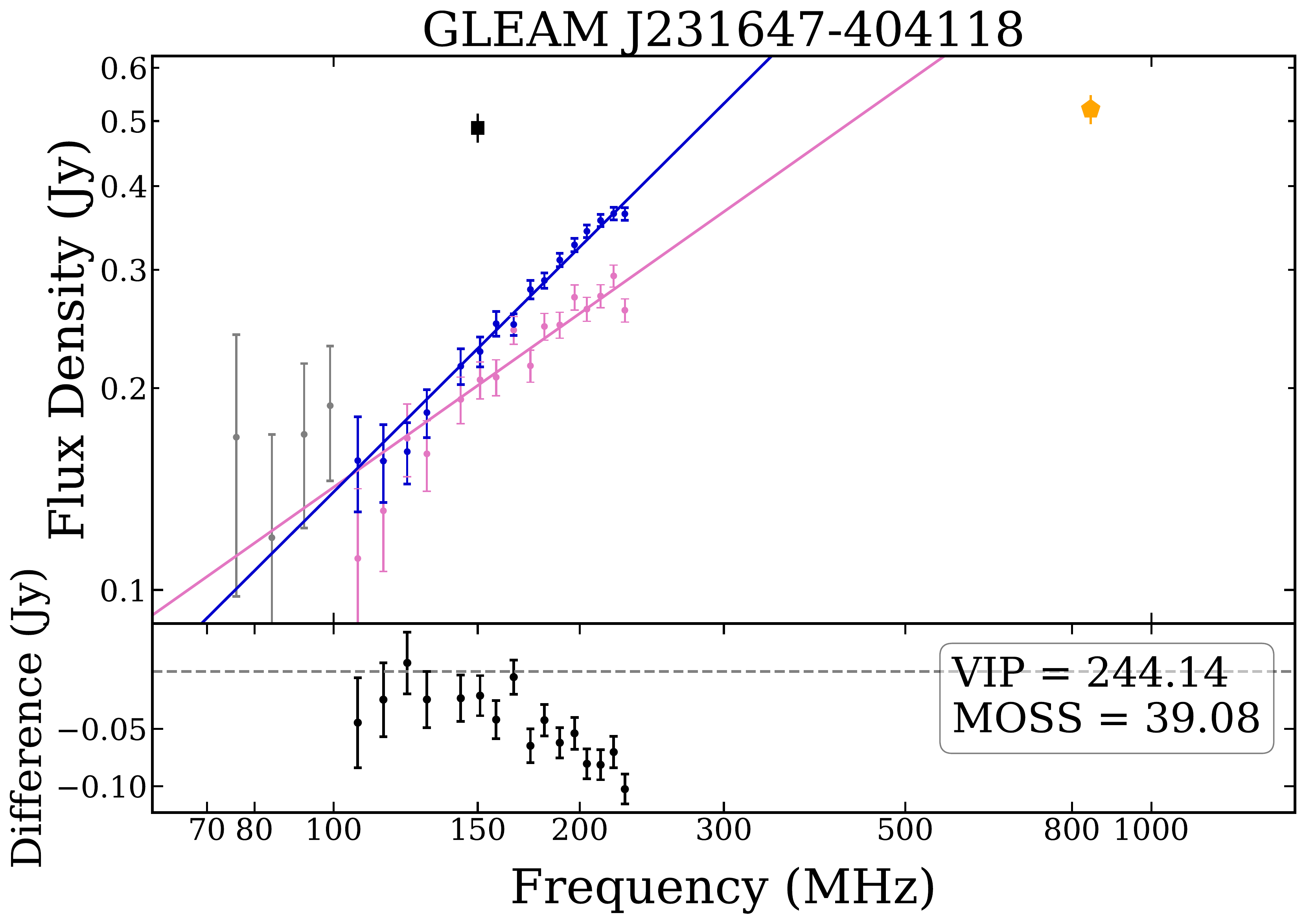} &
\includegraphics[scale=0.15]{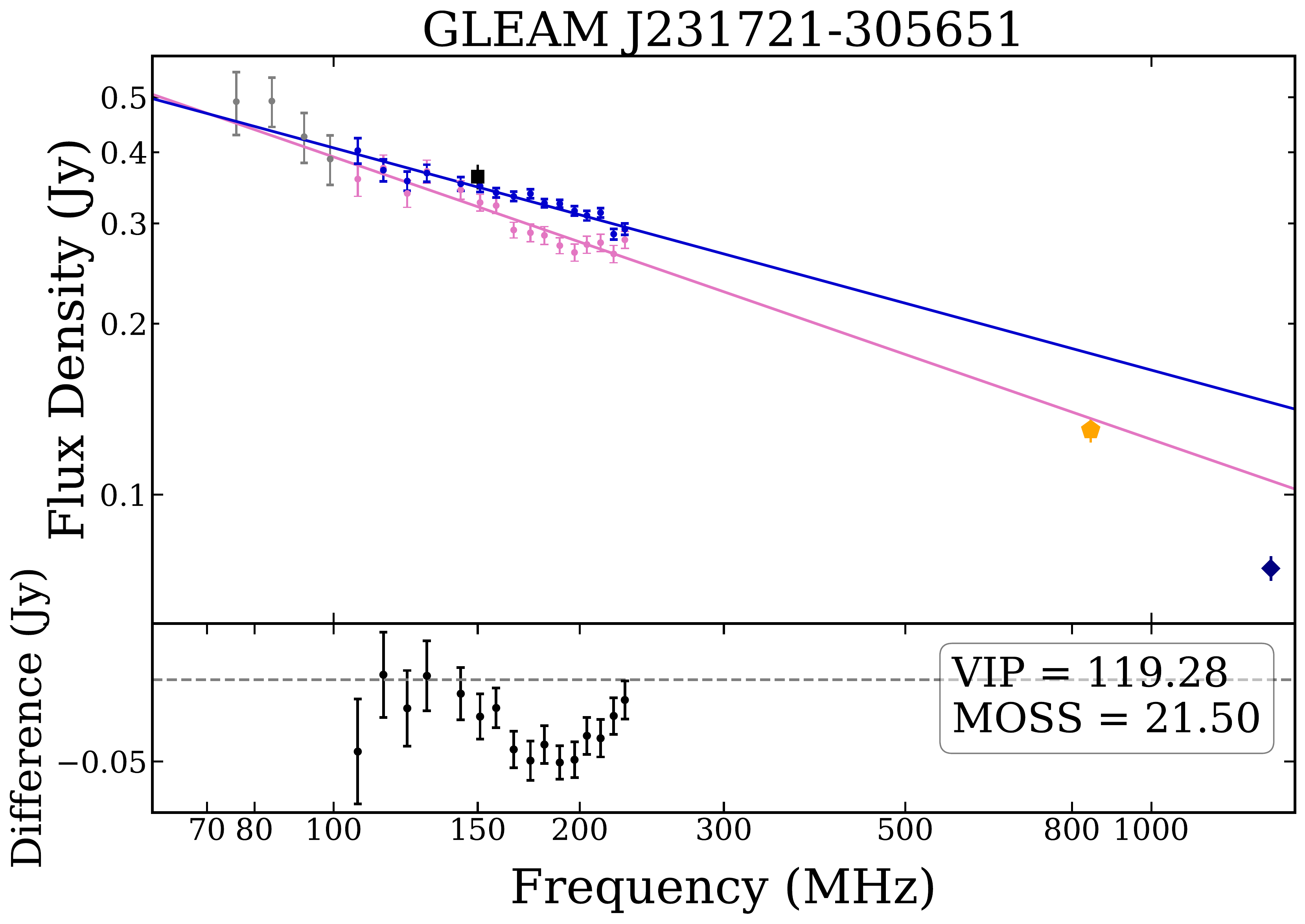} \\
\includegraphics[scale=0.15]{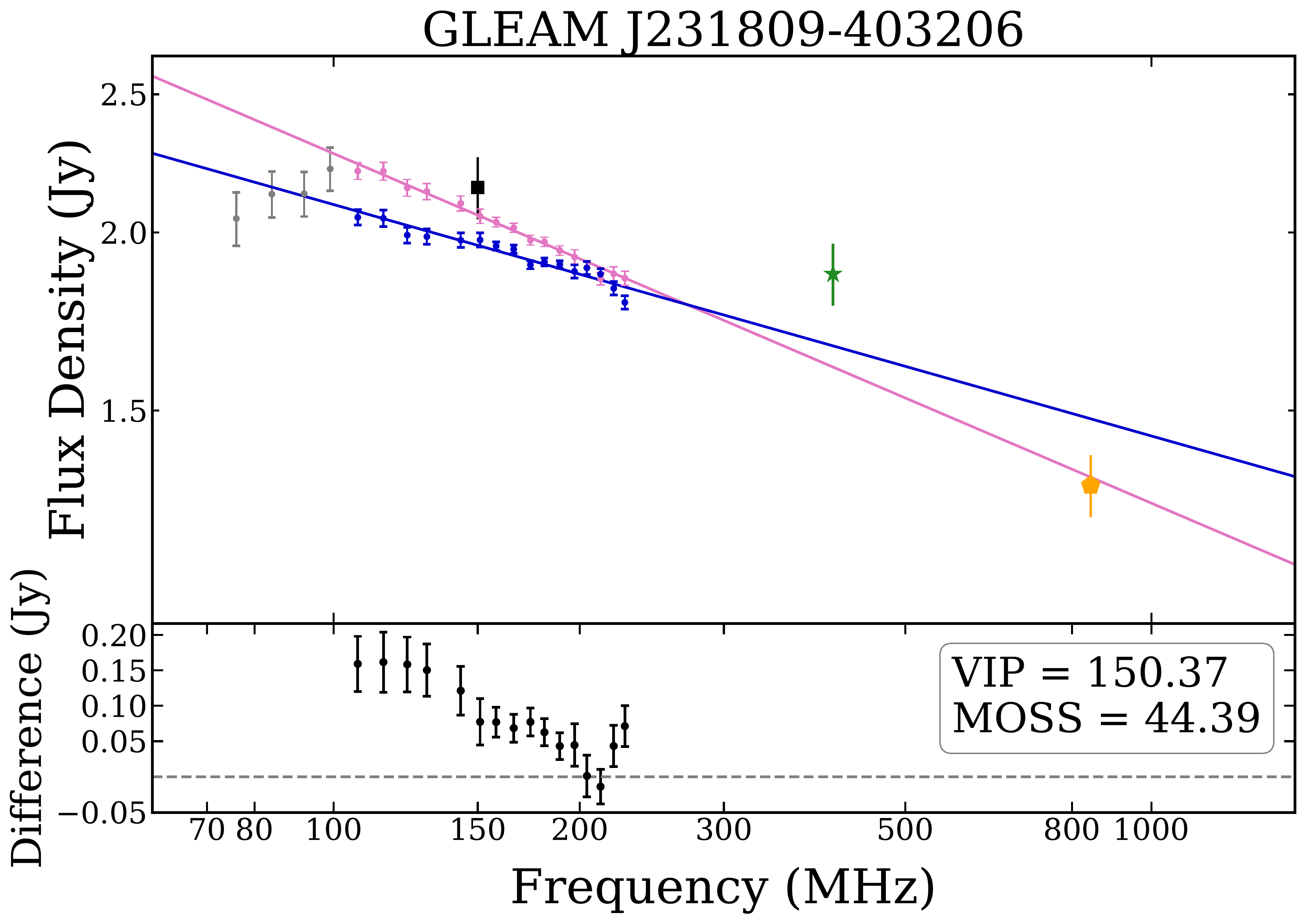} &
\includegraphics[scale=0.15]{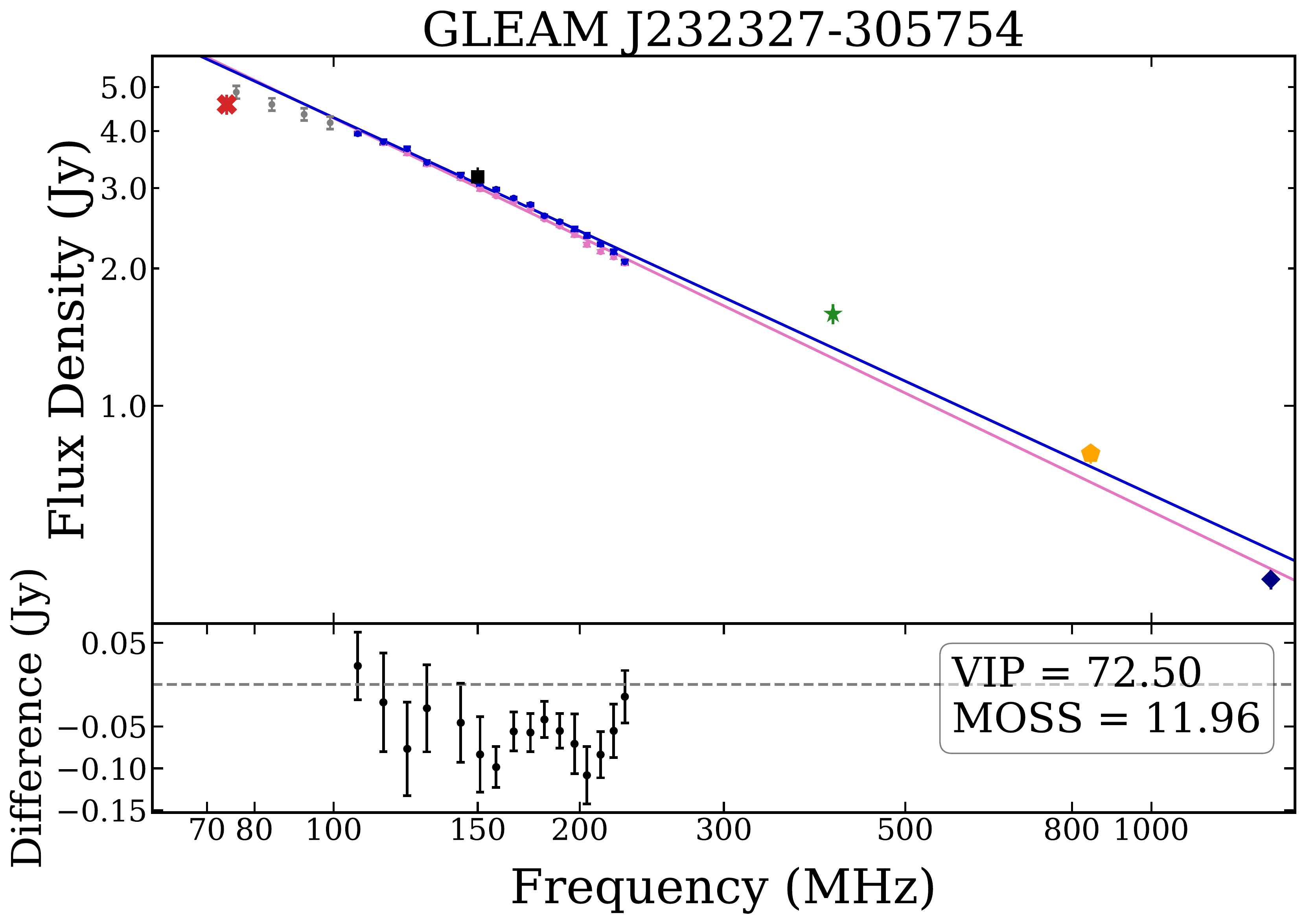} &
\includegraphics[scale=0.15]{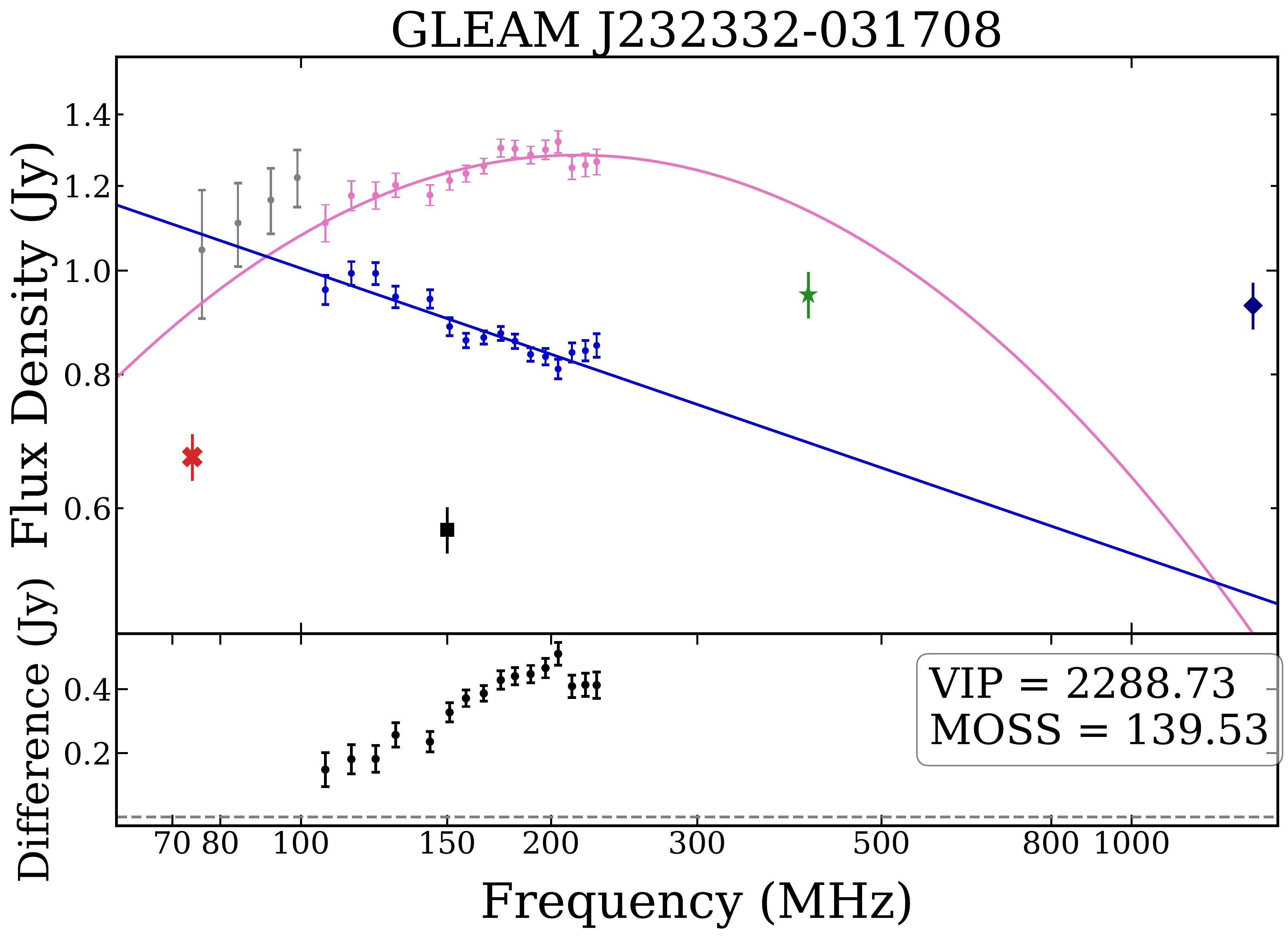} \\
\includegraphics[scale=0.15]{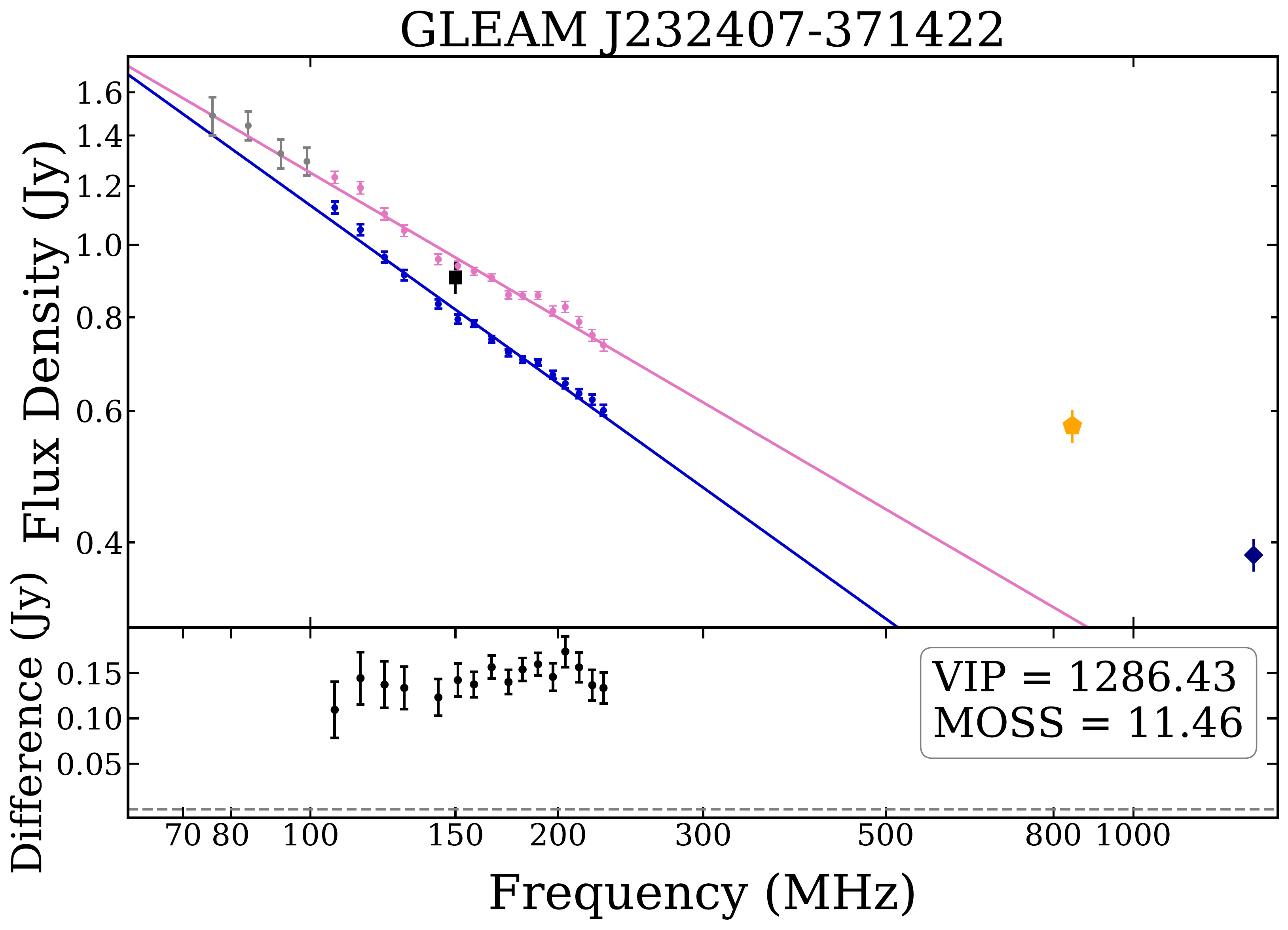} &
\includegraphics[scale=0.15]{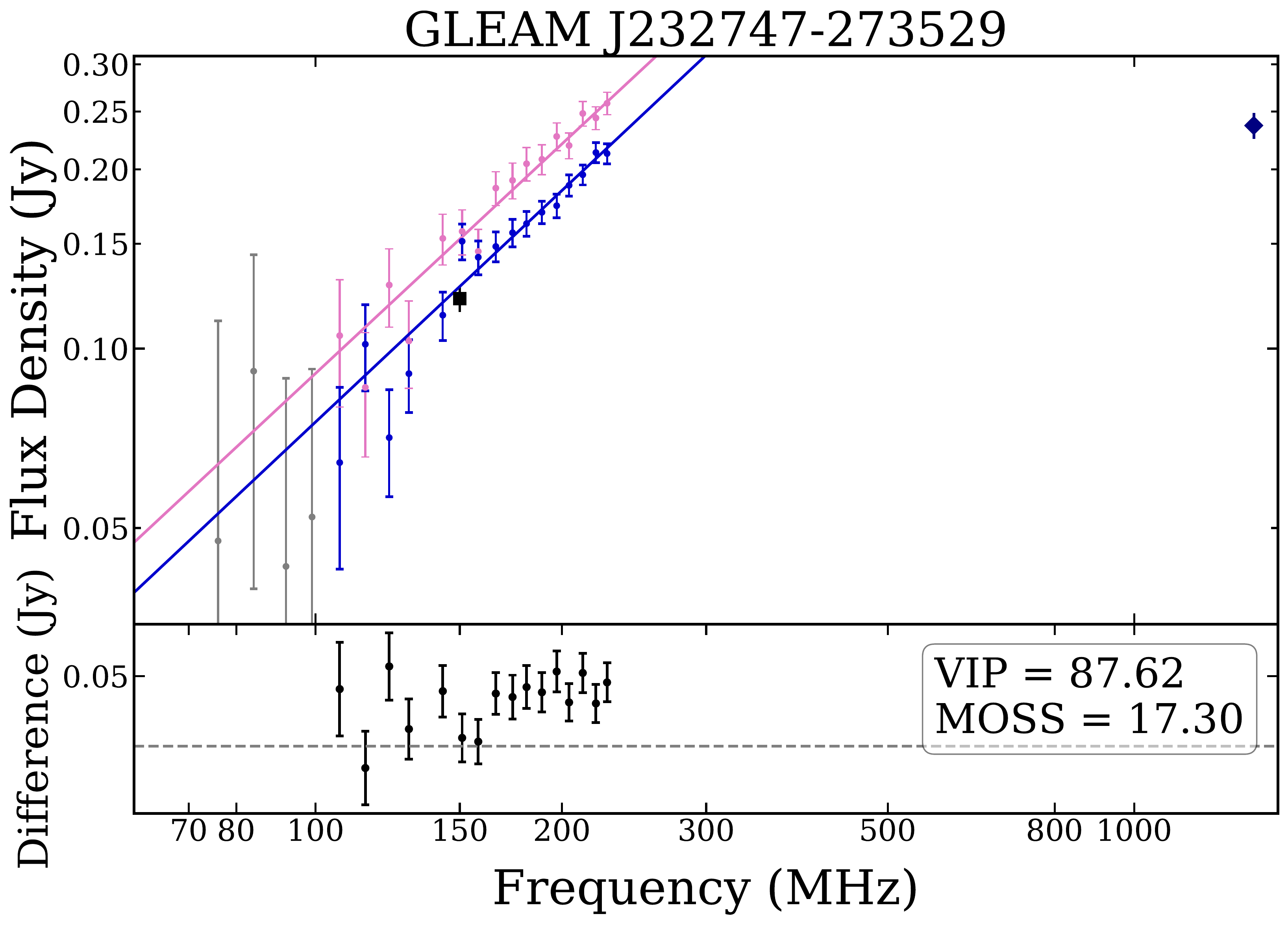} &
\includegraphics[scale=0.15]{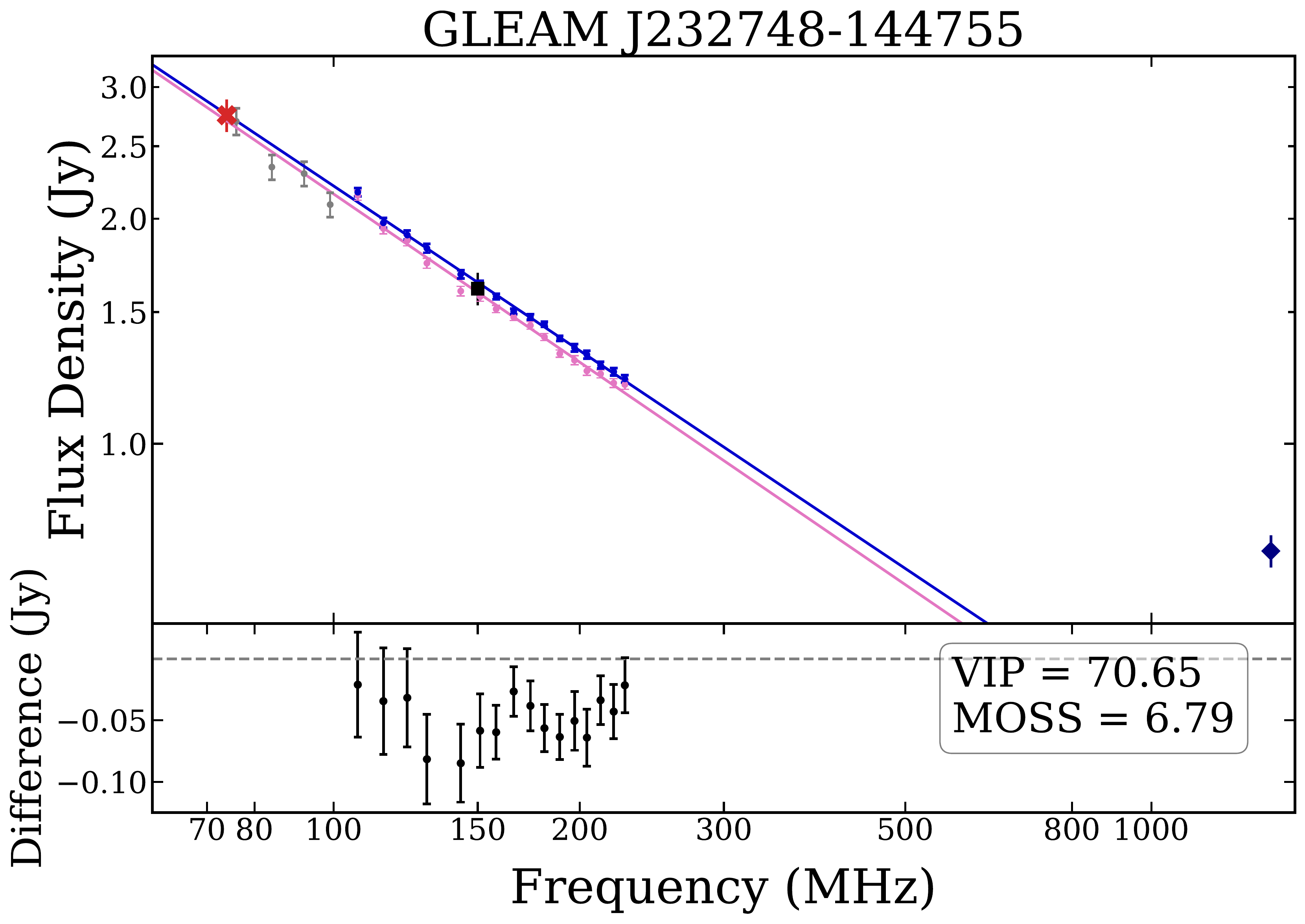} \\
\includegraphics[scale=0.15]{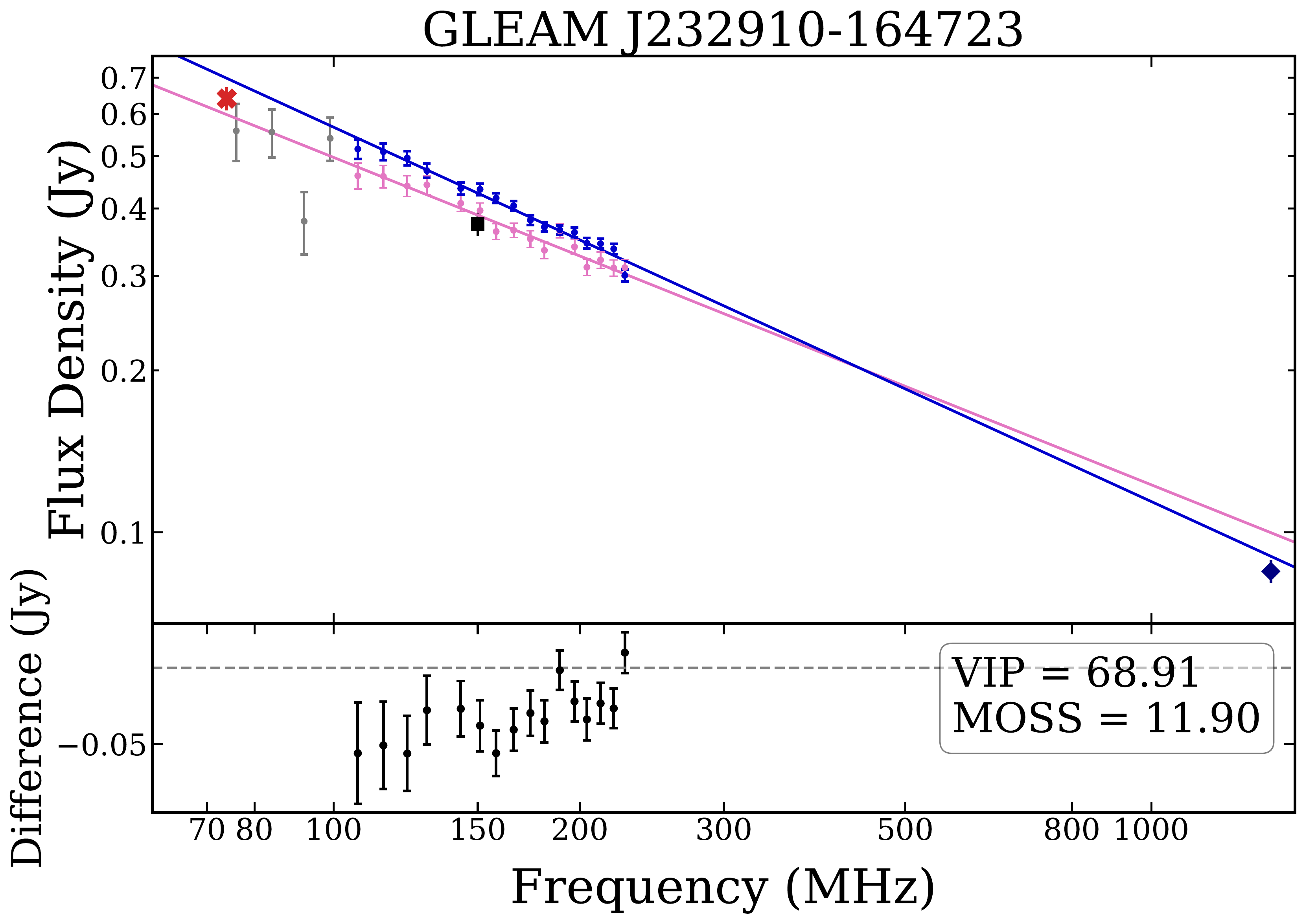} &
\includegraphics[scale=0.15]{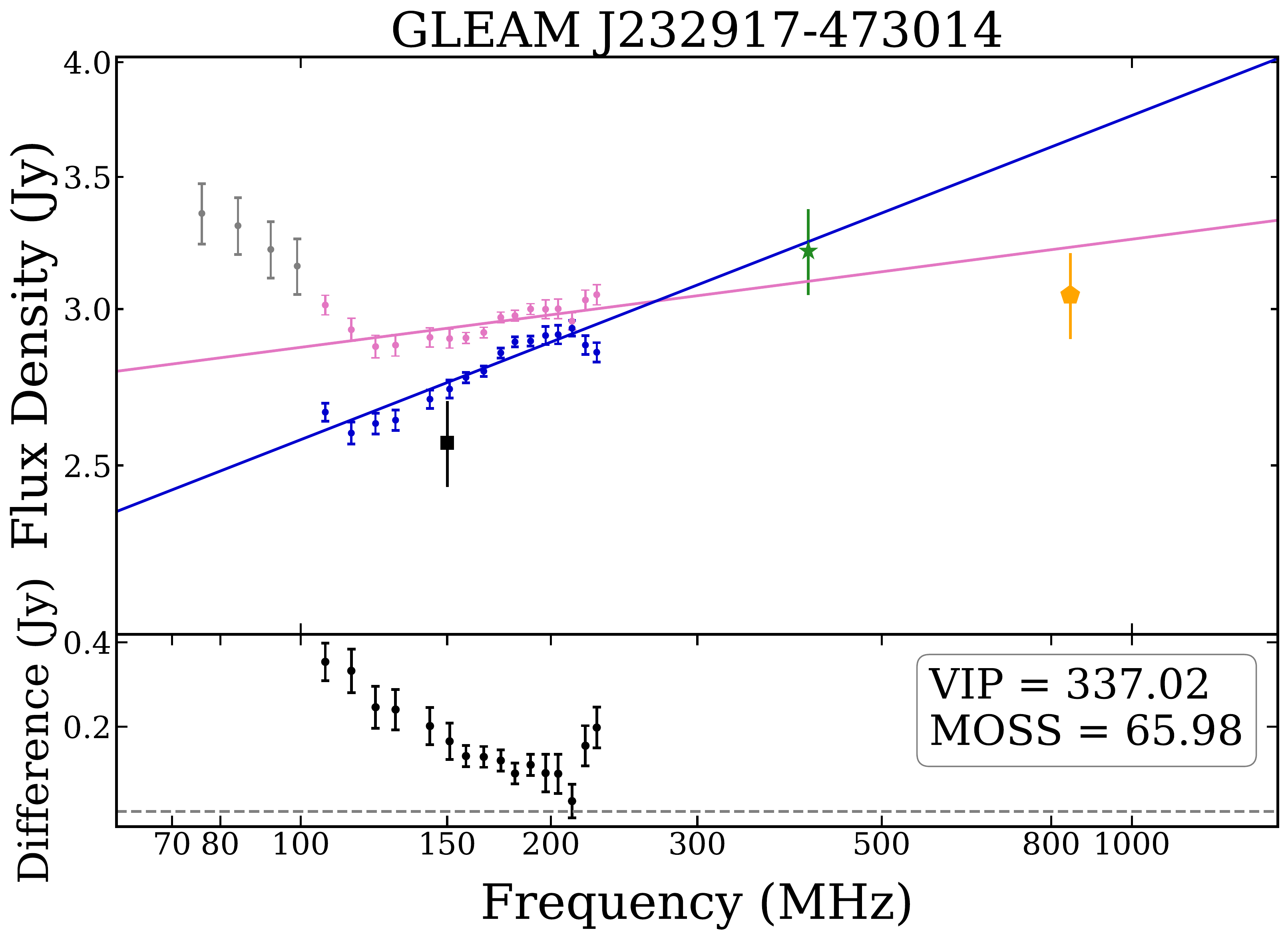} &
\includegraphics[scale=0.15]{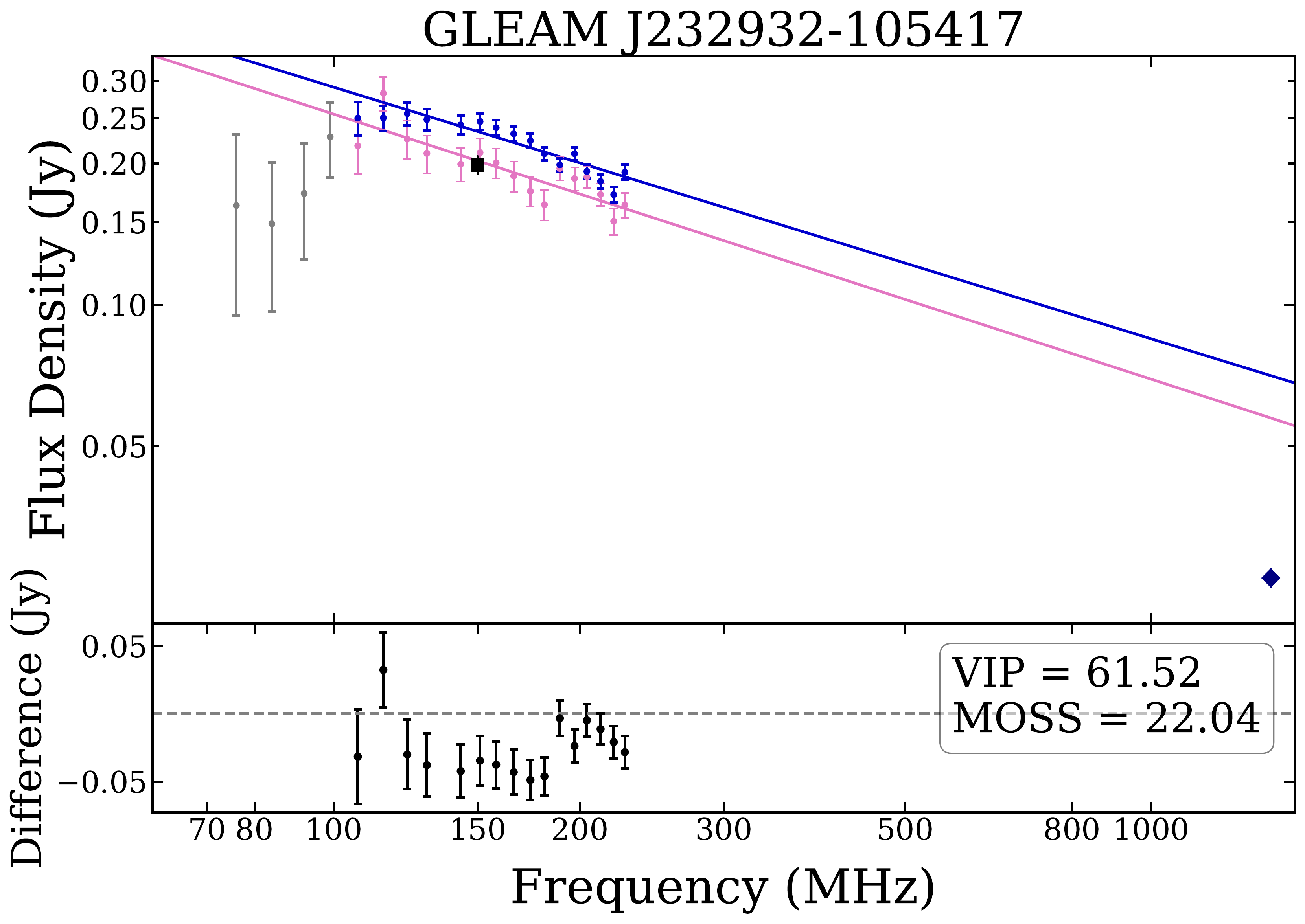} \\
\end{array}$
\caption{(continued) SEDs for all sources classified as variable according to the VIP. For each source the points represent the following data: GLEAM low frequency (72--100\,MHz) (grey circles), Year 1 (pink circles), Year 2 (blue circles), VLSSr (red cross), TGSS (black square), MRC (green star), SUMSS (yellow pentagon), and NVSS (navy diamond). The models for each year are determined by their classification; a source classified with a peak within the observed band was modelled by a quadratic according to Equation~\ref{eq:quadratic}, remaining sources were modelled by a power-law according to Equation~\ref{eq:plaw}.}
\label{app:fig:pg17}
\end{center}
\end{figure*}
\setcounter{figure}{0}
\begin{figure*}
\begin{center}$
\begin{array}{cccccc}
\includegraphics[scale=0.15]{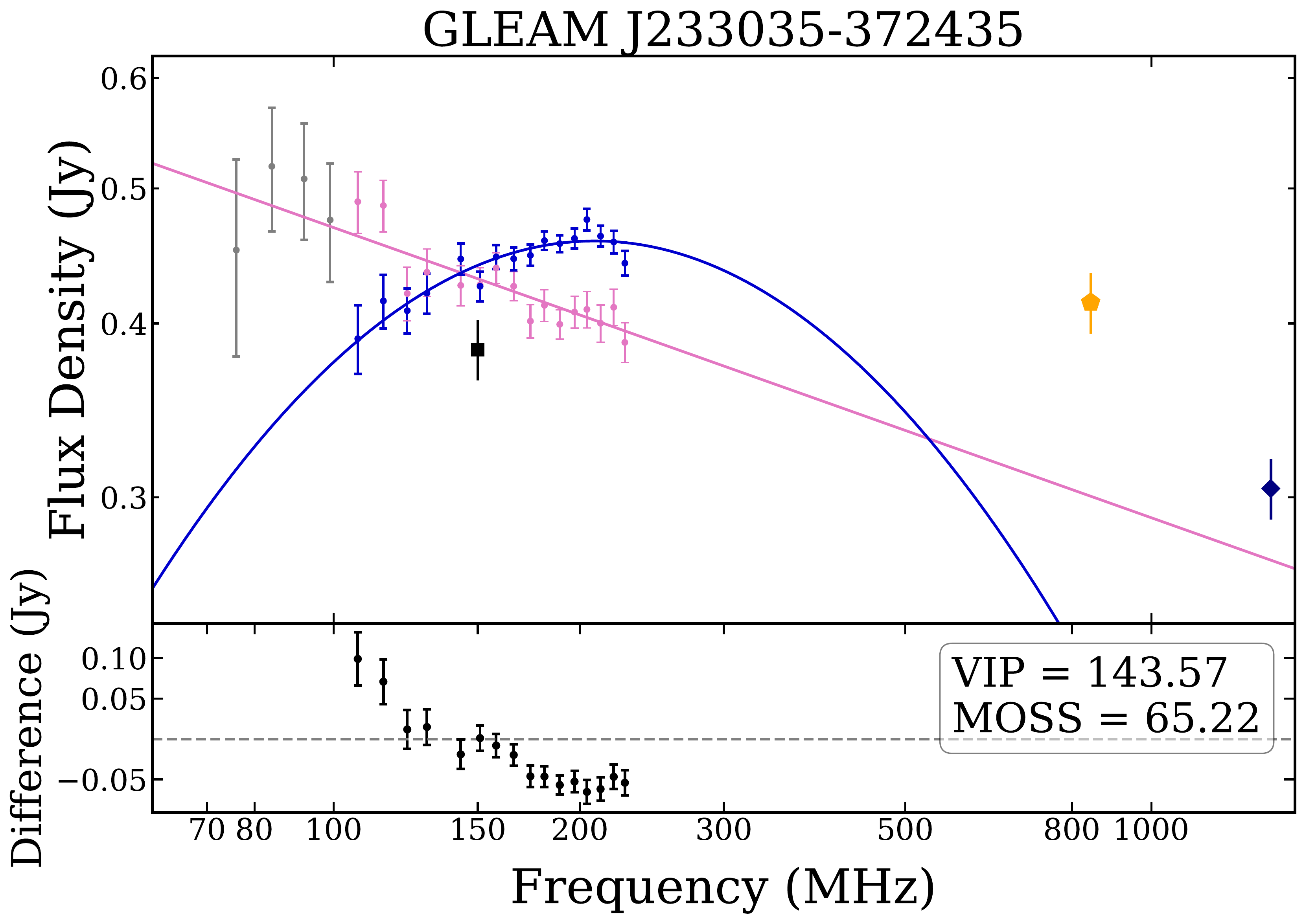} &
\includegraphics[scale=0.15]{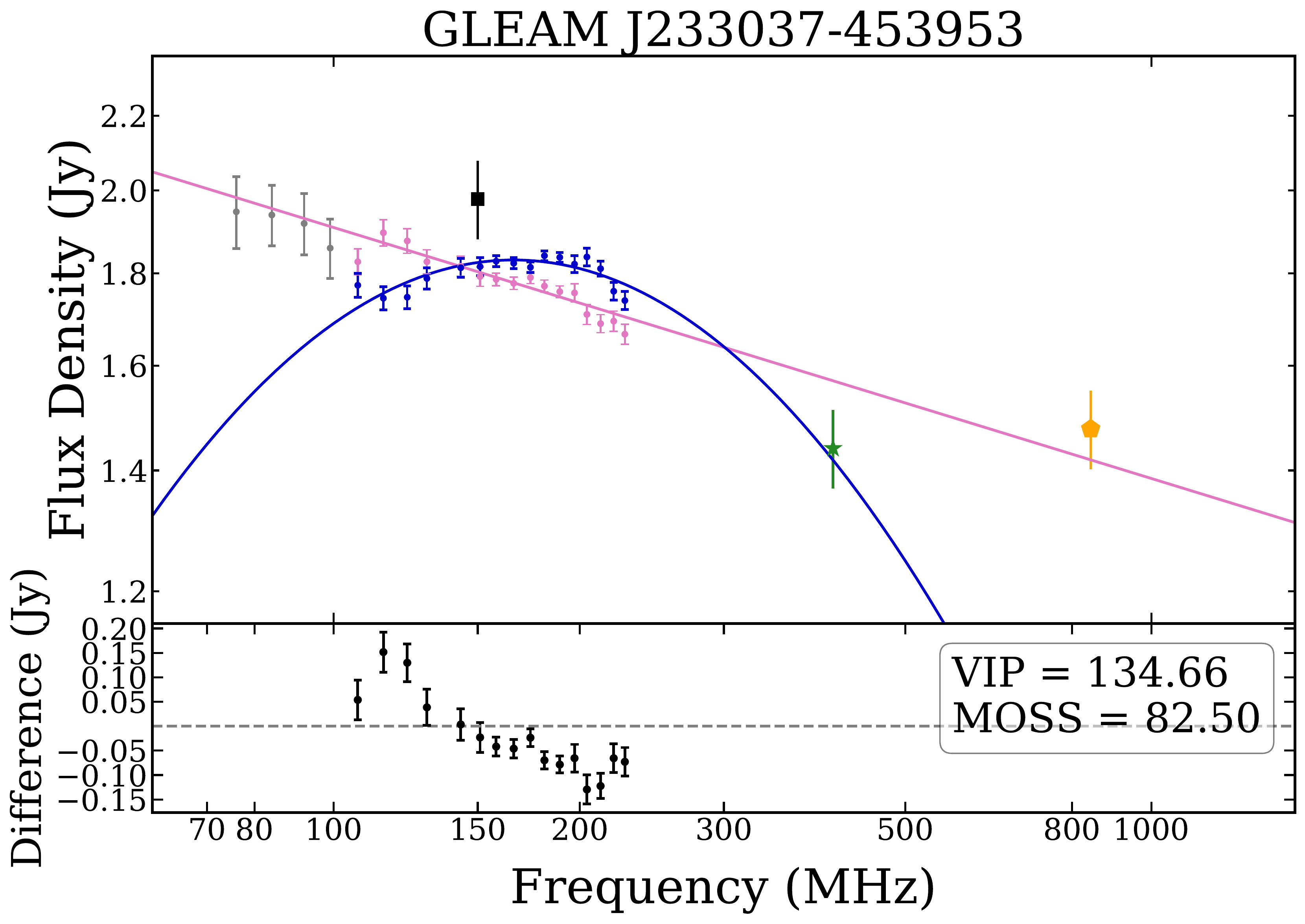} &
\includegraphics[scale=0.15]{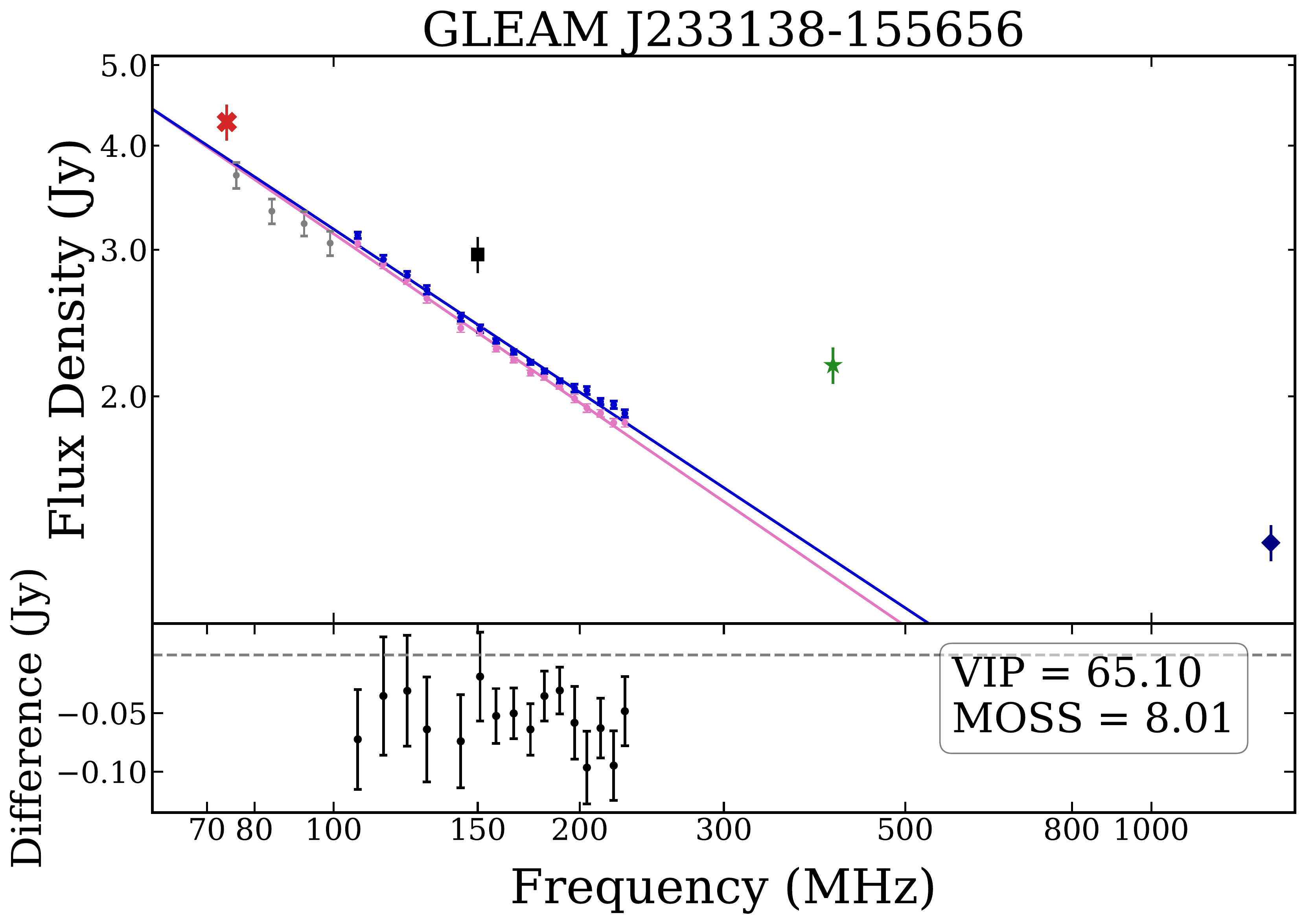} \\
\includegraphics[scale=0.15]{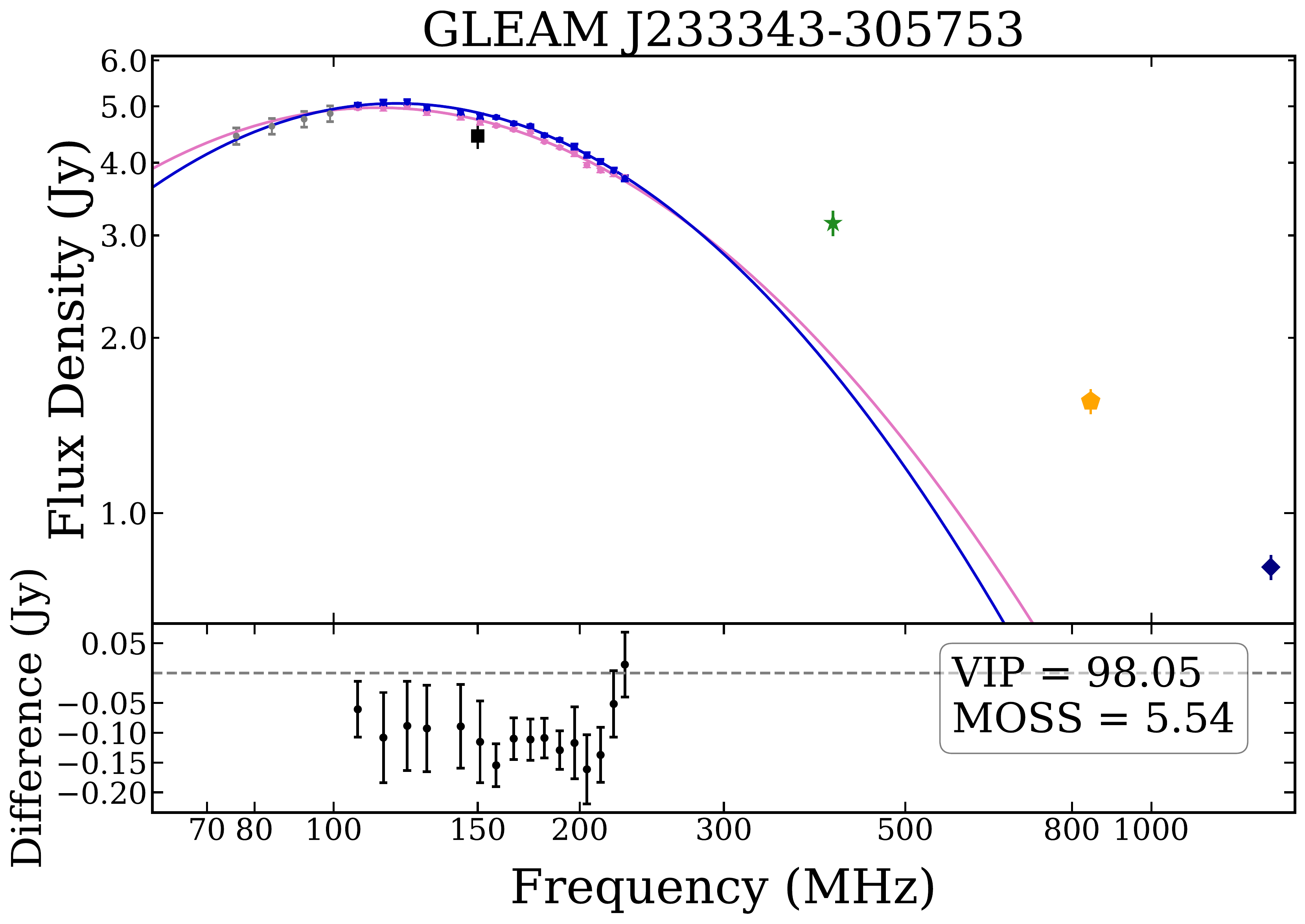} &
\includegraphics[scale=0.15]{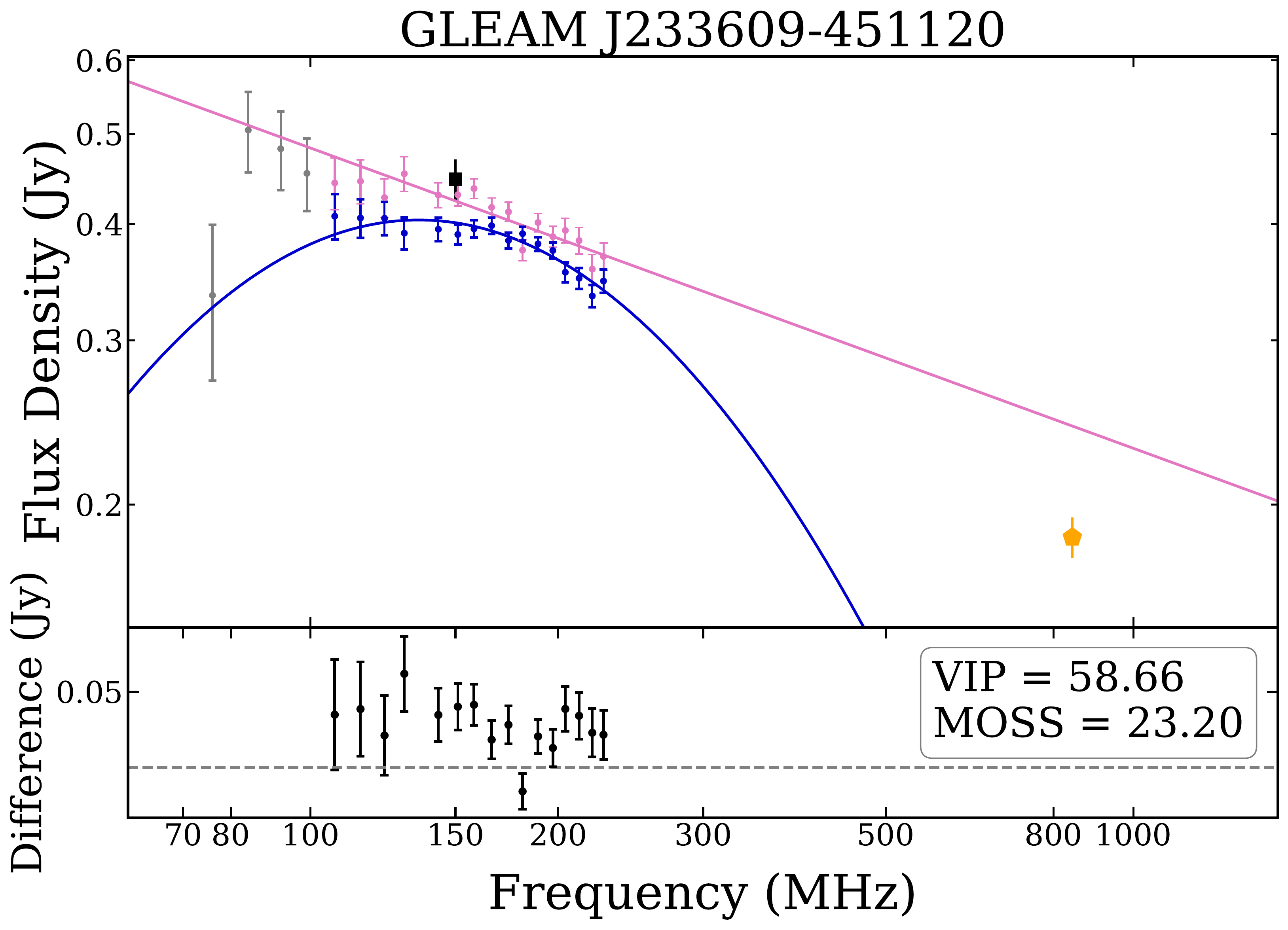} &
\includegraphics[scale=0.15]{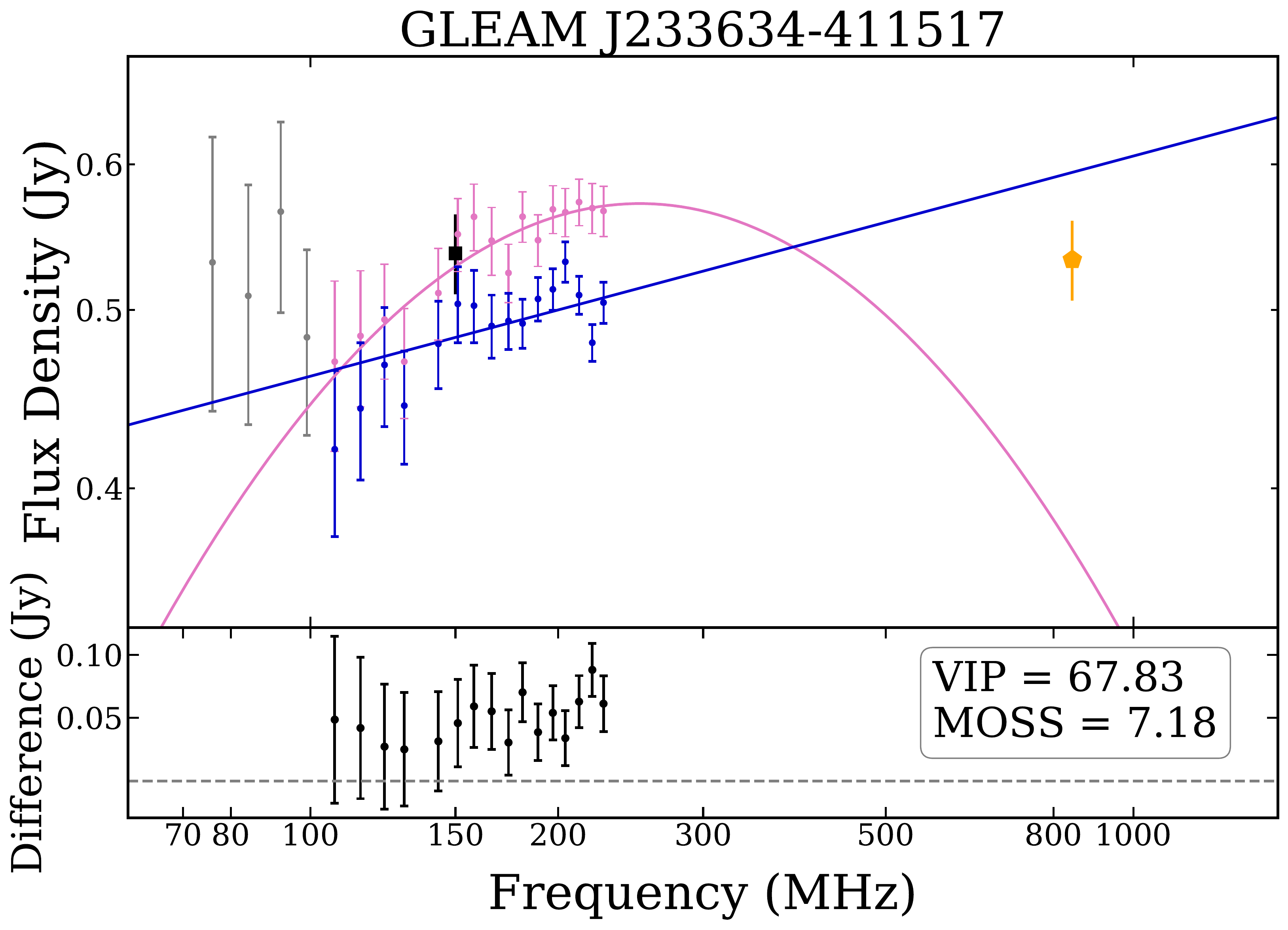} \\
\includegraphics[scale=0.15]{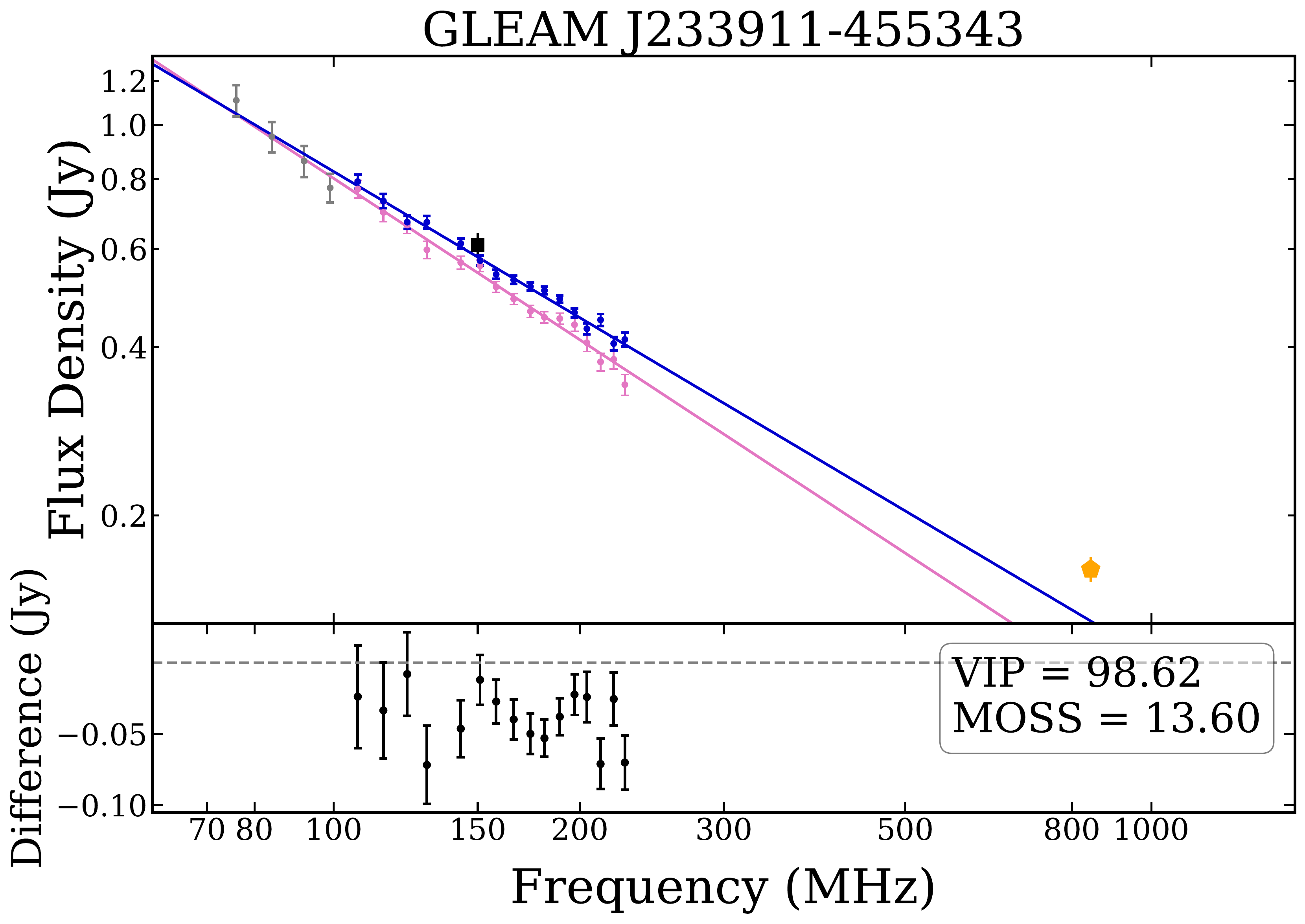} &
\includegraphics[scale=0.15]{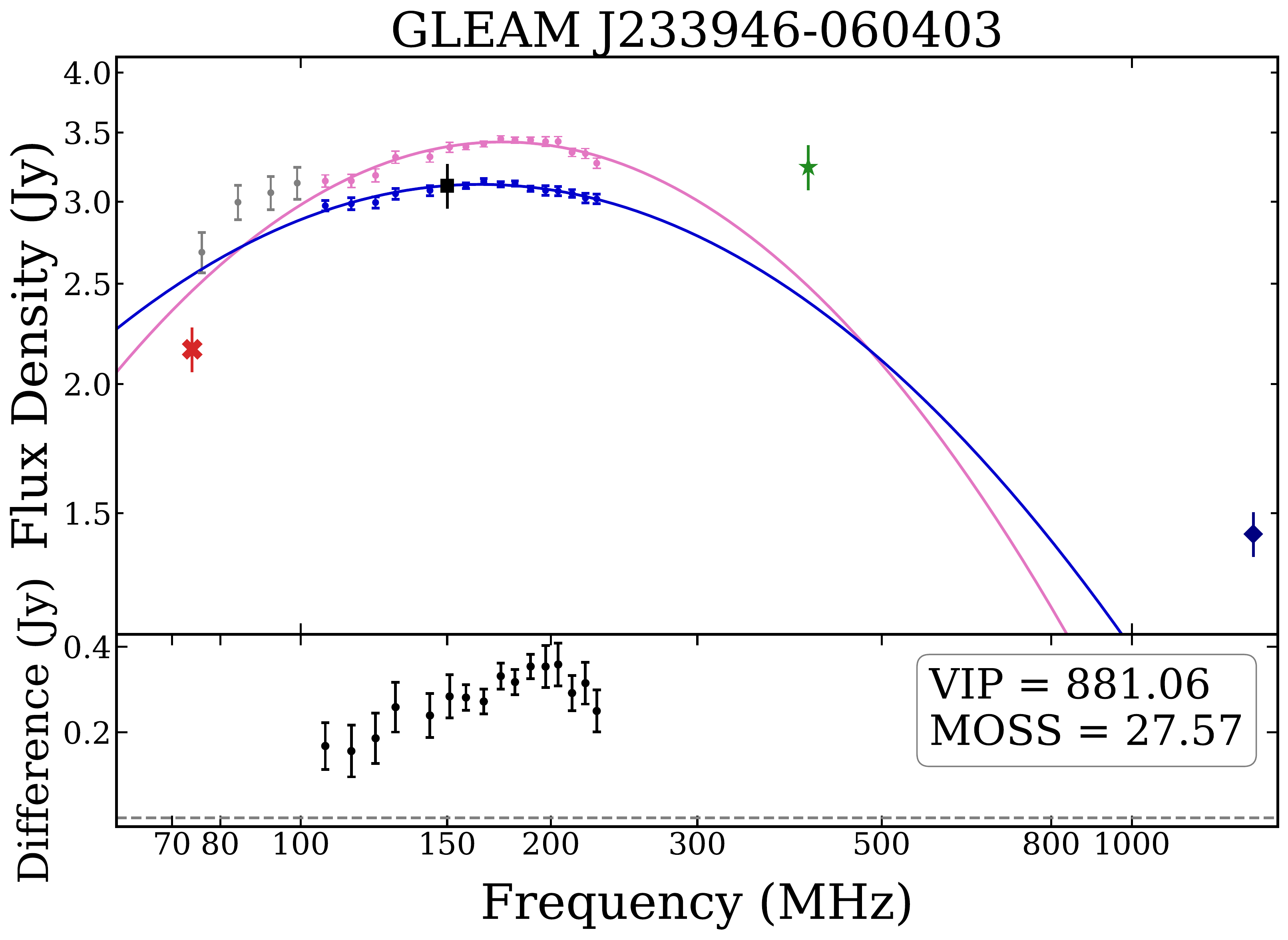} &
\includegraphics[scale=0.15]{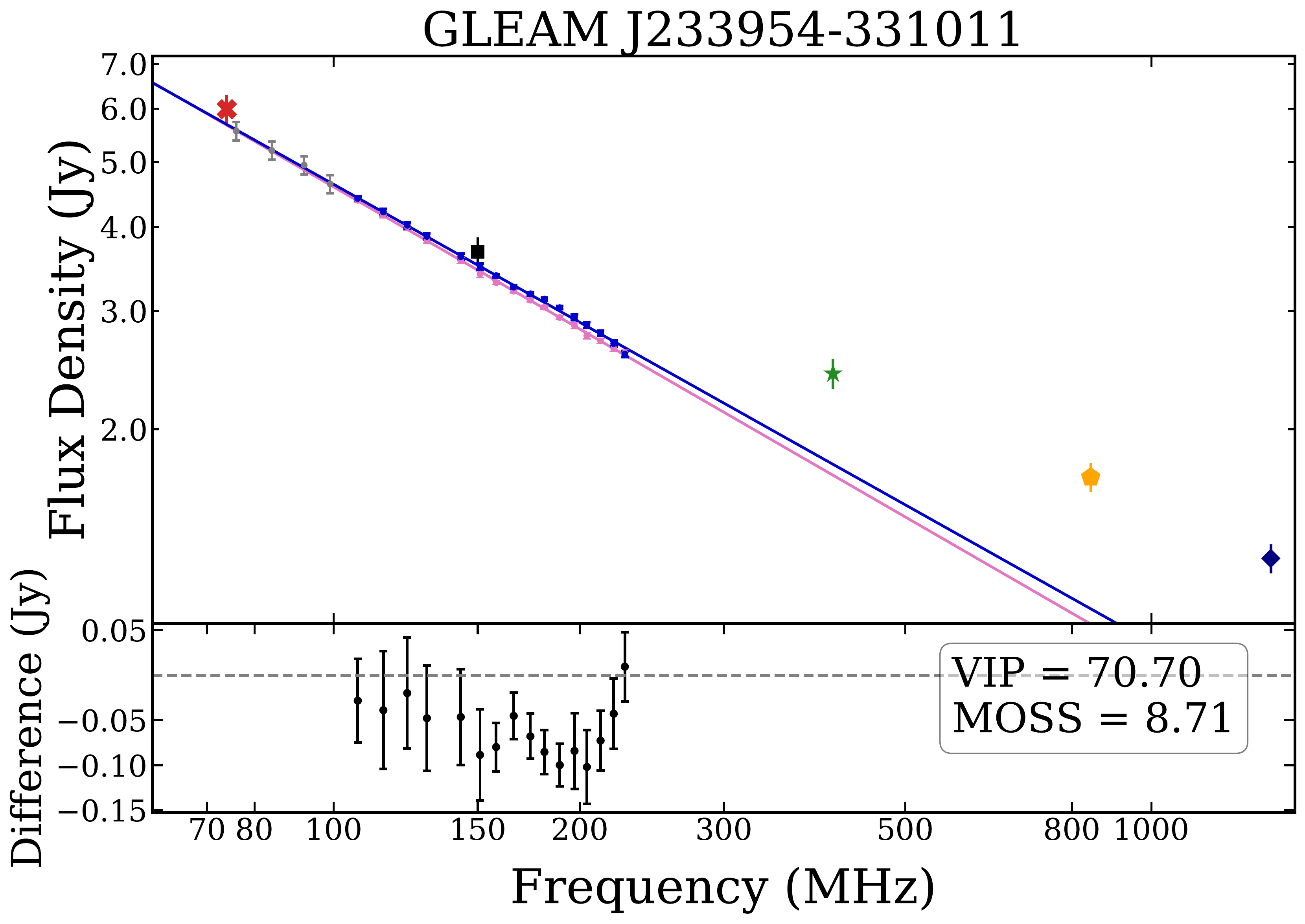} \\
\includegraphics[scale=0.15]{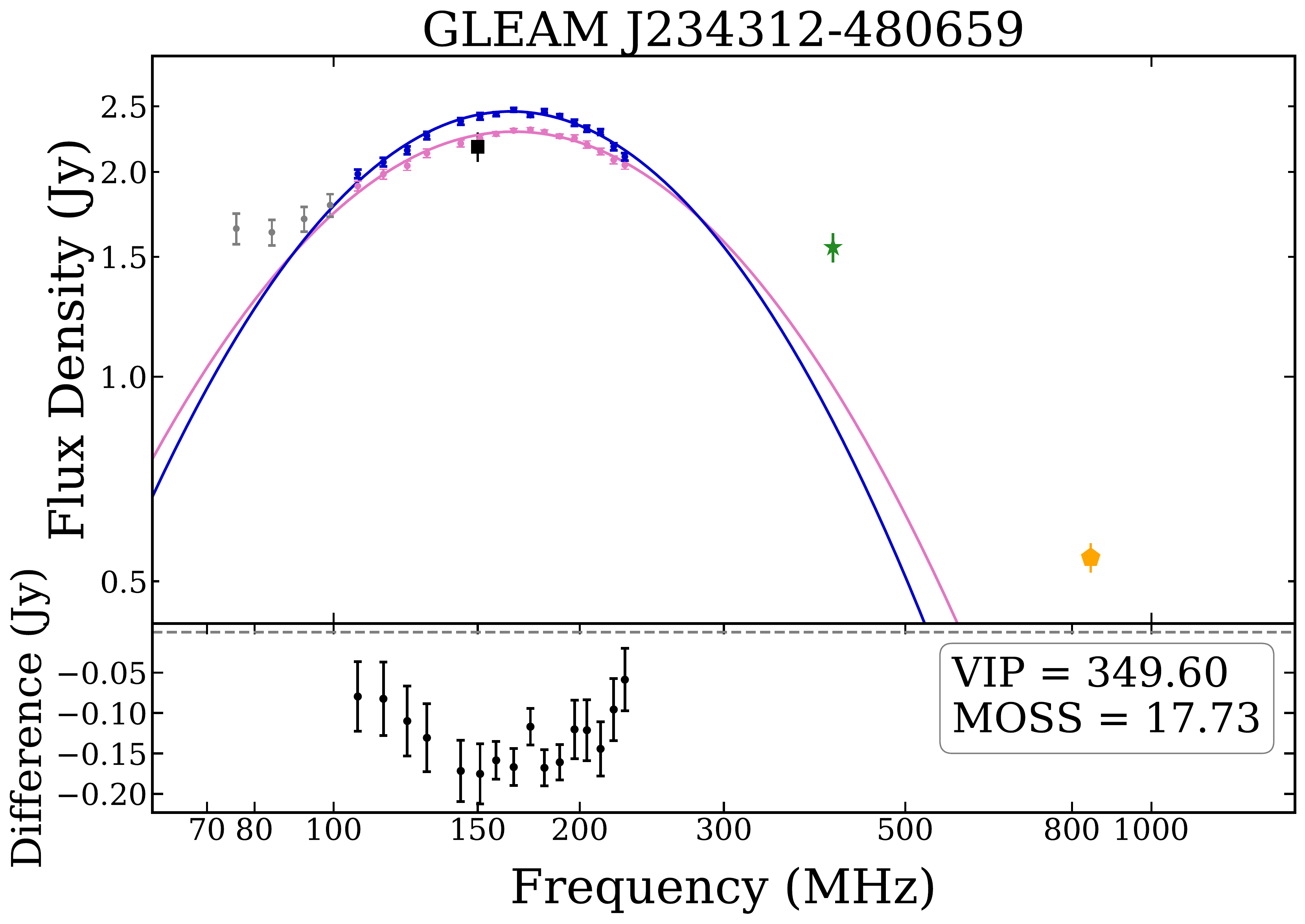} &
\includegraphics[scale=0.15]{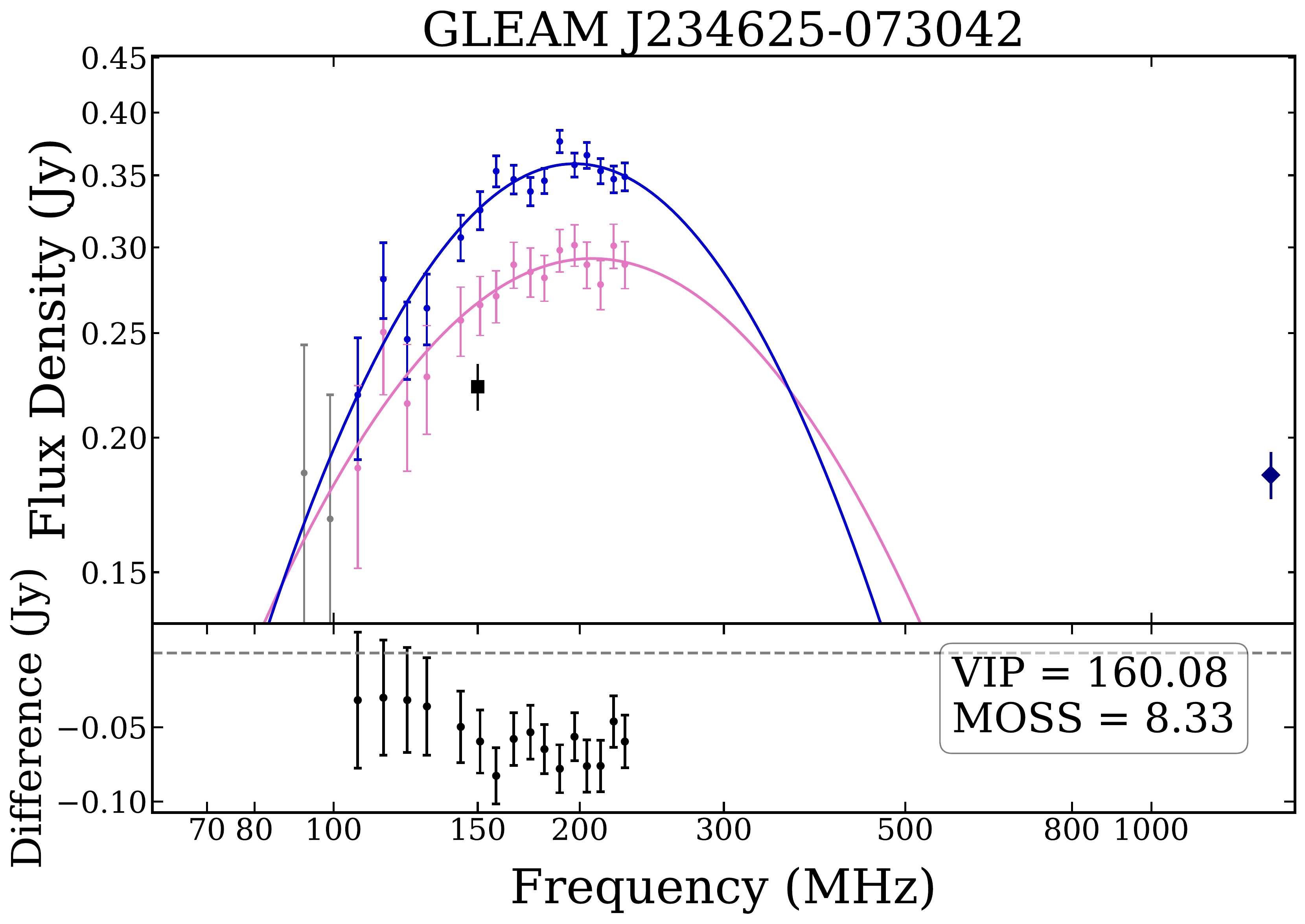} &
\includegraphics[scale=0.15]{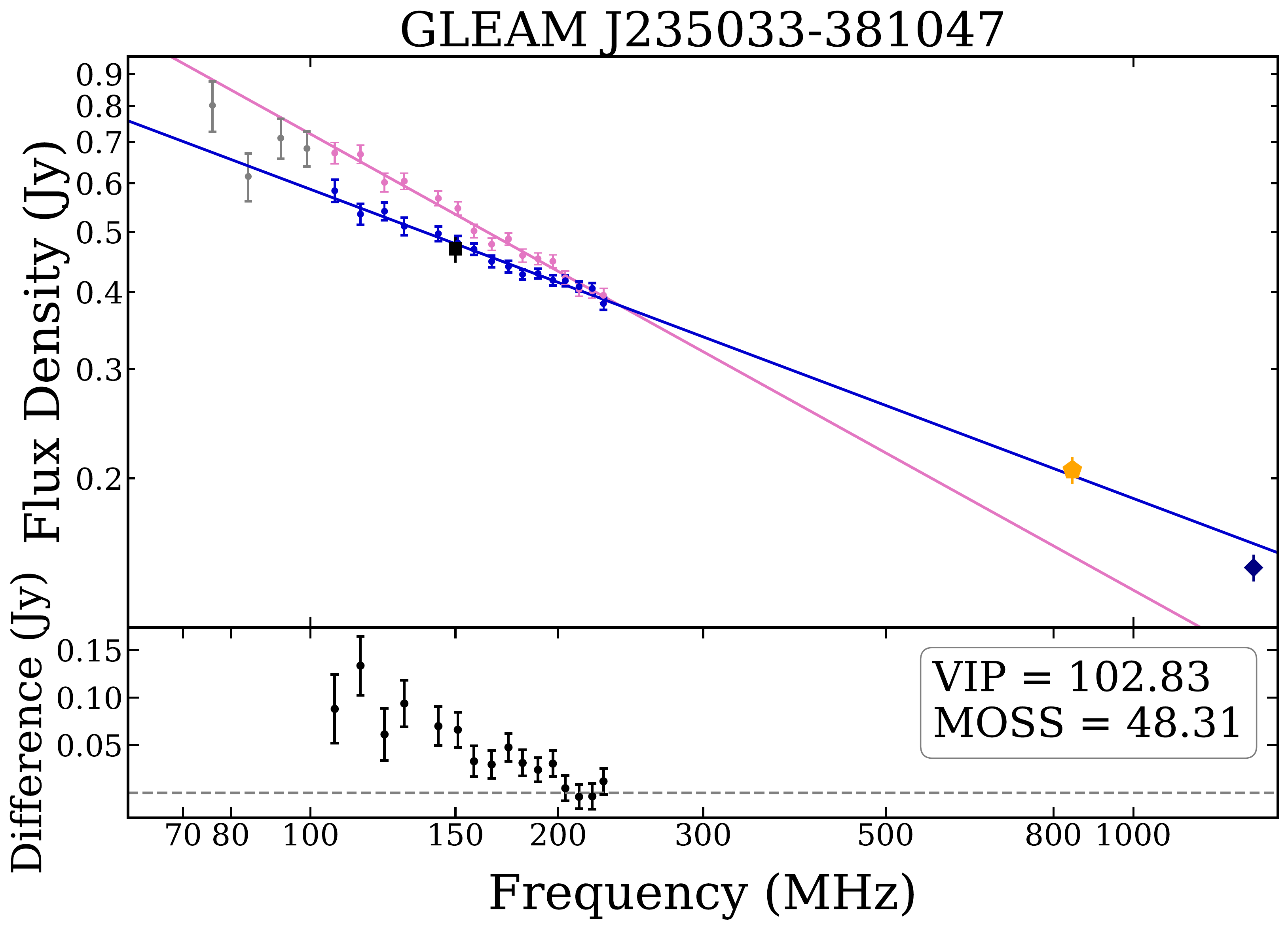} \\
\includegraphics[scale=0.15]{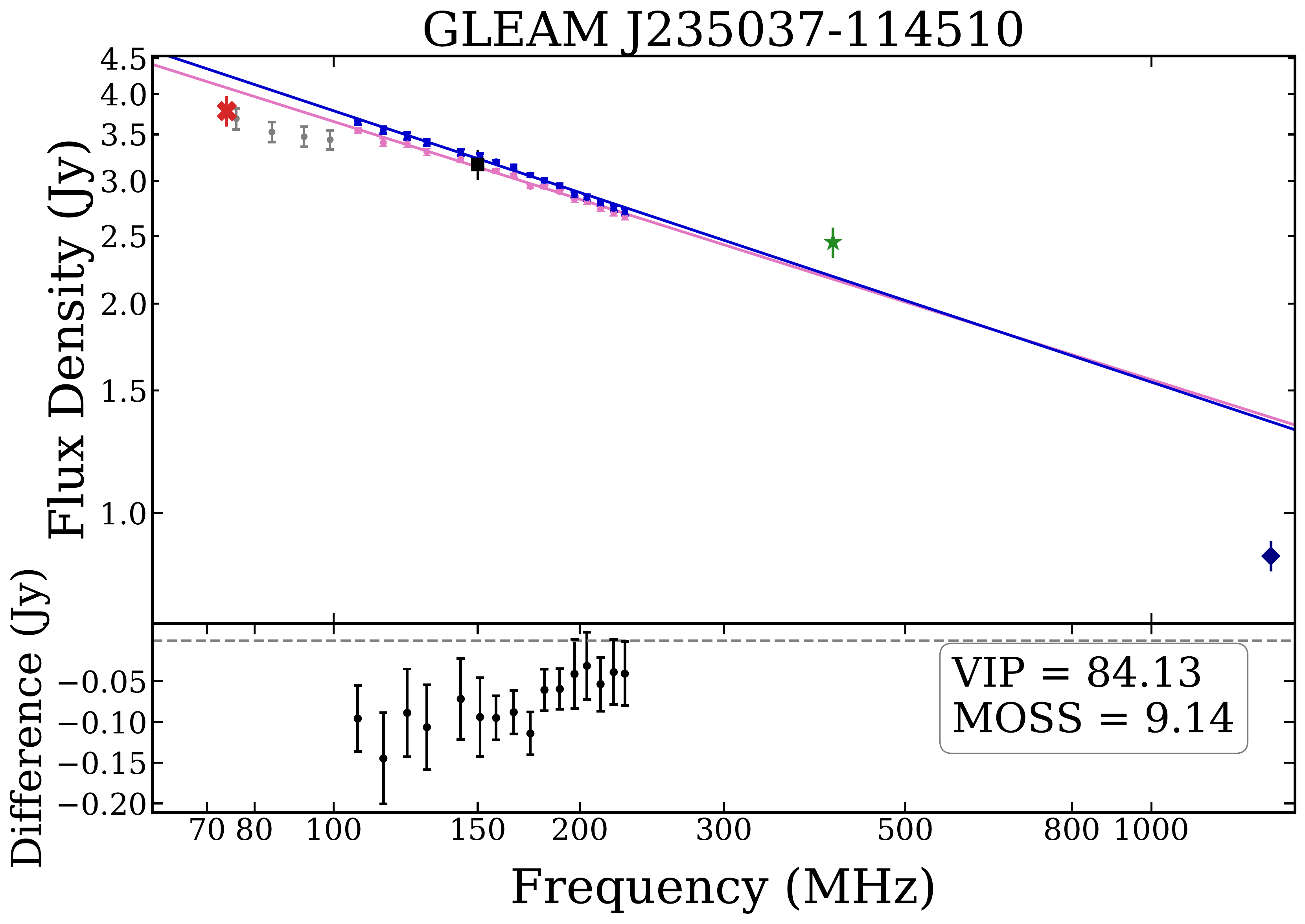} &
\includegraphics[scale=0.15]{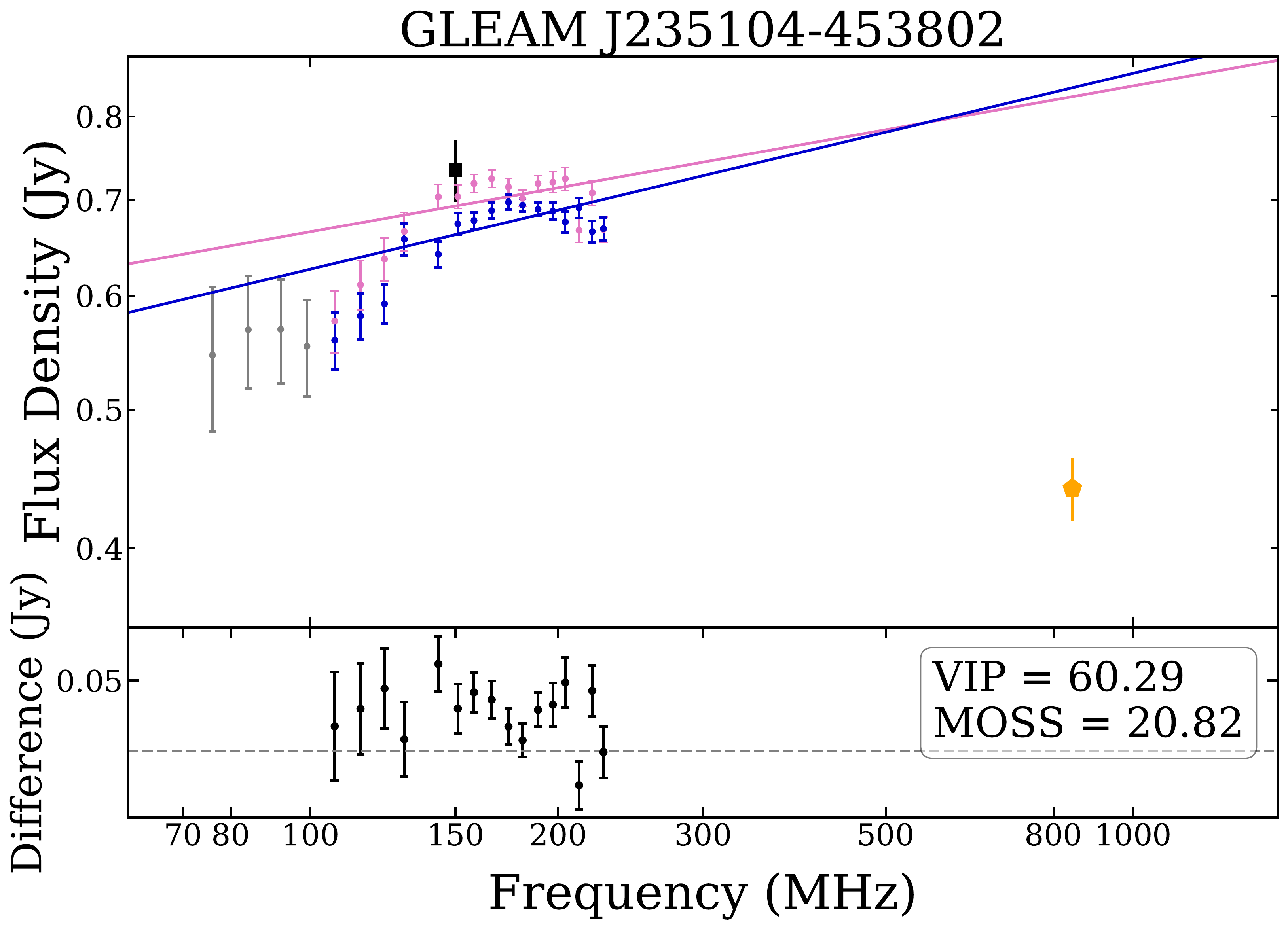} &
\includegraphics[scale=0.15]{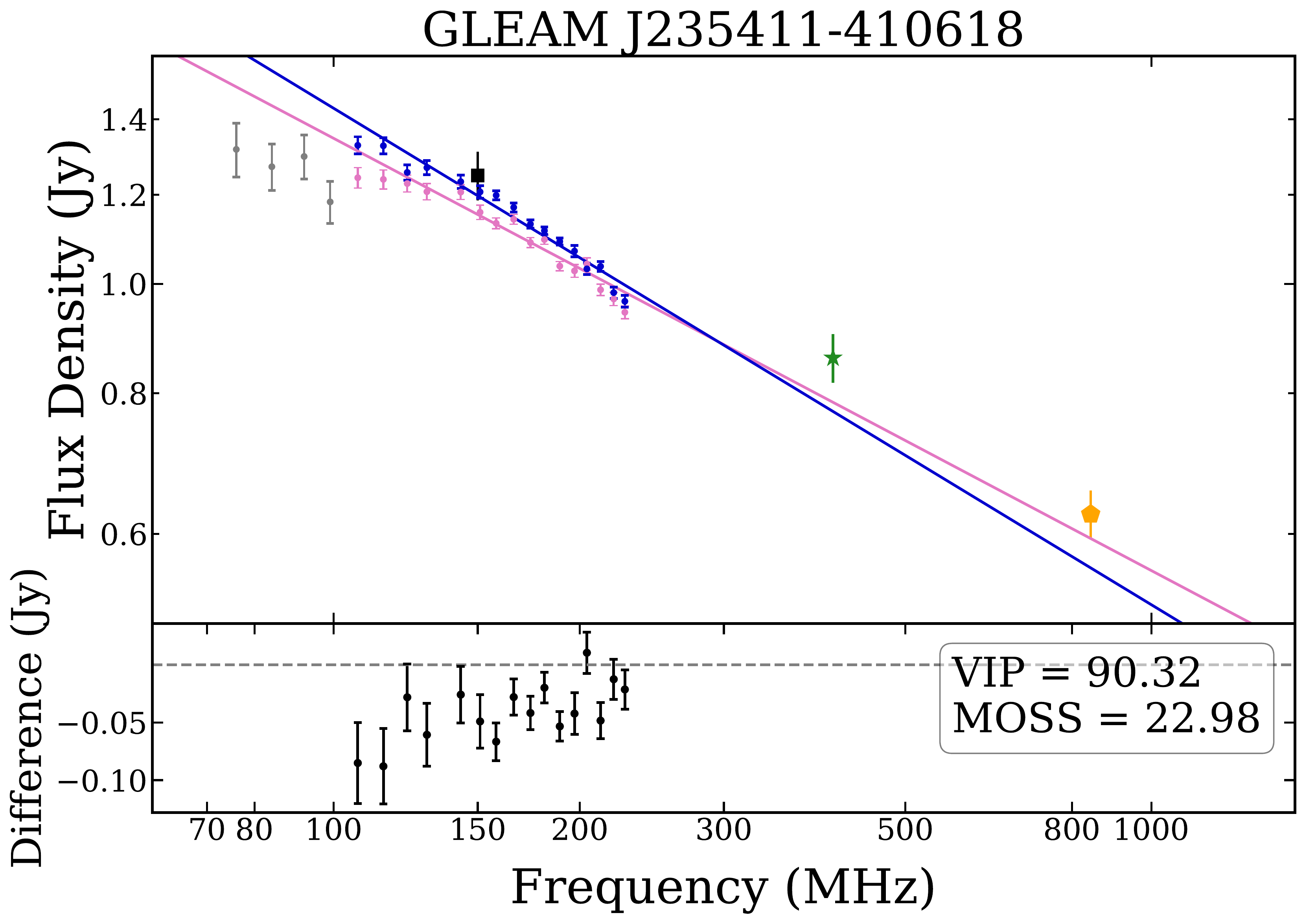} \\
\includegraphics[scale=0.15]{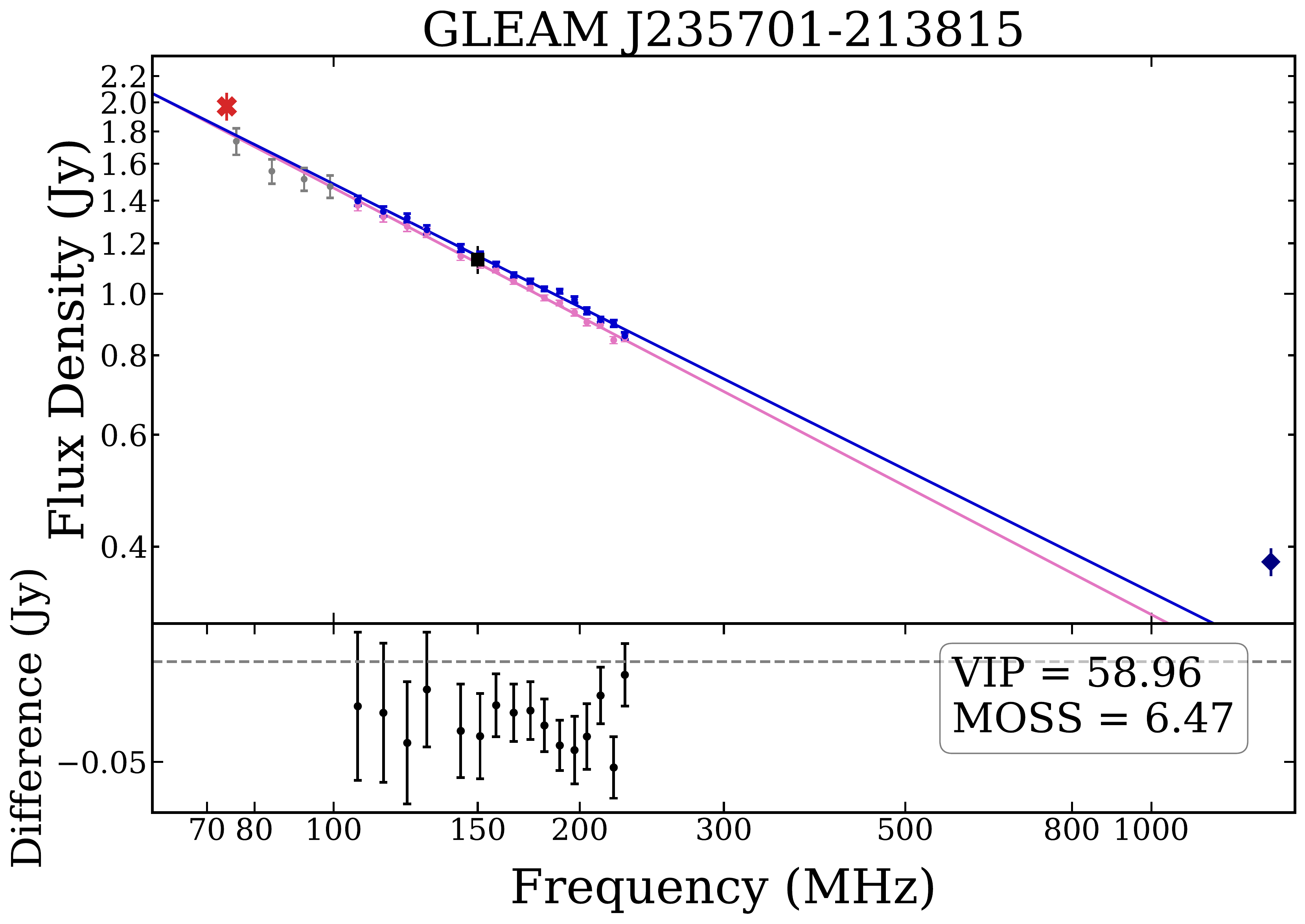} &
\includegraphics[scale=0.15]{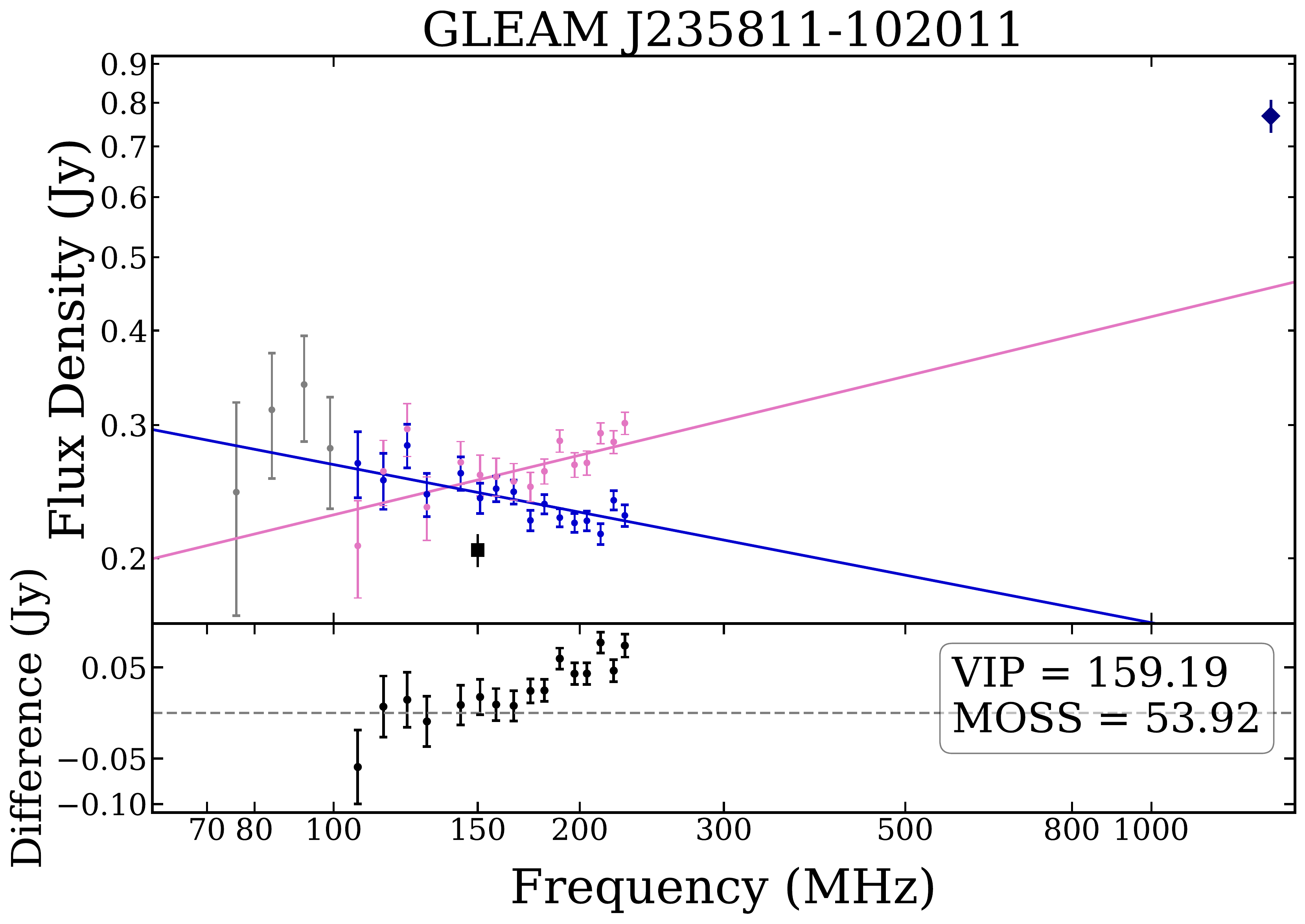} &
\end{array}$
\caption{(continued) SEDs for all sources classified as variable according to the VIP. For each source the points represent the following data: GLEAM low frequency (72--100\,MHz) (grey circles), Year 1 (pink circles), Year 2 (blue circles), VLSSr (red cross), TGSS (black square), MRC (green star), SUMSS (yellow pentagon), and NVSS (navy diamond). The models for each year are determined by their classification; a source classified with a peak within the observed band was modelled by a quadratic according to Equation~\ref{eq:quadratic}, remaining sources were modelled by a power-law according to Equation~\ref{eq:plaw}.}
\label{app:fig:pg18}
\end{center}
\end{figure*}

\bsp	
\label{lastpage}
\end{document}